\documentclass[pdf,12pt]{cernyrep}
\pdfoutput=1

\usepackage{subfigure}
\usepackage{multirow}
\usepackage{amsmath}
\usepackage{amssymb}
\usepackage[english]{babel}
\usepackage{axodraw}
\usepackage{epsfig}
\usepackage{booktabs}
\usepackage{cite}
\usepackage{graphicx}
\usepackage{color}
\usepackage{slashed}
\usepackage{titling}
\usepackage{verbatim}
\usepackage{amsfonts}
\usepackage{mathrsfs}
\usepackage{epstopdf}

\addto\captionsenglish{}

\makeatletter
\def\myaddcontentsline@toc#1#2#3{%
   \addtocontents{#1}{\protect\thispagestyle{empty}}%
   \addtocontents{#1}{\protect \vspace{-.5cm}}
   \addtocontents{#1}{\protect\contentsline{#2}{#3}{}}
   \addtocontents{#1}{\protect \vspace{-.3cm}}}
\def\myaddcontentsline#1#2#3{%
  \expandafter\@ifundefined{addcontentsline@#1}%
  {\addtocontents{#1}{\protect \vspace{-.5cm}}
  \addtocontents{#1}{\protect\contentsline{#2}{#3}{}}
  \addtocontents{#1}{\protect \vspace{-.3cm}}}
  {\csname addcontentsline@#1\endcsname{#1}{#2}{#3}}}
\makeatother

\makeatletter
\def\mysuperaddcontentsline@toc#1#2#3{%
   \addtocontents{#1}{\protect\thispagestyle{empty}}%
   \addtocontents{#1}{\protect \vspace{.1cm}}
   \addtocontents{#1}{\protect\contentsline{#2}{#3}{\thepage}}
   \addtocontents{#1}{\protect \vspace{-.1cm}}}
\def\mysuperaddcontentsline#1#2#3{%
  \expandafter\@ifundefined{addcontentsline@#1}%
  {\addtocontents{#1}{\protect \vspace{.1cm}}
  \addtocontents{#1}{\protect\contentsline{#2}{#3}{\thepage}}
  \addtocontents{#1}{\protect \vspace{-.1cm}}}
  {\csname addcontentsline@#1\endcsname{#1}{#2}{#3}}}
\makeatother

\newcommand{\superpart}[1]{\clearpage \thispagestyle{empty}  \vphantom{blabla}\vspace{5cm} \centerline{\huge #1} 
\mysuperaddcontentsline{toc}{part}{\protect\numberline{} \hspace{-2cm}#1}%
\clearpage}
   
\newcommand{\AddToContent}[1]{\myaddcontentsline{toc}{chapter}{\protect  \textit{\hspace{1.5cm} \small#1}}}

\renewcommand{\thefigure}{\arabic{figure}}
\renewcommand{\thetable}{\arabic{table}}
\renewcommand{\theequation}{\arabic{equation}}

\def\etmiss{E_T^{miss}}
\def\lsp{{\tilde\chi_1^0}}
\def\dofig#1#2{\includegraphics[width=#1]{#2}}
\def\dofigs#1#2#3{\includegraphics[width=#1]{#2}\includegraphics[width=#1]{#3}}

\catcode`\@=11
\font\manfnt=manfnt
\def\Watchout{\@ifnextchar [{\W@tchout}{\W@tchout[1]}}
\def\W@tchout[#1]{{\manfnt\@tempcnta#1\relax%
  \@whilenum\@tempcnta>\z@\do{%
    \char"7F\hskip 0.3em\advance\@tempcnta\m@ne}}}
\let\foo\W@tchout
\def\dubious{\@ifnextchar[{\@dubious}{\@dubious[1]}}

\def\@dubious[#1]{%
  \setbox\@tempboxa\hbox{\@W@tchout#1}
  \@tempdima\wd\@tempboxa
  \list{}{\leftmargin\@tempdima}\item[\hbox to 0pt{\hss\@W@tchout#1}]}
\def\@W@tchout#1{\W@tchout[#1]}
\catcode`\@=12

\def\describeRec{Provide a clear, explicit description of the analysis in publications. 
In particular, the most crucial information such as basic object definitions 
and event selection should be clearly displayed in the publications, preferably 
in tabular form, and kinematic variables utilised should be unambiguously defined. 
Further information necessary to reproduce the analysis should be provided 
on a suitable common platform.}
\def\databaseRec{Provide a common analysis database where all the experimental
  results are stored together with all necessary information about the
  analyses, including well-encapsulated functions, such as multivariate
  analysis (MVA)  functions if they are needed.}
\def\analysisNumbersRec{Provide all crucial numbers regarding the results of the analysis,
     preferably in tabulated form in the publication itself.  Further relevant information, like fit 
     functions or distributions, should be provided as auxiliary material.}
\def\multiBinRec{For multi-bin results, provide an ensemble of sets of the numbers $B$,
  $\delta B$, $\cal{L}$, $\delta\cal{L}$, $Q$, $k$, etc in the auxiliary
  information.
These would be 
created by sampling from the various experiment-specific systematic effects,
such as the jet energy scale, jet energy resolution, etc. 
Systematic uncertainties external to the experiment, such as PDF uncertainties, 
need not be included because they induce correlations across measurements.} 
\def\efficiencyRec{
Provide histograms or functional forms of efficiency maps wherever possible in the auxiliary information, along with precise definitions of the efficiencies, and preferably provide them in 
standard electronic forms that can easily be interfaced with simulation or analysis software.}
\def\simulatorRec{Provide and maintain a public simulator developed by the
  collaboration, or provide official support of an existing one. The public
  simulator would provide the mapping from the pre-detector data to the
  post-reconstruction data.}
\def\likelihoodRec{
Provide a mathematical description of the \underline{final} likelihood
  function in which experimental data and parameters are clearly distinguished, 
  either in the publication or the auxiliary information.}
\def\roostatsRec{Additionally provide an digitized implementation of the
likelihood that is consistent with the mathematical description.}
\def\interpretRec{In the interpretation of experimental results,  
preferably provide the final likelihood function (following 3b or 3c), 
or provide a grid of confidence levels over the parameter space.  
The expected constraints should be given in addition to the observed ones, and 
whatever sensitivity measure is applied must be be precisely defined. 
Modeling of the acceptance needs to be precisely described.}
\def\higgsRec{For Higgs searches, provide all relevant information on a channel-by-channel basis.}
\def\designRec{When relevant, 
design analyses and signal regions that are based on  disjoint sets of events.}

\renewcommand{\eg}[0]{\textit{e.g.}}
\renewcommand{\ie}[0]{\textit{i.e.}}

\newcommand{\eq}[1]{Eq.~(\ref{#1})}
\newcommand{\eqs}[3]{Eqs~(\ref{#1}), (\ref{#2}), and (\ref{#3})}
\newcommand{\eqsfromto}[2]{Eq.s~(\ref{#1}) to (\ref{#2})}

\newcommand{\leer}[1]{}

\newcommand{\URxUBL}[0]{\ensuremath{U(1)_{R} \times U(1)_{B-L}}}
\newcommand{\UYxUBL}[0]{\ensuremath{U(1)_{Y} \times U(1)_{B-L}}}

\newcommand{\Da}[0]{\Delta m^2_{\textsc{A}}}
\newcommand{\seesaw}[0]{see-saw}
\newcommand{\Seesaw}[0]{See-saw}
\newcommand{\PDGIX}[0]{PDG.IX}
\newcommand{\RE}[0]{\text{Re}}
\newcommand{\IM}[0]{\text{Im}}

\newcommand{\numentry}[2]{
\begin{minipage}[t]{1.2cm}\flushright\texttt{#1}\end{minipage}
\hspace{2mm}:
\begin{minipage}[t]{11cm}\noindent
#2\end{minipage}\\[1mm]
}

\renewcommand{\o}[1]{\overline{#1}}
\newcommand{\grandO}[1]{O\mathopen{}\left(#1\right)}
\newcommand{\mt}[1]{\ensuremath{m_{t'}}}
\newcommand{\mb}[1]{\ensuremath{m_{b'}}}
\newcommand{\me}[1]{\ensuremath{m_{\tau'}}}
\newcommand{\mv}[1]{\ensuremath{m_{\nu'}}}
\newcommand{\thetam}[1]{\ensuremath{\theta_{#1}}}
\newcommand{\FP}{\ensuremath{\lambda_{\mathrm{FP}}}}
\newcommand{\mgut}{\ensuremath{M_{\gut}}}
\newcommand{\gut}{{\textsc{gut}}}

\begin{document}

\setcounter{tocdepth}{0}
\thispagestyle{empty}

\vspace{1cm}

\begin{center}
{\Large {\bf LES HOUCHES 2011: PHYSICS AT TEV COLLIDERS \\[4mm]}}
{\Large {\bf NEW PHYSICS WORKING GROUP REPORT}}
\end{center}

\vspace{0.1cm}
\begin{center}
\textbf{G.~Brooijmans}$^{1}$, 
\textbf{B.~Gripaios}$^{2}$, %
\textbf{F.~Moortgat}$^{3}$, %
\textbf{J.~Santiago}$^{4}$  %
\textbf{and}
\textbf{P.~Skands}$^{5}$ %
\textbf{(convenors)}\\
D.~Albornoz~V\'asquez$^6$,  
B.C.~Allanach$^{7}$,
A.~Alloul$^8$,
A.~Arbey$^{5,9,10}$,
A.~Azatov$^{11}$
% ...
\end{center}

 \vspace{1cm}
\begin{center}
{\large {\bf Abstract}}\\[.2cm]
\end{center}
We present the activities of the 
``New Physics'' working group for the ``Physics at TeV Colliders"
workshop (Les Houches, France, 30 May--17 June, 2011).  Our 
report includes new agreements on formats for interfaces between
computational tools, new tool developments,
important signatures for searches at the LHC,
recommendations for presentation of LHC search results, 
as well as additional phenomenological studies.
\vspace{1cm}
\begin{center}
{\bf Acknowledgements}\\[.2cm]
\end{center}
We would like to heartily thank the funding bodies, the organisers 
(G.~B\'elanger, F.~Boudjema, S.~Gascon, C.~Grojean, J.P.~Guillet, S.~Kraml, G.~Moreau, E.~Pilon, G.~Salam, P.~Slavich and D.~Zerwas), the staff and the other
participants of the Les Houches workshop for providing a stimulating and
lively environment in which to work.

\newpage

\vspace{1cm}

\thispagestyle{empty}
\setcounter{page}{2}

\begin{center}
{\footnotesize
$^1$ Physics Department, Columbia University, New York, NY 10027, USA\\ 
$^2$ Cavendish Laboratory, University of Cambridge, Cambridge CB3 0HE, UK \\
$^3$ Institute for Particle Physics, ETH Z\"urich, Z\"urich, Switzerland \\
$^4$ CAFPE and Dpto. de F\'{\i}sica Te\'orica y del Cosmos,
Universidad de Granada, E-18071, Granada, Spain \\
$^5$ Physics Department, CERN, CH-1211 Geneva 23, Switzerland\\
$^6$~Institut d'Astrophysique de Paris, UMR 7095 CNRS, Universit\'e Pierre et
Marie Curie, Paris 75014, France\\
$^7$ DAMTP, CMS, University of Cambridge, Cambridge, CB3 0WA,
UK\\
$^8$ Institut Pluridisciplinaire Hubert Curien/D\'epartement 
    Recherches Subatomiques, Universit\'e de Strasbourg/CNRS-IN2P3, 
    F-67037 Strasbourg, France\\
$^9$ Universit\'e Claude Bernard Lyon 1, CNRS/IN2P3, Institut
de Physique Nucl\'eaire de Lyon, F-69622~Villeurbanne Cedex, France\\
$^{10}$ Centre de Recherche Astrophysique de Lyon, Observatoire de Lyon,
 Saint-Genis Laval Cedex, F-69561, France; CNRS, UMR 5574; Ecole Normale
 Sup\'erieure de Lyon, Lyon, France\\ 
$^{11}$ Dipartimento di Fisica, Universit\`a di Roma
 ``La Sapienza'', INFN Sezione, 00185 Roma, Italy\\
% ...
%
%
}

\end{center}

\vspace{3cm}

\newpage

\tableofcontents

\newpage


\noindent {\Large {\bf Introduction}}
\vspace{.5cm}

{\it G.~Brooijmans, B.~Gripaios, F.~Moortgat, J.~Santiago, P.~Skands}
\vspace{.5cm}

This document is the report of the New Physics
session of the 2011 Les Houches Workshop ``Physics at TeV Colliders''.  During the 
workshop, which brings together theorists and experimenters, a substantial number of 
ideas were discussed and for a number of these, in-depth studies were initiated.  This
report describes the results of those studies.

A first section presents progress specific to the software tools crucial in predicting 
new physics model signatures and interpreting experimental results.  The first contribution
builds on the ``Les Houches Accords'' tradition in proposing a ``Dark Matter Les Houches
Agreement,'' a first step in standardizing the interface between different dark matter 
codes.  Two other contributions propose further developments to the existing Flavor and
SUSY Les Houches Accords.  Additional contributions in the tools Section describe new 
additions to {\sc FeynRules} and {\sc LanHEP}, and the development of {\sc SuSpect3}.

While (unfortunately) no new physics has been discovered yet, results from the 2010 and 2011 
LHC runs have substantial implications for most models of new physics.  In the first three 
contributions in the ``Higgs Boson and Top Quark Physics'' Section, the implications of 
the LHC results on Higgs searches in a variety of scenarios, the CMSSM mass spectrum, and
dark matter are investigated.  The next two contributions study Higgs boson production 
mechanisms in association with top quarks in different models.  Two additional 
contributions use effective theory approaches to study LHC implications of the top quark charge 
asymmetry observed at the Tevatron and multi-top signatures of 
sgluon production.  A two lasts contributions to this Section 
study some cases of new particles leading to single top-like signatures, 
and show the importance of polarization measurements in establishing 
discovery.

In a third Section, scenarios with long-lived particles are
studied. These arise in a variety of new physics models and present
special challenges and opportunities for LHC phenomenology. In a first case, 
particles with a lifetime sufficiently long that decays happen outside detectors are 
considered, with an MSSM point with a long-lived chargino as benchmark
model.  A second
class of signatures studied involves particles that decay after
travelling up to a meter away from 
the interaction point, in the context of scenarios with either heavy
(and long-lived)
neutrinos, or in supersymmetric models in which the observed cosmic
dark matter abundance is ``frozen in'' by the late decay of long-lived
Next-to-Lightest Supersymmetric Particles (NLSPs) to dark matter LSPs.

The fourth Section considers models in which the symmetry that stabilizes
dark matter is $Z_N$, rather than the usual $Z_2$, leading to a
quite different phenomenology.
One important new feature is the presence of dark matter
semi-annihilations, in which a pair of 
dark matter particles interacts to produce a dark matter and a
Standard Model particle, and the impact of these on dark matter phenomenology, including the freeze-out process and
dark matter abundance, is studied.

An important issue with the experimental results involves how to present them, since the 
implications for models not considered as benchmarks by the experimental collaborations
are generally non-trivial (if not impossible) to derive.  The fifth Section of this report
contains a first set of ``recommendations for the presentation of LHC results'' which 
emerged from extensive discussions between theorists and experimenters
at the workshop and elsewhere. The recommendations are split into
those that are considered to be essential and those that are
desiderata and it is hoped that they will stimulate
further discussion, refinement, and implementation across the community. A second contribution in this Section
presents HEPMDB, a storage environment for HEP models which can accommodate, via web 
interface to an HPC cluster, the validation of models, evaluation of LHC predictions 
and an event generation-simulation chain.

The meeting in Les Houches fostered a large number of discussions between theorists and
experimenters, but, as mentioned above, in-depth studies could only be completed for a
number of the generated ideas on the required timescale. Moreover,
many of the discussions begun at the workshop continue to evolve in `real time' as the LHC data
stream in, making it difficult to document them in these proceedings in
a way in which they may remain useful to the community in the longer
term. Still, we expect that many future results will 
benefit from the discussions held at the workshop.

\addtocontents{toc}{\protect\contentsline{part}{\protect\numberline{} \hspace{-2cm}Introduction}{\thepage}}
\AddToContent{G.~Brooijmans, B.~Gripaios, F.~Moortgat, J.~Santiago, P.~Skands}

\superpart{ Tools }


\chapter{DLHA: Dark Matter Les Houches Agreement}

{\it {\rm Coordinator}: Csaba Bal\'azs, \\
David G. Cerde\~no,
Rebecca Leane,
Mitsuru Kakizaki,
Sabine Kraml,
Christopher Savage,
Pat Scott,
Sezen Sekmen,
\\{\bf AMIDAS}:
Chung-Lin Shan,
\\{\bf DarkSUSY}: 
Lars Bergstr\"om,
Torsten Bringmann,
Gintaras Duda,
Joakim Edsj\"o,
Paolo Gondolo,
\\{\bf DMFIT and DMMW}:
Stefano Profumo,
Tesla Jeltema,
\\{\bf DRAGON}: 
Carmelo Evoli,
Daniele Gaggero,
Dario Grasso,
Luca Maccione,
\\{\bf FeynRules}: 
Claude Duhr,
Benjamin Fuks,
\\{\bf GALPROP}:  
Igor Moskalenko,
Elena Orlando,
Troy Porter,
Andrew Strong,
\\{\bf ISAJET}:
Howie Baer,
\\{\bf micrOmegas}:  
Genevieve B\'elanger,
Fawzi Boudjema,
Alexander Pukhov,
\\{\bf PPPC4DMID}: 
Marco Cirelli,
Mario Kadastik,
Paolo Panci,
Martti Raidal,
\\{\bf SLHA}:  
Sven Kreiss,
Frank S. Thomas,
\\{\bf Semi-analytical propagation models}:
Timur Delahaye,
Fiorenza Donato,
Julien Lavalle,
David Maurin,
Pierre Salati,
Richard Taillet,
\\{\bf SuperIso Relic}:
Alexandre Arbey,
Farvah Mahmoudi
}

\begin{abstract}
This work presents a set of conventions and numerical structures that aim to provide a universal interface between computer programs calculating dark matter related observables.  It specifies input and output parameters for the calculation of observables such as abundance, direct and various indirect detection rates.  These parameters range from cosmological to astrophysical to nuclear observables.  The present conventions lay the foundations for defining a future Les Houches Dark Matter Accord. 
\end{abstract}

\section{INTRODUCTION}


Over the last decade a burst of activity has surrounded various dark matter related problems.  As a byproduct of this activity, a number of robust numerical tools have been created by particle, astro-particle and astrophysicists.  These tools predict values for observables that, when contrasted with observation, can shed light on the properties of dark matter.  The dark matter related computer codes address a wide range of physical problems, of which the most important are:
\begin{enumerate}
\item deriving Feynman rules from Lagrangians,
\item calculating scattering amplitudes from Feynman rules,
\item building cross sections or decay rates from amplitudes,
\item computing dark matter abundance from cross sections or decay rates,
\item obtaining dark matter-nucleon scattering rates from cross sections,
\item evaluating cosmic ray yields based on annihilation cross sections or decay rates.
\end{enumerate}


A certainly incomplete but representative list of such computer codes is:
\begin{center}
\begin{tabular}[h]{lll}
 \cite{Pukhov:2004ca}           CalcHEP        (2, 3)          ~~~ & 
 \cite{Gondolo:2005we}          DarkSUSY       (3, 4, 5, 6)    ~~~ &
 \cite{Jeltema:2008hf,DMMWRef}  DMFIT and DMMW (6)             ~~~ \\  
 \cite{DRAGONRef}               DRAGON         (6)             ~~~ & 
 \cite{Christensen:2008py}      FeynRules      (1)             ~~~ & 
 \cite{Strong:2009xj}           GALPROP        (6)             ~~~ \\
 \cite{Paige:2003mg}            ISATools       (4, 5)          ~~~ & 
 \cite{Semenov:2008jy}          LanHEP         (1)             ~~~ & 
 \cite{Belanger:2010pz}         micrOmegas     (2, 3, 4, 5, 6) ~~~ \\
 \cite{PPPC4DMID}               PPPC4DMID      (6)             ~~~ &  
 \cite{Arbey:2011zz}            SuperIso Relic (4)             
\end{tabular}
\end{center}
where the numbers in parentheses after the names indicate what these codes do in the context of the task-list above.  These two lists show that 
today there exists no code that covers all the dark matter related calculations (although some come very close), and
there are significant overlaps between the capabilities of these codes.
%
What these lists do not reveal is how general and sophisticated the various programs are.  Not surprisingly, each of these tools has its strengths and weaknesses.  Necessarily, all these codes contain hard-wired assumptions that potentially limit their capabilities. 


When particle physicists found themselves in a similar situation regarding collider related calculations, they created a series of interfaces that allowed their various tools to communicate with each other \cite{Boos:2001cv, Whalley:2005nh, Skands:2003cj, Allanach:2008qq, Mahmoudi:2010iz}.  These interfaces allow the users of the codes to easily and selectively exploit the features that each of the tools offers.  The success of the Supersymmetric Les Houches Accord (SLHA) and the other accords lies in the fact that they significantly increase ease and flexibility for collider related calculations \cite{Skands:2003cj, Allanach:2008qq}. 


Following the successful path of previous Les Houches accords, the Dark Matter Les Houches agreement (DLHA) proposes a format for storing and exchanging information relevant to dark matter calculations.  In the spirit of the Les Houches accords, DLHA aims to interface various calculators to provide increased flexibility to users of these tools.  Presently these calculators work semi-independently from each other and their consecutive use typically requires tedious interfacing.  However careful this interfacing is, it may jeopardise the integrity of the tools and, in turn, the results of the calculation.  


A related problem DLHA targets is the transparency of the dark matter calculations.  Present codes input and output limited amounts of information while they may contain large amounts of implicit, hard-wired assumptions that affect their results.  The more of this information is accessible, the more control a user has over the calculation.  Additionally, making more assumptions explicit gives the user a chance to change them and this could lead to a more diverse and productive use of these codes.  Since a DLHA file can be part of the input or the output of a numerical code calculating dark matter related observables, it aids in making implicit assumptions explicit and, potentially, it allows for changing some of those assumptions.  


Here we give some examples of how such an agreement enables the user to easily interface various codes with different capabilities and thereby calculate quantities that none of these codes could calculate alone.
\begin{itemize}
\item Assume that code $A$ has the capability to calculate decay rates of a dark matter candidate in an exotic particle model, but code $A$ can only handle standard cosmology when calculating relic abundance of the dark matter.  Assume that code $B$ does not implement the exotic particle model of interest but can calculate relic abundance for non-standard cosmologies.  With the appropriate DLHA interface the decay rate can be communicated from code $A$ to $B$ and relic abundance of an exotic candidate can be calculated in a non-standard cosmology.
\item Assume that code $C$ has the capability to calculate the annihilation cross section of a dark matter candidate with next-to-leading order (NLO) corrections but it has no indirect detection routines built into it.  Meanwhile code $D$ can only do one thing, but it does it excellently: calculate indirect detection rates.  An obvious task for DLHA is to pass the annihilation cross section from $C$ to $D$, thereby making it possible to calculate a cosmic ray yield from annihilating dark matter with NLO corrections.
\end{itemize}
These examples are just a limited sample of the possibilities that DLHA offers for a resourceful user.

\section{CALCULATION OF DARK MATTER RELATED OBSERVABLES}

This section summarizes the relevant details of dark matter related calculations so that we can fix our notation and define the accord in the next section.

\subsection{Relic abundance}

Numerous microscopic models have been proposed to describe the identity of dark matter particles.  Depending on the details of these models, and conditions in the early Universe, dark matter may have been produced in various different ways.  For simplicity, here we only address the thermal production mechanism of weakly interacting massive particles.

\subsubsection{Thermal relic abundance with standard cosmological assumptions}

In this section we recapture some details of the calculation of thermally produced relic abundance of dark matter following standard cosmological assumptions.  According to Big Bang cosmology the Universe cooled from temperatures substantially higher than the mass of a typical dark matter particle.  At those temperatures dark matter particles were in chemical equilibrium with their environment and consequently their (co-moving) number density was unchanged.  At temperatures comparable to the typical weakly interacting dark matter particle, at $T = {\cal O}$(100 GeV/$k_B$), the Universe was radiation dominated.  (Here $k_B$ is the Boltzmann constant.)  As the Universe cooled and expanded further, dark matter particles came in thermal equilibrium with the cooling radiation and consequently their number density decreased exponentially.  Later scattering between dark matter particles became rare until the scattering rate fell below the expansion rate of the Universe for the dark matter particles to stay in equilibrium.  At that stage dark matter particles froze-out, that is their (co-moving) number density ceased to change. 

The change of the average number density of dark matter particles, $n_\chi$, is described by Boltzmann's equation \cite{Goldberg:1983nd,Krauss:1983ik,Srednicki:1988ce,Gondolo:1990dk,Edsjo:1997bg}. This change occurs due to (co-)annihilation to Standard Model particles and to the expansion of the early Universe. Boltzmann's equation reads:
\begin{equation}
\frac{dn_\chi}{dt} = -\langle\sigma v\rangle\left(n_{\chi}^2-n_{\chi,eq} ^2\right) - 3 H n_{\chi} .
\label{DLHA_boltzbest}
\end{equation}
On the right hand side the thermally averaged product of the annihilation cross section and relative velocity of the two annihilating dark matter particles is defined by
\begin{equation}
\langle\sigma v\rangle = \frac{\int{\sigma v\ dn_{1,eq}dn_{2,eq}}}{\int{dn_{1,eq} dn_{2,eq}}}
= \frac{\int{\sigma v \ f(E_1)f(E_2)d^3p_1 d^3p_2}}{\int{f(E_1)f(E_2)d^3p_1 d^3p_2}} .
\label{DLHA_sigmav}
\end{equation}
Here $p_i$ are the three-momenta and $E_i = \sqrt{m_i^2 + p_i^2}$ are the energies of the colliding particles with $i=1,2$%
\footnote{Throughout this section we use natural units with $c = \hbar = 1$, unless stated otherwise.}.
The total (co-)annihilation cross section $\sigma$ depends on the microscopic properties of the dark matter particles and can be calculated assuming a particular dark matter model.  The equilibrium number density is given by thermodynamics
\begin{equation}
n_{\chi,eq} = \frac{g_\chi}{(2\pi)^3}\int{f(E)d^3E} ,
\end{equation}
where $g_\chi$ is the number of internal degrees of freedom of dark matter particles.  Their energy distribution is the familiar
\begin{equation}
f(E)=\frac{1}{\exp(E/k_B T)\pm1} .
\label{DLHA_eq:f(E)}
\end{equation}
In Eq.~(\ref{DLHA_boltzbest}) the Hubble parameter describes the expansion rate, and in a nearly flat Friedmann-Lemaitre-Robertson-Walker (FLRW) Universe it is given by
\begin{equation}
H^2 = \frac{8\pi G_N}{3} \rho .
\label{DLHA_hubs}
\end{equation}
As the Universe was radiation dominated, the total energy density was the standard function of the temperature of the radiation $T$:
\begin{equation}
\rho = \rho_{rad}(T) = g_{eff}(T)\frac{\pi^2}{30}T^4 ,
\label{DLHA_energydens}
\end{equation}
where $g_{eff}(T)$ is the number of internal degrees of freedom of the (effectively) massless particles in the thermal bath\footnote{Here we assume that the various species contributing to the total energy density are all in thermal equilibrium with each other at a common temperature $T$.}. 

The decrease in number density due to expansion of the Universe can be made implicit by rewriting Eq.~(\ref{DLHA_boltzbest}) in terms of the co-moving number density 
\begin{equation}
Y = \frac{n_{\chi}}{s} .
\label{DLHA_y}
\end{equation}
For the radiation dominated Universe the total entropy density is
\begin{equation}
s = s_{rad}(T) = h_{eff}(T)\frac{2\pi^2}{45}T^3 .
\label{DLHA_entropy}
\end{equation}
The number of internal degrees of freedom of the particles contributing to the entropy is $h_{eff}(T)$\footnote{For a relativistic particle with one internal degree of freedom, such as a spin zero boson, $g_{eff} = h_{eff} = 1$.}.
Recasting Boltzmann's equation for $Y$ as a function of $x=m_{\chi}/T$, where $m_\chi$ is the mass of the dark matter particle, we arrive at
\begin{equation}
\frac{dY}{dx} = \frac{1}{3H}\frac{ds}{dx}\langle\sigma v\rangle(Y^2-Y^2_{eq}) .
\label{DLHA_boltznew}
\end{equation}
By utilising Eqs.~(\ref{DLHA_hubs}), (\ref{DLHA_energydens}) and (\ref{DLHA_entropy}) this may also be written as
\begin{equation}
 \frac{dY}{dx} = 
 -\sqrt{\frac{\pi g_\star}{45 G_N}}\frac{m_\chi}{x^2}\langle\sigma v\rangle(Y^2-Y^2_{eq}) .
\label{DLHA_boltznew2}
\end{equation}
The parameter $g_\star$ incorporates the degrees of freedom arising from the energy and entropy densities
\begin{equation}
 g_\star ^{1/2} = \frac{h_{eff}}{g_{eff}^{1/2}}\left(1+\frac{1}{3}\frac{T}{h_{eff}}\frac{dh_{eff}}{dT}\right) .
\label{DLHA_eq:gstar}
\end{equation}
A value for $Y_{eq}$ of a cold gas of dark matter is obtained for (non-)relativistic dark matter particles by taking the (low) high energy limit of the Maxwell-Boltzmann distribution.  For non-relativistic dark matter particles
\begin{equation}
Y_{eq}=\frac{45g_{eff}}{4\pi^4} \frac{x^2 K_2(x)}{h_{eff}(m_\chi/x)} ,
\label{DLHA_yeq}
\end{equation}
where $K_2(x)$ is the modified Bessel function of second order.

In various calculations it can be useful to introduce the freeze-out temperature $T_f$ based on the condition 
\begin{equation}
Y-Y_{eq} = \alpha Y_{eq} ,
\label{DLHA_freezecond}
\end{equation}
where $\alpha > 0$ is a number of order 1. Incorporating this condition into Eq.~(\ref{DLHA_boltznew2}) a statement for freeze-out is obtained
\begin{equation}
 \sqrt{\frac{\pi g_\star}{45G_N}}\frac{m_\chi}{x^2}\langle\sigma v\rangle \alpha(\alpha+2)Y_{eq} = -\frac{d(\ln Y_{eq})}{dx} .
\label{DLHA_boltznew3}
\end{equation}
Upon substitution of Eq.~(\ref{DLHA_yeq}) into Eq.~(\ref{DLHA_boltznew3}) this allows for a numerical solution for $x=x_f$.

Equipped with the co-moving number density after freeze-out, $Y_0$, the relic energy density of dark matter may be calculated.  This is typically given in units of the critical energy density
\begin{equation}
 \Omega_\chi=\frac{\rho_\chi}{\rho_c}=\frac{s_0Y_0m_\chi}{\rho_c} ,
\label{DLHA_reldens}
\end{equation}
where $\rho_c$ is the energy density of a flat FLRW Universe.
Here $s_0$ is the current entropy density of the Universe.  Combining Eqs. (\ref{DLHA_hubs}) and (\ref{DLHA_reldens}), the relic density may also be expressed as
\begin{equation}
 \Omega_\chi h^2 = 
 \frac{8 \pi G_N}{3} s_0 Y_0 m_\chi \times 10^{-4} {\rm (s~Mpc/km)}^2 .
\label{DLHA_omegachi}
\end{equation}
The present day normalized Hubble expansion rate $h$ is determined via 
\begin{equation}
 H_0 = 100~h~{\rm km/(s~Mpc)} .
\label{DLHA_eq:def_h}
\end{equation}

\subsubsection{Thermal relic abundance with non-standard cosmological assumptions}

In this section, we consider the generic scenario described in Refs.~\cite{Arbey:2008kv,Arbey:2009gt} and implemented in {\tt SuperIso Relic} \cite{Arbey:2009gu,Arbey:2011zz} and {\tt AlterBBN} \cite{Arbey:2011nf}. In this scenario the total energy density and entropy density of the Universe are modified: 
\begin{equation}
\rho=\rho_{rad}+\rho_{D} ,  ~~~ s=s_{rad}+s_{D} .
\label{DLHA_update1}
\end{equation}
The temperature dependence of the additional components $\rho_{D}$ and $s_{D}$ can be parametrized as
\begin{equation}
\alpha_{D}(T)=\kappa_\alpha \alpha_{rad}(T_{BBN})\left(\frac{T}{T_{BBN}}\right)^{n_\alpha} ,
\label{DLHA_alphadark}
\end{equation}
where $\alpha=\rho$ or $s$, 
\begin{equation}
\kappa_{\alpha}=\frac{\alpha_{dark}(T_{BBN})}{\alpha_{rad}(T_{BBN})} ,
\end{equation}
and $T_{BBN}$ is the Big Bang Nucleosynthesis temperature.

The time evolution of the total entropy density is given by
\begin{equation}
 \frac{ds}{dt}=-3H s+\Sigma_{D},
 \label{DLHA_eq:dsdt}
\end{equation}
where $\Sigma_{D}$ describes the entropy production of the additional component. Instead of parametrizing the entropy density $s_D$, it is sometimes better (for example for reheating models) to write $\Sigma_{D}$ in the form of Eq.~(\ref{DLHA_alphadark}) with $\alpha=\Sigma$, and it is then related to the entropy density by
\begin{equation}
 \Sigma_{D} = T^2\sqrt{\frac{4\pi^3G}{5}\left(1+\frac{\rho_{D}}{\rho_{rad}}\right)}\left(s_{D} g_{eff}^{1/2}-\frac{1}{3}\frac{h_{eff}}{g_\star^{1/2}}T\frac{ds_{D}}{dT}\right) .
\label{DLHA_en1}
\end{equation}
A possible generalization of the above parameterizations consists of relaxing Eq.~(\ref{DLHA_alphadark}) and letting $\rho_D$, $s_D$, $\Sigma_D$ and/or $H$ be general functions of the temperature.

Introducing the co-moving number density $Y = n_{\chi}/s$, Eq.~(\ref{DLHA_boltzbest}) becomes
\begin{equation}
 \frac{dY}{dx} = 
  -\sqrt{\frac{\pi g_\star}{45}}\frac{m_\chi}{x^2}\left( \frac{1+s_D/s_{rad}}{\sqrt{1+\rho_D/\rho_{rad}}} \right)
  \left( \langle\sigma v\rangle (Y^2-Y^2_{eq}) +
  \frac{Y \Sigma_D}{s_{rad}^2(1+s_D/s_{rad})^2} \right) ,
\label{DLHA_eq:dYdxD}
\end{equation}
where 
\begin{equation}
 Y_{eq}=\frac{45 g_{eff}}{4\pi^4} \frac{1}{1+s_D/s_{rad}} \frac{x^2 K_2(x)}{h_{eff}(m_\chi/x)} .
\label{DLHA_eq:YeqD}
\end{equation}
The relic abundance can then be calculated based on Eq.~(\ref{DLHA_omegachi}).



\subsection{Direct detection}

If dark matter and Standard Model particles interact with strength comparable to that of the electroweak interactions, and if dark matter is massive and fast enough, then it may be detected by observing dark matter scattering on atoms.  This is the aim of the direct dark matter detection experiments.  To assess detection rates we calculate elastic scattering cross sections of dark matter on nuclei.  These cross sections are derived from a Lagrangian describing the effective interaction of dark matter with nuclei.  Nuclear form factors provide a link between the partons, the nucleons and the nucleus of the target species.  
For simplicity, in this first agreement, we only consider elastic dark matter-nucleus scattering. Inelastic scattering may be included in a future edition of this agreement.

The dark matter-nucleus recoil rate per unit detector mass, unit time, and unit recoil energy $E$ is written in the form
\begin{equation}
\frac{dR}{dE} = \frac{\sigma_A(E)}{2m_{\chi} \mu_{A}^2} \, \rho_{\chi} \, \eta(E,t),
\end{equation}
where
\begin{equation}
\mu_{A} = \frac{m_{\chi}m_{A}}{m_{\chi}+m_{A}}
\end{equation}
is the dark matter-nucleus reduced mass with $m_A$ the nucleus mass, $\sigma_A(E)$ is the dark matter-nucleus scattering cross section,  $\rho_{\chi}$ is the local dark matter density, and
\begin{equation}
\eta(E,t) = \int_{|{\bf v}|>v_{\rm min}(E)} \frac{f({\bf v},t)}{|{\bf v}|} \, d^3 v
\label{DLHA_eq:etaEt}
\end{equation}
is an average inverse dark matter speed, also called the velocity integral. Here $f({\bf v},t)$ is the dark matter velocity distribution in the reference frame of the detector, which is expected to be time dependent, and 
\begin{equation}
v_{\rm min}(E) = \sqrt{ \frac{m_A E}{2\mu_A^2} } 
\end{equation}
is the minimum dark matter speed that can impart a recoil energy $E$ to the nucleus. Notice that the dark matter-nucleus differential cross section in recoil energy $E$ for a dark matter particle of initial velocity $v$ with respect to the nucleus has the expression
\begin{equation}
\frac{d\sigma_A}{dE} = \frac{\sigma_A(E)}{E_{\rm max}(v)} \, \Theta\!\left( E_{\rm max}(v) - E\right) ,
\end{equation}
where
\begin{equation}
E_{\rm max}(v) = \frac{2 \mu_A^2 v^2}{m_A}
\end{equation}
is the maximum recoil energy that a particle of velocity $v$ can impart to the nucleus.

Only the non-relativistic limit is relevant for dark matter-nucleus scattering. In this limit, only two kinds of nucleon currents survive: the spin-independent current $\psi^\dagger \psi$ and the spin-dependent current $\psi^\dagger {\bf S} \psi$. Here $\psi$ is the nucleon wave function (a non-relativistic two-component Pauli spinor) and ${\bf S}=\frac{1}{2} \boldsymbol\sigma$ is the nucleon spin operator. 
Therefore one splits the differential dark matter-nucleus scattering cross section $\sigma_{A}(E)$  into its spin-independent (SI) and spin-dependent (SD) contributions,
\begin{equation}
\sigma_{A}(E) = \sigma_{A}^{SI}(E) + \sigma_{A}^{SD}(E).
\label{DLHA_sigman}
\end{equation}
Correspondingly, one often separates the spin-independent and spin-dependent contributions to the  recoil rate $dR/dE$ as
\begin{equation}
\frac{dR}{dE} = \left({\frac{dR}{dE}}\right)^{SI} + \left({\frac{dR}{dE}}\right)^{SD}.
\end{equation}

The spin-independent part $\sigma_{A}^{SI}(E)$ is written as
\begin{equation}
\sigma_{A}^{SI}(E) = \frac{4\mu_{A}^{2}}{\pi}\, {\Big|Z \, f_{\rm p} \, F^{(Z,A)}_{\rm p}(E) + (A - Z) \, f_{\rm n} \, F^{(Z,A)}_{\rm n}(E) \Big|}^{2},
\label{DLHA_sigmanSI}
\end{equation}
where 
$Z$ is the number of protons in the nucleus (atomic number), $A$ is the mass number of the nucleus, $F^{(Z,A)}_{\rm p}(E)$ and $F^{(Z,A)}_{\rm n}(E)$ are proton and nucleon number density form factors for the nucleus $(Z,A)$, normalized to $F^{(Z,A)}_{\rm p}(0)=F^{(Z,A)}_{\rm n}(0)=1$, and finally $2f_{\rm p}$ and $2f_{\rm n}$ are the dimensionless four-particle vertices for the SI dark matter-proton and dark matter-neutron interactions, respectively. 

One also introduces the pointlike dark matter-proton and dark matter-neutron cross sections, which by convention are used when reporting or interpreting experimental results,
\begin{equation}
\sigma_{\rm p}^{SI} = \frac{4\mu_{\rm p}^{2}}{\pi}\, {|f_{\rm p}|}^{2},  
~~~~~~
\sigma_{\rm n}^{SI} = \frac{4\mu_{\rm n}^{2}}{\pi}\, {|f_{\rm n}|}^{2}. \label{DLHA_eq:sigmaSIn}
\end{equation}
Here $\mu_{\rm p}$ and $\mu_{\rm n}$ are the reduced dark matter-proton and dark matter-neutron masses.

\begin{table}
\caption{Four-particle effective vertices for dark matter-proton elastic scattering. Dark matter-neutron vertices are obtained by changing ${\rm p} \to {\rm n}$. $s_{\chi}$ is the spin of the dark matter particle.}
\begin{center}
\setlength{\unitlength}{0.8in}
\begin{tabular}{cccc}
\hbox to 1in{~}& \hbox to 1.5in{\hfil $s_{\chi}=0$ \hfil} & \hbox to 1.5in{\hfil $s_{\chi}=\frac{1}{2}$ \hfil}  &\hbox to 1.5in{\hfil $s_{\chi}=1$ \hfil} \\ &&& \\
\hline
\begin{picture}(1.3,1)
\put(0,.45){spin-independent}
\end{picture}
&
\begin{picture}(1,1)
\put(.2,.2){\line(1,1){0.6}}
\put(.2,.8){\line(1,-1){0.6}}
\put(.09,.1){${\rm p}$}
\put(.825,.1){${\rm p}$}
\put(.09,.82){${\chi}$}
\put(.825,.82){${\chi}$}
\put(.62,.45){$$}
\end{picture}
&
\begin{picture}(1,1)
\put(.2,.2){\line(1,1){0.6}}
\put(.2,.8){\line(1,-1){0.6}}
\put(.09,.1){${\rm p}$}
\put(.825,.1){${\rm p}$}
\put(.09,.82){${\chi}$}
\put(.825,.82){${\chi}$}
\put(.62,.45){$2f_{\rm p}$}
\end{picture}
&
\begin{picture}(1,1)
\put(.2,.2){\line(1,1){0.6}}
\put(.2,.8){\line(1,-1){0.6}}
\put(.09,.1){${\rm p}$}
\put(.825,.1){${\rm p}$}
\put(.09,.82){${\chi}$}
\put(.825,.82){${\chi}$}
\put(.62,.45){$$}
\end{picture}
\\
\begin{picture}(1.3,1)
\put(0,.45){spin-dependent}
\end{picture}
&
\begin{picture}(1.3,1)
\put(0.45,.45){N/A}
\end{picture}
&
\begin{picture}(1,1)
\put(.2,.2){\line(1,1){0.6}}
\put(.2,.8){\line(1,-1){0.6}}
\put(.09,.1){${\rm p}$}
\put(.825,.1){${\rm p}$}
\put(.09,.82){${\chi}$}
\put(.825,.82){${\chi}$}
\put(.62,.45){$2\sqrt{2}G_Fa_{\rm p} \boldsymbol\sigma_{\chi} \!\cdot\!\boldsymbol\sigma_{\rm p}$}
\end{picture}
&
\begin{picture}(1,1)
\put(.2,.2){\line(1,1){0.6}}
\put(.2,.8){\line(1,-1){0.6}}
\put(.09,.1){${\rm p}$}
\put(.825,.1){${\rm p}$}
\put(.09,.82){${\chi}$}
\put(.825,.82){${\chi}$}
\put(.62,.45){$$}
\end{picture}
\\
\hline
\end{tabular}
\end{center}
\end{table}

One often assumes, and we will do so in this first agreement, that 
\begin{equation}
 F^{(Z,A)}_{\rm p}(E) = F^{(Z,A)}_{\rm n}(E) \equiv F_A(E).
\label{DLHA_eq:F_A}
\end{equation}
In this case, one sometimes introduces the pointlike dark matter-nucleus cross section
\begin{equation}
\sigma_{A,0}^{SI} = \frac{4\mu_{A}^{2}}{\pi}\, {\Big|Z \, f_{\rm p} + (A - Z) \, f_{\rm n} \Big|}^{2},
\end{equation} 
which is $\sigma_{A}^{SI}$ with $F^{(Z,A)}_{\rm p}(E)=F^{(Z,A)}_{\rm n}(E)=1$.

The spin-dependent part $\sigma_{A}^{SD}(E)$ is written as
\begin{equation}
\sigma_{(Z,A)}^{SD}(E) = \frac{32\mu_{A}^{2}G_{F}^{2}}{(2J_{A}+1)}\,\big[a_{\rm p}^{2}\,S_{{\rm pp}}^{(Z,A)}(E) + a_{\rm n}^{2}\,S_{{\rm nn}}^{(Z,A)}(E) + a_{\rm p}\,a_{\rm n}\,S_{{\rm pn}}^{(Z,A)}(E)\big].
\label{DLHA_sigmanSD}
\end{equation}
Here 
$G_F=1.16637\times10^{-5}$ $(\hbar c)^3/{\rm GeV}^{2}$ is the Fermi coupling constant, $J_{A}$ is the nucleus total angular momentum in units of $\hbar$, and finally $2\sqrt{2}\,G_{F}a_{\rm p}$ and $2\sqrt{2}\,G_{F}a_{\rm n}$ are the effective four-particle vertices for the SD interaction of DM particles with protons and neutrons.
 
In Eq.~(\ref{DLHA_sigmanSD}), the dimensionless functions $S_{{\rm pp}}^{(Z,A)}(E)$, $S_{{\rm nn}}^{(Z,A)}(E)$, and $S_{{\rm pn}}^{(Z,A)}(E)$ are the proton-proton, neutron-neutron, and proton-neutron nuclear spin structure functions. They can be written in terms of isoscalar and isovector spin structure functions  $S_{00}^{(Z,A)}$, $S_{11}^{(Z,A)}$, $S_{01}^{(Z,A)}$ as 
\begin{eqnarray}
S_{{\rm pp}}^{(Z,A)} & = & S_{00}^{(Z,A)}+S_{11}^{(Z,A)}+S_{01}^{(Z,A)}\,,
\label{DLHA_Ss1}
\\
 S_{{\rm nn}}^{(Z,A)} & = & S_{00}^{(Z,A)}+S_{11}^{A}-S_{01}^{(Z,A)}\,,
\label{DLHA_Ss2}
\\
S_{{\rm pn}}^{(Z,A)} & = & 2(S_{00}^{(Z,A)}-S_{11}^{(Z,A)})\,.
\label{DLHA_Ss3}
\end{eqnarray}
One similarly introduces
\begin{equation}
a_{0} = a_{\rm p} + a_{\rm n}, ~~~~~~
a_{1} = a_{\rm p} - a_{\rm n} .
\end{equation}

When the nuclear spin is approximated by the spin of the odd nucleon only, one finds
\begin{equation}
S_{{\rm pp}}^{(Z,A)} = \frac{\lambda_{A}^{2}\,J_{A}(J_{A}+1)(2J_{A}+1)}{\pi}\,,\,\,S_{{\rm nn}}^{nA} = 0\,,\,\,S_{{\rm pn}}^{(Z,A)} = 0\,,
\label{DLHA_Spp}
\end{equation}
for a proton-odd nucleus, and
\begin{equation}
S_{{\rm pp}}^{(Z,A)} = 0\,\,,\,S_{{\rm nn}}^{A} = \frac{\lambda_{A}^{2}\,J_{A}(J_{A}+1)(2J_{A}+1)}{\pi}\,,\,\,S_{{\rm pn}}^{(Z,A)} = 0\,,
\end{equation}
for a neutron-odd nucleus. Here $\lambda_{A}$ is conventionally defined through the relation $\langle\psi_A|{\bf S}_A|\psi_A\rangle = \lambda_{A}\, \langle \psi_A|{\bf J}_A|\psi_A \rangle$, where $|\psi_A \rangle$ is the nuclear state, ${\bf S}_A$ is the nucleus spin vector, and ${\bf J}_A$ is its total angular momentum vector.

One sometimes introduces dark matter-nucleon pointlike cross sections. For a single proton or a single neutron, $\lambda_{\rm p}=\lambda_{\rm n}=1$, $J_{\rm p}=J_{\rm n}=\frac{1}{2}$, and
\begin{equation}
\sigma_{\rm p}^{SD} = \frac{36\mu_{\rm p}^{2}G_{F}^{2}}{\pi^2}\,|a_{\rm p}|^{2}, 
~~~~~~
\sigma_{\rm n}^{SD} = \frac{36\mu_{\rm n}^{2}G_{F}^{2}}{\pi^2}\,|a_{\rm n}|^{2}. 
\label{DLHA_eq:sigmaSDn}
\end{equation}

We may list the expressions of the effective four-particle dark matter-nucleon couplings $f_{\rm p}$, $f_{\rm n}$, $a_{\rm p}$ and $a_{\rm n}$ in terms of elementary couplings to quarks and gluons for various kinds of dark matter particles, via an effective four-particle lagrangian with spin $s$
\begin{equation}
 {\cal L}_{eff}(s) = 
 \frac{1}{2}\left(\sum_{i=e,o} f_{N,i}(s) {\cal L}_{N,i}^{SI}(s) +
 \sum_{i=e,o} a_{N,i}(s)     {\cal L}_{N,i}^{SD}(s)\right)   .
\label{DLHA_eq:L_eff}
\end{equation}
The operators ${\cal L}_e^{SI}(s)$ and ${\cal L}_o^{SI}(s)$ describe spin independent even and odd interactions, while ${\cal L}_e^{SD}(s)$ and ${\cal L}_o^{SD}(s)$ are their spin dependent counterparts.\footnote{Even and odd couplings are introduced to trace symmetries under particle-antiparticle interchange.  Majorana fermions, for example, have even couplings and non-Majorana fermions have odd couplings.}
Table \ref{DLHA_tableop} shows the explicit forms of these operators for various values of $s$.  
The dark matter-nucleon couplings can be obtained as
\begin{equation}
 f_{\rm N} = \sum_{i=e,o} f_{N,i}(s), ~~~ 
 a_{\rm N} = \sum_{i=e,o} a_{N,i}(s),
\end{equation}

\begin{table}
\begin{center}
\caption{Even and odd operators ${\cal L}_{e,o}^{SI,SD}(s)$ for dark matter interactions with standard quarks $q$.  A scalar field only interacts in a spin independent manner \cite{micromegas22}.}
\vspace{3mm}
\begin{tabular}{|l|c|c|c|}
\hline

 & s & Even operators ($i=e$) & Odd operators ($i=o$) \\ \hline

 Spin independent & 0 & $2M_\chi \chi\chi^\dagger\bar{q}q$ & $i(\partial_\mu\chi\chi^\dagger-\chi\partial_\mu\chi^\dagger)\bar{q}\gamma^\mu q$ \\

 ($SI$) & 1/2 & $\bar{\chi}\chi \bar{q}q$ & $\bar{\chi}\gamma_\mu\chi\bar{q}\gamma^\mu q$ \\
 
 & 1 & $2M_\chi \chi_\mu\chi^\mu\bar{q}q$ & $i(\chi^{\dagger\alpha}\partial_\mu\chi,\alpha-\chi^\alpha\partial_\mu\chi^\dagger_\alpha)\bar{q}\gamma_\mu q$ \\ \hline

 Spin dependent & 1/2 & $\bar{\chi}\gamma_5\gamma_\mu\chi\bar{q}\gamma_5\gamma^\mu q$ & $-\frac{1}{2}\bar{\chi}\sigma_{\mu\nu}\chi\bar{q}\sigma^{\mu\nu}q$ \\
 ($SD$) & 1 & $\sqrt{6}(\partial_\mu\chi^{\dagger\beta}\chi_\gamma-\chi^\dagger_\beta\partial_\mu\chi_\gamma)\epsilon^{\alpha\beta\gamma\mu}\bar{q}\gamma_5\gamma_\mu q$ & $\frac{\sqrt{3}}{2}i(\chi_\mu\chi^{\dagger}_\nu-\chi^\dagger_\mu\chi_\nu)\bar{q}\sigma^{\mu\nu}q$ \\
 
\hline
\end{tabular}\label{DLHA_tableop}
\end{center}
\end{table}

Dark matter-nucleus scattering amplitudes can be calculated based on parton level amplitudes after relating nuclear couplings to dark-matter quark couplings via nucleon form factors
\begin{eqnarray}
 2f_{N,e}=\sum_{q}\frac{m_p}{m_q} {f_q^N s_{q,e}} , ~~~
 2f_{N,o}=\sum_{q}\frac{m_p}{m_q}{f_{V_q}^N s_{q,o}} , \label{DLHA_nuc1form} \\
 2\sqrt{2}G_Fa_{N,e}=\sum_{q}{\Delta_q^N a_{q,e}} , ~~~
 2\sqrt{2}G_Fa_{N,o}=\sum_{q}{\delta_q^N a_{q,o}}  .
\label{DLHA_nuc2form}
\end{eqnarray}
These form factors capture the distribution of quarks within the nucleons.  The light quark flavor scalar form factors are related to the pion-nucleon sigma term and the nucleon and quark masses as
\begin{equation}
 f_u^{p,n} = \frac{m_u}{m_d} \alpha^{+1,-1} f_d^{p,n} , ~~~
 f_d^{p,n} = \frac{2 \sigma_{\pi N}}{(1 + \frac{m_u}{m_d}) m_{p,n}} \frac{\alpha^{0,1}}{1 + \alpha} , ~~~
 f_s^{p,n} = \frac{\sigma_{\pi N} y}{(1 + \frac{m_u}{m_d}) m_{p,n}} \frac{m_s}{m_d} .
\label{DLHA_eq:f_q^N}
\end{equation}
Here $\alpha=B_u/B_d$, $\sigma_{\pi N}=(m_u+m_d)(B_u+B_d)/2$, $y=2B_s/(B_u+B_d)$ and $B_q=\langle N|{\bar q}q|N \rangle$.
The heavy flavor scalar form factors are typically calculated as 
\begin{equation}
 f_Q^N = \frac{2}{27} \left( 1-\sum_{q=u,d,s} f_q^N \right) .
\label{DLHA_eq:f_Q^N}
\end{equation}

\subsection{Indirect detection}

Perhaps the most challenging way to discover dark matter is to detect its annihilation or decay products in the astrophysical environment.  This is called indirect detection and involves the description of an initial (source) distribution of standard particles originating from dense dark matter concentrations.  After the source properties are fixed, the propagation of the secondary particles has to be followed through.  Here we review only the simplest cases: electron, positron, antiproton or photon production via dark matter annihilation and their subsequent propagation to us through our Galaxy using a simplified but effective treatment. We also comment on the modifications which would need to be introduced for a more detailed treatment.

Charged cosmic ray propagation through the Galaxy can be usually described by the diffusion-convection model \cite{Ginzburg:1990sk}. This model assumes homogeneous propagation of charged particles within a certain diffusive region (similar to one of the simplest models of propagation called the leaky box model), but it also takes into account cooling (energy loss) effects. The diffusive region is usually assumed to have the shape of a solid flat cylinder of half-height $L$ and radius $R$ that sandwiches the Galactic plane: inside it, charged cosmic rays are trapped by magnetic fields; outside, they are free to stream away. The (cylindrical) coordinates of the solar system correspond to $\vec r_\odot = (8.33 \ {\rm kpc}, 0  \ {\rm kpc})$ \cite{Gillessen:2008qv}. The phase-space density $\psi_a (\vec r,p,t)$ of a particular cosmic ray species $a$ at an instant $t$, at a Galactic position $\vec r$ and with momentum $p$ can be calculated by solving the cosmic ray transport equation, which has the general form \cite{Strong:2007nh}
\begin{eqnarray}
 \frac{\partial \psi_a (\vec r,p,t)}{\partial t} 
 & = &
 Q_a(\vec r, p, t) 
 + \nabla \cdot ( D_{xx}\nabla\psi_a - \vec V\psi_a ) \nonumber \\
 & + & \frac{\partial}{\partial p}\, \left( p^2 D_{pp} \frac{\partial}{\partial p}\, \frac{1}{p^2}\, \psi_a \right)  
 - \frac{\partial}{\partial p} \left( \dot{p} \psi_a
 - \frac{p}{3} \, (\nabla \cdot \vec V )\psi_a \right)
 - \frac{1}{\tau_f}\psi_a - \frac{1}{\tau_r}\psi_a .
\label{DLHA_eq:transport}
\end{eqnarray}
Usually one assumes that steady state conditions hold, as they do if the typical time scales of the dark matter galactic collapse and of the variation of propagation conditions are much longer than the time scale of propagation itself (which is of the order of 1 Myr at 100 GeV energies). In this case, the l.h.s. can be equated to zero and the dependence on time is dropped for all quantities.

We concentrate on a version of the propagation equation which is sufficient to describe in first approximation the electron or positron and proton or antiproton flux through the Galaxy: 
\begin{equation}
 0 = 
 Q_a(\vec r,E) +
K(E)\ \nabla^2\psi_a +
 \frac{\partial}{\partial E}\Big(b(E)\ \psi_a - \ K_{\rm EE}(E)\ \psi_a\Big) - 
 \frac{\partial}{\partial z}\left({\rm sign}(z)\, V_C \ \psi_a\right),
\label{DLHA_evol_energy}
\end{equation}
where now $E$ is the energy of the secondary particle species $a$. Boundary conditions are imposed such that the CR density vanishes on the outer surface of the cylinder, outside of which the particles are supposed to freely propagate and escape, consistently with the physical picture described above. At $r=0$, one imposes a symmetric condition $\partial \psi_a / \partial r (r=0) = 0$. In momentum space one imposes null boundary conditions. 
The terms containing the spatial diffusion coefficient $K(E)$, the energy loss rate $b(E)$ and the diffusive reacceleration coefficient $K_{\rm EE}(E)$ describe respectively the transport of cosmic ray species through turbulent magnetic fields, their cooling due to different phenomena (such as Inverse Compton scattering (ICS), synchrotron radiation, Coulomb scattering or bremsstrahlung) and their reacceleration due to hits on moving magnetized scattering targets in the Galaxy. The term with the convective velocity $V_C$ describes the characteristics of the Galactic winds emanating vertically from the stars in the disk.  A source term resulting from dark matter annihilation can be written as
\begin{equation}
 Q_a(\vec r,E)=\frac{1}{2}\frac{dN_a}{dE}\langle\sigma_a v\rangle_0 \left(\frac{\rho_g(\vec r)}{m_\chi}\right)^2 .
\label{DLHA_source}
\end{equation}
Here $\langle\sigma_a v \rangle_0$ is the value of the thermally averaged annihilation cross section into the relevant species, and $\rho_g(r)$ is the dark matter energy density in the Galaxy.
The energy distribution of the secondary particle $a$ is $dN_a/dE$, normalized per annihilation.  This formula applies to self-conjugated annihilating dark matter. 
%
In the case of non-self-conjugated dark matter, or of multicomponent dark matter, the quantities in Eq.~(\ref{DLHA_source}) should be replaced as follows, where an index $i$ denotes a charge state and/or particle species (indeed any particle property, collectively called "component") and $f_i=n_i/n$ is the number fraction of the $i$-th component:
\begin{eqnarray}
m_\chi&  \to&   \sum_i f_i m_i \qquad\qquad\qquad\qquad\qquad
\text{(mean mass)},\nonumber \\
\langle \sigma_a v \rangle&  \to&  \sum_{ij} f_i f_j \langle
\sigma_{a,ij} v_{ij} \rangle \qquad\qquad\qquad\,\,\,\text{(mean cross
  section times relative velocity)},\nonumber \\
dN_a/dE&  \to&  \frac{\sum_{ij} f_i f_j \sigma_{a,ij} v_{ij} \, (dN_{a,ij}/dE)}{\sum_{ij} f_i f_j \sigma_{a,ij} v_{ij} } \quad \text{(annihilation spectrum per annihilation).}
\end{eqnarray}

The spatial diffusion coefficient $K(E)$ is generally taken to have the form
\begin{eqnarray}
K(E) =  K_0 \ v^\eta \left(\frac{R}{\rm GeV}\right)^\delta ,
\label{DLHA_eq:KE}
\end{eqnarray}
where $v$ is the speed (in units of $c$) and $R=p/eZ$ is the magnetic rigidity of the cosmic ray particles.  Here $Z$ is the effective nuclear charge of the particle and $e$ is the absolute value of its electric charge (of course the quantities are different from 1 only in the case in which particles other than electrons, positrons, protons or antiprotons are considered). The parameter $\eta$ controls the behavior of diffusion at low energy: recently, in departure from the traditional choice $\eta = 1$, other values (possibly negative) have been advocated. In more detailed treatments, the diffusion coefficient can be considered as space dependent ($K(\vec r, E)$) and a possible dependence on the particle direction of motion, leading to anisotropic diffusion, can be introduced. 

The energy loss rate can be parametrized as
\begin{eqnarray}
b(E) = b_0 \ E^2 .
\label{DLHA_eq:bE}
\end{eqnarray}
This form holds as long as one neglects the fact that energy losses are position dependent in the Galactic halo (e.g.~synchrotron radiation depends on the intensity of the magnetic field, which varies in the Galaxy, and ICS depends on the density of the background light distribution, which also varies in the Galaxy), and as long as one assumes that all energy loss phenomena are proportional to $E^2$ (which is true only if one neglects Coulomb losses and bremsstrahlung and one considers ICS only in the Thomson scattering regime, i.e.~at relatively low electron or positron energy). In a more detailed treatment, the energy loss rate can also be considered as space dependent, $b(\vec r, E)$, and with a more general dependence on the energy. Coulomb losses ($dE/dt \sim$ const) and bremsstrahlung losses ($dE/dt \sim bE$) can also be taken into account.  Some codes compute these using detailed formulae as a function of position and energy, based on the gas, interstellar radiation and magnetic field distributions \cite{Strong:2009xj}.  The ultimate form of the energy loss rate may be a complex function which, as will be described below, can be given under \verb|FUNCTION EnerLoss|. 

Finally, the diffusive reacceleration coefficient $K_{\rm EE}(E)$ is usually parameterized as 
\begin{eqnarray}
K_{\rm EE}(E) = \frac{2}{9} v_{\rm A}^2 \frac{v^4 \, E^2}{K(E)},
\label{DLHA_eq:KEE}
\end{eqnarray}
where $v_{\rm A}$ is the Alfv\'en speed.

A propagator, or Green's function $G$, is used to evolve the flux which originate from the source $Q$ at $\vec r_S$ with energy $E_S$ through the diffusive halo, to reach the Earth at point $\vec r$ with energy $E$.  This allows the general solution for Eq.~(\ref{DLHA_evol_energy}) to be written as
\begin{equation}
 \psi_a(\vec r,E)=\int_E^{m_\chi}dE_S \int{d^3r_S \, G(\vec r,E;\vec r_S,E_S)\, Q_a(\vec r_S,E_S)} .
\label{DLHA_energysoln}
\end{equation}
The differential flux is related to the solution in Eq.~(\ref{DLHA_energysoln}) via
\begin{equation}
 \frac{d\Phi_a}{dE}=\frac{v(E)}{4\pi}\psi_a(\vec r,E) .
\label{DLHA_eq:dPhi_a/dD}
\end{equation}
For proton or antiproton propagation in the Galactic halo, additional terms in Eq.~(\ref{DLHA_evol_energy}) should be introduced to account for spallations on the gas in the disk. 

\medskip

We next turn to photons. In this case, propagation is much simpler since no scattering processes take place. This leads to a straight line propagation without any energy loss, which, in the formalism above, corresponds to a trivial propagator. Thus, for photons $a = \gamma$, by utilising Eqs.~(\ref{DLHA_source}) and (\ref{DLHA_energysoln}), Eq.~(\ref{DLHA_eq:dPhi_a/dD}) can be approximated with
\begin{equation}
 \frac{d\Phi_\gamma}{dE}(\phi) =  \langle\sigma_\gamma v\rangle_0 \frac{dN_\gamma}{dE}\frac{1}{8\pi \, m_\chi^2}\int_0^\infty{\rho_g^2(r(s,\phi))\, ds} .
\label{DLHA_flux}
\end{equation}
Here the square of the Galactic dark matter density profile $\rho_g^2$ is integrated over the line of sight, parameterized by the coordinate $s$.  The angle $\phi$ is the aperture between the direction of the line of sight and the axis connecting the Earth to the Galactic Center. Explicitly, the coordinate $r$, centered on the Galactic Center, reads 
\begin{equation}
 r(s,\phi) = (r_\odot^2+s^2-2\,r_\odot\,s\cos\phi)^{1/2} . 
\label{DLHA_rsphi}
\end{equation}
As for Eq.~(\ref{DLHA_source}), Eq.~(\ref{DLHA_flux}) applies to self-conjugated annihilating dark matter.  
If dark matter is not composed of self-conjugated particles, and $n$ indicates the number density of particles and ${\overline n}$ the number density of antiparticles, the factor $n^2/2$ in the formula above has to be replaced by $n {\overline n}$.

This formula will depend very sensitively on the direction, especially near the galactic center. However, a given gamma-ray instrument will only measure an averaged value, smeared over the angular resolution. Therefore, a more useful quantity to compute is that average value in the direction given by $\phi$ \cite{Bringmann:2008kj},
\begin{equation}
\frac{d\Phi_ \gamma}{dE}(\phi;\Delta\Omega)= \int d\Omega' \frac{d\Phi_ \gamma}{dE}(\varphi',\theta')
R_{\Delta\Omega}(\theta'), 
\end{equation}
where
\begin{equation}
R_{\Delta\Omega}(\theta')= \frac{1}{2\pi\theta_r^2}\exp\left(\frac{-\theta'^2}{2\theta_r^2}\right), 
\end{equation}
describes the angular resolution $\theta_r$ with $d\Omega'=d\varphi' d\cos \theta'$. Here $\phi'$ and $\theta'$ are polar coordinates centered on the direction $\phi$, and $\Delta\Omega = \pi\theta_r^2$ (for small $\theta_r$). For the different, slightly varying lines of sight entering the integral, $\cos\phi$ in Eq.~(\ref{DLHA_rsphi}) is replaced by
$\cos\psi=\cos\phi\cos\theta'-\cos\varphi'\sin\phi\sin\theta'$.

\subsubsection{Dark matter substructures}

If dark matter particles are packed inside dense clumps, their annihilations are enhanced, and so are their indirect signatures at the Earth.  The boost factor by which the yield of a smooth dark matter halo has to be multiplied depends in a simple, but not obvious, way on the spatial distribution of the clumps and on their inner structure.  A population ${\cal P}$ of substructures $i$ generates at the Earth the cosmic ray density 
\begin{equation}
 \psi_a^{sub}({\vec r},E)=
 \left( {\cal S} \equiv \frac{1}{2} \langle\sigma_a v\rangle_0 \left(\frac{\rho_{\odot}}{m_\chi}\right)^2 \right) \,
 {\displaystyle \sum_{i \in {\cal P}}} \; \tilde{G}_{i} \; \xi_{i} \;\; ,
\end{equation}
where the effective propagator $\tilde{G}_{i}$, defined as
\begin{equation}
 \tilde{G}_{i}=\int_E^{m_\chi}dE_S \, G({\vec r},E;{\vec r}_i,E_S) \, \frac{dN_a}{dE} ,
\end{equation}
takes into account the propagation from clump $i$ located
at ${\vec r}_{i}$, and the injection spectrum at the source.
The annihilation volume $\xi_{i}$ would be the volume of
clump $i$ should its density be equal to the Milky Way
dark matter density $\rho_{\odot}$ at the Sun. It is
defined as the integral over the volume of the $i$-th clump
\begin{equation}
\xi_{i}=\int_{\rm clump \, i} d^3r_S \left( \frac{\rho_{\rm DM}(\vec r_S)}{\rho_{\odot}} \right)^2 .
\end{equation}
Because we have no idea of the actual population of
dark matter substructures inside which we are embedded,
a statistical analysis needs to be performed on the
ensemble of all possible realizations of galactic
clump distributions. A population of ${\cal N}_{\rm H}$
substructures inside the Milky Way dark matter halo
yields, on average, the cosmic ray density
\cite{Lavalle:2006vb, Brun:2007tn, Lavalle:1900wn}
\begin{equation}
 \langle \psi_a^{sub}({\vec r},E) \rangle=
 {\cal N}_{\rm H} \, {\cal S} \,
 \int\!d^3r_S \int\!d\xi \; {\cal D}({\vec r}_S,\xi) \, \tilde{G}({\vec r},E;{\vec r}_S) \, \xi .
\end{equation}
The propability to find a dark matter clump at location
${\vec r}_S$ with annihilation volume $\xi$ is denoted
by ${\cal D}({\vec r}_S,\xi)$. The variance associated
to the average substructure signal $\langle \psi_a^{sub}({\vec r},E) \rangle$
can be expressed as
\begin{equation}
 \sigma_{\psi}^{2}({\vec r},E)=
 {\cal N}_{\rm H} \, {\cal S}^{2} \,
 \int\!d^3r_S \int\!d\xi \; {\cal D}({\vec r}_S,\xi) \, \tilde{G}^{2}({\vec r},E;{\vec r}_S) \, \xi^{2}
 \, - \,
 {\displaystyle \frac{\langle \psi_a^{sub}({\vec r},E) \rangle^{2}}{{\cal N}_{\rm H}}} .
\end{equation}

\section{DESCRIPTION OF THE DLHA BLOCKS AND FUNCTIONS}

Information in a DLHA file is organized into blocks.  The general properties of the DLHA blocks follow those of the Supersymmetry Les Houches Accord (SLHA) \cite{Skands:2003cj}.  Similarly to SLHA, the entries within the blocks are identified by the first numerical value(s) within the block.  This feature allows for a flexible order of entries within a block.  Most block entries are optional and can be omitted at writing.  A missing entry typically signals the lack of a calculation within the program that wrote the block.

A DLHA file may contain the following blocks or statements (listed here alphabetically):
\begin{quote}
\verb|BLOCK ABUNDANCE|       \\
\verb|BLOCK ANNIHILATION|    \\
\verb|BLOCK ASTROPROPAG|     \\
\verb|BLOCK COSMOLOGY|       \\
\verb|BLOCK DETECTOR_NUCLEI| \\
\verb|BLOCK DMCLUMPS|        \\
\verb|BLOCK DMSPADIST|       \\
\verb|BLOCK DMVELDIST|       \\
\verb|BLOCK DOFREEDOM|       \\
\verb|BLOCK EFFCOUPLING|     \\
\verb|BLOCK FORMFACTS|       \\
\verb|BLOCK INDIRDETSPECTRUM|\\
\verb|BLOCK MASS|            \\
\verb|BLOCK NDMCROSSSECT|    \\
\verb|BLOCK QNUMBERS|        \\
\verb|BLOCK STRUCTFUN| 
\end{quote}
Blocks \verb|ASTROPROPAG|, \verb|COSMOLOGY|, \verb|DETECTOR_NUCLEI|, \verb|DMCLUMPS|, \verb|DMSPADIST|, \verb|DMVELDIST|, \verb|DOFREEDOM|, \verb|FORMFACTS| and \verb|STRUCTFUN| depend on the cosmological, astrophysical, standard particle and nuclear physics assumptions but are kept independent from the microscopic properties of dark matter.  
The rest of the blocks depend on the microscopic physics describing dark matter.  A given dark matter candidate is identified in the block \verb|MASS| by the PDG number and mass of the particle.  If the PDG code of a particle does not exist then an arbitrary code identifying the candidate can be supplied.  Further microscopic properties of the dark matter candidate are given in block \verb|QNUMBERS|. 

For dark matter models containing multiple dark matter candidates a separate DLHA file has to be created for each candidate.  If multiple dark matter candidates contribute simultaneously to the present abundance and direct or indirect detection signals, multiple sets of blocks may appear for each candidate in separate files.  For example, a DLHA file may contain 
\begin{quote}
\verb|BLOCK MASS| \\
\verb|#  PDG code   mass                particle name | \\
\verb|   1000022    1.29098165E+00   #  1st neutralino| \\
\verb|...|             \\
\verb|BLOCK QNUMBERS|  \\
\verb|...|             \\
\verb|BLOCK ABUNDANCE| \\
\verb|...| 
\end{quote}
while another DLHA file might contain
\begin{quote}
\verb|BLOCK MASS| \\
\verb|#  PDG code   mass                particle name | \\
\verb|   1000039    2.35019093E+00   #  gravitino     | \\
\verb|BLOCK QNUMBERS|  \\
\verb|...|             \\
\verb|BLOCK ABUNDANCE| \\
\verb|...| 
\end{quote}
Here the ellipses denote entries irrelevant to this discussion.  A user of DLHA is responsible for knowing that the two files contain information in the context of multiple dark matter candidates. 

For decaying dark matter particles a standard SLHA decay file can be used to read and write the total decay width of the dark matter particle and its branching ratios into various final states.

\subsection{The FUNCTION object}

Departing from the tradition of previous Les Houches accords, DLHA introduces a new structure that specifies a function.  A function definition is facilitated by DLHA using the following construction
\begin{quote}
\verb|FUNCTION <name> type=<type> args=<number of arguments>| \\
\verb| ...                                                  | \\
\verb|END_FUNCTION| 
\end{quote}
The \verb|FUNCTION| heading denotes the beginning and \verb|END_FUNCTION| the end of the structure.  Each function is identified by a name, which follows the \verb|FUNCTION| heading.  In the various block descriptions below we fix the names of the possible functions and specify their content.  The content of a function can be given in several different ways in DLHA.  The \verb|<type>| variable differentiates between these methods:
\begin{description}
\item[] \verb|type = P| for a predefined function,
\item[] \verb|type = C| for a C function,
\item[] \verb|type = F| for a Fortran function,
\item[] \verb|type = T| for tabular information.
\end{description}
The number of independent variables of the function is given by the numerical value of the last argument \verb|<number of arguments>| of the \verb|FUNCTION| structure.  

While a \verb|FUNCTION| depends on one or more independent variables, it may also carry information about related parameters inside the body of the function.  These parameters can be listed as follows:
\begin{quote}
\verb|FUNCTION <name> type=<type> args=<number of arguments>| \\
\verb| PARAMETERS                                           | \\
\verb|  <parameter name 1>=<value 1>                        | \\
\verb|  <parameter name 2>=<value 2>                        | \\
\verb|  <parameter name 3>=<value 3>                        | \\
\verb|  ...                                                 | \\
\verb| END_PARAMETERS                                       | \\
\verb| <function body>                                      | \\
\verb|END_FUNCTION                                          | 
\end{quote}
Parameter names are fixed by DLHA in the block descriptions similarly to names of functions.  If a parameter value is specified both outside and inside of a function, the value given inside the FUNCTION construction overrides the one appearing outside. 

Predefined functions, typically the most commonly used functions for a given quantity, are specified by DLHA under the description of the various functions.  A predefined function which is listed in the DLHA write-up can be referred to by the following construction following the \verb|FUNCTION| heading:
\begin{quote}
\verb|FUNCTION <name> type=<type> args=<number of arguments>| \\
\verb| DLHA <name> <identifier>                             | \\
\verb| ...                                                  | \\
\verb|END_FUNCTION| 
\end{quote}
Within the block descriptions DLHA fixes the function choices corresponding to various numerical values of identifiers.

As an example, the following structure specifies the use of the Einasto profile for the dark matter galactic halo profile $\rho_g(r)$:
\begin{quote}
\verb|FUNCTION rho_g type=P args=2| \\
\verb| DLHA rho_g 5               | \\
\verb|END_FUNCTION                |
\end{quote}
The Einasto profile depends on two arguments.  One of them may also be fixed as:
\begin{quote}
\verb|FUNCTION rho_g type=P args=1| \\
\verb| DLHA rho_g 5               | \\
\verb| PARAMETERS                 | \\
\verb|  alpha=1                   | \\
\verb| END_PARAMETERS             | \\
\verb|END_FUNCTION                |
\end{quote}

In the present agreement, we only consider functions tabulated on a rectangular (but not necessary equidistant) grid.  In this case the list of the independent variables and the function value is given as an \verb|n|+1 column table.  For tabular functions the line after the function name gives the names of the independent variables and the dimensions of the rectangular grid on which the function is specified.  A schematic example of a function given in tabular format is the following:
\begin{quote}
\verb|FUNCTION rho_g type=T args=2                |\\
\verb| r: 2  alpha: 3                             |\\
\verb|#  r             alpha         rho_g        |\\
\verb|   0.000000E+00  1.000000E+00  2.345678E+00 |\\
\verb|   1.000000E-05  1.000000E+00  1.234567E+00 |\\
\verb|   0.000000E-05  2.000000E+00  4.567890E+00 |\\
\verb|   1.000000E-05  2.000000E+00  3.456789E+00 |\\
\verb|   0.000000E-05  3.000000E+00  6.789012E+00 |\\
\verb|   1.000000E-05  3.000000E+00  5.678901E+00 |\\
\verb|END_FUNCTION                                |
\end{quote}

A \verb|FUNCTION| given in the form of a C language function could appear as:
\begin{quote}
\verb|FUNCTION rho_g type=C args=2              |\\
\verb| #include<math.h>                         |\\
\verb| double Einasto(double r, double alpha)   |\\
\verb| { return exp(-2*(pow(r,alpha)-1)/alpha);}|\\
\verb|END_FUNCTION                              |
\end{quote}

The \verb|FUNCTION| construction also allows passing function names that are included in a pre-compiled library:
\begin{quote}
\verb|FUNCTION <name> type=<type> args=<number of arguments>|\\
\verb| libName=<name of compiled library>                   |\\
\verb| funcName=<name of function in library>               |\\
\verb|END_FUNCTION                              |
\end{quote}
An example of this is shown in Section \ref{DLHA_subsec:BDN}.


\subsection{BLOCK COSMOLOGY}

The \verb|COSMOLOGY| block specifies the values of the cosmological parameters that enter into the calculation of dark matter related observables such as the abundance.  The cosmology block may contain numerical entries such as those below.  In these entries, the temperature is given in units of GeV$/k$, where $k$ is the Boltzmann constant.

\begin{description}
\item[1] The value of the current entropy density of the Universe $s_0$ in units of $({\rm GeV}/\hbar c)^3/k$, appearing in Eq.~(\ref{DLHA_omegachi}).
\end{description}

The cosmology block also accommodates various entries that are defined as functions as described in the previous subsection.  The list of these functions is given below.
 
Non-standard energy density  $\rho_{D}(T)$, as described in Eq.~(\ref{DLHA_update1}), as a function of the temperature and in units of $({\rm GeV}/\hbar c)^4$.
\begin{quote}
\verb|FUNCTION rho_D type=<type> args=1|\\
\verb| ...                             |\\
\verb|END_FUNCTION                     |
\end{quote}

Non-standard entropy density  $s_{D}(T)$, as described in Eq.~(\ref{DLHA_update1}), as a function of the temperature and  in units of $({\rm GeV}/\hbar c)^3$.
\begin{quote}
\verb|FUNCTION s_D type=<type> args=1|\\
\verb| ...                           |\\
\verb|END_FUNCTION                   |
\end{quote}

Non-standard normalized entropy production rate $\Sigma_{D}(T)/\sqrt{G}$, as described in Eq.~(\ref{DLHA_update1}), as a function of the temperature and in units of $({\rm GeV}/\hbar c)^5$.
\begin{quote}
\verb|FUNCTION Sigma_D type=<type> args=1|\\
\verb| ...                               |\\
\verb|END_FUNCTION                       |
\end{quote}

The value of the normalized Hubble expansion rate $H(T)/\sqrt{G}$, as per Eq.~(\ref{DLHA_eq:def_h}), as a function of the temperature.  The normalized Hubble expansion rate is in units of $({\rm GeV}/\hbar c)^2$.
\begin{quote}
\verb|FUNCTION Hubble type=<type> args=1|\\
\verb| ...                              |\\
\verb|END_FUNCTION                      |
\end{quote}

Note that the functions \verb|s_D| and \verb|Sigma_D| are not to be given simultaneously, as they refer to two different parameterizations of the entropy content of the Universe. Also, the functions \verb|rho_D| and \verb|Hubble| do not need to be given simultaneously.

A specific cosmology block, describing a standard cosmological scenario may appear as this:
\begin{quote}
\verb|BLOCK COSMOLOGY| \\
\verb|# identifier(s)  parameter value   comment         | \\
\verb|  1              2.18421585E-03  # s_0 [GeV^3]     | \\
\verb|# -------------------------------------------------| \\
\verb|FUNCTION rho_D type=C args=1     # rho_D(T) [GeV^4]| \\
\verb| ...                                               | \\
\verb|END_FUNCTION                                       | \\
\verb|FUNCTION s_D type=F args=1       # s_D(T) [GeV^3]  | \\
\verb| ...                                               | \\
\verb|END_FUNCTION                                       | 
\end{quote}

\subsection{BLOCK DOFREEDOM}

This block contains the various degrees of freedom entering into the abundance calculation.

The effective degrees of freedom $g_{eff}(T)$ in Eq.~(\ref{DLHA_energydens}) as a function of the temperature are given by   
\begin{quote}
\verb|FUNCTION g_eff type=<type> args=1         |\\
\verb| ...                                      |\\
\verb|END_FUNCTION                              |
\end{quote}
The units of temperature are GeV$/k$.

As an alternative to $g_{eff}(T)$, as a function of the temperature, $g_\star$ can also be given.  The latter is defined in Eq.~(\ref{DLHA_eq:gstar}).  
\begin{quote}
\verb|FUNCTION gstar type=<type> args=1         |\\
\verb| ...                                      |\\
\verb|END_FUNCTION                              |
\end{quote}

The effective degrees of freedom $h_{eff}(T)$ in Eq.~(\ref{DLHA_entropy}) as a function of the temperature (measured in GeV$/k$) are given by
\begin{quote}
\verb|FUNCTION h_eff type=<type> args=1         |\\
\verb| ...                                      |\\
\verb|END_FUNCTION                              |
\end{quote}

This block containing 57 entries of $g_{eff}$ and 48 entries for $h_{eff}$ in tabular format might appear as follows.  
\begin{quote}
\verb|BLOCK DOFREEDOM| \\
\verb|FUNCTION g_eff type=T args=1     | \\
\verb| T: 57                           | \\
\verb|# T [GeV/k]         g_{eff}(T)   | \\
\verb|  1.00000000E-05  2.00000000E-00 | \\
\verb|  5.00000000E-05  2.00000000E-00 | \\
\verb|  1.00000000E-04  4.00000000E-00 | \\
\verb|...                              | \\
\verb|END_FUNCTION                     | \\
\verb|FUNCTION h_eff type=T args=1     | \\
\verb| T: 48                           | \\
\verb|# T [GeV/k]         h_{eff}(T)   | \\
\verb|  2.00000000E-05  1.00000000E-00 | \\
\verb|  4.00000000E-05  1.00000000E-00 | \\
\verb|  6.00000000E-05  2.00000000E-00 | \\
\verb|...                              | \\
\verb|END_FUNCTION                     | 
\end{quote}

\subsection{BLOCK DMSPADIST}

Block \verb|DMSPADIST| contains information regarding the dark matter energy density distribution in various astrophysical objects, such as galaxies or nebulae.  These enter the calculation of direct and indirect detection rates.  
Before discussing the block attributes, we list the most commonly used halo profiles and give their corresponding mathematical expressions.  
%
%
Each profile has a set of parameters which are dependent on the particular halo of interest.  Unless otherwise stated $\rho_0$ sets the normalization of the profile and $r_s$ is the scale radius.

Hernquist-Zhao density profile:
\begin{equation}
\rho_g(r)=\frac{\rho_0}{(r/r_s)^\gamma\left(1+(r/r_s)^{(\alpha-\gamma)/\beta}\right)^\beta} .
\label{DLHA_eq:generalnfw}
\end{equation}

The Navarro-Frenk-White (NFW) distribution is a special case of this with $\alpha = 3$, $\beta = 2$, and $\gamma = 1$~\cite{Navarro:1995iw}.
The Kravtsov et al.~profile can also be obtained from Eq.~(\ref{DLHA_eq:generalnfw}) by setting $\alpha = 3$ and $\beta = 2$~\cite{kra}. 

Modified isothermal profile \cite{isothermal}:
\begin{equation}
\rho_g(r)=\frac{\rho_0}{1+(r/r_s)^2} .
\label{DLHA_eq:isotherm}
\end{equation}

Einasto profile \cite{eina}:
\begin{equation}
 \rho_g(r) = 
 \rho_0 \exp\left( -\frac{2}{\alpha} \left( \frac{r^\alpha}{r_s^\alpha} -1 \right) \right) .
\label{DLHA_eq:Einasto}
\end{equation}

Moore et al.~profile \cite{mooreetal}:
\begin{equation}
\rho_g(r)=\frac{\rho_0}{(r/r_s)^{3/2}\left(1+r/r_s\right)^{3/2}} .
\label{DLHA_eq:Moore}
\end{equation}

Burkert profile~\cite{burkert}:
\begin{equation}
\rho_g(r)=\frac{\rho_0 r_s^3}{(r+r_s)(r^2+r_s^2)} .
\label{DLHA_eq:Burkert}
\end{equation}

\subsubsection{The structure of BLOCK DMSPADIST}

Since dark matter energy density distributions can be specified for various astrophysical objects, such as external galaxies and halo substructures, more than one \verb|BLOCK DMSPADIST| may appear in a DLHA file.  In this case multiple \verb|BLOCK DMSPADIST| blocks carry an index differentiating them from each other.  
For example
\begin{quote}
\verb|BLOCK DMSPADIST_MilkyWay| \\
\verb|... | \\
\verb|    | \\
\verb|BLOCK DMSPADIST_NGC6388| \\
\verb|... | \\
\verb|    | \\
\verb|BLOCK DMSPADIST_VIRGOHI21| \\
\verb|... | 
\end{quote}

The shape of the dark matter density distribution $\rho_g(r)$, featured in Eqs.~(\ref{DLHA_source}) and (\ref{DLHA_flux}), is defined by the function
\begin{quote}
\verb|FUNCTION rho_g type=<type> args=<number of arguments>|\\
\verb| ...                                                 |\\
\verb|END_FUNCTION                                         |
\end{quote}
The galactic halo profile is assumed to be given in units of GeV/$c^2$/cm$^3$, and its radial argument in units of kpc.

Identifiers for numerical parameters featured in this function for predefined density profiles are the following: 
\begin{description}
\item[1] the normalization of the profile $\rho_0$ in units of GeV/$c^2$/cm$^3$,
\item[2] the scale radius $r_s$ in units of kpc,
\item[3] $\alpha$ for the Hernquist-Zhao and Einasto profiles,
\item[4] $\beta$ for the Hernquist-Zhao  profile,
\item[5] $\gamma$ for the Hernquist-Zhao  profile,
\item[6] the value of the dark matter mass density near Earth $\rho_\oplus$ in units of GeV/$c^2$/cm$^3$, 
\item[7] our Galactocentric distance, $R_0$ in units of kpc. 
\end{description}
These variables can also be used within predefined functions as \verb|PARAMETERS| with the following names
\begin{description}
\item[] \verb|rho_0    |: $\rho_0$      ,
\item[] \verb|r_s      |: $r_s$         ,
\item[] \verb|alpha    |: $\alpha$      ,
\item[] \verb|beta     |: $\beta$       ,
\item[] \verb|gamma    |: $\gamma$      ,
\item[] \verb|rho_oplus|: $\rho_\oplus$ ,
\item[] \verb|R_0      |: $R_0$         . 
\end{description}

Predefined values of the most commonly used halo distributions are
\begin{description}
\item[] \verb|DLHA rho_g 1|: Hernquist-Zhao profile, as given in Eq. (\ref{DLHA_eq:generalnfw}),
\item[] \verb|DLHA rho_g 2|: NFW density profile, defined by Eq. (\ref{DLHA_eq:generalnfw}),
\item[] \verb|DLHA rho_g 3|: Kravtsov et al. profile, defined by Eq. (\ref{DLHA_eq:generalnfw}),
\item[] \verb|DLHA rho_g 4|: Modified isothermal profile, as given in Eq. (\ref{DLHA_eq:isotherm}),
\item[] \verb|DLHA rho_g 5|: Einasto profile, as given in Eq. (\ref{DLHA_eq:Einasto}),
\item[] \verb|DLHA rho_g 6|: Moore et al. profile, as given in Eq. (\ref{DLHA_eq:Moore}),
\item[] \verb|DLHA rho_g 7|: Burkert profile, as given in Eq. (\ref{DLHA_eq:Burkert}).
\end{description}

For example, in a DLHA file a predefined NFW profile can be referred to as
\begin{quote}
\verb|# identifier(s)  parameter value    comment               | \\
\verb|  1              0.385954823E+00  # rho_0 [GeV/cm^3]      | \\
\verb|  2              2.000000000E+01  # scale radius r_s [kpc]| \\
\verb|  3              3.000000000E+00  # parameter $\alpha$    | \\
\verb|  4              2.000000000E+00  # parameter $\beta$     | \\
\verb|  5              1.000000000E+00  # parameter $\gamma$    | \\
\verb|# --------------------------------------------------------| \\
\verb|FUNCTION rho_g type=P args=1                              | \\
\verb| DLHA rho_g 2 # NFW profile                               | \\
\verb|END_FUNCTION                                              |
\end{quote}
Alternatively, the following definition can also be used:
\begin{quote}
\verb|FUNCTION rho_g type=P args=1| \\
\verb| DLHA rho_g 2 # NFW profile | \\
\verb| PARAMETERS                 | \\
\verb|  rho_0 = 0.385954823E+00   | \\
\verb|  r_s   = 2.000000000E+01   | \\
\verb|  alpha = 3.000000000E+00   | \\
\verb|  beta  = 2.000000000E+00   | \\
\verb|  gamma = 1.000000000E+00   | \\
\verb| END_PARAMETERS             | \\
\verb|END_FUNCTION                |
\end{quote}

A sample \verb|BLOCK DMSPADIST| for a predefined Kravtsov et al.~halo profile and a tabulated velocity distribution may appear as follows:
\begin{quote}
\verb|BLOCK DMSPADIST_MilkyWay| \\
\verb|# identifier(s)  parameter value    comment              | \\
\verb|  1              0.385954823E+00  # rho_0 [GeV/cm^3]     | \\
\verb|  2              2.000000000E+01  # r_s [kpc]            | \\
\verb|  3              3.000000000E+00  # alpha                | \\
\verb|  4              2.000000000E+00  # beta                 | \\
\verb|  6              0.385954821E+00  # rho_Earth [GeV/cm^3] | \\
\verb|# -------------------------------------------------------| \\
\verb|FUNCTION rho_g type=P args=1      # predefined Kravtsov  | \\
\verb| DLHA rho_g 3                                            | \\
\verb|END_FUNCTION                                             | 
\end{quote}

\subsection{BLOCK DMVELDIST}

This block contains information about velocity distributions of dark matter needed in direct and indirect detection calculations.  
%
%
%
Velocities are given using the Galactic coordinate system, described
below. 
Before discussing the block attributes, we specify common velocity distributions.
%
%

Truncated Maxwellian distribution:
\begin{equation}
\widetilde{f}(v) = \frac{4\pi v^2}{N_{esc}  (2\pi \sigma_v^2)^{3/2}} \, e^{-v^2/2\sigma_v^2} \, \theta(v_{esc}-v) \, 
\label{DLHA_eq:TMaxwellian}
\end{equation}
\indent\indent where $N_{esc} = \mathop{\rm erf}(z)-2ze^{-z^2}/\sqrt{\pi}$ with $z=v_{esc}/\sqrt{2}\sigma_v$.

\indent Subtracted Maxwellian distribution:
\begin{equation}
\widetilde{f}(v) = \frac{4\pi v^2}{N_{esc} (2\pi \sigma_v^2)^{3/2}} \,  \left[ e^{-v^2/2\sigma_v^2} - e^{-v_{esc}^2/2\sigma_v^2}\right] \, \theta(v_{esc}-v) \, 
\label{DLHA_eq:SMaxwellian}
\end{equation}
\indent\indent where $N_{esc} = \mathop{\rm erf}(z)-2z(1+2z^2/3)e^{-z^2}/\sqrt{\pi}$ with $z=v_{esc}/\sqrt{2}\sigma_v$.

Here $N_{esc}$ is a normalization factor such that $\int_0^\infty \widetilde{f}(v) \,dv=1$, $v_{esc}$ is the escape speed (a finite cutoff in the distribution expected due to high-speed dark matter being able to escape from the object's gravitational potential), and $\sigma_v$ is a velocity dispersion parameter.  In the large $v_{esc}$ limit, the distributions in Eqs.~(\ref{DLHA_eq:TMaxwellian}) and (\ref{DLHA_eq:SMaxwellian}) reduce to the same Maxwellian distribution with most probable speed $v_0=\sqrt{2}\, \sigma_v$, 1D velocity dispersion $\sigma_v$, and 3D velocity dispersion $\sqrt{3} \, \sigma_v$.

For direct detection, one needs the local dark matter velocity distribution multiplied by the local dark matter density.  The latter is given in the \verb|DMSPADIST| block.  Here we address the velocity distribution.  Notice that if the local dark matter density is split into different velocity groups, each group has its own block 
\verb|DMVELDIST|.

For direct detection (Eq.~\ref{DLHA_eq:etaEt}), one needs the dark matter velocity distribution $f(\mathbf{v},t)$ in the rest frame of the detector. Its time dependence is expected to arise from the motion of the Earth around the Sun and about its axis. Since these motions are known, $f(\mathbf{v},t)$ can be obtained through a Galilean transformation from the Sun's rest frame,
\begin{equation}
f(\mathbf{v},t) = f_{\odot}(\mathbf{v}_{\rm det} + \mathbf{v}).
\label{DLHA_eq:shiftedfv}
\end{equation}
Here  $f_{\odot}(\mathbf{u})$ is the dark matter heliocentric velocity distribution, which is a function of the dark matter velocity $\mathbf{u}$ with respect to the Sun (heliocentric velocity), and $\mathbf{v}_{\rm det}$ is the detector velocity with respect to the Sun given by $ \mathbf{v}_{\rm det} = \mathbf{v}_{\oplus\rm rev}+{\bf v}_{\oplus\rm rot}$, where $\mathbf{v}_{\oplus\rm rev}$ is the (time-varying) velocity of the Earth revolution relative to the Sun, and ${\bf v}_{\oplus\rm rot}$ is the (time-varying) velocity of the Earth rotation at the location of the detector relative to the Earth's rest frame. 

Enough information should be given in \verb|BLOCK DMVELDIST| to
reconstruct the three-dimensional heliocentric velocity distribution 
$f_{\odot}(\mathbf{u})$. This is achieved by giving the velocity
distribution function $\widetilde{f}(\mathbf{v})$ in some specified but
otherwise arbitrary reference frame $\widetilde{S}$, together with the
velocity $\widetilde{\mathbf{v}}$ of $\widetilde{S}$ with respect to the
Sun, i.e.
\begin{equation}
f_{\odot}(\mathbf{u}) = \widetilde{f}(\mathbf{u}-\widetilde{\mathbf{v}}) .
\label{DLHA_eq:shiftedfvS}
\end{equation}
Though not required (except in a case described below), the reference
frame $\widetilde{S}$ will typically be chosen such that the form for
$\widetilde{f}(\mathbf{v})$ becomes more simplified, as occurs in the
rest frame of an isotropic velocity distribution.

\subsubsection{The structure of BLOCK DMVELDIST}

Similarly to the energy density, dark matter velocity distributions can be specified for various astrophysical objects: galaxies, halo substructures, or even different local velocity components.  Thus, more than one \verb|BLOCK DMVELDIST| may appear in a DLHA file.  Multiple \verb|BLOCK DMVELDIST| blocks are differentiated from each other as
\begin{quote}
\verb|BLOCK DMVELDIST_MilkyWay| \\
\verb|... | \\
\verb|    | \\
\verb|BLOCK DMVELDIST_NGC6388| \\
\verb|... | \\
\verb|    | \\
\verb|BLOCK DMVELDIST_VIRGOHI21| \\
\verb|... | 
\end{quote}


The function $\widetilde{f}(\mathbf{v})$ is given by
\begin{quote}
\verb|FUNCTION fv_g type=<type> args=<number of arguments>|\\
\verb| ...                                                |\\
\verb|END_FUNCTION                                        |
\end{quote}


The units of $\widetilde{f}(\mathbf{v})$  are assumed to be $({\rm km/s})^{-n}$, where $n$ is the number of arguments. 
Two different conventions apply depending on whether $\widetilde{f}$ is presented as a function of one or more than one variable, i.e., depending on whether $n=1$ or $n>1$. 
If $n>1$, $\widetilde{f}(\mathbf{v})$  must be normalized so that $\int \widetilde{f}(\mathbf{v})  \, d^nv=1$. 
%
%
If $n=1$, i.e., if $\widetilde{f}(v)$ is a function of only one variable $v$, it is assumed that (i)  the frame $\widetilde{S}$ coincides with the frame in which the average velocity is zero, (ii)  in this frame the velocity distribution is isotropic, (iii)  the variable $v=|{\bf v}|$ is positive and represents the speed, i.e., the magnitude of the velocity, in the frame $\widetilde{S}$, and (iv)  $\widetilde{f}(v)$ is the speed distribution in $\widetilde{S}$, normalized so that $\int_0^\infty \widetilde{f}(v) \, dv = 1$.

Predefined values of the most commonly used velocity distributions are
\begin{description}
\item[] \verb|DLHA fv_g 1|: Truncated Maxwellian distribution, as given in Eq.~(\ref{DLHA_eq:TMaxwellian}),
\item[] \verb|DLHA fv_g 2|: Subtracted Maxwellian distribution, as given in Eq.~(\ref{DLHA_eq:SMaxwellian}).
\end{description}
Names of numerical \verb|PARAMETERS| featured in these predefined functions are the following: 
\begin{description}
\item[] \verb|sigma_v|: the velocity dispersion parameter $\sigma_v$ in units of km/s,
\item[] \verb|v_esc  |: the escape speed $v_{esc}$ in units of km/s.
\end{description}
These parameters can also be given before the \verb|FUNCTION| definition with the following identifiers
\begin{description}
\item[1] $\sigma_v$ in units of km/s,
\item[2] $v_{esc}$ in units of km/s.
\end{description}

The  $\widetilde{S}$ heliocentric velocity vector $\widetilde{\mathbf{v}}$ is to be given by its components in the Galactic reference frame. The Galactic reference frame is a right-handed Cartesian coordinate system $xyz$ with $x$ axis in the direction of the Galactic Center $(l=0,b=0)$,  $y$ axis  in the direction of the Galactic rotation $(l=90^\circ,b=0)$, and  $z$ axis in the direction of the Galaxy's axis of rotation (North Galactic Pole $b=90^\circ$). The  velocity $\widetilde{\mathbf{v}}$ is thus given by the following parameters.
\begin{description}
\item[11] the  $x$ component $\widetilde{v}_{x}$ of the $\widetilde{S}$ heliocentric velocity in units of km/s,
\item[12] the  $y$ component $\widetilde{v}_{y}$ of the $\widetilde{S}$ heliocentric velocity in units of km/s,
\item[13] the  $z$ component $\widetilde{v}_{z}$ of the $\widetilde{S}$ heliocentric velocity in units of km/s.
\end{description}

A sample \verb|BLOCK DMVELDIST| for a predefined truncated Maxwellian distribution may appear as follows:
\begin{quote}
\verb|BLOCK DMVELDIST_standard_dark_halo| \\
\verb|# identifier(s)  parameter value    comment        | \\
\verb|  1              1.555634918E+02  # sigma_v [km/s] | \\
\verb|  2              6.500000000E+02  # v_esc [km/s]   | \\
\verb|  11             0.000000000E+00  # vframe_x [km/s]| \\
\verb|  12            -2.200000000E+02  # vframe_y [km/s]| \\
\verb|  13             0.000000000E+00  # vframe_z [km/s]| \\
\verb|# -------------------------------------------------| \\
\verb|FUNCTION fv_g type=P args=1 # predefined v distr.  | \\
\verb| DLHA fv_g 1                # truncated Maxwellian | 
\end{quote}

A sample \verb|BLOCK DMVELDIST| for a tabulated velocity distribution may appear as follows:
\begin{quote}
\verb|BLOCK DMVELDIST_MilkyWay| \\
\verb|  1              1.555634918E+02  # sigma_v [km/s]| \\
\verb|  2              6.500000000E+02  # v_esc [km/s]  | \\
\verb|  11             0.000000000E+00  # vframe_x [km/s]| \\
\verb|  12            -2.200000000E+02  # vframe_y [km/s]| \\
\verb|  13             0.000000000E+00  # vframe_z [km/s]| \\
\verb|FUNCTION fv_g type=T args=1       # fv_g tabulated| \\
\verb| v: 57                                            | \\
\verb|# v [km/s]        f(v) [s/km]                     | \\
\verb|  5.00000000E+01  1.20000000E-01                  | \\
\verb|  7.50000000E+01  2.40000000E-01                  | \\
\verb|  1.00000000E+02  4.90000000E-01                  | \\
\verb|  ...                                             | \\
\verb|END_FUNCTION                                      | 
\end{quote}

\subsection{BLOCK FORMFACTS}


The nucleon form factors $f^N$ 
and $\Delta_q^N$ 
as introduced in Eqs.~(\ref{DLHA_nuc1form}) and (\ref{DLHA_nuc2form}).  
The corresponding identifiers are:
\begin{description}
\item[1~~~~~1]      up quark scalar form factor  for the proton $f_u^p$,
\item[1~~~~~2]    down quark scalar form factor  for the proton $f_d^p$,
\item[1~~~~~3] strange quark scalar form factor  for the proton $f_s^p$,
\item[1~~~~~4]   heavy quark scalar form factors for the proton $f_Q^p$,
\item[2~~~~~1]      up quark scalar form factor  for the neutron $f_u^n$,
\item[2~~~~~2]    down quark scalar form factor  for the neutron $f_d^n$,
\item[2~~~~~3] strange quark scalar form factor  for the neutron $f_s^n$,
\item[2~~~~~4]   heavy quark scalar form factors for the neutron $f_Q^n$,
\item[3~~~~~1]      up quark vector form factor for the proton $f_{V_u}^p$,
\item[3~~~~~2]    down quark vector form factor for the proton $f_{V_d}^p$,
\item[4~~~~~1]      up quark vector form factor for the neutron $f_{V_u}^n$,
\item[4~~~~~2]    down quark vector form factor for the neutron $f_{V_d}^n$,
\item[5~~~~~1]      up quark axial-vector form factor for the proton $\Delta_u^p$,
\item[5~~~~~2]    down quark axial-vector form factor for the proton $\Delta_d^p$,
\item[5~~~~~3] strange quark axial-vector form factor for the proton $\Delta_s^p$,
\item[6~~~~~1]      up quark axial-vector form factor for the neutron $\Delta_u^n$,
\item[6~~~~~2]    down quark axial-vector form factor for the neutron $\Delta_d^n$,
\item[6~~~~~3] strange quark axial-vector form factor for the neutron $\Delta_s^n$,
\item[7~~~~~1]      up quark $\sigma_{\mu\nu}$ form factor for the proton $\delta_u^p$,
\item[7~~~~~2]    down quark $\sigma_{\mu\nu}$ form factor for the proton $\delta_d^p$,
\item[7~~~~~3] strange quark $\sigma_{\mu\nu}$ form factor for the proton $\delta_s^p$,
\item[8~~~~~1]      up quark $\sigma_{\mu\nu}$ form factor for the neutron $\delta_u^n$,
\item[8~~~~~2]    down quark $\sigma_{\mu\nu}$ form factor for the neutron $\delta_d^n$,
\item[8~~~~~3] strange quark $\sigma_{\mu\nu}$ form factor for the neutron $\delta_s^n$.
\end{description}

A specific \verb|FORMFACTS| block for a spin-zero dark matter candidate may look like this:
\begin{quote}
\verb|BLOCK FORMFACTS| \\
\verb|# identifier(s)  value              comment| \\
\verb|  1     1        2.30000000E-02  #  f_u^p  | \\
\verb|  1     2        3.30000000E-02  #  f_d^p  | \\
\verb|  1     3        2.60000000E-01  #  f_s^p  | \\
\verb|  2     1        1.80000000E-02  #  f_u^n  | \\
\verb|  2     2        4.20000000E-02  #  f_d^n  | \\
\verb|  2     3        2.60000000E-01  #  f_s^n  | 
\end{quote}

\subsection{BLOCK STRUCTFUN}

The nuclear structure functions $F_A(E)$ and $S_{ij}(E)$ are defined in Eqs.~(\ref{DLHA_sigmanSI}) and (\ref{DLHA_eq:F_A}) for SI interactions, and Eqs.~(\ref{DLHA_sigmanSD}) and (\ref{DLHA_Ss1}-\ref{DLHA_Ss3}) for SD interactions.  They may be given in the form of functions of the recoil energy $E$, such as
\begin{quote}
\verb|FUNCTION F_A type=<type> args=1| \\
\verb|...| \\
\verb|END_FUNCTION| 
\end{quote}
and 
\begin{quote}
\verb|FUNCTION S_ij type=<type> args=1| \\
\verb|...| \\
\verb|END_FUNCTION| 
\end{quote}
The nuclear form factors should carry no unit, while the transferred energy $E$  is in units of keV. For reference, the transferred energy $E$ is related to the momentum transfer $q$ through $E=q^2/(2m_A)$, where $m_A$ is the nucleus mass.

The following common parametrizations of $F_A$ are used.  The exponential form factor defined as
\begin{equation}
 F_A(q r_n) = e^{-\alpha (q r_n)^2 / 2} ~~~ {\rm with} ~~~ 
 r_n = a_n A^{1/3} + b_n .
\label{DLHA_eq:F_A_exp}
\end{equation} 
The names of the corresponding parameters that appear in the \verb|PARAMETERS| list are
\begin{description}
\item[] \verb|A    |: $A$ the mass number of the nucleus   ,
\item[] \verb|Z    |: $Z$ the atomic number of the nucleus ,
\item[] \verb|alpha|: $\alpha$                             ,
\item[] \verb|a_n  |: $a_n$ in units of fm                 ,
\item[] \verb|b_n  |: $b_n$ in units of fm                 .
\end{description}

The Helm form factor
\begin{equation}
 F_A(q r_n) = 3 \frac{j_1(q r_n)}{q r_n} e^{-(q s)^2/2} ~~~ {\rm with} ~~~
 r_n^2 = c^2 + \frac{7}{3} \pi^2 a^2 - 5 s^2 ~~~ {\rm and} ~~~
 c = c_0 A^{1/3} + c_1 .
\end{equation} 
\label{DLHA_eq:F_A_Helm}
Helm specific names for \verb|PARAMETERS| are
\begin{description}
\item[] \verb|A  |: $A$ the mass number of the nucleus  ,
\item[] \verb|Z  |: $Z$ the atomic number of the nucleus ,
\item[] \verb|c_0|: $c_0$ in units of fm                ,
\item[] \verb|c_1|: $c_1$ in units of fm                ,
\item[] \verb|a  |: $a  $ in units of fm                ,
\item[] \verb|s  |: $s  $ skin thickness in units of fm .
\end{description}

The Fermi distribution
\begin{equation}
 F_A(q) =  \int \rho(r) e^{ i {\mathbf q} \cdot {\mathbf r} } d^3r 
 ~~~ {\rm with} ~~~
 \rho(r) = \rho_0 ( 1 + e^{(r-c)/a} )^{-1} ~~~ {\rm and} ~~~
 c = c_0 A^{1/3} + c_1 .
\label{DLHA_eq:F_A_Fermi}
\end{equation} 
Corresponding names of \verb|PARAMETERS| are
\begin{description}
\item[] \verb|A    |: $A$ the mass number of the nucleus ,
\item[] \verb|Z    |: $Z$ the atomic number of the nucleus ,
\item[] \verb|rho_0|: $\rho_0$                           ,
\item[] \verb|c_0  |: $c_0$ in units of fm               ,
\item[] \verb|c_1  |: $c_1$ in units of fm               ,
\item[] \verb|a    |: $a  $ in units of fm               .
\end{description}

Corresponding to these in DLHA the following predefined $F_A$ form factors can be used:
\begin{description}
\item[] \verb|DLHA F_A 1|: exponential form factor,
\item[] \verb|DLHA F_A 2|: Helm form factor       ,
\item[] \verb|DLHA F_A 3|: Fermi distribution     . 
\end{description}

Examples of specifying these form factors are:
\begin{quote}
\verb|# Exponential form factor:                   |\\
\verb|#   F(q r_n) = e^{-\alpha (q r_n)^2 / 2)}    |\\
\verb|#   r_n = a_n A^{1/3} + b_n                  |\\
\verb|FUNCTION F_A type=P args=1                   |\\
\verb| DLHA F_A 1                # Exponential form factor|\\
\verb| PARAMETERS                                  |\\
\verb|  Z = 11                                     |\\
\verb|  A = 23                                     |\\
\verb|  alpha = 0.20000000E+00   #                 |\\
\verb|  a_n   = 1.15000000E+00   # [fm]            |\\
\verb|  b_n   = 0.39000000E+00   # [fm]            |\\
\verb| END_PARAMETERS                              |\\
\verb|END_FUNCTION                                 |\\
\verb|                                             |\\
\verb|# General spin-independent form factor (Helm):        |\\
\verb|#  F(q r_n) = 3 \frac{j_1(q r_n)}{q r_n} e^{-(qs)^2/2}|\\
\verb|#  r_n^2 = c^2 + \frac{7}{3} \pi^2 a^2 - 5 s^2        |\\
\verb|#  c = c_0 A^{1/3} + c_1                              |\\
\verb|FUNCTION F_A type=P args=1                            |\\
\verb| DLHA F_A 2             # Helm form factor            |\\
\verb| PARAMETERS                                           |\\
\verb|  Z = 11                                     |\\
\verb|  A = 23                                     |\\
\verb|  c_0 =  1.23000000E+00 # [fm]                        |\\
\verb|  c_1 = -0.60000000E+00 # [fm]                        |\\
\verb|  a   =  0.52000000E+00 # [fm]                        |\\
\verb|  s   =  0.90000000E+00 # [fm] skin thickness         |\\
\verb| END_PARAMETERS                                       |\\
\verb|END_FUNCTION                                          |\\
\verb|                                             |\\
\verb|# Fermi distribution:                        |\\
\verb|FUNCTION F_A type=P args=1                   |\\
\verb| DLHA F_A 3             # Fermi distribution |\\
\verb| PARAMETERS                                  |\\
\verb|  Z = 11                                     |\\
\verb|  A = 23                                     |\\
\verb|  c_0 =  1.23000000E+00 # [fm]               |\\
\verb|  c_1 = -0.60000000E+00 # [fm]               |\\
\verb|  a   =  0.52000000E+00 # [fm]               |\\
\verb| END_PARAMETERS                              |\\
\verb|END_FUNCTION                                 |\\
\verb|                                             |\\
\verb|# Sodium (spin-dependent, tabulated)|\\
\verb|FUNCTION S_00 type=T args=1         |\\
\verb| PARAMETERS                         |\\
\verb|  Z = 11                            |\\
\verb|  A = 23                            |\\
\verb| END PARAMETERS                     |\\
\verb| E: 101                             |\\
\verb|#  E             S_00               |\\
\verb|   0.000000E+00  1.000000E+00       |\\
\verb|   1.000000E+00  0.997000E+00       |\\
\verb|   2.000000E+00  0.994000E+00       |\\
\verb|   ...                              |\\
\verb|END FUNCTION                        |
\end{quote}

\subsection{BLOCK DETECTOR\_NUCLEI \label{DLHA_subsec:BDN}}

An alternative way to specify nuclear structure functions $S_{ij}$ is the following.

\begin{quote}
\verb|BLOCK DETECTOR_NUCLEI  # DAMA                                |\\
\verb|# Num  Fraction  A    Z   J    FSD  S00      S01      S11    |\\
\verb|  1    0.153     23   11  1.5  STD  S00Na23  S00Na23  S11Na23|\\
\verb|  2    0.847     127  53  2.5  STD  S00I127  S00I127  S11I127|\\ 
\verb|                                                             |\\
\verb|FUNCTION S00Na23 type=C args=1                                |\\
\verb| libName=libmicromegas.so                                    |\\
\verb| funcName=S00Na23                                            |\\
\verb|END_FUNCTION                                                 |
\end{quote}

\subsection{BLOCK MASS}

Block \verb|MASS| is part of SLHA.  It is used by DLHA to identify the dark matter particle and to specify its mass.  Specific examples of \verb|BLOCK MASS| appear in the introduction of this section.

\subsection{BLOCK QNUMBERS}

This block is part of the Les Houches BSM Generator Accord and defined in Ref.~\cite{Alwall:2007mw}.  The block contains information on the spin, self-conjugate nature, the standard $(SU(3)_C$, $SU(2)_W$, $U(1)_Y$) and exotic charges of the particle.  It may be extended to also contain information about quantum numbers corresponding to discrete symmetries such as R-parity, KK-parity, T-parity, Z-parities.  

\subsection{BLOCK ABUNDANCE}

Abundances of any species of dark matter particles receive their own block, detailing their properties.  The elements of the abundance block are the following. 

\begin{description}
\item[1] The freeze-out temperature $T_f$ in units of GeV$/k$ as defined by the condition in Eq.~(\ref{DLHA_freezecond}).
\item[2] The chosen value of $\alpha$ for Eq.~(\ref{DLHA_freezecond}).
\item[3] The thermal average of the total (co-)annihilation cross section times velocity at freeze-out $\langle\sigma v\rangle(T_f)$ in units of cm$^3$/s as defined in Eq.~(\ref{DLHA_sigmav}).
\item[4] The average energy density of the dark matter particle $\Omega_\chi h^2$ due to thermal production, in units of the critical density as given in Eq.~(\ref{DLHA_omegachi}).
\item[5] Same as above but for non-thermal production.
\item[6] Percentage contribution to the total cross section by (co-)annihilation channels.  This line will appear multiple times for a cross section with multiple (co-)annihilation channels.  Each line lists the PDG codes of the two final state particles along with the percent at which it contributes to the total (co-)annihilation cross section.
\end{description}

An example of \verb|BLOCK ABUNDANCE| for a thermally produced dark matter candidate with an abundance of $\Omega_\chi h^2 = 0.11018437$ is the following:
\begin{quote}
\verb|BLOCK ABUNDANCE| \\
\verb|# identifier(s)  parameter value    comment                | \\
\verb|  1              5.16392660E+00  #  T_f [GeV/k]            | \\
\verb|  2              1.50000000E+00  #  alpha                  | \\
\verb|  3              3.18452057E-26  #  <sigma v>(T_f) [cm^3/s]| \\
\verb|  4              0.11018437E+00  #  Omega h^2 thermal      | \\
\verb|# annihilation channel contribution to <sigma v>(T_f) [%]  | \\
\verb|# identifier(s) PGD code 1  PGD code 2    %                | \\
\verb|  6             ...         ...           ...              | 
\end{quote}

\subsection{BLOCK EFFCOUPLING}

The entries of this block are the effective dark matter-nucleon couplings as defined in Eqs.~(\ref{DLHA_nuc1form}) and (\ref{DLHA_nuc2form}).
\begin{description}
\item[1] Spin-independent scalar coupling $f_p$ for $s_\chi=1/2$ and the proton,
\item[2] spin-independent scalar coupling $f_n$ for $s_\chi=1/2$ and the neutron,
\item[3] spin-dependent axial-vector coupling $a_p$ for $s_\chi=1/2$ and the proton,
\item[4] spin-dependent axial-vector coupling $a_n$ for $s_\chi=1/2$ and the neutron.
\end{description}

An example for an \verb|EFFCOUPLING| block is the following
\begin{quote}
\verb|BLOCK EFFCOUPLING| \\
\verb|# identifier(s)  parameter value   comment | \\
\verb|  1              0.10000000E+00  # f_p     | \\
\verb|  2              0.10000000E+00  # f_n     | \\
\verb|  3              0.00000000E+00  # a_p     | \\
\verb|  4              0.00000000E+00  # a_n     | 
\end{quote}

\subsection{BLOCK NDMCROSSSECT}

This block contains the spin-independent and spin-dependent nucleon-dark matter elastic scattering cross sections $\sigma^{SI,SD}_N$ found using Eqs.~(\ref{DLHA_eq:sigmaSIn}) and (\ref{DLHA_eq:sigmaSDn}).  
\begin{description}
\item[1] The spin-independent  proton-dark matter elastic scattering cross section $\sigma^{SI}_{p}$ in units of cm$^2$, 
\item[2] The spin-independent neutron-dark matter elastic scattering cross section $\sigma^{SI}_{n}$ in units of cm$^2$, 
\item[3] The   spin-dependent  proton-dark matter elastic scattering cross section $\sigma^{SD}_{p}$ in units of cm$^2$, 
\item[4] The   spin-dependent neutron-dark matter elastic scattering cross section $\sigma^{SD}_{n}$ in units of cm$^2$.
\end{description}

An example for a \verb|NDMCROSSSECT| block is the following
\begin{quote}
\verb|BLOCK NDMCROSSSECT                                      | \\
\verb|# identifier(s)  parameter value   comment              | \\
\verb|  1              0.12345678E-41  # \sigma^{SI}_p [cm^2] | \\
\verb|  2              0.23456789E-41  # \sigma^{SI}_n [cm^2] | \\
\verb|  3              0.34567890E-41  # \sigma^{SD}_p [cm^2] | \\
\verb|  4              0.45678901E-41  # \sigma^{SD}_n [cm^2] | 
\end{quote}

\subsection{BLOCK ASTROPROPAG}

This and the following blocks describe aspects of the calculations relevant to indirect detection experiments.  \verb|BLOCK ASTROPROPAG| contains parameters and functions of the propagation model for charged cosmic rays.  The parameters in this block are the following:

\begin{description}
\item[1] normalization of the spatial diffusion coefficient $K_0$, in Eq.~(\ref{DLHA_eq:KE}),
\item[2] coefficient $\eta$, controlling the dependence on $v$ of the spatial diffusion coefficient in Eq.~(\ref{DLHA_eq:KE}),
\item[3] coefficient $\delta$ parameterizing the steepness of the spatial diffusion coefficient in energy, in Eq.~(\ref{DLHA_eq:KE}),
\item[4] normalization of the energy loss coefficient $b_0$, in Eq.~(\ref{DLHA_eq:bE}),
\item[5] half-height $L$ of the diffusion box in units of kpc,
\item[6] radius $R$ of the diffusion box in units of kpc,
\item[7] re-acceleration parameter $v_A$ in units of km/s, in Eq.~(\ref{DLHA_eq:KEE}),
\item[8] Galactic wind parameter $V_C$ in units of km/s, in Eq.~(\ref{DLHA_evol_energy}).
\end{description}

A spatial diffusion coefficient with generic energy or rigidity dependence can be defined by the function
\begin{quote}
\verb|FUNCTION SpatDiff type=<type> args=1| \\
\verb| ...                                | \\
\verb|END_FUNCTION                        | 
\end{quote}
A predefined spatial diffusion coefficient dependence is given by Eq.~(\ref{DLHA_eq:KE}) and can be indicated by 
\begin{quote}
\verb|FUNCTION SpatDiff type=P args=1| \\
\verb| DLHA SpatDiff 1               | \\
\verb|END_FUNCTION                   | 
\end{quote}

The energy loss coefficient for charged cosmic rays, $b(E)$, can also be defined by 
\begin{quote}
\verb|FUNCTION EnerLoss type=<type> args=1| \\
\verb| ...                                | \\
\verb|END_FUNCTION                        | 
\end{quote}
Its predefined form, given by Eq.~(\ref{DLHA_eq:bE}), is indicated as
\begin{quote}
\verb|FUNCTION EnerLoss type=P args=1| \\
\verb| DLHA EnerLoss 1               | \\
\verb|END_FUNCTION                   | 
\end{quote}

\subsection{BLOCK DMCLUMPS}

This is a recommended block and it is only sketched in this version of DLHA.  It should store information regarding dark matter substructures.  It stores the relevant information on the distribution of dark matter substructures inside the Milky Way halo.  

The inner structures of clumps are similar to
the dark matter distributions inside galactic halos. Whatever the
favourite density profile, code builders need to compute the
annihilation volume $\xi$ as a function of the substructure mass
$M_{\rm cl}$. For this, they need to compute the virial radius
$R_{\rm vir}$ of a given clump. This radius encompasses an average
density
\begin{equation}
 \bar{\rho}(R_{\rm vir})={\displaystyle \frac{M_{\rm cl}}{(4\pi/3)R_{\rm vir}^{3}}}=
 \Delta_{\rm vir} \, \Omega_{\rm M} \, \rho_{C}^{0} ,
\end{equation}
which is $\Delta_{\rm vir}$ times larger than the average cosmological
matter density $\Omega_{\rm M}\rho_{C}^{0}$. The concentration of the
clump is defined as the ratio
\begin{equation}
c_{\rm vir}=\frac{R_{\rm vir}}{r_{-2}} ,
\end{equation}
where the clump density profile has slope $-2$ at radius $r=r_{-2}$.
The mass-to-concentration relation can be parameterized as
\cite{Lavalle:1900wn}
\begin{equation}
 \ln(c_{\rm vir})={\sum_{i=0}^{4}} \, C_{i} \,
 \left\{ \ln \left({\displaystyle \frac{M_{\rm cl}}{M_{\odot}}}\right) \right\}^{i} .
\end{equation}
The output is the annihilation volume $\xi$ as a function of the clump
mass $M_{\rm cl}$.
%

The substructure probability distribution
${\cal D}({\vec r}_S,\xi)$ and the total number of clumps
${\cal N}_{\rm H}$ inside the Milky Way halo can be inferred
from the space and mass distribution function
${d^{4}{\cal N}_{\rm cl}}/{dM_{\rm cl \,}}{d^{3}r_S}$. As an illustration,
we can assume the space distribution of clumps to be isotropic and independent
of the mass spectrum, so that
\begin{equation}
 {\displaystyle \frac{d^{2}{\cal N}_{\rm cl}}{{4{\pi}r_S^{2}dr_S}\,{dM_{\rm cl}}}}=
 {\displaystyle \frac{d{\cal N}_{\rm cl}}{dM_{\rm cl}}} \times {\cal P}(r_S) .
\end{equation}
The mass distribution is taken in general to be the power law
\begin{equation}
 {\displaystyle \frac{d{\cal N}_{\rm cl}}{dM_{\rm cl}}}=
 {\displaystyle \frac{C}{{M_{\rm cl}}^{\alpha}}} ,
\end{equation}
where the normalization constant $C$ is calculated by requiring that
the Milky Way halo contains a certain number of substructures
with mass in the range $M_{1}$ to $M_{2}$. In Ref.~\cite{Lavalle:1900wn}
for instance, there are 100 clumps between $10^{8}$ and $10^{10} M_{\odot}$.
The mass spectrum of the clumps extends from $M_{\rm inf}$ to $M_{\rm sup}$
which are inputs of the code.
%

The last input is the space distribution function
${\cal P}(r_S)$. Substructures do not follow the smooth dark matter
distribution. In particular, they are tidally disrupted near the Galactic
center, hence a deficit with respect to the smooth component. As an
illustration, a possible clump distribution is given by the isothermal
profile with core radius $a$
\begin{equation}
 {\cal P}(r_S)={\displaystyle \frac{\kappa}{a(a^2+r_S^2)}} .
\end{equation}
Assuming that the integral of ${\cal P}(r_S)$ from $r=r_{\rm min}$
to $r=r_{\rm max}$ is normalized to unity leads to the constant
\begin{equation}
 \kappa={\displaystyle \frac{{1}/{4{\pi}}}{(u_2 - u_1) - \left\{\arctan(u_2) - \arctan(u_1)\right\}}} ,
\end{equation}
where $u_1={r_{\rm min}}/a$ and $u_2={r_{\rm max}}/a$. This is just an
example. What is actually needed is the input function ${\cal P}(r_S)$.

\subsection{BLOCK ANNIHILATION}

Input block (from various particle physics model codes) that can be used by other codes to calculate the source spectra from annihilation.  This block contains the total annihilation cross section and partial cross sections into different Standard Model final states.

\begin{description}
\item[1] Annihilation cross section times velocity, $\sigma v$ in the $v\rightarrow 0$ limit and in units of cm$^3$/s, followed by a table of annihilation channels with branching fractions with the following columns:
\item[column 1] branching fraction (BR),
\item[column 2] number of annihilation products (NDA),
\item[column 3] PDG code of annihilation product 1,
\item[column 4] PDG code of annihilation product 2.
\end{description}

An example entry could look like
\begin{quote}
\verb|BLOCK ANNIHILATION| \\
\verb|      1          3.000000E-26     # sigma v (v->0) [cm^3/s]|\\
\verb|#     BR         NDA   ID1    ID2   comment                |\\
\verb| 9.50000000E-01   2     24    -24 # X X -> W+ W-           |\\
\verb| 4.00000000E-02   2      5     -5 # X X -> b b-bar         |\\
\verb| 1.00000000E-02   2     25     36 # X X -> H_1^0 H_3^0     |
\end{quote}

%
%
%

\subsection{BLOCK INDIRDETSPECTRUM $i$ $n$}
The spectra of end products (such as positrons, gamma rays, antiprotons, etc.) in the halo (or vacuum) or in an environment (like the Sun) can be calculated from the information given in the blocks above. These spectra can then be used as input for programs that solve the cosmic ray propagation equations, or calculate the fluxes from particular sources in the sky in case of neutral particles. If for example the block \verb|ANNIHILATION| is given at the same time as any of the spectrum blocks given below, the blocks below should take precedence, i.e.~the codes should \emph{not} recalculate the spectra if a spectrum is already given in the DLHA file.

The block definition is
\begin{quote}
\verb|BLOCK INDIRDETSPECTRUM i n| 
\end{quote}
where $i$ is the dark matter particle and $n$ is the spectrum type:
\begin{description}
\item[~] $n=1$: positron spectrum in vacuum (or the halo),  
\item[~] $n=2$: antiproton spectrum in vacuum (or the halo),  
\item[~] $n=3$: antideutron spectrum in vacuum (or the halo),  
\item[~] $n=4$: gamma-ray spectrum in vacuum (or the halo),  
\item[~] $n=5$: electron neutrino spectrum in vacuum (or the halo),  
\item[~] $n=6$: muon neutrino spectrum in vacuum (or the halo),  
\item[~] $n=7$: tau neutrino spectrum in vacuum (or the halo),
\item[~] $n=101$: electron neutrino spectrum at the Earth from annihilations in the Sun,
\item[~] $n=102$: electron anti-neutrino spectrum at the Earth from annihilations in the Sun,
\item[~] $n=103$: muon neutrino spectrum at the Earth from annihilations in the Sun,
\item[~] $n=104$: muon anti-neutrino spectrum at the Earth from annihilations in the Sun,
\item[~] $n=105$: tau neutrino spectrum at the Earth from annihilations in the Sun,
\item[~] $n=106$: tau anti-neutrino spectrum at the Earth from annihilations in the Sun,
\item[~] $n=113$: $\mu^-$ spectrum at the Earth (coming from muon neutrino nucleon interactions) from annihilations in the Sun,
\item[~] $n=114$: $\mu^+$ spectrum at the Earth (coming from muon neutrino nucleon interactions) from annihilations in the Sun,
\item[~] $n=201$: electron neutrino spectrum at the Earth from annihilations in the Earth,
\item[~] $n=202$: electron anti-neutrino spectrum at the Earth from annihilations in the Earth,
\item[~] $n=203$: muon neutrino spectrum at the Earth from annihilations in the Earth,
\item[~] $n=204$: muon anti-neutrino spectrum at the Earth from annihilations in the Earth,
\item[~] $n=205$: tau neutrino spectrum at the Earth from annihilations in the Earth,
\item[~] $n=206$: tau anti-neutrino spectrum at the Earth from annihilations in the Earth,
\item[~] $n=213$: $\mu^-$ spectrum at the Earth (coming from muon neutrino nucleon interactions) from annihilations in the Earth,
\item[~] $n=214$: $\mu^+$ spectrum at the Earth (coming from muon neutrino nucleon interactions) from annihilations in the Earth.
\end{description}

The content of the block is the energy distribution $dN/dE$ as the function of the energy $E$:
\begin{quote}
\verb|E        dN/dE | 
\end{quote}
The spectrum $dN/dE$ is the yield per annihilation or decay at that energy (for $n<100$). For spectra from the Sun/Earth ($n>100$), the units of $dN/dE$ are per annihilation per m$^2$.

A typical spectrum block could then look like
\begin{quote}
\verb|BLOCK INDIRDETSPECTRUM 1 4 # gamma-ray spectrum | \\
\verb|#      E        dN/dE                           | \\
\verb|  1.000000E-02  1.000000E-4                     | \\
\verb|  2.000000E-02  1.200000E-4                     | \\
\vdots
\end{quote}

\subsection{DECAY files}

For decaying dark matter particles a standard SLHA decay file can be used to read and write the total decay width of the dark matter particle and its branching ratios into various final states.  In SLHA DECAY entries are possible for decay channels of various particles, including the dark matter particle. They look similar to the \verb|ANNIHILATION| block above. To be able to calculate the complete spectrum from annihilation and decay of a dark matter particle, we also need the partial decay widths (or branching ratios) for other new physics particles, e.g.\ new Higgs bosons. These DECAY structures should follow the same format as in SLHA2. For example, the decay of a Higgs boson might look like
\begin{quote}
\verb|#      PDG       Width                     | \\
\verb|DECAY  1000039   3.287443E+35              | \\
\verb|#      BR        NDD   ID1   ID2    channel| \\
\verb| 9.50000000E-01    2    24   -24  # W+ W-  | \\
\verb| 5.00000000E-02    2     5    -5  # b b-bar| \\
\verb|# -----------------------------------------| \\
\verb|#      PDG       Width                     | \\
\verb|DECAY  35        3.287443E+00              | \\
\verb|#      BR        NDA   ID1   ID2           | \\
\verb|   9.000000E-01    2     5    -5  # b b-bar| \\
\verb|   6.000000E-02    2    24   -24  # W- W+  | \\
\verb|   4.000000E-02    2    23   -23  # Z0 Z0  | 
\end{quote}

\section{OPEN ISSUES}

A partial list of open issues is addressed in various degrees of detail.

\subsection{Cosmology related open issues}

\vspace{3mm}
\begin{itemize}
\item Other generalizations of the standard cosmological equations would be useful. 
\item The standard inflation scenario should be discussed as well.
\item What about a decaying inflaton scenario? Can be presented as an example. 
\end{itemize}

\subsection{Astrophysics related related open issues}

\vspace{3mm}
\begin{itemize}

\item The notation in BLOCK DMSPADIST is confusing.  $\rho_0$ is not the density at $R_0$.  The local density could be called $\rho_\odot$ and the Galactocentric distance $R_\odot$.  It should be clearly stated that either the density is defined with respect to the change of slope (or the interior mass) and then parameters (1, 2, 3, 4, and 5) are needed, or with respect to the local density, which requires parameters (2, 3, 4, 5, 6, and 7). Giving all the parameters is also possible but one needs to check consistency.

\item Ideally, DLHA should be able to accommodate innately anisotropic distributions.

\item A capability for non-spherical and/or clumped halo distributions is also desirable.
\end{itemize}

\subsection{Direct detection related open issues}

\vspace{3mm}
\begin{itemize}
\item For non-self conjugate dark matter particles there is a need to present nuclear cross sections for antiparticles.  This is not possible within the present setup.
\end{itemize}

\subsection{Indirect detection related open issues}

\vspace{3mm}
\begin{itemize}
\item Photon energy losses need to be improved.
\item Solar modulation needs a unified treatment.
\item The quantities that enter the transport equation (for example, the diffusion coefficient, the re-acceleration term, \ldots) should be standardized, in such a way that free parameters are identified. 
\end{itemize}

\subsection{Other open issues}

\vspace{3mm}
\begin{itemize}
\item Kinetic decoupling of DM particles should be discussed. This sets the small-scale cutoff in the spectrum of density perturbations, viz.~the mass of smallest dark matter halos, and can have impact on, e.g., the anisotropy spectrum and the 'boost factor' for indirect searches. 
\item Concerning the calculation of the relic density, the importance of the QCD phase transition should be stressed.  This may impact strongly the calculation if the dark matter candidate is light (10 GeV/c$^2$ or so).  The QCD phase transition temperature should be a parameter to put in BLOCK DOFREEDOM.
\item BLOCK DMPDGCODE: suggestion for a new block more explicitly identifying the DM candidate(s). 
\begin{quote}
\verb| 1   0  1000022 # neutralino|\\
\verb| 1   1  1000012 # sneutrino |\\
\verb| ...                        |\\
\verb| 2   0  9999999 # axion     |
\end{quote}
\item With large tables in large parameter scans, I/O may take a long time.
\item In scenarios/models with $Z_3$, $Z_4$ etc., different interactions and processes appear.
\item What about semi-annihilations?
\end{itemize}

\section*{ACKNOWLEDGEMENTS}

The authors thank Xerxes Tata for useful comments.
C.B. and R.L. were funded in part by the Australian Research Council under Project ID DP0877916 and in part by the ARC Centre of Excellence for Particle Physics at the Terascale.
C.B. was partially funded by Project of Knowledge Innovation Program (PKIP) of the Chinese Academy of Sciences, Grant No. KJCX2.YW.W10. 
G.B., F.B. and A.P. were supported in part by the GDRI-ACPP of CNRS.
T.B. acknowledges support from the German Research Foundation (DFG) through 
Emmy Noether grant BR 3954/1-1.
D.G.C. is supported by the Ram\'on y Cajal program of the Spanish MICINN. T.D. was supported by the Spanish MICINN's  Consolider-Ingenio 2010 Programme under grant CPAN CSD2007-00042. D.G.C. and T.D. also thank the support of the Spanish MICINNs Consolider-Ingenio 2010 Programme under grant MultiDark CSD2009-00064, the Spanish MICINN under grant FPA2009-08958, the Community of Madrid under grant HEPHACOS S2009/ESP-1473, and the European Union under the Marie Curie-ITN program PITN-GA-2009-237920.
The work of M.C. in supported in part by the French national research agency ANR under contract ANR 2010 BLANC 041301 and by the EU ITN network UNILHC.
J.E. and C.S. are grateful to the Swedish Research Council (VR) for
financial support.
P.G. was supported by NSF award PHY-1068111 and acknowledges support by the Oskar Klein Centre during his sabbatical visit.
S.K. was supported by the US National Science Foundation grants PHY-0854724 and PHY-0955626.
A.P. was supported by the Russian foundation for Basic Research, grant RFBR-10-02-01443-a. 
S.P. is partly supported by the US Department of Energy, Contract DE-FG02-04ER41268, and by NSF Grant PHY-0757911. 
P.S. is supported by the Lorne Trottier Chair in Astrophysics and an Institute for Particle Physics Theory Fellowship.
The development of GALPROP is partially supported by NASA grants NNX09AC15G and NNX10AE78G, and by the Max-Planck Society.



\AddToContent{C.~Bal\'azs et al.}
\renewcommand{\thesection}{\arabic{section}}




\chapter{Further developments in the Flavour Les Houches Accord}

{\it F.~Mahmoudi, S.~Heinemeyer, A.~Arbey,
  A.~Bharucha, 
  T.~Goto, U.~Haisch, S.~Kraml, M.~Muhlleitner, 
  J.~Reuter, P.~Slavich} 




\begin{abstract}
The Flavour Les Houches Accord (FLHA) specifies a standard set of conventions
for flavour-related parameters and observables such as Wilson coefficients,
form factors, decay tables, etc, using the generic SUSY Les Houches Accord
(SLHA) file structure. The accord provides a model-independent interface
between codes evaluating and/or using flavour-related observables. \\ 
We present here a few clarifications and improvements to the accord. In
addition, we provide instructions for a new block concerning the electric and
magnetic dipole moments. 
\end{abstract}

\section{INTRODUCTION}
The FLHA \cite{Mahmoudi:2010iz} exploits the existing organisational structure
of SLHA \cite{Skands:2003cj,Allanach:2008qq} and defines an accord for the
exchange of flavour related quantities. In brief, the purpose of this accord
is to present a set of generic definitions for an input/output file structure
which provides a universal framework for interfacing flavour-related
programs. Furthermore, the standardised format provides the users with clear
and well-structured results. 

The FLHA project has started in Les Houches 2009 where a first proposal was
prepared \cite{Butterworth:2010ym}. This proposal was further completed in the
form of an official and published write-up in \cite{Mahmoudi:2010iz}. FLHA is
fully compatible with SLHA and can contain the SLHA blocks relevant for
flavour physics such as \texttt{SMINPUTS}, \texttt{VCKMIN}, \texttt{UPMNSIN},
\texttt{VCKM}, \texttt{IMVCKM}, \texttt{UPMNS}, \texttt{IMUPMNS} and
\texttt{MODSEL}. The FLHA specific
block names start with ``F'' in order to emphasise their belonging to
FLHA. The blocks which are already defined include: \texttt{FCINFO},
\texttt{FMODSEL}, \texttt{FMASS}, \texttt{FLIFE}, \texttt{FCONST},
\texttt{FCONSTRATIO}, \texttt{FBAG}, \texttt{FWCOEF}, \texttt{IMFWCOEF},
\texttt{FOBS}, \texttt{FOBSERR}, \texttt{FOBSSM} and \texttt{FPARAM}. 

In the following, we provide more details about the definition of the meson
mixings in \texttt{BLOCK FOBS} and introduce \texttt{BLOCK FDIPOLE} for the
electric and magnetic dipole moments.

\section{MESON MIXINGS}
Meson mixing is assigned observable type 7 in \texttt{BLOCK FOBS}. In the
FLHA, we assume the standard definition for the meson mixings. 
The oscillation frequency of $Q^0_q$ and $\bar Q^0_q$ mixing is characterised by the mass difference of the heavy and light mass eigenstates~\cite{Buras:1998raa}:
\begin{equation}
\Delta M_q\equiv M_{\rm H}^{q}-M_{\rm L}^{q} = 2 {\rm Re}\, \sqrt{(M_{12}^q-\frac{i}{2}\Gamma_{12}^q)(M_{12}^{q*}-\frac{i}{2}\Gamma_{12}^{q*})},
\end{equation}
where $M_{12}$ and $\Gamma_{12}$ are the transition matrix elements from virtual and physical intermediate states respectively.
For the kaon systems this gives:
\begin{equation}
\Delta M_K = 2 {\rm Re}\, M_{12}.
\end{equation}
In the case of $\Delta B=2$ transitions since $|\Gamma_{12}| \ll |M_{12}|$, one can write:
\begin{equation}
\langle B_q^0| {\cal H}^{\Delta B=2}_{\rm eff} | \bar B_q^0\rangle = 2 M_{B_q}
M_{12}^q \,,
\end{equation}
where $M_{B_q}$ is the mass of $B_q$ meson and
\begin{equation}
\Delta M_q = 2 |M_{12}^q|.
\end{equation}
The quantity given in the block is $\Delta M_q$ and the unit is fixed to
1/ps. We use the PDG number of the oscillating mesons for the parent and the
flipped sign for the daughter.%
\footnote{
Exceptions to this rule are possible, for instance, in the case of 
$K$-$\bar K$ mixing, where the states $K_{\rm long}$ and $K_{\rm short}$
with their PDG numbers 130 and 310, respectively, can be used.
}
~The number of daughters is fixed to 1. For
example, the $B_s - \bar B_s$ oscillation frequency is given as: 

\begin{verbatim}
Block FOBS  # Flavour observables
# ParentPDG type  value      q        NDA  ID1  ID2  ID3 ... comment
   531      7     1.9e01     0        1    -531 # Delta M_s
\end{verbatim}

Similarly, the corresponding SM values and the errors can be given in
\texttt{BLOCK FOBSSM} and \texttt{BLOCK FOBSERR}, respectively. 
Note that the matrix elements (which are not physical observables) cannot be
given in this block. Such quantities can be expressed in terms of Wilson
coefficients, decay constants, bag parameters, etc. which are defined in the
FLHA, or they can be given in a user defined block.

\section{DIPOLE MOMENTS}
We define \texttt{BLOCK FDIPOLE} which contains the electric and magnetic
dipole moments. The standard for each line in the block should correspond to
the FORTRAN format  
\begin{center} 
\texttt{(1x,I10,3x,I1,3x,I1,3x,1P,E16.8,0P,3x,'\#',1x,A)},
\end{center} 
where the first ten-digit integer should be the PDG code of a particle, the
second integer the type (electric or magnetic), the next integer the model
(SM, NP or SM+NP, as in \texttt{FWCOEF}) and finally the last double precision
number the value of the moment. The electric dipole moments must be given in
e.cm unit. 

The PDG codes for the nuclei follows the PDG particle numbering scheme
\cite{Nakamura:2010zzi}: ``Nuclear codes are given as 10-digit numbers
$\pm10LZZZAAAI$. For a (hyper)nucleus consisting of $n_p$ protons, $n_n$
neutrons and $n_\Lambda$ $\Lambda$'s, $A = n_p + n_n + n_\Lambda$ gives the
total baryon number, $Z = n_p$ the total charge and $L = n_\Lambda$ the total
number of strange quarks. $I$ gives the isomer level, with $I = 0$
corresponding to the ground state ... To avoid ambiguities, nuclear codes
should not be applied to a single hadron, like $p$, $n$ or $\Lambda^0$, where
quark-contents-based codes already exist.'' As an example, the PDG code of the
deuteron is 1000010020.  

The types of the moments are defined as follows:\\[2mm]
\numentry{1}{electric}
\numentry{2}{magnetic}

The electric and magnetic moments can be given for the SM only, the New
Physics (NP) only or the total contributions of both SM and NP, using the
following definitions:\\[2mm] 
\numentry{0}{SM}
\numentry{1}{NP}
\numentry{2}{SM+NP}

For example the muon anomalous magnetic moment can be given as:

\begin{verbatim}
Block FDIPOLE  # Electric and Magnetic dipole moments
# PDG_code type  M    value           comment
    13     2     1    3.02e-09  # 1/2 (g-2)_mu
\end{verbatim}

\section*{CONCLUSIONS}
The Flavour Les Houches Accord is a helpful tool for interfacing flavour
physics and high $p_T$ physics codes in order to exploit maximal information
from both collider and flavour data. We presented here a few clarifications
and improvements viz.\ the published write-up \cite{Mahmoudi:2010iz}. The FLHA
will continue to improve as the number of codes and interfaces grows in order
to accommodate the specific needs of the relevant programs.



\AddToContent{F.~Mahmoudi, S.~Heinemeyer, A.~Arbey,
  A.~Bharucha, 
  T.~Goto, U.~Haisch, S.~Kraml, M.~Muhlleitner, 
  J.~Reuter, P.~Slavich} 
\renewcommand{\thesection}{\arabic{section}}




\chapter{Extension of the SUSY Les Houches Accord 2 for see-saw mechanisms}
{\it L.~Basso, A.~Belyaev, D.~Chowdhury,
  D.K.~Ghosh, M.~Hirsch,
S.~Khalil,
S.~Moretti,
B.~O'Leary,
W.~Porod,
F.~Staub} 



\begin{abstract}
The SUSY Les Houches Accord (SLHA) $2$ extended the first SLHA to include
 various generalisations of the Minimal Supersymmetric Standard Model (MSSM) as
 well as its simplest next-to-minimal version.
 Here, we propose further extensions to it, to include the most general and
 well-established \seesaw\ descriptions (types I/II/III, inverse, and linear) in
 both an effective and a simple gauged extension of the MSSM framework.

\end{abstract}

\section{INTRODUCTION}
If neutrinos are {\em Majorana} particles, their mass at low energy is 
described by a unique dimension-5 operator \cite{Weinberg:1979sa}
\begin{equation}\label{eq:dim5}
m_{\nu} = \frac{f}{\Lambda} (H L) (H L).
\end{equation}
Using only renormalizable interactions, there are exactly three tree-level 
models leading to this operator \cite{Ma:1998dn}. The first one is the 
exchange of a heavy fermionic singlet, called the right-handed neutrino. 
This is the celebrated \seesaw\
mechanism~\cite{Minkowski:1977sc,GellMann:1980vs,Yanagida:1979as,
S.L.Glashow,Mohapatra:1979ia}, nowadays called 
\seesaw\ type I. The second possibility is the exchange of a scalar $SU(2)_L$
triplet~\cite{Schechter:1980gr,Cheng:1980qt}. This is commonly known 
as \seesaw\ type II. And lastly, one could also add one (or more) fermionic
triplets to the field content of the SM~\cite{Foot:1988aq}. This is 
known as \seesaw\ type III. The \seesaw\ mechanism provides a rationale for 
the observed smallness of neutrino masses, by the introduction of the 
inverse of some large scale $\Lambda$. In \seesaw\ type I, for example, 
$\Lambda$ is equal to the mass(es) of the right-handed neutrinos. Since 
these are $SU(2)_L$ singlets, their masses can take any value, and with 
neutrino masses as indicated by the results from oscillation experiments 
$m_{\nu} \sim \sqrt{\Da} \sim 0.05$ eV, where $\Da$ is the atmospheric 
neutrino mass splitting, and couplings of order ${\cal O}(1)$, the scale 
of the \seesaw\ is estimated to be very roughly $m_{SS} \sim 10^{15}$ GeV. 
This value is close to, but slightly lower than, the scale of grand 
unification. In addition there exist \seesaw\ models with large couplings at the
 electroweak scale, such as the linear
 \cite{Akhmedov:1995vm} and inverse \cite{Mohapatra:1986bd} \seesaw\ models.

In the MSSM the gauge couplings unify nearly perfectly at an 
energy scale close to $m_{G} \simeq 2 \times 10^{16}$ GeV. Adding 
new particles which are charged under the SM group at a scale below $m_G$ 
tends to destroy this attractive feature of the MSSM, unless the 
new superfields come in complete $SU(5)$ multiplets. For this 
reason, within supersymmetric \seesaw\ models, one usually realizes 
the type II \seesaw\ by adding $\bf 15$-plets~\cite{Ma:1998dn,
Rossi:2002zb} and the type III \seesaw\ by the addition of 
$\bf 24$-plets. 

Just like with any other extensions of the SM based on supersymmetry (SUSY),
 those implementing \seesaw\ realisations have seen several different
 conventions used over the years, many of which have become widespread. Such a
 proliferation of conventions has some drawbacks though from a calculational
 point of view: results obtained by different authors or computing tools are not
 always directly comparable. Indeed, to enable this comparison, a consistency
 check of all the relevant conventions and the implementation of any necessary
 translations thereof must first be made. Needless to say, this is a
 time-consuming and rather error-prone task.

To remedy this problem, the original SUSY Les Houches Accord (SLHA1) was
 proposed \cite{Skands:2003cj}. The SLHA1 uniquely defined a set of conventions
 for SUSY models together with a common interface between codes. The latter can
 be broadly categorised in terms of four different kinds of tools:
 ($i$) spectrum calculators (which calculate the SUSY mass and coupling
 spectrum, assuming some SUSY-breaking terms and a matching of SM parameters to
 known data); ($ii$) observables calculators (packages which calculate one or
 more of the following: inclusive cross sections, decay partial widths, relic
 dark matter densities and  indirect/precision  observables); ($iii$) Monte
 Carlo (MC) event generators (which calculate exclusive cross sections through
 explicit  simulation of high-energy particle collisions, by including resonance
 decays, parton showering, hadronisation and underlying-event effects);
($iv$) SUSY fitting programs (which fit SUSY model parameters to data).
(See http://www.ippp.dur.ac.uk/montecarlo/BSM/ for an up-to-date collection and
 description of such tools.) Further, SLHA1 provided users with input and output
 in a common format, which is more readily comparable and transferable. In
 short, the basic philosophy was to specify a unique set of conventions for SUSY
 extensions of the SM together with generic file structures to be communicated
 across the four types of codes above, $(i)-(iv)$, based on the transfer of
 three different ASCII files: one for model input, one for spectrum calculator
 output and one for decay calculator output.  

The original protocol, SLHA1, was strictly limited to the MSSM with real
 parameters and R-parity conservation neglecting generation mixing.  An expanded version was proposed in 
 \cite{Allanach:2008qq} (see also \cite{Allanach:2005kk} in
 Ref.~\cite{Allanach:2006fy}), known as SLHA2, whereby various MSSM
 generalisations were included: \ie, those involving CP, $R$-parity and flavour
 violation as well as the simplest extension of the MSSM, the so-called
 next-to-MSSM (NMSSM). Herein, we further develop this protocol, by including
 the most general and well-established \seesaw\ descriptions in both an
 effective and a simple gauged extension of the MSSM. The new conventions and
 control switches described here comply with those of SLHA2 (and,
 retrospectively, also SLHA1) unless explicitly mentioned in the text.  
Our effort here is paralled by other generalisations of the previous accords
 documented elsewhere in these proceedings, altogether eventually contributing
 to the definition of a future release of the original protocol.

\section{THE SEE-SAW MECHANISM}
\label{sec:seesawmechanisms}

In this section, we discuss different implementations of the \seesaw\
 mechansim. As already stated in the introduction, the aim of the \seesaw\
 mechansim is to explain the neutrino masses and mixing angles. This is done by
 linking the tiny masses to other parameters which are of the naturally expected
 order. The general idea can be summarized by writing down the most general mass
 matrix combining left-handed neutrino ($L$), right-handed neutrino ($R$) and
 additional singlet fields carrying lepton number ($S$): 
\begin{equation}
 \left(\begin{array}{ccc}
m_{LL} & m_{LR} & m_{LS} \\
m^T_{LR} & m_{RR} & m_{RS} \\
m^T_{LS} & m^T_{RS} & m_{SS}
\end{array}  \right)
\end{equation}
Looking at specific limits of this matrix, we can recover the different \seesaw\
 realizations: $m_{LL} = m_{LS} = m_{RS} = 0$ leads to type I, type III is
 obtained in the same limit but with $m_{RR}$ stemming from $SU(2)_L$ triplets.
 $m_{LL} = m_{RR} = m_{LS} = 0$ is the characteristic matrix for inverse
 \seesaw\ , while $m_{LL} = m_{RR} = m_{SS} = 0$ is the standard parametrization
 of the linear \seesaw\ . What all these different  \seesaw\ models have in
 common is
the way the tiny neutrino masses are recovered, just by suppressing them with
very high scales for the new fields. This is strictly true for the type I/II/III
models. The linear and inverse \seesaw\ versions work slightly differently: the
 heaviness of the
 new fields is reduced at the price of introducing a relatively small
dimensionless parameter, usually connected to an explicit violation of the
lepton number.

In the following subsection we discuss models which can explain the origin of
 the distinct neutrino mass matrices.




\subsection{Type I/II/III}
The simplest see-saw models describe neutrino masses with an effective operator
arising after integrating out heavy superfields.
While one generation of $15$-plets
is sufficient to explain the entire neutrino data, this is not the case with
just one $24$-plet if $SU(5)$-invariant boundary conditions are assumed on the
new parameters and more generations have to be included. We will therefore treat
the number of generations of singlets, 15- and 24-plets as free parameter.
Bearing this in mind,
the models will be described with minimal addition of superfields, in the basis
after $SU(5)$ symmetry breaking (see table~\ref{eff_SS_new_chiral} \footnote{We use 
always the convention that all given 
$U(1)$  charges are those appearing in the covariant derivative, i.e. $\partial_\mu - i g Q A_\mu$.}). 
All these
fields are integrated out during the RGE evaluation, such that, at the SUSY
scale, only the particle content of the MSSM remains. 
 
We give in the following the unified equations which would lead to a mixed
scenario of \seesaw\ types I, II, and III. The
 \eqs{eff_SS_T1_potential}{eff_SS_T1_softbreaking_W}{eff_SS_T1_softbreaking_phi}
 are specific to type I,
 \eqs{eff_SS_T2_potential}{eff_SS_T2_softbreaking_W}{eff_SS_T2_softbreaking_phi}
 refer to type II and
 \eqs{eff_SS_T3_potential}{eff_SS_T3_softbreaking_W}{eff_SS_T3_softbreaking_phi}
 refer to type III.

\begin{table}[!ht]
\begin{center}
\begin{tabular}{|c|c|c|c|c|c|} 
\hline \hline 
\multicolumn{6}{|c|}{Type I}\\
\hline \hline 
SF & Spin 0 & Spin \(\frac{1}{2}\) & Generations & \(U(1)\otimes\,
\text{SU}(2)\otimes\, \text{SU}(3)\) &  $R$-parity of fermion \\ 
\hline 
\(\hat{\nu}^c\) & \(\tilde{\nu}^c\) & \(\nu^c\) & $n_{1}$
 & \((0,{\bf 1},{\bf 1}) \) & $+$\\ 
\hline \hline
\multicolumn{6}{|c|}{Type II}\\
\hline \hline 
SF & Spin 0 & Spin \(\frac{1}{2}\) & Generations & \(U(1)\otimes\,
\text{SU}(2)\otimes\, \text{SU}(3)\) & $R$-parity of fermion \\ 
\hline 
\(\hat{T}\) & \(\tilde{T}\) & \(T\) & $n_{15}$ & \((1,{\bf 3},{\bf 1}) \) & $-$\\ 
\(\hat{\bar{T}}\) & \(\tilde{\bar{T}}\) & \(\bar{T}\) & $n_{15}$ & \((-1,{\bf 3},
{\bf 1}) \) & $-$\\ 
\(\hat{S}\) & \(\tilde{S}\) & \(S\) & $n_{15}$
 & \((-\frac{2}{3},{\bf 1},{\bf 6}) \) & $-$\\ 
\(\hat{\bar{S}}\) & \(\tilde{\bar{S}}^*\) & \(\bar{S}^*\) & $n_{15}$ &
\((\frac{2}{3},{\bf 1},{\bf \overline{6}}) \) & $-$\\ 
\(\hat{Z}\) & \(\tilde{Z}\) & \(Z\) & $n_{15}$
 & \((\frac{1}{6},{\bf 2},{\bf 3}) \) & $-$\\ 
\(\hat{\bar{Z}}\) & \(\tilde{\bar{Z}}\) & \(\bar{Z}\) & $n_{15}$
 & \((-\frac{1}{6},{\bf 2},{\bf \overline{3}}) \) & $-$\\ 
\hline \hline
\multicolumn{6}{|c|}{Type III}\\
\hline \hline 
SF & Spin 0 & Spin \(\frac{1}{2}\) & Generations & \(U(1)\otimes\,
\text{SU}(2)\otimes\, \text{SU}(3)\) &  $R$-parity of fermion \\ 
\hline 
\(\hat{W}_M\) & \(\tilde{W}_M\) & \(W_M\) & $n_{24}$ & \((0,{\bf 3},{\bf 1}) \)
 & $+$\\ 
\(\hat{G}_M\) & \(\tilde{G}_M\) & \(G_M\) & $n_{24}$ & \((0,{\bf 1},{\bf 8}) \)
 & $+$\\ 
\(\hat{B}_M\) & \(\tilde{B}_M\) & \(B_M\) & $n_{24}$ & \((0,{\bf 1},{\bf 1}) \)
 & $+$\\ 
\(\hat{X}_M\) & \(\tilde{X}_M\) & \(X_M\) & $n_{24}$ & \((\frac{5}{6},{\bf 2},
{\bf \overline{3}}) \) & $+$\\ 
\(\hat{\bar{X}}_M\) & \(\tilde{\bar{X}}_M\) & \(\bar{X}_M\) & $n_{24}$ &
\((-\frac{5}{6},{\bf 2},{\bf 3}) \) & $+$\\ 
\hline \hline
\end{tabular} 
\caption{\label{eff_SS_new_chiral} 
New chiral superfields appearing in the effective type I/II/III \seesaw\
models. While $n_{15} = 1$ is sufficient to explain neutrino data, 
$n_1$ and $n_{24}$ must be at least 2. }
\end{center} 
\end{table}

The combined superpotential of all three types can be written as
\begin{displaymath}
W -  W_{\text{MSSM}} = W_{I} + W_{II} + W_{III}
\end{displaymath}
where
\begin{align} \label{eff_SS_T1_potential}
W_{I} = & Y_{\nu} \, \hat{\nu}^c\,\hat{L}\,\hat{H}_u\,+\frac{1}{2} M_{\nu^c}
\,\hat{\nu}^c\,\hat{\nu}^c \\ 
W_{II} = & \frac{1}{\sqrt{2}} \nonumber
Y_T \,\hat{L}\,\hat{T}\,\hat{L}\,+\frac{1}{\sqrt{2}} Y_S
\,\hat{d}^c\,\hat{S}\,\hat{d}^c\,+Y_Z\,\hat{d}^c\,\hat{Z}\,\hat{L}\,\\ 
                 \label{eff_SS_T2_potential}
        & +\frac{1}{\sqrt{2}} \lambda_1
\,\hat{H}_d\,\hat{T}\,\hat{H}_d\,+\frac{1}{\sqrt{2}} \lambda_2
\,\hat{H}_u\,\hat{\bar{T}}\,\hat{H}_u\,+M_T\,\hat{T}\,\hat{\bar{T}}\,+M_Z\,
\hat{Z}\,\hat{\bar{Z}}\,+M_S\,\hat{S}\,\hat{\bar{S}}\, \\ \nonumber
 W_{III} = & \sqrt{\frac{3}{10}} Y_B
\,\hat{H}_u\,\hat{B}_M\,\hat{L}\,+Y_W\,\hat{H}_u\,\hat{W}_M\,\hat{L}\,+Y_X\,
\hat{H}_u\,\hat{\bar{X}}_M\,\hat{d}^c\,\\ \label{eff_SS_T3_potential}
 & +M_X\,\hat{X}_M\,\hat{\bar{X}}_M\,+\frac{1}{2} M_W
\,\hat{W}_M\,\hat{W}_M\,+\frac{1}{2} M_G \,\hat{G}_M\,\hat{G}_M\,+\frac{1}{2}
M_B \,\hat{B}_M\,\hat{B}_M\,
\end{align} 

The soft-breaking terms can be split into three categories: terms stemming from
the superpotential couplings when replacing the fermions with their scalar
superpartners ($L_{SB,W}$), the scalar soft-breaking masses for each chiral
superfield ($L_{SB,\phi}$) and the soft-breaking masses for the gauginos
($L_{SB,\lambda}$). Since the gauge sector is not modified, $L_{SB,\lambda}$
reads as in the MSSM. The soft-breaking terms stemming from the superpotential
are

\begin{displaymath}
L_{\text{SB},W} - L_{\text{SB},W,\text{MSSM}}
= L_{\text{SB},W}^{I} + L_{\text{SB},W}^{II} + L_{\text{SB},W}^{III}
\end{displaymath}
where
\begin{align} \label{eff_SS_T1_softbreaking_W}
L_{\text{SB},W}^{I} = \, & \, T_{\nu}\,\tilde{\nu}^c\,\tilde{L}\,H_u\,
+\frac{1}{2} B_{\nu^c} \,\tilde{\nu}^c\,\tilde{\nu}^c\,+ \mbox{H.c.}\\ \nonumber
L_{\text{SB},W}^{II} = \, & \frac{1}{\sqrt{2}}
T_T \,\tilde{L}\,\tilde{T}\,\tilde{L}\,+\frac{1}{\sqrt{2}} T_S \,\tilde{d}^{c}\,
\tilde{S}\,\tilde{d}^{c}\,+T_Z\,\tilde{d}^{c}\,\tilde{Z}\,\tilde{L}\,
+\frac{1}{\sqrt{2}} T_1\,H_d\,\tilde{T}\,H_d\,\\
\label{eff_SS_T2_softbreaking_W}
 & +\frac{1}{\sqrt{2}} T_2 \,H_u\,\tilde{\bar{T}}\,H_u\,+ B_T\,\tilde{T}\,
\tilde{\bar{T}}\,+ B_Z\,\tilde{Z}\,\tilde{\bar{Z}}\,+ B_S\,\tilde{S}\,
\tilde{\bar{S}}\, + \mbox{h.c.}\\  \nonumber
L_{\text{SB},W}^{III} = \, & \sqrt{\frac{3}{10}} T_B \,H_u\,\tilde{B}_M\,
\tilde{L}\,+T_W\,H_u\,\tilde{W}_M\,\tilde{L}\,+T_X\,H_u\,\tilde{\bar{X}}_M\,
\tilde{d}^{c}\,+B_X\,\tilde{X}_M\,\tilde{\bar{X}}_M\\
\label{eff_SS_T3_softbreaking_W}
 & \,+\frac{1}{2} B_W \,\tilde{W}_M\,\tilde{W}_M\,+\frac{1}{2} B_G \,
\tilde{G}_M\,\tilde{G}_M\,+\frac{1}{2} B_B \,\tilde{B}_M\,\tilde{B}_M\,
+\mbox{H.c.} 
\end{align} 
while the soft-breaking scalar masses read
\begin{displaymath}
L_{\text{SB},\phi} - L_{\text{SB},\phi,MSSM} = L_{\text{SB},\phi}^{I}
+ L_{\text{SB},\phi}^{II} + L_{\text{SB},\phi}^{III}
\end{displaymath}
where
\begin{align} \label{eff_SS_T1_softbreaking_phi}
L_{\text{SB},\phi}^{I} = & - (\tilde{\nu}^c)^\dagger {m_{\nu^c}^{2}}
\tilde{\nu}^c \\ \label{eff_SS_T2_softbreaking_phi}
L_{\text{SB},\phi}^{II} = & -m_S^2 \tilde{S}^* \tilde{S} -m_{\bar{S}}^2
\tilde{\bar{S}}^* \tilde{\bar{S}} -m_T^2 \tilde{T}^* \tilde{T} -m_{\bar{T}}^2
\tilde{\bar{T}}^* \tilde{\bar{T}} -m_Z^2 \tilde{Z}^* \tilde{Z} -m_{\bar{Z}}^2
\tilde{\bar{Z}}^* \tilde{\bar{Z}} \\         \nonumber
L_{\text{SB},\phi}^{III} = \, & - \tilde{B}_M^\dagger m^2_B \tilde{B}_M
-\tilde{W}_M^\dagger m^2_W \tilde{W}_M  - \tilde{G}_M^\dagger m^2_G \tilde{G}_M
\\ \label{eff_SS_T3_softbreaking_phi}
& \, - \tilde{X}_M^\dagger m^2_X \tilde{X}_M  -  \tilde{\bar{X}}_M^\dagger
m^2_{\bar{X}} \tilde{\bar{X}}_M 
\end{align}

\paragraph{GUT conditions and free parameters}
Since the new interactions in
\eqsfromto{eff_SS_T2_potential}{eff_SS_T3_potential} are the result of
$SU(5)$-invariant terms, it is natural to assume a unification of the different
couplings at the GUT scale. 
\begin{align}
& M_T = M_Z = M_S \equiv M_{15}, \hspace{1cm} 
Y_S = Y_T = Y_Z \equiv Y_{15} \\
& Y_B = Y_W = Y_X  \equiv Y_{24}, \hspace{1cm}
 M_X = M_W = M_G = M_B  \equiv M_{24}
\end{align}
In the same way the bi- and trilinear soft-breaking terms unify and they are
connected to the superpotential parameters by
\begin{align}
& B_{\nu^c} \equiv B_0 M_{\nu^c}, \hspace{1cm} T_{\nu} \equiv A_0 Y_{\nu} \\
& B_{15} \equiv B_0 M_{15}, \hspace{1cm}  T_{15} \equiv A_0 Y_{15}\\
& B_{24} \equiv B_0 M_{24}, \hspace{1cm} T_{24} \equiv A_0 Y_{24} 
\end{align}
 In case of
CMSSM-like boundary conditions, this leads to the following free parameters
\begin{equation}
B_0, \thickspace M_{\nu^c}, \thickspace Y_{\nu} \thickspace M_{15}, \lambda_1,
 \lambda_2, \thickspace Y_{15}, \thickspace M_{24}, \thickspace Y_{24}
\end{equation}
in addition to the well-known MSSM parameters
\begin{equation}
 m_0, \thickspace M_{1/2}, \thickspace A_0, \thickspace \tan\beta, \thickspace
\text{sign}(\mu) \label{eq:CMSSM}
\end{equation}
In principle, this $B_{0}$ is not the same as the $B$ for the Higgs, though in a
minimal case they may be defined to be equal at the GUT scale. Furthermore,
 $T_i = A_0 Y_i$ holds at the GUT scale. 


\paragraph{Effective neutrino masses.}
The effective neutrino mass matrices appearing in type I/II/III at SUSY scale
are
\begin{align}
 \label{eq:mnuI}
m^I_\nu = &\, - \frac{v^2_u}{2} Y^T_\nu M^{-1}_R Y_\nu \\
\label{eq:ssII}
m^{II}_\nu = &\, \frac{v_u^2}{2} \frac{\lambda_2}{M_T}Y_T. \\
\label{eq:mnu_seesawIIIa}
m^{III}_\nu = &\,- \frac{v^2_u}{2} 
\left( \frac{3}{10} Y^T_B M^{-1}_B Y_B + \frac{1}{2} Y^T_W M^{-1}_W Y_W \right).
\end{align}


\subsection{Inverse and linear \seesaw}
\label{sec:inverse_linear}
The inverse and linear \seesaw\ realisations are obtained in models that provide
three generations of a further gauge singlet carrying lepton number in addition
to three generations of the well-known right-handed neutrino superfields, here
\(\hat{\nu}^c\) (see table~\ref{tab:chiral_lin_inv}).

\begin{table}[!ht]
\begin{center}
\begin{tabular}{|c|c|c|c|c|c|c|c|} 
\hline \hline 
SF & Spin 0 & Spin \(\frac{1}{2}\) & Generations & \(U(1)\otimes\,
\text{SU}(2)\otimes\, \text{SU}(3)\) &  lepton number & $R$-parity of fermion\\
\hline 
\(\hat{\nu}^c\) & \(\tilde{\nu}^c\) & \(\nu^c\) & $n_{\nu^c}$
 & \((0,{\bf 1},{\bf 1}) \) & $+1$ & $+$\\ 
\({\hat{N}}_{S}\) & \({\tilde{N}}_{S}\) & \(N_{S}\) & $n_{N_S}$
 & \((0,{\bf 1},{\bf 1}) \) & $-1$ & $+$\\ 
\hline \hline
\end{tabular} 
\caption{\label{tab:chiral_lin_inv}
New chiral superfields appearing in models with inverse and linear
\seesaw.}
\end{center} 
\end{table}

The only additional terms in the superpotential which are allowed by
conservation of gauge quantum numbers are 
\begin{equation}
\label{eq:W_lin_inv}
 W - W_{\text{MSSM}} = Y_{\nu}\,\hat{\nu}^c\,\hat{L}\,\hat{H}_u\,
+M_R\,\hat{\nu}^c\,{\hat{N}}_{S}\,
+ \left\{ \begin{array}{c c}
\frac{1}{2} {\mu}_{N}\,{\hat{N}}_{S}\,{\hat{N}}_{S} & \mbox{inverse \seesaw}\\
Y_{LN} \,{\hat{N}}_{S}\,\hat{L}\,\hat{H}_u\, & \mbox{linear \seesaw}
\end{array} \right. .
\end{equation}
It is important to note that the last term in each model breaks lepton number
explicitly, but is expected for different reasons.

The soft-breaking terms read
\begin{align}
L_{\text{SB},W} = \, & L_{\text{SB},W,\text{MSSM}}\, + T_{\nu}\,\tilde{\nu}^c\,
\tilde{L}\,H_u\,+B_R\,\tilde{\nu}^c\,{\tilde{N}}_{S}\,
+ \left\{ \begin{array}{c c}
\frac{1}{2} B_N \,{\tilde{N}}_{S}\,{\tilde{N}}_{S} & \mbox{inverse \seesaw}\\
T_{LN} \,{\tilde{N}}_{S}\,\tilde{L} H_u & \mbox{linear \seesaw}
\end{array} \right\} + \mbox{H.c.} \\
L_{\text{SB},\phi} = \, & L_{\text{SB},\phi,\text{MSSM}}
- (\tilde{\nu}^c)^\dagger {m_{\nu^c}^{2}} \tilde{\nu}^c - {\tilde{N}}_{S} m_N^2
{\tilde{N}}_{S}^{\ast} \, .
\end{align}
while $L_{\text{SB},\lambda}$ is again the same as for the MSSM. 
It is necessary to split the sneutrinos and the singlets into their scalar and
pseudoscalar components:
\begin{equation} 
\label{eq:decomposition_sneutrinos}
\tilde{\nu}_L =  \, \frac{1}{\sqrt{2}} \left(\sigma_L  + i \phi_L\right),
\thickspace 
\tilde{\nu}^c =  \, \frac{1}{\sqrt{2}} \left(\sigma_R  + i  \phi_R \right),
\thickspace 
{\tilde{N}}_{S} = \, \frac{1}{\sqrt{2}} \left(\sigma_S  + i  \phi_S \right)\, .
\end{equation} 
In comparison to the MSSM, additional mixings between fields take place: the
left- and right-handed scalar components mix with the scalar component of the
singlet fields. The same holds for the pseudoscalar components. Furthermore, the
neutrinos mix with the fermionic singlet fields to build up $9$ Majorana
fermions. All three appearing $9 \times 9$ mass matrices can be diagonalised by
unitary matrices. We define the basis for the mass matrices as
\begin{itemize}
 \item Scalar sneutrinos: \( \left(\sigma_{L}, \sigma_{R}, \sigma_{S}\right)^T\)
 \item Pseudoscalar sneutrinos:
       \( \left(\phi_{L}, \phi_{R}, \phi_{S}\right)^T\)
 \item Neutrinos: \( \left(\nu_L, \nu^c, N_{S}\right)^T\)
\end{itemize}
The neutrino mass matrix then reads
\begin{equation}
\label{eq:InvLinMM}
 \left(\begin{array}{ccc}
0 & \frac{v_u}{\sqrt{2}}Y_\nu & 0 \\
\frac{v_u}{\sqrt{2}}Y^T_\nu & 0 & M_R \\
0 & M^T_R &  \mu_N
       \end{array} \right)
\mbox{ (inverse) or }
 \left(\begin{array}{ccc}
0 & \frac{v_u}{\sqrt{2}}Y_\nu & \frac{v_u}{\sqrt{2}} Y_{LN} \\
\frac{v_u}{\sqrt{2}}Y^T_\nu & 0 & M_R \\
\frac{v_u}{\sqrt{2}} Y^T_{LN} & M^T_R & 0
       \end{array} \right)
\mbox{ (linear),}
\end{equation}
Note, the presence of $v_u$ in all terms of the first column and row is just
 coincidence caused by the given, minimal particle content. For more general
 models different VEVs can appear. 

\paragraph{Free parameters}
If CMSSM-like boundary conditions are assumed, the following new free parameters
arise:
\begin{equation}
M_R, \thickspace Y_{LN}, \thickspace \mu_N, \thickspace B_0
\end{equation}
in addition to those given in \eq{eq:CMSSM}.

Calculating the eigenvalues of the above mass matrices, it can be seen that the
 light neutrino masses are linear functions of $Y_{LN}$ in the linear
 \seesaw\ models, while the neutrino masses are linearly proportional to
 $\mu_N$, as in the inverse \seesaw\ models. The neutrino masses in the two
 models read
\begin{align}
m^{LS}_\nu \simeq &\, \frac{v_u^2}{2} \left(Y_\nu (Y_{LN} M_R^{-1})^T +
(Y_{LN} M_R^{-1})Y_\nu^T \right) \, ,\\
m^{IS}_\nu \simeq &\, \frac{v_u^2}{2} Y_\nu (M^T_R)^{-1} \mu_N M^{-1} Y^T_\nu \,.
\end{align}
Hence we propose that both models, and any combination of the two, be specified
 by extending $Y_{\nu}$ to a $3 \times ( n_{\nu^c} + n_{N_S} )$ Yukawa matrix,
 incorporating $Y_{LN}$ as $Y_{\nu}^{ij}$ with $i$ running from $n_{\nu^c} + 1$
 to $n_{\nu^c} + n_{N_S}$. Implementation of each model consists of zeros being
 specified in the appropriate entries in the relevant matrices (\textit{e.g.}
 specifying that the elements of $Y_{\nu}$ corresponding to $Y_{LN}$ are zero
 recovers the inverse \seesaw\ model, while specifying that ${\mu}_{N}$ is zero
 recovers the linear \seesaw\ model).


\subsection{\Seesaw\ in models with \URxUBL\ gauge sector}
\label{sec:RxB-L}
Models based on a $SO(10)$ GUT theory can lead to a gauge sector containing the
product group \URxUBL, through the breaking pattern 
\begin{align}
SO(10) & \rightarrow SU(3)_C \times SU(2)_L \times SU(2)_R \times U(1)_{B-L}
& \rightarrow SU(3)_C \times SU(2)_L \times U(1)_R \times U(1)_{B-L}\, .
\end{align}
The \URxUBL\ factors will be subsequently broken to the hypercharge $U(1)_Y$ of
the SM. However, it is possible that this final breaking scale is just around
the TeV scale without spoiling gauge unification \cite{Malinsky:2005bi}. This
can therefore lead to interesting phenomenology and can have an important impact on
the Higgs sector \cite{Hirsch:2011hg}. The first version of these models
included a linear \seesaw\ mechanism, but it has been shown that also the
 inverse \seesaw\ can be included \cite{DeRomeri:2011ie}. Since the realization
 of the linear model is rather involved, we describe here only the version with
 inverse \seesaw.

Notice that, in general,
these models contain not only gauge couplings per each Abelian gauge group, but
also so-called `off-diagonal couplings', as discussed in
appendix~\ref{sec:kineticmixing}. The minimal particle content for such model
extending the MSSM, leading to the spontaneous breaking of \URxUBL\ and to
neutrino masses, is given in table~\ref{tab:chiral_RxBL}. This particle content
consists of $3$ generations of $16$-plets of $SO(10)$, $2$ additional Higgs
fields and $3$ generations of a singlet field. The vector superfields are given
in table~\ref{tab:vector_RxBL}.
\begin{table}[!ht]
\begin{center} 
\begin{tabular}{|c|c|c|c|c|c|}
\hline \hline 
SF & Spin 0 & Spin \(\frac{1}{2}\) & Generations & \( U(1)_{B-L}\otimes\,
\text{SU}(2)\otimes\, U(1)_R\otimes\, \text{SU}(3)\)\\ 
\hline 
\hline
\multicolumn{5}{|c|}{Matter fields (fermionic components have positive
$R$-parity)}
\\
\hline \hline
\(\hat{Q}\) & \(\tilde{Q}=\left(\begin{array}{c} \tilde{u}_L \\ \tilde{d}_L
\end{array} \right) \) & \(Q=\left(\begin{array}{c} u_L \\ d_L \end{array}
\right)\) & 3 & \((\frac{1}{6},{\bf 2},0,{\bf 3}) \) \\ 
\(\hat{L}\) & \(\tilde{L}=\left(\begin{array}{c} \tilde{\nu}_L \\ \tilde{e}_L
\end{array} \right)\) & \(L=\left(\begin{array}{c} \nu_L \\ e_L \end{array}
\right)\) & 3 & \((-\frac{1}{2},{\bf 2},0,{\bf 1}) \) \\ 
\(\hat{u^c}\) & \(\tilde{u}^c\) & \(u^c\) & 3 & \((-\frac{1}{6},{\bf 1},
-\frac{1}{2},{\bf \overline{3}}) \) \\ 
\(\hat{d^c}\) & \(\tilde{d}^c\) & \(d^c\) & 3 & \((-\frac{1}{6},{\bf 1},
\frac{1}{2},{\bf \overline{3}}) \) \\ 
\( {\hat \nu}^c \) & \(\tilde{\nu}^c\) & \( \nu^c\) & 3 & \((\frac{1}{2},
{\bf 1},-\frac{1}{2},{\bf 1}) \) \\ 
\(\hat{e^c}\) & \(\tilde{e}^c\) & \(e^c\) & 3 & \((\frac{1}{2},{\bf 1},
\frac{1}{2},{\bf 1}) \) \\ 
\( {\hat{N}}_{S}\) & \( {\tilde{N}}_{S} \) & \( N_{S} \) & $n_{N_S}$ &
\((0,{\bf 1},0,{\bf 1}) \) \\
\hline 
\hline
\multicolumn{5}{|c|}{Higgs fields (scalar components have positive
$R$-parity)}\\
\hline \hline
\(\hat{H}_d\) & \(H_d = \left(\begin{array}{c} H_d^0 \\ H_d^- \end{array}
\right)\) & \(\tilde{H}_d = \left(\begin{array}{c} \tilde{H}_d^0 \\
\tilde{H}_d^- \end{array} \right)\) & 1 & \((0,{\bf 2},-\frac{1}{2},{\bf 1}) \)
\\ 
\(\hat{H}_u\) & \(H_u= \left(\begin{array}{c} H_u^+ \\ H_u^0\end{array}
\right)\) & \(\tilde{H}_u= \left(\begin{array}{c} \tilde{H}_u^+ \\ \tilde{H}_u^0
\end{array} \right)\) & 1 & \((0,{\bf 2},\frac{1}{2},{\bf 1}) \) \\ 
\(\hat{\xi}_R\) & \(\xi_R^0\) & \(\tilde{\xi}_R^0\) & 1 & \((-\frac{1}{2},
{\bf 1},\frac{1}{2},{\bf 1}) \) \\
\(\hat{\bar{\xi}}_R\) & \(\bar{\xi}_R^0\) & \(\tilde{\bar{\xi}}_R^0\) & 1 &
\((\frac{1}{2},{\bf 1},-\frac{1}{2},{\bf 1}) \) \\ 
\hline \hline
\end{tabular} 
\end{center} 
\caption{\label{tab:chiral_RxBL}
Chiral Superfields appearing in models with \URxUBL\ gauge sector which
incorporate linear and inverse \seesaw\ mechanisms.
}
\end{table}
\begin{table}[!ht]
\begin{center} 
\begin{tabular}{|c|c|c|c|c|c|} 
\hline \hline 
SF & Spin \(\frac{1}{2}\) & Spin 1 & \(SU(N)\) & Coupling & Name \\ 
 \hline 
\(\hat{B}'\) & \(\lambda_{\tilde{B}'}\) & \(B'\) & \(U(1)\) & \(g_{BL}\) &
\text{B-L}\\ 
\(\hat{W}_L\) & \(\lambda_{L}\) & \(W_L\) & \(\text{SU}(2)\) & \(g_L\) &
\text{left}\\ 
\({\hat B}_R \) &\(\lambda_{\tilde{B}_R}\) & \(B_R\) & \(U(1)\) & \(g_R\) &
\text{right}\\ 
\(\hat{g}\) & \(\lambda_{\tilde{g}}\) & \(g\) & \(\text{SU}(3)\) & \(g_s\) &
\text{color}\\ 
\hline \hline
\end{tabular} 
\end{center} 
\caption{\label{tab:vector_RxBL}
Vector superfields appearing in models with \URxUBL\ gauge sector.}
\end{table}
\paragraph{Superpotential}
We assume for the following discussion that the superpotential can contain the
following terms:
\begin{align} 
\label{eq:W_RxBL}
W -W_{MSSM} = & \, Y_{\nu}\,\hat{\nu^c}\,\hat{L}\,\hat{H}_u\, - \mu_{\xi} \,
\hat{\bar{\xi}}_R\,\hat{\xi}_R\, +Y_{N\nu^c} {\hat{N}}_{S}\hat{\nu}^c\hat{\xi}_R
+ \mu_N {\hat{N}}_{S}{\hat{N}}_{S} \, .
\end{align} 
Notice that, again, the ${\hat{N}}_{S}$ superfield carries lepton number and
that therefore the term $\mu_N$ provides its explicit violation. 

\paragraph{Soft-breaking terms} 
\begin{align} 
L_{\text{SB},W} - L_{\text{SB},W,\text{MSSM}}
= \, & T_{\nu}\,\tilde{L}\,H_u\,\tilde{\nu}^c\,
- B_{\xi} \,\bar{\xi}_R\,\xi_R, +T_{N\nu^c} {\tilde{N}}_{S} \tilde{\nu}^c \xi_R
 + B_N {\tilde{N}}_{S}{\tilde{N}}_{S}
 + \mbox{H.c.} \\ 
L_{\text{SB},\phi} - L_{\text{SB},\phi,\text{MSSM}}
= \,& - {\tilde{N}}_{S}^\dagger m_{N}^2 \tilde{S}
- m_{\xi}^2 |\xi_R|^2 - m_{\bar{\xi}}^2 |\bar{\xi}_R|^2
- (\tilde{\nu}^c)^\dagger {m_{\nu^c}^{2}} \tilde{\nu}^c \\ 
L_{\text{SB},\lambda} = \, & \frac{1}{2}\left(- \lambda_{\tilde{B}'}^{2} M_{B-L}
- 2 \lambda_{\tilde{B}'} \lambda_{\tilde{B}_R} {M}_{R B}
- M_2 \lambda_{{\tilde{W}},{i}}^{2} - M_3 \lambda_{{\tilde{g}},{\alpha}}^{2}
- \lambda_{\tilde{B}_R}^{2} {M}_{R} + \mbox{H.c.}
\right) 
\end{align} 
The term $\lambda_{\tilde{B}'} \lambda_{\tilde{B}_R} {M}_{R B}$ is a consquence
of the presence of two Abelian gauge groups,
see section~\ref{sec:kineticmixing}.

\paragraph{Symmetry breaking}
Since it is assumed that in these models the scale of spontaneous symmetry
breaking to the SM gauge group is near the TeV scale, it is possible to restrict
ourselves to a direct one-step breaking pattern
(\textit{i.e.}, $SU(2)_L \times \URxUBL \rightarrow U(1)_{\text{EM}}$). This
breaking pattern takes place when the Higgs fields in the left and right sectors
receive VEVs. We can parametrise the scalar fields as follows: 
\begin{align} 
H^0_d = & \, \frac{1}{\sqrt{2}} \left(v_d  +  \sigma_d  +i \phi_d\right),
\thickspace 
H^0_u =  \, \frac{1}{\sqrt{2}} \left(v_u + \sigma_u  +  i \phi_u \right) \\
\xi_R = & \, \frac{1}{\sqrt{2}} \left(v_{\xi_R}  +  \sigma_R  +i \phi_R\right),
\thickspace 
\bar{\xi}_R =  \, \frac{1}{\sqrt{2}} \left(v_{\bar{\xi}_R} + \sigma_{\bar{R}}
+ i \phi_{\bar{R}} \right)\, .
\end{align}
It is useful to define the quantities $v_R = v_{\xi_R}^2 + v_{{\bar{\xi}_R}}^2$
and $\tan\beta_R = \frac{v_{{\bar{\xi}_R}}}{v_{\xi_R}}$, in analogy to
$v^2 = v_d^2 + v_u^2$ and to $\tan\beta = \frac{v_u}{v_d}$.

\paragraph{Particle mixing}
Additional mixing effects take place in the gauge and Higgs sectors due to the
additional gauge fields considered, besides the neutrino and sneutrino cases.
The three neutral gauge bosons $B'$, $B_R$ and $W^3$ mix to form three mass
eigenstates: the massless photon, the well-known $Z$ boson, and a $Z'$ boson.
This mixing can be parameterised by a unitary $3 \times 3$ matrix which
diagonalises the mass matrix of the gauge bosons, such as
\begin{equation}
 (\gamma, Z, Z')^T = U^{{\gamma}ZZ^{\prime}} (B', B_R, W^3)^T\, .
\end{equation}
Similarly, this model contains $7$ neutralinos which are an admixture of the
three neutral gauginos, of the two neutral components of the Higgsino doublets
and of the two additional fermions coming from the right sector. The mass
matrix, written in the basis $\left(\lambda_{\tilde{B}'}, \tilde{W}^0,
\tilde{H}_d^0, \tilde{H}_u^0, \lambda_{\tilde{B}_R}, \tilde{\xi_R},
\tilde{\bar{\xi}}_R\right)$, can be diagonalised by a unitary matrix, here
denoted with $Z^N$. In the Higgs sector we choose the mixing basis and rotation
matrices to be, respectively,
\begin{itemize}
 \item Scalar Higgs fields: \( \left(\sigma_{d}, \sigma_{u}, \sigma_{R},
                            \sigma_{\bar{R}}\right)^T\) and $Z^H$
 \item Pseudoscalar Higgs fields: \( \left(\phi_{d}, \phi_{u}, \phi_{R},
                                  \phi_{\bar{R}}\right)^T\) and $Z^A$\, .
\end{itemize}
The neutrino and sneutrino sectors are similar to the case discussed in
section~\ref{sec:inverse_linear}: the scalar fields are decomposed into their
{\it{CP}}-even and odd components according to
eq.~(\ref{eq:decomposition_sneutrinos}). The mass matrices are defined in the
same basis. If all terms of \eq{eq:W_RxBL} are present, the resulting
masses of the light neutrinos are a result of
an inverse \seesaw. The mass matrix is analog to the left matrix given in \eq{eq:InvLinMM}
with $M_R$ replaced by $\frac{1}{\sqrt{2}} Y_{N{\nu}^{c}} v_{\xi_R}$.

\paragraph{Free parameters}
If CMSSM-like boundary conditions are assumed, the following new free parameters
arise:
\begin{equation}
\thickspace Y_{N{\nu}^{c}},
 \thickspace \mu_N,\thickspace B_0, \thickspace \tan\beta_R, \thickspace
 \text{sign}(\mu_{\xi}) \thickspace M_{Z'}
\end{equation}
in addition to those given in \eq{eq:CMSSM}. Here we have assumed that
the parameters $\mu_\xi$ and $B_\xi$ are fixed by the tadpole equations. The
 relationships of the soft trilinear terms to the Yukawa couplings are as
 before, and  $B_N = B_0 \mu_N$.


\subsection{\Seesaw\ in models with \UYxUBL\ gauge sector}
\label{sec:YxB-L}
The final category of models considered here includes an additional $B-L$ gauge
group tensored to the SM gauge groups, \textit{i.e.}
$SU(3)_C \times SU(2)_L \times \UYxUBL$. The corresponding vector superfields
are given in table~\ref{tab:vector_YxBL}. The minimal version of these models
\cite{Khalil:2007dr,FileviezPerez:2010ek,O'Leary:2011yq} extends the MSSM particle content with three generations of
right-handed superfields. Two additional scalars, singlets with respect to SM
gauge interactions but carrying $B-L$ charge, are added to break $U(1)_{B-L}$,
as well as allowing for a Majorana mass term for the right-handed neutrino
superfields. Furthermore, two new lepton fields per generation can be included to
specifically implement the inverse \seesaw\ mechanism \cite{Elsayed:2011de} . All particles and their
quantum numbers are given in table~\ref{tab:chiral_YxBL}. This table contains
also the charge assignment under a $Z_2$ symmetry which is just present in case
of the inverse \seesaw\ model\footnote{Notice that in comparison to Ref.~\cite{Elsayed:2011de}, the charge assignments of the new particles in the inverse \seesaw\ model, as well as the $Z_2$ symmetry, have been redifined for consistency with the similar minimal model of Ref.\cite{Khalil:2007dr}.}. 

\begin{table}[!ht]
\begin{center} 
\begin{tabular}{|c|c|c|c|c|c|} 
\hline \hline 
SF & Spin \(\frac{1}{2}\) & Spin 1 & \(SU(N)\) & Coupling & Name \\ 
 \hline 
\(\hat{B}\) & \(\lambda_{\tilde{B}}\) & \(B\) & \(U(1)\) & \(g_1\)
&\text{hypercharge}\\ 
\(\hat{W}\) & \(\lambda_{\tilde{W}}\) & \(W^-\) & \(\text{SU}(2)\) & \(g_2\)
&\text{left}\\ 
\(\hat{g}\) & \(\lambda_{\tilde{g}}\) & \(g\) & \(\text{SU}(3)\) & \(g_3\)
&\text{color}\\ 
\(\hat{B}'\) & \(\lambda_{\tilde{B}{}'}\) & \(B'\) & \(U(1)\) & \(g_{B}\)
&\text{B-L}\\ 
\hline \hline
\end{tabular} 
\end{center} 
\caption{\label{tab:vector_YxBL}
Vector superfields appearing in models with \UYxUBL\ gauge sector.}
\end{table}

\begin{table}[!ht]
\begin{center} 
\scalebox{0.91}{
\begin{tabular}{|c|c|c|c|c|c|c|c|} 
\hline \hline 
SF & Spin 0 & Spin \(\frac{1}{2}\) & Generations & \((U(1)_Y\otimes\,
\text{SU}(2)\otimes\, \text{SU}(3)\otimes\, U(1)_{B-L}\) & $Z_2$ inverse SS\\ 
\hline 
\hline
\multicolumn{6}{|c|}{Matter fields (fermionic components have positive
$R$-parity)}
\\
\hline \hline
\(\hat{Q}\) & \(\tilde{Q}=\left(\begin{array}{c} \tilde{u}_L \\ \tilde{d}_L
\end{array} \right) \) & \(Q=\left(\begin{array}{c} u_L \\ d_L \end{array}
\right)\) & 3 & \((\frac{1}{6},{\bf 2},{\bf
3},\frac{1}{6}) \)  & $+$\\ 
\(\hat{L}\) & \(\tilde{L}=\left(\begin{array}{c} \tilde{\nu}_L \\ \tilde{e}_L
\end{array} \right)\) & \(L=\left(\begin{array}{c} \nu_L \\ e_L \end{array}
\right)\) & 3 & \((-\frac{1}{2},{\bf 2},{\bf
1},-\frac{1}{2}) \)  & $+$\\ 
\(\hat{d}^c\) & \(\tilde{d}^c\) & \(d^c\) & 3 & \((\frac{1}{3},{\bf
1},{\bf \overline{3}},-\frac{1}{6}) \)  & $+$\\ 
\(\hat{u}^c\) & \(\tilde{u}^c\) & \(u^c\) & 3 & \((-\frac{2}{3},{\bf
1},{\bf \overline{3}},-\frac{1}{6}) \)  & $+$\\ 
\(\hat{e}^c\) & \(\tilde{e}^c\) & \(e^c\) & 3 & \((1,{\bf 1},{\bf
1},\frac{1}{2}) \)  & $+$\\ 
\(\hat{\nu}^c\) & \(\tilde{\nu}^c\) & \(\nu^c\) & 3 & \((0,{\bf 1},{\bf
1},\frac{1}{2}) \) & $+$\\ 
\hline 
\hline
\multicolumn{6}{|c|}{Higgs fields (scalar components have positive
$R$-parity)}\\
\hline \hline
\(\hat{\eta}\) & \(\eta\) & \(\tilde{\eta}\) & 1 & \((0,{\bf 1},{\bf
1},-1) \)  & $-$\\ 
\(\hat{\bar{\eta}}\) & \(\bar{\eta}\) & \(\tilde{\bar{\eta}}\) & 1 &
\((0,{\bf 1},{\bf 1},+1)\)  & $-$\\ 
\(\hat{H}_d\) & \(H_d\) & \(\tilde{H}_d\) & 1 & \((-\frac{1}{2},{\bf 2},
{\bf1},0) \)  & $+$\\ 
\(\hat{H}_u\) & \(H_u\) & \(\tilde{H}_u\) & 1 & \((\frac{1}{2},{\bf 2},
{\bf1},0) \) & $+$\\ 
\hline 
\hline
\multicolumn{6}{|c|}{Additional field for inverse \seesaw\ (fermionic components
have positive $R$-parity)}\\
\hline \hline
\({\hat{N}}_{S}\) & \({\tilde{N}}_{S}\) & \(N_{S}\) & $n_{N_S}$ &
\((0,{\bf 1},{\bf 1},-\frac{1}{2}) \) & $-$\\ 
\({\hat{N}}_{S}^{\prime}\) & \({\tilde{N}}_{S}^{\prime}\) & \(N_{S}^{\prime}\)
& $n_{N_S}$ & \((0,{\bf 1},{\bf 1},\frac{1}{2}) \) & $+$\\ 
\hline \hline
\end{tabular}}
\end{center} 
\caption{\label{tab:chiral_YxBL}
Chiral Superfields appearing in models with \UYxUBL\ gauge sector. The
minimal particle content is needed for \seesaw\ type I, the additional fields
 can be used to incorporate an inverse \seesaw. The $SU(2)_L$ doublets are named as
 in table~\ref{tab:chiral_RxBL}. The $Z_2$ in the last column is not present in
 the minimal model but just in the model with inverse \seesaw. The number of
 generations of ${\hat{N}}_{S}^{\prime}$ must match those of ${\hat{N}}_{S}$ for
 anomaly cancellation.}
\end{table}

The additional terms for the superpotential in comparison to the MSSM read 
\begin{align} 
\label{eq:W_YxBL}
\nonumber W - W_{\text{MSSM}} =
& \, Y_{\nu}{\hat{l}}{\hat{H}}_{u}{\hat{\nu}}^{c}
+ \left\{ \begin{array}{c c}
Y_{{\eta}{\nu}^{c}}{\hat{\nu}}^{c}{\hat{\eta}}{\hat{\nu}}^{c}
- {\mu}_{{\eta}}{\hat{\eta}}{\hat{{\bar{\eta}}}} & \mbox{minimal \seesaw}\\
Y_{IS}{\hat{\nu}}^{c}{\hat{\eta}}{\hat{N}}_{S}
+ {\mu}_{N}{\hat{N}}_{S}{\hat{N}}_{S} & \mbox{inverse \seesaw}
\end{array} \right. .
\end{align}

The extra parity is not applicable to the minimal case, thus the \seesaw\ term
$Y_{{\eta}{\nu}^{c}}{\hat{\nu}}^{c}{\hat{\eta}}{\hat{\nu}}^{c}$ is allowed,
whereas it is forbidden in the inverse \seesaw\ model. Instead, in the inverse
\seesaw\ model the ${\mu}_{N}{\hat{N}}_{S}{\hat{N}}_{S}$ term plays an important
role, even though since ${\hat{N}}_{S}$ carries $B-L$ charge, this term
violates $B-L$ charge conservation similarly to the terms in
\eq{eq:W_lin_inv} and \eq{eq:W_RxBL} which break lepton number.
This term is assumed to come from higher-order effects. (A similar term for the
${\hat{N}}_{S}^{\prime}$ field is possible but not relevant, since the
$N_{S}^{\prime}$ does not take part in the mixing with the neutrinos.) A
possible bilinear term ${\hat{N}}_{S} {\hat{N}}_{S}^{\prime}$ is forbidden by
the $Z_2$ symmetry given in the last column of Table~\ref{tab:chiral_YxBL}. This
discrete symmetry also forbids terms like ${\hat{N}}_{S} {\hat{N}}_{S} \eta$ or
${\hat{N}}_{S}^{\prime} {\hat{N}}_{S}^{\prime} \bar{\eta}$ as well as the
${\mu}_{{\eta}}$-term that is necessary to obtain a pure inverse \seesaw\
scenario. The
additional soft-breaking terms are written as follows:
\begin{align} 
L_{\text{SB},W} - L_{\text{SB},W,\text{MSSM}}
 = & T_{\nu}{\tilde{l}} H_{u} {\tilde{\nu}}^{c}
+ \left\{ \begin{array}{c c}
T_{{\eta}{\nu}^{c}}{\tilde{\nu}}^{c}{\eta}{\tilde{\nu}}^{c}
- B_{{\eta}}{\eta}{\bar{{\eta}}} & \mbox{minimal see-saw}\\
T_{IS}{\tilde{\nu}}^{c}{\eta}{\tilde{N}}_{S}
+ B_{N}{\tilde{N}}_{S}{\tilde{N}}_{S} & \mbox{inverse see-saw}
\end{array} \right\} + \mbox{H.c.}\\
L_{\text{SB},\phi} - L_{\text{SB},\phi,\text{MSSM}} = & \, - m_{\eta}^2 |\eta|^2
- m_{\bar{\eta}}^2
|\bar{\eta}|^2 - ( {\tilde{N}}_{S}^{\ast} m_{N}^{2} {\tilde{N}}_{S}
+ {\tilde{N}}_{S^{\prime}}^{\ast} m_{N^{\prime}}^{2} {\tilde{N}}_{S}^{\prime}
\mbox{ in inverse see-saw} ) \\
L_{\text{SB},\lambda} = \, & \frac{1}{2}\left(- \lambda_{\tilde{B}}^{2} M_1
- \lambda_{\tilde{B}} \lambda_{\tilde{B}{}'} {M}_{B B'} - M_2
\lambda_{{\tilde{W}},{i}}^{2} - M_3 \lambda_{{\tilde{g}},{\alpha}}^{2}
- \lambda_{\tilde{B}{}'}^{2} {M}_{BL} + \mbox{H.c.} \right) 
\end{align}

To break $SU(2)_L \times \UYxUBL$ to $U(1)_{em}$, the neutral
MSSM Higgs fields and the new SM scalar singlets acquire VEVs:
\begin{align} 
H^0_d = & \, \frac{1}{\sqrt{2}} \left(v_d + \sigma_d + i \phi_d\right),
\thickspace 
H^0_u = \, \frac{1}{\sqrt{2}} \left(v_u + \sigma_u + i \phi_u \right) \\
\eta = & \, \frac{1}{\sqrt{2}} \left(\sigma_{\eta} + v_{\eta} + i \phi_{\eta}
\right), \thickspace 
\bar{\eta} =  \, \frac{1}{\sqrt{2}} \left( \sigma_{\bar{\eta}} + v_{\bar{\eta}}
+ i \phi_{\bar{\eta}} \right)
\end{align} 
We also define here 
\begin{equation}
 \label{eq:v_tb}
v^2 = v_d^2 + v_u^2, \hspace{1cm} \tan\beta = \frac{v_u}{v_d}
\end{equation}
as well as 
\begin{equation}
v_x = v_{\eta_R}^2 + v_{\bar{\eta}_R}^2, \hspace{1cm}
\tan\beta' = \frac{v_{\bar{\eta}}}{v_{\eta}}
\end{equation}
The left- and right-handed sneutrinos are decomposed into their scalar and
pseudoscalar components according to \eq{eq:decomposition_sneutrinos}.
Similarly, in the inverse \seesaw\ model, the scalar component of
 ${\hat{N}}_{S}$ reads
\begin{equation}
{\tilde{N}}_{S} = \frac{1}{\sqrt{2}} \left( {\sigma}_{S} + i {\phi}_{S} \right)
\end{equation}
The additional mixing effects which take place in this model are similar to the
case of \URxUBL\ discussed in section~\ref{sec:RxB-L}. In the
gauge sector three neutral gauge bosons appear which mix to give rise to the
massless photon, the $Z$ boson, and a $Z'$ boson:
\begin{equation}
 (\gamma, Z, Z')^T = U^{\gamma Z} (B,W^3,B')^T\, .
\end{equation}
Notice that when the kinetic mixing is neglected, the $B'$ field decouples and
the mass matrix of the gauge bosons becomes block diagonal, where the upper
$2\times 2$ block reads as in the SM. In the matter sector we choose the basis
and the mixing matrices which diagonalise the mass matrices respectively as
follows:
\begin{itemize}
 \item Neutralinos: $\left(\lambda_{\tilde{B}}, \tilde{W}^0, \tilde{H}_d^0,
                    \tilde{H}_u^0, \lambda_{\tilde{B}{}'}, \tilde{\eta},
                    \tilde{\bar{\eta}}\right)^T$ and $Z^N$
 \item Scalar Higgs fields: \( \left(\sigma_{d}, \sigma_{u}, \sigma_{\eta},
                            \sigma_{\bar{\eta}}\right)^T\) and $Z^H$
 \item Pseudoscalar Higgs fields: \( \left(\phi_{d}, \phi_{u}, \phi_{\eta},
                                  \phi_{\bar{\eta}}\right)^T\) and $Z^A$
 \item Scalar sneutrinos: $\left(\sigma_{L},\sigma_{R}, \sigma_{S}\right)^T$ and
                          $Z^{\sigma v}$
 \item Pseudoscalar sneutrinos: $\left(\phi_{L},\phi_{R}, \phi_{S}\right)^T$ and
                                $Z^{\phi v}$
 \item Neutrinos: $\left(\nu_L,\nu^c, N_{S}\right)^T$ and $U^V$
\end{itemize}
where the ${\sigma}_{S}, {\phi}_{S}$, and $N_{S}$ are only present in the
inverse \seesaw\ case. The neutrino mass matrices for the minimal and inverse
\seesaw\ cases can be written as
\begin{equation}
 \left(\begin{array}{cc}
0 & \frac{v_u}{\sqrt{2}}Y_\nu \\
\frac{v_u}{\sqrt{2}}Y^T_\nu & \frac{v_{\eta}}{\sqrt{2}} Y_{{\eta}{\nu}^{c}}
\end{array} \right) \mbox{ or }
\left(\begin{array}{ccc}
0 & \frac{v_u}{\sqrt{2}}Y_\nu & 0 \\
\frac{v_u}{\sqrt{2}}Y^T_\nu & 0 & \frac{v_{\eta}}{\sqrt{2}} Y_{IS} \\
0 & \frac{v_{\eta}}{\sqrt{2}} Y^T_{IS} &  \mu_N \end{array} \right)
\end{equation}
respectively.
In the minimal case, the \seesaw\ of type I is recovered. The light neutrino
 mass matrix can be approximated in this case by
\begin{equation}
 m_\nu \simeq -\frac{v_u^2}{\sqrt{2}v_\eta} Y_\nu Y^{-1}_{{\eta}{\nu}^{c}} Y_\nu
\,.
\end{equation}
In the inverse \seesaw\ case, the light neutrino masses can be written as
\begin{equation}
 m_\nu \simeq -\frac{v_u^2}{2} Y^T_\nu (Y^T_{IS})^{-1} \mu_N Y_{IS} Y_\nu
\end{equation}

\paragraph{Free parameters}
If the minimum conditions for the vacuum are solved with respect to $\mu$,
$B_\mu$, ${\mu}_{{\eta}}$ and $B_{\eta}$, the following parameters can be
treated as free in addition of those given in \eq{eq:CMSSM}:
\begin{equation}
Y_{{\eta}{\nu}^{c}}, \thickspace Y_{IS}, \thickspace \mu_N, \thickspace B_0,
\thickspace \tan\beta', \thickspace \text{sign}({\mu}_{{\eta}}),
\thickspace M_{Z'}\, 
\end{equation}
where we have used again $B_N = B_0 \mu_N$.

\section{TOWARDS SLHAv3}

The purpose of this paper is to define the extensions of the SUSY Les Houches
Accords required to incorporate the models described above. In this context and
with respect to the implementation of further models in future, it is helpful to
propose some general rules for the naming of blocks and the allocation of PDG
particle codes, which we present below.

We would like to point out that the SLHA conventions allow for redundant
information to be contained in additional blocks. Hence, in the interests of
backward compatibility, we propose that if the following names for the blocks
are used, but in cases where the old SLHA1/2 blocks would also suffice, they
should also be written in addition, \eg\ if there are only four
 neutralinos, \texttt{NMIX} should be written as well as \texttt{NEUTRALINORM},
 both containing the same information.

\paragraph{Block names for input and output}
Blocks used to give parameters as input end with ``\texttt{IN}'', while the
corresponding values as output are written in blocks without the ending
``\texttt{IN}''. For example, the gauge couplings at the output scale are given
in the \texttt{GAUGE} block, while one could define them as input at the input
scale with the \texttt{GAUGEIN} block.

\paragraph{Block names for mixing matrices}
In SLHA 2 the neutralino and Higgs mixing matrices of the MSSM and NMSSM were
named differently. However, the information about the current model is already
given in the block {\tt MODSEL} and the renaming is thus redundant. Therefore,
in order to prevent a confusing amount of names for mixing matrices in different
models we propose to use always the same names for the following mixing matrices
regardless of their dimension:
\begin{itemize}
 \item Scalar Higgs mixing matrix: {\tt SCALARRM} 
 \item Pseudoscalar Higgs mixing matrix: {\tt PSEUDOSCALARRM} 
 \item Charged Higgs mixing matrix: {\tt CHARGEDSCALARRM} 
 \item Neutralino mixing matrix: {\tt NEUTRALINORM} (corresponds to {\tt NMIX}
       in SLHA1/2)
 \item Chargino mixing matrices: {\tt CHARGINOPLUSRM} and {\tt CHARGINOMINUSRM}
       (corresponds to {\tt VMIX} and {\tt UMIX} respectively in  SLHA1/2)
 \item Up-squark mixing matrix: {\tt UPSQUARKRM} (corresponds to {\tt USQMIX} in
        SLHA1/2)
 \item Down-squark mixing matrix: {\tt DOWNSQUARKRM} (corresponds to
       {\tt DSQMIX} in  SLHA1/2)
 \item Charged slepton mixing matrix: {\tt CHARGEDSLEPTONRM} (corresponds to
       {\tt SELMIX} in  SLHA1/2)
 \item Sneutrino mixing matrix: {\tt SNEUTRINORM} (corresponds to {\tt SNUMIX}
       in the MSSM). In the case of splitting the sneutrinos into real and
       imaginary parts, {\tt SNEUTRINOEVENRM} for the CP-even and
       {\tt SNEUTRINOODDRM} for the CP-odd states should be used
       (corresponding to {\tt SNSMIX} and {\tt SNAMIX} in SLHA1/2) 
\end{itemize}
To distinguish these names from the SLHA1/2-specific names we have always used
the suffix {\tt RM} for {\tt R}otation {\tt M}atrix.

\paragraph{Block names for couplings}
For the naming of blocks which correspond to couplings and soft-breaking terms
we propose to use names which already give information about the meaning of the
parameter: the new names contain abbreviations for the involved fields and start
with a prefix to assign the meaning of the parameters. The different prefixes
should be
 \begin{itemize}
   \item {\tt Y}: trilinear superpotential coupling
   \item {\tt T}: trilinear softbreaking coupling
   \item {\tt M}: bilinear superpotential coupling
   \item {\tt B}: bilinear softbreaking coupling
   \item {\tt L}: linear superpotential coupling
   \item {\tt S}: linear softbreaking coupling
   \item {\tt M2}: soft breaking scalar mass-squared
 \end{itemize}
Because of the prefixes which already give information about the spin of the
involved particles, it is not necessary to distinguish between fermions and
scalars when naming the blocks. However, this requires the convention that all
Majorana mass terms for fermions from chiral superfields appear in the
superpotential, leaving only bilinear mass terms for the scalars in the soft
SUSY-breaking Lagrangian. We propose the following names:
 \begin{itemize}
  \item MSSM: $H_u$: {\tt HU}, $H_d$: {\tt HD}, $d^c$: {\tt D}, $u^c$: {\tt U},
              $q$: {\tt Q}, $e^c$: {\tt E}, $l$: {\tt L}
  \item \Seesaw\ types I-III: $\nu^c$: {\tt NUR}, $S$: {\tt 15S},
                              $\bar{S}$: {\tt 15SB}, $Z$: {\tt 15Z},
                              $\bar{Z}$: {\tt 15ZB}, $T$: {\tt 15T},
                              $\bar{T}$: {\tt 15TB}, $G$: {\tt 24G},
                              $W$: {\tt 24W}, $X$: {\tt 24X},
                              $\bar{X}$: {\tt 24XB}, $B$: {\tt 24B}
  \item inverse and linear \seesaw: $S$: {\tt SL}
  \item \URxUBL\ models: $\xi_R$: {\tt CR}, $\bar{\xi}_R$: {\tt CRB}
  \item \UYxUBL\ models: $\eta$: {\tt BIL}, $\bar{\eta}$: {\tt BILB},
                         $N_{S}$: {\tt NS}, $N_{S}^{\prime}$: {\tt NSP} 
 \end{itemize}
Using these conventions, the block {\tt YU} for the up-type Yukawa coupling in
the MSSM would be replaced by {\tt YHUQU} and the block name for the $\mu$-term
would be {\tt MHUHD}.

\paragraph{EXTPAR and MSOFT}
 In the MSSM the input parameters for the gaugino masses and the Higgs
soft-breaking masses are given in {\tt EXTPAR}. However, their output is given
in {\tt MSOFT}. Similary, the singlet couplings in the NMSSM are defined by
{\tt EXTPAR} but their output is in {\tt NMSSMRUN}. This is in some conflict
with the general rule to use always the the input name plus {\tt IN} as output.
Therefore, we propose that all new one-dimensional soft-breaking parameters and
couplings involved in the \seesaw\ models presented here are not given in
{\tt EXTPAR} as input, but rather in the given output with the prefix {\tt IN}.

\paragraph{PDG particle numbering scheme}
We would like to introduce a consistent numbering scheme for the additional
elementary particles introduced by such models, such that further particles may
be added without worrying about accidentally using a pre-existing code, which
also allows one to gather some information about the particle. While the
proposed scheme would give new numbers to particles which already have codes,
we restrict ourselves just to giving our new particles unique codes, while
hoping that codes compliant with the new standard would be able to read either
the old code or the new code for particles. The proposal is a signed nine digit
integer for each mass eigenstate, which should easily fit in a 32-bit integer.

We note that the existing scheme is already mildly inconsistent, insofar as
adding a fourth generation already has a PDG numbering scheme (17 for the extra
neutrino and 18 for the $\tau^{\prime}$), yet SLHA2 uses 100017 and 100018 for
CP-odd sneutrinos which are degrees of freedom from the first three generations.
We note also that the Flavor Les Houches Accord \cite{Mahmoudi:2010iz} uses 17
and 18 for summing over the three Standard Model generations.

Since the particle code is a signed integer, pairs of conjugate particles are
 assigned the same code with different signs. We propose a convention for which
 particle gets the positive sign:
\begin{itemize}
\item If the particle is self-conjugate, there is only the positive sign. If the
      fermion is Majorana, it only has positive sign. If the scalar part of the
      superfield can be written in terms of self-conjugate CP eigenstates, it
      should be. If it cannot, the scalar field should be assigned a sign
      according to the rules below.

\item If the particle has non-zero electric charge, the state with positive
      electric charge is taken as the particle (hence the positron is taken as
      the particle).

\item If the particle is electrically neutral, but has baryon number $B$ or
      lepton number $L$, the state with positive $B - L$ is taken as the
      particle (hence antineutrinos are taken as the particle).

\item If the particle has $B - L = 0$ according to traditional assignment, but
      is still baryon- or lepton-like, a ``temporary'' $B - L$ is assigned. If
      the particle has color charge a temporary $B$ is first assigned by finding
      the combination of triplets and antitriplets which could combine to form
      the particle's representation: adding $+1/3$ for each triplet and $-1/3$
      for each antitriplet, the combination which has the {\em lowest magnitude}
      of temporary $B$ is assigned; \eg\ an octet may be formed by a
      triplet with an antitriplet, giving $B = 0$, or by three triplets, giving
      $B = 1$, or three antitriplets, giving $B = -1$: in this case, $B = 0$ as
      the lowest $|B|$ is assigned. This temporary $B - L$ is only for the
      purposes of determining which state is taken to be the particle with
      positive code.

\item If the neutral, colorless particle has no natural assignment by
      (temporary) $B - L$, there are still a few cases:
 \begin{itemize}
 \item the fermion is massless: the left-handed fermion is given the positive
       code, thus in supersymmetric models, the scalars left-chiral superfields
       also are given positive codes if their fermions are massless.
 \item the fermion has a Dirac mass: this does not occur in the MSSM, NMSSM, or
       any of the extended models described here. However it is conceivable. We
       propose that the model builder is responsible for deciding to assign a
       temporary lepton number to one of the fields and thus fix the convention.
 \end{itemize}
\end{itemize}

Under this scheme, quarks and squarks would have the same signs as they already
 have in the PDG conventions, leptons would have the opposite sign (\eg\ the
 muon would have a negative code), and charginos and $W$ bosons would have the
 same signs.

Once the sign is fixed, the digits are as follows:
\begin{itemize}
\item[1st digit:] 1 if it is a mass eigenstate which has an admixture of a
                  Standard Model gauge eigenstate (including the Higgs doublet),
                  2 if it does not mix with the
                  Standard Model particles. 
\item[2nd digit:] twice the spin of the particle. 
\item[3rd digit:] the CP nature: complex bosons (spin 0, 1, or 2) and Dirac
                  fermions have 0, while (massless or massive) scalar bosons,
                  massless vector bosons, massless tensor bosons, and
                  Majorana fermions have 1. Pseudoscalar bosons (massless or
                  massive) and massive vector and tensor bosons have 2.
\item[4th and 5th digits:] a 2-digit number for the $SU(3)$ representation;
                           $SU(3)$ singlets have 00; representations up to
                           dimension 64 have either the Dynkin labels for an
                           unbarred representation, or 99 minus the Dynkin
                           labels for the barred representation. These
                           representations are enumerated in
                           \cite{Slansky:1981yr}. Any particles with $SU(3)$
                           representation that does not fit into this scheme are
                           assigned 99 (for instance those of dimension 65 or
                           greater).
\item[6th and 7th digits:] a 2-digit number for the electric charge; six times
                           the absolute value of the electric charge, relative
                           to the electron, up to a maximum of $98/6$. Any
                           particle with a charge that does not fit into
                           this scheme is assigned 99. For example, an electron
                           would have 06, while a down quark would have 02, and
                           a doubly-charged scalar would have 12.
\item[8th and 9th digits:] a generation number; 01 should be given to the
                           lightest particle of any group which share the same
                           first seven digits, 02 to the 2nd-lightest, and so
                           on.
\end{itemize}

Consequently what might be considered to be the same model with or without
$R$-parity will have different codes for some of its mass eigenstates. In the
MSSM, with $R$-parity there would be three charged antileptons 110000601,
110000602, and 110000603, and two charginos 210000601 and 210000602, while
without $R$-parity there would be five charged antileptons 110000601, 110000602,
110000603, 110000604 and 110000605, and no 210000601 or 210000602.

Additionally, the codes for the neutrinos will depend on whether they are Dirac
or Majorana in the considered model.

\section{EXTENSIONS TO SLHA}


In this section we describe the implementation of the models presented in 
section~\ref{sec:seesawmechanisms} using the conventions defined in the previous section. Note that all
parameters can be implemented in complex forms and the
corresponding information can be passed by using the corresponding blocks
starting with ``{\tt IM}'' \cite{Allanach:2008qq}.

\subsection{Block format}
\label{sec:blockformat}
In the following, all blocks with a single index are to be written in the
FORTRAN format\\ (1x,I5,3x,1P,E16.8,0P,3x,'\#',1x,A) (the same format as the
 SLHA1 blocks \texttt{HMIX}, \texttt{GAUGE}, \textit{etc.}). These blocks will
 be denoted as ``rank one''. All blocks with two indices are to be written in
 the FORTRAN format (1x,I2,1x,I2,3x,1P,E16.8,0P,3x,'\#',1x,A) (the same format
 as the SLHA1 blocks \texttt{NMIX}, \texttt{UMIX}, \textit{etc.}). These blocks
 will be denoted as ``rank two''. All blocks with two indices are to be written
 in the FORTRAN format (1x,I2,1x,I2,1x,I2,3x,1P,E16.8,0P,3x,'\#',1x,A) (the same
 format as the SLHA2 blocks \texttt{RVLAMLLE}, \texttt{RVLAMLQD},
 \textit{etc.}). These blocks will be denoted as ``rank three''.

\subsection{Blocks required for each model}
\label{sec:blocksformodels}
Each model requires the presence of certain blocks. Some blocks are common to
several models. We summarize the blocks needed for each of the models
described in section \ref{sec:seesawmechanisms} under the entry for selecting
this model in the \texttt{MODSEL} block \ref{sec:blockmodsel}.

\subsection{Extra Flags In Existing Blocks}
\label{sec:extendingexistingblocks}

\subsubsection{Block \texttt{MODSEL}}
\label{sec:blockmodsel}
Flag 3 (particle content) has further switches, arranged in groups: 11X is for
 effective realizations, 12X non-effective models, 13X for \URxUBL\ models, and
 14X is for \UYxUBL\ models, the initial digit 1 indicating \seesaw. 2XY might
 be used for fourth-generation models, 3XY for another set of models, and so on.
 In addition to the sets of required blocks listed for each model below, the
 \texttt{SEESAWGENERATIONS} block must also be given, and the appropriate flags
 set.
\begin{itemize}
\item Effective models
 \begin{itemize}
 \item[110:] Most general $SU(5)$-invariant effective \seesaw\ (combination of
             types I-III); all the blocks required for types I-III below are
             required.
 \item[111:] type I \seesaw; the following blocks are required:
             \texttt{MNURNURIN, BNURNURIN, M2NURNURIN, YNURLHUIN, TNURLHUIN}.
 \item[112:] type II \seesaw\ ($SU(5)$ version); the following blocks are
             required:
             \texttt{M15S15SBIN, B15S15SBIN, M215SIN, M215SBIN,
             M15T15TBIN, B15T15TBIN, M215TIN, M215TBIN, M15Z15ZBIN,
             B15Z15ZBIN, M215ZIN, M215ZBIN, YD15SDIN, TD15SDIN, YL15TLIN,
             TL15TLIN, YD15ZLIN, TD15ZLIN, YHD15THDIN, THD15THDIN, YHU15TBHUIN,
             THU15TBHUIN}.
             Any or all of the blocks
             \texttt{M15S15SBIN, M15T15TBIN, M15Z15ZBIN}
             may be left absent if the block \texttt{M15IN} is present, and the
             missing terms are taken as copies of \texttt{M15IN}. Similarly for
             \texttt{YD15SDIN, YL15TLIN, YD15ZLIN} and \texttt{Y15IN}, and
             \texttt{TD15SDIN, TL15TLIN, TD15ZLIN} and \texttt{Y15IN} times
             $A_{0}$.
 \item[113:] type III \seesaw\ ($SU(5)$ version); the following blocks are
             required:
             \texttt{M24W24WIN, B24W24WIN, M224WIN, M24G24GIN, B24G24GIN,
             M224GIN, M24B24BIN, B24B24BIN, M224BIN, M24X24XBIN,
             B24X24XBIN, M224XIN, M224XBIN, YHU24BLIN, THU24BLN,
             YHU24WLIN, THU24WLN, YHU24BLIN, THU24BLN, YHU24XBDIN, THU24XBDIN}.
             Any or all of the blocks
             \texttt{M24W24WIN, M24G24GIN, M24B24BIN, M24X24XBIN} may be left
             absent if the block \texttt{M24IN} is present, and the missing terms
             are taken as copies of \texttt{M24IN}. Similarly for
             \texttt{B24W24WIN, B24G24GIN, B24B24BIN, B24X24XBIN} and
             \texttt{B24IN}; \texttt{YHU24BLIN, YHU24WLIN, YHU24XBDIN} and
             \texttt{Y24IN}; and \texttt{THU24BLIN, THU24WLIN, THU24XBDIN} and
             \texttt{Y24IN} times $A_{0}$.
 \item[114:] type II \seesaw\ (minimal version, only triplets); the following
             blocks are required:
             \texttt{M15T15TBIN, B15T15TBIN, M215TIN, M215TBIN, YL15TLIN,
             TL15TLIN, YHD15THDIN, THD15THDIN, YHU15TBHUIN, THU15TBHUIN}.
             If \texttt{TL15TLIN} is absent, \texttt{YL15TLIN} times $A_{0}$
             is used in its place.
 \item[115:] type III \seesaw\ (minimal version, only triplets); the following
             blocks are required:
             \texttt{M24W24WIN, B24W24WIN, M224WIN, YHU24WLIN, THU24WLN}.
             If \texttt{THU24WLN} is absent, \texttt{YHU24WLIN} times $A_{0}$
             is used in its place.
 \end{itemize}
\item Linear and inverse \seesaw:
 \begin{itemize}
  \item[120]: combined inverse and linear \seesaw; the following blocks are
              required: \texttt{MNURSIN, BNURSIN, MNSNSIN, BNSNSIN, M2NSIN,
              M2NURIN, YNURLHUIN, TNURLHUIN}.
              Purely inverse or linear \seesaw\ is specified by appropriate
              zeroes in the block entries.
 \end{itemize}
\item \URxUBL:
 \begin{itemize}
 \item[131]: reserved for minimal \seesaw
 \item[132]: inverse \seesaw; the following blocks are required in
             addition to those required by the \URxUBL\ minimal \seesaw:
             \texttt{MNSNSIN, BNSNSIN, M2NSIN, YNSNURCRIN, TNSNURCRIN}.
 \item[133]: reserved for linear \seesaw
 \end{itemize}
\item \UYxUBL:
 \begin{itemize}
 \item[141]: minimal \seesaw; the following blocks are required:
             \texttt{MBILBILBIN, BBILBILBIN, M2NURIN, M2BILIN,
             M2BILBIN, YNURLHUIN, TNURLHUIN, YNURBILNURIN, TNURBILNURIN}.
 \item[142]: inverse \seesaw; the following blocks are required in
             addition to those required by the \UYxUBL\ minimal \seesaw:
             \texttt{MNSNSIN, BNSNSIN, M2NSIN, YNSNURBILIN, TNSNURBILIN}.
 \end{itemize}
\end{itemize}
In addition, the \URxUBL\ and \UYxUBL\ models require that the appropriate
entries in \texttt{GAUGEIN, MSOFTIN, MINPAR}, and \texttt{EXTPAR} are set
correctly.

\subsubsection{Block \texttt{MINPAR}}
\label{sec:blockminpar}
We extend the \texttt{MINPAR} block to include the input parameters necessary
for the addition of an extra Abelian gauge group. We do not reproduce the
existing entries here since there are quite a few. We only show those that are
new and those that have new interpretations.
\begin{itemize}
\item[2:] The common soft mass term for all the the gauginos.

\item[6:] The cosine of the phase of the ${\mu}^{\prime}$ parameter.

\item[7:] The ratio of the two vacuum expectation values that either break
          \URxUBL to $U(1)_{Y}$ or break $U(1)_{B-L}$ leaving $U(1)_{Y}$ intact.

\item[8:] The mass of the $Z^{\prime}$ boson.

\item[9:] The common bilinear mass parameter $B_0$.
\end{itemize}

\subsubsection{Block \texttt{GAUGE}}
\label{sec:blockgauge}
We extend the \texttt{GAUGE} block to include the couplings necessary for the
addition of an extra Abelian gauge group. We reproduce the existing entries here
for completeness.
\begin{itemize}
\item[1:] The coupling $g_{1}(Q)$ (also known as $g^{\prime}(Q)$ in some
          conventions) of the $U(1)_{Y}$ gauge in models which have it, or the
          coupling $g_{R}(Q)$ of the $U(1)_{R}$ gauge in such models where the
          $U(1)_{R}$ in combination with another $U(1)$ gauge breaks down to
          $U(1)_{Y}$.

\item[2:] The coupling $g_{2}(Q)$ (also known as $g(Q)$ in some conventions) of
          the $SU(2)_{L}$ gauge.

\item[3:] The coupling $g_{3}(Q)$ of the $SU(3)_{c}$ gauge.

\item[4:] The coupling $g_{BL}(Q)$ of the $U(1)_{B-L}$ gauge.

\item[14:] The off-diagonal coupling $g_{T}(Q)$ of the two $U(1)$ gauges in the
           triangle basis.
\end{itemize}

\subsubsection{Block \texttt{MSOFT}}
\label{sec:blockmsoft}
We extend the \texttt{MSOFT} block to include the soft mass terms necessary for
the addition of an extra Abelian gauge group. We do not reproduce the existing
entries here since there are quite a few. We only show those that are new and
those that have new interpretations.
\begin{itemize}
\item[1:] The soft mass term for the gaugino of the $U(1)_{Y}$ gauge in models
          which have it, or the gaugino of the $U(1)_{R}$ gauge in such models
          where the $U(1)_{R}$ in combination with another $U(1)$ gauge breaks
          down to $U(1)_{Y}$.

\item[4:] The soft mass term for the gaugino of the $U(1)_{B-L}$ gauge.

\item[5:] The soft mass term mixing the gauginos of the two $U(1)$ gauges.
\end{itemize}

\subsubsection{Block \texttt{EXTPAR}}
\label{sec:blockextpar}
We extend the \texttt{EXTPAR} block to include the parameters necessary for
the addition of an extra Abelian gauge group. We do not reproduce the existing
entries here since there are quite a few. We only show those that are new and
those that have new interpretations. Bear in mind that just as in the MSSM,
 contradictory inputs should not be given.
\begin{itemize}
\item[124:] The tree-level mass-squared $m_{{\phi}_{R}}^{2}$ of the pseudoscalar
            formed by the doublets breaking $SU(2)_R$.
\item[126:] The pole mass $m_{{\phi}_{R}}$ of the pseudoscalar formed by the
            doublets breaking $SU(2)_R$.
\end{itemize}

\subsection{Extended blocks superseding existing blocks}
\label{sec:supersedingexistingblocks}

\subsubsection{Block \texttt{XPMNSRM} (rank two), superseding \texttt{UPMNS}
 (rank one)}
\label{sec:blockneutrinorm}
In the super-PMNS basis the mass matrices of the charged
leptons and the three light neutrinos are diagonal and the
relevant generation mixing information is given in the PMNS matrix.
In the SLHA2 the input block \texttt{UPMNSIN} was defined in terms
of three mixing angles and three phases \cite{Allanach:2008qq}
whereas for the output the complete $3\times 3$ mixing has to be
given in the block \texttt{UPMNS}. In general more than 3 neutrinos contribute
and thus we propose that similarly the generalized PMNS matrix, which
is rectangular, shall be given for both input and output.
In the basis where, as above, the charged lepton and the extended neutrino
mass matrices are diagonal, this matrix corresponds to the coupling
between the left-handed charged leptons with the $W$ boson and the neutrinos
divided by $g/\sqrt{2}$.

\subsection{New blocks}
\label{sec:newblocks}
Blocks used to give parameters as input end with ``\texttt{IN}'', while the
corresponding values as output are written in blocks without the ending
``\texttt{IN}''.

\subsubsection{Block \texttt{SEESAWGENERATIONS} (rank one)}
This block specifies the number of generations of the extra fields of the
 \seesaw\ models.
\begin{itemize}
\item[1:] The number $n_{{\nu}^{c}}$ of right-handed neutrino generations in the
          type I, inverse, linear, and \UYxUBL\ \seesaw\ models.
\item[2:] The number $n_{N_{S}}$ of neutrino-like singlet generations in the
          inverse, linear, and \UYxUBL\ \seesaw\ models, and also the number of
          parity-odd neutrino-like singlet generations in \UYxUBL\ inverse
          \seesaw\ models.
\item[15:] The number $n_{15}$ of 15-plet generations in the type II model.
\item[24:] The number $n_{24}$ of 24-plet generations in the type III model.
\end{itemize}

\subsubsection{Block \texttt{SCALARRM} (rank two)}
This block specifies the mixing matrix of the neutral scalar Higgs bosons. The
gauge eigenstates that are rotated into the mass-ordered mass eigenstates are
ordered as $H_{d}^{0}, H_{u}^{0}$, then a model-dependent ordering. For \seesaw\
types II and III, there are no further entries, because these fields are assumed
to be integrated out. For \URxUBL, they are
${\xi}_{R}^{0}, {\bar{\xi}}_{R}^{0}$. For \UYxUBL, they are
${\eta}, {\bar{\eta}}$.

\subsubsection{Block \texttt{PSEUDOSCALARRM} (rank two)}
This block specifies the mixing matrix of the neutral pseudoscalar Higgs bosons.
They are ordered analogously to how \texttt{SCALARRM} is ordered.


\subsubsection{Block \texttt{GAUGEIN} (rank one)}
This block specifies the gauge couplings at the input scale. The entries are
analogous to those of the \texttt{GAUGE} block.

\subsubsection{Block \texttt{MSOFTIN} (rank one)}
This block specifies the soft mass terms at the input scale. The entries are
analogous to those of the \texttt{MSOFT} block.

\subsubsection{Superpotential mass matrix blocks (rank two)}
The superpotential mass matrix blocks are \texttt{MNURNUR(IN)} for
$M_{{\nu}^{c}}$, \texttt{M15S15SB(IN)} for $M_{S}$, \texttt{M15T15TB(IN)} for
$M_{T}$, \texttt{M15Z15ZB(IN)} for $M_{Z}$, \texttt{M24W24W(IN)} for $M_{W}$,
\texttt{M24G24G(IN)} for $M_{G}$, \texttt{M24B24B(IN)} for $M_{B}$,
\texttt{M24X24XB(IN)} for $M_{X}$, \texttt{MNURS(IN)} for $M_{R}$,
\texttt{MNSNS(IN)} for ${\mu}_{N}$, \texttt{MCRCRB(IN)} for ${\mu}_{{\xi}}$, and
\texttt{MBILBILB(IN)} for ${\mu}_{{\eta}}$. \texttt{M15(IN)} for $M_{15}$ is
used when $M_{S}$, $M_{T}$, and $M_{Z}$ are set to a common value.
\texttt{M24(IN)} for $M_{24}$ is used when $M_{W}$, $M_{B}$, $M_{G}$, and
$M_{X}$ are set to a common value. Though some of the models are described with
only a single generation of some types of field, we allow for extra generations,
and thus define all the mass matrix blocks as being rank two, even though the
minimal case would use only the $( 1, 1 )$ entry of some of them.

\subsubsection{Soft SUSY-breaking mass matrix blocks (rank two)}
The soft SUSY-breaking mass matrix blocks are \texttt{M2NUR(IN)} for
$m_{{\tilde{{\nu}}}^{c}}^{2}$, \texttt{M215S(IN)} for $m_{S}^{2}$,
\texttt{M215SB(IN)} for $m_{\bar{S}}^{2}$, \texttt{M215T(IN)} for $m_{T}^{2}$,
\texttt{M215TB(IN)} for $m_{\bar{T}}^{2}$, \texttt{M215Z(IN)} for $m_{Z}^{2}$,
\texttt{M215ZB(IN)} for $m_{\bar{Z}}^{2}$, \texttt{M224W(IN)} for $m_{W}^{2}$,
\texttt{M224G(IN)} for $m_{G}^{2}$, \texttt{M224B(IN)} for $m_{B}^{2}$,
\texttt{M224X(IN)} for $m_{X}^{2}$, \texttt{M224XB(IN)} for $m_{{\bar{X}}}^{2}$,
\texttt{M2NS(IN)} for $m_{N}^{2}$, \texttt{M2CR(IN)} for $m_{{\xi}}^{2}$,
\texttt{M2CRB(IN)} for $m_{{\bar{{\xi}}}}^{2}$, \texttt{M2BIL(IN)} for
$m_{{\eta}}^{2}$, \texttt{M2BILB(IN)} for $m_{{\bar{{\eta}}}}^{2}$, and
\texttt{M2NSP(IN)} for $m_{N^{\prime}}^{2}$. Though some of the models are
described with only a single generation of some types of field, we allow for
extra generations, and thus define all the mass matrix blocks as being rank two,
even though the minimal case would use only the $( 1, 1 )$ entry of some of
them.

\subsubsection{Soft SUSY-breaking bilinear matrix blocks (rank two)}
The soft SUSY-breaking bilinear matrix blocks are \texttt{BNURNUR(IN)} for
$B_{{\tilde{{\nu}}}^{c}}$, \texttt{B15S15SB(IN)} for $B_{S}$,
\texttt{B15T15TB(IN)} for $B_{T}$, \texttt{B15Z15ZB(IN)} for $B_{Z}$,
\texttt{B24W24W(IN)} for $B_{W}$, \textit{B24G24G(IN)} for $B_{G}$,
\texttt{B24B24B(IN)} for $B_{B}$, \texttt{B24X24XB(IN)} for $B_{X}$,
\texttt{BNURS(IN)} for $B_{R}$, \texttt{BNSNS(IN)} for $B_{N}$,
\texttt{BCRCRB(IN)} for $B_{\xi}$, and \texttt{BBILBILB(IN)} for $B_{\eta}$.
\texttt{B15(IN)} for $B_{15}$ is used when $B_{S}$, $B_{T}$, and $B_{Z}$ are set
to a common value. \texttt{B24(IN)} for $B_{24}$ is used when $B_{W}$, $B_{B}$,
$B_{G}$, and $B_{X}$ are set to a common value.

\subsubsection{Yukawa coupling matrix blocks (rank three)}
The first digit is the generation index of the 15-plet or 24-plet field in type
II and III models respectively, and is 1 for the other models. The 2nd pair of
indices correspond to the usual indices of the minimal cases.
The Yukawa coupling matrix blocks are \texttt{YNURLHU(IN)} for $Y_{{\nu}}$,
\texttt{YL15TL(IN)} for $Y_{T}$, \texttt{YD15SD(IN)} for $Y_{S}$,
\texttt{YD15ZL(IN)} for $Y_{Z}$, \texttt{YHD15THD(IN)} for ${\lambda}_{1}$,
\texttt{YHU15TBHU(IN)} for ${\lambda}_{2}$, \texttt{YHU24BL(IN)} for $Y_{B}$,
\texttt{YHU24WL(IN)} for $Y_{W}$, \texttt{YHU24XBD(IN)} for $Y_{X}$,
\texttt{YNSLHU(IN)} for $Y_{LN}$, \texttt{YNSNURCR(IN)} for $Y_{N{\nu}^{c}}$,
\texttt{YNURBILNUR(IN)} for $Y_{{\eta}{\nu}^{c}}$, and \texttt{YNSNURBIL(IN)}
for $Y_{IS}$. \texttt{Y15(IN)} for $Y_{15}$ is used when $Y_{S}$, $Y_{T}$, and
$Y_{Z}$ are set to a common value. \texttt{Y24(IN)} for $Y_{24}$ is used when
$Y_{B}$, $Y_{W}$, and $Y_{X}$ are set to a common value.

\subsubsection{Soft SUSY-breaking trilinear matrix blocks (rank three)}
There is a soft SUSY-breaking trilinear matrix block for each Yukawa coupling
matrix block, with the same name but with the initial ``\texttt{Y}'' replaced by
 ``\texttt{T}'', corresponding to the ``$T_{\text{blah}}$'' associated with the
``$Y_{\text{blah}}$'' matrices. However, there is no \texttt{T15(IN)} for
$T_{15}$ or \texttt{T24(IN)} for $T_{24}$; instead, in such constrained
parameter sets, $A_{0} \times Y_{15/24}$ is used.

\subsubsection{Block \texttt{GAMZZPRM} (rank two)}
This is the gauge boson rotation matrix $U^{{\gamma}ZZ^{\prime}}$ given at the
SUSY scale.

\section*{CONCLUSIONS AND OUTLOOK}
In this contribution we propose an extension of the exisiting SLHA accords to
 include several \seesaw\ models.
Firstly, this requires new blocks to be defined 
for the additional couplings and masses needed within the various models. In
 this connection, we do not restrict ourselves to individual versions of the
 latter but also allow for combinations of such models in order to be as general
 as possible. 
Secondly, several new particles have to be postulated. For these, we propose
 a 9-digit scheme for the corresponding PDG-codes, which could in fact be of
 more general use than for \seesaw\ models only, yet it needs to be tested
 extensively against the properties of existing and possibly new SUSY models
 before widespread adoption. 
One issue that has to be addressed as next step is the proper definition of the
 additional parameters in the so-called super-PMNS basis. Moreover, also
 $SU(2)_R$ models are not yet covered in this proposal.

\appendix
\section{Kinetic mixing}
\label{sec:kineticmixing}
It is well known that in models with several \(U(1)\) gauge groups,
kinetic mixing terms
\begin{equation}
\label{eq:offfieldstrength}
- \chi_{ab}  \hat{F}^{a, \mu \nu} \hat{F}^b_{\mu \nu}, \quad a \neq b
\end{equation}
between the field-strength tensors are allowed by gauge and Lorentz
invariance \cite{Holdom:1985ag}, as $\hat{F}^{a, \mu \nu}$ and 
$\hat{F}^{b, \mu \nu}$ are gauge invariant quantities by themselves,
see \textit{e.g.}\  \cite{Babu:1997st}. Even if these terms are absent at tree
level at a particular scale, they might be generated by RGE effects
\cite{delAguila:1988jz,delAguila:1987st}. This happens usually if the
two Abelian gauge groups cannot be embedded in a larger
gauge group simultaneously or if incomplete gauge multiplets of the
fundamental theory are integrated out.
It is easier to work with  non-canonical covariant
derivates instead of off-diagonal field-strength tensors such as in eq.~(\ref{eq:offfieldstrength}). 
The equivalence of both approaches has been shown in \cite{delAguila:1988jz,Fonseca:2011vn}. 
We show here the special case of two Abelian gauge groups $U(1)_A \times U(1)_B$. The 
covariant derivatives has the form
\begin{equation}
\label{eq:kovariantDerivative}
 D_\mu  = \partial_\mu - i Q_{\phi}^{T} G  A 
\end{equation}
where \(Q_{\phi}\) is a vector containing the charges of the field $\phi$ with
respect to the two Abelian gauge groups, $G$ is the gauge coupling matrix
\begin{equation}
\label{eq:Gmatrix}
 G = \left( \begin{array}{cc} g_{AA} & g_{AB} \\ g_{BA} & g_{BB} \end{array} \right)
\end{equation}
and $A$ contains the gauge bosons $A = ( A^A_\mu, A^B_\mu )^T$. As long as the two 
Abelian gauge groups are unbroken there is freedom to rotate the gauge bosons. It is convenient
to choose a basis in which $G$ gets a triangle form
\begin{equation}
\label{eq:Gmatrixtri}
 G' = \left( \begin{array}{cc} g & \tilde{g} \\ 0 & g' \end{array} \right)
\end{equation}
Mixing effects of Abelian gauge groups appear not only in the gauge sector but also for the gauginos 
because also terms of the form 
\begin{equation}
 M_{AB} \lambda_A \lambda_B
\end{equation}
are allowed by gauge and Lorentz invariance \cite{Fonseca:2011vn,Braam:2011xh}.

\section{PDG CODES AND EXTENSIONS}\label{SLHA_sect_PDG-tables}

We summarise here our codes for existing and for the extra particle content of the \seesaw\ models. In the first column of Tables~\ref{slha_pdg1} to \ref{slha_pdg3}, the conventional PDG number is shown.

\begin{table}[hbt]
\begin{center}
\subfigure{
\begin{tabular}{|l|c|c|}
\hline
\multicolumn{3}{|c|}{\bf Colored Fermions} \\
\hline
 PDG & Particle & \PDGIX \\
\hline
2      & $u$          & 110100401 \\
4      & $c$          & 110100402 \\
6      & $t$          & 110100403 \\
-1     & $\bar{d}$    & 110890201 \\
-3     & $\bar{s}$    & 110890202 \\
-5     & $\bar{b}$    & 110890203 \\
\hline
1000021 & $\tilde{g}$ & 211110001 \\
\hline
%

%

\multicolumn{3}{c}{(a)} \\
\multicolumn{3}{c}{} \\
\hline
\multicolumn{3}{|c|}{\bf Colorless, Charged Fermions} \\
\hline
 PDG & Particle & \PDGIX \\
\hline 
-11      & $e^+$                 & 110000601 \\
-13      & $\mu^+$               & 110000602 \\
-15      & $\tau^+$              & 110000603 \\
 1000024 & $\tilde{\chi}_1^+$    & 210000601 \\
 1000037 & $\tilde{\chi}_2^+$    & 210000602 \\
\hline
\multicolumn{3}{c}{(b)} \\
\end{tabular}
 }
 \hspace{1cm}
\subfigure{
\begin{tabular}{|l|c|c|}
\hline
\multicolumn{3}{|c|}{\bf Neutral Fermions} \\
\hline
 PDG & Particle & \PDGIX \\
\hline 
-12/12      & $\bar{\nu}^{D}_1$ / $\nu^{M}_1$ & 110000001 /111000001  \\
-14/14      & $\bar{\nu}^{D}_2$ / $\nu^{M}_2$ & 110000002 /111000002  \\
-16/16      & $\bar{\nu}^{D}_3$ / $\nu^{M}_3$ & 110000003 /111000003  \\
          & $\bar{\nu}^{D}_4$ / $\nu^{M}_4$ & 110000004 /111000004  \\
          & $\bar{\nu}^{D}_5$ / $\nu^{M}_5$ & 110000005 /111000005  \\
          & $\bar{\nu}^{D}_6$ / $\nu^{M}_6$ & 110000006 /111000006  \\
          & $\bar{\nu}^{D}_7$ / $\nu^{M}_7$ & 110000007 /111000007  \\
          & $\bar{\nu}^{D}_8$ / $\nu^{M}_8$ & 110000008 /111000008  \\
          & $\bar{\nu}^{D}_9$ / $\nu^{M}_9$ & 110000009 /111000009  \\
 1000022 & $\tilde{\chi}^0_1$       & 211000001  \\
 1000023 & $\tilde{\chi}^0_2$       & 211000002  \\
 1000025 & $\tilde{\chi}^0_3$       & 211000003  \\
 1000035 & $\tilde{\chi}^0_4$       & 211000004  \\
 1000045 & $\tilde{\chi}^0_5$       & 211000005  \\
         & $\tilde{\chi}^0_6$       & 211000006  \\
         & $\tilde{\chi}^0_7$       & 211000007  \\
\hline
\multicolumn{3}{c}{(c)} \\
\end{tabular}
}
\end{center}
\caption{\label{slha_pdg1}(a) The down-type quarks are considered antiparticles due to having
 negative electric charge, hence the anti-downs are the defining states, and
 since they are color antitriplets, their color digits are 89 ($99-10$).
(b) In the case of $R$-parity violation, these fields mix to eigenstates
 with codes {\tt 1100006XY}.
(c) $\nu^D$ are dirac neutrinos while $\nu^M$ are Majorana neutrinos.
 In the case of $R$-parity violation, the fields mix forming (Majorana) mass
 eigenstates with \PDGIX\ codes {\tt 1110000XY}.}
\end{table}


\begin{table}[hbt]
\begin{center}
\subfigure{
\begin{tabular}{|l|c|c|}
\hline
\multicolumn{3}{|c|}{\bf Colored Scalars} \\
\hline
 PDG & Particle & \PDGIX \\
\hline 
 1000002 & $\tilde{u}_1$   & 200100401  \\ 
 1000004 & $\tilde{u}_2$   & 200100402  \\
 1000006 & $\tilde{u}_3$   & 200100403  \\
 2000002 & $\tilde{u}_4$   & 200100404  \\
 2000004 & $\tilde{u}_5$   & 200100405  \\
 2000006 & $\tilde{u}_6$   & 200100406  \\
-1000001 & $\tilde{d}^*_1$ & 200890201  \\ 
-1000003 & $\tilde{d}^*_2$ & 200890202  \\
-1000005 & $\tilde{d}^*_3$ & 200890203  \\
-2000001 & $\tilde{d}^*_4$ & 200890204  \\
-2000003 & $\tilde{d}^*_5$ & 200890205  \\
-2000005 & $\tilde{d}^*_6$ & 200890206  \\
\hline
\multicolumn{3}{c}{(a)} \\
\end{tabular}}
 \hspace{1cm}
\subfigure{
\begin{tabular}{|l|c|c|}
\hline
\multicolumn{3}{|c|}{\bf Vector Bosons} \\
\hline
 PDG & Particle & \PDGIX \\
\hline 
21  & $g$                &  121110001 \\
22  & $\gamma$           &  121000001 \\
23  & $Z$                &  122000001 \\
32  & $Z'$               &  122000002 \\
33  & $Z^{\prime\prime}$ &  122000003 \\
24  & $W^+$              &  120000601 \\
34  & ${W'}^+$           &  120000602 \\
\hline
\multicolumn{3}{c}{(b)} \\
\end{tabular}}
\end{center}
\caption{\label{slha_pdg2}
(a) New PDG code for colored scalars. 
(b) The $W^{+}$ is considered to be the particle (and hence the $W^{-}$ the
 antiparticle) since it has positive electric charge.}
\end{table}

\begin{table}[htb]
\begin{center}
\begin{tabular}{|l|c|c|}
\hline
\multicolumn{3}{|c|}{\bf Colorless, Charged Scalars} \\
\hline
 PDG & Particle & \PDGIX \\
\hline
25    & $h_1$   & 101000001 \\
35    & $h_2$   & 101000002 \\
45    & $h_3$   & 101000003 \\
      & $h_4$   & 101000004 \\
36    & $A^0_1$ & 102000001 \\
46    & $A^0_2$ & 102000002 \\
-1000012/1000012 & $\tilde{\nu}^*_1$ / $\RE(\tilde{\nu}_1)$&  200000001 /  201000001  \\
-1000014/1000014 & $\tilde{\nu}^*_2$ / $\RE(\tilde{\nu}_2)$&  200000002 /  201000002  \\
-1000016/1000016 & $\tilde{\nu}^*_3$ / $\RE(\tilde{\nu}_3)$&  200000003 /  201000003  \\
1000017 &  $\IM(\tilde{\nu}_1)$ & 202000001 \\
1000018 &  $\IM(\tilde{\nu}_2)$ & 202000002 \\
1000019 &  $\IM(\tilde{\nu}_3)$ & 202000003 \\
        & $\tilde{\nu}^*_4$ / $\RE(\tilde{\nu}_4)$ / $\IM(\tilde{\nu}_4)$ & 200000004 /  201000004 / 202000004 \\
        & $\tilde{\nu}^*_5$ / $\RE(\tilde{\nu}_5)$ / $\IM(\tilde{\nu}_5)$ & 200000005 /  201000005 / 202000005 \\
        & $\tilde{\nu}^*_6$ / $\RE(\tilde{\nu}_6)$ / $\IM(\tilde{\nu}_6)$ & 200000006 /  201000006 / 202000006 \\
        & $\tilde{\nu}^*_7$ / $\RE(\tilde{\nu}_7)$ / $\IM(\tilde{\nu}_7)$ & 200000007 /  201000007 / 202000007 \\
        & $\tilde{\nu}^*_8$ / $\RE(\tilde{\nu}_8)$ / $\IM(\tilde{\nu}_8)$ & 200000008 /  201000008 / 202000008 \\
        & $\tilde{\nu}^*_9$ / $\RE(\tilde{\nu}_9)$ / $\IM(\tilde{\nu}_9)$ & 200000009 /  201000009 / 202000009 \\
\hline
\end{tabular}
\caption{\label{slha_pdg3} In the case of $R$-parity violation but CP conservation, the CP-even
 and CP-odd components mix separately to form eigenstates with \PDGIX\ codes
 {\tt 1010000XY} and {\tt 1020000XY}, respectively. If CP violation is present,
 the numbers are {\tt 1000000XY} and {\tt 2000000XY} respectively (no $R$-parity
 violation) or just {\tt 1000000XY} (with $R$-parity violation).}
 \end{center}
\end{table}

\section*{ACKNOWLEDGEMENTS}
We want to congratulate the Les Houches organisers for a great and fruitful
workshop atmosphere, good food, and the amazing scenery of the French alps. 
We also want to thank Nazila Mahmoudi for the enlighting comments about FLHA.
LB, AB and SM thank the NExT Institute and Royal Society 
for partial financial support. LB has also been partially supported by the Deutsche Forschungsgemeinschaft through
the Research Training Group GRK\,1102
\textit{Physics of Hadron Accelerators}.
D.K.G. acknowledges partial support from the Department of Science and
Technology, India under the grant SR/S2/HEP-12/2006.
BOL and WP are supported by the  by the
 German Ministry of Education and Research (BMBF) under contract
 no.\ 05H09WWEF.



\AddToContent{L.~Basso,
A.~Belyaev,
D.~Chowdhury,
D.K.~Ghosh,
M.~Hirsch,
S.~Khalil,
S.~Moretti,
B.~O'Leary,
W.~Porod,
F.~Staub}
\renewcommand{\thesection}{\arabic{section}}
\renewcommand{\thesubsection}{\thesection.\arabic{subsection}}
\renewcommand{\thesubsubsection}{\thesubsection.\arabic{subsubsection}}
\renewcommand{\thefigure}{\arabic{figure}}
\renewcommand{\thetable}{\arabic{table}}
\renewcommand{\theequation}{\arabic{equation}}


\chapter{Automatic computation of supersymmetric renormalization group equations with FeynRules}

{\it Adam Alloul, Neil Christensen, Claude Duhr, Benjamin Fuks,
  Michel Rausch de Traubenberg}


\begin{abstract}
We discuss the progress on the automatic computation of renormalization group equations with {\sc FeynRules}. We also present a
new interface to the {\sc SuSpect} 3 package in order to automatically calculate supersymmetric spectra .
\end{abstract}

\section{INTRODUCTION}
\label{sec:feynrules_introduction}

Supersymmetry, and in particular its minimal version dubbed the Minimal
Supersymmetric Standard Model (MSSM) \cite{Nilles:1983ge, Haber:1984rc}, is
among the most popular extensions of the Standard Model (SM) of particle
physics. Linking fields with opposite statistics, it matches bosonic (fermionic)
partners to the SM fermions (bosons). The MSSM addresses a set of
conceptual problems of the SM, such as stabilizing the large hierarchy between the
electroweak scale and the Planck scale, unifying the gauge couplings at high energy or
proposing a candidate for the dark matter in the universe. 
However, if supersymmetry were an exact symmetry, both the SM
particles and their superpartners would have the same mass. As this is not observed
experimentally, supersymmetry must be broken at low energies. In 
order to remain a viable solution to the hierarchy problem, this breaking must
be soft, resulting in superpartners with masses around the TeV scale. 
Therefore, the quest for supersymmetric particles is one of the main
experimental topics at the Large Hadron Collider (LHC) at CERN. The latest
results of the general purpose LHC experiments ATLAS and CMS are currently
pushing the limits on the masses of the superpartners to a higher and higher range
\cite{atlas,cms}, thereby  severely constraining the supersymmetric parameter space.
However, most limits only hold
in the context of the so-called constrained MSSM framework, where the 105 free
parameters of the MSSM are reduced to a set of four parameters and a sign,
assuming some organizing principle based on unification at high energy. 
In contrast, there is a
much broader class of supersymmetric theories which may be valuable to be
investigated from a phenomenological point of view.

Even if all of these non-minimal models contain many free parameters, an organizing
principle governing the bulk of the parameters is strongly favored. Indeed, such a principle allows to reduce the number of degrees of freedom (of the parameter space) and renders the model 
accessible to phenomenological studies. Furthermore, it is also
motivated by the current experimental data, coming from \emph{e.g.} flavor physics. 
In their most general version, supersymmetric theories predict large rates for
flavor violating processes, which must however, in order to be in
agreement with the observations, be suppressed within the framework of the SM. An
organizing principle would then either forbid or suppress the relevant terms in
the supersymmetric Lagrangian, restoring agreement with the data.  Inspired by
scenarios such as supergravity scenarios with a minimal K\"ahler potential, 
where \textit{e.g.} all the scalar particles get a
common mass at the Planck scale, this organizing principle may however only be
valid at some high energy scale. Therefore, in order to study the phenomenology of 
supersymmetric models at collider experiment scales, all the parameters must be run down from the
high input scale (where, \emph{e.g.}, they are supposed to unify)
 to low energies using renormalization group equations.

In the following, we describe an extension of the superspace module of
{\sc FeynRules}~\cite{Christensen:2008py,Duhr:2011se} which allows to automatically derive the renormalization group
equations related to the parameters of any supersymmetric theory from the
superfield content of the
theory and the implementation of the model dependent pieces of the Lagrangian,
{\it i.e.}, the superpotential and the soft 
supersymmetry breaking Lagrangian. Furthermore, we also show how these 
equations can be automatically exported to the supersymmetric spectrum generator 
{\sc SuSpect} 3 for a numerical extraction of the model spectrum, the {\sc C++}
version of the {\sc SuSpect} 2 package \cite{Djouadi:2002ze}. 

The outline of this contribution is as follows: in Section~\ref{sec:feynrules_suspect} we present a review on 
how to obtain renormalization group equations for generic supersymmetric theories.
In Section~\ref{sec:feynrules_rge} we show how to extract these equations at the one-loop level automatically with {\sc FeynRules}, and in Section~\ref{sec:feynrules_suspect_interface} we describe the new interface to the {\sc SuSpect} 3 program. Our conclusions are drawn
in Section \ref{sec:feynrules_concl}.

\section{RENORMALISATION GROUP EQUATIONS FOR A GENERIC SOFTLY BROKEN SUPERSYMMETRIC MODEL}
\label{sec:feynrules_suspect}
A generic supersymmetric theory can be described by fixing the superfield content 
and the (gauge) symmetries leaving the corresponding Lagrangian invariant. We introduce a
gauge group $G$ and representations ${\cal R}$ of the corresponding algebra
spanned by the hermitian matrices $T^a$ and we denote the gauge coupling constant
by $g$. The chiral sector of the theory consists
in a set of left-handed chiral superfields $\Phi^i$ lying in the representation
${\cal R}^i$ of $G$. The gauge sector contains a vector superfield $V$
of the gauge group. The supersymmetric and gauge invariant Lagrangian, together
with the associated soft supersymmetry breaking Lagrangian, is given by
\begin{equation}\label{eq:feynrules_lsusy}
  {\cal L} = {\cal L}_{\rm chiral} + {\cal L}_{\rm Yang-Mills} + {\cal L}_{\rm
    superpotential} + {\cal L}_{\rm soft} \ ,
\end{equation}
where ${\cal L}_{\rm chiral}$ and ${\cal L}_{\rm Yang-Mills}$ contain the kinetic terms
as well as the gauge interactions of the different particles, which are entirely
fixed by supersymmetry and gauge invariance. Their explicit form in terms of component
fields can be found, {\it e.g.}, in Refs.\ \cite{Duhr:2011se,wb,FuksRausch}.
In contrast, ${\cal L}_{\rm
superpotential}$ is model-dependent (but gauge-invariant) and is derived from the 
superpotential describing the
interactions among the different chiral supermultiplets. 
Assuming renormalizability in addition, the most general superpotential 
takes the form
\begin{equation}
  W(\Phi) = \frac16 y_{ijk} \Phi^i \Phi^j \Phi^k + \frac12 \mu_{ij} \Phi^i \Phi^j +
    \alpha_i \Phi^i \ ,
\end{equation}
where $y$, $\mu$ and $\alpha$ are the coupling strengths dictating the
supersymmetric (non-gauge) interactions of the theory.
Finally, the last part of the Lagrangian of Eq.\ \eqref{eq:feynrules_lsusy}, ${\cal L}_{\rm
soft}$, is also model-dependent (and gauge-invariant as well) 
and contains the soft supersymmetry 
breaking terms {\it i.e.} terms of dimension
two and three breaking supersymmetry. The latter can be generically written as
\begin{equation}
  {\cal L}_{\rm soft} = -\frac16 h_{ijk} \phi^i \phi^j\phi^k - \frac12 b_{ij}
    \phi^i \phi^j - a_i \phi^i - \frac12 M \lambda \cdot \lambda -\frac12
    \phi^\dag_i (m^2)^i{}_j \phi^j+ {\rm h.c.} \ ,
\end{equation}
where $\phi^i$ denotes the scalar component  of the chiral superfield $\Phi^i$ and
$\lambda$ the (two-component fermionic) gaugino field embedded within 
the vector superfield $V$. This Lagrangian
contains hence mass terms for gauginos and scalar fields $M$ and $m^2$ 
as well as linear, bilinear and trilinear scalar interactions derived
from the form of the superpotential. The coupling strengths of the interactions are given by the
$h$, $b$ and $a$ parameters. In the rest of this section, all the supersymmetry-breaking and
superpotential parameters are assumed to be generic non-zero complex numbers.

We now turn to the renormalization group equations linking the values of the
parameters at two different energy scales. Even if the results have been known
at the one-loop \cite{Derendinger:1983bz, Falck:1985aa}
and two-loop levels for a long time \cite{Martin:1993zk, Yamada:1993uh,
Yamada:1993ga, Yamada:1994id, Jack:1994rk}, and could even be derived by other
means (\emph{e.g.}, supergraph techniques), we only focus here on the one-loop
contributions.
An extension to the two-loop level is foreseen in a near future.
A generic renormalization group equation for a parameter $x$ takes the form
\begin{equation}\label{eq:feynrules_genrge}
  \frac{{\rm d}}{{\rm d} t } x =  \beta_x \ .
\end{equation}
where we have introduced the $\beta$-function of the parameter $x$. As
previously mentioned, this quantity can be expanded perturbatively,
\begin{equation}
  \beta_x = \frac{1}{16 \pi^2} \beta_x^{(1)}  + \frac{1}{(16 \pi^2)^2}
\beta_x^{(2)} + \ldots\ ,
\end{equation}
where $\beta_x^{(1)}$ and $\beta_x^{(2)}$ are the one-loop and two-loop
contributions, respectively.
In the following, we then provide the first coefficient of
the $\beta$-functions of all the parameters introduced above.

Starting with the gauge sector, the first coefficients of the
$\beta$-functions of the gauge coupling constant $g$ and of the
gaugino mass parameter $M$ read
\begin{equation}\label{eq:feynrules_beta1}\begin{split}
  \beta_g^{(1)} =&\ g^3 \Big[ S({\cal R}) - 3 C(G) \Big] \ , \\
  \beta_M^{(1)} =&\ 2 g^2 M \Big[ S({\cal R}) - 3 C(G) \Big] \ .
\end{split}\end{equation}
In the expressions above, we have introduced the quadratic Casimir invariant
$C(G)$ of the adjoint representation and $S({\cal R})$, the total Dynkin index
for the representation ${\cal R}$. Let us note that this last quantity must be
summed over all the chiral superfields lying in this representation, accounting
in addition for their multiplicity. In the case of an abelian group, we define $S({\cal
R})$ as the sum of the squared charges of the whole chiral content of the
theory, accounting again for the multiplicity of each superfield. 
Turning to the superpotential parameters, the first coefficient of the
$\beta$-functions of the linear, bilinear and trilinear interaction
parameters are given by,
\begin{equation}\label{eq:feynrules_beta2}\begin{split}
  \beta_{\alpha_i}^{(1)} =&\ \alpha_p\ (\gamma^{(1)})^p{}_i \ , \\ 
  \beta_{\mu_{ij}}^{(1)} =&\ \mu_{ip}\ (\gamma^{(1)})^p{}_j + 
    \mu_{pj}\ (\gamma^{(1)})^p{}_i  \ , \\ 
  \beta_{y_{ijk}}^{(1)} =&\ y_{ijp}\ (\gamma^{(1)})^p{}_k + 
    y_{ipk}\ (\gamma^{(1)})^p{}_j + y_{pjk}\ (\gamma^{(1)})^p{}_i  \ ,
\end{split}\end{equation}
respectively. In the expression above, the first coefficient of the anomalous
dimensions $\gamma$ are defined as 
\begin{equation}
  (\gamma^{(1)})^j{}_i =  \frac12 y_{ipq} y^{\ast jpq}  -2 \delta^j{}_i g^2 C(i)
   \ ,
\end{equation}
where $C(i)$ refers to the quadratic Casimir invariant of the representation of
$G$ in which the chiral superfield $\Phi^i$ lies. The first coefficients of the
beta functions of the soft supersymmetry breaking parameters,
\textit{i.e.}, of the linear, bilinear and trilinear scalar interactions, are
given by
\begin{equation}\label{eq:feynrules_beta3}\begin{split}
  \beta_{a_i}^{(1)} =&\ \frac12 y_{ipq} y^{\ast rpq} a_r +  y^{\ast rpq}  h_{ipq} 
     \alpha_r + \mu_{ir} y^{\ast rpq} b_{pq} + 2 y_{ipq} (m^2)^q{}_r \mu^{\ast
      pr} + h_{ipq} b^{\ast pq} \ , \\
  \beta_{b_{ij}}^{(1)} =&\ \frac12 b_{ip} y^{\ast pqr} y_{qrj} + \frac12 y_{ijr}
     y^{\ast rpq} b_{pq} + \mu_{ip} y^{\ast pqr} h_{qrj}  - 2 \big[ b_{ij} - 2 M
   \mu_{ij} \big] g^2 C(i) + \big( i \leftrightarrow j\big) \ , \\ 
  \beta_{h_{ijk}}^{(1)} =&\ \frac12 h_{ijr} y^{\ast rpq} y_{pqk} + y_{ijr}
    y^{\ast rpq} h_{pqk} - 2 \big[ h_{ijk} - 2 M y_{ijk}\big] g^2 C(k) + \big( i
    \leftrightarrow k\big) + \big( k \leftrightarrow j\big) \ .
\end{split}\end{equation}
Finally, the first term in the beta function for the scalar mass
terms reads
\begin{equation}\label{eq:feynrules_beta4}\begin{split}
  \beta^{(1)}_{ (m^2)^i{}_j} =&\ \frac12 y^{\ast ipq} y_{pqr} (m^2)^r{}_j +
    \frac12 y_{jpq} y^{\ast pqr} (m^2)^i{}_r + 2 y^{\ast ipq} y_{jpr}
    (m^2)^r{}_q + h_{jpq} h^{\ast ipq}\\ &\ - 8 \delta^i{}_j M M^\dag g^2 C(i) + 2
     g^2 (T_a)^i{}_j {\rm Tr}\big[ T^a m^2 \big] \ . 
\end{split}\end{equation}
We note that the terms in this equation depending explicitly on the
representation matrices of the gauge group vanish identically for non-abelian
gauge groups. In the abelian case, these matrices must be read
as the squared abelian charge of the considered superfield.

\section{EXTRACTING RENORMALIZATION GROUP EQUATIONS WITH FEYNRULES}
\label{sec:feynrules_rge}
In order to be able to extract the renormalization group
equations of a given supersymmetric theory in an automated way with {\sc FeynRules}, 
the latter must be
implemented using the superspace module~\cite{Duhr:2011se}. However, as discussed in the
previous section, the derivation of the coefficients of the beta-functions
requires the computation of the multiplicity of each superfield. In order to
allow {\sc FeynRules} to perform this task efficiently, 
both the superpotential and the soft supersymmetry breaking Lagrangian must be
implemented using the function \texttt{SUDot} for 
$SU(N)$-invariant products of $N$ superfields. Hence, taking the example of the
MSSM, the $\mu$-term of the superpotential would be implemented as
\begin{verbatim}
   MUH SUDot[HU[aa] , HD[aa], aa]
\end{verbatim}
where the \texttt{SUDot} function receives as arguments the sequence of the
contracted superfields with all indices explicit. The contracted $SU(N)$
indices are set to a single given value for all superfields, the latter being
repeated as the last
argument of the \texttt{SUDot} function. In our example, it is labeled
\texttt{aa}. As another example, the up-type squark trilinear soft supersymmetry
breaking interaction terms would be implemented as
\begin{verbatim}
  -tu[ff1,ff2] URs[ff1,cc1] SUDot[QLs[aa,ff2,cc1] , hus[aa], aa]
\end{verbatim} 
We refer
to Ref.\ \cite{Duhr:2011se} for the definition of the superfields, fields and parameters.
After the model file has been loaded,
the renormalization group equations can be extracted via the command 
\begin{verbatim}
  RGE[LSoft, SuperW]
\end{verbatim}
where \texttt{LSoft} and \texttt{SuperW} denote the variables in which the soft supersymmetry breaking
Lagrangian have been stored. The \texttt{RGE} command computes the renormalization group
equations according to Eq.\ \eqref{eq:feynrules_genrge}, using the beta-functions provided
in Eqs.\ \eqref{eq:feynrules_beta1}, \eqref{eq:feynrules_beta2}, \eqref{eq:feynrules_beta3} and
\eqref{eq:feynrules_beta4}. 

We have validated our implementation against the well-known results from the
literature both in the context of the MSSM \cite{Martin:1993zk} and the
Next-to-Minimal Supersymmetric Standard Model \cite{uli}. We give, as
examples, two renormalization group equations given by {\sc FeynRules} in the
framework of the general MSSM. For notations, we refer to Ref.\
\cite{Duhr:2011se}. The evolution of the $\mu$-parameter of the superpotential
and the one of the corresponding $b$-parameter in the soft supersymmetry breaking
Lagrangian are driven by 
\begin{equation}\begin{split}
\frac{{\rm d}\mu}{{\rm d} t} = &\ \mu \bigg[ 
   -\frac{3 g^{\prime 2}}{80  \pi^2} 
   - \frac{3 g_w^2}{16 \pi^2}  
   +  \frac{3}{16 \pi^2} {\rm Tr} \big[ \mathbf{y^d}^\dag \mathbf{y^d} \big]
   +  \frac{3}{16 \pi^2} {\rm Tr} \big[ \mathbf{y^u}^\dag \mathbf{y^u} \big]
   +  \frac{1}{16 \pi^2} {\rm Tr} \big[ \mathbf{y^e}^\dag \mathbf{y^e} \big]
   \bigg]\ , \\
\frac{{\rm d} b}{{\rm d} t} = &\ b \bigg[ 
   - \frac{3 g^{\prime 2}}{80  \pi^2} 
   - \frac{3 g_w^2}{16 \pi^2} 
   +  \frac{3}{16 \pi^2} {\rm Tr} \big[ \mathbf{y^d}^\dag \mathbf{y^d} \big]
   +  \frac{3}{16 \pi^2} {\rm Tr} \big[ \mathbf{y^u}^\dag \mathbf{y^u} \big]
   +  \frac{1}{16 \pi^2} {\rm Tr} \big[ \mathbf{y^e}^\dag \mathbf{y^e} \big]
  \bigg] \\ +&\ \mu \bigg[ 
     \frac{3 g^{\prime 2} M_1}{40  \pi^2} 
   + \frac{3 g_w^2 M_2}{8  \pi^2} 
   + \frac{3}{8 \pi^2} {\rm Tr} \big[ \mathbf{y^d}^\dag \mathbf{T^d} \big]
   + \frac{3}{8 \pi^2} {\rm Tr} \big[ \mathbf{y^u}^\dag \mathbf{T^u} \big]
   + \frac{1}{8 \pi^2} {\rm Tr} \big[ \mathbf{y^e}^\dag \mathbf{T^e} \big]
   \bigg]  \ . 
\end{split}\end{equation}

\section{AN INTERFACE TO THE SUSPECT 3 SUPERSYMMETRIC SPECTRUM GENERATOR}
\label{sec:feynrules_suspect_interface}
In collaboration with the {\sc SuSpect} 3 team, A.\ Djouadi, J.L.\ Kneur, 
G.\ Moultaka, M.\ Ughetto and D. Zerwas, we have designed an interface exporting
the renormalization group equations derived by {\sc FeynRules} into a format
which can be used by the spectrum calculator {\sc SuSpect}~3
presented in Section \ref{SuSpect3_chapter}.
Within {\sc Mathematica}, this interface is called as any other {\sc
FeynRules} interface,
\begin{verbatim}
  WriteSuSpectOutput[LSoft, SuperW]
\end{verbatim} 
where again, the variables \texttt{LSoft} and \texttt{SuperW} contain the soft
supersymmetry breaking Lagrangian and the superpotential, respectively.
When this command is issued, {\sc FeynRules} internally calls the function {\tt RGE}
and exports the results in a header and source file
\texttt{DerivativeExternal.h} and \texttt{DerivativeExternal.cxx}, respectively. 
The header file contains the class
definitions of {\sc SuSpect} 3 and the source file contains the 
renormalization group equations themselves.
The two files have to be copied into the {\sc SuSpect}~3 directories
(\texttt{inc} and \texttt{src}). Once {\sc SuSpect}~3 has been recompiled, the
renormalization group equations provided by {\sc FeynRules} can be used either
if the counter 0 of the block \texttt{SUSPECT\textunderscore CONFIG} of the
input SLHA file is set to 99 (external renormalization group equations) or
directly in memory as described in Section \ref{SuSpect3_chapter}.

\section{CONCLUSION}\label{sec:feynrules_concl}
In this contribution we have presented a new module in the {\sc FeynRules} package that allows to compute one-loop renormalization group equations for supersymmetric theories in an automated way. {\sc Feynrules} extracts all the one-loop $\beta$ functions of the model directly from the superfield spectrum and the superpotential and soft-supersymmetry breaking terms of the Lagrangian, which allows to obtain the renormalization group equations for any supersymmetric model. The renormalization group equations obtained in this way can then be written to file in a format that can be linked against the {\sc SuSpect}~3 program, thus allowing to perform the renormalization group running numerically and to obtain the spectrum for arbitrary non-minimal supersymmetric theories.

\section*{ACKNOWLEDGEMENTS}
AA, BF and MRT are particularly grateful to D. Zerwas for enlightening discussions on
{\sc SuSpect} 3. CD and BF would like to thank CERN and LAPTH for their
hospitality during the workshop
during which some of the work contained herein was performed. AA, BF and MRT are
supported by the Theory-LHC France-initiative of the CNRS/IN2P3. This work was supported by the Research Executive Agency (REA) of the European Union under the Grant Agreement number PITN-GA-2010-264564 (LHCPhenoNet).

\AddToContent{A.~Alloul, N.~Christensen, C.~Duhr, B.~Fuks, 
  M.~Rausch~de~Traubenberg}
\renewcommand{\thesection}{\arabic{section}}


\chapter{Implementation and automated validation of the minimal $Z'$ model in FeynRules}

{\it Lorenzo Basso, Neil D. Christensen, Claude Duhr, Benjamin
Fuks, Christian Speckner}




\begin{abstract}
We describe the implementation of a well-known class of $U(1)$ gauge models,
 the ``minimal'' $Z'$ models, in FeynRules.  We also describe a new
automatized validation tool for FeynRules models which is controlled
by a web interface and allows the user to run a complete set of
$2\to2$ processes on different matrix element generators, different
gauges, and compare between them all. If existing, the comparison with independent implementations is also possible.  This tool has been used to
validate our implementation of the ``minimal'' $Z'$ models.
\end{abstract}

\section{INTRODUCTION}

The Large Hadron Collider (LHC) has begun exploring the TeV scale of
physics.  
To determine the full set of properties of the new physics expected to appear,
we will have to implement new theoretical models for TeV scale physics into LHC
simulation software such as
CalcHEP\cite{Pukhov:1999gg,Boos:2004kh,Pukhov:2004ca},
MadGraph\cite{Stelzer:1994ta,Maltoni:2002qb,Alwall:2007st,Alwall:2008pm,Alwall:2011uj}
and Whizard\cite{Moretti:2001zz,Kilian:2007gr}. 

Among the various possible extensions of the Standard Model (SM),
those containing an extra
$U(1)$ gauge group play a special role
(see~\cite{Langacker:2008yv,Erler:2009jh,Blumenhagen:2005mu} and references
therein). The associated $Z'$ boson(s) may be among the first objects to
be probed at the LHC, due to the simplicity of their signals at colliders. The phenomenology of $Z'$
bosons has been intensively studied (for some 
reviews, see~\cite{Rizzo:1996ce,Cvetic:1997wu,Appelquist:2002mw,
Carena:2004xs,Accomando:2010fz}). 
Despite the important role of $Z'$ physics, we note the
absence of commonly accessible code for its analysis with the
existing Monte Carlo tools.  In the past, each analysis has started with a new, private
implementation of a particular $Z'$ model, a process which is
inefficient and error prone.  
%
%

For this reason, we have implemented a class of $U(1)$ models in
FeynRules\cite{Christensen:2008py,Christensen:2009jx,Christensen:2010wz} which
covers the complete set of ``minimal'' $Z'$
models~\cite{Appelquist:2002mw,Carena:2004xs,Salvioni:2009mt,Basso:2011na}.  In
this implementation, the various ``minimal'' models are chosen by setting a
parameter which continuously varies among the many possibilities.  By
implementing this model in FeynRules, we allow its implementation to be
used in CalcHEP, MadGraph and Whizard without any modification of those codes.
Furthermore, we have made this model implementation public on the FeynRules
website so that anyone can use it.

Although this approach of automatically generating the different model implementations
from a single ``master'' implementation greatly reduces the risk of
creating faulty model implementation, possible issues can still arise from either
faulty FeynRules input or software bugs. While it is virtually impossible to
prove the correctness of a model implementation, confidence can be gained
from comparing the model between different codes supported by
FeynRules, between different gauges and, if
available, by comparing to existing, independent implementations. Still, doing
such a comparison by hand is very cumbersome as it requires expertise
in each simulation package, setting up each package, matching the respective
setups (in particular the cuts and
model input) and then running and comparing a set of processes as large as
possible. This procedure shows a lot of potential for automatization.
In order to facilitate this, we have developed an easy-to-use validation framework
driven by a web application. The framework allows registered users to upload a
model which is automatically run through FeynRules in order to check for
syntactical correctness and generate model implementations for the different
Monte Carlo tools supported by the framework (at the moment MadGraph, CalcHEP
and Whizard). After this sanity check, a list of all possible $2\longrightarrow
2$ processes in the model (restrictions can be applied to the process list) is
generated. For each process on the list and each different generator, the cross
section is automatically evaluated and compared between the different tools.

Using this web validation tool, we have validated our class of
``minimal'' $Z'$ models by comparing over a thousand $2\to2$ processes
between CalcHEP and Whizard and by comparing between Feynman and
unitary gauge.  This allowed us to find the remaining bugs in this
model implementation and fix them.  The model we have made public
shows agreement between CalcHEP and Whizard and between Feynman and
unitary gauges.


\section{THE MINIMAL $Z'$ MODELS IN FEYNRULES\label{sec:BL}}

In this section we discuss the implementation of the Minimal $Z'$
Model~\cite{Basso:2011na,Basso:2010jm}. This model is based on a $SU(3)_C\times
SU(2)_L\times U(1)_Y\times U(1)_{B-L}$ gauge symmetry, i.e., the Standard Model
(SM) gauge group is augmented by a $U(1)$ factor related to the gauged Baryon
minus Lepton ($B-L$) number. The model is ``minimal'' as it extends the SM with
a minimum amount of modifications in all the sectors. In addition to the new
$U(1)_{B-L}$ gauge sector, the model includes three right-handed neutrinos
(designed to cancel anomalies) and an additional complex Higgs boson ($\chi$)
which is a singlet under the SM gauge groups but is charged under $U(1)_{B-L}$
and is responsible for giving mass to the additional $Z'$ gauge boson.  The
classical, gauge invariant Lagrangian can be decomposed as
$\mathscr{L}=\mathscr{L}_{YM} + \mathscr{L}_S + \mathscr{L}_f + \mathscr{L}_Y +
\mathscr{L}_{ghost}$.

The non-Abelian field strengths in $\mathscr{L}_{YM}$ are the same as in the SM
whereas the Abelian ones can be written as $\mathscr{L}^{\rm Abel}_{YM} =
-\frac{1}{4}F^{\mu\nu}F_{\mu\nu}-\frac{1}{4}F^{\prime\mu\nu}F^\prime _{\mu\nu}$
(where $F_{\mu\nu} = \partial _{\mu}B_{\nu} - \partial _{\nu}B_{\mu}$ and
$F^\prime_{\mu\nu} = \partial _{\mu}B^\prime_{\nu} - \partial
_{\nu}B^\prime_{\mu}$).  In this field basis, the covariant derivative reads:
\begin{equation}\label{cov_der}
D_{\mu}\equiv \partial _{\mu} + ig_S T^{\alpha}G_{\mu}^{\phantom{o}\alpha} 
+ igT^aW_{\mu}^{\phantom{o}a} +ig_1YB_{\mu} +i(\widetilde{g}Y + g_1'Y_{B-L})B'_{\mu}\, .
\end{equation}
Since $\widetilde{g}$ is a free parameter of the theory (assuming a
fixed $g'_1$), this model describes a one-dimensional class of $Z'$ theories. 
A discrete set of popular $Z'$ models
(see, e.g.~\cite{Carena:2004xs,Appelquist:2002mw,Salvioni:2009mt}) can be
 recovered by a suitable definition of both $\widetilde{g}$ and $g_1'$.
 A few of them are summarised in Table~\ref{models}.
\begin{table}[!ht]
\begin{center}
\begin{tabular}{|c|l|}
\hline
Model & Parameterization \\
\hline
$U(1)_{B-L}$ & $\widetilde{g}=0$ \\
\hline
$U(1)_\chi$ & $\widetilde{g}=-4/5 g'_1$ \\ 
\hline
$U(1)_R$ & $\widetilde{g}=-2 g'_1$  \\ 
\hline
\end{tabular}
\end{center}
\caption{Specific parametrisations of the Minimal $Z'$ Models:
$U(1)_{B-L}$, $U(1)_\chi$ and $U(1)_R$.}
\label{models}
\end{table}

We now describe a few important aspects of our
FeynRules implementation of this model. First of all, we note that the
model files are set to Feynman gauge by default.  If unitary gauge is
desired, the variable \verb|$FeynmanGauge| must be set to \verb|False|
inside the Mathematica session after loading the model, but before
calculating Feynman rules.  

Unless stated otherwise, all particles and parameters that exist in the SM are
the same as in the default FeynRules SM implementation.  We have renamed the
tauon to \verb|l| and its antiparticle to \verb|L| in order to match the CalcHEP
conventions.  Since we have added new gauge bosons, scalar fields and neutrinos
which mix with the SM fields, their mixing partners have been mostly renamed.
The new names are given in Table~\ref{tab:FR_MinZp_conv} together with the new
parameters.
For an overall description of their meaning and some useful relations, see
Refs.~\cite{Basso:2011na,Basso:2010jm}.
The columns marked FR give the names that are used in the matrix
element generators.

\begin{table}[!h]
\begin{center}
\begin{tabular}{|c|c|c|c|}
\multicolumn{4}{c}{Particles}\\
\hline
& \multicolumn{2}{|c|}{Symbol} & \mbox{PDG code}\\ 
&$\mathscr{L}$&FR&\\
\hline 
Light Neutrino 1&  $\nu_{l_1}$ 	& \verb|n1| 	& $12$\\  
Light Neutrino 2&  $\nu_{l_2}$ 	& \verb|n2| 	& $14$\\  
Light Neutrino 3&  $\nu_{l_3}$ 	& \verb|n3| 	& $16$\\  
Heavy Neutrino 1&  $\nu_{h_1}$ 	& $\sim$\verb|n1| 	& $9910012$\\  
Heavy Neutrino 2&  $\nu_{h_2}$ 	& $\sim$\verb|n2| 	& $9910014$\\  
Heavy Neutrino 3&  $\nu_{h_3}$ 	& $\sim$\verb|n3| 	& $9910016$\\  
Light Higgs&  $h_1$ 	& \verb|H1| 	& $9900025$\\  
Heavy Higgs&  $h_2$ 	& \verb|H2| 	& $9900026$\\  
$Z'$ boson&  $Z'$ 	& \verb|Zp| 	& $9900032$\\   
  \hline
\end{tabular}
\hspace{0.25in}
\begin{tabular}{|c|c||c|c|}
\multicolumn{4}{c}{Parameters}\\ 
\hline
\multicolumn{2}{|c||}{Independent} & \multicolumn{2}{|c|}{Dependent}\\
$\mathscr{L}$&FR&$\mathscr{L}$&FR\\
\hline 
$g'_1$ & \verb|g1p| & $M_Z$   & \verb|MZ| \\  
$\widetilde{g}$ & \verb|gt| & $\sin{\theta'}$   & \verb|Sp|\\  
$M_{Z'}$ & \verb|MZp| & $x$   & \verb|x| \\  
$m_{h_1}$ & \verb|MH1| & $\lambda_1$   & \verb|lam1| \\  
$m_{h_2}$ & \verb|MH2| & $\lambda_2$   & \verb|lam2|\\  
$\sin{\alpha}$ & \verb|Sa| & $\lambda_3$   & \verb|lam3| \\  
$m_{\nu_l}$ & \verb|MnL| & $y^D_i$   & \verb|yndi|\\  
$m_{\nu_h}$ & \verb|MnH| & $y^M_i$   & \verb|ynmi|\\  
$M_W$ & \verb|MW|  & $\sin{\alpha_i}$   & \verb|Sani|\\   
  \hline
\end{tabular}
\end{center}
\caption{\label{tab:FR_MinZp_conv} New particles and parameters and the related
ASCII symbols exported by FeynRules to the matrix element generators.}
\end{table}

In order to use the resulting matrix element generator code, it is
important to know which parameters are independent and can be adjusted
by the user.  
In the neutral gauge boson sector, 
the choice we made is to keep the new gauge couplings ($g'_1$ and $\widetilde{g}$) as
 independent parameters, thus allowing to easily
 identify the  various benchmark models,
and to express the remaining quantities in terms of the $Z'$ mass, using the
following relation, independent of the $Z$ mass:
%
\begin{equation}
x=\frac{M_{Z'}}{2g'_1} \sqrt{1-\frac{\widetilde{g}^2 v^2}{4 M_{Z'}^2-v^2(g_w^2+g_1^2)}}\, ,
\end{equation}
which generalises a similar formula for the pure $B-L$ model (where $\widetilde{g}=0$).
This choice implies that the $Z$ boson mass is now treated as an
internal parameter, function of $M_{Z'}$, despite the usual conventions. 
%
It is left to the user to
choose a $Z'$ mass that satisfies existing experimental limits.
%
%
%
On the other hand, we have defined the $W$ boson mass as an
independent parameter.
Finally, in the scalar sector, the scalar masses and mixing angle $\alpha$ are
kept as input parameters instead of the $\lambda_i$ parameters appearing in the potential. (See Ref.~\cite{Basso:2010jm} for details.)
%

\section{WEB VALIDATION\label{sec:web}}

Although FeynRules greatly simplifies the process of implementing new models
and reduces the propability of errors during implementation considerably, there is still the
possibility of mistakes.  We have taken great pains to validate the
FeynRules package \cite{Christensen:2009jx} and we plan to continually extend and
improve these tests in order to make the FeynRules package ever more dependable.
On the other hand, when a new model is implemented, it has been up to the author
of that new model to validate it.  We have proposed guidelines for this
validation \cite{Butterworth:2010ym}, but, in practice, the level of model validation has been
irregular.  
The problem is
that, as a result of there being thousands of Feynman rules, testing a handful
of them (which is achievable by a human) is not sufficient to fully validate a
model.  It would be far better if all the vertices could be tested.  This
requires automation.  Some tests that can be automated are the following:
\begin{itemize}
\item comparison of all $1\to2$ and $1\to3$ partial widths between matrix element
generators,
\item comparison of all $2\to2$ and $2\to3$ cross sections between
matrix element generators,
\item comparison between gauges in these processes,
\item comparison of these processes with an independent implementation of the model.
\end{itemize}
Even though these tests cannot guarantee a model implementation is
error-free, our experience shows that they can help to discover remaining bugs in
many cases.

We have begun a project to create an automated validation tool which 
implements the tests described in the previous paragraph for FeynRules models.
Although not yet finalized, it has achieved sufficient maturity to be useful for
model validation and has, in fact, been used for the validation of the ``minimal'' $Z'$ model here
described, of which we will say more in the next section.  This
validation tool is implemented as a web (HTML / Javascript)
application and should, therefore, work on any
web enabled device.  A user who wishes to validate a model using this package,
would point his or her browser to the url of the web validation website and
would control the validation of his or her model via the web interface.
Approved, authenticated users can upload FeynRules model files for testing.
They can also upload implementations of their models done by independent
parties for comparison.  Once the model files have passed some basic sanity
tests, the user is able to run the full suite of validation tests on his or
her model.  The model and the results of the validation will remain private until
explicitly made public by the author, when ready.  In the future, we plan to add
the ability to export the results of the validations to other useful formats,
such as \LaTeX.

After uploading the model files and passing the sanity tests, the user can
generate a list of processes for comparison.  We currently support $2\to2$
processes, but plan to add support for others in the future.   By default, the
web interface generates all $2\to2$ processes allowed by the symmetries of the
model, up to permutations of the external particles.  The user can reduce the
number of generated processes by entering further restrictions.  The
restrictions that are currently supported are the following:
\begin{itemize}
\item restricting the number of particles of each spin,
\item limiting the
number of particles of each gauge group index defined in the model,
\item restricting the number of particles of each charge type defined in the
model.
\end{itemize}
In a future version, the user will be able to fine tune this list by
manually adding or removing processes.  

After generating a list of processes to study, the user can choose which matrix
element generators to use and in which gauges.  Currently available choices are
CalcHEP in Feynman and unitary gauge, MadGraph in unitary gauge and Whizard in
Feynman and unitary gauge.  Support for other matrix element generators is
planned for the future.  The user can also choose whether to include an
independently implemented model in the validation.  Typically, multiple tests
are performed:
\begin{itemize}
\item in one test, multiple matrix element generators are run, all
in unitary gauge;
\item in another test, gauge invariance is tested by running
CalcHEP in both gauges;
\item in yet another test, gauge invariance is again tested by
running Whizard in both gauges\footnote{We note that testing gauge invariance in
CalcHEP is slightly different than testing gauge invariance in Whizard.  In
CalcHEP, both Goldstone bosons and Faddeev-Popov ghosts are included in the
external states to cancel unphysical polarizations of gauge bosons in the
analytic polarization sum.  Whizard on the other hand calculates helicity
amplitudes and sums over physical helicities only. As a result,
they test slightly different aspects of gauge invariance which can be
useful for finding bugs. }.
\end{itemize}

Once the process list, the matrix element generators and the gauges are chosen,
the calculations can be performed.  The web validation organizes the
calculations and submits them to a Condor \cite{condor-practice} scheduler for running them on
a computer cluster.  For each
process, it collides the incoming particles at an energy which is greater than
threshold and far from any resonance in order to not be sensitive to particle
widths (each matrix element generator deals with the widths differently).  In
addition, a $p_T$ cut is applied in order to remove T-channel singularities.  The
web validation only submits a small number of jobs to the cluster at a time for
each test.  When those jobs are finished, it submits more.  In this way, if
multiple users are running tests, the jobs are mixed together on the Condor
cluster and each sees progress at roughly the same rate.  Once a job finishes,
the web validation software stores the resulting cross section and Monte-Carlo
uncertainty in a MySQL database for later analysis.  Comparisons of the
differential cross section at an individual phase space point is planned for a
later version.

After the jobs are done, the validation software analyzes $\chi^2$ for each
process which is given by:
\begin{equation}\label{eq: chi2 binned}
\chi^2 = \sum_i \left(\frac{\sigma_i-\sigma_{best}}{\Delta\sigma_i}\right)^2\ ,
\end{equation}
where $\sigma_i$ is the cross section obtained from matrix element generator and
gauge $i$, $\sigma_{best}$ is the best value for the cross section and
$\Delta\sigma_i$ is the Monte Carlo uncertainty for the cross section obtained
from matrix element generator and gauge $i$.  The value of $\sigma_{best}$ is
obtained by minimizing $\chi^2$.
%
The resulting values for $\chi^2$ are then histogrammed and a $\chi^2$
distribution is plotted and presented on the web page for the test (see examples
in the next section) along with the theoretical $\chi^2$ distribution
theoretically expected from Poisson statistics.
%

Typically, a correctly implemented model will give a $\chi^2$ distribution as
good as or better than the theoretical $\chi^2$ curve\footnote{
We find that if the model is implemented correctly, then the $\chi^2$
distribution is often shifted towards smaller $\chi^2$ than would be expected
from pure Poisson statistics.}.
Furthermore, the results of the processes are listed in order of
$\chi^2$ starting with the highest value.  When the histogrammed $\chi^2$ values
are worse than the theoretical $\chi^2$ curve, it is likely that there is
something wrong with the model.  In this case, there are typically one or more
processes with very large values of $\chi^2$ listed at the top of the process
list.  By analyzing the details of these discrepant processes, the user can
often find the mistake in his or her implementation.  

Validating simulation software is extremely important in order to avoid wrong
physics results coming from faulty software.
We, again, emphasize that this automated validation software should complement
other tests.
 The user should still check that the
Feynman rules given by FeynRules, as well as various physical distributions,
 are correct and agree with published vertices/distributions where possible.
This automated
validation software should be used as an additional test that increases the
confidence in a model implementation.

\section{VALIDATION OF THE MINIMAL $Z'$ MODEL\label{sec:BLval}}

Before running it through the automated validation suite described in the previous
section, the minimal $Z^\prime$ model had already been checked in the following
ways:
\begin{itemize}
\item the limit cases of the SM and the minimal $B-L$ model~\cite{Basso:2008iv} are correctly obtained,
\item the $Z'$ width and all its partial decay widths have been
  compared to calculations by hand and agree,
\item we have compared our production cross section 
for the $Z'_\chi$ with the similar $SO(10)$ inspired model in
  the literature \cite{Accomando:2010fz} and found agreement to within the uncetainties of the
  PDF's and matrix element generators used to do the calculations,
\item we have checked that certain vanishing vertices are not
  spuriously produced by our implementation (e.g., no
 interactions between a photon ghost and other
  neutral gauge bosons or neutral ghosts are allowed).
\end{itemize}

After completing these initial tests,
the model was uploaded to the validation server and a series of tests was run.
We used the validation software to generate all $2\to2$ processes allowed by the
symmetries of our model, which gave 1424 processes on which we ran the following tests:
\begin{itemize}
\item Test 1: For each process, we compared the Feynman and unitary gauges in
  CalcHEP,
\item Test 2: For each process, we compared the Feynman and unitary gauges in
  Whizard,
\item Test 3: For each process, we compared CalcHEP and Whizard in unitary
  gauge.
\end{itemize}

\begin{figure}
\begin{center}
\includegraphics[scale=1]{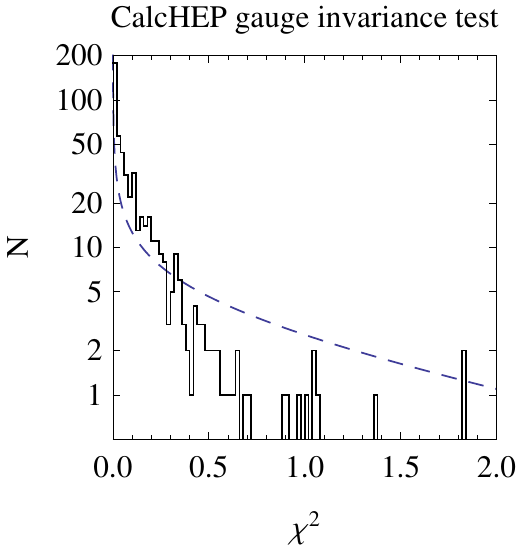}
\includegraphics[scale=1]{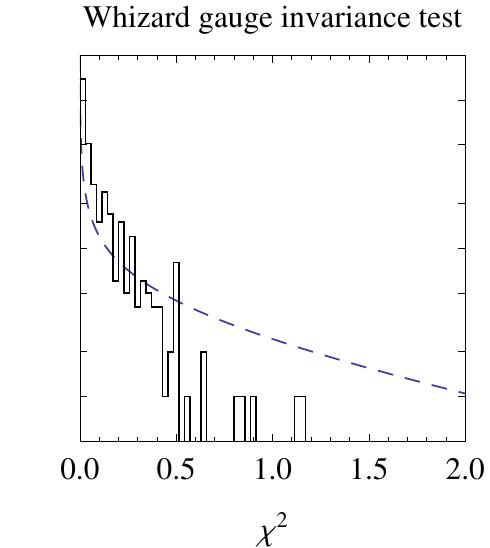}
\includegraphics[scale=1]{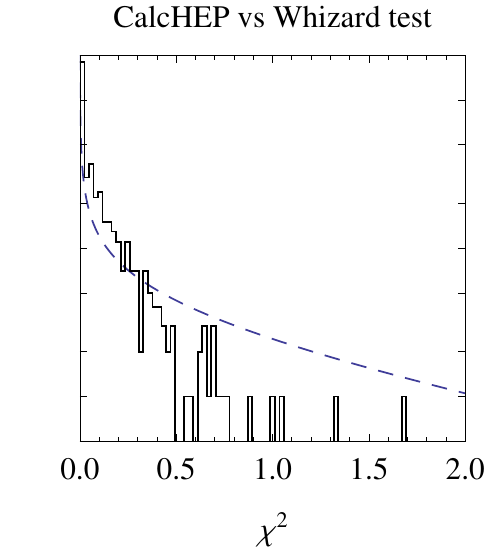}
\caption{Plots of the binned (solid) and theoretical (dashed)
  $\chi^2$.  The results of the CalcHEP gauge invariance test, Whizard
gauge invariance test and CalcHEP versus Whizard (in unitary gauge)
test are shown on the left, middle and right respectively. Actual data is solid,
while the theoretical $\chi^2$ distribution is given by the dashed line.}
\label{fig: chi2}
\end{center}
\end{figure}

Plots of the resulting $\chi^2$ distribution (c.f. Eq. \ref{eq: chi2 binned})
are shown in Fig. \ref{fig: chi2}.  
 %
The curves show that
\begin{itemize}
\item No large discrepancy was found.
\item The binned results from our tests did better than expected from
  a theoretical $\chi^2$ distribution.  
\end{itemize}

\section{CONCLUSIONS\label{sec:conc}}

We have implemented a class of $U(1)$ gauge models that spans the space of
``minimal'' $Z'$ models with a continuous parameter interpolating among
them. 
The FeynRules imlementation of the model has been made publicly available
through the FeynRules website.

We have also implemented a new automatized validation package.  This
package is accessible on the internet and allows a registered user to
upload their models and run a series of tests on them.  The tests
which are currently supported are a comparison of all $2\to2$
processes between different matrix element generators and between
Feynman and unitary gauges.  After calculation of each cross section,
the validation package calculates $\chi^2$ for each process and presents the
resulting $\chi^2$ distribution to the user together with the test results.  

Using this web validation package we have validated the
implementation of the ``minimal'' $Z'$ models and found full consistency.

\section*{ACKNOWLEDGEMENTS}
We would like to congratulate the Les Houches organisers for a great and
fruitful workshop atmosphere, good food, and the amazing scenery of the French
 alps.
LB thanks the NExT Institute for partial financial support.
LB and CS have been supported by the Deutsche Forschungsgemeinschaft through
the Research Training Group GRK\,1102
\textit{Physics of Hadron Accelerators}.
NC was partially supported by the United States National Science
Foundation under grant NSF-PHY-0705682 and partially by
the PITTsburgh Particle Physics, Astrophysics and Cosmology Center
(PITT PACC).
The server that ran the web validation software was bought with funds
from IISN 4.4517.08. We would like to thank Vincent Boucher and the
Center for Cosmology, Particle Physics and Phenomenology (CP3) at the
Universit\'{e} catholique de Louvain / Louvain-la-Neuve for administering
the server.



\AddToContent{L.~Basso, N.~Christensen, C.~Duhr, B.~Fuks, C.~Speckner}
\renewcommand{\thesection}{\arabic{section}}



\chapter{Implementation of scalar and vector resonances into FeynRules and LanHEP}


\begin{abstract}
In this summary report we discuss the implementation of a set of operators in {\sc FeynRules} and {\sc LanHEP}
that can be used to study in a bottom-up approach
new scalar and vector resonances coupling to quarks.
\end{abstract}

\section{INTRODUCTION}
\label{sec:AFB_introduction}

Monte Carlo event generators play an important role in making reliable predictions for events to be observed in collider experiments, both to describe the backgrounds and possible candidate signals. 
In particular, the simulation of a hadronic collision requires not only an accurate description of the underlying hard scattering process, but also of the parton showering and hadronization as efficiently provided by programs such as {\sc Herwig}~\cite{Corcella:2000bw,Gieseke:2011na}, {\sc Pythia}~\cite{Sjostrand:2006za, Sjostrand:2007gs} and {\sc Sherpa}~\cite{Gleisberg:2003xi,Gleisberg:2008ta}. As regards the generation of the hard matrix element itself, a lot of effort has gone into the development of several multipurpose matrix element generators,  such as
{\sc CompHEP}/{\sc CalcHEP} \cite{Pukhov:1999gg,  Boos:2004kh,Pukhov:2004ca}, {\sc MadGraph}/{\sc MadEvent} \cite{Stelzer:1994ta,Maltoni:2002qb,Alwall:2007st,Alwall:2008pm,Alwall:2011uj}, {\sc Sherpa} and {\sc Whizard}~\cite{Kilian:2007gr, Moretti:2001zz}. Even though these programs are in principle able to generate the (tree-level) matrix element for any process in a renormalizable quantum field theory built on scalar, vector and fermion fields, the implementation of a full Beyond the Standard Model (BSM) theory can be a tedious and error-prone task, often requiring the implementation of one vertex at the time following the conventions specific to each code. Recently, this task has been alleviated by the introduction of new tools  that allow to obtain the Feynman rules directly from a Lagrangian and to export them to various matrix element generators in an automated way.

The aim of this contribution is to present the summary report on the implementation of a certain class of operators that allow to study possible BSM contributions such as the
forward-backward asymmetry measured by the CDF and D\O\ collaborations
at the Fermilab Tevatron collider~\cite{Aaltonen:2011kc,:2007qb} into various matrix element generators. The goal is to provide a framework in which the anomaly observed at the Tevatron can be studied in a bottom-up approach. More precisely, we have implemented all the operators classified in Refs.~\cite{delAguila:2010mx,AguilarSaavedra:2011vw} into both {\sc FeynRules}~\cite{Christensen:2008py,Christensen:2009jx,Duhr:2011se} and 
{\sc LanHEP}~\cite{Semenov:1998eb,Semenov:2002jw,Semenov:2008jy,Semenov:2010qt}.

{\sc FeynRules } is a {\sc Mathematica}\textregistered\footnote{{\sc Mathematica}\ is a registered trademark of Wolfram Research, Inc. } package that allows computation of Feynman rules from a Lagrangian and can export them to various matrix element generators in an automated way, thus allowing implementation of the model into these tools. For the moment, interfaces to {\sc CompHEP}/{\sc CalcHEP}, {\sc FeynArts}/{\sc FormCalc}~\cite{Hahn:1998yk, Vermaseren:2000nd, Fliegner:1999jq, Fliegner:2000uy, Tentyukov:2004hz, Tentyukov:2007mu}, {\sc GoSam}~\cite{Cullen:2011ac}, {\sc MadGraph}/{\sc MadEvent}~\cite{deAquino:2011ub,Degrande:2011ua}, {\sc Sherpa} and {\sc Whizard}~\cite{Christensen:2010wz} are available.

The {\sc LanHEP} package is also designed to automatically generate Feynman rules for a given   Lagrangian.
{\sc LanHEP} is written in  the {\tt C} programming language
and can be easily compiled and installed on any platform. Its first version was released  in 1996 and was effectively used
to derive Feynman rules for the  Minimal Supersymmetric Standard Model for the {\sc CompHEP}  package.
{\sc LanHEP}  reads the Lagrangian written as a text file in a compact form, close to the one used in publications. 
Lagrangian terms can be written with summation over indices of broken symmetries
and using special symbols for complicated expressions, such as covariant derivatives
and strength tensors for gauge fields. {\sc LanHEP} creates an output file with  Feynman rules
in the format  of {\sc CompHEP}, {\sc CalcHEP} and  {\sc FeynArts}  model files as well as has an option to generate the 
output file in the form of a \LaTeX\ table.  Also, {\sc LanHEP} can generate one-loop counterterms in the {\sc FeynArts}
format. One should also add that  the latest {\sf LanHEP v.3.15} has an option (under test) for
outputting the model in UFO format which
provides a way to interface it with  {\sc MadGraph/MadEvent}.

The outline of this contribution is as follows: In Section~\ref{sec:AFB_afb} we review the operators and the spectrum of the resonances we implemented. In Sections~\ref{sec:AFB_feynrules} and~\ref{sec:AFB_lanhep} we discuss the implementation of all these operators into the {\sc FeynRules} and {\sc LanHEP} packages. We compare the two implementations in Section~\ref{sec:AFB_comparison} by computing a selection of decay rates and 2-to-2 cross section with {\sc MadGraph} 5 and {\sc CalcHEP}. Our conclusions are drawn in Section~\ref{sec:AFB_conclusion}.

\section{SCALAR AND VECTOR RESONANCES COUPLING TO QUARK PAIRS}
\label{sec:AFB_afb}
In this section we briefly discuss the various renormalizable operators of scalar or vector resonances coupling to quark pairs we have implemented into {\sc FeynRules} and {\sc LanHEP}. These operators are particularly interesting to study possible BSM origins of the anomaly in the forward-backward asymmetry in top-quark pair production measured at the Tevatron. 
In Refs.~\cite{delAguila:2010mx,AguilarSaavedra:2011vw} a classification was made of possible operators describing scalar and vector resonances carrying charges under the SM gauge group and contributing to the processes $u\,\bar{u}, d\,\bar{d} \to t\,\bar{t}$. These resonances thus provide a (non-exhaustive) list of operators that can be used to study potential BSM corrections to the asymmetry in a bottom-up approach.

The Lagrangian of the ``model'' containing all these new scalar and vector resonances can be written as a sum of three terms
\begin{equation}
\begin{cal}L\end{cal} = \begin{cal}L\end{cal}_{SM} + \begin{cal}L\end{cal}_{BSM, kin} + \begin{cal}L\end{cal}_{BSM, int}\,.
\end{equation}
While $\begin{cal}L\end{cal}_{SM}$ denotes the SM Lagrangian, $\begin{cal}L\end{cal}_{BSM, kin}$ and $\begin{cal}L\end{cal}_{BSM, int}$ describe the resonances' kinetic terms and couplings to quarks. The kinetic terms for the new resonances can schematically be written in the form
\begin{equation}\begin{split}
\begin{cal}L\end{cal}_{BSM, kin} =&\, {1\over2}D_\mu\varphi_i\,D^\mu\varphi_i -{1\over 2}m_{\varphi_i}^2\varphi_i^2
+D_\mu\Phi^\dagger_i\,D^\mu\Phi_i -m_{\Phi_i}^2\Phi_i^\dagger\Phi_i\\
&\, -{1\over4}(D_\mu v_{i\nu} - D_\nu v_{i\mu})(D^\mu v_{i}^{\nu} - D^\nu v_{i}^{\mu}) - {1\over 2}m_{v_i}^2v_{i\mu}v^\mu_i\\
&\, -\frac12 (D_\mu V^\dagger_{i\nu} - D_\nu V^\dagger_{i\mu})(D^\mu V_{i}^{\nu} - D^\nu V_{i}^{\mu}) - m_{V_i}^2V^\dagger_{i\mu}V^\mu_i\,,
\end{split}\end{equation}
where $\varphi_i$ ($\Phi_i$) and $v_{i\mu}$ ($V_{i\mu}$) denote generic real (complex) scalar and vector fields transforming in some irreducible representation of the SM gauge group, and $D_\mu$ denotes the $SU(3)_c\times SU(2)_L\times U(1)_Y$ covariant derivative. The explicit list of all scalar and vector fields considered in Refs.~\cite{delAguila:2010mx,AguilarSaavedra:2011vw} is given in Tables~\ref{tab:AFB_afb_scalar} and~\ref{tab:AFB_afb_vector}. We note that, while all gauge interactions of the new resonances are taken into account via the covariant derivative, we do not include possible mixing effects among the new resonances or with SM particles\footnote{It would in principle be possible to consider for example kinetic mixing between the SM gauge bosons and the new vector fields. We do not however include these terms into the Lagrangian.}. Besides interacting with the SM gauge bosons, the new resonances also couple to quark pairs via three-point interactions described by $\begin{cal}L\end{cal}_{BSM, int}$ and summarized in Tables~\ref{tab:AFB_afb_scalar} and~\ref{tab:AFB_afb_vector}.

\begin{center}
\begin{table}[!t]
\begin{center}
\begin{tabular}{clcc}
\hline\hline
Field & Charges & Operator & Coupling \\
\hline\hline
$\phi$ & $(1,2)_{-1/2}$ & $-\overline{Q}_{L}\,g^u\,u_{R}\,\phi -\overline{Q}_{L}\,g^d\,d_{R}\,\tilde\phi + \textrm{h.c.}$ & --\\
$\Phi$ & $(8,2)_{-1/2}$ & $-\overline{Q}_{L}\,g^u\,T^a\,u_{R}\,\Phi^a -\overline{Q}_{L}\,g^d\,T^a\,d_{R}\,\tilde\Phi^a + \textrm{h.c.}$ & --\\
$\omega^1$ & $(3,1)_{-1/3}$ & $-\epsilon_{ijk}\,\overline{d}_{Rj}\,g\,u^c_{Rk}\,\omega_i^{1\dagger} + \textrm{h.c.}$ & --\\
$\Omega^1$ & $(\bar{6},1)_{-1/3}$ & $-{1\over2}\,\overline{K}_{ij}^\alpha\,[\overline{d}_{Ri}\,g\,u^c_{Rj} + \overline{d}_{Rj}\,g\,u^c_{Ri}]\,\Omega_\alpha^{1\dagger} + \textrm{h.c.}$ & --\\
$\omega^4$ & $(3,1)_{-4/3}$ & $-\epsilon_{ijk}\,\overline{u}_{Rj}\,g\,u^c_{Rk}\,\omega_i^{4\dagger} + \textrm{h.c.}$ & $g^T=-g$\\
$\Omega^4$ & $(\bar{6},1)_{-4/3}$ & $-{1\over2}\,\overline{K}_{ij}^\alpha\,[\overline{u}_{Ri}\,g\,u^c_{Rj} + \overline{u}_{Rj}\,g\,u^c_{Ri}]\,\Omega_\alpha^{4\dagger} + \textrm{h.c.}$ & $g^T=g$\\
$\sigma$ & $(3,3)_{-1/3}$ & $-\epsilon_{ijk}\,\overline{Q}_{Li}\,g\,\tau^I\,\epsilon Q^c_{Lj}\,\sigma_{kI}^{\dagger} + \textrm{h.c.}$ & $g^T=-g$\\
$\Sigma$ & $(\bar{6},3)_{-1/3}$ & $-{1\over2}\,\overline{K}_{ij}^\alpha\,[\overline{Q}_{Li}\,g\,\tau^I\,\epsilon Q^c_{Lj} + \overline{Q}_{Lj}\,g\,\tau^I\,\epsilon Q^c_{Li}]\,\Sigma_{\alpha I}^{\dagger} + \textrm{h.c.}$ & $g^T=g$\\
\hline\hline
\end{tabular}
\caption{\label{tab:AFB_afb_scalar}Operators describing the couplings of the scalar resonances to quarks~\cite{delAguila:2010mx,AguilarSaavedra:2011vw}. $g$, $g^u$ and $g^d$ are coupling matrices in flavor space, $\tau^I$ are the Pauli matrices and $\overline{K}_{ij}^\alpha$ is the Clebsch-Gordan coefficient linking two triplet representations to a sextet representation. The last column gives the symmetries of the coupling matrices $g$ (if any).}
\end{center}
\end{table}
\end{center}

\begin{center}
\begin{table}[!t]
\begin{center}
\begin{tabular}{clcc}
\hline\hline
Field & Charges & Operator & Coupling \\
\hline\hline
$\begin{cal}B\end{cal}_\mu$ & $(1,1)_{0}$ & $-(\overline{Q}_{L}\,g^Q\,\gamma^\mu\,Q_{L} + \overline{u}_{R}\,g^u\,\gamma^\mu\,u_{R} + \overline{d}_{R}\,g^d\,\gamma^\mu\,d_{R})\,\begin{cal}B\end{cal}_\mu$ & $g^\dagger = g$\\
$\begin{cal}W\end{cal}_\mu$ & $(1,3)_{0}$ & $-\overline{Q}_{L}\,g\,\gamma^\mu\,\tau^I\,Q_{L}\,\begin{cal}W\end{cal}_\mu^I$ & $g^\dagger = g$\\
$\begin{cal}B\end{cal}^1_\mu$ & $(1,1)_{1}$ & $-\overline{d}_{R}\,g\,\gamma^\mu\,u_{R}\,\begin{cal}B\end{cal}_\mu^{1\dagger} + \textrm{h.c.}$ & --\\
$\begin{cal}G\end{cal}_\mu$ & $(8,1)_{0}$ & $-(\overline{Q}_{L}\,g^Q\,\gamma^\mu\,T^a\,Q_{L} + \overline{u}_{R}\,g^u\,\gamma^\mu\,T^a\,u_{R} + \overline{d}_{R}\,g^d\,\gamma^\mu\,T^a\,d_{R})\,\begin{cal}G\end{cal}^a_\mu$ & $g^\dagger = g$\\
$\begin{cal}H\end{cal}_\mu$ & $(8,3)_{0}$ & $-\overline{Q}_{L}\,g\,\gamma^\mu\,\tau^I\,T^a\,Q_{L} \,\begin{cal}H\end{cal}^{aI}_\mu$ & $g^\dagger = g$\\
$\begin{cal}G\end{cal}^1_\mu$ & $(8,1)_{1}$ & $-\overline{d}_{R}\,g\,\gamma^\mu\,T^a\,u_{R}\,\begin{cal}G\end{cal}_\mu^{1a\dagger} + \textrm{h.c.}$ & --\\
$\begin{cal}Q\end{cal}^1_\mu$ & $(3,2)_{1/6}$ & $-\epsilon_{ijk}\,\overline{d}_{Rj}\,g\,\gamma^\mu\,\epsilon Q^c_{Lk}\,\begin{cal}Q\end{cal}_\mu^{1i\dagger} + \textrm{h.c.}$ & --\\
$\begin{cal}Q\end{cal}^5_\mu$ & $(3,2)_{-5/6}$ & $-\epsilon_{ijk}\,\overline{u}_{Rj}\,g\,\gamma^\mu\,\epsilon Q^c_{Lk}\,\begin{cal}Q\end{cal}_\mu^{5i\dagger} + \textrm{h.c.}$ & --\\
$\begin{cal}Y\end{cal}^1_\mu$ & $(\bar{6},2)_{1/6}$ & $-{1\over 2}\,\overline{K}_{ij}^{\alpha}\,[\overline{d}_{Ri}\,g\,\gamma^\mu\,\epsilon Q^c_{Lj} + \overline{d}_{Rj}\,g\,\gamma^\mu\,\epsilon Q^c_{Li}]\,\begin{cal}Y\end{cal}_\mu^{1\alpha\dagger} + \textrm{h.c.}$ & --\\
$\begin{cal}Y\end{cal}^5_\mu$ & $(\bar{6},2)_{-5/6}$ & $-{1\over 2}\,\overline{K}_{ij}^{\alpha}\,[\overline{u}_{Ri}\,g\,\gamma^\mu\,\epsilon Q^c_{Lj} + \overline{u}_{Rj}\,g\,\gamma^\mu\,\epsilon Q^c_{Li}]\,\begin{cal}Y\end{cal}_\mu^{5\alpha\dagger} + \textrm{h.c.}$ & --\\
\hline\hline
\end{tabular}
\caption{\label{tab:AFB_afb_vector}Operators describing the couplings of the vector resonances to quarks~\cite{delAguila:2010mx,AguilarSaavedra:2011vw}. $g$, $g^Q$, $g^u$ and $g^d$ are coupling matrices in flavor space, $\tau^I$ are the Pauli matrices and $\overline{K}_{ij}^\alpha$ is the Clebsch-Gordan coefficient linking two triplet representations to a sextet representation. The last column gives the symmetries of the coupling matrices $g$ (if any).}
\end{center}
\end{table}
\end{center}

\section{IMPLEMENTATION INTO FEYNRULES}
\label{sec:AFB_feynrules}

The Lagrangian implemented into FeynRules follows exactly the classification of operators given in Refs.~\cite{delAguila:2010mx,AguilarSaavedra:2011vw}.
The implementation is completely generic, and allows to choose freely the numerical values for the new masses and couplings, including all possible flavor and/or $CP$-violating effects in the coupling constants. In this way the implementation allows for a maximum of flexibility when studying the resonances. 
We have implemented all the operators described in the previous section into {\sc FeynRules}. The implementation is based on the implementation of the Standard Model (SM) in {\sc FeynRules} shipped together with the package. The additional files needed for the new interactions can be downloaded from the {\sc FeynRules} website~\cite{FRwebpage}. In the rest of this section we give a very brief account on the implementation and how to use the model files. For a full documentation we refer to the corresponding website.

In order to allow for the study of only a subset of the new resonances, we have implemented each of the entries in Tables~\ref{tab:AFB_afb_scalar} and~\ref{tab:AFB_afb_vector} into a separate model file. A given subset of files corresponding to a given set of resonances of interest can then be loaded at run time into FeynRules. More precisely, the implementation consists of the following files,
\begin{itemize}
\item {\tt Representation.fr}: The SM implementation in {\sc FeynRules} does not contain any definitions regarding the sextet representation of the QCD gauge group. The file {\tt Representation.fr} extends the definition of the SM gauge group in {\sc FeynRules} by adding the corresponding representation.
\item {\tt SC}$xy${\tt .fr}, $x,y$ being integers: These files contain all the declarations and interactions of a new scalar particle 
transforming in the representation $(x,y)$ of $SU(3)_c\times SU(2)_L$ (\emph{e.g.}, {\tt SC12.fr} contains the definitions of the resonance $\phi$ transforming in the representation $(1,2)_{-1/2}$ defined in Table~\ref{tab:AFB_afb_scalar}).
\item {\tt V}$xy${\tt .fr}, $x,y$ being integers: Similar to {\tt SC}$xy${\tt .fr}, but for the vector resonances.
\end{itemize}
The files can then be loaded into {\sc FeynRules} and combined at leisure at run time. As an example, the SM extended by the scalar resonances $\phi$ and $\sigma$ defined in Table~\ref{tab:AFB_afb_scalar} can be loaded into {\sc FeynRules} via the command
\begin{verbatim}
LoadModel[ "SM.fr", "Representation.fr", "SC12.fr", "SC33.fr"].
\end{verbatim}
We emphasize that the file {\tt SM.fr} should be loaded first, in order to set up correctly the SM part of the model.
Furthermore, in order to allow for a maximum of flexibility, all the new numerical input parameters (masses and coupling matrices) are kept completely generic. In particular the masses of all new resonances are given a default value of 1 TeV, and all the coupling matrices are implemented as generic complex matrices (subject to the symmetry constraints given in the last column in Tables~\ref{tab:AFB_afb_scalar} and~\ref{tab:AFB_afb_vector}), initialized to the identity matrix in the model files. Note that, as the widths of the resonances are functions of the masses and the couplings, it is important to recompute the widths of the particles every time the numerical input parameters are changed. 

After the model has been loaded into {\sc FeynRules}, the Feynman rules can be obtained and exported to matrix element generators in the usual way. However, some of the interactions appearing in these operators have rather unusual color structures (color sextets, $\epsilon_{ijk}$), which are in general not supported by all the matrix element generators. As a consequence not all the interactions can be implemented into all the matrix element generators for which a {\sc FeynRules} interface exists. To our knowledge, the only matrix element generator currently able to deal with all the color structures involved is {\sc MadGraph} 5.

\section{IMPLEMENTATION INTO LANHEP}
\label{sec:AFB_lanhep}

The Lagrangian implemented into {\sc LanHEP} also corresponds exactly to the classification of operators given in Refs.~\cite{delAguila:2010mx,AguilarSaavedra:2011vw}.
One should note that {\sc LanHEP} as well as {\sc CalcHEP} need to be extended to deal with colour sextet representations
as well as  with interactions involving the epsilon colour tensor $\epsilon_{ijk}$ connecting three colour-triplet particles.
So, the respective interactions are not implemented at the moment into {\sc LanHEP}.
The request to extend {\sc LanHEP} to include these interactions has been sent to its author.

The implementation of the rest interactions from Tables~\ref{tab:AFB_afb_scalar} and~\ref{tab:AFB_afb_vector}
is done in the form of a separate model which includes also SM interactions.
All models are available  for download at the
High Energy Model Database website~\cite{HepMDB}.

As in the case of {\sc FeynRules}, the {\sc LanHEP} implementation of these models is  generic
and written  in the form of module files with a naming scheme similar to {\sc FeynRules}.
For example the term with $\mathcal{H}_\mu (8,3)_0$ interactions
is implemented in the \verb|v830.mdl| file which is included in the 
\verb|sm+v830.mdl| model. Similarly, one can add several new interactions to the SM in the {\sc LanHEP}
format.

\section{COMPARISON OF THE TWO IMPLEMENTATIONS}
\label{sec:AFB_comparison}
To verify the {\sc FeynRules} and {\sc LanHEP} implementations, we have confronted the two implementations by computing a selection of decay rates and 2-to-2 cross sections with {\sc MadGraph} 5 and {\sc CalcHEP}. The input files for {\sc MadGraph} 5 were generated with {\sc FeynRules}, whereas the input files for {\sc CalcHEP} were generated with {\sc LanHEP}. In the rest of this section we present a brief summary of the comparison. For all the processes presented in this section, the masses of all the resonances are set to 1 TeV and the coupling matrices are assumed to be the identity matrix (including the CKM matrix).

As a first check, we computed the decay rates of a resonance into light quarks. It is easy to show that, in the scenario where all the resonances have the same mass, and all the coupling matrices are proportional to the identity matrix\footnote{Note that some of the coupling matrices in Table~\ref{tab:AFB_afb_scalar} are antisymmetric. In this case we assume that the entries in the upper triangular half are equal to unity.}, the decay rates  into light quarks are linked and only one of them is independent, which we choose to be
\begin{equation}
\begin{split}
\gamma &\,\equiv \Gamma(\phi^0 \to u\,\bar{u}) = {|g|^2\,N_c\,M\over 16\pi }\,,\\
\end{split}
\end{equation}
where $\phi^0$ denotes the weak isospin $(+1/2)$ component of the scalar field $\phi$, $\phi^T = \left(\phi^0, \phi^-\right)$, $M$ is the mass of the decaying particle and $N_c$ is the number of colors.
The decay rates of all other scalar resonances are related to $\gamma$ via the relations
\begin{equation}
\begin{split}
\Gamma(\phi^- \to d\,\bar{u}) &\,= 2\,\gamma\,, \\
\Gamma(\Phi^0\to u\,\bar{u}) &\,= {1\over 6}\,\gamma\,,\\
\Gamma(\Phi^-\to d\,\bar{u}) &\,=  \Gamma(\Omega^1\to \bar{u}\,\bar{d}) =  
\Gamma(\Sigma_{2/3}\to \bar{d}\,\bar{d}) = \Gamma(\Sigma_{-1/3}\to \bar{u}\,\bar{d}) = \\
&\, = \Gamma(\Sigma_{-4/3}\to \bar{u}\,\bar{u}) =
{1\over 3}\,\gamma\,,\\
\Gamma(\omega^1\to \bar{u}\,\bar{d}) &\,= \Gamma(\Omega^4 \to \bar{u}\,\bar{u}) = \Gamma(\sigma_{-1/3}\to \bar{u}\,\bar{s})= {2\over 3}\,\gamma\,,\\
\Gamma(\sigma_{2/3}\to \bar{d}\,\bar{s}) &\,= \Gamma(\sigma_{-4/3}\to \bar{u}\,\bar{c}) = {4\over3}\,\gamma\,,\\
\Gamma(\omega^4\to \bar{u}\,\bar{c}) &\,= {8\over3}\,\gamma\,.
\end{split}
\end{equation}
Similarly, the decay rates of the vector resonances are related to $\gamma$ as well:
\begin{equation}
\begin{split}
\Gamma({\cal B}\to u\,\bar{u}) &\,= \Gamma({\cal W}^+\to u\,\bar{d}) = {4\over 3} \gamma\,,\\
\Gamma({\cal W}^3\to u\,\bar{u}) &\,= \Gamma({\cal B}^1\to u\,\bar{d}) = {2\over 3}\,\gamma\,,\\
\Gamma({\cal G}\to u\,\bar{u}) &\,= \Gamma({\cal H}^+\to u\,\bar{d})  =
\Gamma({\cal Y}^1_{-1/3}\to \bar{u}\,\bar{d}) = \Gamma({\cal Y}^1_{2/3}\to \bar{d}\,\bar{d})= \\
&\,=
\Gamma({\cal Y}^5_{4/3}\to \bar{u}\,\bar{u}) = \Gamma({\cal Y}^5_{-1/3}\to \bar{u}\,\bar{d})
= {2\over 9}\,\gamma\,,\\
\Gamma({\cal G}^1\to u\,\bar{d})  &\,=\Gamma({\cal H}^3\to u\,\bar{u}) = 
 {1\over 9}\,\gamma\,,\\
 \Gamma({\cal Q}^1_{2/3}\to \bar{d}\,\bar{d})&\, = \Gamma({\cal Q}^1_{-1/3}\to \bar{u}\,\bar{d})= 
 \Gamma_{{\cal Q}^5_{-1/3}\to \bar{u}\,\bar{d}} = \Gamma_{{\cal Q}^5_{-4/3}\to \bar{u}\,\bar{u}}=
 {4\over 9}\,\gamma\,.
 \end{split}
 \end{equation}
We have computed all the decay rates into light quarks numerically via {\sc MadGraph} 5 and {\sc CalcHEP}, and we have checked that all the above relations are verified. The results are summarized in Table~\ref{tab:Afb_decays}. Note that the {\sc LanHEP} implementation currently does not include $\epsilon_{ijk}$ or sextet color structures.

\begin{table}
\begin{center}
\begin{tabular}{ccc}
\hline\hline
 & {\sc FR + MG5} &LH + CH\\
 \hline\hline
$\phi^0 \to u\,\bar{u}$ & 59.683 & 59.683\\
$\phi^- \to d\,\bar{u}$ & 119.37 & 119.37\\
$\Phi^0 \to u\,\bar{u}$ & 9.9472 &  9.9472\\
$\Phi^- \to d\,\bar{u}$ & 19.894 & 19.894\\
$\omega^1 \to \bar{u}\,\bar{d}$ & 39.789 & --\\
$\Omega^1 \to \bar{u}\,\bar{d}$ & 19.894 & --\\
$\omega^4 \to \bar{u}\,\bar{c}$ & 159.15 & -- \\
$\Omega^4 \to \bar{u}\,\bar{u}$ & 39.789& --\\
$\sigma_{-1/3} \to \bar{u}\,\bar{s}$ & 39.789& -- \\
$\sigma_{2/3} \to \bar{d}\,\bar{s}$ & 79.578& --\\
$\sigma_{-4/3} \to \bar{u}\,\bar{c}$ & 79.578& -- \\
$\Sigma_{-1/3} \to \bar{u}\,\bar{d}$ & 19.894& -- \\
$\Sigma_{-4/3} \to \bar{u}\,\bar{u}$ & 19.894& -- \\
$\Sigma_{2/3} \to \bar{d}\,\bar{d}$ & 19.894& -- \\
\hline\hline
\end{tabular}
\hspace{10mm}
\begin{tabular}{ccc}
\hline\hline
 & {\sc FR + MG5} &LH + CH\\
 \hline\hline
${\cal B} \to u\,\bar{u}$ & 79.577 &	79.577\\ 
${\cal W}^3 \to u\,\bar{u}$ & 39.789&	39.789 \\ 
${\cal W}^+ \to u\,\bar{d}$ & 79.577&	79.577\\ 
${\cal B}^1 \to u\,\bar{d}$ & 39.789 &  39.789\\ 
${\cal G} \to u\,\bar{u}$ & 13.263  &  13.263\\ 
${\cal H}^3 \to u\,\bar{u}$ & 6.632 &  6.632\\ 
${\cal H}^+ \to u\,\bar{d}$ & 13.263 & 13.263\\ 
${\cal G}^1 \to u\,\bar{d}$ &  6.632 & 6.632\\ 
${\cal Q}^{1}_{2/3} \to \bar{d}\,\bar{d}$ & 26.526 & --\\ 
${\cal Q}^{1}_{-1/3} \to \bar{u}\,\bar{d}$ & 26.526 & --\\ 
${\cal Q}^{5}_{-1/3} \to \bar{u}\,\bar{d}$ & 26.526 & --\\ 
${\cal Q}^{5}_{-4/3} \to \bar{u}\,\bar{u}$ &  26.526& --\\ 
${\cal Y}^{1}_{-1/3} \to \bar{u}\,\bar{d}$ &  13.263& --\\ 
${\cal Y}^{1}_{2/3} \to \bar{d}\,\bar{d}$ &  13.263& --\\ 
${\cal Y}^{5}_{-4/3} \to \bar{u}\,\bar{u}$ &  13.263& --\\ 
${\cal Y}^{5}_{-1/3} \to \bar{u}\,\bar{d}$ &  13.263& --\\ 
\hline\hline
\end{tabular}
\caption{\label{tab:Afb_decays}Decay widths (in GeV) for the scalar (left) and vector (right) resonances computed with {\sc FeynRules} and {\sc MadGraph} 5 (FR+MG5), and {\sc LanHEP} and {\sc CalcHEP} (LH+CH). The {\sc LanHEP} implementation currently does not include $\epsilon_{ijk}$ or sextet color structures.}
\end{center}
\end{table}

As the main purpose of our implementation is to study possible BSM contributions to the $t\,\bar{t}$ asymmetry measured at the Tevatron, we have also compared a selection of cross sections for $u\,\bar{u}, d\,\bar{d} \to t\,\bar{t}$. We generated the signal diagrams with {\sc MadGraph} 5 and {\sc CalcHEP} for a (partonic) center-of-mass energy $\sqrt{s} = 2$ TeV, and we neglected widths in the propagators. We found perfect agreement between the {\sc FeynRules} and {\sc LanHEP} implementations. The results are summarized in Table~\ref{tab:Afb_xsecs}. In particular, the results satisfy the following relations (which follow simply from color algebra considerations):
\begin{equation}
\begin{split}
\sigma(u\,\bar{u} \to \Phi^0 \to t\,\bar{t})  &\, = {N_c^2-1\over 4N_c^2}\,\sigma(u\,\bar{u} \to \phi^0 \to t\,\bar{t}) = {2\over9}\,\sigma(u\,\bar{u} \to \phi^0 \to t\,\bar{t})\,,\\
\sigma(u\,\bar{u} \to {\cal G} \to t\,\bar{t})  &\, = {N_c^2-1\over 4N_c^2}\,\sigma(u\,\bar{u} \to {\cal B} \to t\,\bar{t}) = {2\over9}\,\sigma(u\,\bar{u} \to {\cal B} \to t\,\bar{t})\,,\\
\sigma(u\,\bar{u} \to {\cal H}^3 \to t\,\bar{t})  &\, = {N_c^2-1\over 4N_c^2}\,\sigma(u\,\bar{u} \to {\cal W}^3 \to t\,\bar{t}) = {2\over9}\,\sigma(u\,\bar{u} \to {\cal W}^3 \to t\,\bar{t})\,.
\end{split}
\end{equation}

\begin{table}
\begin{center}
\begin{tabular}{ccc}
\hline\hline
 & FR+MG5 & LH+CH \\
 \hline\hline
$u\,\bar{u} \to \phi^0, (\phi^0)^\dagger \to t\,\bar{t}$ & 1.6633 & 1.6707\\
$d\,\bar{d} \to \phi^0, (\phi^0)^\dagger \to t\,\bar{t}$ & 1.6633 & 1.6707\\
\hline
$u\,\bar{u} \to \Phi^0,(\Phi^0)^\dagger \to t\,\bar{t}$ & 0.3696& 0.3713
 \\
$d\,\bar{d} \to \Phi^0,(\Phi^0)^\dagger \to t\,\bar{t}$ & 0.3696& 0.3713 \\
\hline
$u\,\bar{u} \to {\cal B} \to t\,\bar{t}$ & 4.5618 & 4.5890\\
$d\,\bar{d} \to {\cal B} \to t\,\bar{t}$ & 4.5618 & 4.5890\\
\hline
$u\,\bar{u} \to {\cal W}^3 \to t\,\bar{t}$ & 1.1173 & 1.1122\\
$d\,\bar{d} \to {\cal W}^3 \to t\,\bar{t}$ & 1.1173 & 1.1122\\
\hline
$u\,\bar{u} \to {\cal G} \to t\,\bar{t}$ & 1.0137 & 1.0198 \\
$d\,\bar{d} \to {\cal G} \to t\,\bar{t}$ & 1.0137 & 1.0198\\
\hline
$u\,\bar{u} \to {\cal H}^3 \to t\,\bar{t}$ & 0.2483 &0.2494\\
$d\,\bar{d} \to {\cal H}^3 \to t\,\bar{t}$ & 0.2483 &0.2494 \\
\hline\hline
\end{tabular}
\caption{\label{tab:Afb_xsecs}Cross sections (in pb) computed with {\sc FeynRules} and {\sc MadGraph} 5 (FR+MG5), and {\sc LanHEP} and {\sc CalcHEP} (LH+CH). No cuts are applied to the final state.}
\end{center}
\end{table}

\section{CONCLUSION}\label{sec:AFB_conclusion}
In this contribution we have discussed the implementation into {\sc FeynRules} and {\sc LanHEP} of all the scalar and vector resonances with SM quantum numbers coupled to quark pairs presented in Refs.~\cite{delAguila:2010mx,AguilarSaavedra:2011vw}. These operators are relevant to study in a bottom-up approach the possible BSM contributions to the anomaly in the forward-backward asymmetry in top pair production measured by the CDF and D\O\ experiments. The implementation was done in a modular way, allowing the user to decide which resonances to consider, and gives the possibility to include all possible flavor and or $CP$-violating effects. 
The {\sc FeynRules} model files are publicly available and can be downloaded from the {\sc FeynRules} website~\cite{FRwebpage}. 
The {\sc lanHEP} model files will be  publicly available soon for download
at the HEPMDB website~\cite{HepMDB}.

\section*{ACKNOWLEDGEMENTS}
CD and BF would like to thank CERN and LAPTH for their
hospitality during the workshop
during which some of the work contained herein was performed. BF is supported by the Theory-LHC France-initiative of the CNRS/IN2P3. This work was supported by the Research Executive Agency (REA) of the European Union under the Grant Agreement number PITN-GA-2010-264564 (LHCPhenoNet).
A.B.  thanks the NExT Institute and Royal Society 
for partial financial support. 



\AddToContent{A.~Belyaev, N.~Christensen, C.~Duhr, B.~Fuks}
\renewcommand{\thesection}{\arabic{section}}






\chapter{SuSpect3 \label{SuSpect3_chapter}}

{\it A.Djouadi, J-L. Kneur, G. Moultaka, M. Ughetto, D. Zerwas}


\begin{abstract}
  We present here an overview of the SuSpect3 project, ultimately aimed 
to be a major upgrade of the present SuSpect2 code for calculation of 
spectra in various supersymmetric models.
\end{abstract}

\section{INTRODUCTION}
The public Fortran code {\bf SuSpect} \cite{Djouadi:2002ze} 
calculates the supersymmetric and Higgs particle spectrum in the Minimal
Supersymmetric Standard Model (MSSM). In its present version (latest 2.41), 
it can deal with specific supersymmetry-breaking models with universal 
boundary conditions at high scales, such as
the gravity (mSUGRA), anomaly (AMSB) or gauge (GMSB) mediated supersymmetry
breaking models, as well as non-universal MSSM (restricted however to $R$--parity and $CP$
conservation). Input and Output can be driven from the standard SLHA format files\cite{Skands:2003cj}.
The algorithm includes the main mandatory ingredients such as the renormalization group evolution (RGE) 
of parameters between low and high energy scales, the consistent implementation of 
radiative electroweak symmetry breaking, and the calculation of the physical masses 
of the Higgs  bosons and supersymmetric particles including the full one-loop and dominant 
two-loop radiative corrections. In addition a control of important theoretical 
and experimental features is available, such as the absence of non-physical minima, 
the amount of fine-tuning in the electroweak symmetry breaking condition, or the agreement with 
some precision observables. Although SuSpect2 is still considered essentially up-to-date and will continue to 
be maintained in the future, a major upgrade is timely for several reasons.

Given the experimental prospects for supersymmetry, the available tools for the 
interpretation of the data to come at the LHC should allow high flexibility in implementing
new models and/or new theoretical improvements and variants of the existing ones.
It is obviously necessary to keep up with the state of the art regarding any eventual new and more precise theoretical 
calculation that can affect the predictions.
It is also desirable to be able to encode in the same tool a large variety of supersymmetric models,
non-minimal extensions, more general flavor structure in the (s)quark and (s)lepton sectors,
$R$-parity violation, $CP$-violating phases, new energy thresholds related to low or high
scale extended gauge groups, modified GUT or universality conditions, etc. 
The task becomes less formidable than it may seem if one recognizes properly the 
generic features of the various extensions. An object oriented programming approach
is particularly suitable in this context. SuSpect3, which we describe below, is the first
step in this project. It is a C++ version of the latest SuSpect2 version, fully rewritten 
with an object-oriented architecture but containing the same algorithms as SuSpect2 (though details 
of the implementation have changed).
In the following the structure and the implementation will be described first. 
Then a comparison between SuSpect2 and SuSpect3 for an  mSUGRA benchmark point as well 
as a scan will be described.

\section{STRUCTURE AND IMPLEMENTATION}

C++ was chosen for the implementation of SuSpect3. The use of this language is widespread
in the experimental high energy physics community, where ATLAS and CMS rely on it for a major 
part of their software,
as well as in the theoretical community, where the event generators PYTHIA8~\cite{Sjostrand:2007gs} 
and Herwig++~\cite{Bahr:2008pv} have 
moved from FORTRAN to this language.

In writing SuSpect3 care was taken not to resort to a line-by-line technical 
conversion of SuSpect2 as this would have been a change of syntax with no added benefit.
The rewrite limits the inheritance structure to essentially three layers. 
Additionally the possibility of using other codes for parts of the calculations usually performed
by SuSpect2 is also foreseen. An example is the use of the renormalization group equations provided by 
tools such as SARAH~\cite{Staub:2008uz} or Feynrules~\cite{LesHouchesBSM2011}.

As a first test of the flexibility and the robustness of the new implementation, the first 
and second generation of sfermions were separated. In SuSpect2 these were hard-coded to be equal. 
The change took less than a day (testing included), giving a first indication of the robustness of the
new structure. 

The internal communication between different objects is assured by
an implementation of the SLHA structure in memory. The SLHA object
contains all SLHA1 standard blocks as well as supplementary blocks 
used internally.

The top object \texttt{suspect} has three methods: Initialize, Execute and Finalize. 
For the initialization polymorphism is used: one can either use the default settings
of SuSpect (SPS1a), read an SLHA file by passing the name of the file as argument 
or create and initialize an SLHA object in memory and pass it as argument to the object \texttt{suspect}. 
After the calculation, the output is given either on screen or written to a file. 

The requested model is configured in the \texttt{suspect} object:
the MSSM (defined at either a high scale or at the EWSB scale), mSUGRA, AMSB, mGMSB and CompressedSUSY  
have been implemented so far. 
In these objects only the initialization
of the model as well as the definition of the boundary conditions 
have to be implemented. The next layer consists of \texttt{Model4Scales}, \texttt{Model3Scales} 
or \texttt{Model2Scales}. These classes implement the logic of the sequence 
of the calculation of the spectrum. The separation is driven by the number of 
scales involved in the definition of the model.
All \texttt{Model2/3/4Scales} are objects which inherite from a common baseclass \texttt{ModelBase}. 
This class holds the data members common to all classes: 
an RGE solver, determination of the $\overline{\text{DR}}$ parameters, 
the calculators of the particle masses, 
the implementation of electroweak symmetry breaking (\texttt{EWSBBase}) etc. 
As an example: the object \texttt{ModelmSUGRA} (created in the object \texttt{suspect}) 
inherits from \texttt{Model3Scales} as its parameters
are defined at the GUT scale, RGE-evolved to the EWSB scale and corrections
to the Standard Model particles are calculated at the scale of the Z boson 
mass. \texttt{ModelAMSB} is derived from \texttt{Model3Scales}. As GMSB has an 
additional intermediate scale, it is derived from the \texttt{Model4Scales}, whereas 
\texttt{ModelLowScaleMSSM} is derived from \texttt{Model2Scales}.

The RGE evolution from one scale to another is steered by the object \texttt{Model2/3/4Scales}.
The \texttt{RgeEvolution} object can be called multiple times in a calculation. 
The object fetches the information in the SLHA object, solves the RGE equations
and writes the final result back into the SLHA object. 

Each particle to be calculated is implemented as an object which can be configured
to include radiative corrections or not. All particle objects inherite 
from a common baseclass. The calculation of the particle
masses from the parameters is steered by the object \texttt{ModelBase}.
The result of the calculations is stored in the SLHA object.  

If the environment variable \texttt{ROOTSYS} is defined, ROOT will 
be linked automatically. If the variable is not defined, the compilation
will be performed without in a transparent way. The ROOT output file
contains a tree \texttt{suspect3} with the content of the SLHA blocks. 

\section{Results}

\begin{center}
\begin{figure}[!ht]
\begin{minipage}[h]{0.5\linewidth}
\centering
\includegraphics[width=0.98\linewidth]{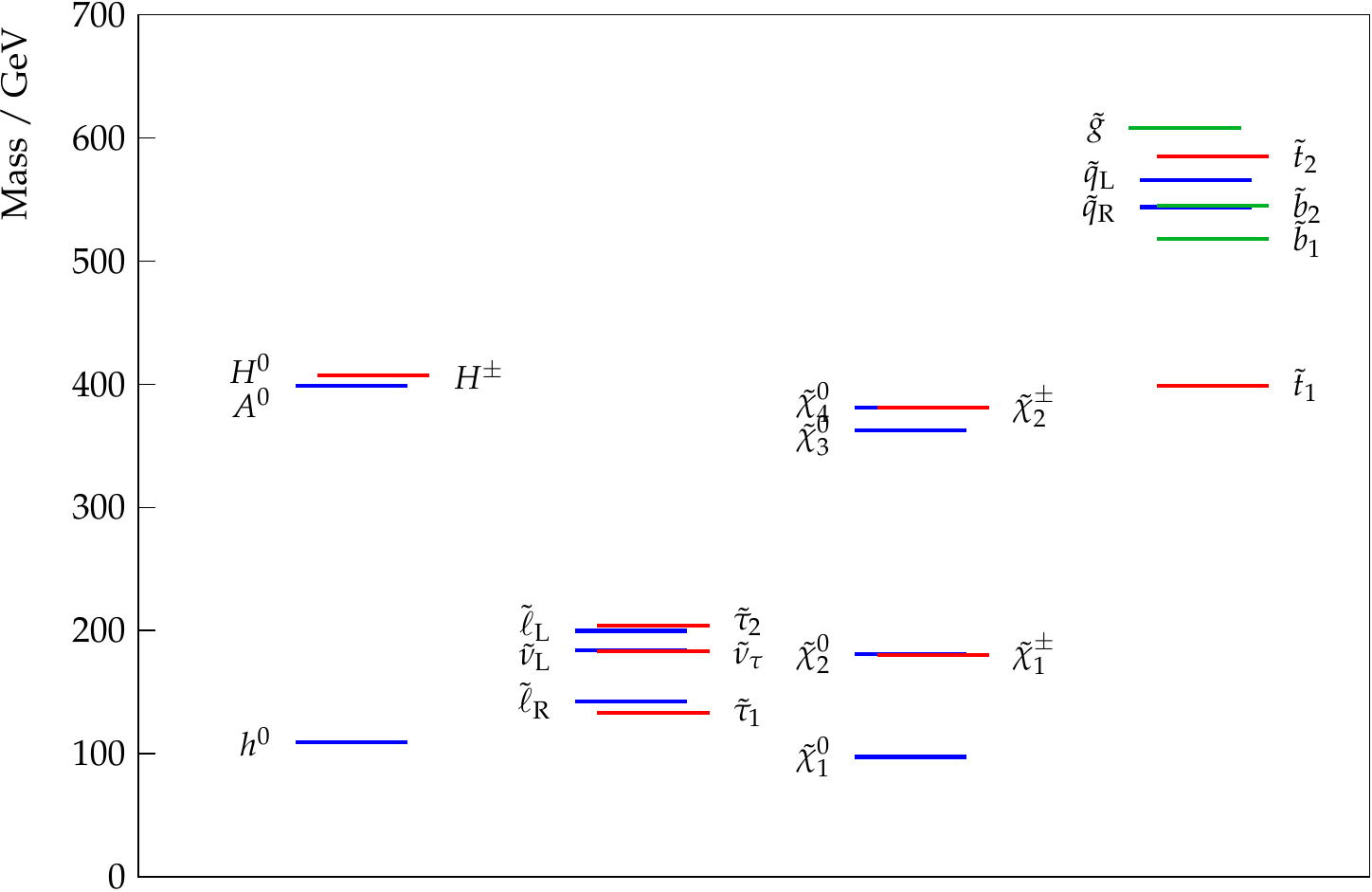}
\vspace{8mm}
\end{minipage}
\hspace{-0.2cm}
\begin{minipage}[h]{0.7\linewidth}
\centering
\includegraphics[width=\linewidth]{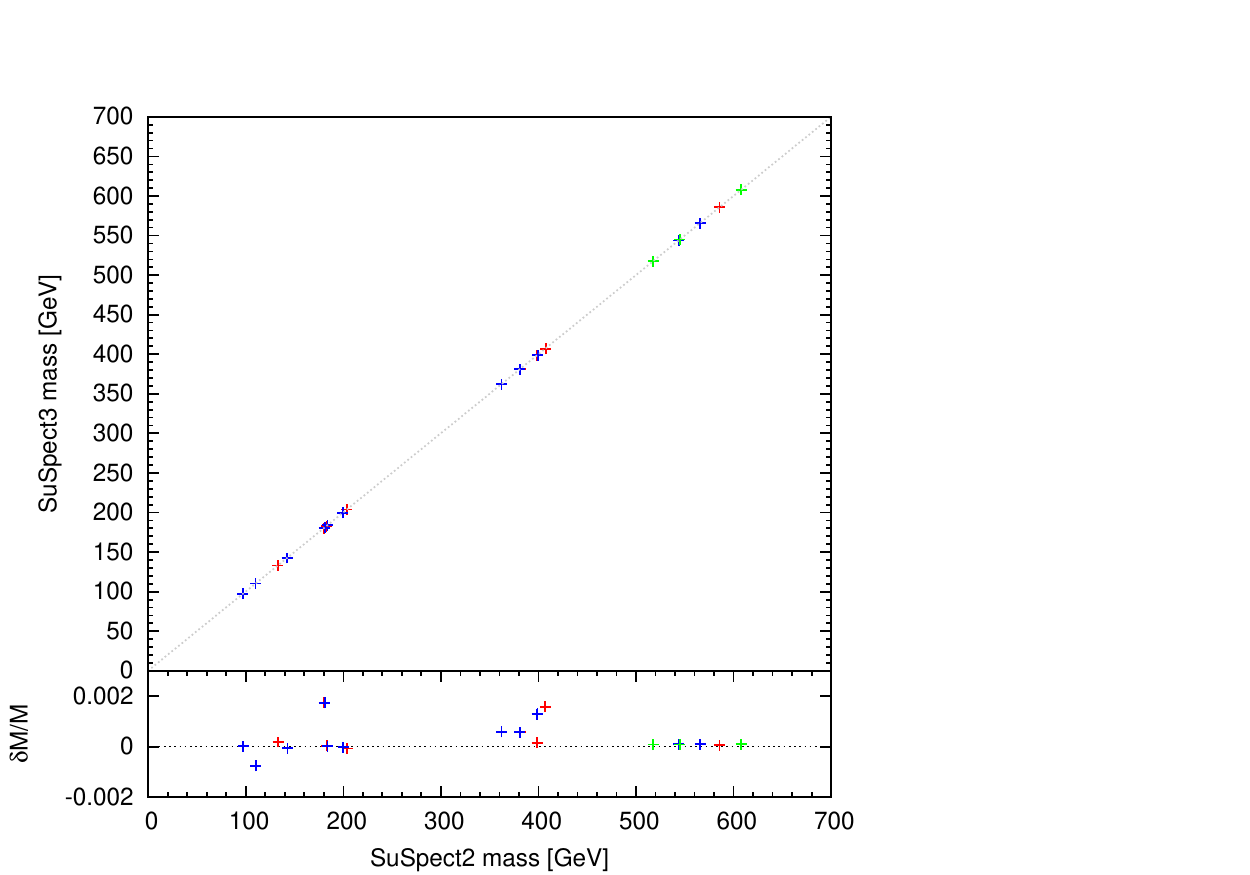}
\end{minipage}
\caption{\label{SuSpect3_sps1a} On the left, the SPS1a spectrum is shown as 
computed by SuSpect3. On the right, the comparison of the masses calculated by SuSpect2 and SuSpect3 is shown.}
\end{figure}
\end{center}

As a benchmark point the well known mSUGRA point SPS1a~\cite{Allanach:2002nj} has been used
to compare SuSpect3 and SuSpect2.
SPS1a has the GUT scale boundary conditions: $A_0=-100$~GeV, $\tan\beta=10$, $m_0=100$~GeV and $m_{1/2}=250$~GeV. The 
Higgs mass parameter $\mu$ was chosen to be positive.
The electroweak symmetry breaking scale (EWSB) is set to $1$~TeV as suggested in Ref.~\cite{AguilarSaavedra:2005pw}.
At each step of the algorithm, comparisons were performed to make sure that all details agree. 
As an example, the spectrum predicted by SuSpect3 as well as the comparison of the mass of the particles
with respect to SuSpect2 are shown in Figure~\ref{SuSpect3_sps1a}. The two predictions are in good agreement 
at the per mil level.

\begin{figure}[!ht]
\includegraphics[width=0.5\linewidth]{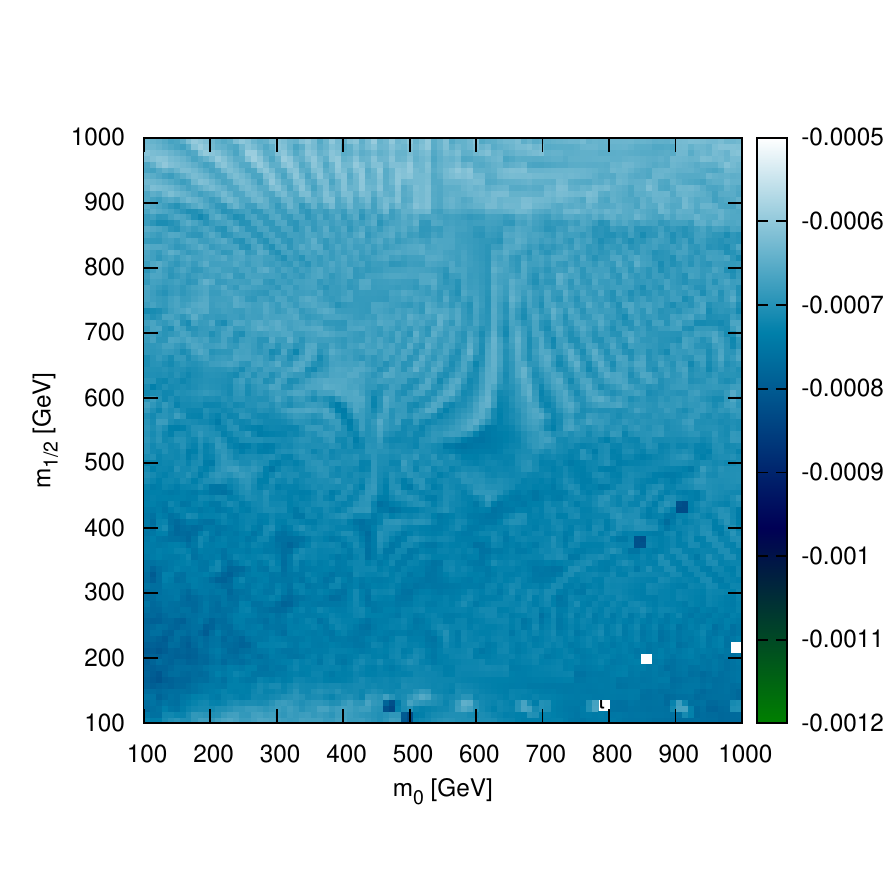}\includegraphics[width=0.5\linewidth]{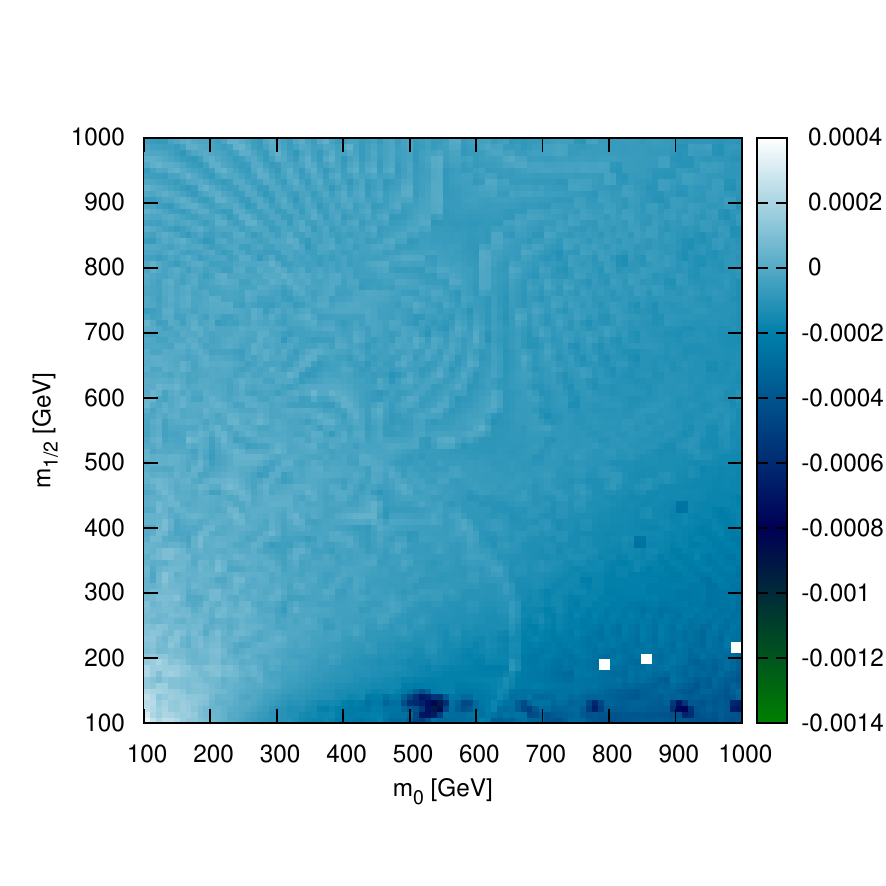}
\caption{\label{SuSpect3_higgs-stop1-scans} The two plots show relative differences between SuSpect2 and SuSpect3 with respect to SuSpect3. The left one is for $M_h$, and the right one for $M_{\tilde{t}_1}$. Scan on $m_0$ and $m_{1/2}$ has been performed for mSUGRA with $A_0=-100$, $\tan\beta=10$ and $\mu>0$.}
\end{figure}

As a further illustration a scan was performed in mSUGRA. Three parameters ($A_0=-100$~GeV, $\tan\beta=10$, $\mu>0$) were fixed.
The common scalar and sfermion masses were scanned from $100$~GeV to $1$~TeV in steps of $9$~GeV.
In Figure~\ref{SuSpect3_higgs-stop1-scans} the relative deviation of SuSpect2 and SuSpect3 is shown. 
As an illustration the mass of the lightest Higgs boson and the mass of the lightest stop quark are used. The first
is sensitive to the radiative corrections and the second is sensitive to the mixing effects. Thus both are good
sensitive indicators whether all ingredients of the calculation agree.
With the exception of 3 points each for the stop and the Higgs mass the agreement is at the per mil level
for the 10201 points analyzed.
For the stop quark mass the RMS of the relative difference is one tenth of a per mil. 

\section{OUTLOOK}

The rewriting of SuSpect3 in C++ with added features and a more flexible design to accomodate the use of 
calculations by other tools is on its way. All models currently implemented in SuSpect2 have been implemented 
in SuSpect3. All major options allowed by SuSpect2 have also been implemented in SuSpect3. 
The first comparisons for SPS1a show very good agreement. In the next months SuSpect3 will 
be tested intensely before making it available to a selected public for tests and later on releasing
it publicly as an addition to SuSpect2.

\section*{ACKNOWLEDGEMENTS}
D.Z. would like to thank the Aspen Center for Physics for its hospitality.
Part of this work was developed in the GDR Terascale (CNRS).



\AddToContent{A.~Djouadi, J-L.~Kneur, G.~Moultaka, M.~Ughetto, D.~Zerwas}
\renewcommand{\thesection}{\arabic{section}}


\superpart{ Higgs Boson and Top Quark Physics }





\chapter{Implications of new physics for Higgs searches at the LHC }

{\it Daniel Albornoz V\'asquez,   
Alexandre Arbey, 
Marco Battaglia,
Genevi\`eve B\'elanger,  
Alexander Belyaev, 
C\'eline Boehm,
Fawzi Boudjema,  
Matthew Brown, 
Jonathan Da Silva, 
Abdelhak Djouadi,
Guillaume Drieu La Rochelle, 
B\'eranger Dumont, 
Rohini Godbole, 
Mitsuru Kakizaki,  
Sabine Kraml,  
Farvah Mahmoudi, 
Alexander Pukhov, 
Sezen Sekmen,   
Ak\i{}n Wingerter,
Chris Wymant}

\begin{abstract}
We examine to what extent the Higgs sectors of extensions of the Standard Model (SM) are 
constrained on the one hand by the LHC searches for the SM Higgs boson, on the other hand by 
LHC searches for new particles beyond the SM.  
The models considered include the MSSM and some of its extensions, as well as a minimal UED model 
and the SM with a sequential 4th generation. Among other issues, we discuss expectations for the signal 
strength of a light Higgs in the $\gamma\gamma$ channel in these models. Where relevant we discuss also the $WW/ZZ$ channels.  
Moreover, we assess the implications of the recent hints of a 
Higgs signal at the LHC corresponding to a Higgs mass near 125~GeV. 
\end{abstract}

\section{Introduction}
The search for the Higgs boson is one of the primary goals of the Large Hadron Collider (LHC).  
At present, combined ATLAS analyses with integrated luminosity of up to 4.9~fb$^{-1}$  
exclude, at 95\%~CL, a SM Higgs in the mass ranges $112.7~{\rm GeV}<M_H<115.5~{\rm GeV}$ and 
$131~{\rm GeV}<M_H<453~{\rm GeV}$~\cite{ATLAS-CONF-2011-163}.
CMS excludes at 95\%~CL the SM Higgs in the range $127~{\rm GeV}<M_H<600~{\rm GeV}$ 
with integrated luminosity of 4.6--4.7~fb$^{-1}$~\cite{Chatrchyan:2012tx}. This leaves a window of   
$m_h=115.5-127$~GeV, in good agreement with expectations from electroweak fits.  

Furthermore, ATLAS reported an excess of 2.8$\sigma$ in $H\rightarrow \gamma\gamma$, consistent 
with $M_H\approx 126$~GeV~\cite{ATLAS-CONF-2011-163}. This excess is larger than expected from a SM Higgs 
by $(\sigma\times BR)/(\sigma\times BR)_{\rm SM}=2\pm 0.8$. 
CMS also reported an excess in the same channel but for a 
slightly lower mass of $M_H\approx 124$~GeV---this excess is consistent with a SM Higgs~\cite{Chatrchyan:2012tx}. 
New limits in the channels $H\rightarrow\tau\tau,\, bb$ as well as in $H\rightarrow WW,\, ZZ$ with the 
gauge bosons decaying leptonically were also reported. 
For a Higgs lighter than about 140 GeV, limits on $(\sigma\times BR)/(\sigma\times BR)_{\rm SM}$ range 
from 3--5~\cite{Chatrchyan:2012tx} in the $\tau\tau$ channel and from 3--8 in the $b\bar b$ channel from vector boson fusion. 

These results, however, cannot be applied directly to the Higgs boson(s) in extensions of the SM.
Indeed for both production and decay, the predictions for non-standard Higgses~\cite{CPNSH} 
can differ significantly from the SM ones. 
First, the Higgs couplings to vector bosons or to fermions can differ from the values in the SM.  
In supersymmetry (SUSY), for instance, these couplings (in particular the ones to b-quarks) are modified in the limit 
of low pseudoscalar Higgs mass, $M_A$~\cite{Djouadi:2005gj,Drees:book,Baer:2006rs}. 
Furthermore, the loop-induced couplings $hgg$ and $h\gamma\gamma$ can receive important  
contributions from new fermions and scalars. 
Finally, there can be new decay modes of the Higgs either into stable particles 
(e.g., light dark matter candidates~\cite{Belanger:2000tg,ArkaniHamed:2000bq,Belanger:2001am,Bottino:2003cz,Mambrini:2011ri,Ghosh:2011qc,Andreas:2010dz}, leading to invisible Higgs decays), or into new particles whose decay modes are not searched for, e.g., 
the singlet Higgs of the NMSSM~\cite{Dermisek:2005ar,Chang:2008cw,Ellwanger:2009dp} 
or of models with light scalar singlets~\cite{vanderBij:2006pg}. 
Enhanced invisible decays of the light Higgs of course lead to reduced rates for all visible modes. 
Dedicated searches for the invisible Higgs were advocated at the LHC in either vector boson fusion~\cite{Eboli:2000ze,Davoudiasl:2004aj} or in associated vector boson production~\cite{Godbole:2003it,Davoudiasl:2004aj}.
 
The implications of supersymmetry and other extensions of the SM on Higgs searches have been extensively 
studied over the years. 
In SUSY, the effect of top squarks on loop-induced Higgs processes was found to be large 
especially for light stops~\cite{Djouadi:1998az,Belanger:1999pv,Carena:2002qg,Kinnunen:2005aq,Dermisek:2007fi,Low:2009nj}, while the effect of charginos on the $h\to\gamma\gamma$ width was shown to be more modest~\cite{Djouadi:1996pb,Belanger:2000tg}. Changes in the $h\to b\bar{b}$ partial width due to supersymmetric corrections can enhance or suppress the branching ratio into photons~\cite{Carena:2002qg,Djouadi:2005gj}. 
The modification of the $H\to b\bar{b}$ partial width by introducing an extra singlet that mixes with the doublet 
Higgs can also enhance the $\gamma\gamma$ branching ratio in the NMSSM~\cite{Ellwanger:2010nf} or in models 
with a hidden Higgs sector~\cite{Lebedev:2011iq}.
More generally, in the MSSM with new degrees of freedom (BMSSM), modifications of the Higgs couplings 
can potentially lead to largely suppressed or enhanced signals in all channels~\cite{Carena:2010cs,Carena:2011dm,Boudjema:2011aa}. 
In universal extra-dimension models, little Higgs models and in models with a fourth generation, the new top-like quark in the loop is known to increase the Higgs production from gluons, while the decay into photons is only moderately affected.
On the contrary in extra-dimensional models based on RS (or their composite Higgs 4-d duals), the Higgs production is generally reduced ~\cite{Falkowski:2007hz,Low:2009di }.

Last but not least, the above-mentioned excess of events reported by the LHC collaborations hinting at a 
more or less SM-like Higgs with mass near 125~GeV motivated many interpretations in the contexts of 
extensions of the SM, including e.g.~\cite{Arbey:2011ab,Baer:2011ab,Hall:2011aa,Guo:2011ab,Heinemeyer:2011aa,Draper:2011aa,
Ferreira:2011aa,Carena:2011aa,Ellwanger:2011aa,Burdman:2011ki,Gunion:2012zd,King:2012is,Djouadi:2011aa}. 

In this contribution, we reexamine the case of Higgs production and decays in several extensions of the SM:  the constrained Minimal Supersymmetric Standard Model (CMSSM) and 
the Minimal Supersymmetric Standard Model (MSSM) with parameters defined at the electroweak scale, 
including the so-called phenomenological MSSM (pMSSM) with 19 free parameters 
as well as a similar model with 11 free parameters; the BMSSM with the same field 
content as the MSSM but allowing for higher-dimensional operators; 
the next-to-minimal MSSM (NMSSM); the MSSM with L--R mixed sneutrinos;  
the MSSM with an extended gauge symmetry (UMSSM);  the minimal UED model; and 
the SM with a fourth generation (SM4).   
For each case, we describe the constraints on the parameter space used in the different analyses and examine the consequences for the Higgs sector, both for the mass range and/or for the signal strength in the most relevant 
channels, generally $gg\to H\to \gamma\gamma$. When relevant, other modes, including invisible Higgs decays, 
are  discussed. We also highlight in some cases the implications of the excess corresponding to 
$M_H\approx 125$~GeV observed at the LHC.

\section{The MSSM}

The expectations for the Higgs in the MSSM were explored in a variety of models defined at the GUT scale, such as the CMSSM, AMSB, GMSB, or at the electroweak scale, such as the pMSSM.  In all these models, the light Higgs mass is expected to be less than about 135 GeV (see~\cite{Djouadi:2005gj} and references therein) and above the LEP bound of 114 GeV, thus in the region where the main search channel at the LHC is  $gg\rightarrow h\rightarrow \gamma\gamma$. 
To raise the Higgs mass much above the Z scale requires large contributions from 3rd generation squarks; 
in particular a large mixing in the stop sector is needed to reach the high end of the $M_H$ range possible 
in the MSSM. 
Note, however, that a Higgs with mass near 125 GeV, as favored by recent LHC results, requires some 
fine-tuning~\cite{Hall:2011aa}. 
The parameter space of a given model is subject to various constraints from precision data, searches for new particles at colliders, flavor physics as well as  astroparticle and cosmology constraints. Taking into account all these constraints, the allowed range for the Higgs mass can be much restricted.

The LHC Higgs searches are sensitive not only to the value of the light Higgs mass but also to the strength of the 
Higgs signal in each search channel. 
In the following, we will examine the predictions for the expected light Higgs mass in the MSSM with parameters 
defined at the weak scale, taking into account LHC constraints on SUSY and/or Higgs particles, and/or dark matter constraints. Moreover, we consider 
the signal strength for the main search channel, defined as
\begin{equation}
R_{gg\gamma\gamma}\equiv \frac{\sigma(gg\rightarrow h)_{\rm BSM} BR(h\rightarrow \gamma\gamma)_{\rm BSM}}{\sigma(gg\rightarrow h)_{\rm SM} BR(h\rightarrow \gamma\gamma)_{\rm SM}} \,. 
\end{equation}

Before showing these results,  let us summarize the expectations for the (effective) Higgs couplings to 
SM particles in the MSSM. The $ggh$ coupling is dominated by the top-quark loop, while the $h\gamma\gamma$ 
coupling  is dominated by the contribution from $W$ bosons with a subdominant contribution from top quarks of opposite sign. Both couplings can receive a large contribution from third generation sfermions, in particular from stops~\cite{Djouadi:1998az}. 
The light stop interferes constructively with the top when there is no mixing in the stop sector, while the interference is destructive in the case of large mixing. In the first case $hgg$ will increase, while in the latter it will decrease. 
The contrary is true for the $h\gamma\gamma$ couplings.  Weakly interacting particles such as charginos and 
sleptons  will contribute only to the $h\gamma\gamma$ coupling. In particular,  light staus that are heavily mixed can enhance the $h\to\gamma\gamma$ rate~\cite{Carena:2011aa}.

Apart from the contribution of light superpartners in the loops, other particles that can induce large modifications 
of some of the light Higgs couplings and branching ratios are light pseudoscalars and light LSP's (lightest SUSY particle). 
The latter lead to new invisible decay modes of the Higgs, and thus to reduced rates for all visible modes. 
The light pseudoscalar case is now strongly constrained by LHC searches for a Higgs produced through b-quarks and decaying into $\tau\tau$. Nevertheless values of  $M_A$ for which the tree-level couplings of the Higgs to SM particles can show large deviations as compared to the SM case are still possible. 
For example the ratios of the Higgs couplings to SM particles, $R_{WWh}=\sin(\alpha-\beta)$ and $R_{tth}=\cos\alpha/\sin\beta$ where $\alpha$ is the Higgs mixing angle, are considerably smaller than one only for low values of $M_A$, while   
$R_{bbh}=R_{\tau\tau h}=\sin\alpha/\cos\beta$ is enhanced especially at large values of $\tan\beta$~\cite{Belanger:1999pv,Djouadi:2005gj}. Furthermore, corrections to  the $hb\bar b$ vertex arise from higher-order effects, in particular the $\Delta M_b$ correction, which can lead to a much enhanced $hb\bar b$ coupling at large values of $\tan\beta$~\cite{Djouadi:2005gj}. Because the $b\bar b$ mode is the dominant decay channel of the light Higgs, in this case the total width of the Higgs becomes much larger in the MSSM than in the SM. This means that the branching ratios into other modes, such as $BR(h\to \gamma\gamma)$, can be suppressed even if the $h\gamma\gamma$ coupling is itself SM-like. Supersymmetric contributions can thus lead to modifications in both production and decays as compared to the SM.

\subsection{Constrained MSSM} 
{\it A. Arbey, M. Battaglia, A. Djouadi, F. Mahmoudi} 

\begin{figure}[!t]
\begin{center}
\includegraphics[width=0.5\textwidth]{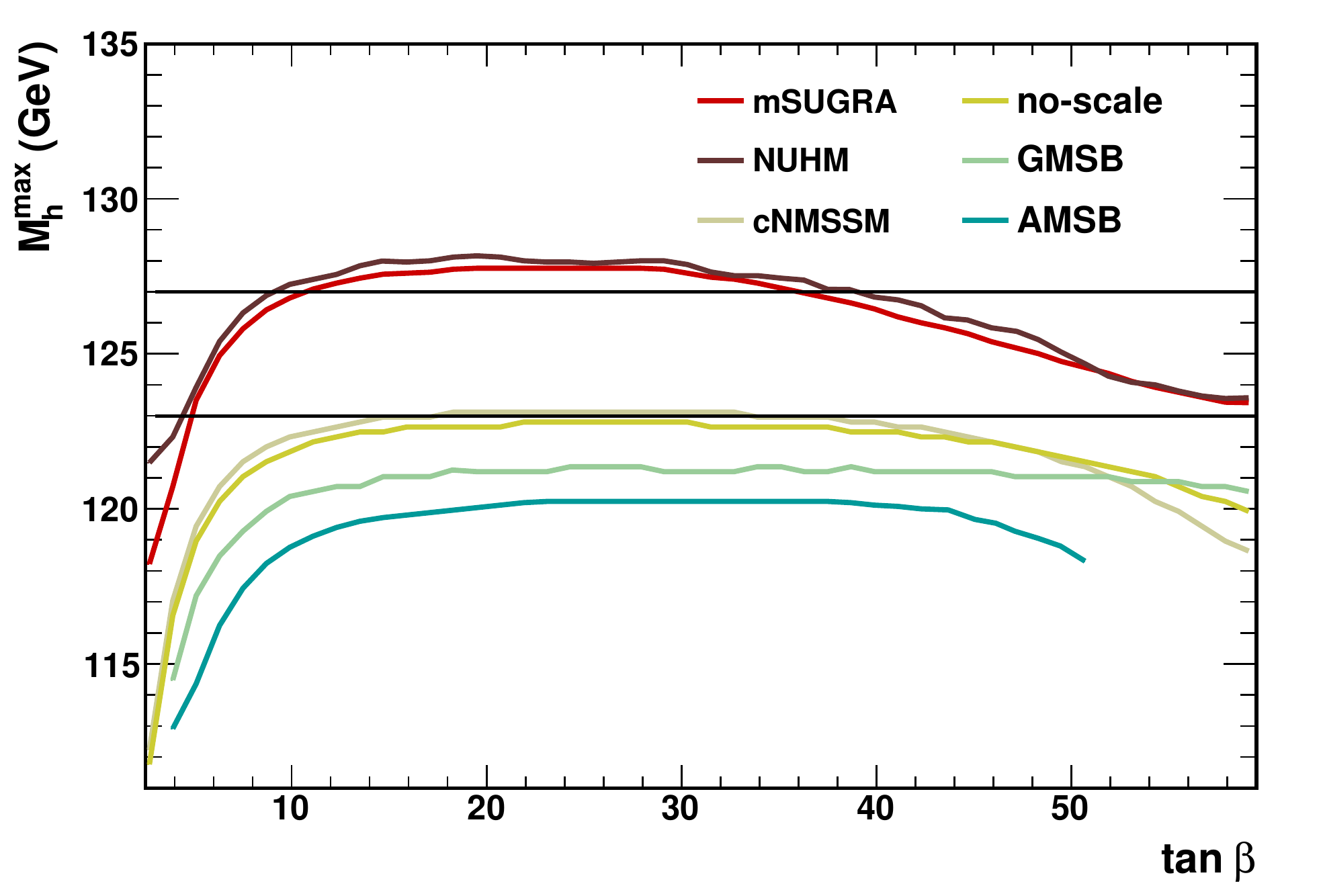}
\end{center}
\caption[]{The maximal value of the h mass defined as the value for
which 99\% of the scan points have a mass smaller than it, shown as a
function of $\tan\beta$ for the various constrained MSSM models.}
\label{Fig:cMSSM}
\end{figure}

In constrained MSSM scenarios the various soft SUSY--breaking parameters
obey a number of universal boundary conditions at a high energy scale
such as the GUT scale, thus reducing the number of basic input
parameters. Since the various parameters which enter the radiative
corrections to the MSSM Higgs sector are not all independent, it is not
possible to freely tune the relevant weak--scale parameters to obtain a
given value of $M_h$. In particular, a Higgs mass of around 125 GeV can
have drastical consequences on the constrained MSSM scenarios
\cite{Arbey:2011ab} as can be seen from Fig. \ref{Fig:cMSSM}.
The upper bound on $M_h$ in these scenarios can be qualitatively
understood by considering in each model the allowed values of the
trilinear coupling $A_t$, which essentially determines the stop mixing
parameter $X_t$ and thus the value of $M_h$ for a given scale $M_S$. In
GMSB, one has $A_t\approx 0$ at relatively low scales and its magnitude
does not
significantly increase in the evolution down to the scale $M_S$. This
implies that we are almost in the no--mixing scenario which gives a low
value of $M_h$. In AMSB, one has a non-zero $A_t$ that is fully
predicted at any renormalisation scale in terms of the Yukawa and gauge
couplings; however, the ratio $A_t/M_S$ with $M_S$ determined from the
overall SUSY breaking scale $m_{3/2}$ turns out to be rather small,
implying again that we are close to the no--mixing scenario. Finally, in
the mSUGRA model, since we have allowed $A_t$ to vary in a wide range as
$|A_0|\leq  9$ TeV, one can get a large $A_t/M_S$ ratio which leads to a
heavier Higgs particle. However, one cannot easily reach $A_t$ values
such that $X_t/M_S \approx \sqrt 6$ so that we are not in the
maximal--mixing scenario.
In turn, in two particular cases of mSUGRA, namely the ``no--scale" and
the ``approximate cNMSSM" scenarios, the upper bound on $M_h$ is much
lower than in the more general mSUGRA case and, in fact, barely reaches
the value $M_h \approx 123$ GeV. The main reason is that these
scenarios involve small values of $A_0$ at the GUT scale, $A_0 \approx
0$ for no--scale and $A_0 \approx -\frac14 m_{1/2}$ for the cNMSSM. One
then obtains $A_t$ values at the weak scale that are too low to generate
a significant stop mixing and, hence, one is again close to the
no--mixing scenario. Thus, only a very small fraction of the parameter
space of these two sub--classes of the mSUGRA model survive (in fact,
the ones leading to the  $M_h^{\rm max}$ value) if one indeed has $M_h
=125 \pm 2$ GeV. These models hence should have a very heavy spectrum as
a value $M_S \gtrsim 3$ TeV is required to increase $M_h^{\rm max}$. In
the VCMSSM, values $M_h \simeq 124.5$ GeV can be reached as $A_0$ can be
large for large $m_0$, $A_0 \approx -m_0$, allowing at least for typical
mixing.
Finally, since the NUHM is more general than mSUGRA as we have two more
free parameters, the $[\tan\beta, M_h]$ area shown in
Fig.~\ref{Fig:cMSSM} is larger than in the mSUGRA case. However, since
we are in the decoupling regime and the value of $M_A$ does not matter
much (as long as it a larger than a few hundred GeV) and  the key
weak--scale parameters entering the determination of $M_h$, i.e. $\tan
\beta, M_S$ and $A_t$ are approximately the same in both models, one
obtains a bound $M_h^{\rm max}$ that is only slightly higher in NUHM
compared to
mSUGRA. Thus, the same discussion above on the mSUGRA scenario, holds
also true in the NUHM case.

\subsection{pMSSM}
{\it A. Arbey, M. Battaglia, A. Djouadi, F. Mahmoudi} 

Studies considering  general MSSM scenarios without implicit correlations between the masses of the supersymmetric particle partners, such as the 19-parameter phenomenological
MSSM (pMSSM), have demonstrated that a wide phase space of solutions, compatible with flavour physics, low energy data and dark matter constraints and beyond the current sensitivity of the LHC experiments, exists~\cite{Conley:2011nn, Sekmen:2011cz, Arbey:2011un}. Even at the end of next year LHC run, with an anticipated integrated luminosity of order of 15~fb$^{-1}$ per experiment, many of these solutions, in particular those with gluino masses above 800~GeV will not be tested. These solutions are compatible with all present bounds and a good fraction of them has acceptable values of the fine tuning parameter. This will not make possible to falsify the MSSM as the model of new physics beyond the SM and the source of relic dark matter in the universe, if no missing $E_T$ signal will be observed at the LHC by the end of 2012. On the other hand, searches for the Higgs bosons can be used as an alternative path to tightly constrain and test the MSSM at the LHC.

\begin{figure}[t!]
\begin{center}
\begin{tabular}{c c}
\hspace*{-2.5mm}\includegraphics[width=0.35\textwidth]{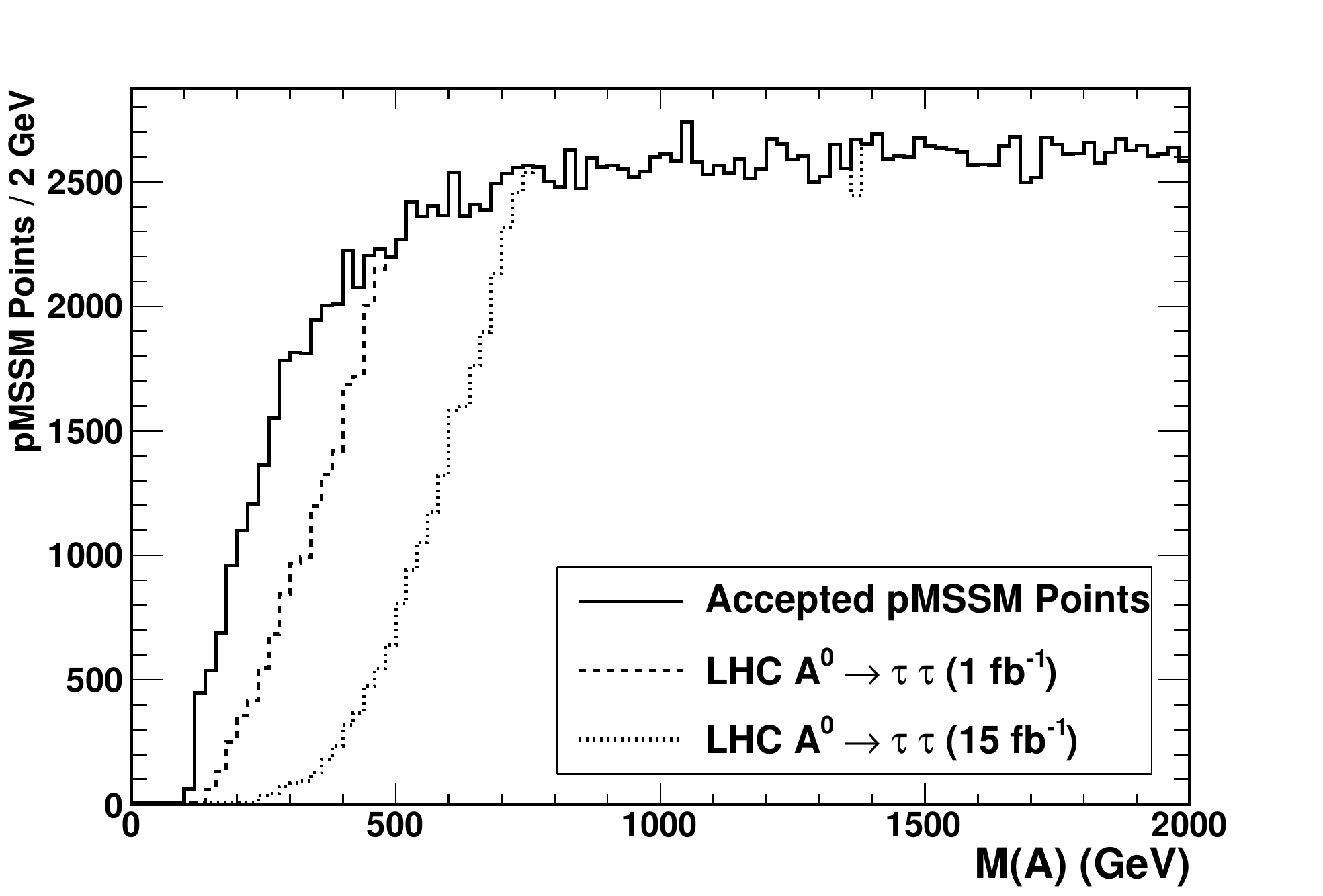}
\hspace*{-2.5mm}\includegraphics[width=0.35\textwidth]{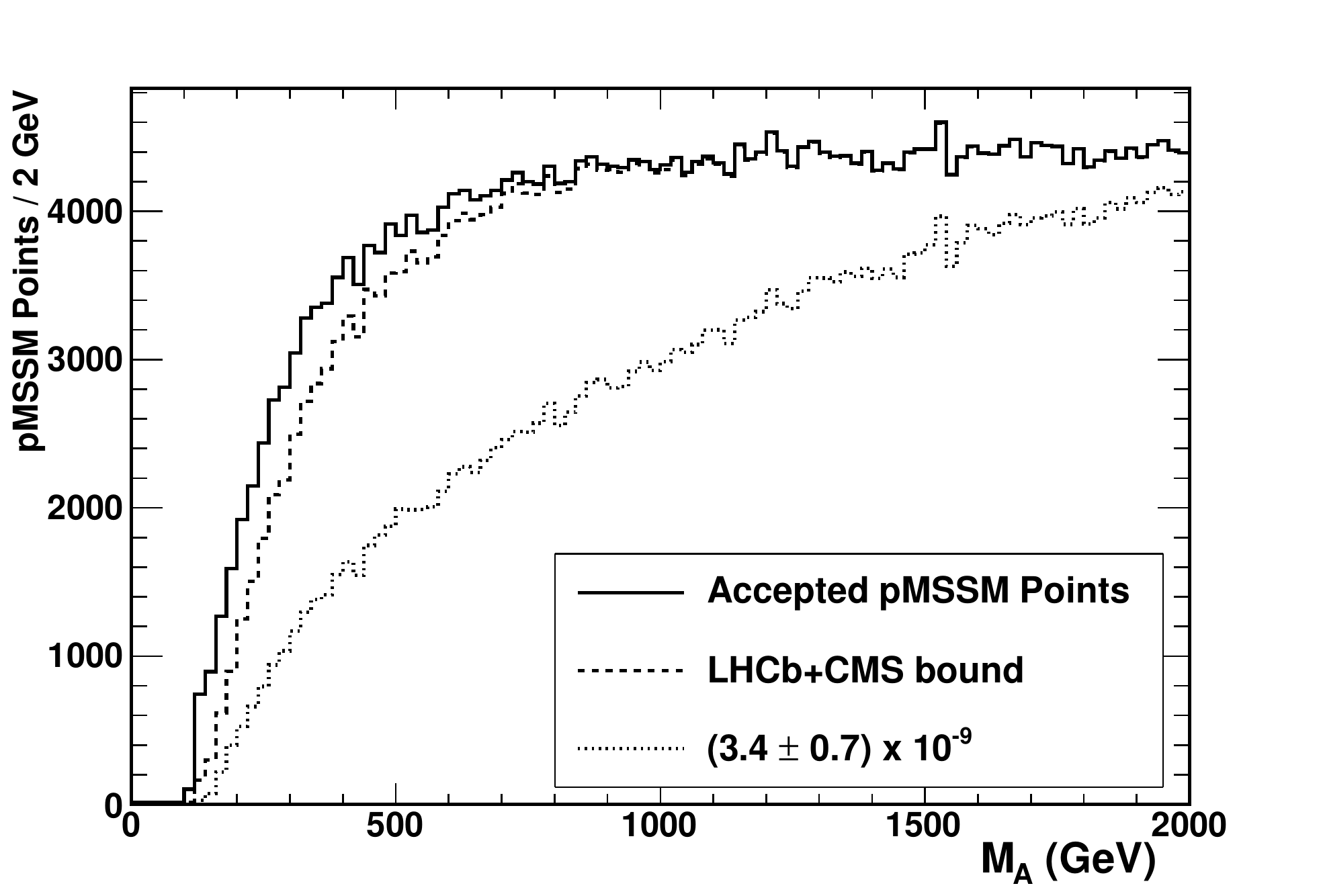}
\hspace*{-2.5mm}\includegraphics[width=0.35\textwidth]{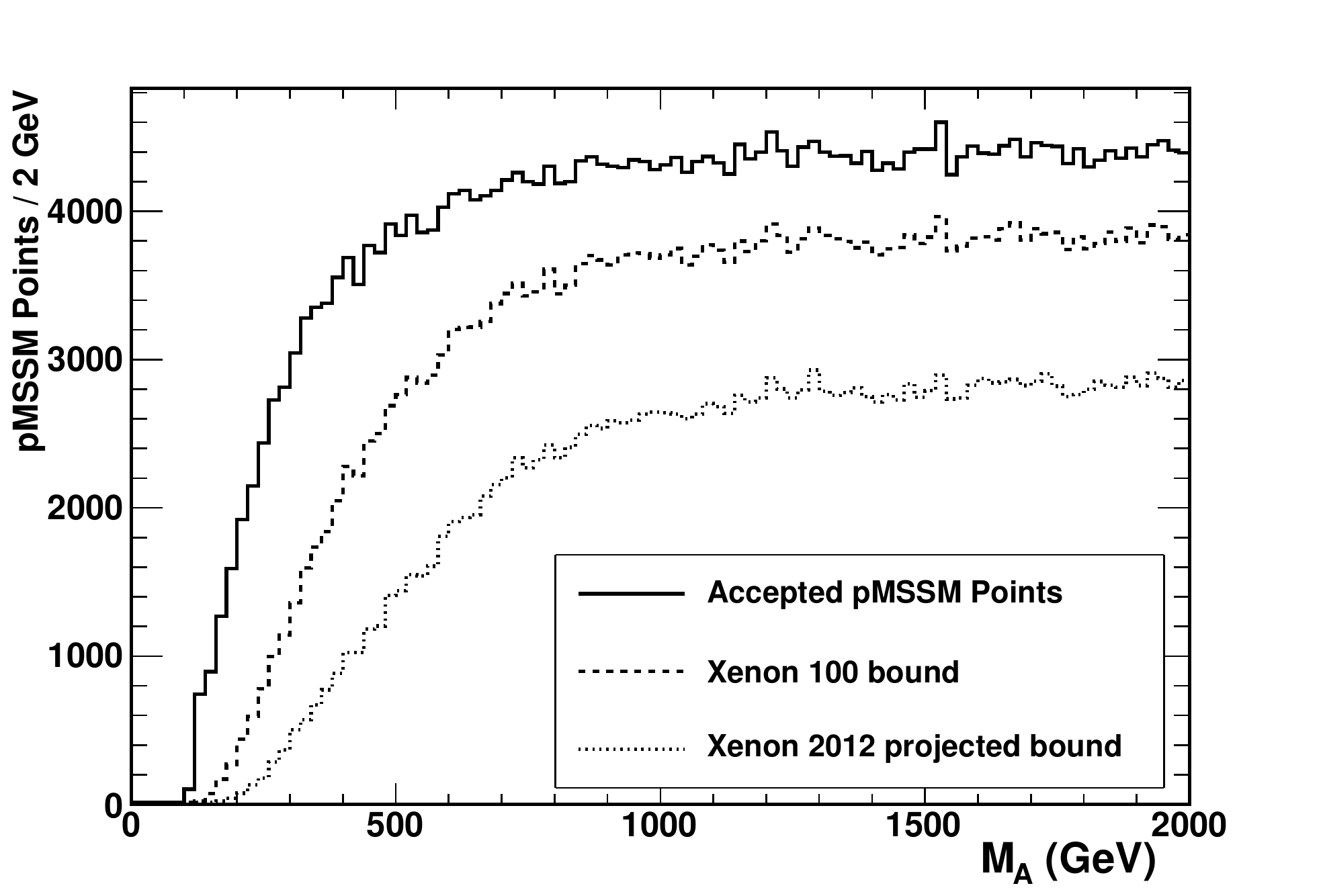}
\end{tabular}
\end{center}
\caption{Distribution of pMSSM points after the $A^0 \rightarrow \tau^+ \tau^-$ search (left), after the $B_s \rightarrow \mu^+ \mu^-$ constraint (middle) and after the dark matter direct detection constraint (right) projected on the $M_A$ for all accepted pMSSM points, and after applying projected limits.}
\label{fig:atau}
\end{figure}

We consider four sets of constraints on SUSY parameters \cite{Arbey:2011aa}.
The result of the direct search for the $A^0$ boson at the LHC is the single most constraining piece of information on the ($M_A , \tan \beta$) plane.
We compute the product of production cross section and decay branching fraction into $\tau$ pairs for the $A^0$ for each accepted pMSSM point. Fig.~\ref{fig:atau} shows the points surviving this selection in function of $M_A$. The 2012 data should severely constrain the low $M_A$ scenario by removing all solutions with $M_A <$ 220~GeV and restricting the region with $M_A <$ 400~GeV to $\tan \beta$ values below 10.

The second important constraint comes from the decay $B_s \to \mu^+ \mu^-$. At large $\tan\beta$, the SUSY contribution to this process is dominated by the exchange of neutral Higgs bosons, and very restrictive constraints can be obtained on the supersymmetric parameters~\cite{Akeroyd:2011kd}. Indeed, the couplings of the neutral Higgs bosons to $b$ quark and muons are proportional to $\tan\beta$, which can lead to enhancement of orders of magnitude as compared to the SM prediction which is helicity suppressed.
We compare our accepted pMSSM points to the combined limit from LHCb and CMS at 95\% C.L. of $\mathrm{BR}(B_s\to\mu^+\mu^-) < 1.1 \times 10^{-8}$~\cite{CMS_plus_LHCb} as well as to the projected constraint in the case of observation of the decay with a SM-like rate of BR($B_s\to\mu^+\mu^-$) = (3.4 $\pm$ 0.7)$\times$10$^{-9}$, to which we have attached a 20\% total relative uncertainty, by the end of the 7 TeV run. The results are presented in Fig.~\ref{fig:atau} in function of $M_A$. The current limit affects the low $M_A$ values up to 700 GeV, excluding large $\tan\beta$ values, below $M_A$=200~GeV. The projected constraint has a stronger impact, with more than half of the spectrum being excluded for $M_A\lesssim 700$ GeV.

A third important sector is that of dark matter direct detection experiments. In particular, the recent XENON~100 result~\cite{Aprile:2011hi}, places a 90\% C.L. upper bound on the spin-independent $\tilde\chi p$ cross section around 10$^{-8}$~pb for $M_{\mathrm{WIMP}}$ = 100~GeV and excludes $\simeq$20\% of the accepted pMSSM points in our scan. By the end of 2012, this bound should be improved by a factor of 7, if no signal is observed.
The $\tilde\chi p$ spin-independent scattering process has contributions from scalar quark exchange and t-channel Higgs exchange~\cite{Jungman:1995df}.
The latter dominates over vast region of the parameter space with the Higgs coupling to the proton depending on its coupling to gluons, through a heavy quark loop and to non-valence quarks. The scattering cross section retains a strong sensitivity on the CP-odd boson mass as highlighted in Fig.~\ref{fig:atau} which shows the pMSSM points retained after the XENON~100 and the projected 2012 sensitivity. The 2012 data should exclude virtually all solutions with $M_A <$ 200~GeV independent on the value of $\tan \beta$, if no signal is detected.

These constraints, originating from different sectors of the theory, are all sensitive to the SUSY parameters most relevant for setting the couplings and decay branching fractions of the light $h^0$ bosons. Their combination provides the boundary conditions for the parameter space where we test the possible suppression of the yields in the LHC Higgs searches.

\begin{figure}[t!]
\begin{center}
\includegraphics[width=0.5\textwidth]{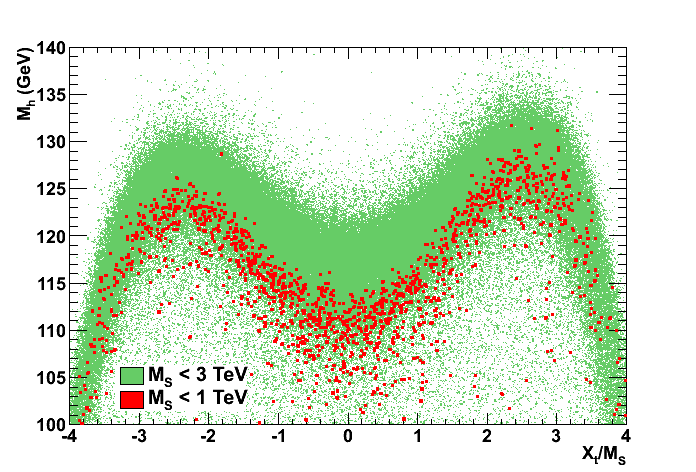}
\end{center}
\caption{The mass of the $h$ boson as a function of $X_t/M_S$ in the pMSSM when all other soft SUSY--breaking parameters and $\tan\beta$ are scanned.}
\label{Fig:pMSSM}
\end{figure}

Finally, the determination of the mass of the lightest Higgs boson with an accuracy of order of 1~GeV places some significant constraints on the SUSY parameters, in particular when its central value corresponds to a mass close to the edge of the range predicted in the MSSM.
In Fig.~\ref{Fig:pMSSM} the obtained $M_h$ in the pMSSM is displayed as a function of the ratio of parameters $X_t/M_S$.

\begin{figure}[h!]
\begin{center}
\includegraphics[width=0.4\textwidth]{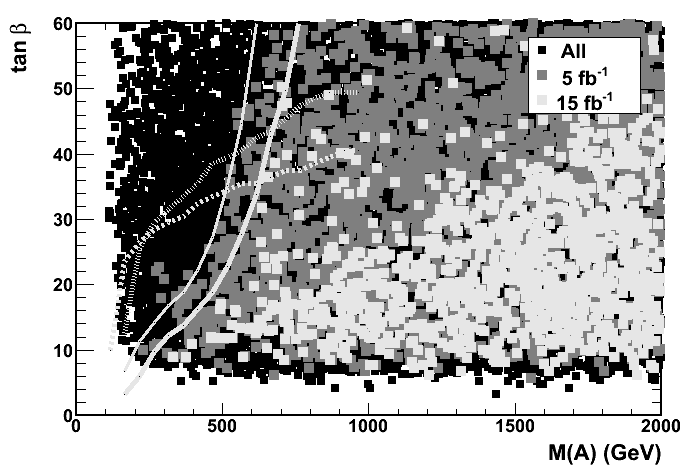}
\end{center}
\caption{pMSSM points in the ($M_A$, $\tan \beta$) parameter space, giving 123 $< M_h <$ 127~GeV. The different shades of grey show the points in the pMSSM without cuts and those allowed by the 2011 data and by the projected 2012 data, assuming no signal beyond the lightest Higgs boson is observed. The lines in the top plot show the regions which include 90\% of the scan points for the $A \rightarrow \tau^+ \tau^-$ and $B_s \rightarrow \mu^+ \mu^-$ decays at the LHC and the dark matter direct detection at the XENON experiment.}
\label{fig:matanb125}
\end{figure}

In order to evaluate the constraints from $M_h$ determination, we select the accepted pMSSM points from our scan, which have 123 $< M_h <$ 127~GeV.
Fig.~\ref{fig:matanb125} shows the points fulfilling these conditions, which are also allowed by the other 2011 data constraints and by the 2012 projection.
We observe that imposing the value of $M_{h}$ selects a wedge in the $(M_A$ - $\tan \beta)$ plane, at rather heavy $A^0$ masses and moderate to large values of $\tan \beta$ and extending beyond the projected sensitivity of the searches in the $A^0 \to \tau^+ \tau^-$ but also that of direct DM detection and would be compatible with a SM-like value for the rate of the $B_s^0 \to \mu^+ \mu^-$ decay.

\subsection{Bayesian analysis of the pMSSM}
{\it S. Kraml and S. Sekmen}

In \cite{Sekmen:2011cz}, we performed a global Bayesian analysis of
the pMSSM in light
of the results of SUSY searches published by the CMS collaboration
based on the first
$\sim$1~fb$^{-1}$ of data taken during the 2011 LHC run at 7~TeV.
Concretely, we used the results from three independent CMS analyses,
the $\alpha_T$~hadronic ~\cite{Collaboration:2011zy},
the same-sign dilepton~\cite{CMS:SS} and
the opposite-sign dilepton~\cite{CMS:OS} analyses, and analyzed whether and how
they impact the probability density distributions of masses and
parameters. To this end,
we compared for each mass and parameter, the ``CMS'' posterior
densities after inclusion
of the above mentioned results to the ``preLHC'' posterior density
based on the constraints from
$BR(b \rightarrow s\gamma)$,  $BR(B_s \rightarrow \mu \mu)$, $R(B_u
\rightarrow \tau \nu)$
and $\Delta a_\mu$, together with Higgs and SUSY mass limits from LEP
as detailed
in Table~1  of \cite{Sekmen:2011cz}.

\begin{figure}[t] \centering
\centering
\subfigure[]
{
   \includegraphics[width=0.3\textwidth]{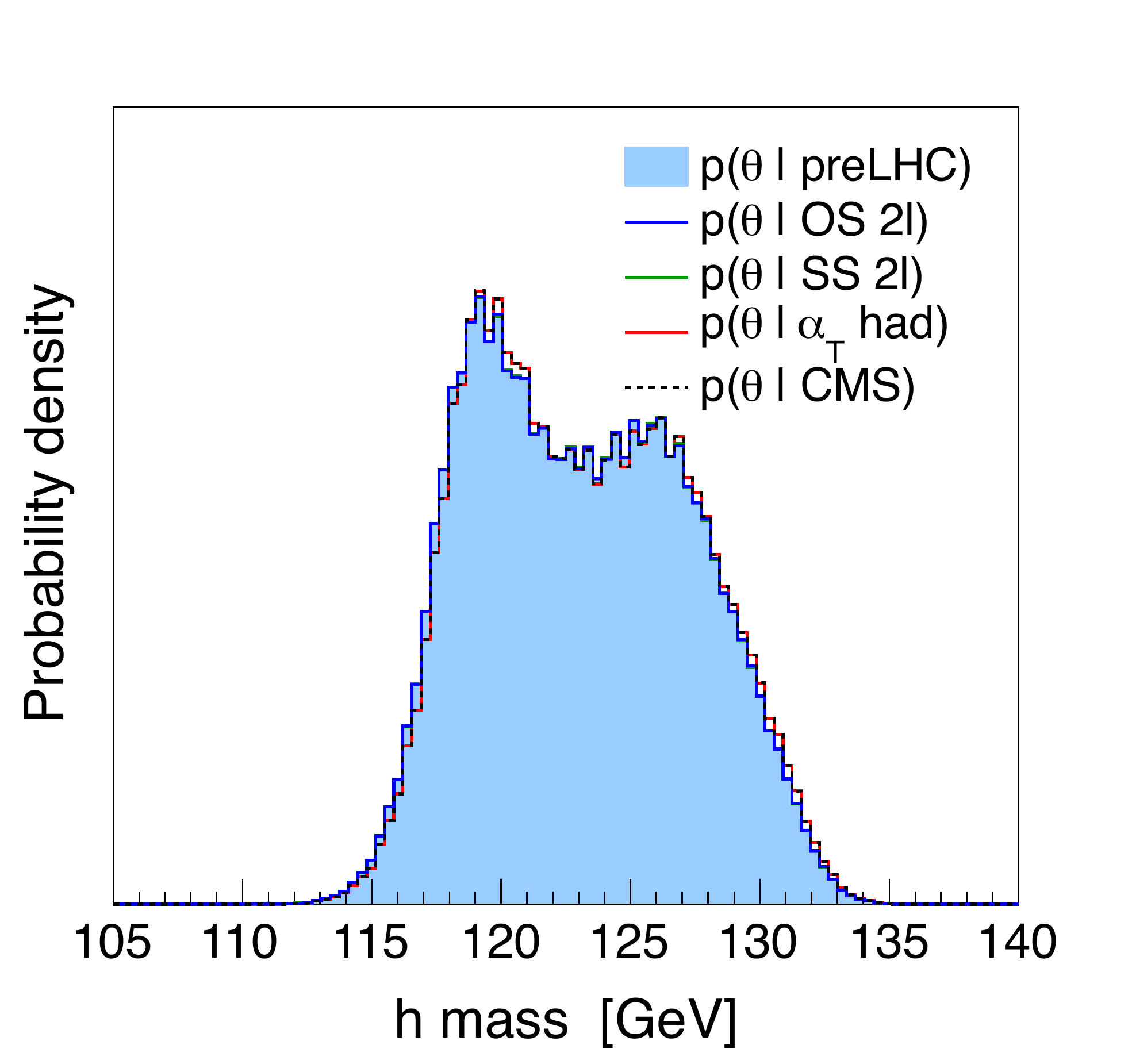}
}
\subfigure[]
{
   \includegraphics[width=0.3\textwidth]{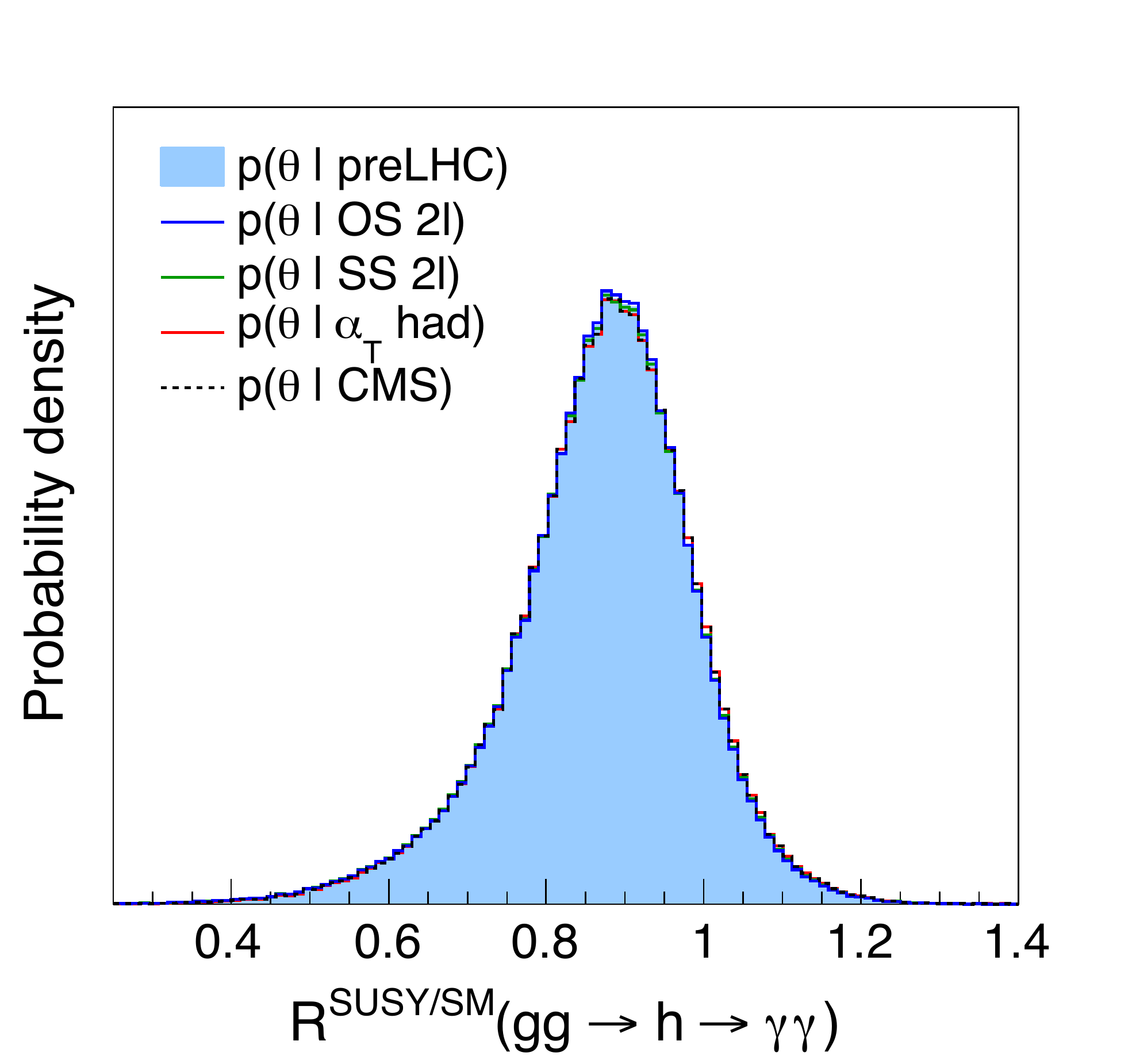}
}
\subfigure[]
{
   \includegraphics[width=0.3\textwidth]{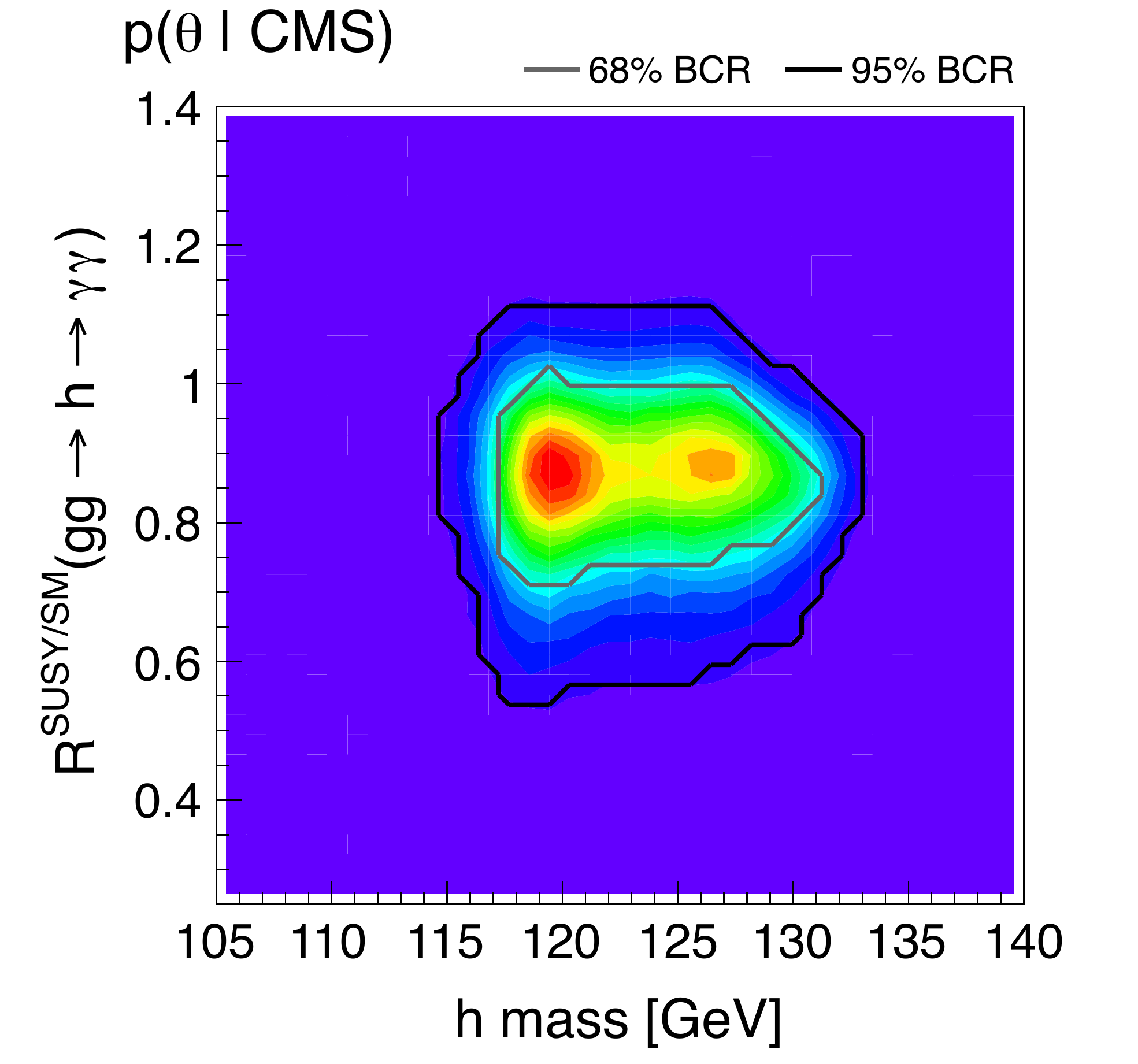}
}
\subfigure[]
{
   \includegraphics[width=0.3\textwidth]{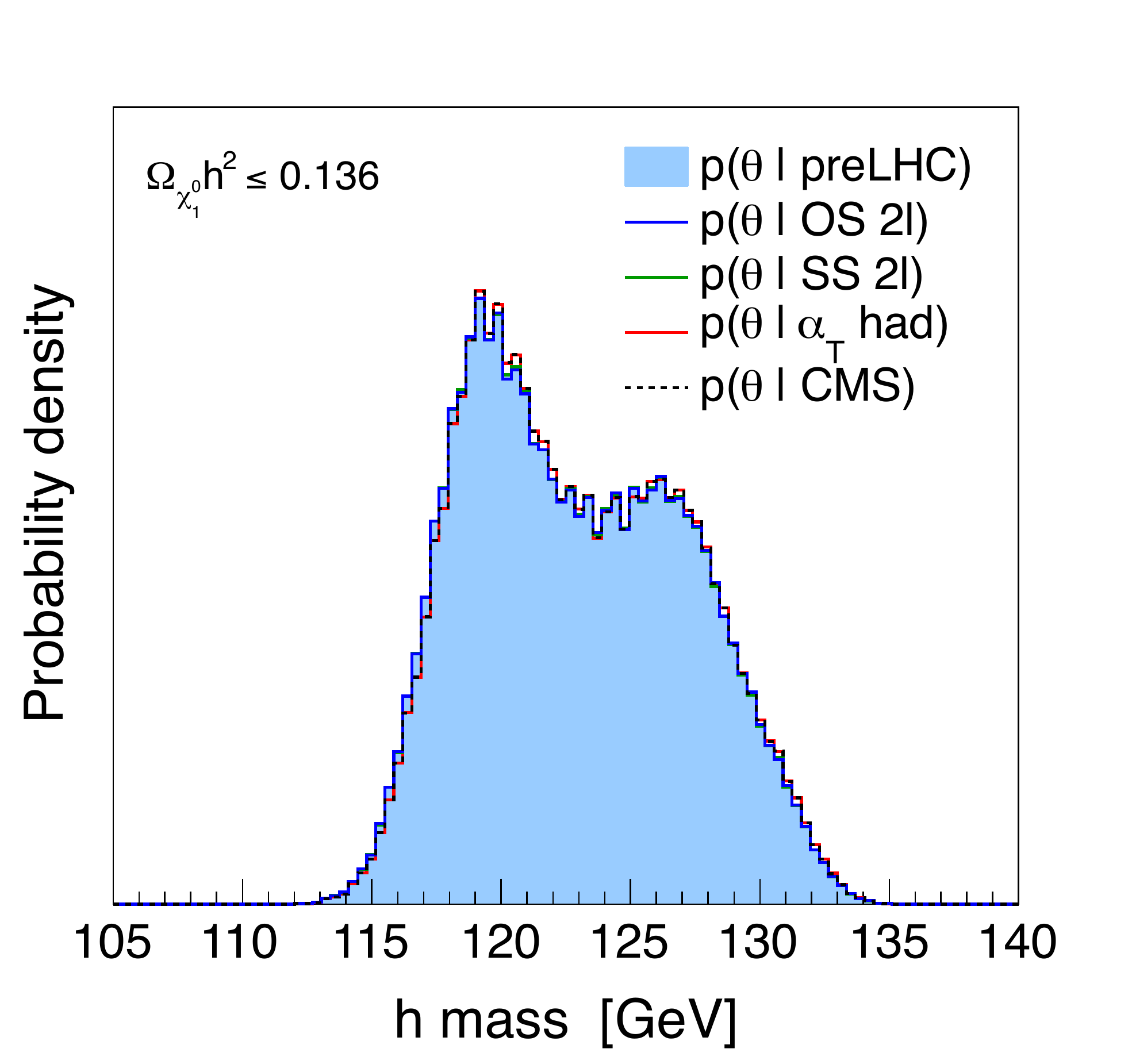}
}
\subfigure[]
{
   \includegraphics[width=0.3\textwidth]{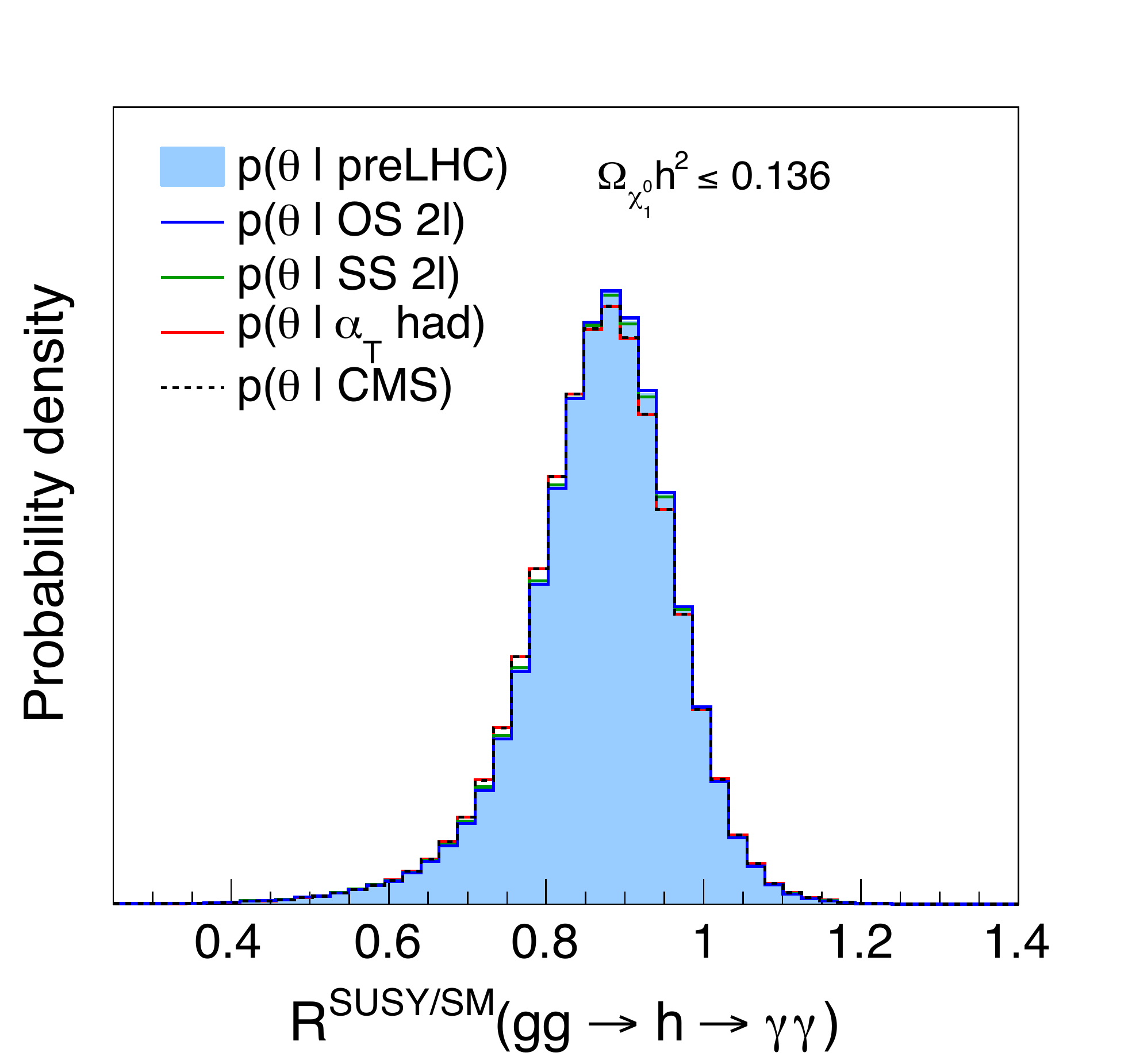}
}
\subfigure[]
{
   \includegraphics[width=0.3\textwidth]{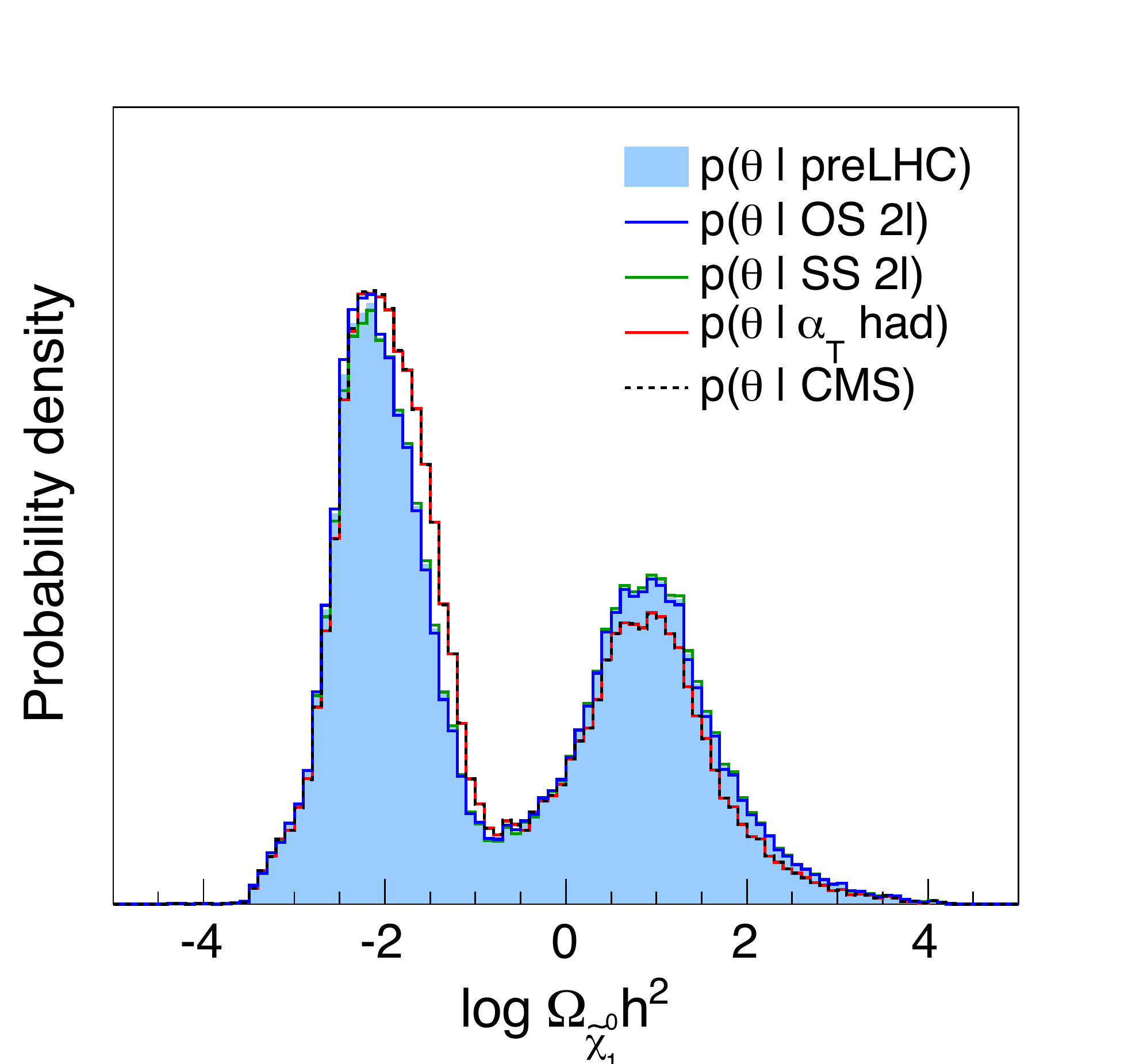}
}
\caption{Marginalized 1D posterior densities for $m_h$ (left) and
$R_{gg\gamma\gamma}$ (middle),
and the 2D posterior in the $R_{gg\gamma\gamma}$ versus $m_h$ plane
(top right).
Also shown is the 1D posterior density of $\Omega h^2$ (bottom right).
In the top row, no constraint is imposed on $\Omega h^2$, while in the
bottom left and bottom-middle plots $\Omega h^2\le 0.136$.
\label{fig:pMSSM-SabineSezen}}
\end{figure}

Our analysis is easily applied to the Higgs sector.  In the 1D distributions of
Fig.~\ref{fig:pMSSM-SabineSezen},
the light blue histograms represent the preLHC probability densities;
the blue, green and red lines show, respectively, the effects of the
OS di-lepton, SS di-lepton and
$\alpha_T$ hadronic CMS analyses; and the dashed black lines show the final
posterior densities after inclusion of the results of all three analyses.
We see that the probability density of $m_h$ peaks at about 120~GeV
and is basically
unaffected by the SUSY search results. When imposing no constraint
on $\Omega h^2$,
the $m_h=123-127$~GeV window has 27.4\% probability.
This decreases only marginally to 26.4\% when requiring $\Omega h^2\le 0.136$.

The $R_{gg\gamma\gamma}$ ratio peaks around 0.9, and does not extend much above 1.
Requiring $\Omega h^2\le 0.136$ slightly narrows the distribution but
does not change it qualitatively.
The probability of finding $R_{gg\gamma\gamma}>1$ is 8.88\% in the general case, and 6.32\%
when requiring $\Omega h^2\le 0.136$.
Here note that we actually approximated
$\sigma(gg\rightarrow h)\times BR(h\rightarrow \gamma\gamma) \approx
  \Gamma_{\rm tot}\times BR(h\rightarrow gg)\times BR(h\rightarrow
\gamma\gamma)$,
computing both the MSSM and the SM widths and branching rations with
{\tt HDECAY\,3.4}~\cite{Djouadi:1997yw}, which is the version
implemented in SUSY-HIT~\cite{Djouadi:2006bz}.
While radiative corrections are somewhat different for $gg\rightarrow h$ and
$h\rightarrow gg$, one can reasonably assume that these differences
largely cancel out when taking the MSSM/SM ratio.

\begin{figure}[t] \centering
\begin{tabular}{lll}
   \includegraphics[width=4.5cm]{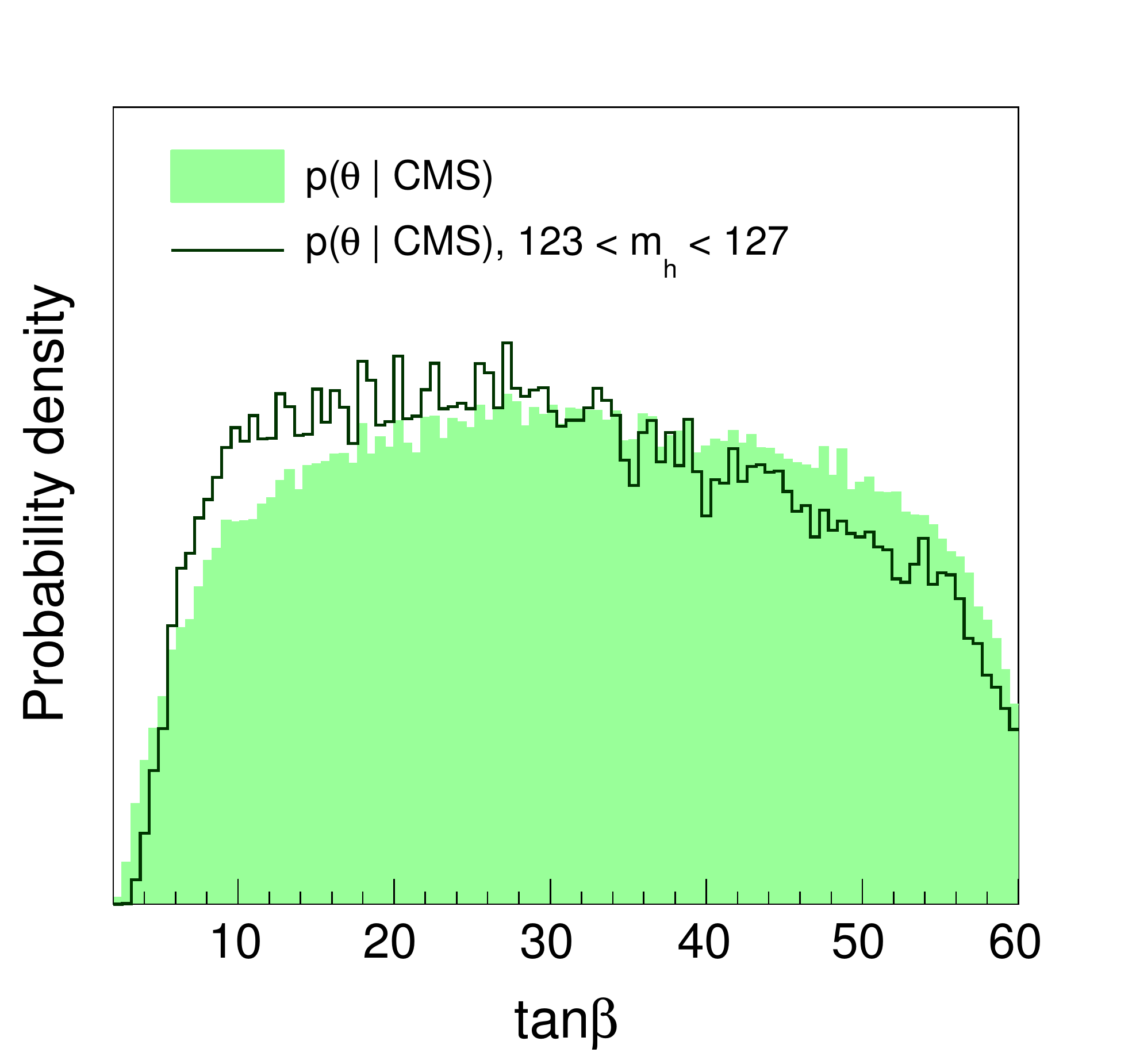} &
   \includegraphics[width=4.5cm]{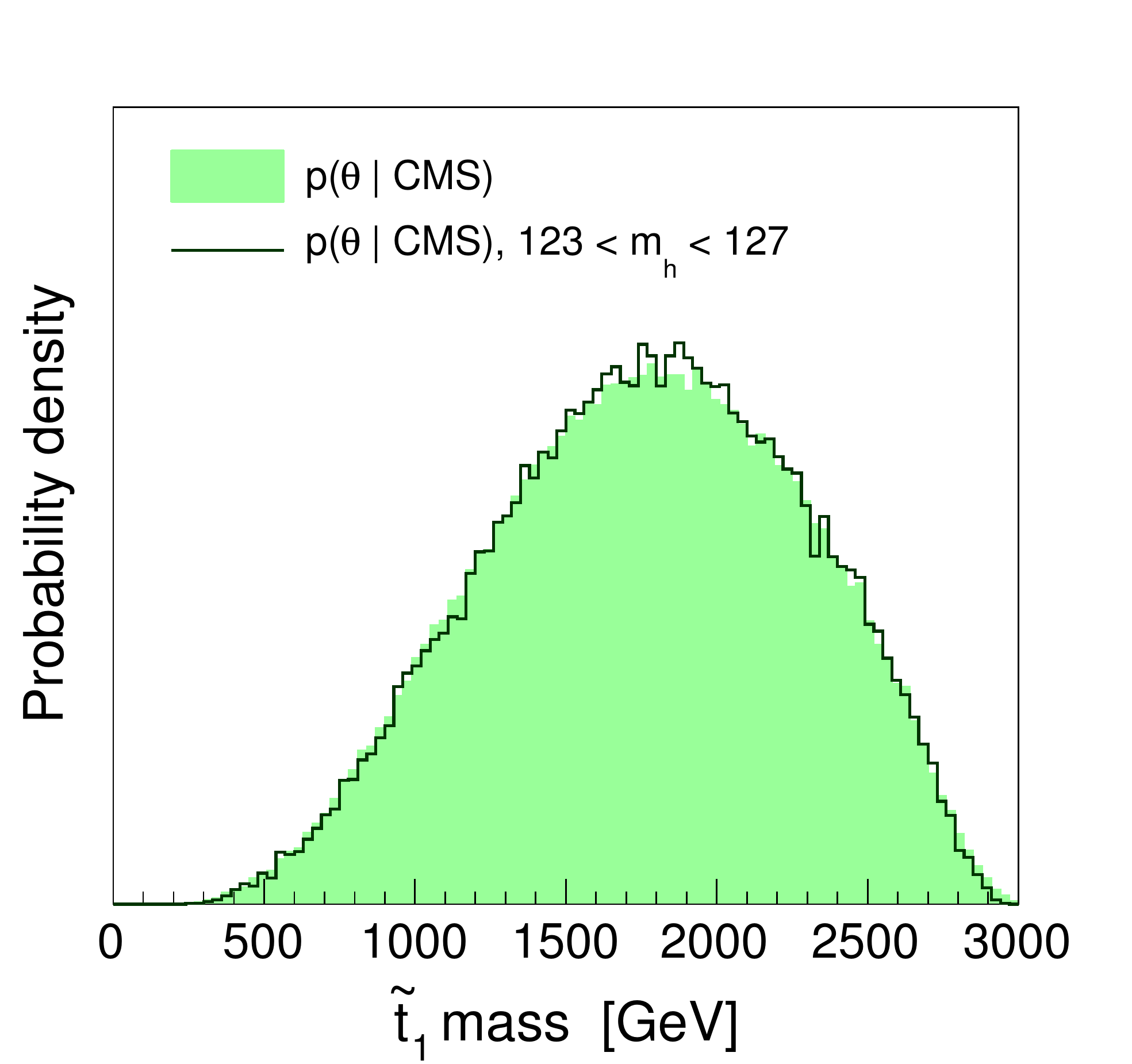}  \\
   \includegraphics[width=4.5cm]{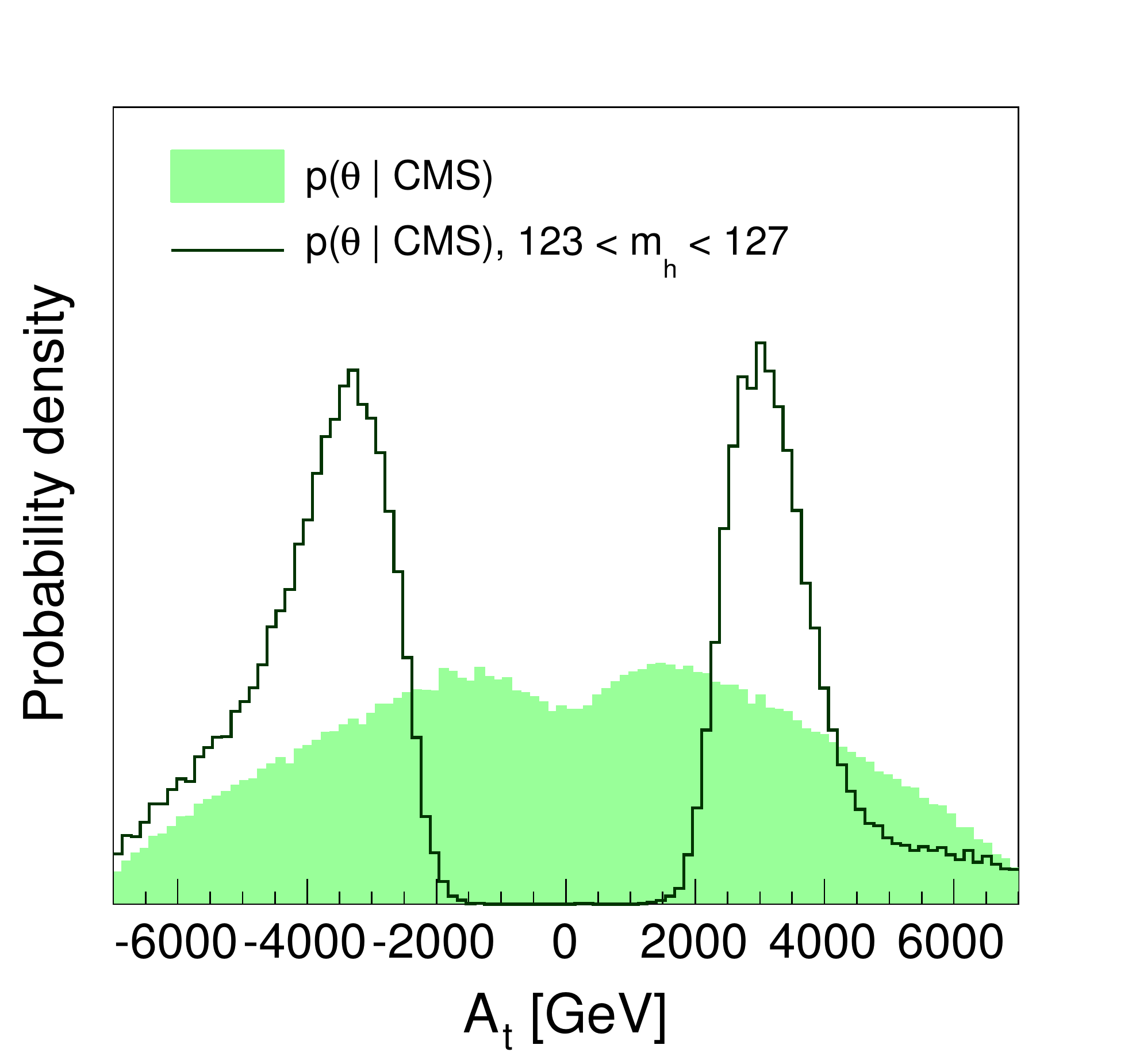}  &
   \includegraphics[width=4.5cm]{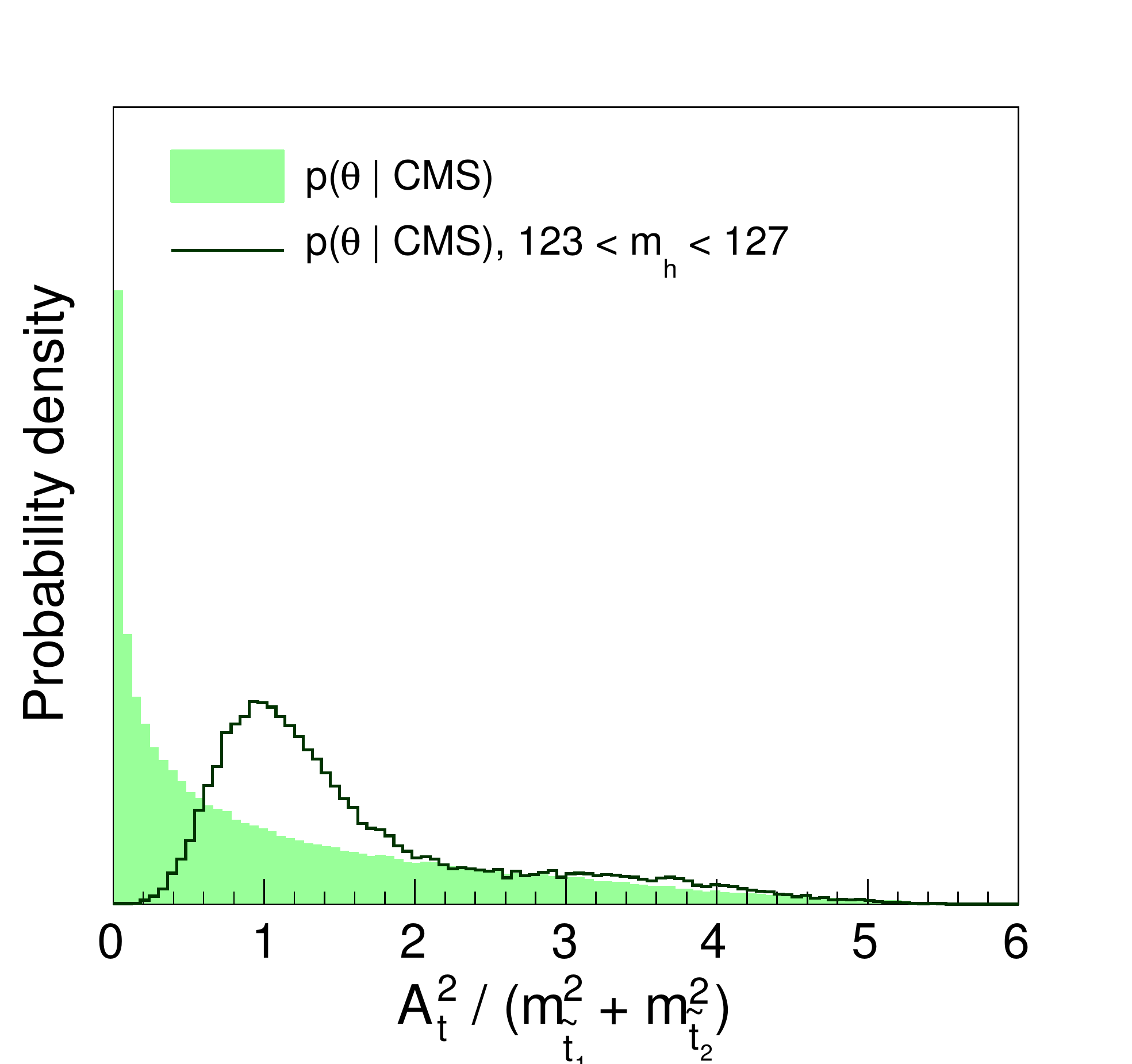} \\
\end{tabular}
\caption{Marginalized 1D posterior densities for $\tan\beta$, $m_{\tilde t_1}$, $A_t$ and 
$A_t^2/(m_{\tilde t_1}^2+m_{\tilde t_2}^2)$ 
after OS, SS and $\alpha_T$ hadronic CMS analyses. 
The effect of imposing $m_h=123-127$~GeV is apparent from comparing the dark green lines to 
the light green filled histograms, which show the probability densities after CMS analyses without 
additional requirements on $m_h$,.
\label{fig:pMSSM-SabineSezen-mh125}}
\end{figure}

It is also interesting to ask what is the effect of requiring a Higgs mass near 125~GeV. 
This is illustrated in Fig.~\ref{fig:pMSSM-SabineSezen-mh125}, where show 
posterior density distributions for some quantities that are relevant for the Higgs sector, 
namely $\tan\beta$, $m_{\tilde t_1}$, $A_t$ and $r_t=A_t^2/(m_{\tilde t_1}^2+m_{\tilde t_2}^2)$. 
The light green filled histograms and the dark green lines compare the probability 
densities after CMS analyses, $p(\theta|{\rm CMS})$, without and with requiring the 
Higgs mass to fall within the $m_h=123-127$~GeV window.
As can be seen, a Higgs mass near 125~GeV somewhat 
favors intermediate values of $\tan\beta$, and strongly favors large $A_t$ with $r_t\approx 1$.  
However, there is basically no effect on the $m_{\tilde t_1}$ distribution. 
We also checked the $m_A$ distribution and found it does not change when requiring
$m_h=123-127$~GeV. 

\subsection{MSSM and light neutralinos}
{\it D. Albornoz Vasquez, G. B\'elanger, R. Godbole, A. Pukhov}

We consider a  constrained version of the pMSSM with only eleven free parameters defined at the electroweak scale, as in~\cite{Vasquez:2011yq,AlbornozVasquez:2011aa}:
\begin{equation}
M_1, \; M_2, \; M_3, \; \mu, \; \tan\beta , \; M_A, \nonumber\\
M_{\tilde{l}_L}, \; M_{\tilde{l}_R}, \; M_{\tilde{q}_{1,\;2}}, \; M_{\tilde{q}_3}, \; A_t. \nonumber
\end{equation}
To explore the parameter space we use 
a Markov Chain Monte-Carlo (MCMC) first presented in~\cite{Vasquez:2010ru} that relies on  micrOMEGAs2.4~\cite{Belanger:2006is,Belanger:2008sj,Belanger:2010gh} for the computation of all observables and on  SuSpect for the calculation of the supersymmetric spectrum~\cite{Djouadi:2002ze}. 
We include limits on B physics observables, on the anomalous magnetic moment of the muon $(g-2)_{\mu}$ and on the sparticles masses obtained from LEP. Limits on the Higgs mass are obtained from  HiggsBounds(3.1.3)
which includes LEP and Tevatron results as well as first LHC results~\cite{Bechtle:2008jh,Bechtle:2011sb}. More recent results from CMS presented in ~\cite{Chatrchyan:2011nx,CMS_higgs_fit} are added a posteriori. We also require that the dark matter relic density be within $10-100\%$ of the WMAP measured value thus allowing for the LSP to be  only a fraction of the dark matter component. Furthermore astroparticle constraints from dark matter direct detection (XENON100) and the photon flux  from dwarf spheroidal galaxies (Fermi-LAT) are added a posteriori.
In this analysis we explore carefully the parameter space which allows for light neutralinos, thus we expect to have some large suppressions of the Higgs signal.

\begin{figure}[!ht] 
\includegraphics[width=8cm,height=6.cm]{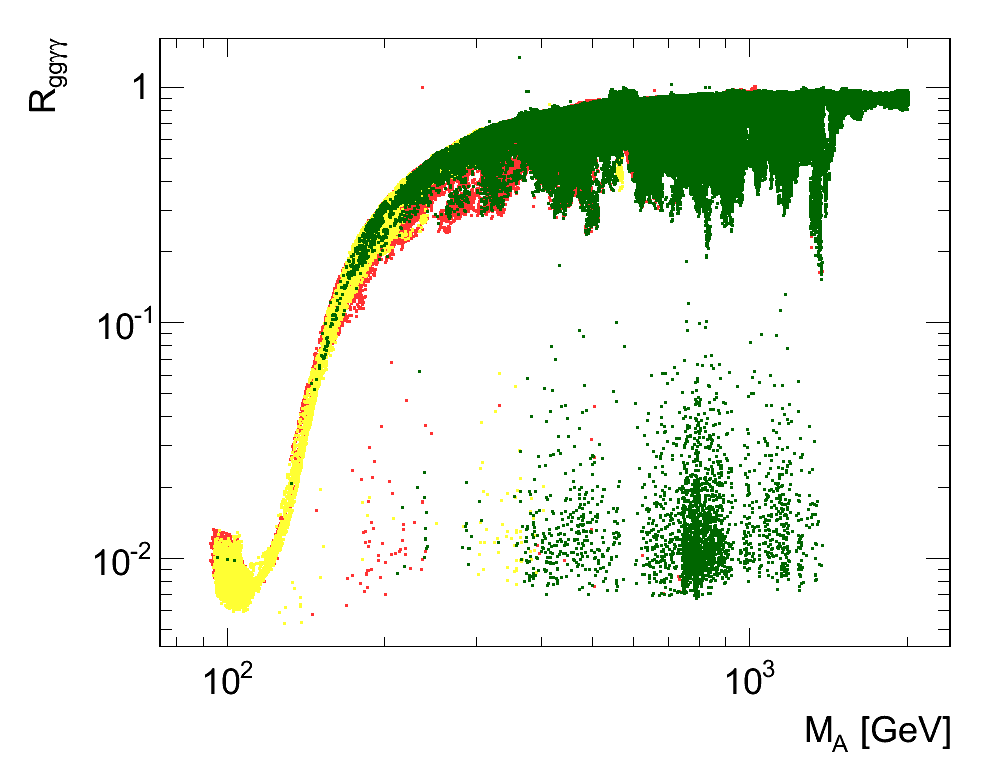} 
\includegraphics[width=8cm,height=6.cm]{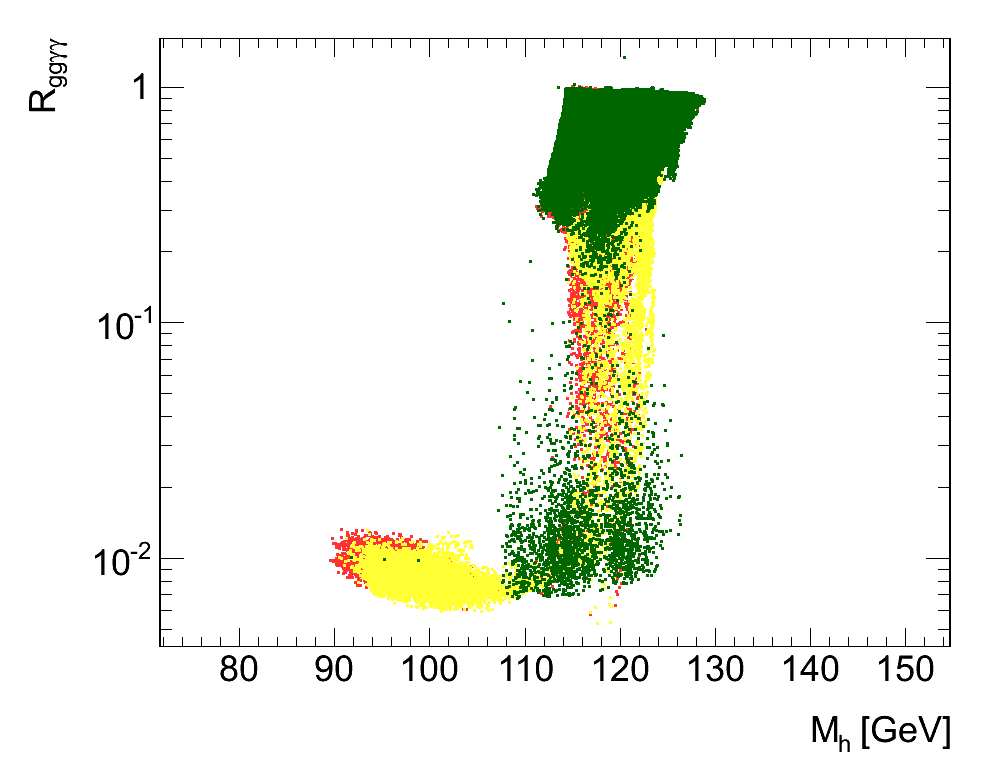} 
 \caption{ $R_{gg\gamma\gamma}$ as a function of the pseudoscalar mass (left) and light  scalar mass (right) for allowed points (green).
 In yellow, points excluded by collider constraints (Higgs and $BR(B_s\rightarrow\mu^+\mu^-)$), in red those excluded by astrophysical constraints (XENON100 and Fermi-LAT). 
  }
 \label{fig:gggg}
\end{figure}

As for the SM Higgs, searches for the light Higgs in the MSSM rely mainly on $gg\rightarrow h\rightarrow \gamma\gamma$. 
By computing the signal strength as compared to the SM we find  that the MSSM Higgs signal can be at most as large as that of the SM Higgs
and is suppressed when $M_A<400{\rm GeV}$ due to shifts in the Higgs couplings as mentioned above. The predictions for $R_{gg\gamma\gamma}$ 
 as a function of the pseudoscalar mass  are displayed in Fig.~\ref{fig:gggg}, left panel. Note that many of the points with a light pseudoscalar are  constrained by either Fermi-LAT, XENON100 (points in red),  latest Higgs searches  at the LHC and/or the new upper limit on $BR(B_s\rightarrow\mu^+\mu^-)$ (points in yellow), nevertheless  there are still allowed points where the Higgs signal is less than 10\% of its SM value. Since  these are associated with a light $M_A$, they can be probed further in searches for the neutral and charged component of the heavy Higgs doublet in CMS and ATLAS and/or a more precise measurement of $BR(B_s\rightarrow\mu^+\mu^-)$. 
When $M_A>400$~GeV and one is in the decoupling limit the suppression in $R_{gg\gamma\gamma}$ is usually more modest. 
Nevertheless it is still possible to have $R_{gg\gamma\gamma}$ as low as 0.2 even when $M_A$ is at the TeV scale. 
This occurs when the LSP is lighter than $M_h/2$ and the invisible width of the Higgs is large.
Note that there are a few points at large values of $M_A$ where $R_{gg\gamma\gamma}\approx 0.01$, these all correspond to points where the LH soft mass for sleptons is below 100 GeV, the charged sleptons are just above the LEP exclusion and the sneutrino is light enough that the dominant decay of the Higgs is into sneutrinos.
Other suppression of $R_{gg\gamma\gamma}$ arises from the stop sector. The effect is usually below 30\%, although for stop masses below 500 GeV we found points where $R_{gg\gamma\gamma}$ can drop as low as 0.4. 

The predictions for $R_{gg\gamma\gamma}$ as a function of the light Higgs mass show that $R_{gg\gamma\gamma}$ has little dependence on the scalar mass provided it lies in the $110-130$~GeV range. However it is important to note that for a Higgs of 126 GeV we find $R_{gg\gamma\gamma}>0.4$, which means that the signal strength would be compatible with the excess of events reported by ATLAS. Also note that for a Higgs in the narrow mass range $110-114$~GeV, we always find $R_{gg\gamma\gamma}<0.8$, because these values of $M_h$ are found only in the non-decoupling limit. 

\begin{figure}[!ht]
\begin{center}
\includegraphics[width=8cm,height=5.5cm]{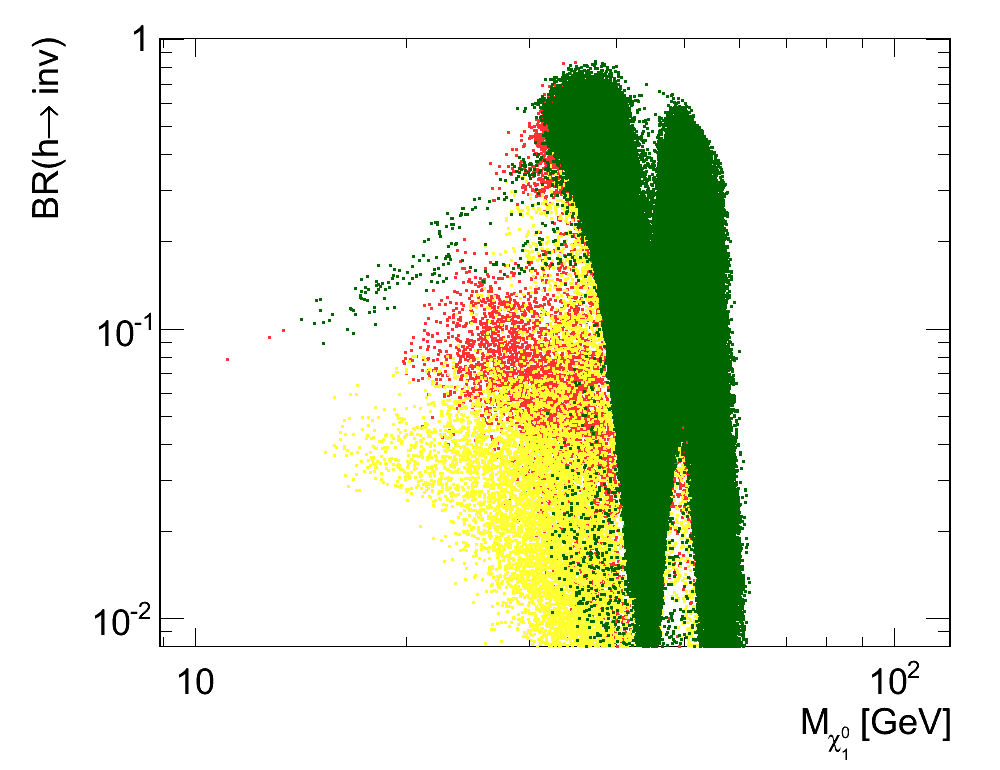} 
\end{center}
\vspace{-0.5cm}
 \caption{ ${\rm BR}(h\rightarrow {\rm inv})$ as a function of the LSP mass, same color code as Fig.~\ref{fig:gggg}.}
 \label{fig:inv_new}
\end{figure}

The Higgs signal can  be more strongly suppressed than found in other studies~\cite{Arbey:2011un,Sekmen:2011cz}. This is due to the large fraction of light neutralinos in our sample allowing for a large branching fraction of the Higgs into invisible particles, see Fig.~\ref{fig:inv_new}.  A large invisible partial width involves special conditions on the parameter space and can therefore be probed by SUSY searches at the LHC. Indeed   invisible decays of the Higgs   requires not only a light LSP but also a significant coupling of the LSP to the Higgs. This means that the light neutralino which is predominantly bino must have also a higgsino component.
Thus $\mu$ cannot be large. This also means that the mass of the lightest chargino, which is driven by $\mu$ is strongly correlated with the Higgs invisible width.
 We find that when the invisible width of the Higgs is larger than 20\%, then the chargino is lighter than 200GeV~\cite{AlbornozVasquez:2011aa}.

\section{Beyond the MSSM}

\subsection{BMSSM} 
{\it F. Boudjema, G. Drieu La Rochelle}

The BMSSM model is the extension of the MSSM obtained by adding to the Higgs sector of the K\"ahler potential and superpotential new higher dimensional operators that are suppressed by a heavy scale $M$ :
\begin{align} K_{\textrm{eff}}&=
K^{(0)}+\frac{1}{M}K^{(1)}+\frac{1}{M^2}K^{(2)}+\grandO{\frac{1}{M^3}}\label{effK}\\
W_{\textrm{eff}}&=
W^{(0)}+\frac{1}{M}W^{(1)}+\frac{1}{M^2}W^{(2)}+\grandO{\frac{1}{M^3}}
\end{align}
Those new operators only involve the MSSM Higgs superfields, so the particle content of the BMSSM is the same as the one of the MSSM. The resulting theory can then be described in the effective field theory approach. This model has been studied in different contexts including Higgs phenomenology (see \cite{Brignole:2003cm,Dine:2007xi,Antoniadis:2007xc,Antoniadis:2008es,Antoniadis:2009rn,Carena:2009gx,Carena:2010cs,Carena:2011dm,Batra:2008rc,Boudjema:2011aa}), vacuum stability~\cite{Blum:2009na}, fine tuning~\cite{Cassel:2009ps,Casas:2003jx}, dark matter~\cite{Cheung:2009qk,Berg:2009mq,Bernal:2009jc,Cassel:2011zd}, cosmology~\cite{Bernal:2009hd}, baryogenesis~\cite{Blum:2008ym,Blum:2010by} and flavour physics~\cite{Bernal:2011pj,Altmannshofer:2011rm}. Regarding the Higgs, its first motivation was to increase the lightest Higgs mass to a large value ($m_h\in[150-250]$ GeV). Here we will not dwell upon the model  itself, but show some of the predictions in the channels searched for at the LHC.

To obtain these predictions we have used the MSSM scenario $m_{h\ max}$ (\cite{Carena:2002qg}), with only two free parameters, $M_{A_0}$ and $\tan\beta$. In the following we have scanned these parameters within the range
\begin{equation}
 M_{A_0}\in[50,450]\ \text{GeV},\qquad\tan\beta\in[2,40].
\end{equation}
We recall that in the $m_{h\ max}$ scenario, the superpartner masses are set to 1 TeV, the gaugino masses are set to $M_2=300$ GeV, $M_1=\frac{5}{3}\tan^2\theta_W\,M_2$ and $M_3=800$ GeV. All trilinear couplings are set to 0 except for $A_b=A_t=2\frac{\mu}{\tan\beta}+2 \text{ TeV}$. The additional parameters brought by the higher-dimensional operators are the new physics scale $M$ (and the supersymmetry breaking mass $m_s$), and the effective coefficients that we name $\zeta_{1i}$ ($i=0,1$) and $a_{ij}$ ($i=1..6\ ,j=0..2$). The scales have been fixed to
\begin{equation}
 M=1.5\ \text{TeV}\qquad\qquad m_s=\mu=300\ \text{GeV},
\end{equation}
and the effective coefficients have been left free. A scan has been performed within
\begin{equation}
 \zeta_{1i},a_{ij}\in[-1,1].
\end{equation}

The phenomenology of this model shows many features different from those of the MSSM, indeed the Higgs scalar sector is dramatically affected so most of the relations between the couplings of the Higgses to the SM particles that held in the MSSM case are no longer true. A detailed analysis can be found in ref. \cite{Carena:2009gx,Antoniadis:2008es} but we will now focus on the application of the LHC constraints that were obtained in \cite{Boudjema:2011aa}.

In all the plots below, the following data sets, corresponding to the December 2011 release, are used to constrain the model.
\begin{itemize}
 \item $Vh\rightarrow V\o{b}b$ : CMS $4.7$ fb$^{-1}$.
 \item $h\rightarrow\o{\tau}\tau$ : CMS $4.6$ fb$^{-1}$ and ATLAS $1.06$ fb$^{-1}$.
 \item $h\rightarrow\gamma\gamma$ : CMS $4.8$ fb$^{-1}$ and ATLAS $4.9$ fb$^{-1}$.
 \item $h\rightarrow ZZ\rightarrow 4l$ : CMS $4.71$ fb$^{-1}$ and ATLAS $4.8$ fb$^{-1}$.
 \item $h\rightarrow WW\rightarrow l\nu l\nu$ : CMS $4.6$ fb$^{-1}$ and ATLAS $2.05$ fb$^{-1}$. 
\end{itemize}
We have used the 95\% CL value for $\sigma_\text{exclusion}$ as given by the collaborations. We have combined all channels by a quadrature sum and we show only points that pass the combined exclusion. Note that the exclusion is done separately on each of the three neutral Higgses (unless they are degenerate in mass, in which case the signals are added) and the test is successful only if each individual test is successful. The search for the charged Higgs $H^+\rightarrow\tau^+\nu_{\tau}$ has also been included.

\begin{figure}[!ht]
\begin{center}
\begin{tabular}{cc}
\includegraphics[scale=0.3,trim=0 0 0 0,clip=true]{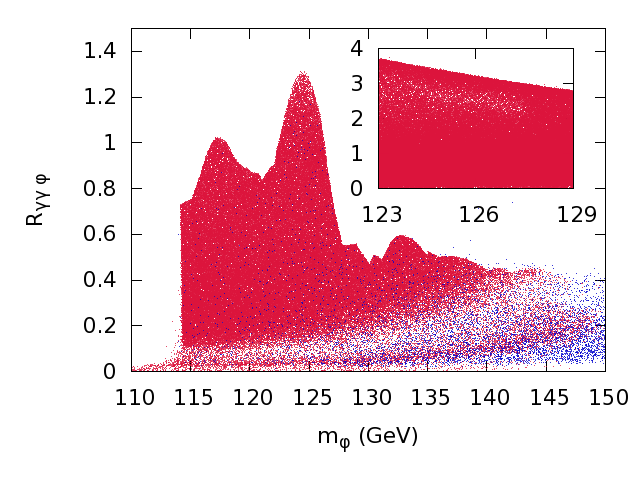} &
\includegraphics[scale=0.3,trim=0 0 0 0,clip=true]{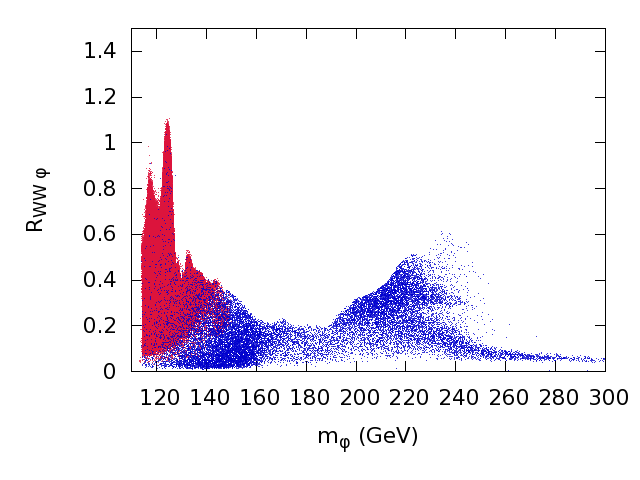} 
\end{tabular}
\end{center}
  \caption{ $R_{\gamma\gamma\ \phi}$ (left) and  $R_{WW\ \phi}$ (right) as a function of the lightest Higgs mass. Red points correspond to signal coming from the lightest Higgs, and blue from the heaviest. We only show points that pass the combination of all LHC searches. On the right corner of the left plot $R_{\gamma\gamma\ h}$ is shown in the case that the 95\% exclusion constraints are switched off.}
\label{fig:gam}
\end{figure}

As compared to the MSSM, the light Higgs of the BMSSM can be much heavier, that is to say $m_h\in[150,250]$ GeV, but only at the price of being SM-like, which means that this case is now ruled out by the LHC. We will then focus on the case of a light Higgs, say $m_h<150$ GeV. The relevant channels are thus $\gamma\gamma$, $ZZ$ and $WW$, but there is also a constraint coming from other Higgses (that may be much heavier) in the $\tau\tau$ channel. Indeed this last one is $\tan\beta$ enhanced, so in some cases it is more constraining than direct searches on the lightest Higgs. We show in figure \ref{fig:gam} the ratio $R_{\gamma\gamma\ \phi}=\frac{\sigma_{pp\rightarrow \phi\rightarrow\gamma\gamma}}{\sigma_{pp\rightarrow h\rightarrow\gamma\gamma\ SM}}$ which tells how much we can lower the prediction of the Standard Model. This ratio is computed by evaluating first which Higgs ($h$ or $H$) has the highest signal given the current sensitivity (that is to say the highest value for $\frac{\sigma}{\sigma_{\text{exclusion}}}$), and then take its mass and its signal strength normalised to the Standard Model. This means that some of the points correspond to the lightest Higgs and others to the heaviest one.

One notices that, as expected, the allowed region is bounded from above by the combination of LHC searches, mainly the $h\rightarrow\gamma\gamma$ with $\sim 5$fb$^{-1}$ from ATLAS and CMS, but also the $WW$ and $ZZ$ channels (since we combine all channels in quadrature). Actually there is only a small range $120-130$~GeV, corresponding to the excesses seen in the $\gamma\gamma$ channels, that allows for points with $R_{\gamma\gamma\ h}>1$. Note that since we are in the no-signal hypothesis we cannot use this graph to discuss on the interpretation of the excess, to this aim, one would have to switch off the constraints from the 95\% exclusion limit on this mass range. This is done in the corner of the plot in fig \ref{fig:gam}, which shows $R_{\gamma\gamma\ \phi}$ when the constraints from $\gamma\gamma$, $ZZ$ and $WW$ channels are not taken into account.

The conclusion from this figure is two-fold: first the signal strength can be much lowered relatively to the SM, which yields a possibly very elusive model, and secondly there are points with signal strength higher than the SM. On the $WW$ side, the signal strength can also be much reduced. Regarding the $R_{\gamma\gamma}$ plot, one notices that the higher part of our range, that is $m_h\in[130-150]$ GeV, is more sensitive to the searches (indeed the upper bound is much lower in this part), so that the model is more constrained on this mass range. This comes from the $WW$ and $ZZ$ channels, that lose sensitivity when the mass decreases.

\subsubsection{Invisible Higgs decays}
\begin{figure}[!ht] 
\begin{center}
\includegraphics[width=8cm,height=6.5cm]{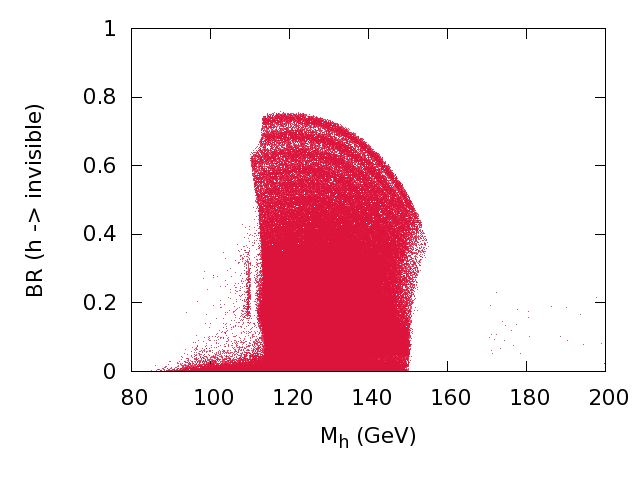} 
 \end{center}
  \caption{ Branching ratio of the light Higgs to invisible in the case where the lightest neutralino is as light as 50 GeV. Only points that pass the combination of all LHC searches are shown.}
 \label{fig:bmssm}
\end{figure}

Our choice of the MSSM benchmark (the $m_{h\ max}$ scenario) has the advantage of leaving the issue of superpartners aside, since the SUSY scale is at 1 TeV, but it also precludes the possibility of rather light neutralinos. If one allows for the lightest neutralino to be as light as 50 GeV, it will modify directly the Higgs signals since a new decay channel opens in this case. In order to have a non-negligible branching ratio, one must ensure that the lightest neutralino has both a bino and a higgsino component: in our case putting $\mu=200$ GeV and $M_1=50$ GeV was enough to see some changes. We have plotted in Fig.~\ref{fig:bmssm}  the branching ratio of the lightest Higgs. While one could hope that it would enhance the robustness of the model towards LHC searches with masses higher than $150$ GeV, it turns out that this branching ratio is quickly overturned by the decays to vector bosons, so that for $m_h>150$ GeV, the branching is lower than 10\% and thus cannot avoid that all points with $m_h>150$ GeV be excluded by the LHC. On the light range however, the branching can go as high as 70\%, which will directly reduce the sensitivity in all channels by the same amount.

We have seen that the BMSSM accounts for a light Higgs that could be quite difficult to find at the LHC: indeed the signal of the lightest Higgs can be reduced significantly with respect to the SM or the MSSM. Though this study has been done in the no-signal hypothesis, the reader would have noticed that the predicted signal can be higher than the SM value, which would be of particular importance in interpreting a possible signal from the latest excesses around $m_h\sim 125$ GeV reported by the ATLAS and CMS collaborations.

\subsection{NMSSM} 
{\it D. Albornoz V\'asquez, G. B\'elanger, C. Boehm, J. Da Silva, C. Wymant}

The NMSSM is a simple extension of the MSSM that
contains an additional gauge singlet superfield. The  VEV of this singlet induces an effective $\mu$ term that is 
 naturally of the order of the electroweak scale, thus providing a solution to the naturalness problem~\cite{Ellwanger:2009dp}.
The model contains  three CP-even  ($H_1,H_2,H_3$) and two CP-odd ($A_1,A_2$) Higgs
bosons, the former result from a  mixing of the singlet with the two neutral CP-even doublets of the MSSM. 
The coupling $\lambda S H_u H_d$ in the superpotential leads to a positive contribution to the mass of the SM-like Higgs boson for 
small values of $\tan\beta$. Furthermore singlet/doublet mixing  can lead to an additional increase in the mass of the SM-like Higgs while the lighter eigenstate, $H_1$, can evade all collider bounds even if its mass  is much below the electroweak scale 
provided its  singlet component is large enough to have much reduced couplings to the SM fields~\cite{Ellwanger:2009dp}. 
A Higgs mass in the range 122-128 GeV is thus obtained naturally in the NMSSM and can correspond either to $H_1$ or $H_2$.  
In the following we discuss the expectations for the main search channel, $gg\rightarrow H_i\rightarrow \gamma\gamma$,
for both scalars.

The model that we consider has input parameters which are defined at the weak scale. 
The free parameters are taken to be the gaugino masses $M_1,M_2=M_3/3$, the Higgs sector parameters $\mu,\tan\beta$, $\lambda,\kappa,A_\lambda,A_\kappa$, two  common soft  masses for the sleptons $M_{\tilde l_L}$, $M_{\tilde l,R}$  and soft masses for the squarks  of the first and second generation $M_{\tilde q_{1,2}}$  and of the third generation, $M_{\tilde q_3}$ as well as only one non-zero trilinear
coupling, $A_t$ (see~\cite{Vasquez:2010ru} for more details). Note that
the model contains an additional neutralino, the singlino. We make two separate analyses. In the first  one we 
take only one common soft mass for sleptons as well as one for squarks and only consider scenarios with a neutralino LSP lighter than 15~GeV, in the second one we relax these conditions. For both analyses we perform an MCMC exploration of the parameter space of the model imposing constraints from Higgs and SUSY searches at colliders,  flavour physics, $(g-2)_\mu$ as well as from dark matter observables such as the  relic density, the spin-independent  direct detection limit from XENON and the photon flux from dwarf spheroidal galaxies~\cite{Vasquez:2010ru}.  The latter are obtained with  micrOMEGAs2.4~\cite{Belanger:2010gh} interfaced to  NMSSMTools~\cite{Ellwanger:2005dv} for the computation of the spectrum and particle physics constraints. 
For all the allowed points we then impose LHC limits from SUSY searches  as well as more recent constraints on the Higgs sector from an interface to {\tt HiggsBounds 3.5.0beta} ~\cite{Bechtle:2008jh,Bechtle:2011sb}.
For SUSY searches we implement the ATLAS jets and missing energy search at $1.04\,{\rm fb}^{-1}$ \cite{Aad:2011ib}.
For each SUSY point, signal events were generated with \textsf{Herwig++ 2.5.1}~\cite{Bahr:2008pv,Gieseke:2011na}. Using \textsf{RIVET 1.5.2}~\cite{Buckley:2010ar} the experimental cuts of each search channel can be applied; for the ATLAS jets and missing $E_T$ search these are included~\cite{Grellscheid:2011ij} in the \textsf{RIVET} package.
For the case with only light neutralinos this SUSY search had hardly any impact on the sample. Firstly the majority of points had a heavy coloured spectrum, giving low production cross-sections. Secondly, allowed points with light squark masses, $300-600$ GeV, had a singlino LSP and rather light sleptons so that leptons were produced in the squark decay chains. The jets and missing $E_T$ SUSY searches that do not veto on leptons, which would show some sensitivity to these configurations, are not included in this analysis at present. For the sample generated allowing for heavier neutralinos, the more familiar result of $m_{\text{squark}}\gtrsim 0.6-1$ TeV (depending on $m_{\text{gluino}}$) was enforced by the search. In this sample the LSP is generally not the singlino thus  allowing for the standard $q\rightarrow q \chi_1^0$ decay.

\begin{figure}	
\centering
\includegraphics[width=0.47\textwidth]{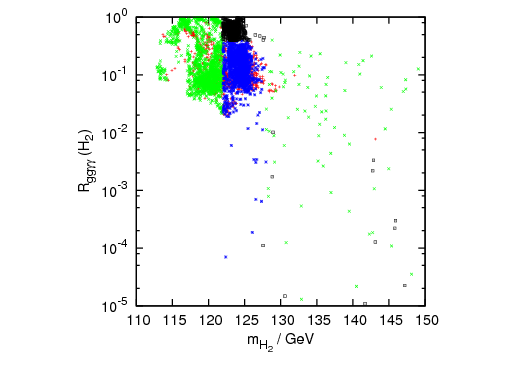}
\includegraphics[width=0.47\textwidth]{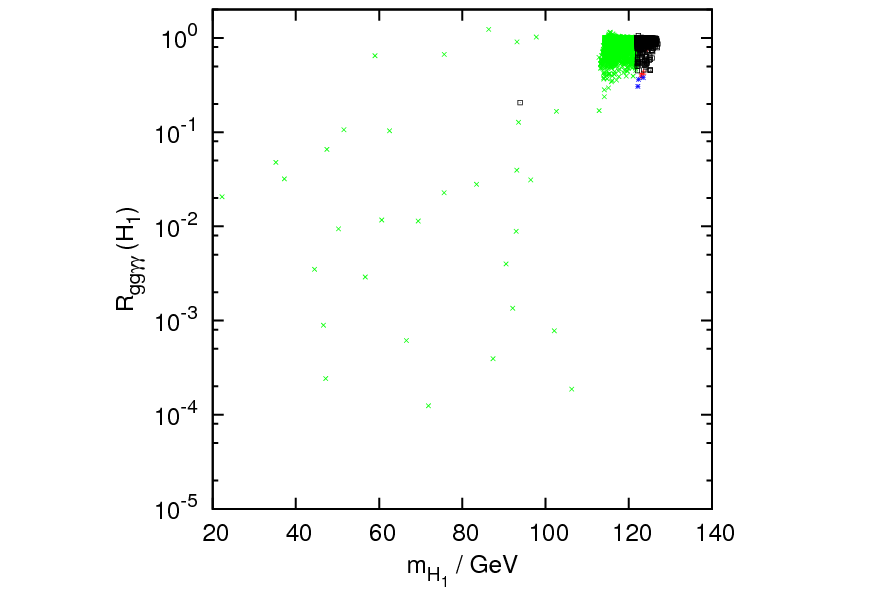}
\caption{$R_{gg\gamma\gamma}$ as a function of $M_{H_2}$ in the light neutralino LSP model (left panel) 
and ${R_{gg\gamma\gamma}}$ as a function of $M_{H_1}$ for arbitrary  neutralino masses (right panel). Blue points have $122~{\rm GeV}<M_{H_i}<128~{\rm GeV}$ and black squares have in addition ${R_{gg\gamma\gamma}}>0.4$.  Red points are ruled out by {\tt HiggsBounds} constraints or SUSY searches.  }
\label{fig:gggg:nmssm}
\end{figure}

We first present the results for the Higgs sector for the scenario with light neutralinos. Imposing this condition means that the model
must contain either a light scalar or pseudoscalar Higgs which is mostly singlet. This field is required  for efficient annihilation
of the neutralino LSP in the  early Universe. Typically we found  $H_1$ to be below the electroweak scale with a large singlet component while $H_2$ was SM-like. In some cases the light singlet Higgs field is a pseudoscalar and $H_1$ is SM-like, its mass barely reaches 122~GeV while  $M_{H_2}$ can take any value up to the TeV scale. It is also possible for both scalars to be heavily mixed and have a mass around 100-130~GeV.
The predictions for $R_{gg\gamma\gamma}$  as a function of the $H_2$ mass  are displayed in Fig.~\ref{fig:gggg:nmssm} (left panel). Here we only display the region where this channel is relevant, that is for a Higgs mass below 150 GeV. 
We find that $R_{gg\gamma\gamma}<1$  and that the maximum rate occurs mostly for a Higgs below 127 GeV.  
The couplings of  $H_2$ are usually SM-like,  nevertheless large suppressions in $R_{gg\gamma\gamma}$ are found.
 This occurs because the width of the Higgs is enhanced by many new non-standard decay channels such as $\chi_1\chi_1,\chi_2\chi_1$ or channels involving light Higgses, thus reducing the branching fraction into photons. We have also computed $R_{gg\gamma\gamma}$ for the lightest Higgs and found that it is usually much below 1 because $H_1$ has a large singlet component. Only a few points with a SM-like $H_1$ with a mass near 122 GeV have  $R_{gg\gamma\gamma}\approx 1$. 

Many points predict a Higgs  in the range compatible with the excess  reported by ATLAS, $122~{\rm GeV}<m_{H_i}<128~{\rm GeV}$, as mentioned above it is usually $H_2$. These points are highlighted in blue  in Fig.~\ref{fig:gggg:nmssm}. For a fraction of these points the strength of the  signal in $\gamma\gamma$ is also compatible with the 2$\sigma$ range reported by ATLAS ($R_{gg\gamma\gamma}>0.4$), these points are denoted by black squares in Fig.~\ref{fig:gggg:nmssm}. Furthermore we found that in the few cases where $H_1$ was heavy enough to be in the preferred range, the corresponding value $R_{gg\gamma\gamma}$ was also  in the desired  range.     

We then relaxed the upper limit on the neutralino mass and repeated
the analysis. In this case we considered only the subset of points
with a relic density  in agreement with the $3\sigma$ value extracted
by WMAP and we did not implement the LHC limits from SUSY searches. As
opposed to the previous case, we found that in general $H_1$ was
SM-like with a mass  near 120 GeV while  $H_2$ was  heavy.  Indeed,
there is no need to fine-tune the Higgs sector to have a light singlet
for efficient neutralino annihilation as other mechanisms are
available. The value of $R_{gg\gamma\gamma}$ for the lightest Higgs,
$H_1$, is displayed in Fig.~\ref{fig:gggg:nmssm} (right panel). 
The points where one Higgs is in the range favored by ATLAS mostly have $R_{gg\gamma\gamma}>0.4$, 
thus they appear as black squares in Fig.~\ref{fig:gggg:nmssm}, the invisible decays of the Higgs being much less important than in the previous case as they are often not kinematically accessible. Note that in this figure  we apply the color-coding for either $H_1$ or $H_2$ so black squares at low values of $M_{H_1}$ correspond to points where $H_2$ is in the ATLAS preferred range  with $R_{gg\gamma\gamma}$ that is not strongly suppressed. 
An analysis of the constrained MSSM~\cite{Gunion:2012zd} also found that after imposing collider, B-physics, dark matter and $(g-2)_\mu$ constraints, only $H_1$ could be in the range preferred by ATLAS and that this Higgs was SM-like.

\subsection{MSSM+RH (s)neutrino}    
{\it B. Dumont and S. Kraml}

A supersymmetric alternative to the see-saw mechanism proposed in~\cite{ArkaniHamed:2000bq,Borzumati:2000mc} can lead to naturally light Dirac neutrinos and generate weak-scale sneutrino A-terms that are not proportional to the small neutrino Yukawa couplings. These large $A_{\tilde{\nu}}$ terms induce a non-negligible mixing between left and right-handed (RH) sneutrinos. 
The lighter sneutrino mass eigenstate, $\tilde{\nu}_1$, can thus become the LSP and a viable cold dark matter candidate. 
A particularly interesting case, on which we concentrate here, is a mainly RH light mixed sneutrino LSP with mass below 
10~GeV~\cite{Belanger:2010cd}. 
Such a light mixed sneutrino LSP has a dramatic effect on SUSY signatures at the LHC~\cite{Belanger:2011ny} 
(see also \cite{Thomas:2007bu}). 
Moreover, since the sneutrino coupling to the Higgs is proportional to $A_{\tilde{\nu}}$, the presence of the mixed sneutrino 
significantly impacts mass and decay modes of the lightest Higgs. First, it is important to notice that $m_h$ is {\it decreased} 
through loop diagrams involving sneutrinos that induce corrections $\propto |A_{\tilde{\nu}}|^4$. Another important consequence 
of the light sneutrino dark matter scenario is a dramatic change of the decay branching fractions of $h$. Indeed, the lightest Higgs decays dominantly invisibly into sneutrinos ($BR\left(h\rightarrow {\rm invisible}\right) \gtrsim 90\%$). In addition, if the decay into neutralinos is kinematically allowed, it adds a small contribution to the invisible width of the lightest Higgs because $BR\left(\tilde{\chi}^0_1 \rightarrow \tilde{\nu} \nu\right) \approx 100\%$.

\begin{figure}[t]\centering
\includegraphics[width=7.6cm,height=6.4cm]{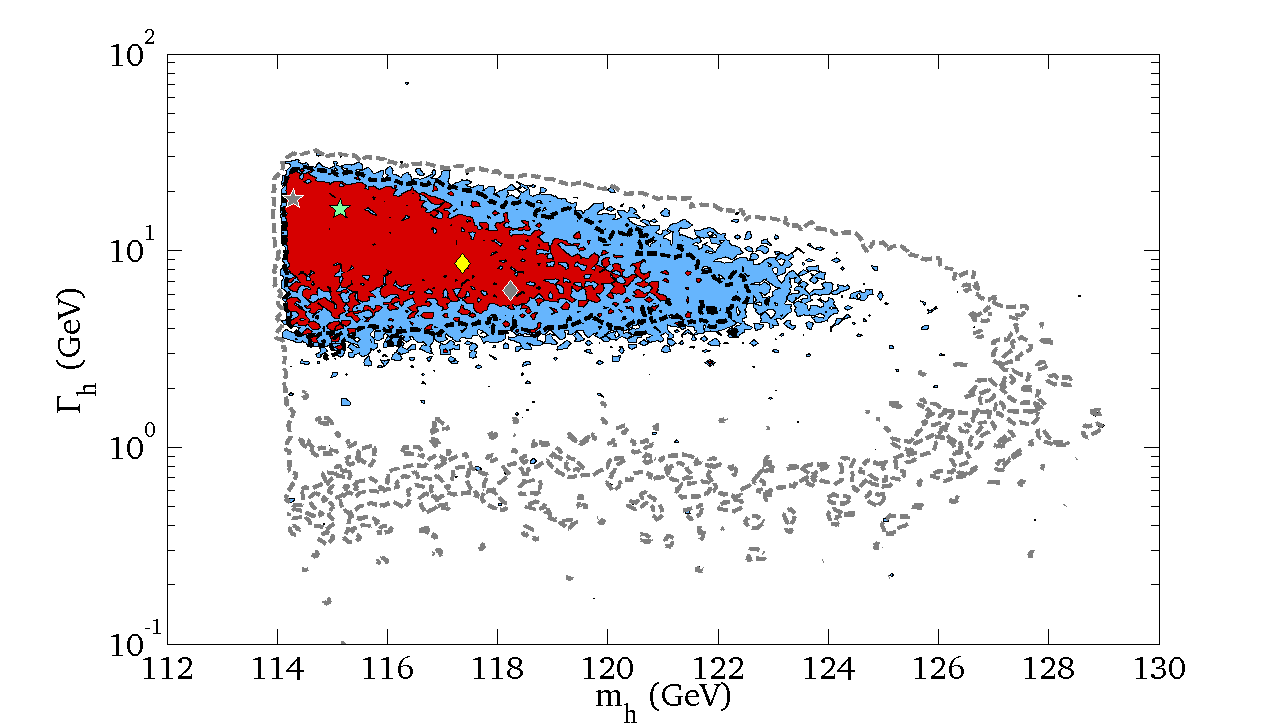}
\includegraphics[width=7.6cm,height=6.4cm]{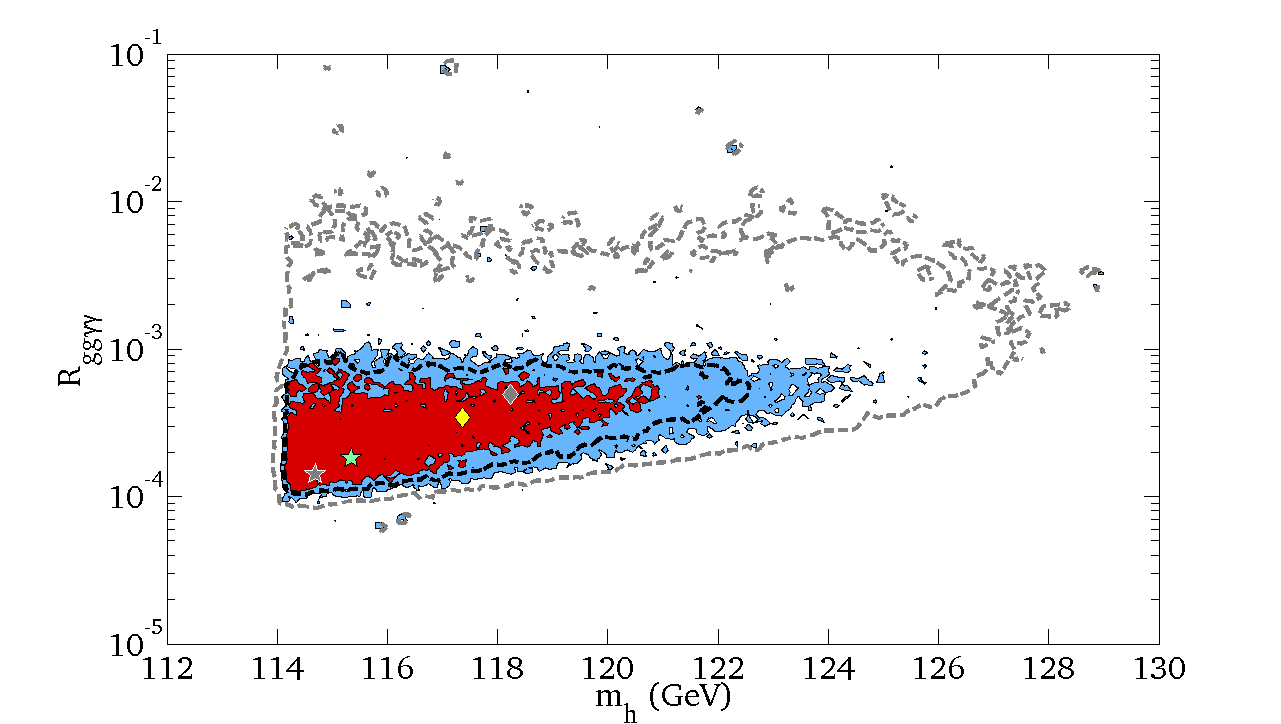}
\caption{Scan results concerning the lightest Higgs. On the left $\Gamma_h$ versus $m_h$, 
on the right $R_{gg\gamma\gamma}$ versus $m_h$. In both cases, the red and blue surfaces are the 68\% and 95\% Bayesian credible regions (BCR), respectively, requiring $m_{\tilde{g}} > 750$~GeV. The green star indicates our best fit point, the yellow rhombus locates the weighted average of the distribution. The black and the grey dotted lines identify the 68\% and 95\% BCR, respectively, in which instead of $m_{\tilde{g}} > 750$~GeV we set
$m_{\tilde{g}} > 500$~GeV. In this latter case, the best fit point and
the weighted average are shown in grey.}
\label{fig:rsnfig}
\end{figure}

To assess the consequences for Higgs phenomenology in a global way, we perform an MCMC analysis of the MSSM+RH 
neutrino parameter space, including all relevant constraints from Higgs and dark matter searches. (Regarding the latter,  
we carefully take into account nuclear and astrophysical uncertainties.) The calculations are done using {\tt micrOMEGAs} linked to appropriately modified versions of {\tt SuSpect} and {\tt HDECAY}. Higgs mass limits are computed with {\tt HiggsBounds3.5.0beta}.
We assume GUT relations for the gaugino masses, $M_1 \simeq M_2/2 \simeq M_3/6$, a common soft mass of 2 TeV for squarks of the first and second generation ($m_{\tilde{q}_{L,R}} \approx 2$~TeV), and of 1 TeV for squarks of the third generation, $m_{\tilde{Q}_3} = m_{\tilde{U}_3} = m_{\tilde{D}_3} = 1$~TeV. The pseudoscalar Higgs mass $m_A$ is also fixed to 1~TeV. We scan over all remaining free parameters, including the sneutrino and charged slepton soft terms of the 1st/2nd (assumed to be degenerate) and the 3rd generation.  
We set a loose 95\% upper bound on the relic density, $\Omega_{\rm DM}h^2 < 0.135$, using a smooth step-like function and we compute the $\chi^2$ for the direct detection limits from XENON100, CDMS and CoGeNT (we thank Thomas Schwetz for providing his private code to this end~\cite{Kopp:2009qt,Schwetz:2011xm}). We combine these constraints, and others (e.g. the value of $(g-2)_{\mu}$), in a global likelihood function.
Regarding SUSY mass limits from the LHC, the CMSSM limit for the case of heavy squarks~\cite{Aad:2011ib, cmsrazor2011} applies to our case. To take this into account we require a posteriori $m_{\tilde{g}} > 750$~GeV.
We present our main results on the properties of the lightest Higgs in this model in Fig.~\ref{fig:rsnfig} and 
Table~\ref{tab:rsnbci}.

The mass of our Higgs is constrained by the invisible searches at LEP~\cite{lep:2001xz,Achard:2004cf} and we find the usual lower bound $m_h > 114$~GeV. We see in Fig.~\ref{fig:rsnfig} that $\Gamma_h$ tends to decrease with increasing $m_h$. Conversely, $R_{gg\gamma\gamma}$ tends to increase with increasing $m_h$. This is because higher masses tend to have a smaller $A_{\tilde{\nu}}$ due to the negative corrections to $m_h$ involving sneutrinos, and then the lightest Higgs-sneutrino-sneutrino coupling, $g_{h\tilde{\nu}\tilde{\nu}} \propto A_{\tilde{\nu}}$, is smaller. For the same reason, 
going from $m_{\tilde{g}} > 500$~GeV to $m_{\tilde{g}} > 750$~GeV has a large effect on the predicted Higgs mass
range: with $M_2\simeq M_3/3$, increasing the gluino mass suppresses the neutralino t-channel exchange contribution 
to the $\tilde \nu_1$ annihilation (i.e.~the $\tilde \nu_1\tilde \nu_1\to \nu\nu$ and $\tilde \nu_1^*\tilde \nu_1^*\to \bar\nu\bar\nu$ channels), which has to be compensated by a larger Higgs-exchange contribution, and hence a larger $A_{\tilde{\nu}}$ term, to still achieve $\Omega_{\rm DM}h^2 < 0.135$. 
Thus the 95\% upper bound on the BCI of $m_h$ changes 
from 124.6~GeV to 124.2~GeV to 122.2~GeV, when going from no limit on $m_{\tilde{g}}$ to $m_{\tilde{g}} > 500$~GeV to  $m_{\tilde{g}} > 750$~GeV. (These numbers will of course somewhat vary if we take the stop soft masses as free parameters in the fit.) 

The left panel of Fig.~\ref{fig:rsnfig} shows that $\Gamma_h \approx 3-30~{\rm GeV} \approx 10^{3-4} \, \Gamma_{h,{\rm SM}}$, due to the predominant decay into two sneutrinos. In the right panel, we see that $R_{gg\gamma\gamma}$ is always $< 0.1$ and more likely of the order of a few $\times 10^{-4}$. Therefore, if the 125 GeV excess in the $\gamma\gamma$ channel is to be confirmed, this model would be ruled out in its present form. If not, this Higgs boson is a very good candidate for the invisible searches at the LHC in the WW fusion channel.

\begin{table}
\begin{center}
\begin{tabular}{|c|c|c|c|c|}\hline
 & \multicolumn{2}{|c|}{68\% BCI} &\multicolumn{2}{|c|}{95\% BCI}\\
Observables   & lower & upper & lower & upper \\
\hline
 $m_h\rm{\ (GeV)}$ & $114.2$ & $118.2$ & $114.1$ & $122.2$ \\
 $\Gamma_h\rm{\ (GeV)}$ & $5.2$ & $14.7$  & $3.5$ & $22.6$ \\
 $1- BR\left(h \rightarrow  {\rm invisible}\right)$ & $2.6 \times 10^{-4}$ & $9.8 \times 10^{-4}$ & $1.5 \times 10^{-4}$ & $2.4 \times 10^{-3}$ \\
 $R_{gg\gamma\gamma} \times 10^4$ & $1.9$ & $5.9$  & $1.2$ & $9.0$ \\
\hline
\end{tabular}
\caption{68\% and 95\% Bayesian credible intervals (BCI) of some Higgs-related observables. \label{tab:rsnbci}}
\end{center}
\end{table}


\subsection{UMSSM} 
{\it G. B\'elanger J. Da Silva}

The UMSSM is an extension of the MSSM  with a gauge group that contains an extra $\mathrm{U(1)}^\prime$ symmetry. Here we will assume that this model is derived from an $E_6$ model and in the numerical example we will choose the $\mathrm{U(1)}^\prime$ to correspond to $U(1)_\psi$, one of the two abelian subgroups of $E_6$.
In addition to the MSSM superfields, the model contains a new vector superfield, $B^\prime$, and a new singlet scalar superfield, $S$, thus leading to a new gauge boson, $Z^\prime$.
The Higgs sector  consists of 3 CP-even scalars, one  CP-odd Higgs and a charged Higgs. 
 The $\mu$  parameter, is generated from the VEV of the singlet field responsible for the breaking of  the $\mathrm{U(1)}^\prime$ symmetry  and is thus naturally at the weak scale. In this model the light Higgs mass receives new contributions from the superpotential term as in the NMSSM as well from $\mathrm{U(1)}^\prime$ D-terms~\cite{Barger:2006dh}. 
 This  means that it is more natural  than in the MSSM to obtain a light Higgs with a mass of 125~GeV. In fact the light Higgs can easily exceed the value already 
 excluded by LHC searches for a Higgs with SM couplings.  
Typically the Higgs spectrum will consist of an SM-like light Higgs, a heavy mostly doublet scalar which is almost degenerate with the pseudoscalar and the charged Higgs, and a predominantly singlet scalar. The singlet state is rather heavy as its mass  is close to that of the $Z^\prime$, which  is constrained by the LHC to be   above the TeV scale. This also implies that
the singlet component of the light state is usually not large enough to significantly relax the bound on the lightest Higgs contrary to what occurs in the NMSSM.
Thus the couplings of the light Higgs are SM-like. 

\begin{figure}[h]\centering
\includegraphics[width=7.6cm,height=6.cm]{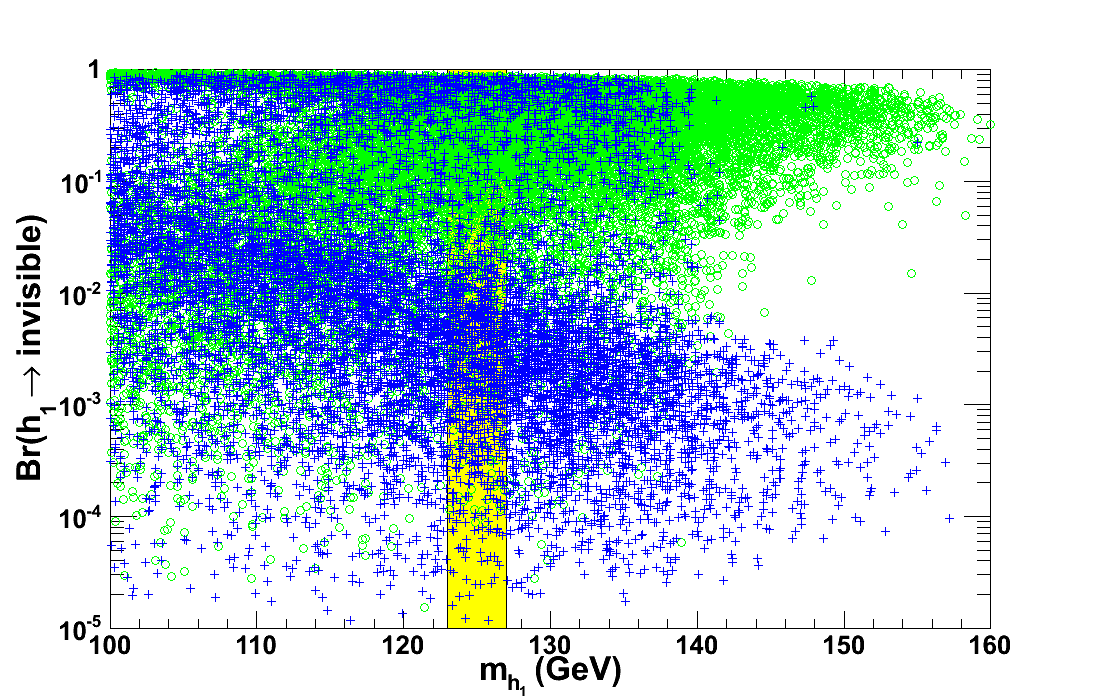} 
\vspace{-0.2cm}
\caption{The invisible branching ratio of the light Higgs as a function of its mass distinguishing the cases where
the sneutrino (circle) or neutralino (cross) is the LSP.  The region with $123~{\rm GeV}<M_h<127~{\rm GeV}$ is  highlighted in yellow.}
\label{fig:umssm}
\end{figure}

Apart from the shift in the mass of the light Higgs, the main impact on the Higgs phenomenology in this model arises from the possible contribution of invisible Higgs decays.  In the UMSSM the dark matter candidate can be either the right-handed sneutrino or the neutralino. Both can be light and contribute to invisible modes of the Higgs. 
In ~\cite{Belanger:2011rs} we have investigated the parameter space of the UMSSM that was consistent with precision measurements, LHC limits on the $Z^\prime$ mass, the upper bound on the relic density, $\Omega h^2< 0.1228$, limits from DM direct detection experiments (XENON100~\cite{Aprile:2011hi})  as well as with the limit from $\Delta M_{d,s}$
 in the B-meson~\cite{Krocker:2011hq} in the case of a sneutrino LSP. Here we allow either dark matter candidate and specifically investigate the invisible branching ratio of the light Higgs. We also vary the common soft squark mass in the range 500-2000~GeV. The results are displayed in Fig.~\ref{fig:umssm} for a light Higgs mass in the range 100-160~GeV considering only the cases when $M_{LSP}<M_h/2$.  
We find that even though the invisible branching ratio can reach one,  in general and even under the condition of a light LSP,  the visible decay modes are not significantly suppressed. Thus the model can accomodate a Higgs in the mass range compatible with  the excess of events observed at the LHC with a signal  strength comparable to the SM Higgs.
 
\section{Non supersymmetric models}
\subsection{UED \\ {\it A. Belyaev, G. B\'elanger, M. Brown, 
 M. Kakizaki, A. Pukhov}} 

Theories with Universal Extra Dimensions (UED) are very promising. In the UED framework {\it all} SM particles propagate in
the bulk, which results in the conservation of Kaluza-Klein (KK) number\footnote{Boundary conditions induce momentum quantization along the extra dimension(s) and the corresponding quantum numbers(s) are called KK number(s).} at tree level and conservation of KK parity at any loop level. This KK
parity conservation is the consequence of the orbifold symmetry (e.g.~$S_1/Z_2$ orbifold and 
$Z_2$ symmetry in case of one universal extra dimension) which insures the conservation of the ``evenness" or ``oddness"
of KK number in an interaction. As a result of this symmetry, UEDs predict a stable
lightest Kaluza-Klein particle (LKP) which would be a 
candidate for dark matter (DM).

The simplest UED theory is the 5D theory known as minimal Universal Extra Dimensions (mUED). The phenomenology of mUED has
been studied very intensively in many publications, but so far none of the experimental papers has set LHC limits on mUED.
 This is not surprising -- the search for mUED is much more difficult than the search of SUSY within the experimentally well-explored mSUGRA scenario the LHC.

The main reason is that mUED provides much smaller missing transverse momentum due to a small mass splitting between KK-partners of SM particles of the same KK level.  Though DM sets an upper limit on the scale of mUED below about 1.6 TeV~\cite{Belanger:2011},
this scale will be very difficult to test even with the 14 TeV LHC. However, the Higgs sector provides a
very powerful constraint on mUED. First of all, the Higgs mass is limited to be below 230 GeV by the simple requirement that the dark
matter candidate should be neutral. Second, as we will
demonstrate, the signals from the Higgs boson from mUED are {\it always} enhanced as compared to those of the SM. Therefore, the excellent LHC
sensitivity to the SM Higgs boson would lead to even better sensitivity to  the Higgs boson within the mUED scenario. For example, by
the end of 2012 the LHC will be able to completely exclude the SM Higgs bosom below 230 GeV, or observe it at least at the 3$\sigma$ level. 
If the hints of a Higgs signal are not confirmed, and a SM Higgs boson is excluded, this   would mean that mUED is excluded completely for all scales!

Here we evaluate and present the effect of KK-particles in the loop for $gg\to h$ production and Higgs decay to $\gamma \gamma$
and $W^+W^-$. This effect is different for different Higgs signatures and this non-universality should be taken into account
when establishing limits on the mUED parameter space. As far as we know, this is the first time the above mentioned 
non-universality is taken into account in the context of UED models.

The matrix elements $\mathcal{M}$ for the $gg\to h$ and $h\to \gamma\gamma$ processes both take the form $\mathcal{M} = \tilde{\mathcal{M}}(m_{h}^{2}g^{\mu\nu}/2 - p^{\nu}q^{\mu})\sum\epsilon_{\mu}\epsilon_{\nu}$, where $p$ and $q$ are the momenta of the external photons or gluons (approximated as on-shell), $m_{h}$ is the Higgs boson mass, and the sum is over gluon or photon polarisations. For $gg\to h$ we have
\[
\tilde{\mathcal{M}}_{ggh} = -\frac{\alpha_{s}}{4\pi v}\sum_{q} F_{ggh,q}
\]
with
\[
F_{ggh,q} =f_{F}(m_{q},m_{h}) + \sum_{n=1}^{N} m_{q}\sin(2a_{q}^{(n)}) \left(\frac{1}{m_{q,1}^{(n)}} f_{F}(m_{q,1}^{(n)},m_{h}) + \frac{1}{m_{q,2}^{(n)}} f_{F}(m_{q,2}^{(n)}, m_{h}) \right)
\]
and
\[
f_{F}(m,m_{h}) = \frac{8m^{2}}{m_{h}^{2}} - \frac{4m^{2}}{m_{h}^{2}}(m_{h}^{2}-4m^{2})C_{0}(m,m_{h}),
\]
where $C_{0}$ is the scalar three-point Passarino-Veltman function
\[
C_{0}(m,m_{h}) =
\begin{cases}
 -\frac{2}{m_{h}^{2}}\left[\arcsin\left(\frac{m_{h}}{2m}\right)\right]^{2} & m^{2} \ge m_{h}^{2}/4
 \\
 \frac{1}{2m_{h}^2}\left[\ln\left(\frac{1+\sqrt{1-4m^{2}/m_{h}^{2}}}{1-\sqrt{1-4m^{2}/m_{h}^{2}}}\right) - i\pi\right]^{2} & m^{2} < m_{h}^{2}/4.
\end{cases}
\]
Note that for each SM quark there are \emph{two} KK partners at each KK level $n$ with masses $m_{q,1}^{(n)}$ and $m_{q,2}^{(n)}$. At tree level these are equal:
\[
m_{q,\text{tree}}^{(n)} = \sqrt{m_{q}^{2}+ \frac{n^{2}}{R^{2}}},
\]
where $R$ is the compactification radius of the extra dimension (and the same goes for $W$ boson and scalar masses used below); but if radiative corrections are included (see e.g.~Ref.~\cite{Cheng:2002iz}) the masses split. Similarly, the mixing angles $a_{q}^{(n)}$ required to diagonalise the KK quark mass matrices differ from the tree-level form
\[
a_{q,\text{tree}}^{(n)} = \frac{1}{2}\arctan\left(\frac{m_{q}R}{n}\right)
\] 
once radiative corrections are included, but the mixings for type 1 and 2 quarks remain equal to each other, in contrast to the masses. In our analysis we used one-loop corrected expressions for all masses and mixings as detailed in Ref.~\cite{Cheng:2002iz}).
One should also note that the higher-order corrections to $gg\to H$ presented here can be substantial, of the order of 50\% or even larger.  However, these  corrections largely cancel out when taking the ratio of mUED to SM rates.

The sum over KK modes $n$ is taken up to a cutoff $N$, corresponding to a momentum cutoff in the extra dimension of $N R^{-1}$. Mild cutoff-dependence is expected in perturbatively non-renormalisable theories such as mUED. In our quantitative analysis we chose $N=20$ and included only $t$ and $b$ in the sum over quark flavours $q$, which is an excellent approximation due to the size of their Yukawa couplings compared to those of the other quarks. One should note that for large $N$ the rest of the sum is proportional to 
$1/N$. Therefore, setting $N=20$ yields about 5\% accuracy as compared to the full sum.

The $h\to \gamma\gamma$ matrix element is given by
\[
\tilde{\mathcal{M}}_{h\gamma\gamma} = -\frac{e^{2}g}{16\pi^{2}m_{W}}F_{h\gamma\gamma},
\]
where
\[
F_{h\gamma\gamma} = F_{h\gamma\gamma}^{\text{SM}} + \sum_{n=1}^{Nt}F_{h\gamma\gamma}^{(n)}
\]
and
\[
F_{h\gamma\gamma}^{\text{SM}} = f_{V}(m_{W},m_{h}) + \sum_{f}n_{c}Q_{f}^{2}f_{F}(m_{f},m_{h}).
\]
The Fermion loop function $f_{F}$ was given earlier and the vector function introduced above is
\[
f_{V}(m,m_{h}) = -2 -12\frac{m^{2}}{m_{h}^{2}} + 12\frac{m^{2}}{m_{h}^{2}}(m_{h}^{2}-2m^{2})C_{0}(m,m_{h}).
\]
The sum could be taken over all SM fermions $f$, each with charge $Q_{f}e$, setting $n_{c}$ to 3 for quarks and 1 for leptons. In our treatment, however, we again only included $t$ and $b$ quarks. The contribution from the $n$th level KK fermions is
\begin{align*}
F_{h\gamma\gamma}^{(n)} &= \frac{m_{W}^{2}}{m_{W,n}^{2}}f_{V}(m_{W,n},m_{h}) + f_{S}^{(n)}(m_{a,n},m_{W,n},m_{h})\\
&\quad+\sum_{f}n_{c}Q_{f}^{2}\left(\frac{m_{f}}{m_{f,1}^{(n)}}\sin 2a_{f}^{(n)} f_{F}(m_{f,1}^{(n)},m_{h}) + \frac{m_{f}}{m_{f,2}^{(n)}}\sin 2a_{f}^{(n)} f_{F}(m_{f,2}^{(n)},m_{h})\right).
\end{align*}
Here we have a new scalar loop function for the charged scalars $a_{\pm}^{(n)}$ that appear in the loop,
\begin{align*}
f_{S}(m_{a,n},m_{W,n},m_{h}) &= \left[\frac{2m_{W}^{2}}{m_{W,n}^{2}}\left(1-\frac{2m_{a,n}^{2}}{m_{h}^{2}}\right) - 2\right] [1-2C_{0}(m_{a,n},m_{h})].
\end{align*}

\begin{figure}[hbt]
\begin{center}
\includegraphics[width=7.6cm,height=6.cm]{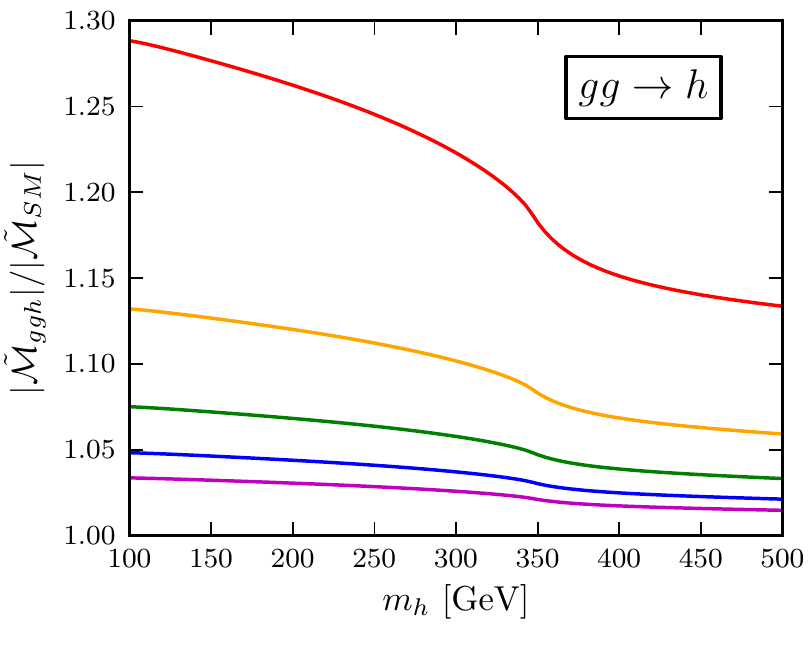}%
\includegraphics[width=7.6cm,height=6.cm]{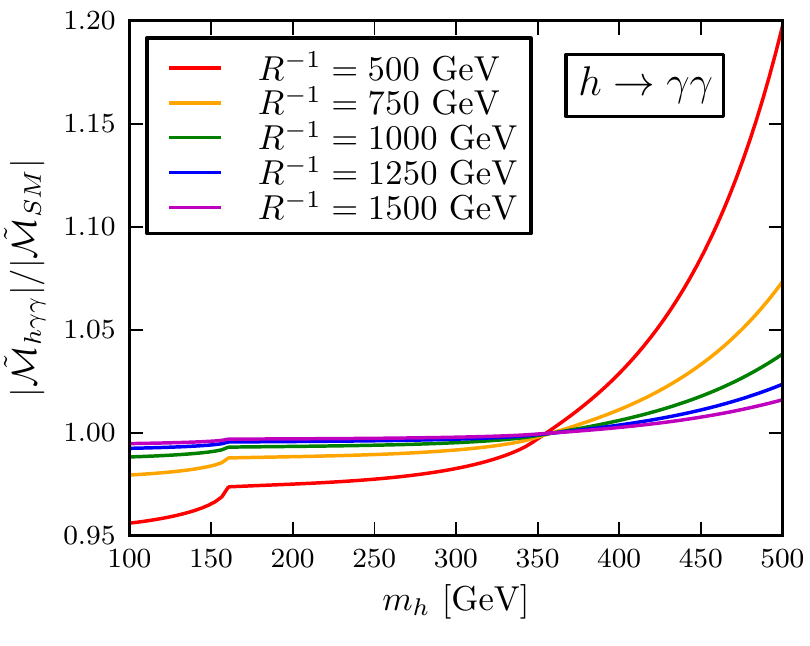}
\end{center}
\caption{Ratio of mUED and SM matrix elements for $ggh$ (left) and $h\gamma\gamma$ (right) processes for the following values of $R^{-1}$ from top to bottom in each plot: 500, 750, 1000, 1250 and 1500 GeV. The number of KK modes is $N=20$.}
\label{fig:mattratios}
\end{figure}

The dependence of the two matrix elements on $m_{h}$ and the inverse compactification radius $R^{-1}$ (these are the two free parameters of mUED, other than the very weak cutoff dependence) is shown in Figure~\ref{fig:mattratios}, and clearly indicates that for a light Higgs boson the $hgg$ coupling is enhanced while $h\gamma\gamma$ is suppressed. The $R^{-1}$ dependence enters  through the KK masses and mixing angles.

\begin{figure}[bt]
\begin{center}
\includegraphics[width=0.5\textwidth]{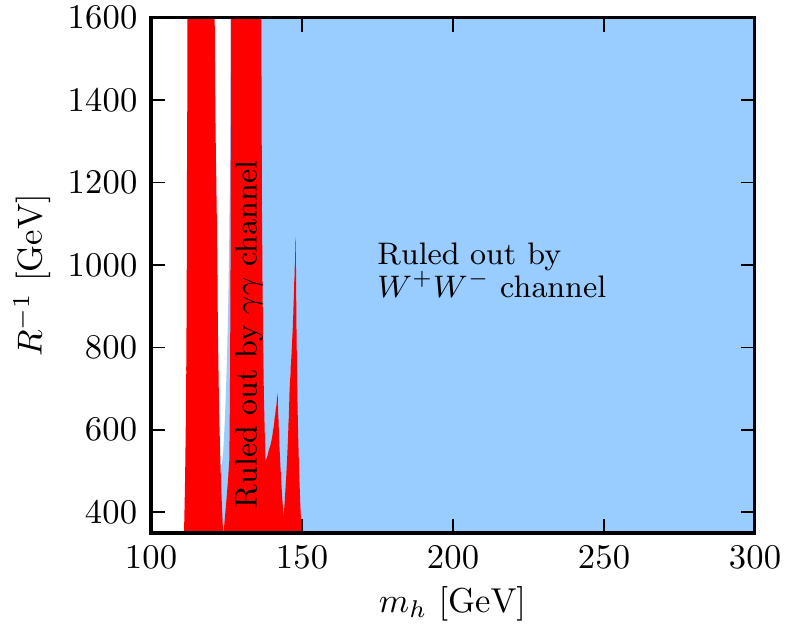}%
\includegraphics[width=0.51\textwidth]{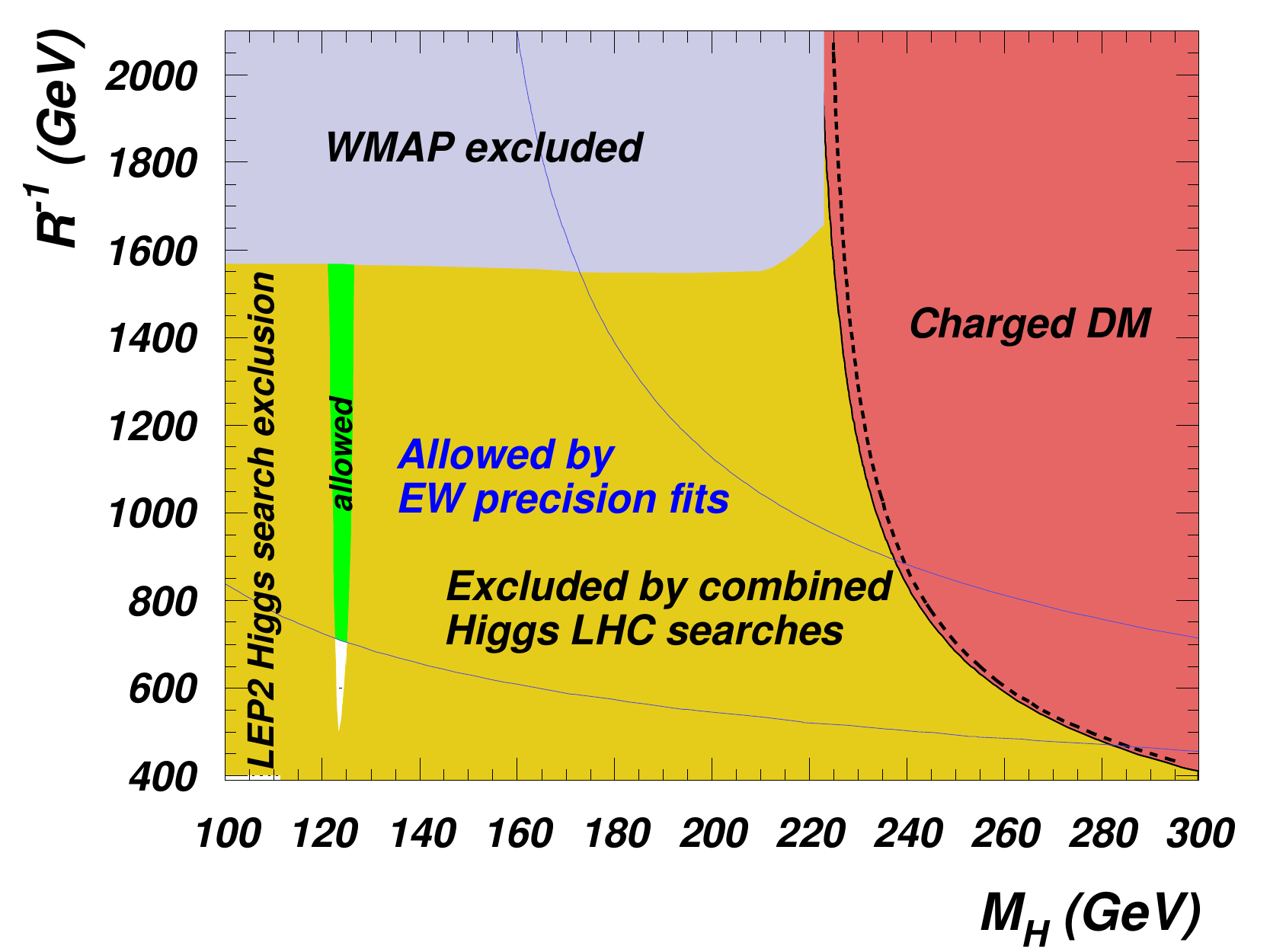}
\end{center}
\caption{Left: exclusion of mUED $(m_h,R^{-1})$ parameter space at 95\% CL from the
Higgs boson search using combined ATLAS and CMS limits in the $\gamma\gamma$ and $W^{+}W^{-}$
channels. Right: combination of Higgs constraints (from the left frame), EW precision and DM relic density limits on the mUED parameter space.}
\label{fig:mued-limits}
\end{figure}

Using our model's predictions for Higgs boson production enhancement for different values of $m_{h}$ and $R^{-1}$ together with the latest CMS~\cite{Chatrchyan:2012tx} and ATLAS~\cite{ATLAS-CONF-2011-163} limits on Higgs boson production, we can exclude regions of the $(m_h,R^{-1})$ plane. We statistically combined data from Fig.~6 (top) of the CMS paper and Fig.~3 of the ATLAS paper in each of the $\gamma\gamma$ and $W^{+}W^{-}$ channels; the resulting limits on mUED are shown in Fig~\ref{fig:mued-limits} (left).These constraints are combined with other constraints from DM relic density~\cite{Belanger:2011} as well as EW precision tests~\cite{Gogoladze:2006br} in Fig.~\ref{fig:mued-limits} (right).

We can see that the Higgs search provides a powerful constraint on the  mUED model
since Higgs boson production is enhanced.
Compared to previous studies \cite{Nishiwaki:2011gk}, 
we  have included mass corrections for the particles in the loop,
providing more realistic predictions of mUED cross sections,
and have accurately combined non-universal enhancement in
$\gamma\gamma$ and $W^+W^-$ matrix elements, which has not been done previously.
This allowed us to find accurate limits on the mUED 
$(m_h,R^{-1})$ parameter space. After combination of ATLAS and CMS limits 
for each individual channel ($\gamma\gamma$ and $W^{+}W^{-}$),
we found that $R^{-1}<500$~GeV is   excluded at 95\%CL, while
for larger  $R^{-1}$ only a very narrow ($\pm 1-3$~GeV) mass window around
$m_h=125$~GeV is left. This is the region where  the excess of the events
in the Higgs search channels is reported by the ATLAS and CMS collaborations and where
the exclusion limit is weaker. 

\subsection{SM4}
{\it A. Wingerter}

\subsubsection{Motivation for a Fourth Chiral Generation}

Introducing a fourth chiral generation is one of the simplest and most straightforward extensions of the SM. There are 9 additional parameters, namely the masses of the fourth generation fermions \mt{4}, \mb{4}, \me{4}, \mv{4}, the mixing angles $\theta_{14}$, $\theta_{24}$, $\theta_{34}$, and 2 new CP phases. The relatively small number of new parameters makes the Standard Model with four generations (SM4) highly predictable and testable.

Before observing the slight excess around 126 GeV~\cite{ATLAS-CONF-2011-163,Chatrchyan:2012tx} that may indicate the presence of a Higgs boson, some physicists were getting increasingly worried about the tension between the lower Higgs boson mass bound of 114.4 GeV set by LEP~\cite{Barate:2003sz} and the best-fit value of $m_H=80^{+30}_{-23}$ GeV~\cite{Nakamura:2010zzi} as obtained from electroweak precision measurements (EWPMs). In the presence of a fourth generation, the EWPMs may be consistent with much larger values for the Higgs mass, if the mass splittings $|\mt{4}-\mb{4}|$ and $|\me{4}-\mv{4}|$ are chosen accordingly (see Fig.~(13) in Ref.~\cite{Baak:2011ze}). Also, the SM does not predict the number of generations, and the constraints from flavor physics are weaker than previously assumed~\cite{Bobrowski:2009ng} and allow for sizable mixing of order the Cabibbo angle between the third and the fourth generation, even after taking the EWPMs into account~\cite{Chanowitz:2009mz}. The \textit{direct}\footnote{Previous determinations of $\left|V_{tb}\right|$ assumed unitarity of the 3-by-3 CKM matrix.} measurement of $\left|V_{tb}\right|=1.02^{+0.10}_{-0.11}$ from single-top production~\cite{Abazov:2011pt} is consistent with $\left|V_{tb}\right| = 1$, but leaves enough room for a fourth generation replica $t'$ of the top quark to couple to the $b$ quark. Basically, this is all in support of a ``Why not?'' type of argument.

Some of the other reasons usually given in favor of the existence of a fourth generation are less convincing. Following Ref.~\cite{Hung:1997zj} many recent publications have maintained that in the presence of a fourth generation the gauge couplings may unify without supersymmetry (SUSY). However, a closer inspection shows that the fourth generation quark and Higgs boson masses required in this scenario are excluded by experiment~\cite{CMS-PAS-EXO-11-051,CMS-PAS-EXO-11-036,ATLAS-CONF-2011-135,CMS_higgs_fit}. Also, it has been discussed in the literature whether the extra CP violation introduced by the 2 additional phases in the 4-by-4 CKM matrix may be sufficient to render electroweak baryogenesis in the SM viable~\cite{Hou:2008xd}. According to Refs.~\cite{Ham:2004xh,Fok:2008yg}, the first order phase transition in the SM is not strong enough to preserve the generated baryon asymmetry from being washed out, but the presence of extra scalars in the MSSM may change the picture.

\subsubsection{Impact of a 126 GeV Higgs Boson on the SM4}

\begin{figure}[h!]
\begin{center}
\subfigure[$\lambda$ approaches an ultraviolet fixed point \FP{}.]{
\includegraphics[width=0.47\textwidth]{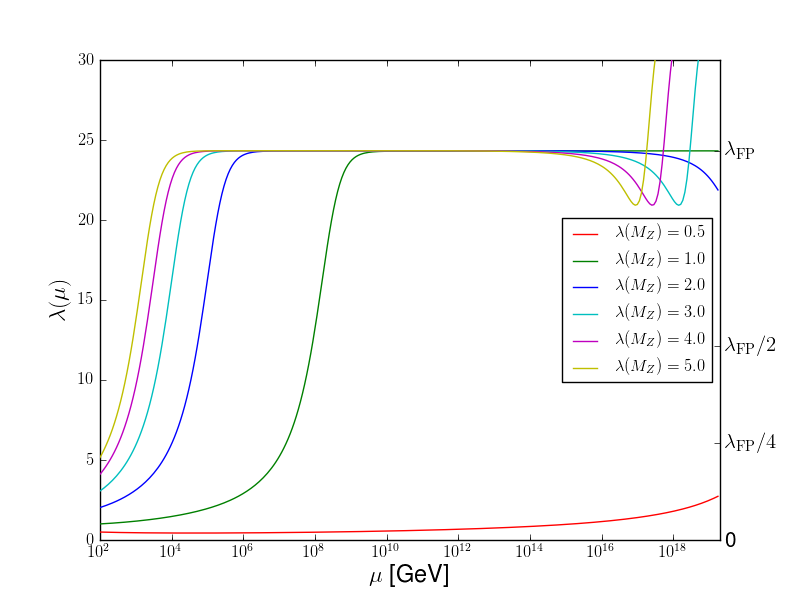}
\label{fig:runninglambda-triv}
}
\subfigure[For small starting values, $\lambda$ becomes negative.]{
\includegraphics[width=0.47\textwidth]{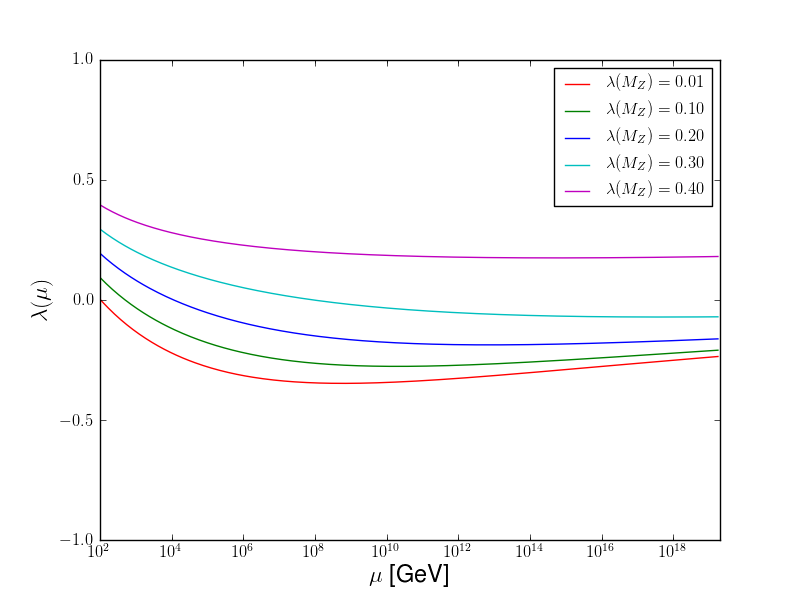}
\label{fig:runninglambda-stab}
}
\caption{The 2-loop running of the quartic Higgs coupling $\lambda(\mu)$ in the SM.}
\label{fig:runninglambda}
\end{center}
\end{figure}

At 1-loop, the running quartic coupling $\lambda(\mu)$ develops a Landau pole for some $\mu=\mu_0$, and this tells us that the SM stops making sense at the scale $\mu_0$ for the corresponding Higgs mass $m_H^2=\lambda\,v^2$. This fact can be interpreted in either of two ways: If we assume that the theory is valid beyond $\mu_0$, the Higgs mass must be less than $m_H$. On the other hand, if we assume (or know) that the Higgs mass is $m_H$, the triviality bound gives us the maximal scale $\mu_0$ up to which the SM can be \textit{perturbatively} valid. At 2-loop, the quartic coupling does not develop a Landau pole, but approaches an ultraviolet fixed point \FP{} (see Fig.~\ref{fig:runninglambda} (a)). The SM ceases to be perturbative long before $\lambda$ reaches the fixed point~\cite{Riesselmann:1996is}, and usually one adopts either $\lambda<\FP{}/2$ (loose) or $\lambda<\FP{}/4$ (tight) as a condition for perturbativity. 

A lower bound on the Higgs mass can be derived from the requirement of vacuum stability~\cite{Cabibbo:1979ay}. For too small starting values, $\lambda(\mu)$ is driven negative (see Fig.~\ref{fig:runninglambda} (b)), and to a good approximation~\cite{Altarelli:1994rb,Casas:1996aq} that is tantamount to the effective Higgs potential becoming unbounded from below. As before, the scale where this happens indicates where the SM breaks down. Alternatively, one obtains a lower limit on the Higgs mass. It is important to note that in this 2-loop analysis (both for the triviality and the stability bound), the radiative corrections to the tree-level relations between the Higgs boson mass $m_H$ and the quartic coupling $\lambda$ on the one hand, and the top quark mass $m_t$ and the Yukawa coupling $y_t$ on the other, cannot be neglected~\cite{Hambye:1996wb}. The generalization of these so-called Higgs and top matching corrections to the case of the SM4 are given in Ref.~\cite{Wingerter:2011dk}.

\begin{figure}[h!]
\begin{center}
\subfigure[The Standard Model.]{
\includegraphics[width=0.47\textwidth]{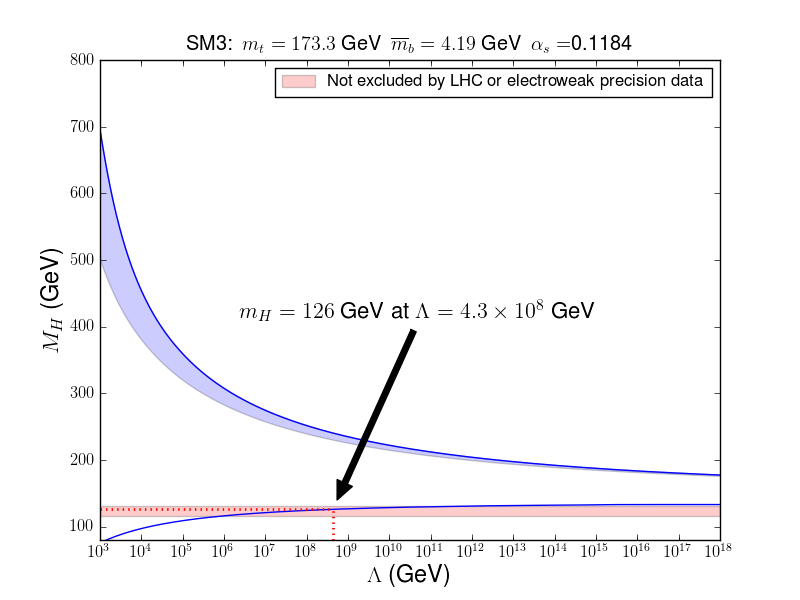}
\label{fig:SM3-trivstab}
}
\subfigure[The Standard Model with four generations.]{
\includegraphics[width=0.47\textwidth]{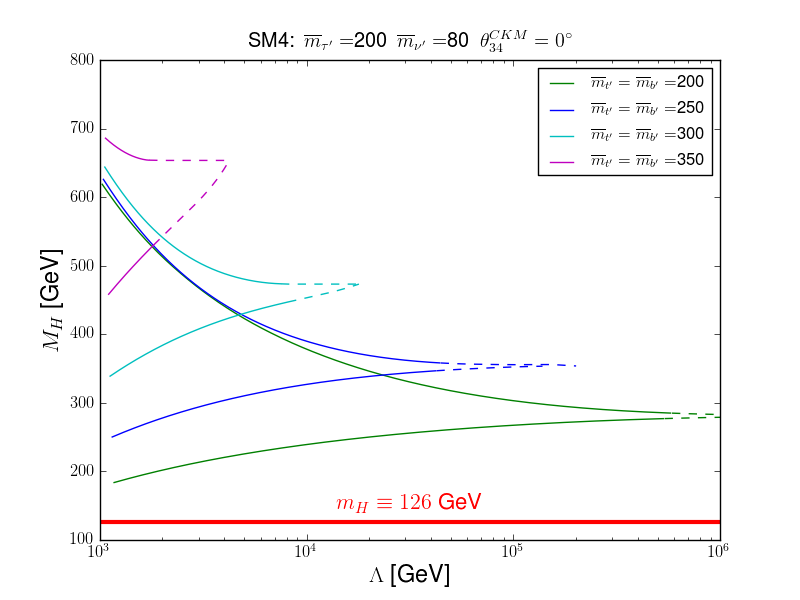}
\label{fig:SM4-trivstab}
}
\caption{The triviality and stability bounds in the Standard Model with three and four generations, respectively~\cite{Wingerter:2011dk}. The blue shaded area below the triviality bound in the left panel corresponds to the uncertainty introduced by choosing $\lambda<\FP{}/2$ or $\lambda<\FP{}/4$, respectively, as the condition for perturbativity of the quartic coupling. The dashed lines in the right panel indicate that at least one of the Yukawa couplings has become non-perturbative, i.e.~$y_{f}\geq\sqrt{4\pi}$ for $f=t,t',b',\tau',\nu'$.}
\label{fig:trivstab}
\end{center}
\end{figure}

In Fig.~\ref{fig:trivstab} (a) we show the triviality and stability bounds for the SM. The recent LHC results~\cite{ATLAS-CONF-2011-163,Chatrchyan:2012tx} together with LEP data~\cite{Barate:2003sz} exclude the mass range $m_H<115.5$ GeV and $127\text{ GeV} < m_H < 600\text{ GeV}$ at 95\% C.L. Electroweak precision measurements set 95\% C.L.~upper bounds for the Higgs mass of 185 GeV and 158 GeV, respectively, depending on whether one includes the LEP direct search results or not~\cite{aleph:2010vi}. This basically leaves a small window of $115.5\text{ GeV} < m_H < 127\text{ GeV}$ that is not yet excluded (red band in Fig.~\ref{fig:trivstab} (a)). One immediate consequence is that the SM cannot be perturbatively valid up to some grand unification scale\footnote{Since the stability bound is asymptotically flat, choosing \mgut $\sim10^{15}$ GeV or $10^{16}$ GeV does not change the conclusions.} \mgut{}: The stability bound crosses 127 GeV at $\Lambda = 1.3\times 10^{9}$ GeV. If the slight excess that points to a Higgs boson with mass $m_H=126$ GeV can be substantiated with more data, this would correspond to a cut-off scale for the SM around $\Lambda_c = 4.3\times10^{8}$ GeV (red dotted line in Fig.~\ref{fig:trivstab} (a)). Albeit beyond the reach of the LHC, this is one of the lowest scales so far for which we have hints for physics beyond the Standard Model (BSM). Of course, this does not mean that there is no new physics before $\Lambda_c$, but rather this sets an upper bound. One easy way to evade that conclusion is e.g.~to assume that the vacuum is metastable or to introduce a second Higgs doublet.

In Fig.~\ref{fig:trivstab} (b) we show the triviality and stability bounds for the SM4. Perhaps the most striking feature (as compared to the case of the SM) is that the curves now intersect and bound a finite region. As such, there is a priori a highest scale $\Lambda_c$ beyond which the theory does not make sense. Generally, the Yukawa couplings become non-perturbative before reaching $\Lambda_c$ (indicated by the dashed line style in Fig.~\ref{fig:trivstab} (b)) so that the real cut-off scale is lower. Note that the stability bound is much more sensitive to the new particles than the triviality bound and sets the stronger constraints. 

We have indicated in Fig.~\ref{fig:trivstab} (b) the slight excess around 126 GeV and see that it is clearly excluded by the stability bound for all values of the fourth generation fermion masses. More plots where \mt{4}, \mb{4}, \me{4}, \mv{4} and \thetam{34} are varied within sensible limits can be found in Ref.~\cite{Wingerter:2011dk} and lead to the same conclusion. The other two mixing angles \thetam{14} and \thetam{24} are small~\cite{Bobrowski:2009ng}, and their effect on the stability and triviality bounds is negligible. Even without taking this excess into account, the Higgs exclusion limits in the context of the SM4~\cite{ATLAS-CONF-2011-135,CMS_higgs_fit} cannot be reconciled with the stability and triviality bounds and therefore forbid a fourth chiral generation of fermions for any choice of masses and mixing angles. Of course, this statement is subject to the assumptions we used to calculate the stability and triviality bounds such as absolute stability of the vacuum, no more than one Higgs doublet, etc. For the sake of completeness, though, it should be noted that the Higgs searches do not cover the region $m_H>600$ GeV, and there exists a small window $600\text{ GeV}<m_H<700\text{ GeV}$ that is allowed by the stability and triviality bounds, and by the EWPMs. However, as Fig.~\ref{fig:trivstab}  and Fig.~2 in Ref.~\cite{Wingerter:2011dk} show, the cut-off scale in this case is at most 2 TeV.

Also, the experimental limits on the fourth generation quark~\cite{CMS-PAS-EXO-11-051,CMS-PAS-EXO-11-036} and lepton~\cite{Nakamura:2010zzi} masses are now closing in on the unitarity bounds $m_{t'}$, $m_{b'}$ $\lesssim500$ GeV and $m_{\tau'}$, $m_{\nu'}$ $\lesssim1000$ GeV~\cite{Chanowitz:1978mv} and leave very little room for a fourth generation. The bounds we obtained from theory by considering the stability and triviality bounds are complementary in the sense that they also apply in situations where the direct search limits do not hold, and at the same time set stronger constraints on the lepton masses than currently available from direct search experiments. For a critical discussion of the direct search limits, especially the assumption $\mt{4}-\mb{4}<M_W$, see Ref.~\cite{Wingerter:2011dk}.

\section{Conclusions}

The properties  of non-standard Higgs bosons can differ significantly from those of the SM one 
for each production and decay process at the LHC. While for most of the MSSM parameter space the most relevant search channel is 
a Higgs produced in gluon fusion decaying into two photons, in extensions of the MSSM or in UED the Higgs decay into vector boson pairs
is also an important channel. Interpreting LHC results therefore requires careful combination of the limits from each production times decay channel 
independently.  An exclusion of the SM Higgs by the end of this year at 95\% C.L. would rule out the whole parameter space of mUED while SUSY models have enough flexibility to avoid this limit. 
We have also shown that the excess of events corresponding to a Higgs with a mass of 125 GeV reported by ATLAS and CMS could be compatible with any of the extensions to the Standard Model considered,  provided the invisible Higgs decay mode is not too large.

\section{Acknowledgements}

The work of GB, FB, AP was supported in part by the GDRI-ACPP of CNRS.
AB and MB thank  the Royal Society for  partial financial support. AB also
thanks  the NExT Institute and SEPnet for financial support.
JD acknowledges the CMIRA 2011 EXPLO'RA DOC program of the French region Rh\^one-Alpes for financial support.
RMG  wishes to acknowledge the Department of Science and Technology of India,
for financial support  under the J.C. Bose Fellowship scheme under grant no.
SR/S2/JCB-64/2007.
The work of AP was supported by the Russian foundation for Basic Research, 
grant RFBR-10-02-01443-a.

\AddToContent{D.~Albornoz V\'asquez et al.}  
\renewcommand{\thesection}{\arabic{section}}


\chapter{CMSSM mass spectrum  at the LHC after the 125 GeV Higgs boson results}

{\it M. Kadastik, K. Kannike, A. Racioppi, M. Raidal}

\begin{abstract}
We show that  if the Higgs boson mass is $125\pm 1$~GeV, the CMSSM sparticle spectrum that is testable at the LHC is
restricted only to two distinct possibilities -- either of the lightest slepton or stop are degenerate in mass with the lightest
supersymmetric particle that is the dark matter. This implies that the CMSSM spectrum is now known.  
However, the two possible options both represent  very difficult spectra for the LHC because very soft particles are produced in sparticle decays.
We encourage the  LHC experiments to perform detailed studies of the two possible CMSSM sparticle spectra.

\end{abstract}

\section{INTRODUCTION}

The consistent ATLAS~\cite{ATLAS-CONF-2011-163} and CMS~\cite{Chatrchyan:2012tx} 
hints for the existence of a $M_H\approx 125$~GeV Higgs boson have profound
implications~\cite{Hall:2011aa,Arbey:2011ab,Baer:2011ab,Heinemeyer:2011aa,Draper:2011aa,Carena:2011aa,Kadastik:2011aa}
for the sparticle mass spectrum in the context of the minimal supersymmetric standard model (MSSM).
In particular, the versions of the MSSM with unification constraints on supersymmetry (SUSY)
breaking parameters, such as the CMSSM or mSUGRA, becomes severely 
fine-tuned~\cite{Kadastik:2011aa,Buchmueller:2011ab,Akula:2011aa,Strege:2011pk,Cao:2011sn}. The reason is that
such a high value of the Higgs boson mass requires unusually large scalar masses to generate the indicated Higgs
mass at loop level. 

At the same time, a  $M_H\approx 125$~GeV Higgs boson mass implies that the possible LHC phenomenology of
new CMSSM particles becomes highly predictive. This is because  the global fits 
of the CMSSM parameter space are dominated by two phenomena~\cite{Farina:2011bh}. 
The first one is the production of the correct dark matter (DM) relic density via very finely tuned freeze-out processes.
The second is explaining the measured value of mthe uon anomalous magnetic moment $(g-2)_\mu$ with extra contributions from 
sparticles. All other phenomenological constraints, summarized in Table~\ref{susysig:tab}, are less constraining and, at present,
play a role only in some particular corner of the CMSSM parameter space. As a result, the DM freeze-out processes alone 
determine the sparticle spectrum that is potentially observable at the LHC.

In this work we point out that the 125~GeV Higgs implies only two possible options for the CMSSM phenomenology at the LHC --
the lightest stable SUSY particle, the DM,  is (almost) degenerate with either the lightest slepton or stop. This is because for $M_H\approx 125$~GeV
only the slepton and stop coannihilation processes can produce the correct DM density and have light sparticles.
All other DM freeze-out processes imply unobservable sparticle spectra at the LHC. If, in addition, one also requires generation of 
the measured $(g-2)_\mu,$, at $3\sigma$ level only the slepton coannihilation region of the CMSSM parameter space survives 
(with a poor fit).  This may be testable at the LHC.

The implications of those results for the LHC are twofold. On the one hand, the CMSSM spectra testable at the LHC are now known 
and the experiments can concentrate on detailed studies of those spectra. On the other hand, those spectra are really
difficult because the transverse momenta of the particles produced in sparticle decays are predicted to be very small due to the
sparticle degeneracies.
The aim of this work is to encourage the LHC experiments to analyze the slepton and stop degenerate spectra in detail.

\section{THE DIFFICULT CMSSM SPECTRA}
\label{sec:cmssm}

The CMSSM is one of the most thoroughly studied SUSY models and  the CMSSM parameter space was rather 
fine-tuned already before the 125~GeV Higgs hint~\cite{Farina:2011bh,Buchmueller:2011ki,Buchmueller:2011sw,Bertone:2011nj,Fowlie:2011mb}.  
Naturally, if the Higgs boson is discovered with mass $M_H=125\pm 1$~GeV, one would like to know what the 
implication of this discovery is for the LHC phenomenology of this model. 
At the GUT scale the parameter space of the CMSSM is described by five parameters,
\begin{eqnarray}
m_0,\; M_{1/2},\; A_0, \; \tan\beta,\; {\rm sign} (\mu),
\end{eqnarray}
the common scalar mass, the common gaugino mass, the common trilinear coupling, the ratio of the two Higgs 
vacuum expectation values (vevs) and the 
sign of the higgsino mass parameter. To scan over the CMSSM parameter space we randomly generate parameters 
in the following ranges: $300< m_0,\; M_{1/2}< 10000$~GeV, $|A_0|<5 m_0,$ $3<\tan\beta <60$, $ {\rm sign}(\mu)=\pm.$
We use the MicrOMEGAs package \cite{Belanger:2006is,Belanger:2010gh} to compute the electroweak scale sparticle mass spectrum, the Higgs boson masses, the DM relic abundance $\Omega_{\rm DM}$,
the spin-independent DM-nucleon direct detection cross section $\sigma_{\rm SI}$ and the other observables in Table~\ref{susysig:tab}.
In addition, we require $M_H=125\pm 1$~GeV. 
There is a few GeV theoretical uncertainty in the computation of SUSY Higgs masses
in the available codes. 
Therefore, to select the phenomenologically acceptable parameter space  we impose $3\sigma$ hard cuts for the observables in  Table~\ref{susysig:tab}.
Our approach should be regarded as an example study of the CMSSM parameter space
for a heavy Higgs boson; qualitatively similar results should hold if the real Higgs boson mass deviates from 125~GeV by a few GeV.
We first  study the parameter space that induces correct $M_H$ and $\Omega_{\rm DM}$.
We discuss the implications of the $(g-2)_\mu$ constraint later.

\begin{table}[t]
\begin{center}
\begin{tabular}{c|cc}
quantity & experiment & Standard Model\\ \hline
$\alpha_3(M_Z)$~\cite{Bethke:2009jm} &  $0.1184 \pm 0.0007$ & parameter\\
$m_t$~\cite{Lancaster:2011wr} & $173.2\pm 0.9$&parameter\\ 
$m_b$~\cite{Nakamura:2010zzi} & $4.19\pm 0.12$&parameter\\ \hline
$\Omega_{\rm DM} h^2$~\cite{Larson:2010gs} & $0.112 \pm 0.0056$ & 0\\
$\delta a_\mu$~\cite{Davier:2010nc} & $(2.8\pm0.8)\times10^{-9}$ & 0\\
BR$(B_d\to X_s\gamma)$~\cite{Misiak:2006zs} & $(3.50\pm0.17)\times10^{-4}$ & $(3.15\pm0.23)\,10^{-4}$ \\
BR$(B_s\to \mu^+\mu^-)$~\cite{CMS-PAS-BPH-11-019} & $<1.1\times10^{-8}$ at 95\%C.L.& $(0.33\pm 0.03)\,10^{-8}$\\
BR$(B_u\to \tau\bar\nu)$/SM~\cite{Buchmueller:2009fn} & $1.25\pm0.40$ & 1\\
\end{tabular}
\end{center}
\caption{\label{susysig:tab}\em Constraints used for the CMSSM analyses.}
\end{table}

\begin{figure}
\begin{center}
\includegraphics[width=0.9\textwidth]{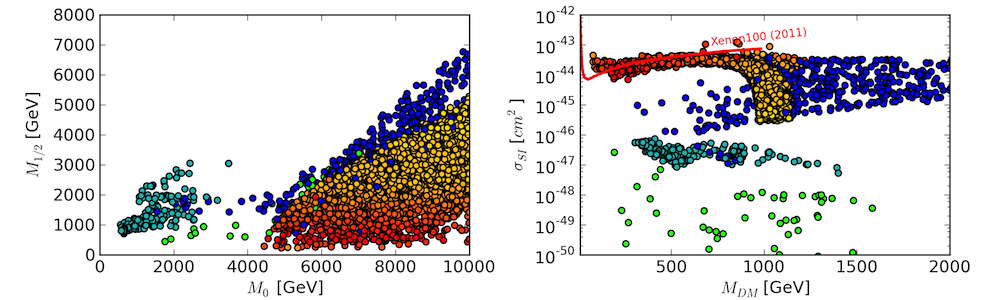}
 \caption{
Points in the CMSSM parameter space yielding $M_H=125\pm 1$~GeV. Colours represent different 
dominant DM freeze-out processes.
Light blue: slepton co-annihilation; green: stop co-annihilation; red to orange: well-tempered neutralino;
yellow: higgsino;  dark blue: heavy Higgs resonances. No $(g-2)_\mu$ constraint is imposed.
}
\label{sig_fig1}
\end{center}
\end{figure}

In Fig.~\ref{sig_fig1} we present our results in scatter plots without the $(g-2)_\mu$ constraint.
In the  left panel the results are presented in the $(m_0, M_{1/2})$ plane and
in the right panel  in the $(M_{\rm DM}, \sigma_{\rm SI})$ plane, where
the first 100 days XENON100 constraint~\cite{Aprile:2011hi} is also shown.
One can identify five distinctive parameter space regions according to the dominant DM annihilation processes~\cite{Kadastik:2011aa}.
\begin{itemize}
\item The light blue points with small  $m_0$ and $ M_{1/2}$
represent the slepton co-annihilation region. They  feature very large values of $\tan\beta.$
Those points represent the best fit value of the CMSSM~\cite{Farina:2011bh}
and have low enough sparticle masses to allow potential SUSY discovery at the LHC. However, their spin-independent direct detection cross section
is predicted to be below $10^{-46}$~cm$^2$ and remains unobservable by XENON100. The present XENON100 experimental bound is 
plotted in the right panel by a solid red line. This is the only parameter region that survives at $3\sigma$ level after the  $(g-2)_\mu$ constraint is
 imposed.

\item The green dots  represent the stop co-annihilation region. Consequently those points have the lowest possible stop mass and, due to the 
mass degeneracy with DM, stops can be long lived and seen as stable very slow particles ($R$-hadrons) at the LHC. 
The feature of those points is an enormous trilinear coupling and very large mixing. In addition, the gluino mass can be 
in reach of the LHC. The spin-independent direct detection cross section
is, unfortunately,  unobservable.

\item The dots represented by continuous colour code from red to orange 
represent the so called well-tempered neutralino, i.e., neutralinos with large bino-higgsino mixing.
The colour varies according to the higgsino component from red (predominantly bino) to yellow (pure higgsino).
Therefore those points can simultaneously have small DM mass  and large DM-nucleon scattering cross sections 
 that can be well tested by XENON100. However, apart from the DM,
all other sparticle masses are predicted to be too heavy to be directly produced at the LHC.

\item The yellow dots around $M_{\rm DM}\sim 1$~TeV represent the pure higgsino DM that is almost degenerate in mass with the chargino.
 The sparticle mass spectrum is predicted to be even heavier than 
in the previous case because the DM scale is fixed to be high.
These points represent the most general and most abundant bulk of the  $M_H=125$~GeV Higgs scenario --
apart from the light DM and heavy Higgs boson there are no other observable consequences because stops can completely decouple.
In our case the 10~TeV bound on stops is imposed only because we did not generate larger values of $m_0.$

\item The dark blue points represent heavy Higgs resonances. Those points are featured by very large values of $\tan\beta$ and  give the heaviest mass spectrum. In essence those points are just smeared out higgsino points due to additional Higgs-mediated processes.
\end{itemize}

\begin{figure}
\begin{center}
\includegraphics[width=0.9\textwidth]{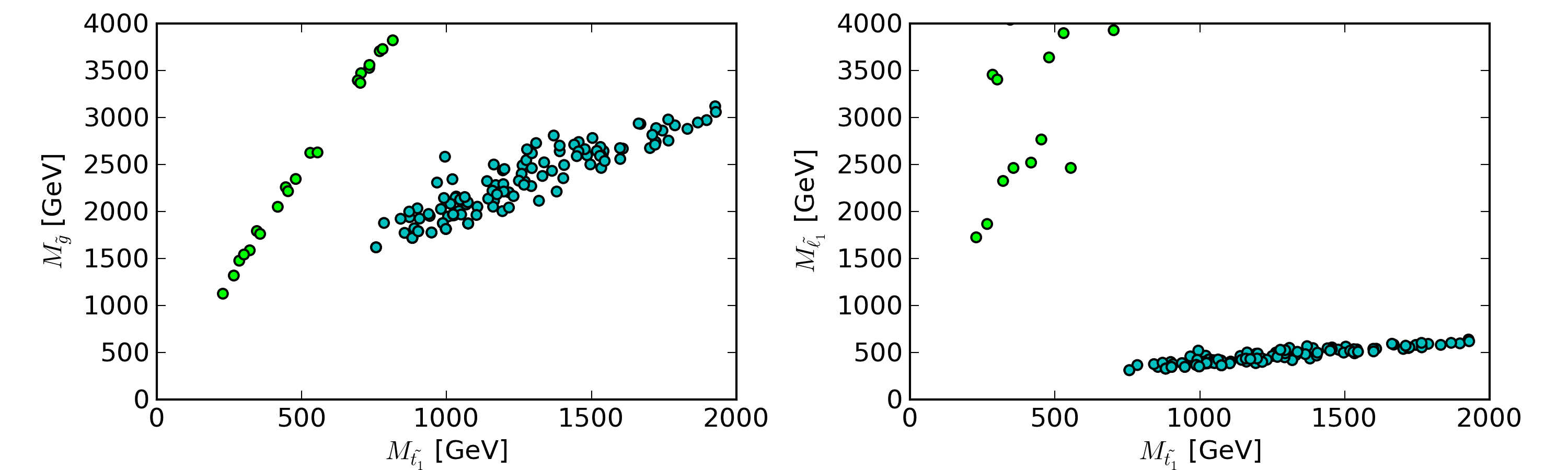}
 \caption{
Scatter plots presenting correlations between physical lightest stop and gluino and lightest slepton  masses. 
}
\label{sig_fig2}
\end{center}
\end{figure}

In order to study the testability of those parameter space regions at the LHC we plot in Fig.~\ref{sig_fig2} 
the physical gluino mass against the lightest stop mass and the lightest slepton mass against the lightest stop mass.
Clearly,  the slepton and stop co-annihilation regions are the only two regions that are of interest for the LHC phenomenology.  
According to Ref.~\cite{Baer:2011aa}
sparticles with those masses may be discovered already at the 7~TeV LHC. Interestingly, due to the stop mass degeneracy with DM
the stops can be long-lived. In this case one must search for $R$-hadrons at the LHC experiments.

\begin{figure}
\begin{center}
\includegraphics[width=0.9\textwidth]{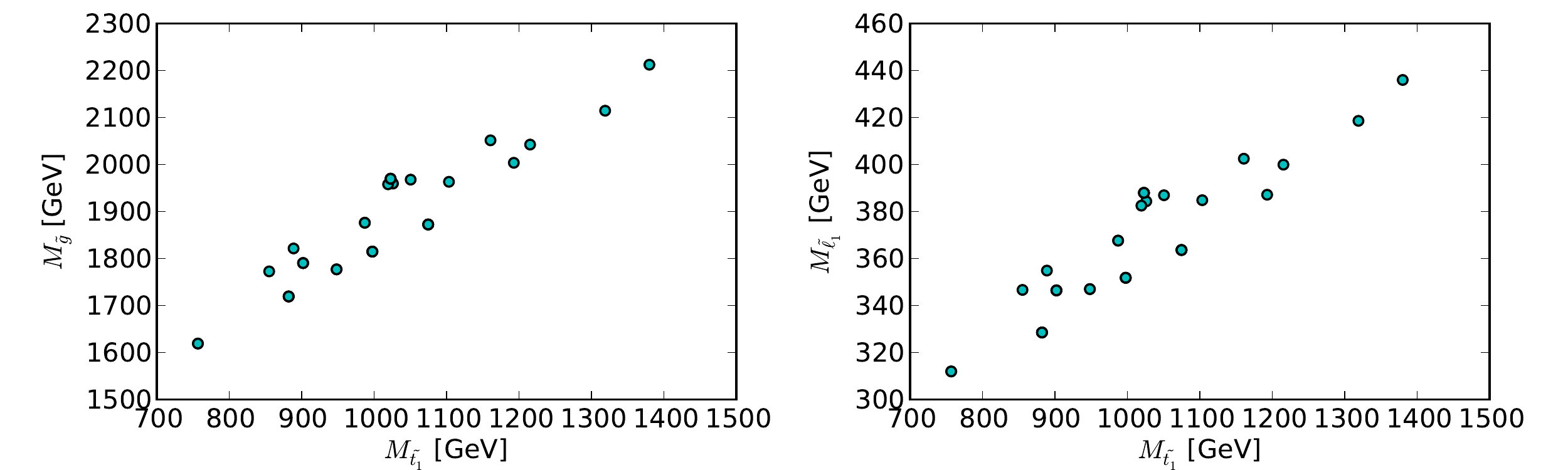}
 \caption{The same as  in Fig.~\ref{sig_fig2} after imposing a $3\sigma$ constraint on the $(g-2)_\mu$ prediction.
}
\label{sig_fig3}
\end{center}
\end{figure}

So far we have ignored the $(g-2)_\mu$ constraint. If we impose a hard $3\sigma$ cut 
on the generated parameter space, only the slepton co-annihilation region survives. The result is plotted in Fig.~\ref{sig_fig3}
where we repeat the content of  Fig.~\ref{sig_fig2} but with the additional $(g-2)_\mu$ constraint. As expected, the observed deviation
in $(g-2)_\mu$ from the SM prediction is hard to explain in SUSY models with heavy spectrum. 
Therefore the two measurements,  $(g-2)_\mu$ and $M_H=125$~GeV,  are essentially in conflict in the CMSSM. 
The conflict is mildest in the slepton co-annihilation case
because of large $\tan\beta$ and the lightest sparticle spectrum. 
 Therefore, for the $M_H=125$~GeV Higgs boson, we predict definite sparticle masses and correlations between them, 
shown in  Fig.~\ref{sig_fig3}, for the LHC. If the CMSSM is realized in Nature and if it contributes significantly to $(g-2)_\mu$,
 the sparticle spectrum is essentially fixed and potentially observable at the LHC.

\section{DISCUSSION AND CONCLUSIONS}
\label{sec:concl}

We studied the CMSSM by scanning over its parameter space allowing the sparticle mass parameters to be very large. 
We required the Higgs boson mass to be in the range $125\pm 1$~GeV.
The first considered case was without attempting to explain $(g-2)_\mu$ in the context of the CMSSM.
 We confirmed that for very large $A$-terms there exists a stop co-annihilation region 
where all DM,  stop and gluino are preferably light. Due to the mass degeneracy between stop and DM the stops can be long lived
resulting in non-trivial LHC phenomenology. The second parameter region that is potentially testable at the LHC is the slepton
co-annihilation region. For all other cases the $M_H\approx 125$~GeV Higgs boson implies very heavy sparticle masses. The exception is, of course,
the DM that can be light due to bino-higgsino mixing even if other sparticles are as heavy as 10~TeV. In this case the CMSSM 
cannot be tested at the LHC but there still is a chance to see the DM scattering off nuclei in the XENON100. 
Those results imply that from the point of view of the LHC phenomenology, the CMSSM sparticle spectrum is known and the LHC
experiments should perform detailed studies of the stop and slepton co-annihilation spectra. Those, however, are the difficult ones
at the LHC because of soft leptons produced in sparticle decays to the DM sparticles.

If, in addition, one attempts to explain  also $(g-2)_\mu$ in this framework, there is immediate tension between the high SUSY scale and the 
large value of the needed $(g-2)_\mu$ contribution. We found that imposing the $(g-2)_\mu$ constraint, only the slepton
co-annihilation region survived at $3\sigma$ level. In this case the CMSSM has a definite predictions of the sparticle masses and spectrum
to be tested at the LHC experiments.

\section*{ACKNOWLEDGEMENTS}
We thank A. Strumia for several discussions.
Part of this work was performed in the  Les Houches 2011 summer institute. This work was supported by the ESF grants 8090, 8499, 8943, MTT8, MTT59, MTT60, MJD140, JD164, by the recurrent financing SF0690030s09 project
and by the European Union through the European Regional Development Fund.

\AddToContent{M. Kadastik, K. Kannike, A. Racioppi, M. Raidal}
\renewcommand{\thesection}{\arabic{section}}



\chapter{Improved vacuum stability in scalar dark matter models}

{\it M. Kadastik, K. Kannike, A. Racioppi, M. Raidal}

\begin{abstract}
If the Higgs boson mass is $125\pm 1$~GeV, the standard model cannot be a fundamental theory up to the GUT scale due to
the vacuum stability arguments. We show that if the dark matter of the Universe consists of stable scalar particles, the
vacuum stability constraints are modified and the theory can be valid up to a high scale. We argue that the present
LHC data may favour such a scenario. 

\end{abstract}

\section{INTRODUCTION}

The scalar potential of the standard model (SM) Higgs boson~\cite{Englert:1964et,Higgs:1964ia,Higgs:1964pj,Guralnik:1964eu}
is one of the best studied objects in particle physics. If the Higgs boson mass is $M_H\approx 125$~GeV as indicated by the recent
experimental results by ATLAS~\cite{ATLAS-CONF-2011-163} and CMS~\cite{Chatrchyan:2012tx} experiments,
the SM cannot be valid up to the Grand Unification (GUT) or Planck scales~\cite{EliasMiro:2011aa,Kadastik:2011aa}.
For such a low Higgs boson mass its quartic self-coupling $\lambda$
runs to negative values at energy scales much below the GUT scale,
causing vacuum instability at higher scales $\Lambda.$  This result is
presented in the left panel of Fig.~\ref{hv_fig1} in which the renormalization
group equation running of the SM Higgs self-coupling is presented at the two loop level for different values of the Higgs mass.
Consecuently, the SM alone cannot be valid up to the scales required by gauge coupling unification and proton decays arguments.

On the other hand, any realistic model of new physics beyond the SM must  explain the existence of dark matter (DM)  in the Universe.
By far the simplest extension of the SM is  obtained by adding  a real~\cite{McDonald:1993ex,Burgess:2000yq,Barger:2007im} 
or complex~\cite{Barger:2008jx,Kadastik:2009dj} singlet scalar field to  the SM scalar potential. 
In addition to minimality, such a SM extension is motivated by GUT arguments~\cite{Kadastik:2009dj}.
In those models the DM and Higgs sectors are related via the Higgs portal and the scalar potentials are in general rather complicated. 
Due to  new self-interactions in the scalar sector, the SM Higgs quartic coupling renormalization is modified, and  one might expect that 
the triviality $\lambda (\Lambda)=0$ may be achieved for much higher values of $\Lambda.$

In addition to the vacuum stability arguments, the scalar DM may also be motivated by the LHC data.
In the light of recent XENON100 and LHC results, this possibility has been studied in Refs.~\cite{Farina:2011bh,Mambrini:2011ik,Raidal:2011xk}.
It was argued in Ref.~\cite{Raidal:2011xk} that the LHC data may
contain indications for Higgs boson invisible decays consistent with the direct DM 
searches in XENON100. Thus it is necessary to study the SM-like Higgs boson properties in every possible production and decay channel.

The aim of this note is to point out that, indeed, if the DM of the Universe consists of scalar singlets, the stability of scalar potential of the 
model may be guaranteed up to the GUT scale even for the Higgs boson mass 125~GeV.  Thus, in this case, there is no 
reason to introduce any intermediate scale in the theory and the known physics can be considered to be fundamental 
until Grand Unification takes over.

\section{VACUUM STABILITY IN SCALAR DARK MATTER MODELS}
\label{sec:hv_cmssm}

\begin{figure}
\begin{center}
\includegraphics[width=0.45\textwidth]{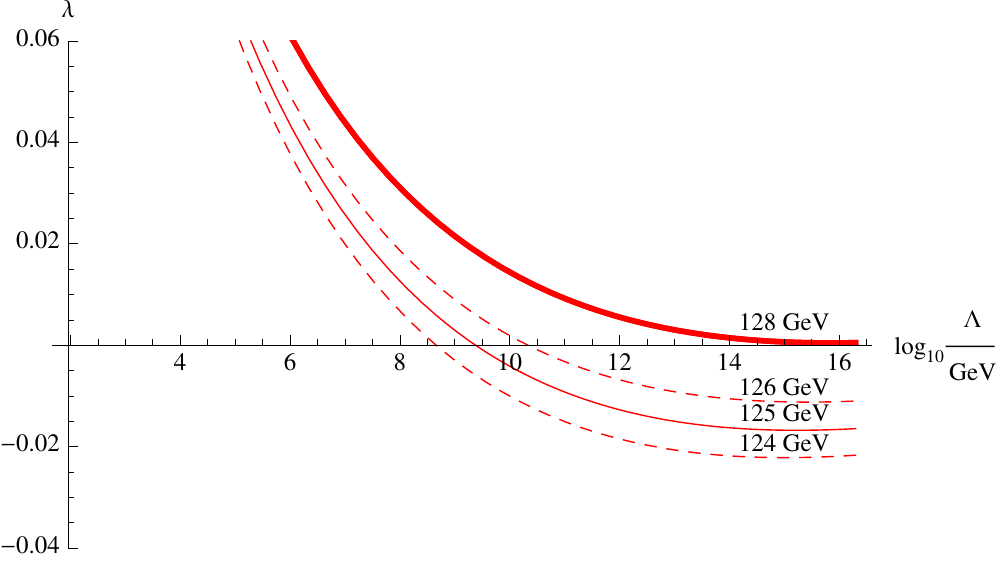}
\includegraphics[width=0.45\textwidth]{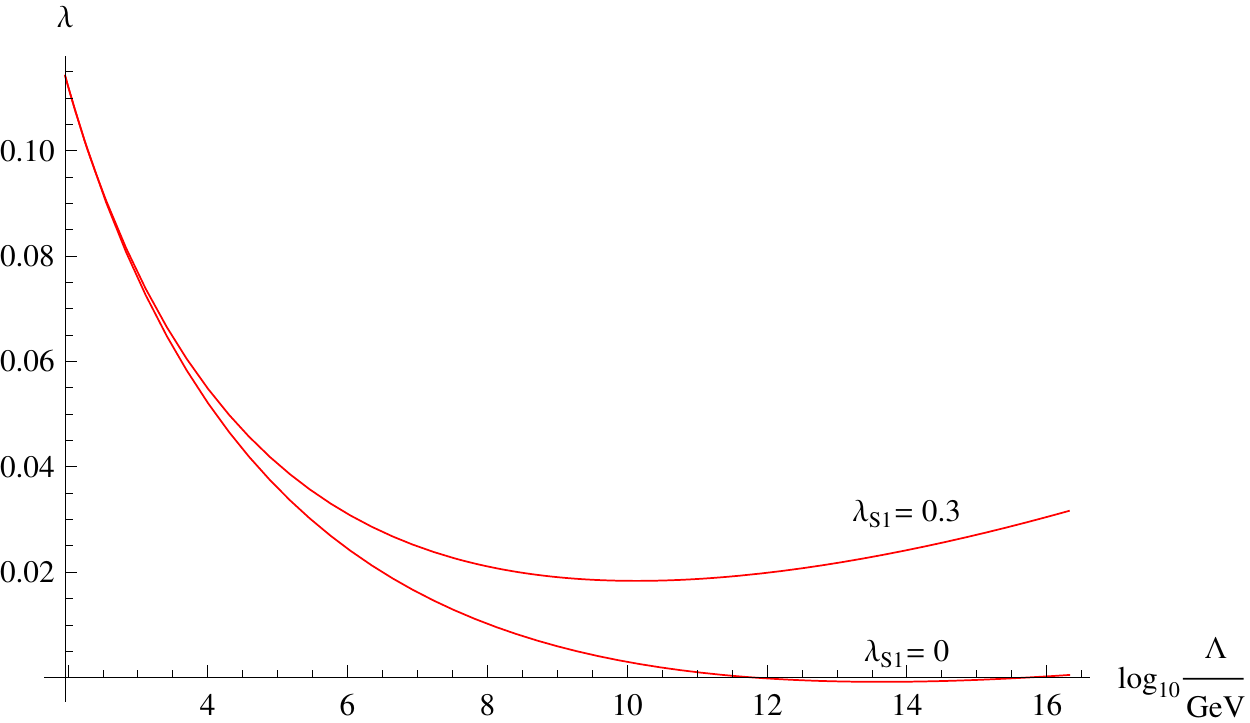}
 \caption{
Left: running of the SM Higgs boson self-coupling $\lambda$ for
different Higgs boson masses at the two loop level.
Right: running of the Higgs self-coupling in the complex singlet model for two different values of $\lambda_{S1}$ at one loop level. 
}
\label{hv_fig1}
\end{center}
\end{figure}

The natural question to ask is that what happens to the vacuum stability in models with extended scalar sector?
Here we study the case of   complex singlet scalar $S = (S_H + i S_A)/\sqrt{2}$, but the phenomenology in the real singlet case is similar. 
 The vacuum stability of the real singlet model has been previously studied also in \cite{Gonderinger:2009jp}. 
 
 Denoting the SM Higgs boson with $H_1,$ the most general Lagrangian invariant under the $Z_2$ transformations $H_1 \to H_1$, $S \to -S$  is
 given by 
\begin{equation}
\begin{split}
    V &= \mu_{1}^{2} H_{1}^\dagger  H_{1} + \lambda_{1} (H_{1}^\dagger  H_{1})^{2}
    + \mu_{S}^{2} S^\dagger  S + \frac{\mu_{S}^{\prime 2}}{2} \left[ S^{2} + (S^\dagger )^{2} \right]
    + \lambda_{S} (S^\dagger  S)^{2} + \frac{ \lambda'_{S} }{2} \left[ S^{4} + (S^\dagger )^{4} \right] \\
    &+ \frac{ \lambda''_{S} }{2} (S^\dagger  S) \left[ S^{2} + (S^\dagger )^{2} \right] + \lambda_{S1}( S^\dagger  S) (H_{1}^\dagger  H_{1}) + \frac{ \lambda'_{S1} }{2} (H_{1}^\dagger  H_{1}) \left[ S^{2} + (S^\dagger )^{2} \right].
\end{split}
\label{eq:Vsing}
\end{equation}
The vacuum stability conditions for the complex singlet model with a global $U(1)$ are given in \cite{Barger:2008jx}. However, those conditions are not 
applicable here because this model is far too simple compared to the general case \eqref{eq:Vsing}.  For the general model the full vacuum stability 
conditions are rather complicated and have been addressed previously in Ref.~\cite{Kadastik:2009cu}.  
However, the conditions of  \cite{Kadastik:2009cu} turn out to be too restrictive because they are derived by requiring the matrix of 
quartic couplings to be positive.  This is required only if the coefficients of biquadratic terms are negative and, in general, cut out some
of the allowed parameter space.  The milder conditions arising from pure quartic terms of the potential \eqref{eq:Vsing} 
are
\begin{equation}
	\lambda_1 \geq 0, \quad \lambda_{S} + \lambda'_S \geq |\lambda''_S|.
\label{eq:vacuum:stability:S:quartic}
\end{equation}
For simplicity we consider in addition only the case that the coefficents of the terms biquadratic in real fields (\emph{e.g.} the coefficient of $S_H^2 S_A^2$) are all non-negative, giving
\begin{equation}
	\lambda_S - 3 \lambda'_S \geq 0, \quad \lambda_{S1} 
	-  |\lambda'_{S1}| \geq 0.
\label{eq:vacuum:stability:S:all:biquadr:pos}
\end{equation}
Doing this, we exclude a part of the points that would be allowed by the full vacuum stability conditions. However, this is sufficient for our purposes because
our aim is  to show that  regions of the parameter space exist  that lower the SM Higgs boson mass vacuum stability bound.

The one-loop renormalization group equations (RGEs) can be obtained from those in \cite{Kadastik:2009cu} by setting all couplings of the inert doublet to zero.
The RGEs show that nonzero $\lambda_{S1}$ or $\lambda'_{S1}$ give a positive contribution to the $\beta$-function of $\lambda_1$, pushing the scale where $\lambda_1 \equiv \lambda = 0$ higher. For qualitative understanding of the model, we let $\lambda_S = \lambda'_S = \lambda''_S = \lambda'_{S1} = 0$. 

In the right panel of Fig.~\ref{hv_fig1} we plot the  one loop level
running for a $125$~GeV Higgs boson quartic coupling for
$\lambda_{S1} = 0$ (the SM case) and for $\lambda_{S1} = 0.3$. In the
latter case, the minimum bound on the Higgs boson mass from the vacuum stability argument is lowered and the vacuum can be stable up to the GUT or Planck scale. This simple example demonstrates that the 125~GeV Higgs boson excess at the LHC may call for
the extended scalar sector that simultaneously solves the vacuum stability and DM problems.

\section{CONCLUSIONS}
\label{sec:hv_concl}

 We have considered the implications of  a $M_H\approx 125$~GeV Higgs boson for the vacuum stability in scalar DM models.
 In the SM the 125~GeV Higgs boson implies that the model cannot be consistent with the gauge coupling unification scale nor with the 
 scale required by proton stability.
 We have shown here that in the case of non-SUSY scalar DM models  the vacuum stability bounds are modified due to more complicated 
scalar potentials. Therefore, unlike the SM,   the scalar DM models can be consistent models up to the GUT or Planck scales
even if the present LHC excess at $M_H\approx 125$~GeV is be confirmed.
If this is the case Nature has chosen,  the Higgs boson interactions and possibly also its branching fractions at the LHC will be modified due to the
coupling to DM.

\section*{ACKNOWLEDGEMENTS}
Part of this work was performed in the  Les Houches 2011 summer institute. This work was supported by the ESF grants 8090, 8499, 8943, MTT8, MTT59, MTT60, MJD140, JD164, by the recurrent financing SF0690030s09 project
and by the European Union through the European Regional Development Fund.

\AddToContent{M. Kadastik, K. Kannike, A. Racioppi, M. Raidal}
\renewcommand{\thesection}{\arabic{section}}


\chapter{Higgs boson production via new heavy vector-like top
decays at the LHC}

{\it A.~Azatov, O.~Bondu, A.~Falkowski,
M.~Felcini, S.~Gascon-Shotkin,
D.~K.~Ghosh, G.~Moreau, A.~Y.~Rodr\'iguez-Marrero, S.~Sekmen}



\begin{abstract}
Motivated by the recent severe constraints over a wide mass
range from the direct Higgs boson searches at LHC,
we first build the minimal model of vector-like quarks where the
rate for the gluon fusion mechanism can be greatly suppressed.
Within the obtained model, compatible with the present LHC
constraints on the direct Higgs search, we demonstrate that the
indirect
Higgs production, $pp \to t' \bar t'$, $t'\to t h$, allows to
discover the Higgs boson and measure several of its possible
masses
through the decay channels $h\to \gamma\gamma$, $h\to WW, ZZ$.
We also comment on the recent hint in LHC data
from a possible $\sim 125$~GeV Higgs, in the presence of $t'$
quarks.
\end{abstract}

\section{INTRODUCTION}

The primary issue of the present Large Hadron Collider (LHC) is
certainly the direct search for
the still unobserved cornerstone of the Standard Model (SM),
namely the Higgs boson or any signal from alternative
Electro-Weak Symmetry Breaking (EWSB) mechanisms. The SM is
probably not the ultimate model of nature and
it is clear that new channels for the Higgs production,
that can arise in extensions of the SM, would have a profound
impact on the discovery potential of the Higgs boson at the LHC.
An attractive possibility is the Higgs production in decays of
additional heavy colored particles as those can be copiously
pair produced
at the LHC via strong interactions.

Within the well-motivated theories beyond the SM, there are some
candidates for such new heavy colored states,
extra quarks with vector-like couplings, whose existence is
predicted by most of the alternatives to supersymmetry.
In this context, to maintain a naturally
light Higgs boson, divergent quantum corrections from loops of
the top quark are often canceled by top-partner
contributions~\cite{ArkaniHamed:2001nc,ArkaniHamed:2002qx,ArkaniHamed:2002qy}.
Let us describe important examples here. 
In the so-called little Higgs scenarios, the vector-like quarks
arise as partners of the SM fields being promoted to larger
multiplets. 
In composite
Higgs~\cite{Contino:2003ve,Agashe:2004rs,Agashe:2005dk,Contino:2006qr,Burdman:2007sx}
and composite
top~\cite{Contino:2003ve,Agashe:2004rs,Agashe:2005dk,Contino:2006qr,Burdman:2007sx,Hill:1991at}
models, the vector-like quarks
are excited resonances of the bound states constituting the SM
particles.
In extra-dimensional models with (SM) quarks in the bulk,
the vector-like quarks are Kaluza-Klein (KK)
excitations of these
bulk fields~\cite{Cheng:1999bg} like in the Gauge-Higgs
unification mechanism (see e.g.
Ref.~\cite{Carena:2006bn,Carena:2007ua})
or in the Randall-Sundrum (RS)
scenario~\cite{Randall:1999ee,Gogberashvili:1998vx,Gherghetta:2000qt}
-- where some of those KK excitations, the so-called custodians,
can be as light as a few hundred's of
GeV~\cite{Agashe:2003zs,Bouchart:2008vp,Djouadi:2006rk,Djouadi:2007eg,Agashe:2006at,Bouchart:2009vq}.
Another example is a gauge coupling unification theory where
vector-like quarks
are embedded into the simplest ${\rm SU(5)}$
representations~\cite{Kilic:2010fs}.

Vector-like quarks with same electric charge as the up-type quarks are often called top-partners (noted
$t'$)
as these new heavy states mix in general predominantly with the
top quark -- due to the large top mass and to the related
feature that the top
quark is in general more intimately connected to ultraviolet
physics, like e.g. in composite Higgs models.
A $t'$ can also be called a top-partner in the sense that it is
contained in the same group representation
as the top quark with respect to symmetries, like the
approximate global symmetry of the little Higgs
models~\cite{ArkaniHamed:2001nc,ArkaniHamed:2002qx,ArkaniHamed:2002qy},
the gauge unification symmetry~\cite{Kilic:2010fs} or
the custodial symmetry of RS versions with bulk
matter~\cite{Agashe:2003zs,Bouchart:2008vp,Djouadi:2006rk,Djouadi:2007eg,Agashe:2006at,Bouchart:2009vq}
(explaining the SM fermion mass
hierarchies~\cite{Huber:2000ie,Huber:2001ug,Huber:2002gp,Huber:2003sf,Chang:2005ya,Moreau:2006np,Moreau:2005kz,Agashe:2004cp,Agashe:2004ay,Agashe:2006wa,Agashe:2006iy,Grossman:1999ra,Appelquist:2002ft,Gherghetta:2003he,Moreau:2004qe}).

At this level, one must mention that
searches for direct production of
vector-like quarks at the LHC have been considered from
a model-independent point of view in
Ref.~\cite{Cacciapaglia:2010vn,Cacciapaglia:2011fx,Gopalakrishna:2011ef,AguilarSaavedra:2005pv,AguilarSaavedra:2009es}
but also in specific frameworks such as the little Higgs models
(versions sufficiently safe from EW precision
constraints)~\cite{Burdman:2002ns,Han:2003wu,Perelstein:2003wd,Cheng:2005as,Cacciapaglia:2009cu}
or the composite Higgs
hypothesis~\cite{Vignaroli:2011ik,Contino:2008hi}
and the dual RS
context~\cite{Dennis:2007tv,Agashe:2004bm,Brooijmans:2010tn,Carena:2006bn}.
These past searches focus generally on the discovery of the
vector-like quarks, rather than using these extra quarks
to enhance the discovery and identification potential for other
unknown particles such as Higgs scalars.

In relation to Higgs detection, there exist studies utilizing the possible Higgs production through vector-like quark decays,
as described in the following. Indeed, it is well known since some time~\cite{delAguila:1989rq,delAguila:1989ba}
that vector-like quark production could be a copious source of Higgs bosons (Higgs factory). 
\\ Relatively light Higgs bosons produced from the decay of top-partners 
can be highly boosted and good candidates for analyses based
on jet substructure.  This method
has been applied~\cite{Kribs:2010ii} for a $\sim 130$~GeV Higgs
decaying to $b\bar b$ at the $14$~TeV LHC, in order to identify with good efficiency the jets from Higgs decays.
In the simple model considered there,
the $t'$ is a singlet under the ${\rm SU(2)_L}$ gauge group,
which determines the $t'$ couplings and its branching fractions (at tree level) for the decays into the Higgs boson and the two EW gauge bosons $t'\to th$, $t'\to tZ$, $t'\to Wb$.
\\ The top-partner can also be singly produced which leads to
different final states as compared to the pair production; because of the
phase space suppression, the single production becomes competitive with the pair
production at a high $t'$ mass, depending upon the considered model
(since the involved $t'$ couplings to $h,Z^0,W^\pm$ are fixed by
the $t'$ quantum numbers)~\footnote{The single production
becomes dominant
typically around $m_{t'}\sim 700$~GeV as in the scenario with a
singlet $t'$~\cite{AguilarSaavedra:2005pv} or in little Higgs
models~\cite{Azuelos:2004dm}.}.
The reconstruction of the Higgs boson produced in the $t'$
decay, itself singly produced at the $14$~TeV LHC, was studied
in Ref.~\cite{Azuelos:2004dm}
assuming the Higgs mass known (to be $120$~GeV) and focusing on
the channel $h\to b\bar b$ -- with the
combinatorial background only. 
This was performed for a singlet $t'$ in the ``Littlest Higgs''
model with the asymptotic branching ratio values of the high
$m_{t'}$ regime: $B_{t' \to t h}=25\%$, $B_{t' \to t Z}=25\%$,
$B_{t' \to b W} =50\%$ (from the EW equivalence theorem).
\\ Similarly, a vector-like colored $b'$ state produced at the
$14$~TeV LHC can act as a `Higgs factory' thanks to its decay
$b'\to bh$.
It was shown~\cite{Kilic:2010fs} that a Higgs mass
reconstruction can be obtained with a limited accuracy,
concentrating on the decay $h\to WW$
($W\to l\nu$) for $m_h=200$~GeV and assuming the $m_{b'}$ value
to be deduced from a preliminary analysis based on the more
appropriate
channel $b'\to bZ$. Theoretically, the $b'$ was originating from
the upper component of a ${\rm SU(2)_L}$ doublet so there was no
channel $b'\to t W$.   
\\ Higgs mass reconstructions via $t'$ and $b'$ decays were also
studied for the $14$~TeV LHC, based on a light Higgs decaying to
$b\bar b$
in the basic models with a unique extra $t'$ and/or a unique
extra $b'$~\cite{AguilarSaavedra:2009es}.

In the present paper, we use the pair production and decay of a
vector-like top to develop new searches of the Higgs boson at
the  LHC and to study other channels for the Higgs mass reconstruction
like $h\to\gamma \gamma$ (diphoton) and  $h\to ZZ$.
The original theoretical model considered here, including two
top-partners, is constructed to allow interesting
interpretations correlating the
indirect (via vector-like top decay) and direct Higgs production searches at the
LHC, as described in the following.
A few characteristic parameter sets -- with vector-like top mass in the range between $\sim$~400 and 800~GeV
-- are chosen as benchmark points avoiding large $t'$ contributions to the Higgs rates (constrained by present LHC data) and simultaneously allowing for significant branching fraction values ($>10\%$) of the vector-like top decay to the Higgs boson.  
\\ Assuming the presence at low-energy scales only of extra
vector-like quark multiplets containing some $t'$, we have
elaborated a minimal model allowing to strongly suppress the Higgs production via gluon fusion, as compared to the SM. 
In this simple but non-trivial model, the $gg\rightarrow h$ cross section suppression factor possibly reaches values
below $10^{-1}$ at hadron colliders; this is to be put in contrast with 
the $t'$ representations taken usually in the RS scenario 
\cite{Bouchart:2009vq,Djouadi:2007fm,Djouadi:2009nb,Djouadi:2011aj,Azatov:2010pf}~\footnote{$t'$
representations (to which SM fields are promoted)
with analog rate suppression effects can arise with the ${\rm
O(3)}$ subgroup
\cite{Agashe:2006at} implementable in the composite Higgs
model~\cite{Falkowski:2007hz,Espinosa:2010vn}
and RS scenario~\cite{Ledroit:2007ik,Casagrande:2010si} which
can reach strong suppressions,
respectively of $\sigma_{\rm gg\to h}/\sigma_{\rm gg\to h}^{\rm
SM}\sim 35\%$ and $\sim 10\%$.}
and with minimal supersymmetric theories for which such a
suppression is not possible to get (see respectively
Ref.~\cite{Djouadi:2007fm,Bouchart:2009vq} and
Ref.~\cite{Djouadi:1998az}). The chiral case of a fourth quark
generation can even only increase
considerably the gluon fusion rate.
\\ The minimal $t'$ model suggested here is interesting in the
sense that it can easily lead to the following interpretation:
for example, a $255$~GeV Higgs is excluded in the SM by the present LHC
results~\cite{CMSweb,ATLASweb} but can still
exist in the above minimal SM extension with $t'$ where the
reduced Higgs production cross section can be below the LHC
upper limits.
In other words, the Higgs boson would really be light but not
detectable with the present luminosity/energy, via conventional
channels.
A channel that could then allow the Higgs discovery would be 
through the $t'$ pair production and decays, as illustrated in
this paper.
Another possible hypothesis is that an excess of events, like the
recent hint in data explainable by a $125$~GeV Higgs~\cite{ATLASweb},
would be confirmed soon at the LHC.
Then the measured Higgs production cross section times branching
ratios could certainly be exactly reproduced by the present $t'$
model, given the parameter freedom in this model and its capability of
inducing large Higgs rate corrections of both signs.
Then investigating such a new Higgs production channel as the
$t'$ decay developed here, would of course be relevant in particular to
confirm the Higgs existence.
Finally, in the case of a signal from a heavy Higgs, say  
above $500$~GeV~\footnote{Such 
a heavy Higgs would be neither SM-like, as disfavored by the EW
precision tests,
nor belonging to a supersymmetric extension, as forbidden by the
Higgs sector structure.
However, it could perfectly be e.g. in a RS scenario where its
contributions to the oblique T parameter can be
compensated by new KK-induced contributions.}
as we will consider here, the same fit of Higgs data would be
instructive as a test of the present $t'$ model and similarly
the $t'$ decay should be considered as a complementary channel
of Higgs production.

\section{THE THEORETICAL MODEL}

Let us assume the presence at low-energy scales of a unique
additional vector-like quark multiplet including a $t'$
component.
Then, irrespective of the representation of this multiplet under the
${\rm SU(2)_L}$ gauge group (i.e. whether the $t'$ belongs to a
singlet, doublet,\dots), the interference between the next heavier top mass eigenstate $t_2$
[composed of $t$, $t'$, $t''$, with $t_1$ being the SM top quark] and the $t_1$ 
contributions~\footnote{The top quark exchange in the loop is
the dominant contribution in the SM.}
to the triangle loop of the gluon-gluon 
fusion mechanism will be systematically constructive. This is
due to the fact that the physical sign of the Yukawa coupling
and mass insertion involved in this loop -- two chirality flips
are necessary -- will be systematically identical giving rise
to a positive product (for $t_1$ as well as for $t_2$). Hence,
the cross section of the gluon fusion mechanism 
may be increased or slightly decreased 
(because of a possible $t$ Yukawa coupling reduction) relatively to the SM case.
\\ To get the minimal scenario with only additional vector-like
quark multiplet including $t'$ components able to strongly
suppress gluon fusion, one needs to introduce a first
top-partner $t'$ in a ${\rm SU(2)_L}$ doublet as well as
a second top-partner $t''$ in a singlet. In order to simplify the presentation, we
do not consider the doublet including a
$b'$~\footnote{Similar results are expected in such a
situation.} that would also be exchanged
in the triangle loop. So we end up with the doublet
$(q_{5/3},t')$, $q_{5/3}$ being an exotic quark with electric
charge $5/3$
and without self-Yukawa coupling (thus with no contribution in the loop).
Indeed, with this field content, all the possible generic mass
terms and Yukawa couplings appearing in the Lagrangian are,
\begin{equation} 
{\cal L}_{\rm Yuk.} =  
Y \overline{\left ( \begin{array}{c}
t \\  b 
\end{array} \right )}_{L}
H^\dagger t^c_R +
Y' \overline{\left ( \begin{array}{c}
q_{5/3} \\  t' 
\end{array} \right )}_L
H t^c_R +
Y'' \overline{\left ( \begin{array}{c}
q_{5/3} \\  t' 
\end{array} \right )}_{L/R}
H t''_{R/L} + 
\tilde Y \overline{\left ( \begin{array}{c}
t \\ b 
\end{array} \right )}_L
H^\dagger t''_R 
\nonumber
\end{equation}
\begin{equation} 
+ \ Y_b \overline{\left ( \begin{array}{c}
t \\  b 
\end{array} \right )}_{L}
H b^c_R 
+ m \ \bar t''_L t^c_R
+ m' \overline{\left ( \begin{array}{c}
q_{5/3} \\  t' 
\end{array} \right )}_{L}
\left ( \begin{array}{c}
q_{5/3} \\  t' 
\end{array} \right )_{R}
+ m'' \ \bar t''_{L} t''_{R}
+ {\rm H.c.}
\label{VTH:LagDoub}
\end{equation}
where $H$ represents the SM Higgs doublet and $L/R$ the fermion
chiralities.
By construction, the vector-like quarks possess the same quantum
numbers 
for the left-handed and right-handed states.
We have not written the Yukawa couplings for the first two quark
generations as their mixings with the top-partners $t',t''$
are negligible compared to the $t$-$t'$-$t''$ mixing and the CKM
mixing angles are typically small, so that the first two quark generations
are decoupled from $b,t,t',t''$.
Note that a field redefinition rotating $t^c_R$ and $t''_R$ can
allow to eliminate the $m$ term without loss of generality.
A last remark is that the $Y''$ term could be split in two terms
with different chiralities and coupling constants.
The Lagrangian~(\ref{VTH:LagDoub}) gives rise, after EWSB, to
this top mass matrix:
\begin{equation} 
{\cal L}_{\rm mass} =
\overline{\left ( \begin{array}{c} t \\ t' \\ t'' \end{array}
\right )}_L
\left ( \begin{array}{ccc}  
Yv & 0 & \tilde Y v\\  
Y' v & m' & Y'' v\\ 
m & Y'' v & m''
\end{array} \right )
\left ( \begin{array}{c} t^c \\ t' \\ t'' \end{array} \right )_R+ {\rm H.c.}
\label{VTH:MassDoub}
\end{equation}
with $v\simeq 174$~GeV the SM vacuum expectation value of the
Higgs boson.
In our notations of Eq.(\ref{VTH:MassDoub}), the parameters $Y$,
$Y'$, $Y''$ and $\tilde Y$
contain the whole sign (i.e. the combination of the ${\rm
SU(2)_L}$ contraction signs and Yukawa coupling constant signs).
Note that vector-like fermions do not require EWSB to acquire
mass.
The non-trivial consequence of the present $t',t''$ charge
assignment choice is the presence of Yukawa terms in the
block diagonal matrix of Eq.(\ref{VTH:MassDoub}) associated to
the top-partners~\footnote{Namely the last two lines and columns
of this
mass matrix.} (such Yukawa matrix elements would be absent in
the first case of a unique top-partner). This feature of the
mass structure
allows for a strong suppressions of the gluon fusion mechanism. 
In particular, the own top-partner ($t',t''$) Yukawa coupling
($Y''$) sign can be chosen independently of the top ($t$) Yukawa
coupling
($Y$) sign in order to generate a destructive interference
between the top and top-partner loops.

\section{$t_2$ RATES AND DIRECT CONSTRAINTS}

We consider the model described in the previous section and
concentrate on the phenomenology of the lighter top eigenstate
$t_2$; the $t_3$ eigenstate production is subdominant given its higher mass. In a
second stage, one should add the contributions to the Higgs
production
from the $t_3$ decays like $t_3\to t_1 h$ or $t_3\to t_2 Z$. 

In Table~\ref{VTH:DoubletTable}, we define our benchmark points
by the values of the fundamental parameters -- including the
Higgs mass -- and the corresponding $m_{t_2}$, $m_{t_3}$ values.
Those sets of parameters are selected
in particular to have large $B_{t_2\to t_1 h}$ increasing the
studied Higgs signal. Note that in the pure model with a unique
doublet $(q_{5/3},t')$,
$B_{t_2 \to b W}$ is negligible compared to $B_{t_2 \to t_1 h}$
and $B_{t_2 \to t_1 Z}$~\cite{AguilarSaavedra:2009es}. For none
of the considered benchmark points,
the channel $t_2 \to q_{5/3} W$ is open. The table also provides
the theoretical $t_2$ widths and the $\sigma_{\bar t_2t_2}$
cross sections for the $t_2$ pair production at LHC computed
with the HATHOR program~\cite{Aliev:2010zk} at NNLO.
As a comparison, we give in Table~\ref{VTH:DoubletTable} the
expected SM cross
sections~\cite{LHCHiggsCrossSectionWorkingGroup:2011ti}
for Higgs production via gluon fusion $\sigma^{\rm SM}_{\rm gg\to h}$.
It is physically important to note that the branching ratios $B_{t_2 \to t_1 h}$ and $B_{t_2 \to t_1 Z}$ are non-vanishing in contrast
with the case of a fourth generation $t'$ quark so that the observation of such decays (discussed in Section \ref{VTH:exp-analysis}) 
would even prove the vector-like nature of the $t'$ quarks.

The table presents finally the CMS constraints on the
observables $\sigma_{\bar t_2t_2} B^2_{t_2}$
derived from the search for pair production of a heavy top-like quark~\cite{CMSweb,Chatrchyan:2011ay,CMS-PAS-EXO-11-051}
(present bounds from ATLAS are obtained with a lower integrated
luminosity, 37 pb$^{-1}$, and are thus less stringent~\cite{ATLASweb}).
It appears that the corresponding theoretical values, predicted
in the models considered here, respect those experimental limits
for $m_{t_2}$ as low as
$\sim 400$~GeV~\footnote{We have also checked that the Tevatron
constraints are satisfied.}.
Increasing theoretically $B_{t_2\to t_1 h}$ goes in the direction of reducing these
constraints.

Due to these lower constraints on $m_{t_2}$ typically around
$400$~GeV, the $t_2$ pair production suffers from a significant
phase space suppression so that
the whole rate for a single Higgs production through the $t_2$
decay is smaller than for the usual gluon fusion mechanism;
there is e.g. a factor of $\sim 10$ for the
point A at $14$~TeV as shown the Table~\ref{VTH:DoubletTable}.
However, the number of Higgs events coming from the $t_2$ decay
can be significant at high
LHC luminosities; this Higgs production channel can thus be
interesting and especially in the case where the gluon fusion
mechanism is strongly suppressed
by the presence of $t', t"$ states, as occurs for instance with
the point B (see all the rates at $14$~TeV in the table).

\begin{table}[!ht]
\vspace*{.5cm}
\begin{center}
\begin{tabular}{|c|c|c|c|c|}
\hline  
Parameter Set & A & B & C & D  \\
\hline 
\hline
$Y$ / $\tilde Y$  & -1.43 / 2 & 1.15 / 0.4 & 1.12 / -0.5 & 1.05 / -0.3 \\
\hline  
$Y'$ / $Y''$  & 1.85 / -1 & -1.5 / 1.6 & 1.1 / 1.65 & 1.7 / 1.9 \\
\hline  
$m$ / $m'$ (GeV) & 0 / 370 & 0 / 770 & 0 / 810 & 80 / 1100 \\
\hline  
$m''$ (GeV) & 510 & 980 & 850 & 1100 \\
\hline 
\hline  
$m_{t_3}$ (GeV) & 722 & 1181 & 1125 & 1454 \\
\hline  
$m_{t_2}$ (GeV) & 403 & 626 & 572 & 788 \\
\hline  
$m_{h}$ (GeV) & 125 & 255 & 320 &  540 \\
\hline  \hline  
$\sigma^{\rm NNLO}_{\rm gg\to h}$ (pb) @ 7 TeV & 15.31 & 3.18 &
2.25 & 0.58 \\
\hline 
$\sigma^{\rm NNLO}_{\rm gg\to h}$ (pb) @ 14 TeV & 49.85 & 13.50
& 10.59 & 3.85 \\
\hline
$\sigma^{\rm t'}_{\rm gg\to h}/\sigma_{\rm gg\to h}^{\rm SM}$ &
1.27 & 0.45 & 0.40 & 0.65 \\
\hline
$\sigma^{\rm NLO}_{\bar t_1t_1h}$ (pb) @ 7 TeV & 0.0194 & 0.0037
& 0.0016 & $7 \ 10^{-4}$ \\
\hline 
$\sigma^{\rm NLO}_{\bar t_1t_1h}$ (pb) @ 14 TeV & 0.138 & 0.036
& 0.021 & 0.015 \\
\hline 
$\sigma^{\rm NNLO}_{\bar t_2t_2}$ (pb) @ 7 TeV & 1.361 & 0.0709
& 0.136 & 0.0115 \\
\hline 
$\sigma^{\rm NNLO}_{\bar t_2t_2}$ (pb) @ 14 TeV & 13.53 & 1.164
& 1.975 & 0.284 \\
\hline    
$B_{t_2 \to t_1 h}$ (\%) & 62.6 & 60.8 & 13.5 & 43.0 \\ 
\hline  
$B_{t_2 \to t_1 Z}$ (\%) & 28.6 & 25.0 & 46.1 & 40.3 \\
\hline
$B_{t_2 \to b W}$ (\%) & 8.8 & 14.2 & 40.4 & 16.6 \\
\hline
$\Gamma_{t_2}$ (GeV) & 4.4 & 19.8 & 6.5 & 8.8  \\
\hline \hline
$\sigma_{\bar t_2t_2} B^2_{t_2 \to b W}$ (pb) & 0.01 & 0.001 &
0.022 & 0.000(3) \\
LHC bound & $<$ 0.6 & $\times$ & $<$ 0.597 & $\times$    \\  
\hline
$\sigma_{\bar t_2t_2} B^2_{t_2 \to t_1 Z}$ (pb) & 0.11 & 0.004 &
0.029 & 0.002 \\
LHC bound & $<$ 0.45 & $\times$ & $\times$  & $\times$   \\  
\hline \hline
$S$ / $T$ & 0.05 / 0.05 & -0.01 / 0.23 & -0.01 / 0.30 & -0.01 / 0.28\\ 
\hline
\end{tabular}
\end{center}
\caption{Cross sections for the $t_2$ pair production at the
$7$~TeV and $14$~TeV LHC together with the $t_2$ widths
$\Gamma_{t_2}$ and
branching ratios, for characteristic sets of the parameters
[with the induced $m_{t_2}$, $m_{t_3}$ values].
The LHC upper limits and theoretical predictions
for the observables $\sigma_{\bar t_2t_2} B^2_{t_2 \to b W}$,
$\sigma_{\bar t_2t_2} B^2_{t_2 \to t_1 Z}$ are shown just above
the last line
(the crosses indicate the absence of experimental limit at the
associated $m_{t_2}$ values). In the last line are given the
oblique parameters $S$ and $T$ [after subtraction of the SM
contributions so that those only include new physics effects].
For the point A, we have used the indicated $Y''$ value for the
$t''_R$ coupling and $Y''=-0.3$ for the $t''_L$ vertex. Finally,
we give the cross section in the $t'^{(\prime)}$ model for the
$t_1\bar t_1h$ production which is the subleading reaction able
to contribute to the
studied signal with tops in the final state (Higgs production in
$t_2$ decays), given our cuts on the selected events. For
comparison,
we present the cross sections for the main gluon fusion
mechanism of Higgs production at LHC in the SM at NNLO, as well
as the ratio with the rate
in our scenario.}
\label{VTH:DoubletTable}
\end{table}

\section{CONSTRAINTS FROM DIRECT HIGGS BOSON SEARCHES}

\noindent Set {\bf A} - At $m_{h}=125$~GeV, all the sensitive
channels for searching the Higgs boson at hadron colliders are
the decays $h\to \gamma\gamma$, $h\to WW$ (with $W\to \ell\nu$),
$h\to ZZ$ (with $Z\to \ell\bar \ell$), $h\to\tau\bar \tau$ and
$h\to b\bar b$.
The latest bounds on the Higgs boson rates obtained at the LHC
read as,
$\sigma_{\rm gg\to h}B_{\rm h\to \gamma\gamma}/\sigma_{\rm gg\to
h}^{\rm SM}B_{\rm h\to \gamma\gamma}^{\rm SM}$
$< 2$ \cite{CMSweb}, 
$\sigma_{\rm gg\to h}B_{\rm h\to WW}/\sigma_{\rm gg\to h}^{\rm
SM}B_{\rm h\to WW}^{\rm SM}$
$< 1.3$ \cite{CMSweb}, 
$\sigma_{\rm gg\to h}B_{\rm h\to ZZ}/\sigma_{\rm gg\to h}^{\rm
SM}B_{\rm h\to ZZ}^{\rm SM}$
$< 2.2$ \cite{CMSweb},  
$\sigma_{\rm gg\to h}B_{\rm h\to \tau\tau}/$
$\sigma_{\rm gg\to h}^{\rm SM}B_{\rm h\to \tau\tau}^{\rm SM}$
$< 3.2$ \cite{CMSweb} and \\
$\sigma_{\rm gg\to h}B_{\rm h\to bb}/\sigma_{\rm gg\to h}^{\rm
SM}B_{\rm h\to bb}^{\rm SM}$
$< 5$ \cite{CMSweb}. 
These bounds are compatible with the rates calculated taking
into account the $t-t'-t''$ mixing effect on the top quark
Yukawa coupling as
well as the $t_2$ and $t_3$ eigenstate contributions in the
triangular loop of the gluon fusion mechanism,
for our parameter set A. This parameter set yields indeed:   
\begin{equation}
\sigma_{\rm gg\to h}^{\rm t'}B_{\rm h\to \gamma\gamma}^{\rm
t'}/\sigma_{\rm gg\to h}^{\rm SM}B_{\rm h\to \gamma\gamma}^{\rm
SM}=1.16 \ \ ; \ \
\sigma_{\rm gg\to h}^{\rm t'}B_{\rm h\to WW}^{\rm
t'}/\sigma_{\rm gg\to h}^{\rm SM}B_{\rm h\to WW}^{\rm SM}=1.25 \
\ [A]
\label{VTH:SigB1}
\end{equation}
\begin{equation}
\sigma_{\rm gg\to h}^{\rm t'}B_{\rm h\to ZZ}^{\rm
t'}/\sigma_{\rm gg\to h}^{\rm SM}B_{\rm h\to ZZ}^{\rm SM}=1.25 \
\ ; \ \
\sigma_{\rm gg\to h}^{\rm t'}B_{\rm h\to \tau\tau}^{\rm
t'}/\sigma_{\rm gg\to h}^{\rm SM}B_{\rm h\to \tau\tau}^{\rm
SM}=1.25 \ \ [A]
\label{VTH:SigB2}
\end{equation}
\begin{equation}
\sigma_{\rm gg\to h}^{\rm t'}B_{\rm h\to bb}^{\rm
t'}/\sigma_{\rm gg\to h}^{\rm SM}B_{\rm h\to bb}^{\rm SM}=1.25 \
\ [A]
\label{VTH:SigB3}
\end{equation}
The cross section for the Higgs production is enhanced,
$$\sigma^{\rm t'}_{\rm gg\to h}/\sigma_{\rm gg\to h}^{\rm SM}=
1.27 \ \ [A],$$
due to the combination of two possible effects: 
the increase of the $t_1$ Yukawa coupling and the constructive
interferences between the $t_1$ contribution and the $t_2$,
$t_3$ ones.
In contrast, the branching ratio for the decay channel into
diphoton is slightly decreased,
$$B_{\rm h\to \gamma\gamma}^{\rm t'}/B_{\rm h\to
\gamma\gamma}^{\rm SM}= 0.91 \ \ [A].$$ But the resulting
product
$\sigma_{\rm gg\to h}^{\rm t'}B_{\rm h\to \gamma\gamma}^{\rm
t'}$ is increased relatively to the SM case as shown in 
Eq.(\ref{VTH:SigB1}).
\\ Such parameters might reproduce the excesses in the
$\gamma\gamma$, $ZZ$ and $WW$
channels observed by the ATLAS \cite{ATLASweb} and CMS~\cite{CMSweb}) Collaborations.
In particular, it is interesting to note that the present
theoretical model allows for either an increase of $\sigma^{\rm
t'}_{\rm gg\to h}$
compared to the SM, as here, or a decrease as with the following
parameter sets.

\noindent Sets {\bf B,C} - For these sets of parameters where
$m_{h}=255$~GeV or $320$~GeV, all the Higgs decays have
negligible
widths relatively to the dominant channels $h\to ZZ$ and $h\to
WW$, as in the SM case.
Hence the branching ratios $B_{\rm h\to ZZ}$ and $B_{\rm h\to
WW}$ remain unchanged in the present model
with vector-like top quarks where only the decay widths for
$h\to gg$, $h\to \gamma \gamma$ and $h\to \gamma
Z$ are modified.
As a consequence, the experimental limits on $\sigma_{\rm gg\to
h}/\sigma_{\rm gg\to h}^{\rm SM}<0.50$ ($0.40$) [for $m_{h}\simeq 255
\ (320)$~GeV]
issued from the LHC combined investigations using the $h\to
ZZ,WW$ channels exclusively
\cite{CMSweb,ATLASweb} can be applied directly to our framework
where one gets
\begin{equation}
\sigma_{\rm gg\to h}^{\rm t'}/\sigma_{\rm gg\to h}^{\rm SM}
=0.45 \ \ [B] \ \ ; \ \ 0.40 \ \ [C]
\end{equation}
which does not conflict with the above LHC limits.
\\ Note that for the point C, $\sigma_{\rm gg\to h}^{\rm t'}$ is
strongly reduced compared to SM. A factor $1/10$ could even be
achieved in the
present theoretical model but variants of the multiplet choice
(non-minimal in term of field content),
allowing coupling correction cancellations, should 
then be used instead to pass the indirect constraints discussed
in Section~\ref{VTH:oblique}

\noindent Set {\bf D} - For  $m_{h}=540$~GeV, the Higgs boson is
searched only through its decays into $ZZ$ and $WW$.
The strongest bounds on the Higgs rates from the LHC read as \\
$\sigma_{\rm gg\to h}B_{\rm h\to ZZ}/\sigma_{\rm gg\to h}^{\rm
SM}B_{\rm h\to ZZ}^{\rm SM}<1$~\cite{CMSweb} and
$\sigma_{\rm gg\to h}B_{\rm h\to WW}/\sigma_{\rm gg\to h}^{\rm
SM}B_{\rm h\to WW}^{\rm SM}<1.6$~\cite{CMSweb}.
These upper limits are clearly in good agreement with the rates
calculated in the presence of the $t'$ and $t''$ states (that
modifies
$B_{\rm h\to t\bar t}$) for the set D, namely,
\begin{equation}
\sigma_{\rm gg\to h}^{\rm t'}B_{\rm h\to ZZ}^{\rm
t'}/\sigma_{\rm gg\to h}^{\rm SM}B_{\rm h\to ZZ}^{\rm SM} = 0.69
\ \ \ \ \ \
\sigma_{\rm gg\to h}^{\rm t'}B_{\rm h\to WW}^{\rm
t'}/\sigma_{\rm gg\to h}^{\rm SM}B_{\rm h\to WW}^{\rm SM} = 0.69
\\
\end{equation}
where $$B_{\rm h\to ZZ}^{\rm t'}/B_{\rm h\to ZZ}^{\rm SM} = 1.06
\ \ \ \ \ \ B_{\rm h\to WW}^{\rm t'}/B_{\rm h\to WW}^{\rm SM} =
1.06 \ \ [D].$$


\section{INDIRECT CONSTRAINTS AND OBLIQUE PARAMETERS}
\label{VTH:oblique}

Given the absence of a precise measurement for the $Zt\bar t$
vertex (coupling directly modified by the $t-t'-t''$ mixing),
the main indirect constraints to the present model come from the corrections to the gauge boson vacuum
polarizations induced by the loops of $q_{5/3}$,$t'$,$t''$
states. The values of the oblique parameters $S,T$
that we obtain, according to the preliminary calculations of
Ref.~\cite{Barbieri:2006bg,Lavoura:1992np}, are given in
Table~\ref{VTH:DoubletTable}.
They appear to belong to the $1\sigma$ regions induced by the
long list of EW precision observables measured mainly at the LEP
collider
\cite{Nakamura:2010zzi}. 
\\ Let us remark that the input parameters of
Table~\ref{VTH:DoubletTable} (i.e. the theoretical values in the
first four lines) have been chosen to
fix a panel of characteristic benchmark points for $m_{t_2}$
that pass the indirect constraints as well as the bounds from
direct Higgs search described in previous section; however those
two types of constraints allow large domains of the parameter
space (varying also $m_h$). The precise setting of the $Y$
coupling reflects mainly
the experimental precision on the top quark mass measurement (and not any fine-tuning).

The $t',t''$ states could contribute to Flavor Changing Neutral Current (FCNC) reactions which are experimentally well constrained; from the theoretical point of view,  
these FCNC contributions rely precisely on the whole set of Yukawa coupling constants for the entire quark sector. The treatment of such an high degree of freedom 
in the parameter space is beyond the scope of the present study.

Finally, given the relative heavyness of the $t_2$ quark, we have checked that the predicted value for the $V_{tb}$ CKM matrix element 
is in agreement with the experimental measurement close to unity obtained (without assuming $3\times 3$ unitarity) through the single top
quark production cross section at Tevatron \cite{Nakamura:2010zzi}.


\section{HIGGS SIGNAL RECONSTRUCTION IN $t_2\bar{t}_2\rightarrow th+X$ EVENTS}
\label{VTH:exp-analysis}

We have studied the sensitivity, at the 14 TeV LHC, of a search
for $pp \to t_2 \bar t_2$ production, with one of the $t_2$
decaying to $th$, and the other decaying to $Wb$ or $tZ$ or
$th$, resulting into $thWb$, $thtZ$ and $thth$ final states, respectively.
For the signal event generation, we have implemented the
couplings of the afore described $t_2$ model in {\small \tt
FeynRules}~\cite{Christensen:2008py}~\footnote{We thank Claude
Duhr and Benjamin Fuks for their precious help in this
implementation.}
interfaced with {\small \tt MadGRAPH}~\cite{Herquet:2008zz} for
the {\it Monte Carlo} generation, {\small \tt
PYTHIA}~\cite{Sjostrand:2006za}
for the hadronization part and {\small \tt
DELPHES}~\cite{Ovyn:2009tx} for the fast simulation of the CMS
detector response.
The main backgrounds were generated with {\small \tt
ALPGEN}~\cite{Mangano:2002ea} interfaced to {\small \tt PYTHIA}
and {\small \tt DELPHES}, as for the signal events. Physics
objects used for the analysis (photons, leptons and jets) were 
defined emulating the requirements used in real CMS Higgs
searches in the 2011 data ~\cite{CMSweb}. Reconstructed jets and leptons (electrons or
muons) are required to have a transverse momentum larger than 30
GeV, while for photons a transverse momentum lower cut of 20 GeV
is applied.
In particular, we followed closely the physics object definition and selection cuts used for the  real data 7 TeV Higgs analysis in the diphoton channel~\cite{CMS-PAS-HIG-11-030} and in the four lepton channel~\cite{CMS-PAS-HIG-11-025}. We did not attempt to do a dedicated cut optimization for the 14 TeV signals of interest, in order to be able to extrapolate reliably some measured background rejection factors (in particular for the high cross-section background processes, such as diphoton plus jets and $t\bar t$ plus jets) from 7 TeV, where experimental measurements are available, to the 14 TeV case, where we only dispose so far of limited statistics MC samples.

\subsection{Search for $th+Wb/tZ/th$ signal in the diphoton plus
multijets channel}
To study the sensitivity of a search for $thtZ,\ thWb$ and $thth$ final states, 
when the Higgs is relatively light, as for point A with
$m_{h}=125$~GeV, we exploit the $h\to
\gamma\gamma$ decay channel. As for a relatively light Higgs in the SM, in spite of the relatively small expected B$(h\to\gamma\gamma)$ value, 
this is expected to be a channel with very good sensitivity (high signal-over-background ratio), due to the excellent diphoton mass resolution,
that allows to identify the Higgs mass signal over a background that can be precisely measured in the side-bands.  

In our analysis, photon identification and isolation requirements
are applied that emulate the photon selection used in the CMS
search for the SM Higgs boson to diphoton final states~\cite{CMS-PAS-HIG-11-030}. 
Signals from   $th+tZ/Wb/th$, $h\to\gamma\gamma$, final states are characterized by large number of energetic jets,
from top and heavy vector boson decays, in addition to the two
photons from the Higgs decay. Thus, we apply also a requirement
on the minimum number of jets.
The numbers of events expected for the signal and the
backgrounds, after diphoton selection and the requirements of
more than 6 or 8 jets, are given in Table~\ref{tabVTH:dibosona}. The background contributions considered for this study are from 
physics processes where a real photon pair is produced. Background from fake photons is not included. Based on preliminary studies with real data at 7 TeV, it is expected that for the selection requirement proposed here, in particular the large jet multiplicity and the diphoton mass cut,  the background from fake photons plus multijets contributes a small fraction of the total background. 

\begin{table}[htbp]
\begin{center}
\begin{tabular}{|c|c|c|c|c|}
\hline
Signal (point A) & $\sigma \times B$ (fb)	& $N(>$ 6 jets) & $N(>$ 8 jets) & $N(115$ GeV$< m^{\gamma\gamma}< 135$ GeV)\\
\hline
	$thbW,\ h\rightarrow \gamma\gamma$	&3.1	&	$	4.0$	&	$	1.4$	&	$	1.1	$	\\
	$thtZ,\ h\rightarrow \gamma\gamma$	&10.1	&	$	12.5	$	&	$	6.2$	&	$	5.1	$	\\
	$thth,\ h\rightarrow \gamma\gamma$		&22.1	&	13.5  &	$	7.3	$	&	$	5.8	$	\\
\hline
	Background	& $\sigma$(fb)	& $N(>$ 6 jets) & $N(>$ 8 jets) & $N(115$ GeV$< m^{\gamma\gamma}< 135$ GeV)\\
\hline
	$W + \gamma\gamma$ + jets	&450	&	$	110	$	&	$	20.0	$	&	$	4.0 $	\\
	$t\bar{t}$ + $\gamma\gamma$ + jets	&15.5	&	$	11.0	$	&	$	8.0	$	&	$	1.2	$	\\
$t\bar{t}W + \gamma\gamma$ + jets &0.0678	& $ 0.05$ & $ 0.03 $ & $ 0.0020$\\
\hline
\end{tabular}
\end{center}
\caption{Search for $thbW,$ $thtZ$ and  $thth$ final states, with $h\rightarrow \gamma\gamma$, in events with  two photons plus
multijets, at the 14 TeV LHC. 
The number of produced  signal events, per fb$^{-1}$ of integrated luminosity, is given as  $\sigma \times B$ (fb),  
for the signals  final states and  for benchmark point A  (see Table~\ref{VTH:DoubletTable}), while for the 
backgrounds of interest, the cross-sections $\sigma$ are given. 
The additional columns show the number of expected events, in 20 fb$^{-1}$ at 14 TeV,  for the signal final states, compared to the contributions from 
background processes, after requiring two high transverse momentum, isolated  photons, high jet
multiplicity and  the diphoton invariant mass to be in a 20 GeV window around the generated Higgs mass value of 125 GeV.}

\label{tabVTH:dibosona}
\end{table}

\begin{figure}[htbp]
\begin{center}
\begin{tabular}{cc}
\includegraphics[width=0.5\textwidth]{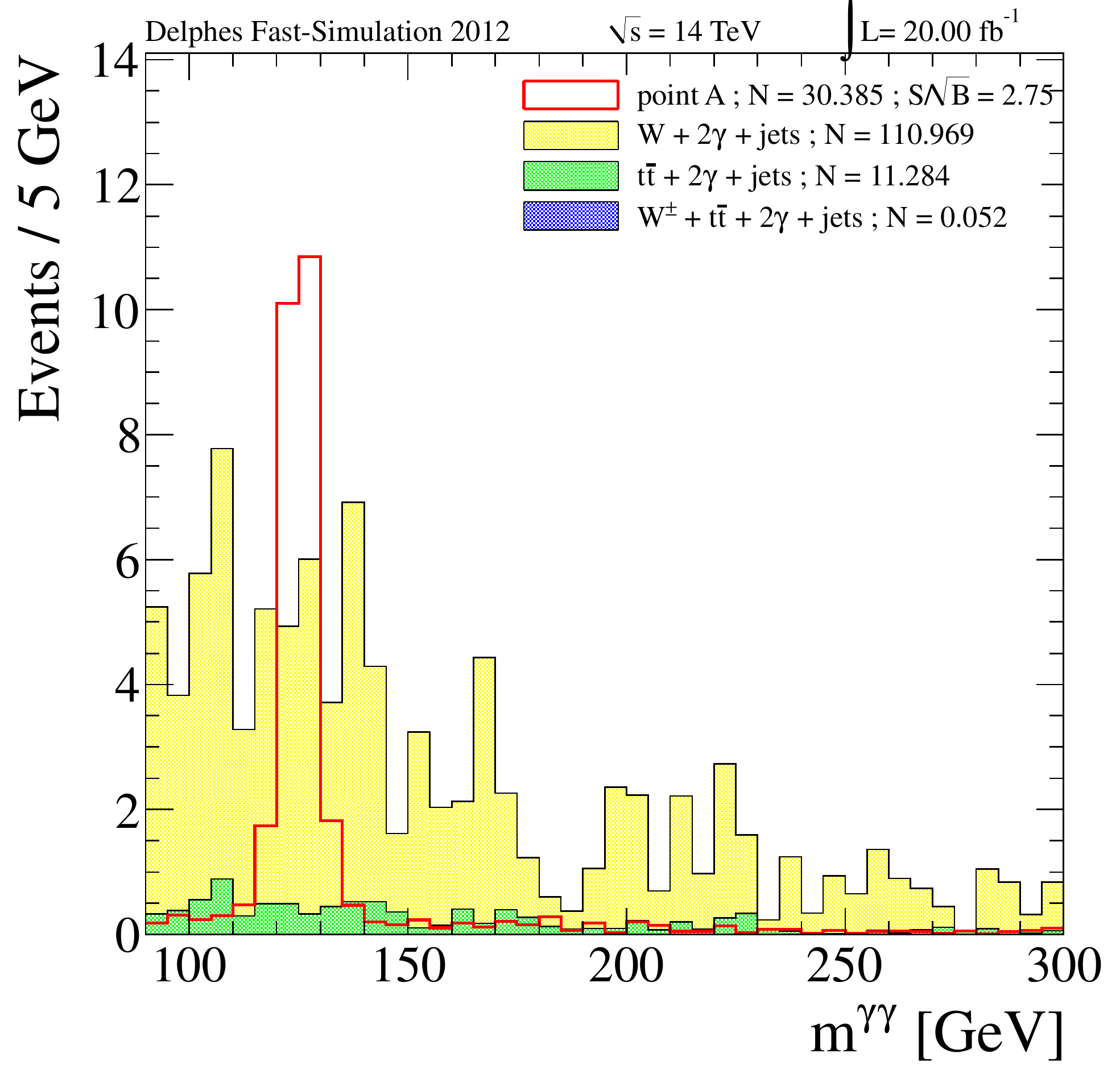}
&
\includegraphics[width=0.5\textwidth]{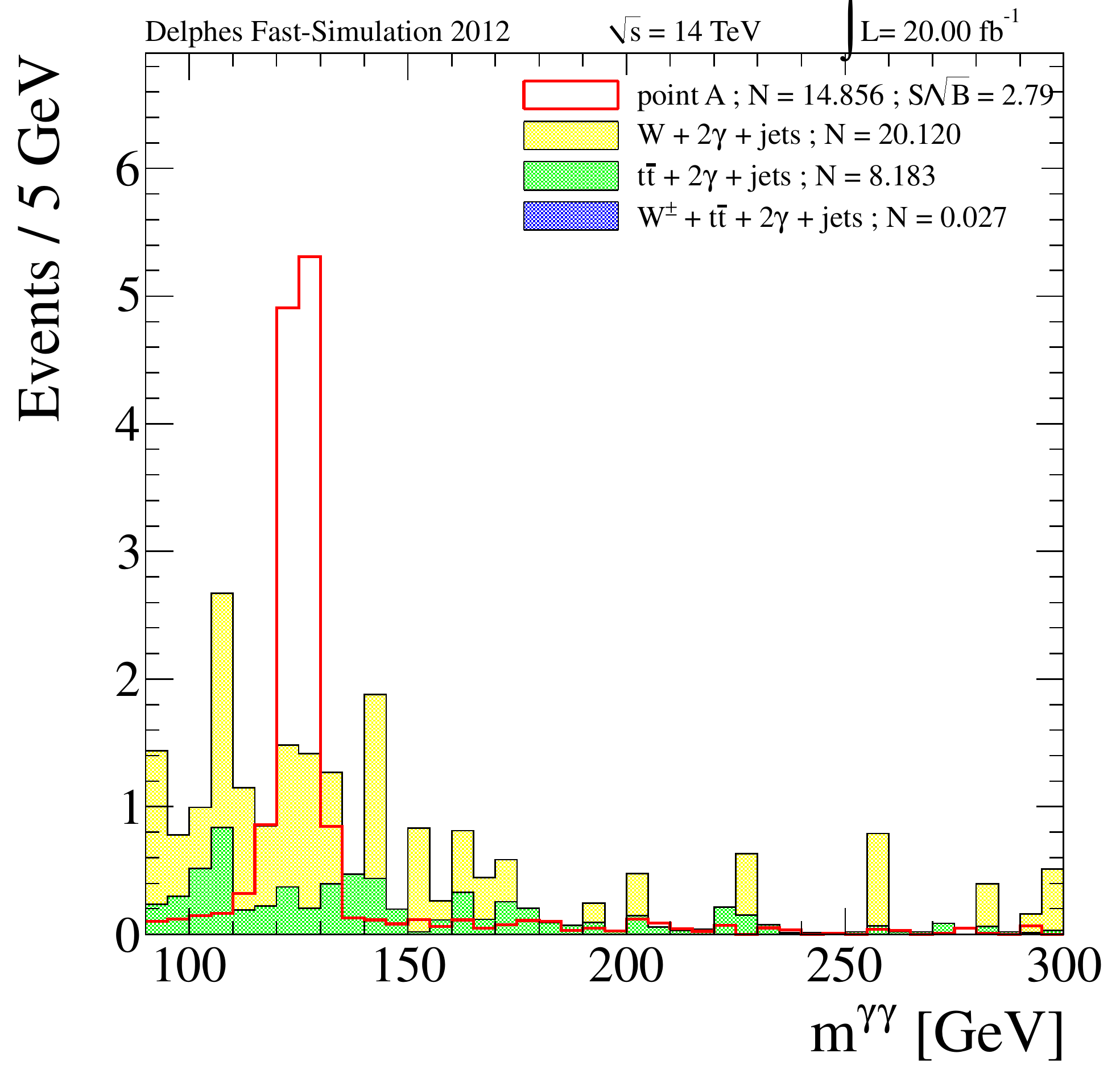}
\\
\end{tabular}
\caption{Reconstructed diphoton mass $m^{\gamma\gamma}$  for simulated events at 14 TeV,  for the sum of the contributions from  $thbW$, $thtZ$ and  $thth$, $h\rightarrow \gamma\gamma$,  signal final states (point A, see 
Table~\ref{VTH:DoubletTable}) compared to the sum of contributions from 
background processes, after requiring in the simulated events at least
two high transverse momentum, isolated photons and more than 6 jets, left panel,
or more than 8 jets, right panel. Event numbers are normalized to 20 fb$^{-1}$ of integrated luminosity according to the expected production cross-section times branching fraction values given in  Table~\ref{tabVTH:dibosona}.}
\label{VTH:dibosona}
\end{center}
\end{figure}

In virtue of the good diphoton mass resolution (1-2\% for a
110 to 130 GeV Higgs) of the detector, the best discriminating
quantity after selection is the reconstructed diphoton mass. The
diphoton invariant mass distribution for signal and background events,
after the requirement on the number of jets to be larger than
six or eight, is shown in Figure~\ref{VTH:dibosona}. The number of
events with more than eight jets, with a diphoton mass in the
window between 115 GeV and 135 GeV, is also given in
Table~\ref{tabVTH:dibosona}. 
We estimate that for an integrated luminosity of 20 fb$^{-1}$, a signal, corresponding to the point A benchmark, is
detectable with a significance $S/\sqrt{B}\sim 5$.

\subsection{Search for $th+Wb/tZ/th$ signal in the  four leptons plus multijets channel}
In order to study the sensitivity of a search for $thWb,\ thtZ$ and $thth$ final states, when the Higgs boson is relatively heavy, as expected for the points B, C and D, with $m_{h}=225, 320, 540$~GeV, respectively, we exploit
the decay channel into four charged leptons $h\to ZZ\rightarrow 4l$. 
Lepton identification and isolation
requirements are applied to emulate the lepton selection used
in the CMS search for the SM Higgs boson to four lepton final
states~\cite{CMS-PAS-HIG-11-025}. 

The numbers of events expected after the four lepton requirement, as well as after the additional requirement
of a number of jets above 5, are shown for signal and
backgrounds in Table~\ref{tabVTH:dibosonb}. The main background is provided by the $t\bar t$+jets process, because of its huge cross-section. The requirement of four high momentum, isolated leptons reduces this background by several orders of magnitude, as shown in Table~\ref{tabVTH:dibosonb}, where  a reduction factor due to four lepton requirements  is applied, as measured in real data at 7 TeV for the SM Higgs search in the four lepton channel~\cite{CMS-PAS-HIG-11-025}. 

\begin{table}[htbp]
\begin{center}
\begin{tabular}{|c|c|c|c|c|}
\hline
Process & $\sigma \times B$ (fb)	& $N(4l)$ & $N(>$ 5 jets) & $N(m^{4l}>200$ GeV) \\ \hline
Signal             & $thbW / thtZ / thth$       &  $thbW / thtZ / thth$   &  $thbW / thtZ / thth$ &   $thbW / thtZ / thth$ \\ 
                      &   $h\rightarrow ZZ$         &   $h\rightarrow ZZ$      &   $h\rightarrow ZZ$    &   $h\rightarrow ZZ$ \\ \hline
 Point B & 59.9 / 105 / 256    &  9.0 / 26.2 / 47.4 & 4.5 / 15.1 / 29.0 &  2.9 / 10.6 / 19.2 \\
 Point C & 66.6 / 76.0 / 22.2 &  8.7 / 17.5 / 3.9  & 4.1 / 10.1 / 2.4  &  2.5 / 6.9  / 1.5  \\
 Point D & 10.7 / 26.1 / 27.8  &  1.6 / 7.0  / 5.7  & 0.8 / 4.1  / 3.5  &  0.6 / 3.1  / 2.5  \\
\hline
Background          & $\sigma$(fb) & $N(4l)$ & $N(>$ 5 jets) & $N(m^{4l}>200$ GeV) \\ \hline
$t\bar{t}$ + jets            &   920 $10^{3}$& 22.2      & 8.5   &  2.7       \\
$t\bar{t}W$ + jets         &   90                   & $<$1     & $<$1  &  $<$1 \\
\hline
\end{tabular}
\end{center}
\caption{Search for $thbW,$ $thtZ$ and  $thth$ final states, with $h\rightarrow ZZ$, in events with  four leptons plus
multijets, at the 14 TeV LHC. 
The number of produced  events, per fb$^{-1}$ of integrated luminosity, is given as  $\sigma \times B$ (fb),  
for the signals  final states,  for benchmark points B, C, D (see 
Table~\ref{VTH:DoubletTable}), while for the backgrounds of interest, the cross-sections $\sigma$ are given. 
The additional columns show the number of expected events, in 20 fb$^{-1}$ at 14 TeV,  for the signal final states, compared to the contribution from background processes, after requiring 
four high transverse momentum, isolated leptons, $N(4l)$, more than five high $p_T$ jets,  $N(>$ 5 jets), and the four lepton 
mass to be larger than 200 GeV,  $N(m^{4l}>200$ GeV).}
\label{tabVTH:dibosonb}
\end{table}

\begin{figure}[htbp]
\begin{center}
\begin{tabular}{cc}
\includegraphics[width=0.5\textwidth]{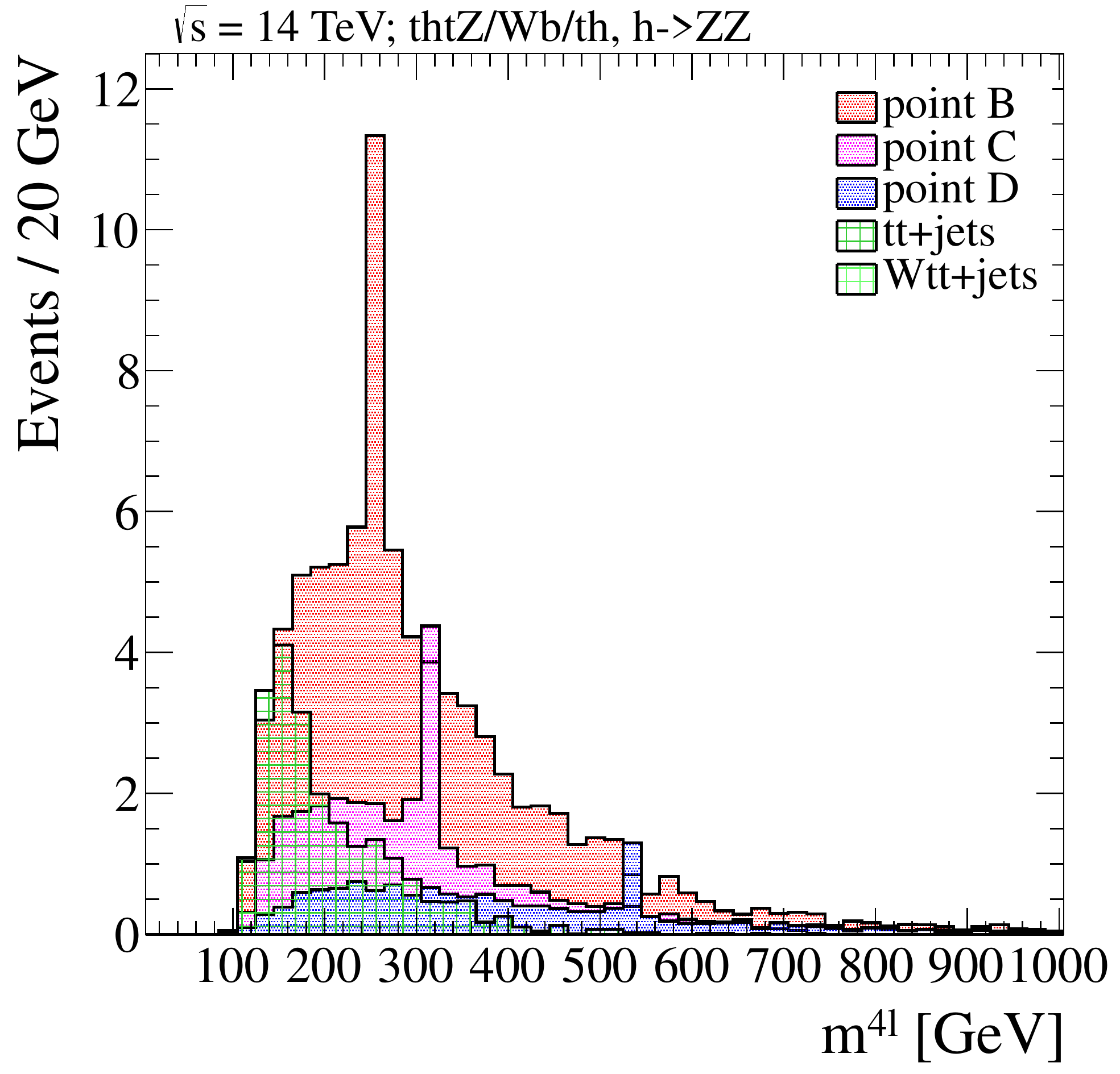}
&
\includegraphics[width=0.5\textwidth]{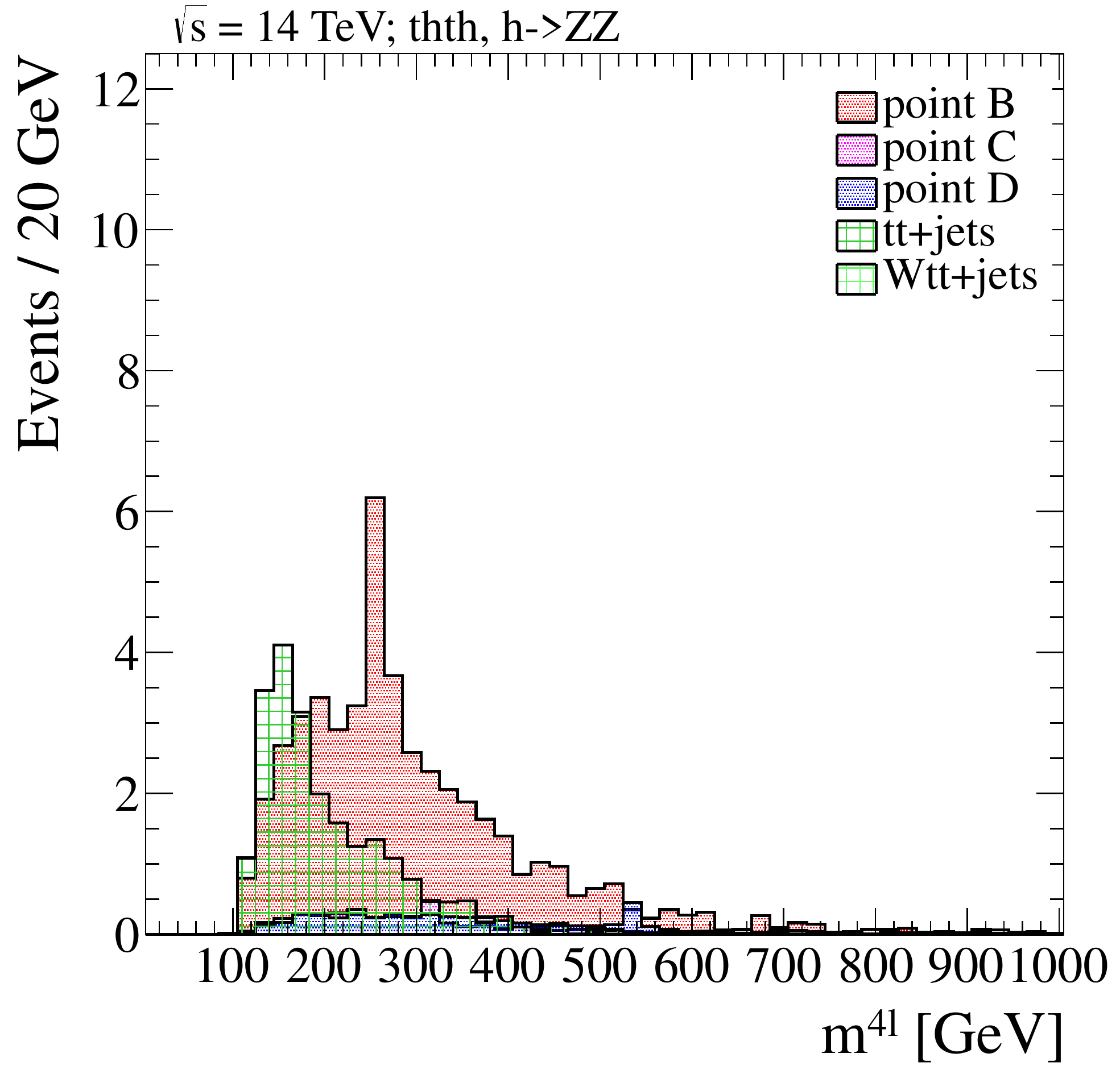}
\\
\includegraphics[width=0.5\textwidth]{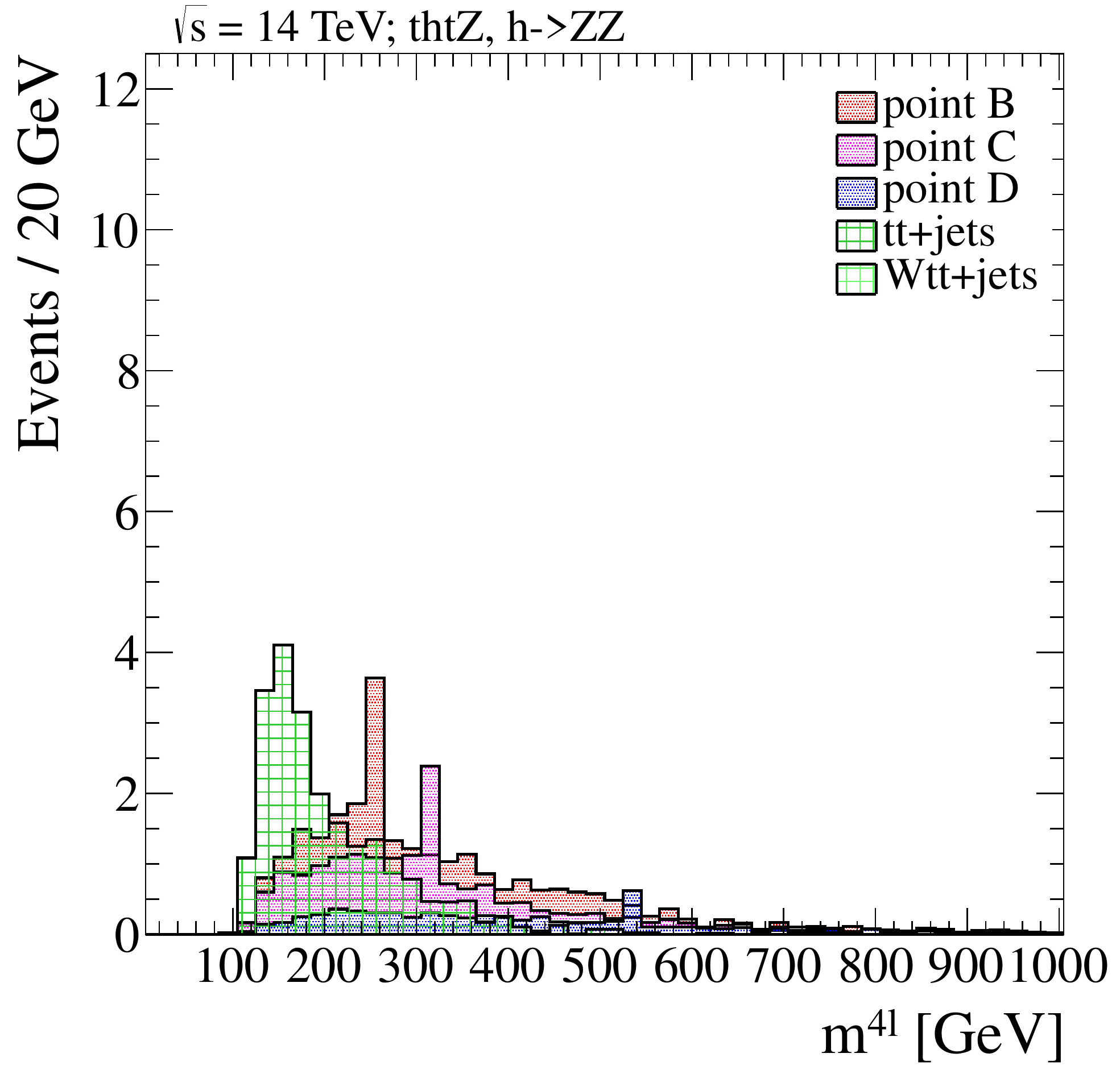}
& 
\includegraphics[width=0.5\textwidth]{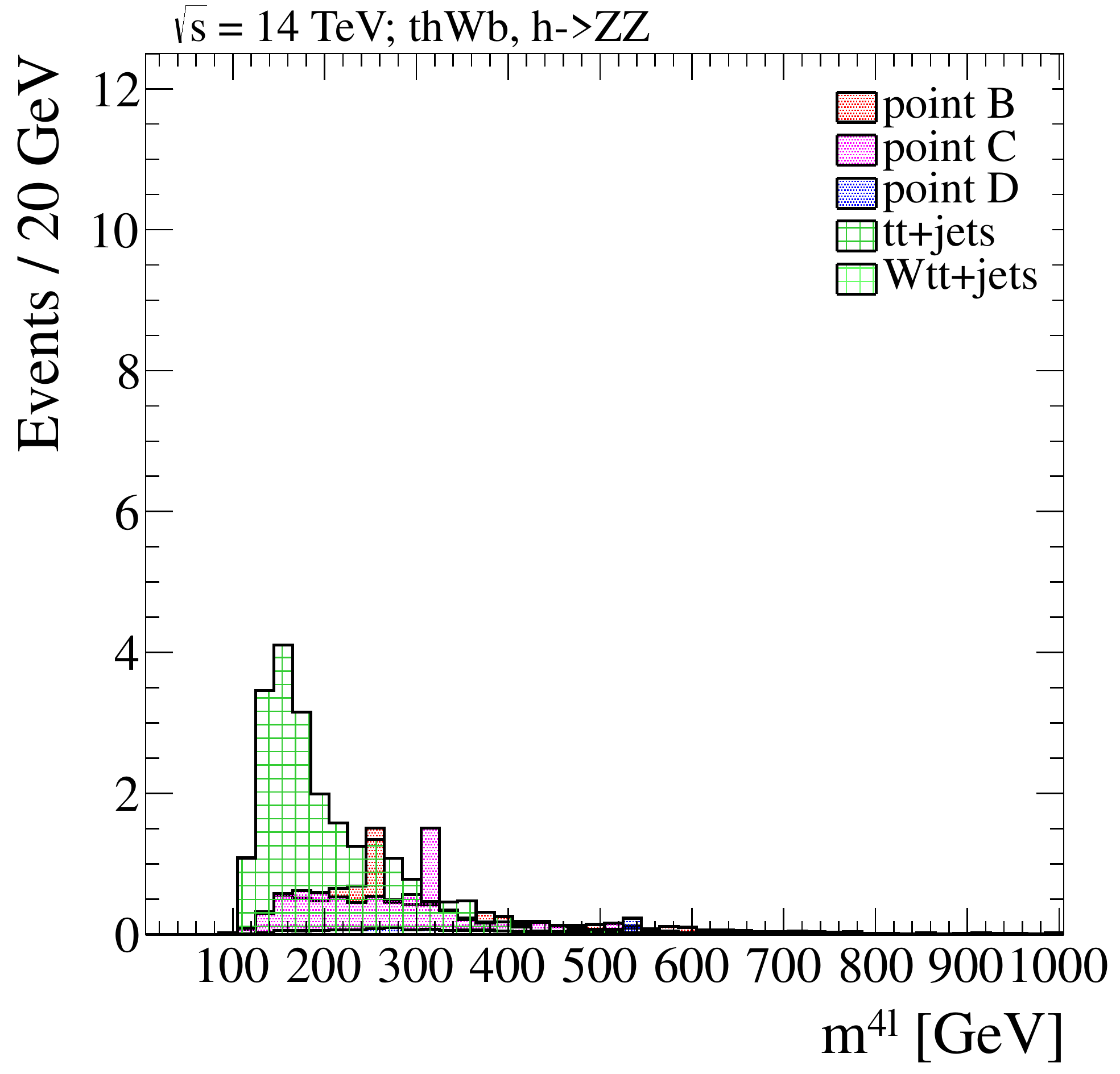}
\\
\end{tabular}
\caption{Reconstructed four leptonn mass, $m^{4l}$,  for simulated events at 14 TeV,  for the sum of the contributions from  $thbW$, $thtZ$ and  $thth$,  $h\rightarrow ZZ$,  signal final states (points B, C, D, see Table~\ref{VTH:DoubletTable}) compared to the sum of the contributions from 
background processes, after requiring four high transverse momentum, isolated leptons and high jet multiplicity. Event numbers are normalized to 20~fb$^{-1}$ of integrated luminosity according to the expected production 
cross section times branching fraction values given in the Table~\ref{VTH:DoubletTable}. The
reconstructed four lepton invariant mass distributions, for three different Higgs mass values,  $m_{h}=255, 320, 540$~GeV, are shown summed over the three signal final states (top left right hand side panel) and separately for each signal final state.}
\label{VTH:dibosonb}
\end{center}
\end{figure}

The distributions of the four lepton invariant mass, $m^{4l}$, for signal and background
events, after the requirement on the number of jets to be larger
than five, are shown in Figure~\ref{VTH:dibosonb}. The number of events in these distributions are normalized to an integrated luminosity of 20 fb$^{-1}$ using the signal cross sections and final state branching ratios expected for the specific benchmark scenarios (B, C and D), detailed in Table~\ref{VTH:DoubletTable}. The $m^{4l}$ signal distributions are shown separately for the sum of the normalized contributions from the different final states, $thbW$, $thtZ$,  $thth$, as well as for the individual normalized contributions.
We also note from the distributions given independently for each final state that all of the three final states can lead to a Higgs discovery, thanks to the  good  experimental resolution of the $h\rightarrow ZZ\rightarrow 4 l$ mass reconstruction.
 Signal events, with reconstructed mass significantly lower, or higher, than the generated Higgs mass
value, are in general events where one lepton from the $h\rightarrow ZZ\rightarrow 4 l$ decay  does not pass the lepton selection criteria,   while another lepton from an accompanying $W$ or$ Z$ decays is selected
to reconstruct the four lepton mass. The background, mainly from the $t\bar t$ + jets process,  is mostly
concentrated in the  $m^{4l}$ region below 200 GeV. Applying a cut  $m^{4l}>200$ GeV  removes a large fraction of the background and improves the signal significance, as shown in Table~\ref{tabVTH:dibosonb}.
We estimate that, for an integrated luminosity of 20 fb$^{-1}$, a
signal, corresponding to the B and C benchmark scenarios, is detectable with  significance $S/\sqrt{B}\gtrsim 5$, 
after the simple selection cuts described above. About a factor of two larger integrate  luminosity, or a more performant analysis, would be needed for 
a five sigma detection of a signal corresponding to the D scenario, with a  $t_2$ mass of about 800 GeV and a pair production cross section 
of about  0.3 pb. 

\section{CONCLUSIONS}

\section{CONCLUSIONS}

We have shown that there exist regions of parameter space in beyond SM scenarios, allowed by present
phenomenological and experimental constraints,
in particular from the direct Higgs searches, where vector-like
top partner production at the LHC may give rise to new Higgs
production channels, over a large Higgs mass range, extending
from 120 GeV to more than 500 GeV. Indeed the SM Higgs mass
exclusion, in the range between 130 and 600 GeV, does not hold
in these beyond SM scenarios. We have shown that for some relevant characteristic top-partner and Higgs mass values, the expected production cross sections and decay branching ratios lead to a
signal of Higgs production, from the decay of vector-like top-partners, which is detectable at the 14 TeV LHC over a large Higgs
mass range (125-540 GeV) using the diphoton and four lepton mass distributions.
The Higgs mass peak arising in those distributions can be reconstructed with a good mass resolution ($\sim$ 1-2\%). Further
studies~\cite{thisteam:2012} are in progress to fully explore
the discovery potential of the LHC in these scenarios.

\section*{ACKNOWLEDGEMENTS}
The authors would like to thank `Les Houches' School for
hospitality offered and the organizers of this Workshop
(fruitful as always)
during which the work contained herein was initiated.
D.K.G. acknow\-ledges partial support from the Department of Science and Technology, India under the grant
SR/S2/HEP-12/2006.

\AddToContent{A.~Azatov, O.~Bondu, A.~Falkowski,
M.~Felcini, S.~Gascon-Shotkin,
D.~K.~Ghosh, G.~Moreau, A.~Y.~Rodr\'iguez-Marrero, S.~Sekmen}
\renewcommand{\thesection}{\arabic{section}}


\chapter{Composite Higgs Models and the $t\bar{t}H$ Channel}

\author{\it Adri\'an Carmona, Mikael Chala, Jos\'e Santiago}

\begin{abstract}
Despite its suppressed couplings to Standard Model particles, 
a composite Higgs with mass $m_H=125$ GeV and a moderate 
degree of compositeness can be consistent
with current Higgs searches, including a sizable enhancement in the
$H\to \gamma \gamma$ channel. Heavy resonances common to many
composite Higgs models can mediate new Higgs production mechanisms. 
In particular, the $t\bar{t}H$
channel can be accesible at the LHC in these models through the exchange of
colored vector and fermion resonances. In this case, the
$t\bar{t}H$ channel is not a direct measure of the top Yukawa coupling.
\end{abstract}

\section{INTRODUCTION}

Composite Higgs models~\cite{Kaplan:1983fs,Kaplan:1983sm}
are arguably among the best motivated models of
natural electroweak symmetry breaking. 
Despite the expected presence new strongly coupled resonances,
current LHC searches have not significantly constrained the parameter
space of composite Higgs models. The main reason is that precision
electroweak and flavor constraints already require a moderate degree
of compositeness and relatively heavy, $M \geq 2$ TeV, new
resonances, beyond the reach of current searches (for a summary 
see~\cite{Santiago:2011dn} and references there in). 
Proper tests of the composite nature of the Higgs, based on
longitudinal gauge boson and Higgs scattering, will require even more
energy or luminosity to have any discriminating power (see for
instance~\cite{Contino:2010mh}). Thus, it makes
sense to investigate whether the tremendous effort that is being put
on the search for the Higgs can give us information on the composite
sector. Indeed, current Higgs searches are beginning to constraint
some of the parameter space of the simplest composite Higgs
models~\cite{Espinosa:2010vn,Azatov:2012bz,Espinosa:2012qj,Carmi:2012yp,Espinosa:2012ir}.  
In this note
we argue that the ``Higgs effort'' the LHC community is going
through could give valuable information on other aspects of the
composite sector not directly related to the Higgs itself. In
particular, new vector and fermion resonances are typically present in
these models. New color-octet vector resonances can be
produced with a sizable cross section and predominantly decay
into a Standard Model (SM) quark and a new fermionic resonance, The
latter will decay, with a relatively large branching ratio, into a SM
quark and the Higgs.
This is a new source of Higgs bosons in composite models that can give
information not only on the Higgs itself but also on other composite
states, like the color-octet vector and the fermion resonances. 
Note that this mechanism is intrinsically different to previous
studies of new vector-like quark pair production (through their QCD
interactions) followed by Higgs decays as discussed elsewhere in these
proceedings. In our case, the
presence of an intermediate resonance with relatively large couplings
and the reduced phase space suppression due to single production allow
us to investigate much higher energies and new sectors of the theory.

In this note we consider the minimal composite Higgs model, based on
an $SO(5)/SO(4)$ symmetry breaking pattern, with composite fermionic
matter transforming on the vector representation of $SO(5)$
(MCMH$_5$~\cite{Agashe:2004rs,Agashe:2006at,Contino:2006qr}). We take 
as an example a composite Higgs mass of $125$ GeV 
with a moderate degree of compositeness
and focus on the $t\bar{t} H$ channel. Despite
the suppressed tree-level 
Higgs couplings to all SM particles due to the composite
nature of the Higgs, there can be enhancements in certain channels
with respect to the SM, as currently hinted by LHC
data~\cite{:2012si,Chatrchyan:2012tx}.  
Many other options in
terms of Higgs masses and decay channels are possible even in the
minimal composite Higgs model and will be discussed elsewhere~\cite{inpre}.

\section{The Model\label{model}}

\begin{figure}[ht]
\begin{center}
\includegraphics[width=0.43\textwidth,clip=]{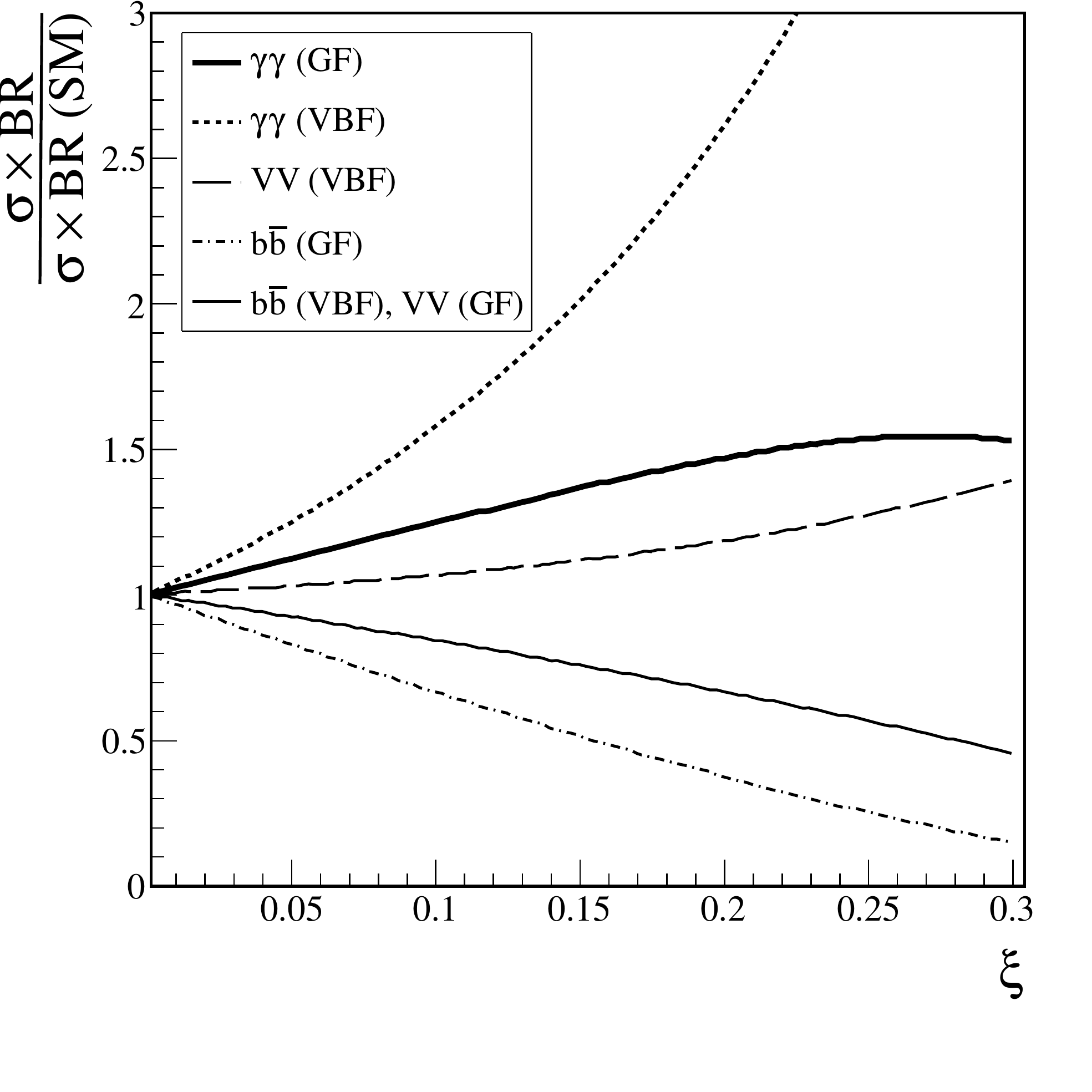} 
\hfil 
\includegraphics[width=0.43\textwidth]{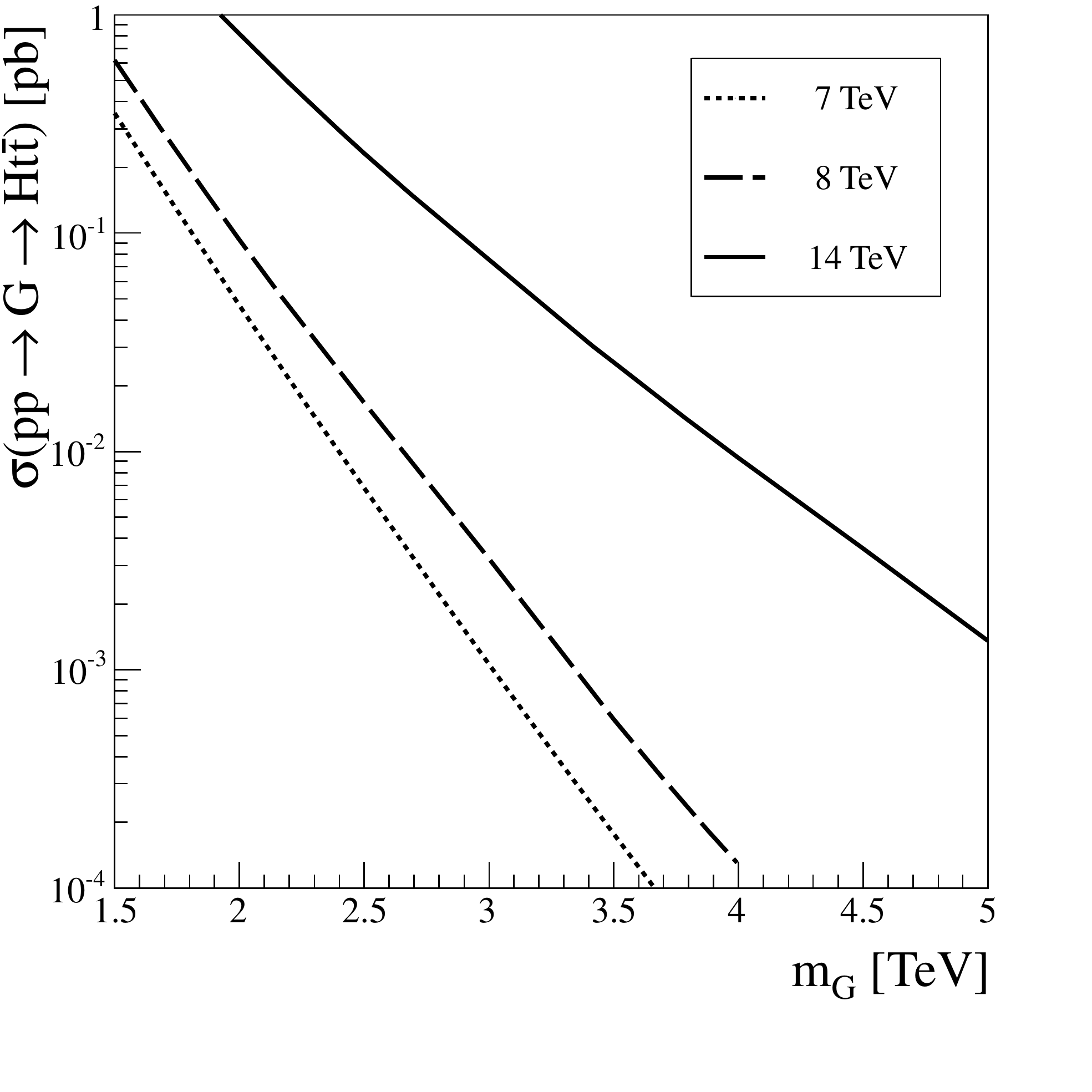}
\vspace{-0.3cm}
\caption{Left panel: ratio of the Higgs production cross section times branching
  ratio into different channels with respect to the SM as a function
  of the degree of 
  compositeness $\xi$ ($V$ stands for either $W$ or $Z$). 
Right panel: $t\bar{t}H$ production cross
  section in the benchmark composite model mediated by a color octet
  vector resonance with decay into a fermionic resonance and a top
  quark (see text for details).}
\label{channels}
\end{center}
\end{figure}

For the sake of concreteness we focus on the minimal $SO(5)/SO(4)$
composite Higgs model based on the $5$ representation of $SO(5)$ for
the fermion composite states (MCHM$_5$~\cite{Contino:2006qr}). 
In this model the relevant tree level Higgs couplings to SM particles are given,
assuming heavy enough new resonances so that one can neglect the
effect of mixing, by (see for
instance~\cite{Contino:2010mh,Espinosa:2010vn}): 
\begin{equation}
R_{hVV}\equiv \frac{g_{hVV}}{g^{\mathrm{SM}}_{hVV}}=\sqrt{1-\xi},\quad 
R_{hff}\equiv \frac{g_{hff}}{g^{\mathrm{SM}}_{hff}}=\frac{1-2\xi}{\sqrt{1-\xi}},
\end{equation}
where $V$ and $f$ stand for any electroweak gauge boson and SM fermion
respectively. The production cross
section receives a suppression proportional to these same factors:
\begin{equation}
\frac{\sigma(gg\to H)}
{\sigma(gg\to H)_{SM}}=R_{hff}^2,\quad
\frac{\sigma(qq\to qq H)}
{\sigma(qq\to qqH)_{SM}}(VBF)=R_{hVV}^2.
\end{equation}
The different decay widths scale with the corresponding couplings squared
except for the $h\to \gamma \gamma$ channel that reads
\begin{equation}
\Gamma(H\to \gamma\gamma)
=\frac{(R_{hff} I_\gamma + R_{hVV}
  J_\gamma)^2}
{(I_\gamma+J_\gamma)^2}
\Gamma^{SM}(H\to \gamma\gamma),
\end{equation}
where
\begin{equation}
I_\gamma=-\frac{8}{3} x_t[1+(1-x_t)f(x_t)],
\quad
J_\gamma=2+3x_W[1-(2-x_W)f(x_W)],
\end{equation}
with $x_t=4 m_t^2/m_h^2$, $x_W=4m_W^2/m_h^2$ and
\begin{equation}
f(x)=\left\{\begin{array}{ll}
\arcsin[1/\sqrt{x}]^2, & x\geq 1, \\
-\frac{1}{4}\left[\log
  \frac{1+\sqrt{1-x}}{1-\sqrt{1-x}}-\mathrm{i}\pi\right]^2, &
x<1.
\end{array}\right.
\end{equation}
We show in Fig.~\ref{channels} (left panel) the ratio of the Higgs
production cross 
section times branching ratio in several different channels with
respect to the corresponding 
SM one as a function of the degree of compositeness of
the Higgs, defined by $\xi=v^2/f^2$ with $v=246$ GeV and $f$ the decay
constant associated to the composite sector. 
We have separated the production mechanism in gluon fusion and vector boson
fusion, as they receive different suppression factors. 
As we see in the figure, there is a
sizeable increase in the $\gamma \gamma$ channel, both in gluon fusion
and vector boson fusion. This is due to a larger reduction of the top
versus the $W$ contribution in the loop (thus resulting in a smaller
cancellation) together with a reduction in the total width.

New vector and fermion resonances are common in composite Higgs
models. It has been recently emphasized that single production of new
quark resonances, mediated by color octet vector resonances, can be a
competitive discovery channel for these
models~\cite{Barcelo:2011vk,Vignaroli:2011ik,Barcelo:2011wu,Bini:2011zb}. Only
decays into electroweak gauge bosons and a SM quark were considered in these
references. However, the heavy quarks will decay with a sizable
branching ratio into the composite Higgs and a SM quark. Thus, they
provide a novel production mechanism in composite Higgs models that
could give further information on the composite sector of the
theory. To be specific, in this note we consider the two-site version
of the MCHM$_5$. Full details of the model can be found
in~\cite{Bini:2011zb}. 
The ingredients relevant for the study in this note are the
presence of new massive gluons of mass $m_G$ and new vector-like
quarks with a relatively large coupling to the top quark. We choose
the masses of the lightest new quarks to be $m_Q=m_G/2$ so that the
decay $G\to Q\bar{Q}$ is kinematically suppressed. 
The leading decay is then $G \to Q \bar{t},\, t \bar{Q}$. A sizable
fraction of the heavy quarks will in turn decay into $H t,~H\bar{t}$, thus
resulting in a $t\bar{t}H$ final state. We show in Fig.~\ref{channels}
(right panel) the production cross section times branching fraction in
this channel for a sample point in parameter
space
as a function of
$m_G$ and for the LHC with $\sqrt{s}=7,~8$ and $14$
TeV, respectively.~\footnote{Technically, we have chosen, following
  the notation  
  in~\cite{Bini:2011zb}, $Y_{\ast U}=Y_{\ast D} = g_{\ast 3}=3$, $\sin
  \phi_L=0.6$ and $s_2=0.1$ and fermion masses such that
  $m_{\tilde{T}}=m_{T_{5/3}}=m_G/2$. Also, we have fixed the Higgs
  mass to $m_H=125$ GeV and its degree of compositeness $\xi=0.2$.}
These cross
sections are of the order of the SM $t\bar{t}H$ production for
$m_G\approx 2-2.5$ TeV and become much smaller for higher gluon
masses. However we will see in the next section that our new
production mechanism has very distinctive kinematics, mainly due to
the large mass of the intermediate particles, and the signal can be
extracted from the background with a high statistical significance.

\section{$t\bar{t}H$ search in composite Higgs models\label{analysis}}

The process we are interested in consists of single production of the
new vector-like quarks mediated by an s-channel exchange of the massive
gluon and followed by a decay into a Higgs and a top quark,
\begin{equation}
pp\to G \to Q \bar{t},\bar{Q}t \to H t \bar{t}.
\end{equation} 
We consider in this note the $H\to b\bar{b}$ channel, which has
$B(H\to b\bar{b})=0.47$ (the decay into
$W W^\ast$ has also a relatively large branching fraction and will be
considered elsewhere~\cite{inpre}).
The decays of the Higgs and top quarks 
result in a very large multiplicity final state that makes, \textit{a
  priori}, full reconstruction difficult. However we do not need full
reconstruction to observe a signal. In fact, it is easy to find a
simple set of cuts that reduce the relevant backgrounds to manageable
levels. The $S_T$ distribution, defined as the scalar sum of the $p_T$
of the relevant jets, leptons and missing energy, then reduces the
remaining background to essentially negligible levels, allowing for a
clean extraction of the signal. The initial set of cuts strongly
depends on the available energy. In the low energy phase ($\sqrt{s}=7$
or $8$ TeV), the masses that can be tested are not too large and the
best strategy is to require a large number of jets and many of them to
be tagged as b-jets. In
the high energy phase ($\sqrt{s}=14$ TeV), however, the available energy
is enough to produce much heavier resonances. As a result the tops and
the Higgs are typically quite boosted and their decay products very
collimated. Thus, we ask for a smaller number of
(fat) jets but require them to have invariant masses close to the top
and the Higgs.

\begin{figure}[t]
\begin{center}
\includegraphics[width=0.43\textwidth]{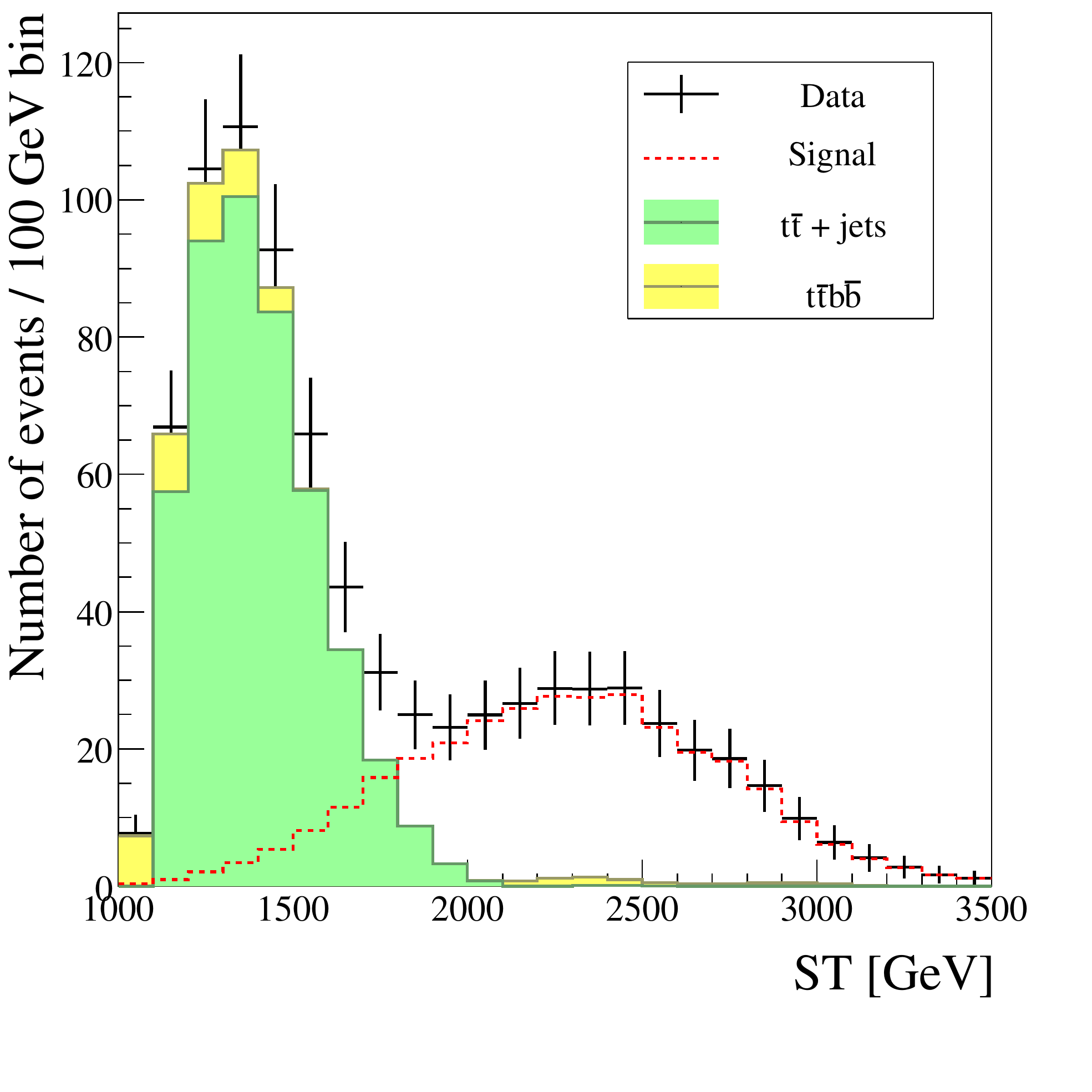}
\hfil
\includegraphics[width=0.43\textwidth]{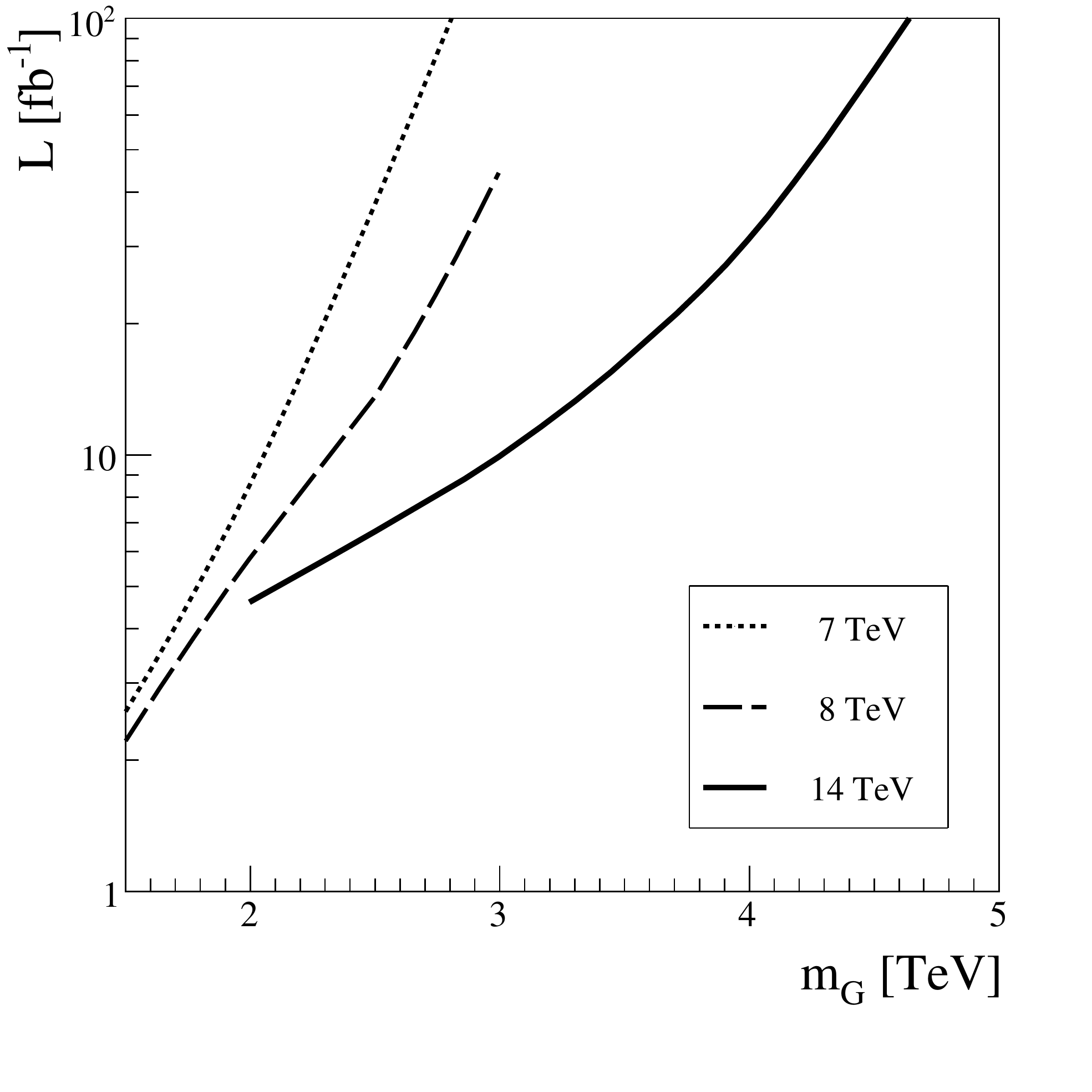}
\vspace{-0.3cm}
\caption{Left panel: $S_T$ distribution for the signal (red dotted) 
and the $t \bar{t}$ (green) and $t\bar{t}
b\bar{b}$ (yellow) backgrounds for $m_G=3$ TeV 
and an integrated luminosity of
  50 fb$^{-1}$ with $\sqrt{14}$ TeV.
Right panel: Luminosity required for a 5 $\sigma$ discovery as a
function of $m_G$ for
$\sqrt{s}=7$ TeV (dotted), 8 TeV (dashed) and 14 TeV (solid).}
\label{results}
\end{center}
\end{figure}

We have simulated the signal and background events with {\tt MADGRAPH
  V4.5.0}~\cite{Alwall:2007st} and {\tt ALPGEN
  V.2.13}~\cite{Mangano:2002ea}, respectively. In both cases the
unweighted events have been passed to {\tt PYTHIA
  6.4}~\cite{Sjostrand:2006za} for 
hadronization and showering and {\tt DELPHES V1.9}~\cite{Ovyn:2009tx}
for detector 
simulation (we have used a tuned version of the ATLAS card to better
agree with published experimental data). The main backgrounds are $t\bar{t}$ and
$t\bar{t}b\bar{b}$ (the former has been simulated with up to $4$ extra jets
properly matched with the parton shower using the MLM method). We have used
the CTEQ6L1 PDFs. 
As we have mentioned, the optimal analyses are different depending on
the available center of mass energy. In all cases jets are defined with
a cone size $\Delta R=0.7$, $p_T(j)> 30$ GeV and
$|\eta_j|<5$. Isolated charged leptons ($e$ or $\mu$) are considered
when $p_T(l)> 20$ GeV and $|\eta_l|<2.5$. A b-tagging
efficiency of 0.7 has been assumed. 
In the analysis for the low energy phase we require
\begin{itemize}
\item At least 4 jets, of which at least 3 must be tagged as b-jets.
\item At least 1 isolated charged lepton.
\end{itemize}
The analysis of the high energy phase requires
\begin{itemize}
\item At least 3 jets, with a minimum of 2 b tags.
\item At least 1 isolated charged lepton.
\item We then order all the jets according to their invariant mass and
  require the two jets with the largest invariant masses to be near the
  top and the Higgs mass, respectively, $|m_{j_1}-m_t|\leq 40$ GeV and
  $|m_{j_2}-m_H|\leq 40$ GeV.
\end{itemize}
The cuts above reduce the background to manageable levels. 
Once the number of background events has been reduced to a reasonable level, we
use $S_T$, defined as the scalar sum of the transverse
momenta of the four (three) hardest jets in the low (high) energy
analysis, 
the hardest charged lepton and missing energy in the event,
as the discriminating variable. We show in the left panel of
Fig.~\ref{results} a sample
distribution of $S_T$ for the main backgrounds and the signal for
$m_G=3$ GeV at the
14 TeV LHC. In order to estimate the statistical
significance of the signal, we introduce the following mass-dependent cut
on $S_T$ 
\begin{itemize}
\item $S_T>0.9,~1.1,~1.5$ TeV for $m_G=1.5,~2,~2.5$ TeV (low energy phase), 
\item $S_T>1.2,~1.5,~1.7,~2$ TeV for 
  $m_G=~2,~2.5,~3,~\geq 3.5$ TeV (high energy phase),  
\end{itemize}
and compute the statistical significance from the formula~\cite{Bayatian:2006zz}
\begin{equation}
\mathcal{S}(s,b)=\sqrt{2\times \left[ (s+b) \ln \left(
    1+\frac{s}{b}\right)
-s \right]}.
\end{equation} 
We show in the right panel of Fig.~\ref{results} the luminosity
required for a 5 $\sigma$ discovery
as a function of $m_G$ for the three energies considered. With the $\sim 5$
fb$^{-1}$ already collected at 7 TeV masses up to $m_G \approx 1.8$
TeV can be probed. This could go up to 2.7 TeV (4.6 TeV) 
for 20 fb$^{-1}$ at 8 TeV 
(100 fb$^{-1}$ at 14 TeV).

\section{Conclusions\label{conclusions}}

We have seen that a composite Higgs with $m_H=125$ GeV and a moderate
degree of compositeness is compatible with current Higgs searches. In
such a model, the Higgs boson can be produced in association with a
$t\bar{t}$ pair through the exchange of a heavy gluon with subsequent
decay into a new fermion resonance (which decays itself into the Higgs
and a SM top) and the top quark. We have shown that simple cuts on the
number of jets and b-jets can reduce the background to manageable
levels. For the highest masses probed, boosted techniques are very
useful. Using then $S_T$ as a discriminating variable allows for a
clean extraction of the signal up to masses $m_G \approx 
1.8$ TeV, 2.7 TeV and 4.6 TeV for the LHC with $\sqrt{s}=7,~8$ and 14 TeV,
respectively and an integrated luminosity of 5, 20 and 100 fb$^{-1}$,
respectively.

\section*{ACKNOWLEDGEMENTS}
We would like to thank the Les Houches school for physics for their
hospitality during the workshop where some of the work contained herein was
performed, as well as the worskhop organizers. This work has been
partially supported by projects FPA2006-05294, FPA2010-17915, FQM 101,
FQM 03048 and FQM 6552 and the FPU programme.

\AddToContent{A.~Carmona, M.~Chala, J.~Santiago}
\renewcommand{\thesection}{\arabic{section}}


\chapter{Top Polarisation in $H^-t$ and $Wt$ production}

{\it R.~M.~Godbole, L.~Hartgring, I.~Niessen, C.~D.~White}


\begin{abstract}
We consider laboratory frame observables obtained from leptonic decay products of
top quarks produced in association with a charged Higgs or $W$ boson. These 
are robust against QCD corrections to top quark decay, and can be used to pin down the parameter 
space of a charged Higgs boson, or reduce backgrounds in $H^-t$ and (Standard 
Model) $Wt$ production.
\end{abstract}

\section{INTRODUCTION}
The proximity of the top quark mass to the scale of electroweak symmetry 
breaking makes the top quark sector a potentially useful window through which
to look for new physics effects. Particularly useful in this regard is single
top production, which in the Standard Model is purely electroweak at leading
order. Single top quarks can be produced in association with other particles,
such as $W$ bosons or other quarks (in the SM), or more exotic species in new
physics scenarios. In general, a singly produced top quark may have a net
degree of longitudinal polarisation $P_t$, in contrast to the case of top 
quark pair production, which in the SM leaves the top quark {\it unpolarised} 
on average ($P_t=0$). Assuming the top quark decays according to 
$t\rightarrow Wb\rightarrow ff' b$ (where $f$ is a quark or lepton, and $f'$ 
the associated antiquark or neutrino), the decay product $f$ is distributed
in the top quark rest frame according to
\begin{equation}
\frac{1}{\Gamma_l}\frac{\mathrm{d}\Gamma_l}{\mathrm{d}\cos\theta_{f,\rm rest}}=\frac{1}{2}\left(1+ \kappa_f P_t\cos\theta_{f,\rm rest}\right).\label{eq:topdecay}
\end{equation}
Here $\Gamma_l$ is the partial decay width, and the polar angle 
$\theta_{f,\rm rest}$ is the angle between the decay product $f$ and the top spin vector in the top quark rest frame. 
$\kappa_f$ is the analysing power of the decay product $f$.
If $f$ is a lepton ($f=l$), it is approximately
independent of corrections to the top quark {\it decay}, thus depending solely
on the {\it production} stage of the top quark. In this case, 
$\kappa_l$ is approximately one. It follows that by studying
leptonic decay products of single top quarks, one may elucidate the net
polarisation of the top quark and, in turn, its coupling to BSM (or SM) particles.

Rather than relying on leptonic distributions in the top quark rest frame, it 
is more practical to consider alternative observables defined in the laboratory (lab) frame,
which require reconstruction of just the top quark direction in the lab and not 
necessarily that of the rest frame. Two such observables
are the azimuthal angle $\phi_l$ between the top quark and its decay lepton
(taking the top quark direction and beam axis to define the $(x,z)$ plane),
and the polar angle $\theta_l$ between the top quark and leptonic directions.
These were defined in Ref.~\cite{Godbole:2010kr} in the context of 
top pair production, and have subsequently been considered in $H^-t$ and $Wt$ 
production~\cite{Huitu:2010ad,Godbole:2011vw}. 

\section{$H^-t$ PRODUCTION}
\begin{figure}
\begin{center}
\includegraphics[width=0.4\textwidth]{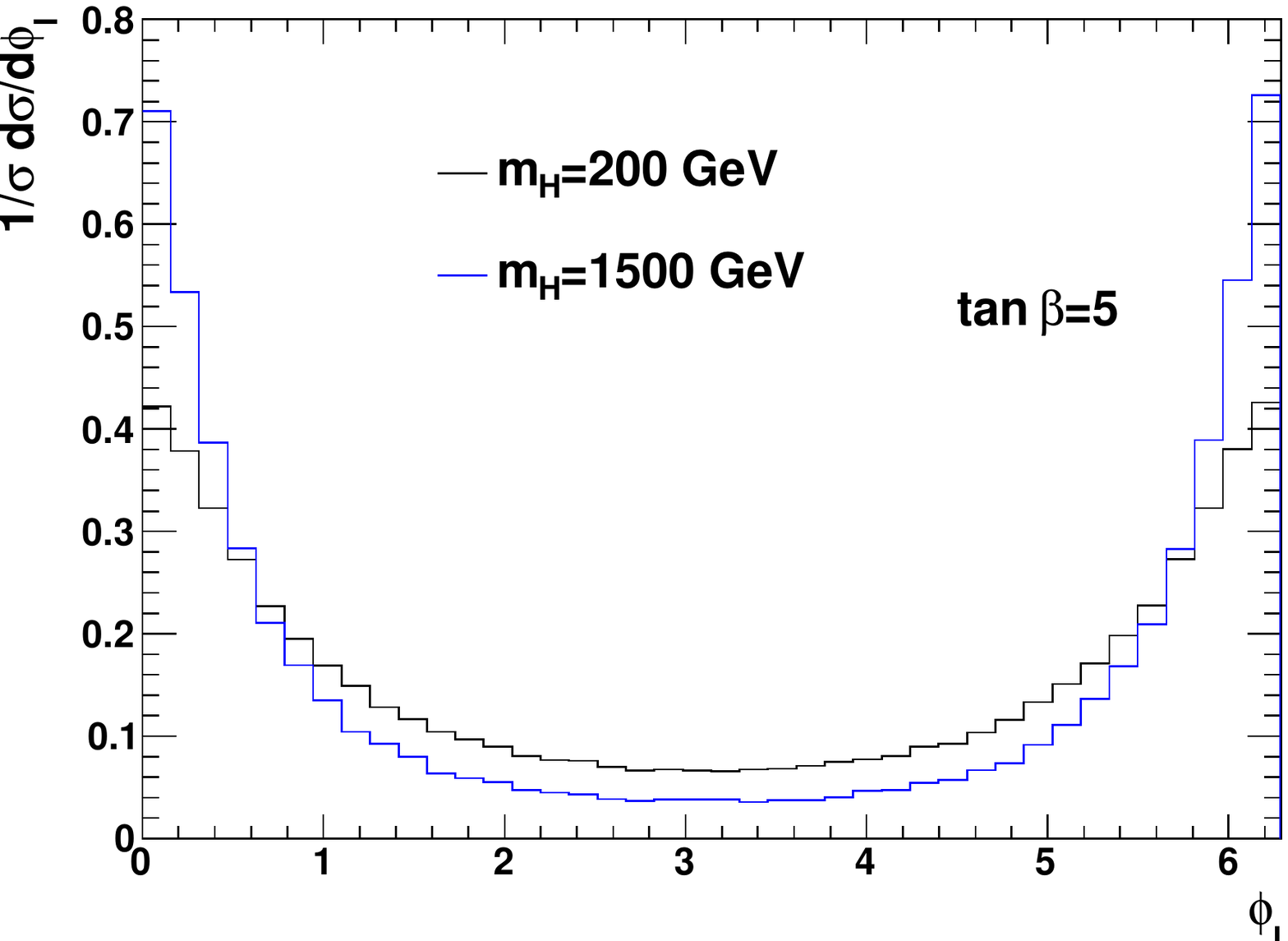}
\includegraphics[width=0.4\textwidth]{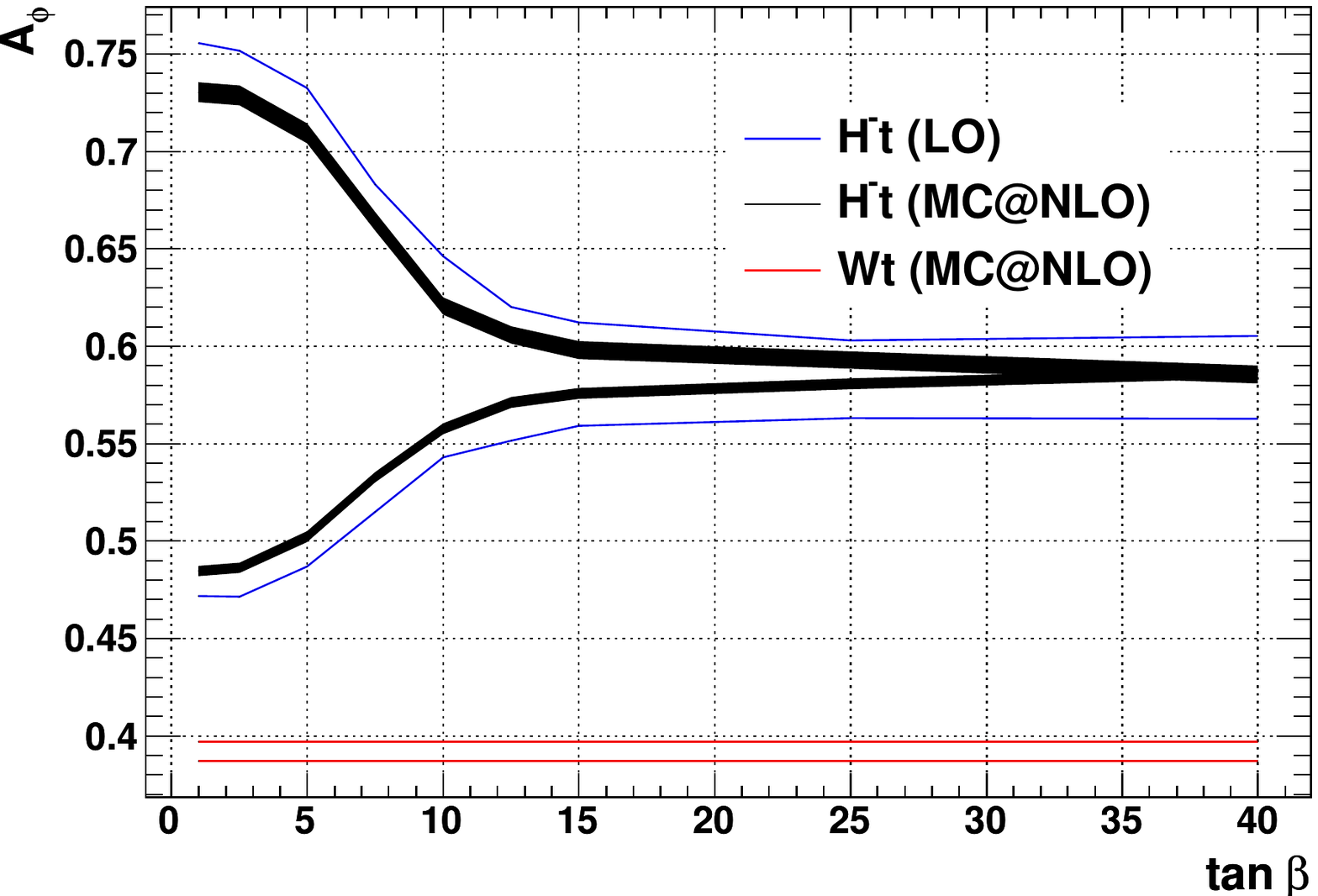}
\caption{Left: the azimuthal angle $\phi_l$ of the top quark decay lepton for 
$\tan\beta=5$ and extremal charged Higgs mass values; right: the azimuthal
asymmetry parameter, for $H^-t$ and $Wt$ production.}
\label{phiplot}
\end{center}
\end{figure}

In the context of a general type II two-Higgs doublet model, we investigated 
the above angular observables 
using recently developed MC@NLO software~\cite{Weydert:2009vr}. 
The polarisation of the top quark depends on the ratio of the vacuum 
expectation values $\tan\beta$, as well
as the charged Higgs boson mass $m_{H^-}$. Representative results for $\phi_l$ are 
shown in Fig.~\ref{phiplot}. 
In addition to the distribution itself, one may also consider the asymmetry 
parameter~\cite{Godbole:2010kr}:
\begin{equation}
A_\phi=\frac{\sigma(\cos\phi_l>0)-\sigma(\cos\phi_l<0)}{\sigma(\cos\phi_l>0)
+\sigma(\cos\phi_l<0)},
\label{Aldef}
\end{equation}
which efficiently distils the difference in distributions at different points
in parameter space. Results for $A_\phi$ are also shown in 
Fig.~\ref{phiplot}, where MC@NLO results (black) are compared with leading order (LO) results
(blue) obtained using MadGraph~\cite{Alwall:2011uj}. Note that there is a 
pronounced difference in $A_\phi$ for different points in $(\tan\beta,m_{H^-})$
space, particularly at low Higgs mass values. Furthermore, the MC@NLO curve
does not differ significantly from the LO curve, and thus the observable is 
highly stable with respect to next-to-leading order (NLO) and parton shower effects. The lower band in
Fig.~\ref{phiplot} constitutes an estimate of the $Wt$ result, using 
MC@NLO~\cite{Frixione:2008yi}. This is very different to the $H^-t$ result,
providing information useful for reducing the $Wt$ background.\\

Further useful information can be gained from considering the polar angle and a suitable asymmetry parameter is defined in Ref.~\cite{Godbole:2011vw}. This asymmetry is
capable of distinguishing different Higgs boson masses at large $\tan\beta$ (where the 
azimuthal asymmetry cannot). In Ref.~\cite{Godbole:2011vw}, we also
studied observables related to the energy of the top quark decay 
products~\cite{Shelton:2008nq}, and further useful asymmetry parameters were 
defined which add complementary information to the angular observables in the 
boosted top quark limit.\\

\section{$Wt$ PRODUCTION}
As well as being a background to $H^-t$ production, $Wt$ production is
of interest as a signal in its own right. Top quark pair
production is a major background, and one may reduce this background using
polarisation information: in pair production the $t$ quark is unpolarised
on average, whereas in $Wt$ production it is fully polarised. 
In Ref.~\cite{Godbole:2011vw} we examined the $\phi_l$ and $\theta_l$ distributions
for $Wt$ and $t\bar{t}$ production, for the $Wt$ signal cuts 
of Ref.~\cite{White:2009yt}, using MC@NLO~\cite{Frixione:2008yi}. The corresponding
asymmetry parameters are $(A_\phi,A_\theta)=(0.33,0.02)$ ($Wt$) and 
$(0.63,0.26)$ ($t\bar{t}$). There is thus a pronounced difference, which is
potentially useful for distinguishing the two processes in e.g. a multivariate
analysis.

\section*{CONCLUSIONS}
Polarisation information can efficiently be used to distinguish single top quark
production processes from their backgrounds, as well as to pin down the
parameters of a charged Higgs boson model. Lab frame leptonic observables
can be defined, which are robust against NLO and parton shower corrections.

\AddToContent{R.~M.~Godbole, L.~Hartgring, I.~Niessen, C.~D.~White}


\chapter{Probing New Physics with Single Top $+$ X}

{\it Devin G.~E.~Walker, Jiang-Hao Yu and C.-P. Yuan}

\begin{abstract}
We investigate single top production in association with a variety of Standard Model final states with a focus on the early LHC run.  Here X can be jets, large missing transverse momentum and/or electroweak gauge bosons.  For this review, we focus on searches for new $W'$ gauge bosons and heavy $B'$ quarks.  A general survey of the possible new physics with associated single top production can be found in Ref.~\cite{walkeryuyuan}.  In essence, the channel probes a variety of new physics, is intricately tied to the new physics at the TeV scale and is essential for maximizing the potential of the early LHC for new physics.  
\end{abstract}

\section{Introduction}
For the first time in history, the TeV scale is being directly probed by the CERN Large Hadron Collider (LHC).  Even at this early stage, hints of the Standard Model (SM) higgs boson~\cite{ATLAS:2012si,Chatrchyan:2012tx} as well as strong constraints on new physics have been produced.  Of particular interest are the constraints on new scalar and gauge bosons (e.g.~resonances~\cite{Aad:2012cg,Collaboration:2011dca,Aad:2011yg,Aad:2011fq,Chatrchyan:2011ns,ATLAS:2011ab,Chatrchyan:2011fq,Aad:2011tq,Aad:2011ch}) as well as new exotic fermions~\cite{Aad:2011yn,Chatrchyan:2011ay,ATLAS:2011ai} which are harbingers of natural new physics scenarios.  Some of the most popular scenarios which exhibit this new physics can be found in Refs.~\cite{Dimopoulos:1981zb,Weinberg:1975gm,Susskind:1978ms,Hill:1980at,Hill:1980sq,Chivukula:1998wd,Hill:2002ap,Kaplan:1983sm,ArkaniHamed:2001nc,ArkaniHamed:2002qy,ArkaniHamed:1998rs,Randall:1999ee,Hewett:1988xc,Chivukula:2005xm,Chivukula:2006cg,Chivukula:2009ck}.  In this review, we show how associated single top production (a top quark produced in conjunction with a variety of SM final states) can provide a sensitive probe to heavy $B'$ quarks and $W'$ gauge bosons.  
\newline
\newline
\textbf{A Motivation from Top Collider Physics:}  The LHC is a ``top factory.''  At 7 TeV, 85,000 single top quark events are expected for 1 fb$^{-1}$ of data~\cite{ATLAS-CONF-2011-101}.  Importantly, the production cross section for these background SM events are precisely known at next-to-leading order (NLO) and next-to-leading log (NLL).  In addition, processes involving single top (quark) production may be unique in their ability to probe the nature of electroweak symmetry breaking.  The top quark mass is of order the electroweak scale thereby suggesting that top quark production may be sensitive to new physics beyond the SM.  Unlike other quarks, SM top quarks decay to a $W$ boson and a bottom quark \textit{before} hadronizing.  This provides a unique opportunity to probe the underlying electroweak interactions as well as efficiently tag the events.  Observation of SM single top production has been experimentally challenging.  One of the major reasons is the jet not associated with the top decay is frequently in the forward regions in the detector.  It is therefore hard to distinguish these single top events from the dominant QCD backgrounds.  The fact that SM single top production has this difficulty is a \textbf{\textit{strength}} when considering single top production from new physics.  When new physics is produced that decays to single top $+ X$, the large mass of the new physics often results in the decay products being in the center of the detector.  
\newline
\newline
 This review paper is organized as follows:   In the next section, we outline the new physics that can be produced from single top $+ X$ production at the LHC.  As mentioned in the abstract, we restrict our focus to new $W'$ bosons as well as new heavy quarks in this review.  Please see Ref.~\cite{walkeryuyuan} for an analysis of all the new physics listed in the next section.  Appendix A includes a brief overview of our reconstruction methodology and any important definitions.  Sections 3, 4 and 5 contain our analysis for the new $W'$s and heavy quarks.  Finally we conclude.

\section{New Physics Searches with Single Top $+$ X}
\label{sec:npsearches}

This review is part of a larger series of papers searching for new physics in the single top $+$ X final state~\cite{walkeryuyuan}.  Here we list the new physics that can be probed.
\vspace{0.4cm}
\begin{center}
\begin{tabular}{c | c } 
Single Top Signatures 				& New Physics Probed  	\\ 
 (partonic level)						& 				  \\ \hline \\
  $t + b$-$\mathrm{jet}/\mathrm{jet}$		& $W'$ Gauge Bosons \\
								& Diquark Resonances \\ 
								& Colored Resonances \\	
								& Charged Higgses \\ & \\	 \hline & \\		
$t + Z/W \,(+\,\, \mathrm{jet})$  	                   & Heavy Top Quarks \\ 									
								& Heavy Bottom Quarks \\ & \\	
$t + h + \mathrm{jet}/W$ 				& SM Higgs  \\ 	& \\
 $t + h^- + b$-$\mathrm{jet}$ 			& Charged Higgses \\  & \\
$t + W + \mathrm{jet}$ 				& Charged $5/3$ Quarks  \\  & \\
$t + E\!\!\!\!\slash_{T} + \mathrm{jet}$ 	& Dark Matter  \\  
								& Top Fermion Partners (e.g. stops) \\ & \\ \hline
\end{tabular}
\end{center}
\vspace{0.4cm}
The above chart is constructed by considering all possible SM particles on the external legs.  Based on the SM symmetries, all possible new physics in the intermediate state(s) are enumerated at tree-level with the simplest tree-level topologies.  We require at least one SM top quark in the final state; and, the initial states are light quarks or gluons.  For this review, we focus on $W'$ bosons and heavy quarks. 

\section{Searching for $W'$ Bosons with  $t + b$-$\mathrm{jet}/\mathrm{jet}$}
\label{sec:Wprime}

\subsection{Theoretical Considerations for $W'$ Bosons}  

Previous studies of $W'$ bosons decaying to top quarks have focused on the LHC at center of mass energies of 14 TeV~\cite{Gopalakrishna:2010xm}.  We examine the potential of the early LHC to identify $W'$ bosons during the early LHC as well as the experimental reach.  Many natural models of new physics beyond the SM have relatively light $W'$ bosons.  They are needed to cancel quadratic divergences induced by SM $W$ bosons on the higgs potential.  However, the tightest exclusion limits come from leptophobic $W'$ bosons which decay to dijets~\cite{Chatrchyan:2011ns,Aaltonen:2008dn}
\begin{eqnarray}
 280	< m_{W'} < 1500 \textrm{ GeV}.
\end{eqnarray}
To ensure a model independent analysis, we write down the following parametrization
\begin{eqnarray}
{\cal L} =  \frac{i\,g_2}{\sqrt{2}}\,~\overline{q}_i \gamma^\mu (f_{L} P_L +  f_{R} P_R)\,V_{ij}~q'_j~W^{\prime +}_\mu + {\rm h.c.}, 
\label{eq:modelindep}
\end{eqnarray}
where $q$ and $q'$ are the up and down-type quarks, respectively.  Here $i = u, c, t$, $j = d, s,b$, and $g_2 = e/\sin\theta_W$.  Also, $V_{ij}$ is the Cabibbo-Kobayashi-Maskawa matrix and $P_{L,R}$ are the chirality projection matrices.  For simplicity in the next section, we consider only the case with a purely left-handed current ($f_{L} = 1$, and $f_{R} = 0$); our study can be extended easily to other cases.  

\subsection{Signal and Background Processes for $W^\prime$ Searches}
The $s$- and $t$-channel signal processes are
\begin{eqnarray}
	p \, p \to t \, \overline{b} + \mathrm{h.c.} &\hspace{3cm}&  p  \, p \to t  \, q\,\,\,\, \,\,\mathrm{or} \,\,\, \to \overline{t}  \, q. 
	\end{eqnarray}
For the $t$-channel processes on the right, the $t$ and $\overline{t}$ final states are equally possible because the initial $b$ and $\overline{b}$ partons occur with equal probability.  For this channel, we only consider top quarks that decay semi-leptonically.  The isolated lepton (with a sufficiently large $p_T$) allows for the event to be effectively tagged.  
The following SM backgrounds are dominant: 
\begin{eqnarray}
p   \,  p \to t   \,  \mathrm{+jet(s)}    &\hspace{2.5cm}& p   \,  p \to W   \,  \mathrm{+jets} \label{eq:Wbkg1}  \\
p  \,  p \to \bar{t}   \,  t  &\hspace{2.5cm}&   p   \,  p \to W   \,  W/Z  \label{eq:Wbkg2} 
\end{eqnarray} 
The SM $p \,p \to t   \,  \mathrm{+jet(s)} $ background is irreducible.  
The other  SM backgrounds in equations~\ref{eq:Wbkg1} and \ref{eq:Wbkg2} are important insofar as they can mimic the $t + \mathrm{jet}$ final state.  

\subsection{Event Simulations for $W^\prime$ Searches}
\begin{figure}
	\begin{center}
	\includegraphics[scale=0.31]{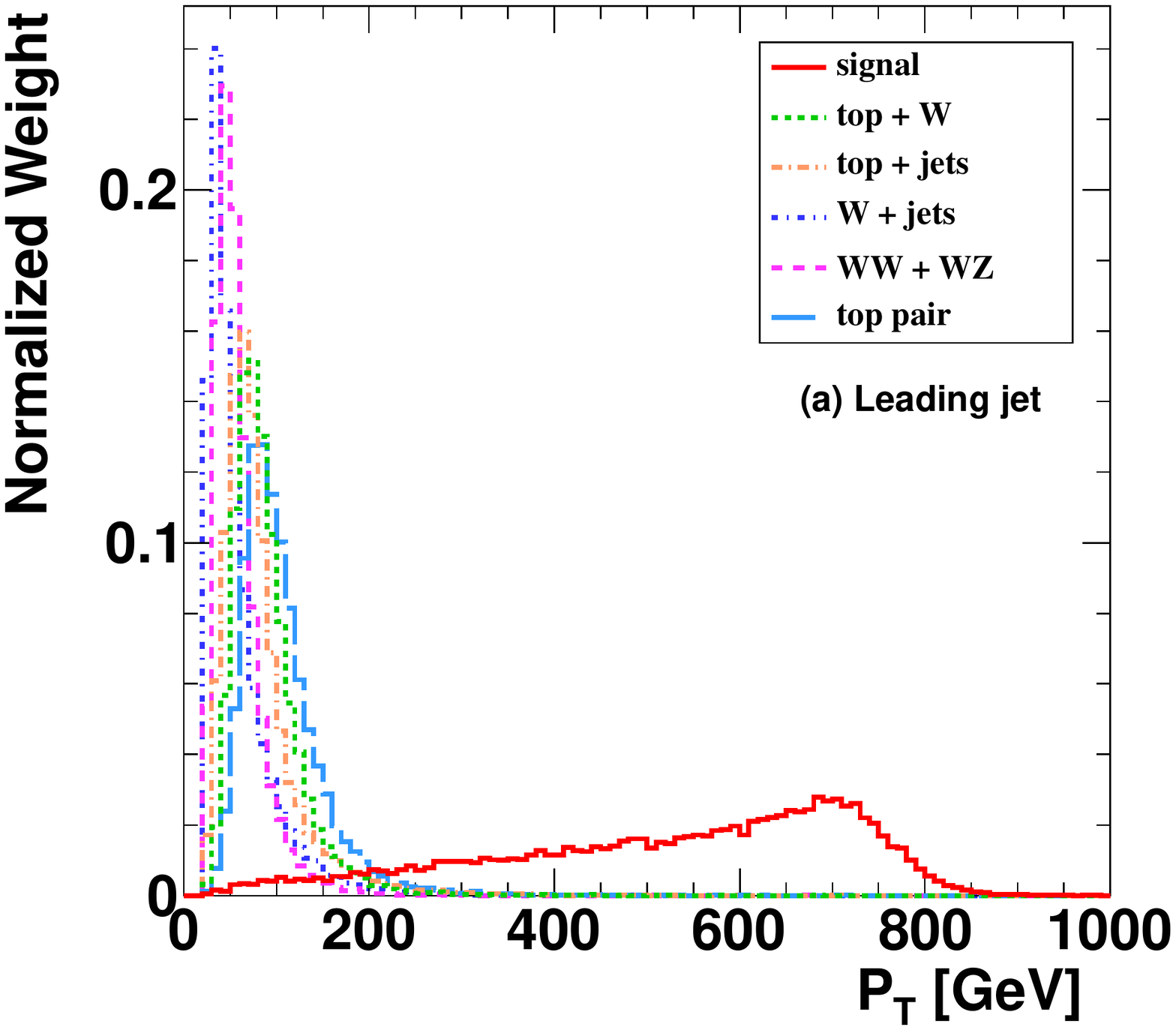} \hspace{0.5cm}
	\includegraphics[scale=0.31]{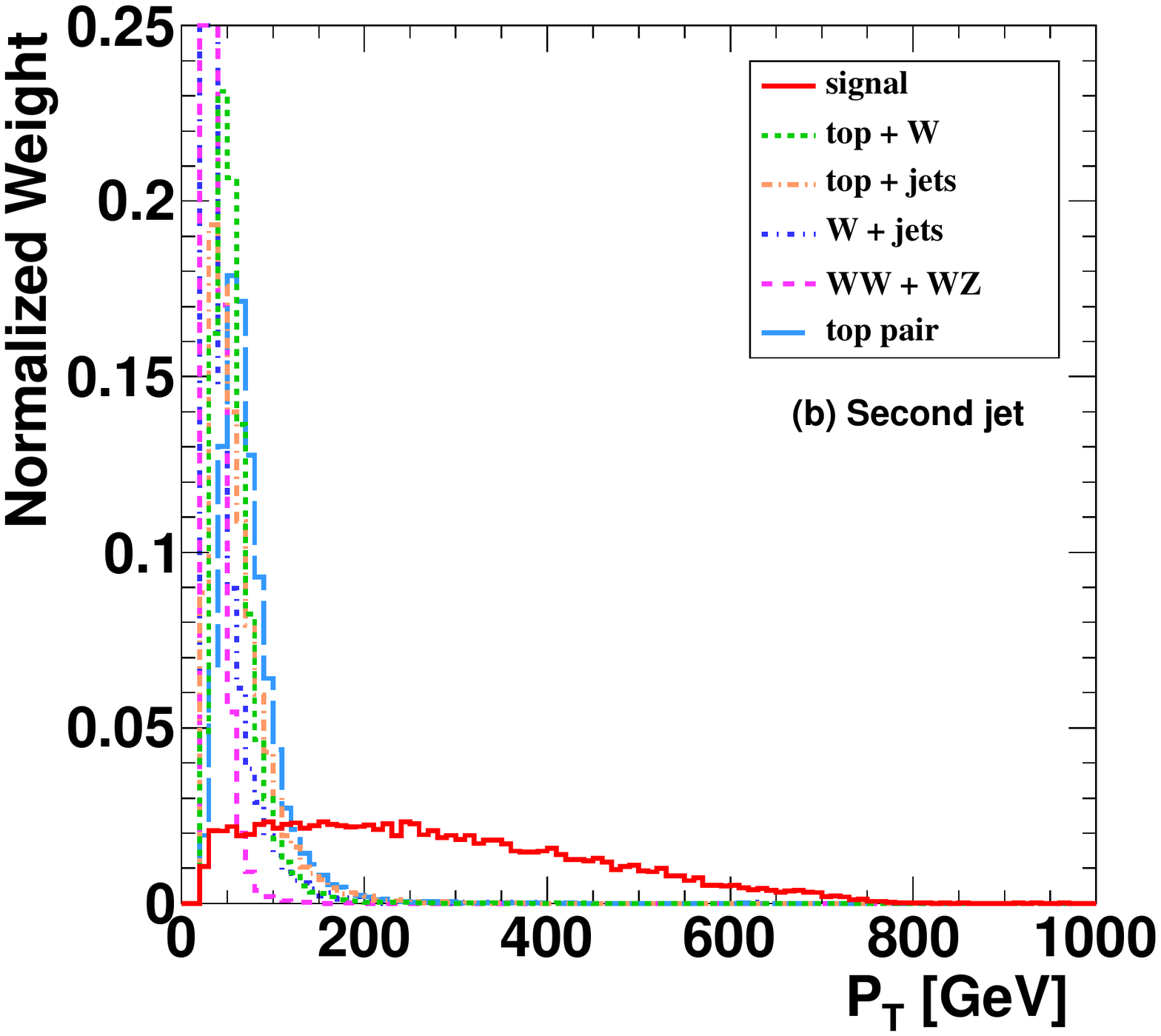} \\ \vspace{1cm}
	\includegraphics[scale=0.31]{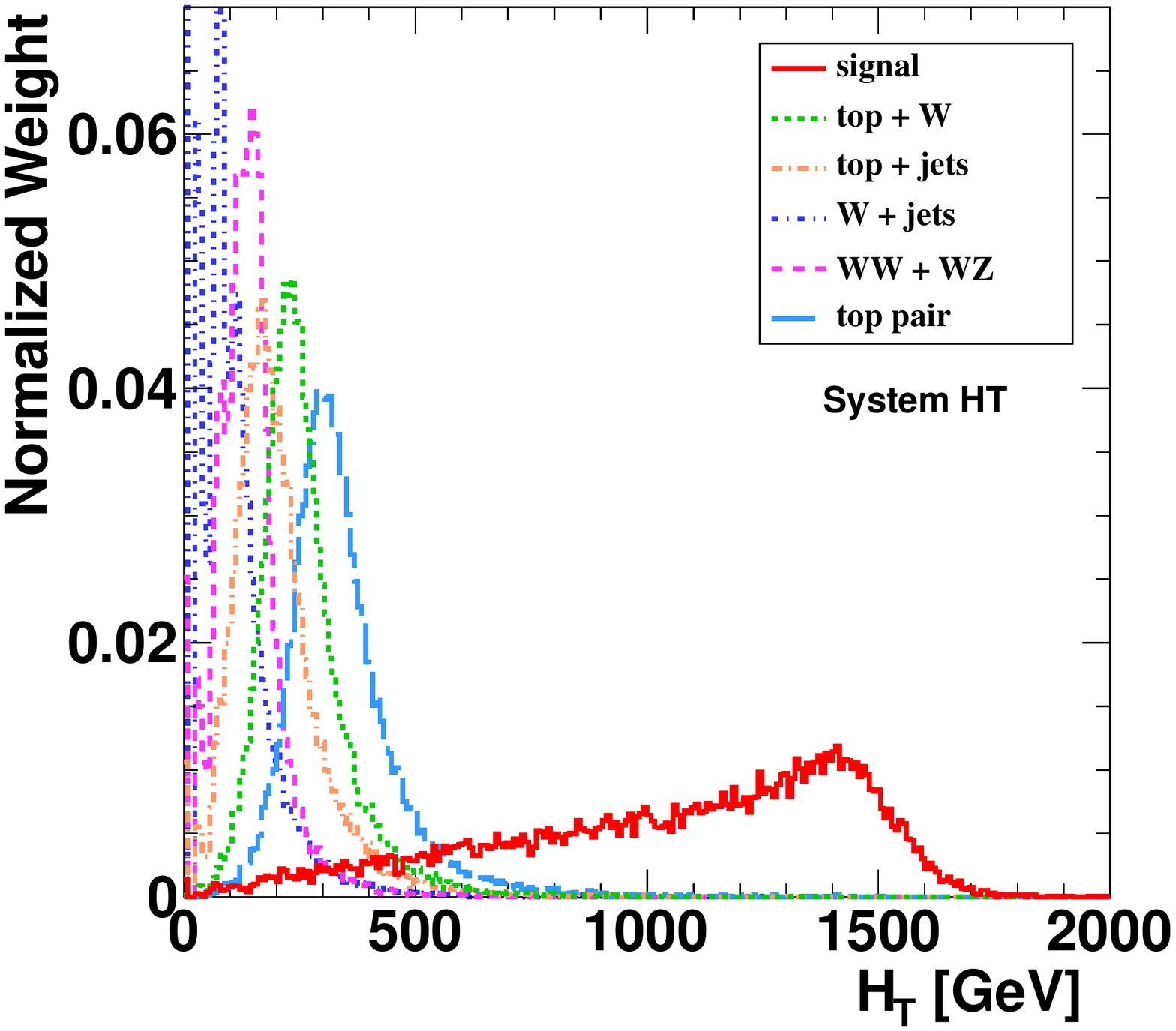} \hspace{0.5cm}
	\includegraphics[scale=0.31]{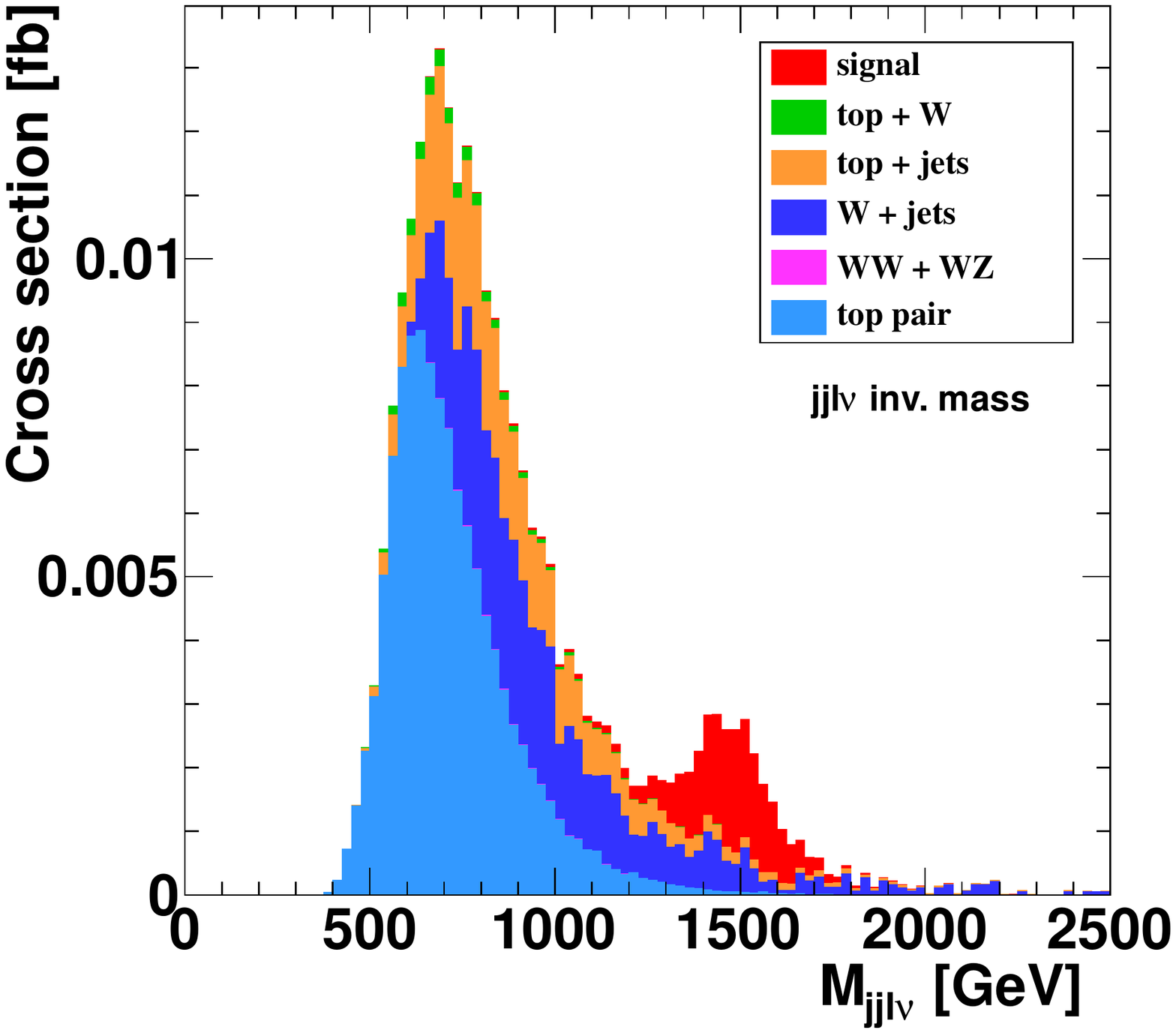}
	\end{center}
\caption{Normalized $p_{T}$ distributions of the leading jet (upper left panel) and sub-leading  jet (upper right panel) for the signal and backgrounds.  The signal process features a 1.5 TeV $W'$ boson with the coupling $f_L = 1$.  See equation~\ref{eq:modelindep} for the definition of $f_L$.  The $H_{T}$ distributions for signal and backgrounds are also plotted (lower left panel).  It is clear an $H_T$ cut can be useful in separating the signal and backgrounds.  After the cuts in equations~\ref{eq:WpTcuts}-\ref{eqW:cut2} and equation~\ref{eq:htcut} are made, the invariant mass distribution  is plotted (lower right panel).} 
\label{figW:jetpt}
\end{figure}

We focus on a 1.5 TeV $W'$ boson with coupling $f_L = 1$.  (See equation~\ref{eq:modelindep}.)  We use CTEQ6L parton distribution functions~\cite{Pumplin:2002vw} in our event simulations.  The renormalization and factorization scales are chosen to be the $W'$ boson mass.  The signal and background events are generated with MadGraph5/MadEvent.  We impose the following basic $p_T$ cuts
\begin{eqnarray}
p_T^j \geq 25\,{\rm GeV} & \hspace{3.2cm} & p_{T}^{\ell}\geq 25\,{\rm GeV}.  \label{eq:WpTcuts}
\end{eqnarray}
We also confine our search to the center of the detector
\begin{eqnarray}
&\left|\eta_{j}\right| \leq 2.5 \hspace{3.2cm} &\left|\eta_{\ell}\right| \leq 2.5.  \label{eq:Wetacuts}
\end{eqnarray}
As mentioned in the introduction for SM single top production, the jet not associated with the top quark decay is often forward in the detector.  Restricting our focus to the center of the detector helps to eliminate this background.  As mentioned before, we also require the leptons and jets to be well separated.  We require the following cone sizes
\begin{eqnarray}
\Delta R_{jj} > 0.4 & \hspace{2.5cm} \Delta R_{j\ell} > 0.4 \hspace{2.5cm} & \Delta R_{\ell\ell} > 0.2. 
\end{eqnarray}
Since we require the top quark to decay semi-leptonically, the resulting neutrino introduces missing transverse momentum ($\not{\!{\rm E}}_T$) in the event.  To further reduce the SM background, we require
\begin{equation}
\not{\!{\rm E}}_T > 25\,\,{\rm GeV}. \label{eq:Wetmiss}
\end{equation}
The $t + \mathrm{jet}$ signature has a final state with one isolated charged lepton (electron or muon), two high $p_T$ jets, and large missing transverse momentum ($E_{T}\!\!\!\!\!\!\!\slash\,\,\,$) from the missing neutrino.  Large $p_T$ cuts on the leading jet can be used to suppress most of the backgrounds.  In the upper panels of Figure~\ref{figW:jetpt}, we show the $p_T$ distributions for the leading and sub-leading jets for the signal and background.  As reminder, we consider the 1.5 TeV $W'$ boson with $f_L = 1$ as the signal; the discriminating power of the $p_T$ cuts is clear in those plots. 
%
We require
\begin{eqnarray}
p_T &\geq& ~200\,\,{\rm GeV}\,\,\,\,\,(\mathrm{leading\,\,jet}),  \label{eqW:cut1} \\
p_T&\geq& ~80\,\,{\rm GeV}\,\,\,\,\,\,(\mathrm{subleading\,\,jet}). \label{eqW:cut2}
\end{eqnarray}
Another useful variable is the scalar sum of the $p_T$'s of all the particles in the final state.  In Appendix A, we define
\begin{eqnarray}
H_T= p_T^{\ell^+} + \not{\!{\rm E}}_T + \sum_{j} p_T^j.
\label{eqW:cut3}
\end{eqnarray}
Here $j$ runs over all of the well-separated jets in the event.  We apply the following cut
\begin{eqnarray}
H_T \geq 600\,\,\mathrm{GeV}.
\label{eq:htcut}
\end{eqnarray}
In Figure~\ref{figW:jetpt}, a slight signal is visible near 1.5 TeV.  To further optimize the signal, we apply b-tagging and perform standard sideband analysis.  We apply the following invariant mass window cut
\begin{equation}
|m_{l\nu jj} - M_{W^\prime}| < 400\,\,\,\mathrm{GeV}.
\label{eq:Wmasscut}
\end{equation}
Here $M_{W^\prime}$ center of the resonance peak in the invariant mass distribution in Figure~\ref{figW:jetpt}.  Of course, the $\nu$ momentum is determined with the top reconstruction in Appendix A.  After all of the cuts, a 1.5 TeV $W'$ boson with a coupling of $f_L = 1$ is clearly visible.  For reference, after all of the cuts, the $s$-channel and $t$-channel signal cross sections are 
\begin{eqnarray}
\sigma (p \,p \to W^{\prime}\to b \, l^\pm  \,\nu \, \bar{b})  &=& \ 13.136 \pm 0.0246   \,\,\mathrm{ pb,} \\
\sigma (p \, p \to t(\bar{t}) \,q \to b (\bar{b})  \, l^\pm \, \nu  \,q) & = & \ 8.5404 \pm 0.0218 \,\,\mathrm{ fb.}
\end{eqnarray}
Table~\ref{stopx_tab1} gives a breakdown of the effect of the cuts on the signal and background cross sections.
\begin{table}[t]
\label{tabW:cs}
\begin{center}
\vspace{0.5cm}
\begin{tabular}{|l|c|c|c|c|c|c|}  \hline
$\sigma$(fb)         	& Signal	& $t + W$	& $t +\rm{jets}$	& $t\bar{t}$& $WV$	& $W+\rm{jets}$\\ \hline\hline
No cuts                      	&  58.96   	&  2861  	&  18877      	& 25840 	& 9888     	& 4018600   \\ \hline
Basic cuts (in eq.~\ref{eq:WpTcuts}-\ref{eq:Wetmiss}) 
				&  33.38   	&  833   	&  4049      	& 7207    	& 2265 	& 284516   \\ \hline
~+ hard $p_T$ cuts (in eq.~\ref{eqW:cut1}-\ref{eqW:cut2})  
				&  28.65  	&  11.1 	&  158.5     	& 232.6	& 3.67	& 7836.2   \\ \hline
~+ $H_T$ cut (in eq.~\ref{eq:htcut})    
				&  26.45   	&  5.69 	&  59.8      	& 137.6	& 3.51   	& 3782   \\ \hline
~+ Tagged $b$-jet               
				&  22.36   &  3.44	&  44.8      & 115.7    & 0.19  & 61.98   \\ \hline \hline 
~+ Mass window  (in eq.~\ref{eq:Wmasscut})      &  21.59   & 0.13       & 7.80      & 3.97   & 0.03   & 16.02   \\ \hline
\end{tabular}
\end{center}
\caption{Cross sections for the signal and various background processes without and with the cuts are listed.}
\label{stopx_tab1}
\end{table}

\subsection{Discovery Potential and Chiral Properties for $W^\prime$ Boson Searches}
\begin{figure}
      \begin{center}
	\includegraphics[scale=0.325]{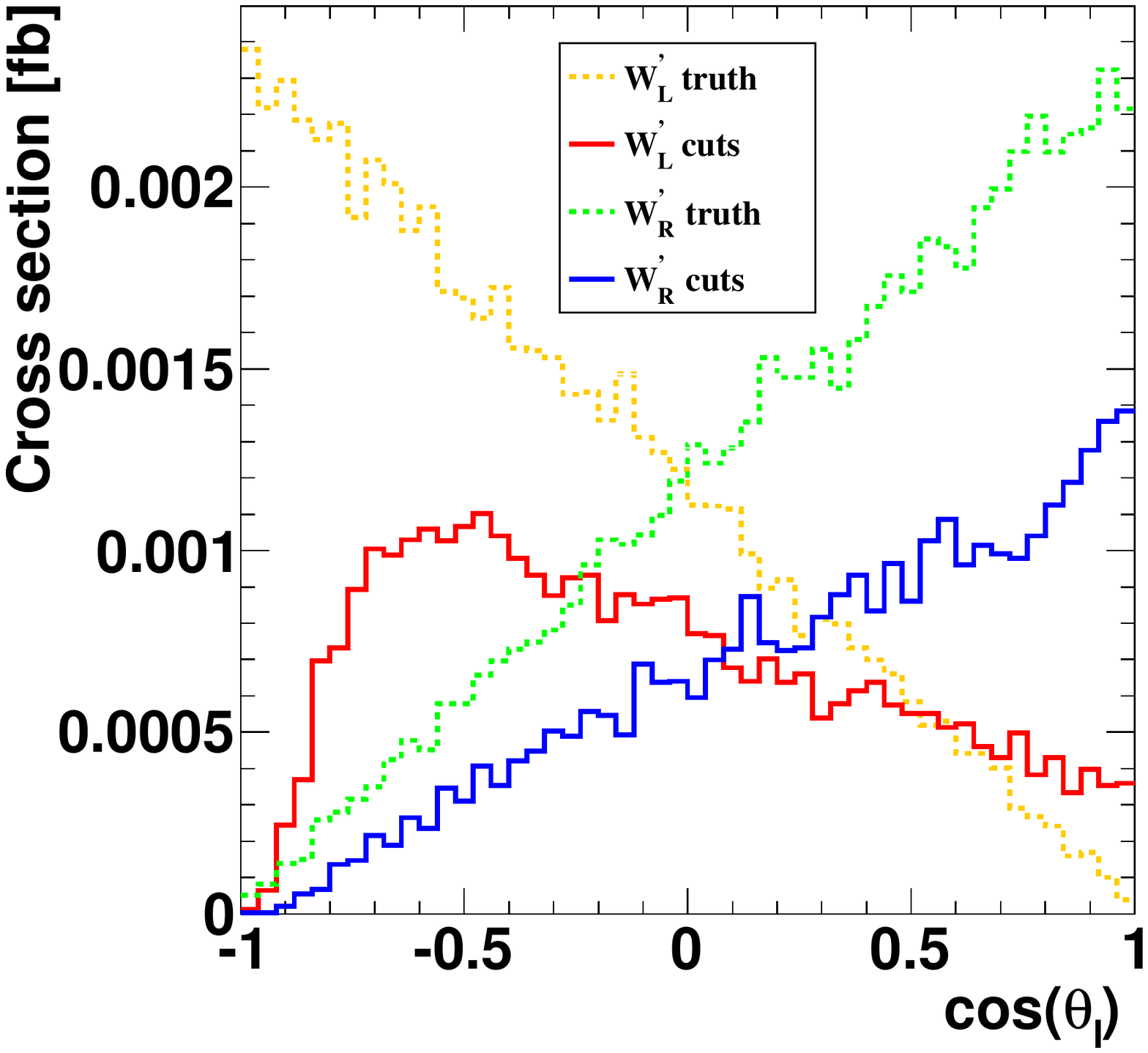}
	\end{center}
\caption{The angular distributions of the final lepton $\cos\theta_l$ for the left-handed $W^\prime$.  See Appendix A for definitions of $\cos\theta_l$.}
	\label{figW:toppol}
\end{figure}
The SM backgrounds are suppressed efficiently such that less than 1 background event survives after cuts with an integrated luminosity of 10 fb $^{-1}$.  We now consider a $1$ GeV $W^\prime$ gauge bosons with $f_L = 1$.  For that amount of luminosity we achieve
\begin{equation}
S/\sqrt{B} \sim 5
\end{equation}
where $S$ and $B$ denotes the number of signal and background events, respectively.
\newline
\newline
After the discovery of the $W^\prime$ boson, one would like to know its mass, spin, and couplings.   Angular distributions of its decay products can be investigated to definitively determine the spin and chiral structure of the $W^\prime$ boson to the SM fermions. The chirality of the $W^\prime$ coupling to SM fermions is best measured from the polarization of the top quark. Among the top quark decay products, the charged lepton from $t \to  b + l + \nu$ is the best analyzer of the top quark spin. For a left-handed top quark, the charged lepton moves preferentially against the direction of motion of the top quark, while for a right-handed top quark the charged lepton moves along the direction of motion of the top quark. The angular correlation of the lepton is $(1 \pm \cos \theta_l)/2$, with the ($+$) choice for right-handed and ($-$) for left-handed top quarks, where $\theta_l$ is  the angle of the lepton in the rest frame of top quark relative to the top quark direction of motion in the center-of-mass (c.m.) frame of the incoming partons.  In Figure~\ref{figW:toppol}, we plot $\cos \theta_l$ for $f_L = 1$, $f_R = 0$ ($W^\prime_L$) as well as $f_L = 0$ and$f_R = 1$ ($W^\prime_R$).  We expect a flat angular distribution for the SM $t\bar{t}$ background because the top quark and anti-top quark are not polarized. Therefore, the angular distributions of the lepton can be used to discriminate $W^\prime$ models in which the chirality of the $W^\prime$ coupling to SM fermions differs.

\section{Searching for Heavy Top and Bottom Quarks with $t + W + \mathrm{jet}$}

\subsection{Theoretical Considerations for Heavy Quarks} 

In many models of natural electroweak symmetry breaking, the following tree-level coupling exists
\begin{eqnarray}
B' \to W \,  t  \label{eq:Tdecay2}
\end{eqnarray}
with $\mathcal{O}(1)$ strength.  Often the heavy bottom quarks are vector-like because of the minimal impact on precision electroweak bounds.  In natural theories, heavy top partners (and the corresponding bottom partners) are \textit{essential} for canceling the quadratic divergences induced by the SM top quark on the higgs potential.  Thus, these particles are expected to be relatively light.  Recent experimental bounds on heavy tops and bottom quarks are~\cite{Chatrchyan:2011em}
\begin{eqnarray}
	m_{T'} \ge 311 \textrm{ GeV}, \hspace{2.5cm}  m_{B'} \ge 338 \textrm{ GeV}\,.
\end{eqnarray}
We show this bound can be improved during the early LHC running.  In this review, we restrict our focus to $B'$ heavy quarks production with a $t + W$ final state.  We parameterize the heavy quark interactions in a manner similar to Ref.~\cite{Atre:2011ae} in order to keep our analysis model independent\footnote{In Ref.~\cite{Atre:2011ae}, the new heavy quarks couple preferentially to first generation quarks.  Thus, the decay of the heavy quarks will involve light quarks instead of the top.}.  After electroweak symmetry breaking, the effective couplings between the electroweak gauge bosons, the third generation SM quarks and the new heavy quarks is
\begin{eqnarray}
	{\cal L} &=& \frac{g_2}{\sqrt{2}}\, W^+_\mu \biggl(\bar{t}\, \gamma^\mu (f_{tB'}^L P_L + f_{tB'}^R P_R) B' \biggr) 
	+ \frac{g_2}{\sqrt{2}} \,W^-_\mu \biggl( 
	\bar{b} \,\gamma^\mu (f_{bT'}^L P_L + f_{bT'}^R P_R) T' \biggr)\nonumber\\
	&+& \frac{g_2}{2\,c_W} Z_\mu \,\biggl(\bar{t} \,\gamma^\mu (f_{tT'}^L P_L + f_{tT'}^R P_R) T' + 
	\bar{b} \,\gamma^\mu (f_{bB'}^L P_L + f_{bB'}^R P_R) B' \biggr) + h.c. 
	\label{eq:effectivequarkcoupling}
\end{eqnarray}	
Here $g_2$ is the $SU(2)_L$ coupling, $c_W$ is the cosine of the Weinberg angle, $P_{L,R}$ are the usual projection operators and $f$ parametrizes the chirality of the couplings.  For the vector-like heavy quarks, the couplings $f_{qQ'}$ in the above Lagrangian can be parametrized in a model independent way as
\begin{eqnarray}
	f_{qQ'}^{L,R} \simeq \frac{v}{m_{Q'}} \kappa_{qQ'}^{L,R}\,,
\end{eqnarray}
where $\kappa_{qQ'}^{L,R}$ is a dimensionless parameter. 
For the chiral heavy quarks, the couplings $f_{qQ}$ are equal to 
\begin{eqnarray}
	f_{qQ'}^{L} = \left|V_{qQ'}^{CKM}\right|\,, \quad f_{qQ'}^{L} = 0\,.
\end{eqnarray}
In order to generate the heavy bottom quark resonance, we require the following magnetic moment operator
\begin{equation}
\mathcal{O}_1 = {g_s\,\lambda_i^2 \over 2 M_{B'}}\,\overline{B'}\,\sigma^{\mu\nu} \,G_{\mu\nu}\,b,
\end{equation}
where $G_{\mu\nu} = G^a\, T^a/2$ is the field strength tensor of the gluon and $g_s$ is the strong coupling constant.  By naive dimensional analysis, we take the suppression scale to be twice the excited resonance mass.

\subsection{Signal and Background Processes for Heavy Bottom Quarks}

As alluded to above, for heavy bottom quarks the signal production processes are
\begin{eqnarray}
	p \,p \to B'  \to t\, W^- 
\end{eqnarray}
with the final decays
\begin{eqnarray}
	W^\pm \to l^\pm\, \nu \ ,&\hspace{3cm}& t(\bar{t}) \to b(\bar{b})l\,^\pm \,\nu.
\end{eqnarray}
which is clean for discovery.
In addition to the backgrounds listed in equations~\ref{eq:Wbkg1}-\ref{eq:Wbkg2}, an additional background to be considered is 
\begin{eqnarray}
p   \,  p &\to& t   \, W.  \label{eq:HQbkg1} 
\end{eqnarray} 
The single top SM background in equation~\ref{eq:HQbkg1} is irreducible.  

\subsection{Event Simulations for Heavy Bottom Quarks ($t + W + \mathrm{jet}$)}

\begin{figure}
\begin{center}
	\includegraphics[scale=0.31]{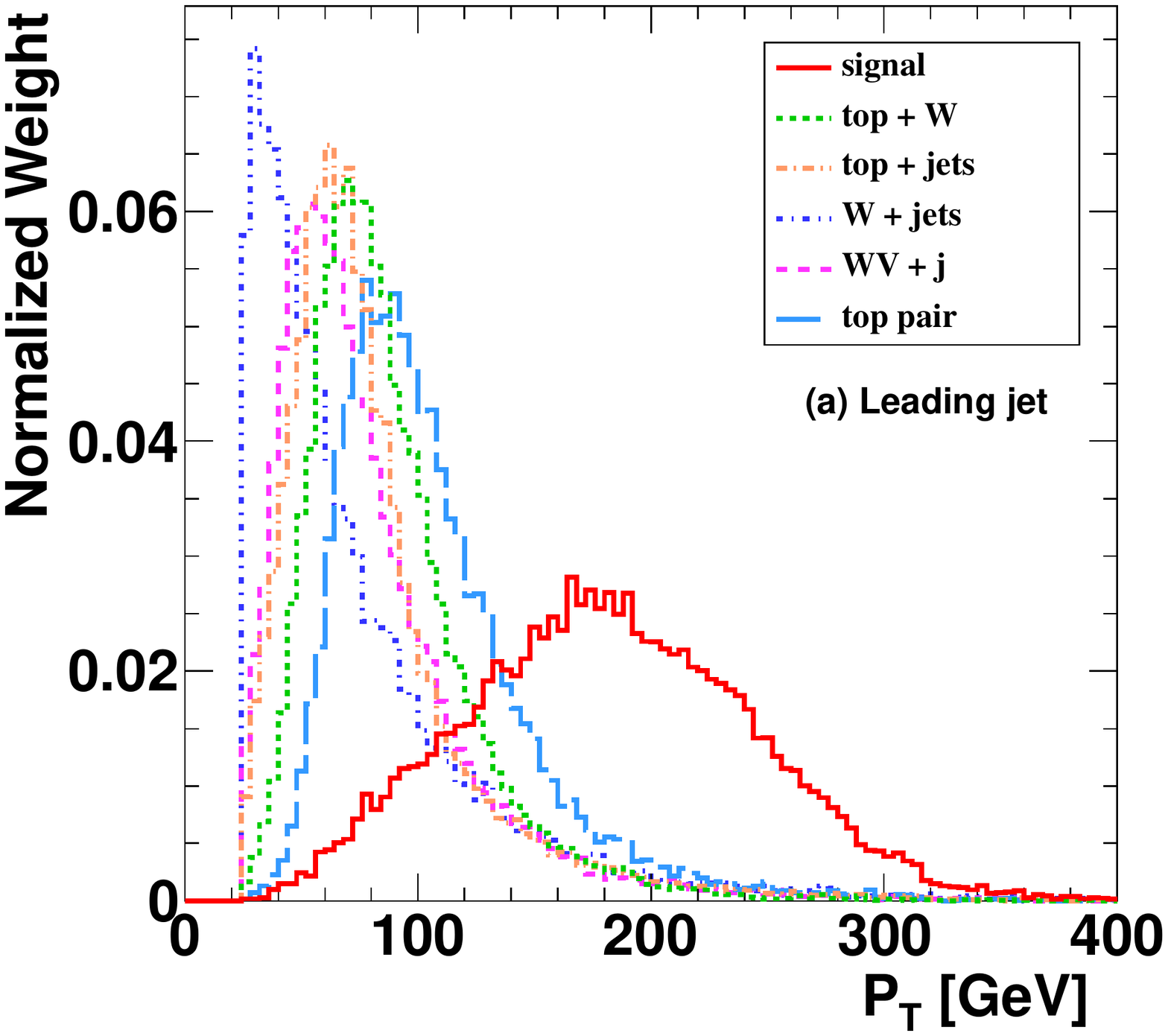}  
	 \includegraphics[scale=0.31]{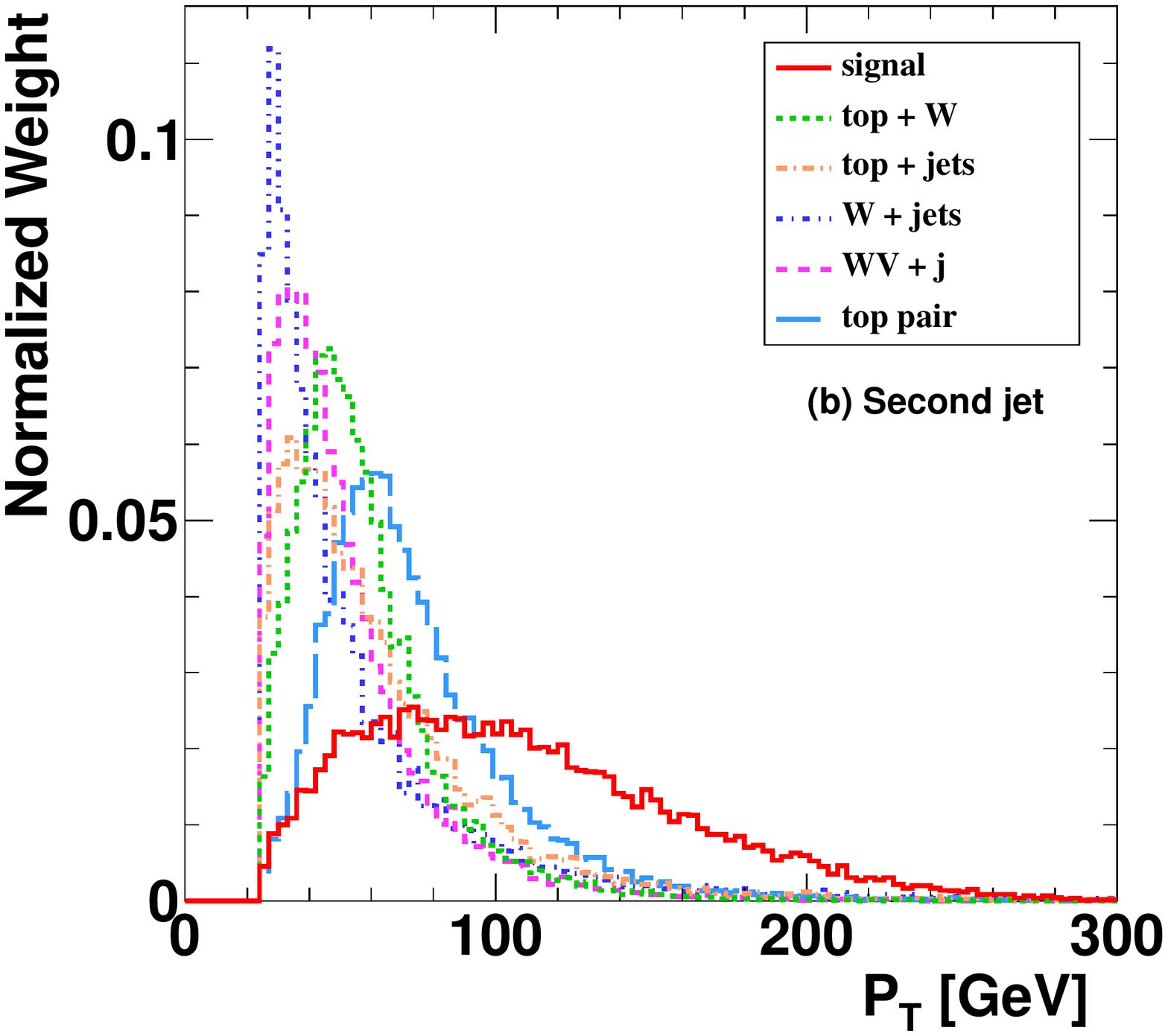} \\ 
	\includegraphics[scale=0.31]{stopx/SysInvHT0.pdf}  
	\includegraphics[scale=0.31]{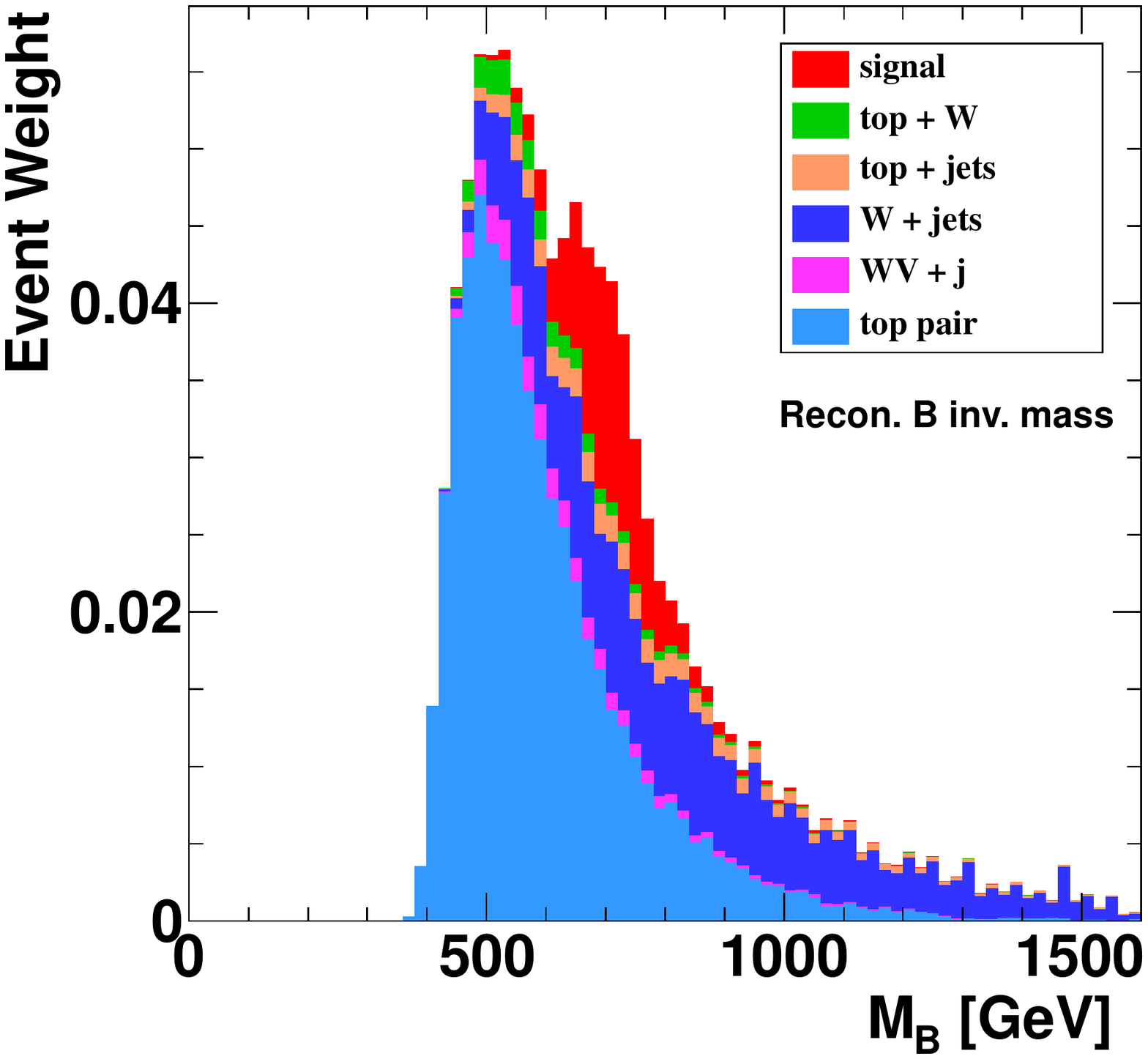}
	\label{figT:jetpt}
\end{center}
\caption{$p_{T}$ distributions from the leading jet (upper right) and subleading jets (upper left).  The $H_T$ distribution is in the lower left panel.  It is clear these variables can be used to separate the signal from the background.  After the cuts, the reconstructed $B'$ invariant mass distributions for signal and backgrounds is plotted in the lower right panel.} \label{stopx:bprimekin}
\end{figure}
\begin{table}[tb]
\label{tabB:cs}
\begin{center}
\vspace{0.5cm}
\begin{tabular}{|l|c|c|c|c|c|c|}  \hline
$\sigma$(fb)                     & Signal     & $t + W$          & $t +\rm{jets}$& $t\bar{t}$ & $WV+\rm{jets}$& $W+\rm{jets}$\\ \hline\hline
No cuts                          &  681.570   &  2861.00         &  18877.0      & 22200.0    & 10007     & 2457400   \\ \hline
Basic cuts  &  326.199   &  833.695         &  4049.12      & 5575.53    & 1721.2    & 88343.5   \\ \hline
~+hard cuts  &  210.469   &  151.776         &  257.671      & 2888.22    & 288.202   & 23713.9   \\ \hline
~+HT cuts (in Eq.~\ref{eqHQ:cut4})    &  196.531   &  52.3563         &  90.6096      & 1247.64    & 125.087   & 12901.4   \\ \hline
~+Tagging one $b$-jet               &  118.559   &  33.4737         &  46.2486      & 580.530    & 32.5228   & 284.724   \\ \hline \hline 
~+Mass window cuts in Eq.\ref{eqB:cut4}&  98.4187   &  12.7315         &  17.9331      & 147.630    & 10.0070   & 81.3498   \\ \hline
\end{tabular}
\end{center}
\caption{Cross sections for the signal and various background processes without and with the cuts are listed.} \label{stopx:tabbp}
\end{table}
The basic selection cuts are all the same as for the $W'$ boson search, up to the requirement 
\begin{equation}
\not{\!{\rm E}}_T > 25\,\,{\rm GeV}.
\end{equation}
In addition to these basic cuts, we now place additional cuts to optimize the signal. 
\newline
\newline
From Fig.~\Ref{stopx:bprimekin}, it is clear the $p_T$ of leading jet for the signal is much larger than the backgrounds.  We require
\begin{eqnarray}
p_T &\geq& ~80\,\,{\rm GeV}\,\,\,\,\,(\mathrm{leading\,\,jet}) \label{eqHQ:cut1} \\
p_T&\geq& ~50\,\,{\rm GeV}\,\,\,\,\,\,(\mathrm{subleading\,\,jet}). \label{eqHQ:cut2} \\
p_T&\geq& ~40\,\,{\rm GeV}\,\,\,\,\,\,(\mathrm{additional\,\,jet}). \label{eqHQ:cut3}
\end{eqnarray}
We again define the scalar sum of the $p_T$'s of all the particles in the final state.
\begin{eqnarray}
H_T= p_T^{\ell^+} + \not{\!{\rm E}}_T + \sum_{j} p_T^j.
\label{eqHQ:cut4}
\end{eqnarray}
Here $j$ runs over all of the well-separated jets in the event.  We apply the following cut
\begin{eqnarray}
H_T \geq 425\,\,\mathrm{GeV}.
\label{eq:HQhtcut}
\end{eqnarray}
We impose the invariant mass window cuts 
\begin{eqnarray}
\left|m_{l\nu bjj} - m_{B'}\right| < 100 \textrm{  GeV}\,. 
 \label{eqB:cut4}
\end{eqnarray}
After these cuts, we show the $B'$ resonance in Fig.~\Ref{stopx:bprimekin}.   For the heavy $B'$ quark with 700 GeV mass, the significance with 1\,fb$^{?1}$ luminosity is 
\begin{equation}
S/\sqrt{B} \sim 6.
\end{equation}
%
Cross sections for the various background processes without cuts are given in Table~\ref{stopx:tabbp}.

\section{Conclusions}

In this review, we focused on searches for new $W'$ gauge bosons as well as $B'$ heavy quarks.  This is done under the rubric of searches for new physics under single top $+ X$.  A general survey of the possible new physics with associated single top production can be found in Ref.~\cite{walkeryuyuan}.  We showed TeV $W'$ bosons and $B'$ heavy quarks are easily accessible during the early LHC running.

\section*{Acknowledgements}
We would like to thank S.~Chivukula J.~Hewett, T.~Rizzo, E.~Simmons and B.~Shuve for useful discussions.  The work of J.-H.Y. and C.-P. Y. is supported in part by the U.S. National Science Foundation under Grant No. PHY-0855561.  This work of D.W. is supported in part by a grant from the National Academies of Science and the LHC Theory Initiative.


\vspace{0.5cm}
\appendix

A few basic definitions were used in the analysis in the previous sections are given here. We place some important definitions not included in the above text; if needed, see Ref.~\cite{Barger:2006hm,Han:2005mu} for additional background.

\section{ Detector Effects}

We often use b-tagging to increase the signal efficiency.  A  tagging efficiency of 60\% is used in our analysis. We take into account a mis-tag rate for a light non-b quark (including the charm quark) to mimic a b jet, with mistag efficiency $\epsilon_{j \to b} = 0.5\%$ 
\newline
\newline
We smear the electron and hadronic energy according to Gaussians given by 
\begin{eqnarray}
{\Delta E^e_j \over E^e_j} &=& {0.05 \over \sqrt{E_j/\mathrm{GeV}}} \oplus 0.0055 \\ 
{\Delta E^h_j \over E^h_j} &=&  {1 \over \sqrt{E_j/\mathrm{GeV}}} \oplus 0.05,
\end{eqnarray}
respectively.  The muon momentum is smeared by
\begin{eqnarray}
{\Delta p^l_T \over p^l_T} &=& 0.36\,{p_T^l \over \mathrm{TeV}}  \oplus {0.013 \over \sqrt{\sin\theta}}.
\end{eqnarray}
where $\theta$ is the polar angle of the lepton with respect to the beam direction in the lab frame.  These smearing parameters are consistent with both the ATLAS and CMS experiments.  Finally, we use FastJet to cluster the final state jets with a cone size $R = 0.4$.  

\section{Top Tagging and Reconstruction}

\textbf{Hadronic Top Reconstruction:}  Here the top quark decays hadronically to a b-jet and a $W$ bosons which subsequently decays to jets.  In order to tag the top simply require the following invariant masses
\begin{eqnarray}
m^2_W = (p_1 + p_2)^2  &\hspace{2.5cm}& m^2_t = (p_1 + p_2 + p_3)^2.
\end{eqnarray}
to be within the ranges
\begin{eqnarray}
 77 < m_W <  83\,\,\,\mathrm{GeV} &\hspace{2.5cm}& 168 < m_t < 176\,\,\,\mathrm{GeV}.
\end{eqnarray}
\newline
\newline
\textbf{Semileptonic Top Reconstruction:}  For simplicity, we consider top decays to electrons or muons only.  With only one neutrino in the final state, the missing momentum generates four unknowns.  The missing transverse momentum
\begin{eqnarray}
\vec{p}_{T}\!\!\!\!\!\!\slash \,\,\,= - \sum_i \vec{p}_{T_i} &\hspace{3cm}& E_{T}\!\!\!\!\!\!\!\slash  \,\,\,\,= - \sum_i E_i,
 \end{eqnarray}
determines only two of the four components of the missing neutrino momentum.  The rest of the components of the momentum can be determined by taking the $W$ boson and top quark mass as an input for on-shell top quark production and decays.  To see this explicitly, we first demand that  $m_{l \nu}^2 = M_W^2$.  The longitudinal momentum of the neutrino is formally expressed as
\begin{equation}
p_{ L}^{(\nu)}  ={1 \over {2\, p_{T}^{(l)\,2}}}
 \left( {A\, p^{(l)}_{L} \pm E^{(l)} \sqrt{A^2  - 4\,{p}^{(l)\,2}_{T}E_{T}\!\!\!\!\!\!\!\slash\,\,\,^2}} \right),
 \label{eq:pLv1}
\end{equation}
where $A = M_W^2 + 2 \,\vec{p}_{T}^{\,(l)} \cdot \,\vec{E}_{T}\!\!\!\!\!\!\!\slash\,\,\, $.  Again $l = e,\mu$.  The two-fold degeneracy of the neutrino longitudinal momentum (above) is resolved by considering the top quark mass.  Thus, if $A^2 - 4 \, {p}^{(l)\,2}_{T}\,E_{T}\!\!\!\!\!\!\!\slash\,\,\,^{\,2} \geq 0$, the value of $p_{\nu L}$ that best yields the known top mass via $m_{l\nu b}^2 = m_t^2$ is selected.  
\newline
\newline
Detector effects can reduce the top reconstruction efficiency and, in particular, force the radical in equation~\ref{eq:pLv1} to be negative. In this case, in order to better recover the correct kinematics~\cite{Barger:2006hm}, we instead first reconstruct the top quark directly by demanding $m_{l\nu b}^2 = m_t^2$.  The longitudinal momentum of the neutrino is expressed as
\begin{eqnarray}
p^{(\nu)}_{L} & = &  {A'\, p_{(bl)\,L} \over 2(E_{(bl)}^{2} - p^{2}_{(bl)\,L})} \ \\
&\pm&  {1\over 2(E_{(bl)}^2 - p_{(bl)\, L}^2)}\,\biggl( {{p_{(bl)\,L}^2 A'^{\,2}  + (E_{(bl)}^2 - p_{(bl)\, L}^2) \,
( A'^{\,2} - 4 E_{(bl)}^2 E_{T}\!\!\!\!\!\!\!\slash\,\,\,^2  )} }\biggr)^{1/2} \nonumber
\end{eqnarray}
where $A' = m_t^2 - M_{(bl)}^2 + 2 \,\vec{p}_{(bl)\, T} \cdot \,\vec{\vec{E}_{T}\!\!\!\!\!\!\!\slash\,\,\,} $.  Here $l$ runs over $e,\mu$.  The two-fold ambiguity is broken by choosing the value that best reconstructs 
$M_W^2 = m_{l \nu}^2 $.  Notice because $A' \sim m_t^2$ in the equation above, the value under the radical has a smaller chance of being negative.

\section{Spin Determination: Lepton Polarization}  

The charged lepton from top quark decay is maximally correlated with top quark 
spin.  The connection between top quark spin and the charged lepton can be found from 
the distribution in $\theta_{\rm hel}$, 
the angle of the lepton in the rest frame of top quark relative to the topquark
direction of motion in the overall c.m.~frame. This is usually named as ``helicity'' basis. 
The angular correlation of the lepton $\ell^+$ is given by 
\begin{eqnarray}
\frac{1}{\sigma}\frac{d\sigma}{d\cos\theta_{\rm hel}}=\frac{1\pm \cos\theta_{\rm hel}}{2},
\label{eq:spin}
\end{eqnarray}
with the ($+$) choice for right-handed and ($-$) for left-handed top quarks; 
Clearly, the charged lepton from a right-handed top quark prefers to move along
the top quark direction .
In the top-quark rest frame, 
75\% (25\%) of charged leptons from $t_R$ ($t_L$) decays follow the top-quark direction.

\AddToContent{D.~G.~E.~Walker, J.-H.~Yu and C.-P.~Yuan}
\renewcommand{\thesection}{\arabic{section}}
\renewcommand{\thesubsection}{\thesection.\arabic{subsection}}
\renewcommand{\thesubsubsection}{\thesubsection.\arabic{subsubsection}}
\renewcommand{\thefigure}{\arabic{figure}}
\renewcommand{\thetable}{\arabic{table}}
\renewcommand{\theequation}{\arabic{equation}}





\chapter{Effective axial-vector coupling of the gluon
and top-quark charge asymmetry at the LHC}

{\it Emidio Gabrielli, Andrea Giammanco,  Antonio Racioppi,
Martti Raidal}

\begin{abstract}
We study the top quark charge asymmetry induced by an effective axial-vector coupling of the gluon at the LHC experiments.
We showed that rapidity cut-dependent asymmetries are more sensitive to the new physics than the independent ones.
We also study the dependence of  the asymmetries and variations of  the total $t\bar t$ cross sections on the invariant mass of the $t\bar t$ system
and show that  it would be necessary to measure those quantities as functions of $m_{tt}$ at the LHC.
If this is done,  in the context of the considered new physics scenario, the 7~TeV LHC has enough sensitivity either to confirm the Tevatron top charge
  asymmetry anomaly or to rule it out.
  In the latter case the LHC will be able to put stringent constraint on the new physics scale $\Lambda$ in this framework.
\end{abstract}

\section{INTRODUCTION}

The $3.4~\sigma$ excess over the standard model (SM) predictions~\cite{Kuhn:1998jr,Kuhn:1998kw,Bowen:2005ap,Antunano:2007da}
in the top-quark charge asymmetry ($A_{FB}^t$)
observed by the CDF collaboration at the Fermilab Tevatron~\cite{Aaltonen:2011kc}, has triggered
numerous theoretical and experimental studies of top quark production at hadron colliders.
An intriguing  property of the measured asymmetry is that it increases with the $t\bar t$ invariant mass $m_{t\bar{t}}.$
At the same time the measured $t\bar t$ production cross section is consistent, within experimental errors,
 with the SM predictions~\cite{Ahrens:2011px,Cacciari:2008zb,Moch:2008qy} both at Tevatron~\cite{Aaltonen:2010bs,Abazov:2009si} and
 at the CERN Large Hadron Collider (LHC)~\cite{ATLAS-CONF-2011-140,CMS-PAS-TOP-11-007}. Motivated by those results, the SM predictions for the
 $t\bar t$ charge asymmetry have been revised~\cite{Kuhn:2011ri,Hollik:2011ps}, showing a moderate $20\%$ increase due to QED and
 electroweak corrections.

 Numerous new physics explanations (see Ref.~\cite{Westhoff:2011tq} for a recent summary review) have been proposed to
explain the observed anomaly, predicting the existence of new particles whose contributions produce the asymmetry.
Those scenarios can be directly tested at the LHC experiments by looking for new
 particle interactions. In the light of the LHC results several popular explanations to the $t\bar t$ charge asymmetry such as
the axigluons~\cite{Ferrario:2008wm,Ferrario:2009bz,Rodrigo:2010gm}, $Z'$~\cite{Jung:2009jz} or $W'$~\cite{Cheung:2009ch} bosons are stringently constrained.

However, effective field theory offers also model-independent tests of the top quark charge asymmetry.
Particularly interesting among those is the one due to an effective axial-vector coupling of the gluon~\cite{Gabrielli:2011jf}.
This scenario does predict the correct sign and the correct $m_{t\bar{t}}$ dependence of the Tevatron anomaly and
does not necessarily require new light resonances. Therefore tests of this scenario  require in particular
the measurement of the top quark charge asymmetry dependence on  $m_{t\bar{t}}$ at the LHC.

Although the LHC is a $pp$ collider with a symmetric initial state, the $t\bar t$ charge asymmetry can be
defined and studied at the LHC in $q \bar q$ collisions using anti-quarks from the sea~\cite{Kuhn:1998jr,Kuhn:1998kw}. Because the sea quark
parton distributions differ from the valence ones, top and anti-top quarks are preferably produced at different
rapidities. Therefore,
studying top quark charge asymmetries at large rapidities and large invariant masses
will enhance the asymmetries both in the SM as well as in scenarios of new physics~\cite{Kuhn:2011ri}.
At present the ATLAS~\cite{ATLAS-CONF-2011-106} and CMS~\cite{CMS-PAS-TOP-11-014} experiments have published only their measurements of rapidity cut-independent
 top quark charge asymmetries. Their results are consistent with the SM predictions. 
The first results on rapidity
and invariant mass-dependent observables for top quark charge asymmetry have been released
by the CMS experiment, comparing data with the SM prediction and with our prediction for the considered new physics scenario~\cite{CMS-PAS-TOP-11-030}.

 The aim of this work is to study rapidity and invariant mass-dependent top quark charge asymmetries at the LHC
 in the scenario of an effective axial-vector coupling of the gluon. We show that the rapidity and invariant mass-dependent
 asymmetries are much more appropriate for testing this type of new physics scenarios than the cut-independent ones used by
 ATLAS~\cite{ATLAS-CONF-2011-106} and CMS~\cite{CMS-PAS-TOP-11-014} so far. We show that the Tevatron observation of large asymmetry can be either confirmed or ruled out already at the
 7~TeV LHC with 10~fb$^{-1}$ data.

\section{EFFECTIVE GLUON COUPLINGS IN FIELD THEORY}

The most general effective Lagrangian for quark-gluon
interactions, containing the lowest dimensional operators, and
compatible with gauge-, CP-, and Lorentz-invariance, is \cite{Gabrielli:2011jf}
\begin{eqnarray}
{\cal L}&=& -ig_S\Big\{ \bar Q T^a\left[\gamma^\mu\left(1+
g_V(q^2,M)
+ \gamma_5 g_A(q^2,M)\right) G^a_\mu\right. \nonumber \\
&+& \left.  g_P(q^2,M)q^{\mu}\gamma_5 G^a_\mu+ g_M(q^2,M)
\sigma^{\mu\nu} G^{a}_{\mu\nu} \right] Q\Big\} ,
\label{effective_ttbar_A_vertex}
\end{eqnarray}
where $g_S$ is the strong coupling constant, $G^a_\mu$ is the gluon field, $T^a$ are the color matrices,
$M$ is some energy scale, $q^2$ is the invariant momentum-squared carried by the gluon and $Q$ denotes a generic quark field.
 From now on we will assume that the dominant contribution
to the $g_{A,P}(q^2,M)$ form factors arises from new physics (NP) with characteristic
scale above the electroweak (EW) scale. In this case the scale $M$ should be identified
with the NP scale.
Model independently, QCD gauge invariance requires that $2m_Q g_A(q^2,M)=q^2 g_P(q^2,M)$,
thus
\begin{eqnarray}
\lim_{q^2\to 0} g_{A,V}(q^2,M)=0\, \label{effective_ttbar_A_ginv}
\end{eqnarray}
since no $1/q^2$ singularities are present in $g_P$. As observed
before~\cite{Gabrielli:2011jf}, Eq.(\ref{effective_ttbar_A_ginv}) does
not impose any additional constraint on the form factors $g_{A,V}$,
which could have different magnitudes at arbitrary $q^ 2$.
Therefore, gauge-invariance does not prevent us to have $g_{V} \ll
g_{A}$ as long as $q^2 \neq 0$.
 If the origin of the large $A_{FB}^t$ is due to NP that has $(V\pm A)$
currents as in the SM, large $g_{V}$ and $g_{A}$ can be generated.
However, this scenario is
phenomenologically unacceptable~\cite{Gabrielli:2011jf} because $g_V$ is strongly
constrained by the total $q\bar q\to t\bar t$ cross section. Thus
we will neglect the contribution of the vectorial form factor
$g_V(q^2,M)$ in Eq.(\ref{effective_ttbar_A_vertex}), and consider
only NP scenarios that generate $g_A$ with the hierarchy $g_V \ll
g_A$. In the limit of $q^2\ll M^2$, it is useful to parametrize
the axial-vector form factor as
\begin{eqnarray}
g_A(q^2,M)=\frac{q^2}{\Lambda^2} F(q^2,\Lambda)\, ,
\label{effective_ttbar_A_gA}
\end{eqnarray}
where  we absorb the NP coupling $\alpha_{NP}$ and loop factor into the NP scale, $\Lambda^2=M^2/(4\pi \alpha_{NP}).$
Notice the $q^2$ dependence that enhances the asymmetry at large invariant masses.

\begin{figure}[t]
\begin{center}
\includegraphics[width=0.18\textwidth]{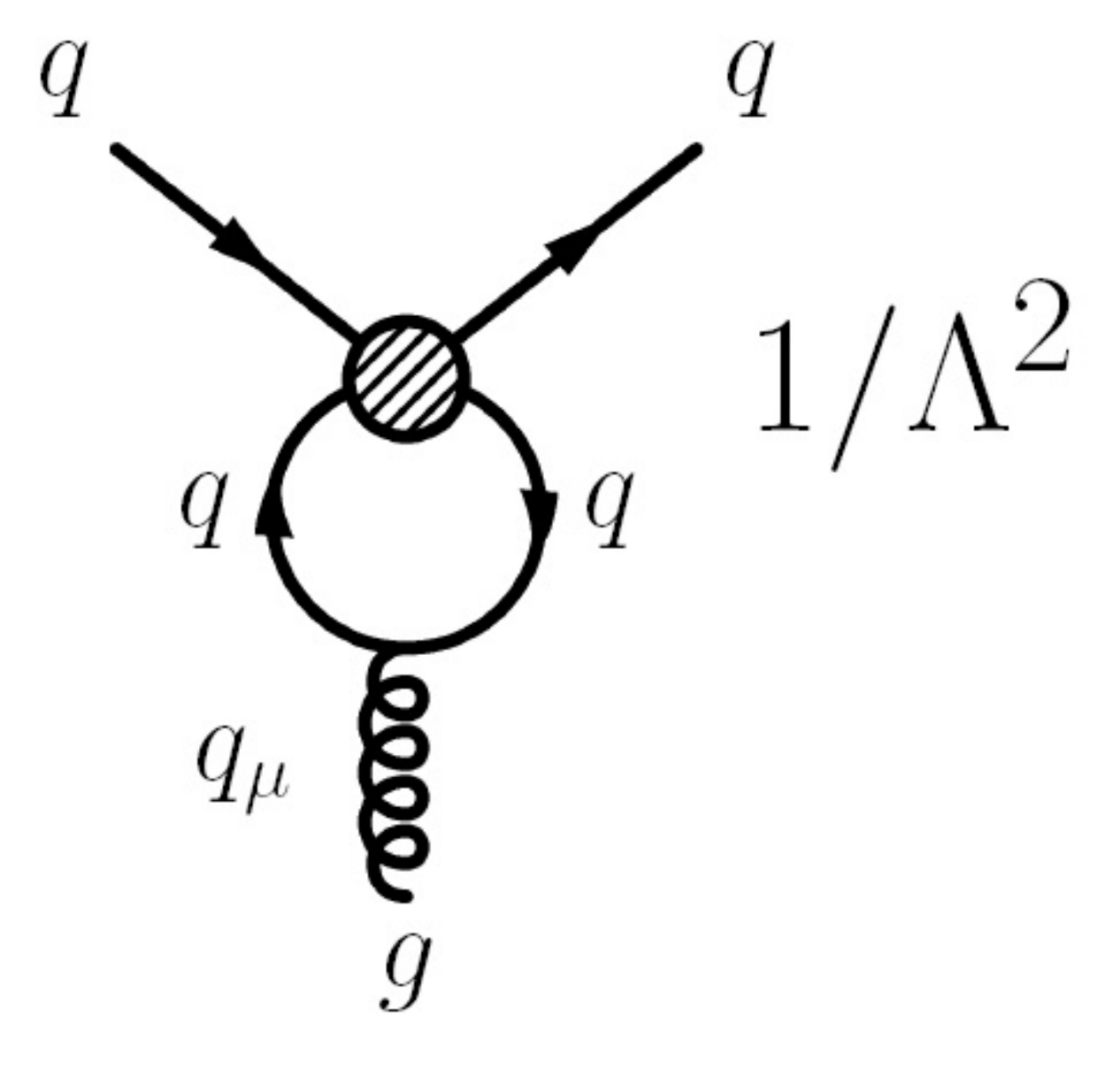}
\vspace{-0.3cm}
\caption{Feynman diagram in the effective low energy theory
that generates the effective axial-vector coupling of gluon.}
\label{effective_ttbar_A_fig:effvert} 
\end{center}
\end{figure}

Assuming that there are perturbative NP above the EW scale,
model independently the effective operators~\cite{Delaunay:2011gv}
\begin{eqnarray}
O^{1,8}_{AV} &=&\frac{1}{{\Lambda}^2}
[\bar Q T_{1,8}\gamma^\mu\gamma_5  Q] [\bar Q T_{1,8} \gamma^\mu Q]\, ,
\label{effective_ttbar_A_OAV} \\
O^{1,8}_{PS}&=&\frac{1}{{\Lambda}^2} [\bar Q T_{1,8}\gamma_5  Q]
[\bar Q T_{1,8} Q] , \label{effective_ttbar_A_OPS}
\end{eqnarray}
 generate $g_A$ via the one-loop diagrams depicted in Fig.~\ref{effective_ttbar_A_fig:effvert}.
Here $T_1=1$ and $ T_8=T^a,$ thus both isoscalar and octet operators contribute. Notice that:
 $(i)$ no $g_V$ is induced due to QCD parity conservation;
   $(ii)$ the one-loop induced $g_A$ is enhanced by $\log(q^2/\Lambda^2)$;
 $(iii)$ the operators $O^{1,8}_{AV}$, $O^{1,8}_{PS}$ do not induce flavor changing (FC) processes; however, there could be different quark flavors in the loop
 in Fig.~\ref{effective_ttbar_A_fig:effvert} (extending the operator basis to $Q\to Q',$ $V\leftrightarrow A,$ $P\leftrightarrow S$ is straightforward);
 $(iv)$  the operators $O^{1,8}_{AV}$, $O^{1,8}_{PS}$ do not
interfere with the corresponding QCD induced 4-quark processes.
 The latter point implies that the stringent LHC constraints~\cite{Khachatryan:2011as,Aad:2011aj} on
 4-quark contact interactions do not apply to our scenario.

Alternatively, large $g_A$ might be generated by new
strongly-coupled parity-violating dynamics related to electro-weak
symmetry breaking (EWSB) at the 1-2 TeV scale. A plethora of new
resonances should occur at the LHC allowing to test this scenario.
As we are not able to make predictions in this case, we shall not
elaborate on this new physics possibility and simply assume the
existence of operators (\ref{effective_ttbar_A_OPS}).

\section{CROSS SECTIONS AT THE LHC}

\begin{figure}[t]
\begin{center}
\includegraphics[width=0.45\textwidth]{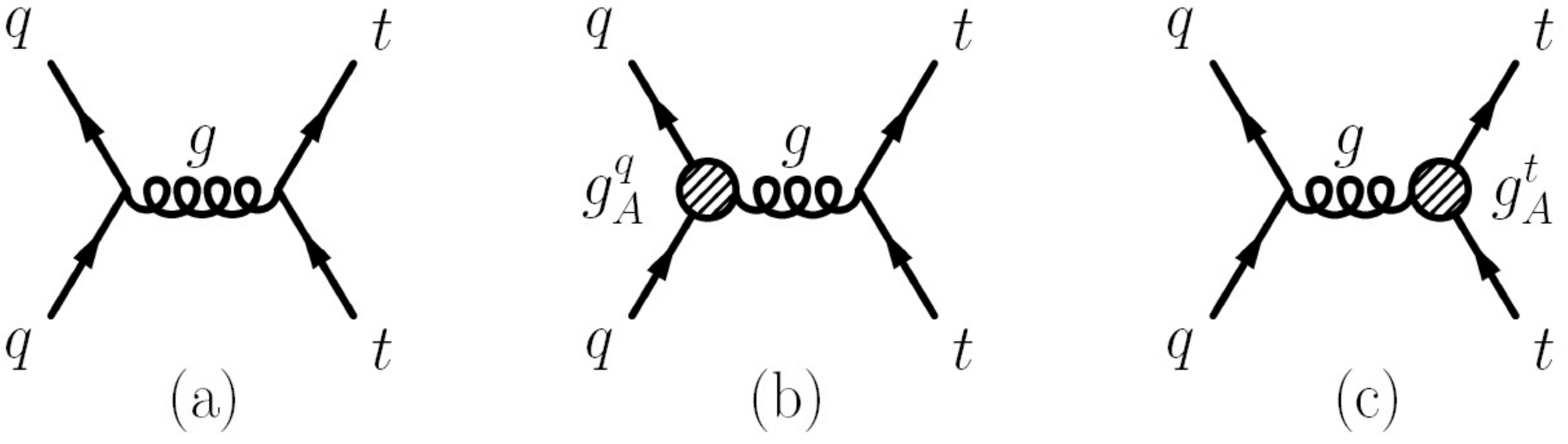}
\vspace{-0.3cm} \caption{Feynman diagrams (a)-(c) for the $q \bar
q \to t \bar t$ process, with the contribution of the gluon
effective axial-vector couplings $g^{q,t}_A$.}
\label{effective_ttbar_A_fig:qqtt}
\end{center}
\end{figure}

\begin{figure}[t]
\begin{center}
\includegraphics[width=0.45\textwidth]{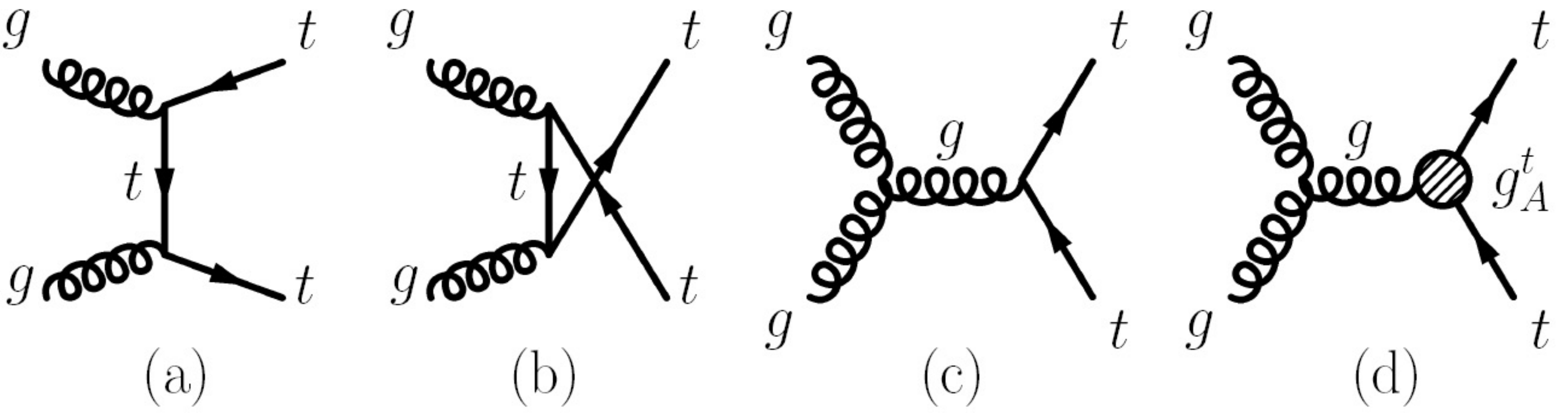}
\vspace{-0.3cm} \caption{Feynman diagrams (a)-(d) for the $g g \to
t \bar t$ process, with the contribution of the gluon effective
axial-vector coupling $g^{t}_A$.}
\label{effective_ttbar_A_fig:ggtt}
\end{center}
\end{figure}

The effective axial-vector coupling of gluon affects the $t\bar t$
cross section at the LHC via the parton level processes
$q\bar{q}\to t\bar t$ and $g g\to t\bar t$ as depicted in
Figs.~\ref{effective_ttbar_A_fig:qqtt} and
\ref{effective_ttbar_A_fig:ggtt}, respectively.
Notice that our scenario modifies not only the subdominant $q\bar
q$ but  also the dominant $gg$ production channel. The effective
axial-vector coupling of the top-quark affects only the s-channel
(diagram \ref{effective_ttbar_A_fig:ggtt}(d)); due to the Ward
identity in Eq.(\ref{effective_ttbar_A_ginv}), the effective
axial-vector contribution vanishes in the $\hat t$- and $\hat
u$-channels, diagrams \ref{effective_ttbar_A_fig:ggtt}(a) and
\ref{effective_ttbar_A_fig:ggtt}(b) respectively, because the
gluons attached to the axial-vector vertex are on-shell. Therefore
detailed measurements of the $t\bar t$ cross sections are needed
at the LHC to test this scenario.

For $q\bar{q}\to t\bar t$ the differential cross section in the massless
light-quarks $q$ limit (summed over all colors) is given by
\begin{eqnarray}
\!\!\!
\frac{d \sigma^{q\bar q}}{d \hat t}&=&
\frac{8 \pi \alpha_S^2}{9\hat{s}^4}
\Big[\Big(\hat{t}^2 +(\hat{s}-2 m_t^2)\hat{t}+\frac{\hat{s}^2}{2}
+m_t^4\Big)\times
\nonumber\\
&&
\left(
1+|g_A^q|^2+|g_A^t|^2 + |g_A^q|^2|g_A^t|^2 \right)
\nonumber \\
&-&2 |g_A^t|^2 \,m_t^2\hat{s} \left(1+|g_A^q|^2\right)
\nonumber \\
&+& 2 {\rm Re}[g_A^q]\, {\rm Re}[g_A^t]\,
\hat{s}(\hat{s}+2\hat{t}-2m_t^2) \Big]\, ,
\label{effective_ttbar_A_dsigmaqq}
\end{eqnarray}
where $g_A^q$ and $g_A^t$ are the corresponding axial-vector form
factors for the light-quark $q$ and top-quark, respectively, and
$\hat  s,\hat t,\hat u$ are the usual Mandelstam variables. After
integrating Eq.(\ref{effective_ttbar_A_dsigmaqq}) over the full
range of $\hat{t}$, the total partonic cross section is given by
\begin{eqnarray}
\sigma^{q\bar q}(\hat s)&=&
\frac{8\pi\alpha^2_S \beta_t}{27 \hat{s}}\Big\{(1+2\frac{m_t^2}{\hat s})
\left(1+|g_A^q|^2\right)+
\nonumber \\
&& \beta_t^2 |g_A^t|^2\left(1+|g_A^q|^2\right)\Big\} ,
\label{effective_ttbar_A_sigmaqq}
\end{eqnarray}
where $\beta=\sqrt{1-\rho}$ and $\rho=4m_t^2/\hat{s}$.

For $g g\to t\bar t$ scattering the differential unpolarized cross section is given by
\begin{eqnarray}
\frac{d \sigma^{gg}}{d\hat{t}}&=&
\frac{\pi \alpha_S^2}{64 \hat{s}^2} \left[
12 M_{ss}+\frac{16}{3} \left(M_{tt}+M_{uu}\right)
-\frac{2}{3} M_{tu} \right. \nonumber \\
&+& \left. 6\left(M_{st}+M_{su}\right)\right],
\label{effective_ttbar_A_dsigmagg}
\end{eqnarray}
with
\begin{eqnarray}
M_{ss}&=&\frac{4}{\hat{s}^2}\left[
\left(\hat{t}-m_t^2\right)\left(\hat{u}-m_t^2\right)
+\left(\hat{t}\hat{u}-m_t^4\right)|g_A^t|^2\right],
\nonumber \\
M_{tt}&=&\frac{2}{(\hat{t}-m_t^2)^2}
\left[\left(\hat{t}-m_t^2\right)\left(\hat{u}-m_t^2\right)
-2m_t^2 \left(\hat{u}+m_t^2\right)\right],
\nonumber \\
M_{tu}&=&\frac{4m_t^2}{(\hat{t}-m_t^2)(\hat{u}-m_t^2)}
\left(\hat{s}-4m_t^2\right),
\nonumber \\
M_{st}&=&\frac{4}{\hat{s}(\hat{t}-m_t^2)}\left[m_t^4-
\hat{t}\left(\hat{s}+\hat{t}\right)\right]\, ,
\label{effective_ttbar_A_M2gg}
\end{eqnarray}
and $M_{uu}=M_{tt}\left\{ t\leftrightarrow u \right\}$,
$M_{su}=M_{st}\left\{ t\leftrightarrow u \right\}$.
This expression is  $SU(3)_c$ gauge invariant.

After integrating Eq.(\ref{effective_ttbar_A_dsigmagg}) over the
full range of $\hat{t}$, we obtain for the total partonic cross
section
\begin{eqnarray}
\hat{\sigma}^{gg}(\hat{s})&=&\frac{\pi\alpha_S^2}{48\hat{s}}
\Big\{\left(16+\rho\left(16+\rho\right)\right)
\log{\left(\frac{1+\beta}{1-\beta}\right)}
 \nonumber\\
&-&
\beta\left(28+31\rho+6|g_A|^2\left(\rho-1\right)\right)\Big\}\, .
\label{effective_ttbar_A_sigmagg}
\end{eqnarray}

Finally, the hadronic cross section  $pp\to t \bar t X$ at LHC is
obtained by convoluting the partonic cross sections in 
Eqs.~(\ref{effective_ttbar_A_sigmaqq}),(\ref{effective_ttbar_A_sigmagg})
with the corresponding parton distribution functions (PDF) for
quarks and gluons, namely
\begin{eqnarray}
\sigma_{p p \to t\bar t X}={\int \left( \sum_qd \mu_q
\sigma_{qq}(\hat s)+ d\mu_{g}\sigma_{gg}(\hat s)\right)},
\label{effective_ttbar_A_xsec}
\end{eqnarray}
where $d \mu_{q}$ and $d \mu_{g}$ indicate the differential
integrations in $dx_1 dx_2$ convoluted with the quark and gluon
PDFs respectively. In the numerical integration of
Eq.(\ref{effective_ttbar_A_xsec}) we have used the CTEQ6L1 parton
distribution function (PDF) \cite{Pumplin:2002vw}, where we set
the PDF scale $\mu=m_t$ with the top quark mass $m_t=172$ GeV.

\section{CHARGE ASYMMETRY AT THE LHC}

The most useful top charge asymmetries at the LHC are defined as~\cite{Kuhn:2011ri}
\begin{itemize}
\item
the {\it cut-independent} charge asymmetry, as measured by ATLAS and CMS,
\begin{eqnarray}
A^y_C= \frac{N(\Delta_y > 0)-N(\Delta_y < 0)}{N(\Delta_y >
0)+N(\Delta_y < 0)}, \label{effective_ttbar_A_CA}
\end{eqnarray}
where $\Delta_y \equiv |y_t|-|y_{\bar t}|$;
\item the {\it integrated} pair charge-asymmetry
\begin{eqnarray}
A^{\rm cut}_C(Y_c)= \frac{N(y_t > y_{\bar t})-N(y_t < y_{\bar t}
)} {N(y_t > y_{\bar t})+N(y_t < y_{\bar t} )},
\label{effective_ttbar_A_CAcut}
\end{eqnarray}
as a function of the cut $Y_c$ on mean rapidity, namely
$(y_t+y_{\bar t})/2 > Y_c$.
\end{itemize}
All the above observables are defined in the laboratory frame.
Due to the symmetry of the initial proton-proton configuration, also $A_C^{\rm cut}(Y_c)$
will vanish if the full rapidity range is integrated, that is when
 $Y_c$ approaches its maximum kinematically allowed value.

Top quark pair production by the gluon fusion mechanism,
which is dominant at the LHC (representing $70\%$
and $90\%$ of the total cross-section at 7 TeV and 14 TeV c.o.m. energy respectively), is charge symmetric
under higher order corrections.
However, the
contributions of the gluon-gluon collisions can be
reduced by imposing a lower cut on the top quark pair invariant mass
$m_{t\bar{t}}$. This has the effect of eliminating
the regions of  lower longitudinal momentum fraction of the colliding partons
where the gluon density is much larger that the quark densities.
In addition, imposing a lower cut on $m_{t\bar{t}}$
also has the advantage of enhancing the $q\bar q \to t \bar t $
contribution to the charge asymmetry, although at the price of
reducing the $t \bar t$ pair statistics.
This requirement has also crucial implications
for our scenario, since it increases
the contribution of the axial-vector coupling of the gluon to the charge
asymmetry. This is due to the fact that the effective coupling
$g_A$ grows as $m^2_{t\bar{t}}/\Lambda^2$ at large $m_{t\bar{t}}$ values (but with
$m_{t\bar{t}}\le \Lambda$). We will restrict our analysis to the
case of real and universal axial-vector gluon couplings. In particular,
we parametrize the axial-vector coupling ( for values of $|q^2|< \Lambda^2 $ )
as follows
\begin{eqnarray}
g_A^t=g_A^q=\frac{q^2}{\Lambda^2}\, \, .
\label{effective_ttbar_A_univ}
\end{eqnarray}
Indeed, in order to explain the Tevatron
anomaly on $A_{FB}^t$, while requiring conservative constraints
on the $t \bar t$ cross sections at the Tevatron,
it was suggested~\cite{Gabrielli:2011jf} that the most favored scenario
is the one where all axial-vector couplings are universal and real,
with the NP scale $\Lambda$ lying in a narrow range
$1~ {\rm TeV} < \Lambda < 1.3~ {\rm TeV}$.
\begin{figure*}[t]
\begin{center}
\includegraphics[width=0.34\textwidth, angle=-90]{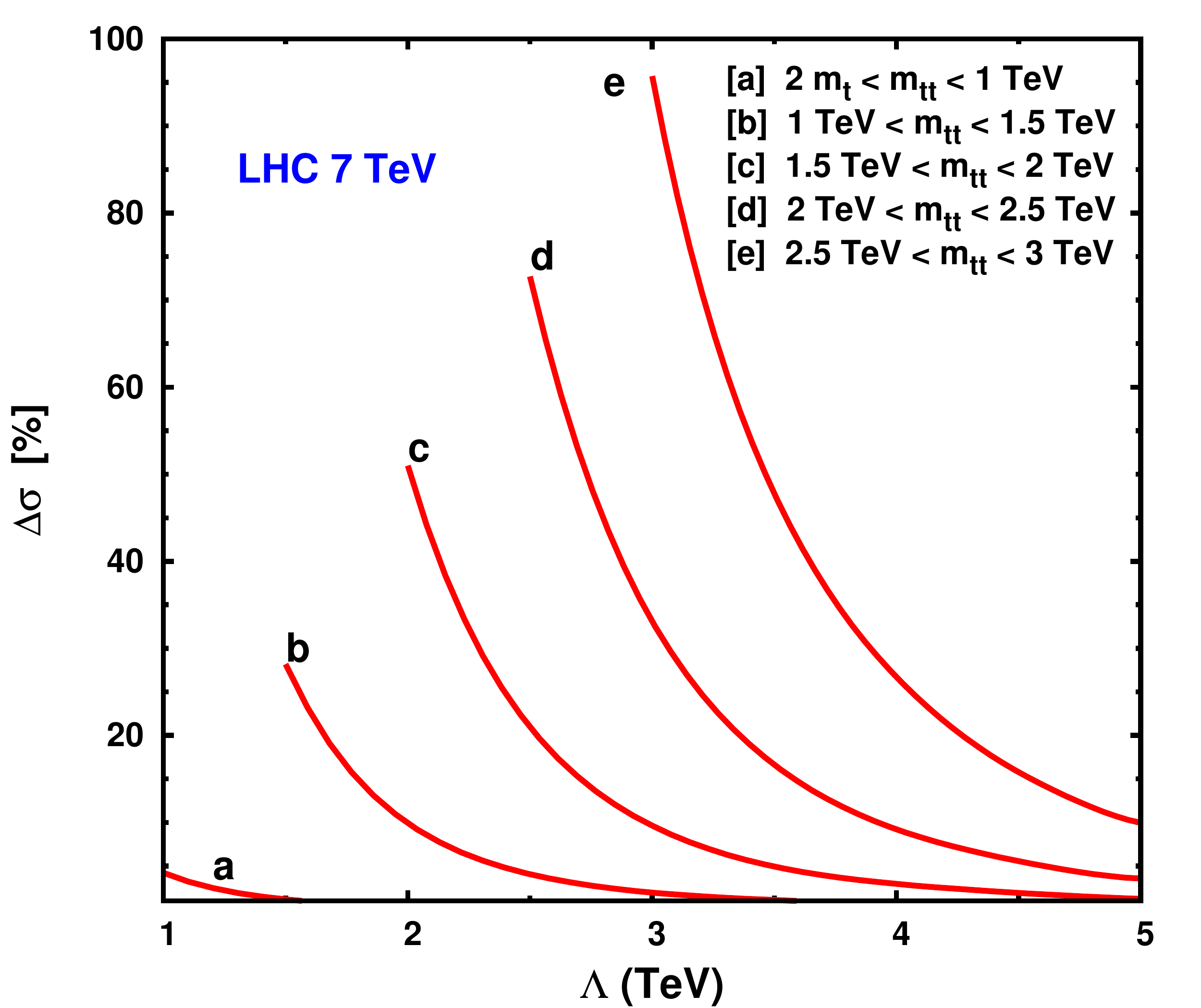}
\includegraphics[width=0.34\textwidth, angle=-90]{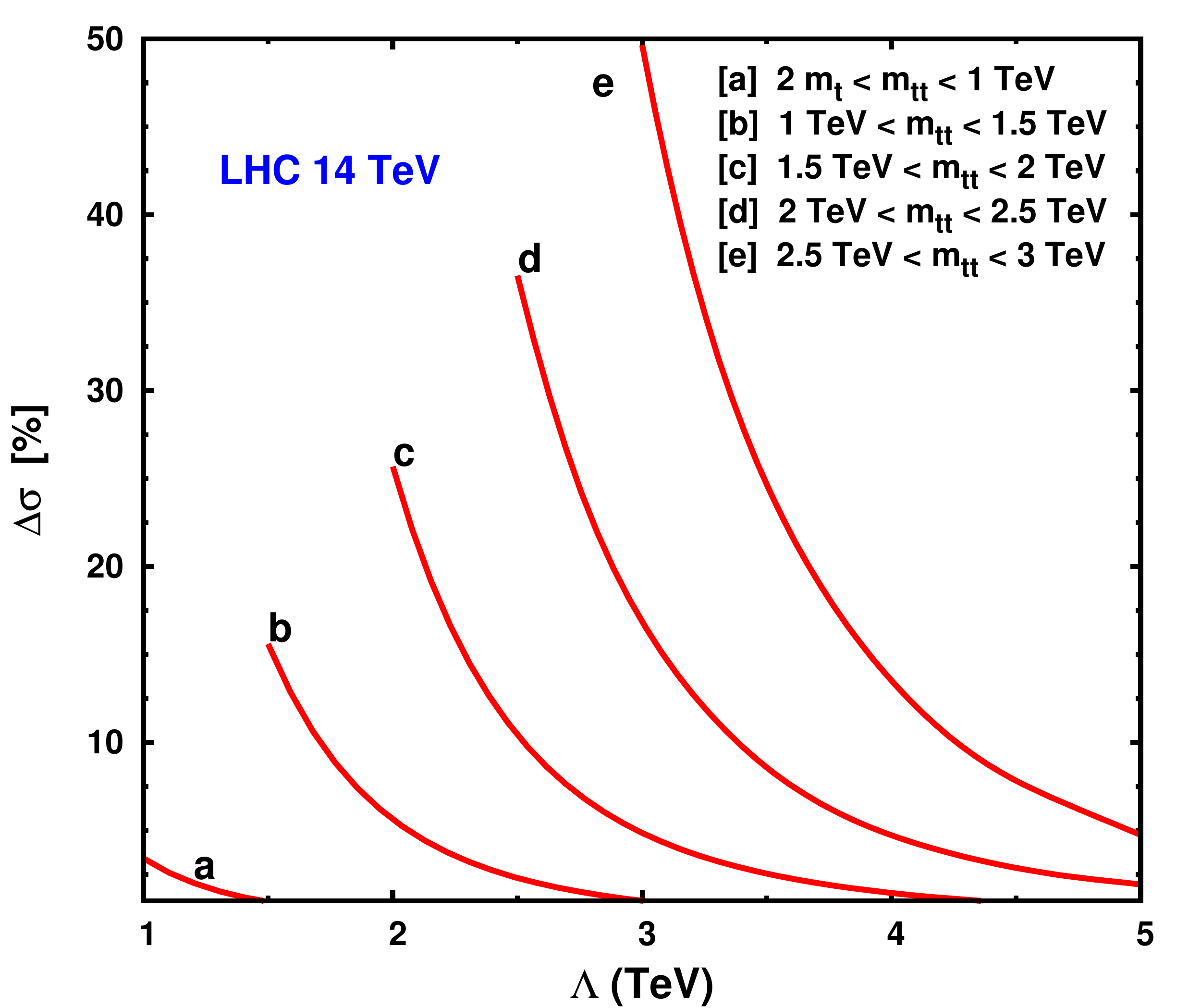}
\vspace{-0.3cm} \caption{Variation $\Delta \sigma$ of the total
cross section for $pp \to t\bar t$ at the LHC with $pp$ center of
mass energies $\sqrt{s}$ = 7 TeV (left panel) and 14
TeV (right panel) for $m_t=172$ GeV, as a function of the scale
$\Lambda$ in TeV, for several regions ([a-e]) of $t\bar t$
invariant mass $m_{t\bar{t}}$.} \label{effective_ttbar_A_fig:deltasigma}
\end{center}
\end{figure*}

In our numerical analysis we have not included the SM contribution to the
charge asymmetry. Indeed, this is almost negligible with respect
to the axial-vector gluon contribution for most of the
kinematic regions considered here. In particular,
we have retained only
the $g_A$ contribution in the numerators of
the asymmetries neglecting the corresponding SM contribution.
However, we have retained the SM effect (at the leading order (LO) in QCD)
in the evaluation of the total number of events entering
the equation for the charge asymmetry.

Then, following the definition of asymmetry,
the exact value of $A_C^{\rm SM+ NP}$, including the SM one
($A^{\rm SM}_C$), are related to the results of $A^{\rm NP}_C$ presented
here by the following equation
\begin{eqnarray}
A_C^{\rm SM+ NP}=A^{\rm NP}_C+\frac{A_C^{\rm SM}}{1+\Delta
\sigma}\, , \label{effective_ttbar_A_asym}
\end{eqnarray}
where  $\Delta \sigma$ is
the percentage variation of the total cross section (defined as
$\Delta \sigma=\sigma^{\rm NP}/\sigma^{\rm SM}$), and $\sigma^{\rm NP~ (SM)}$
represents the pure NP (SM) contributions to the total cross
section evaluated in the same kinematic region of
charge asymmetry. The symbol $A_C$ appearing in the
figures stands for the pure NP contribution to the charge
asymmetry $A^{\rm NP}_C$ as defined above.

 \begin{figure*}[t]
\begin{center}
\includegraphics[width=0.34\textwidth, angle=-90]{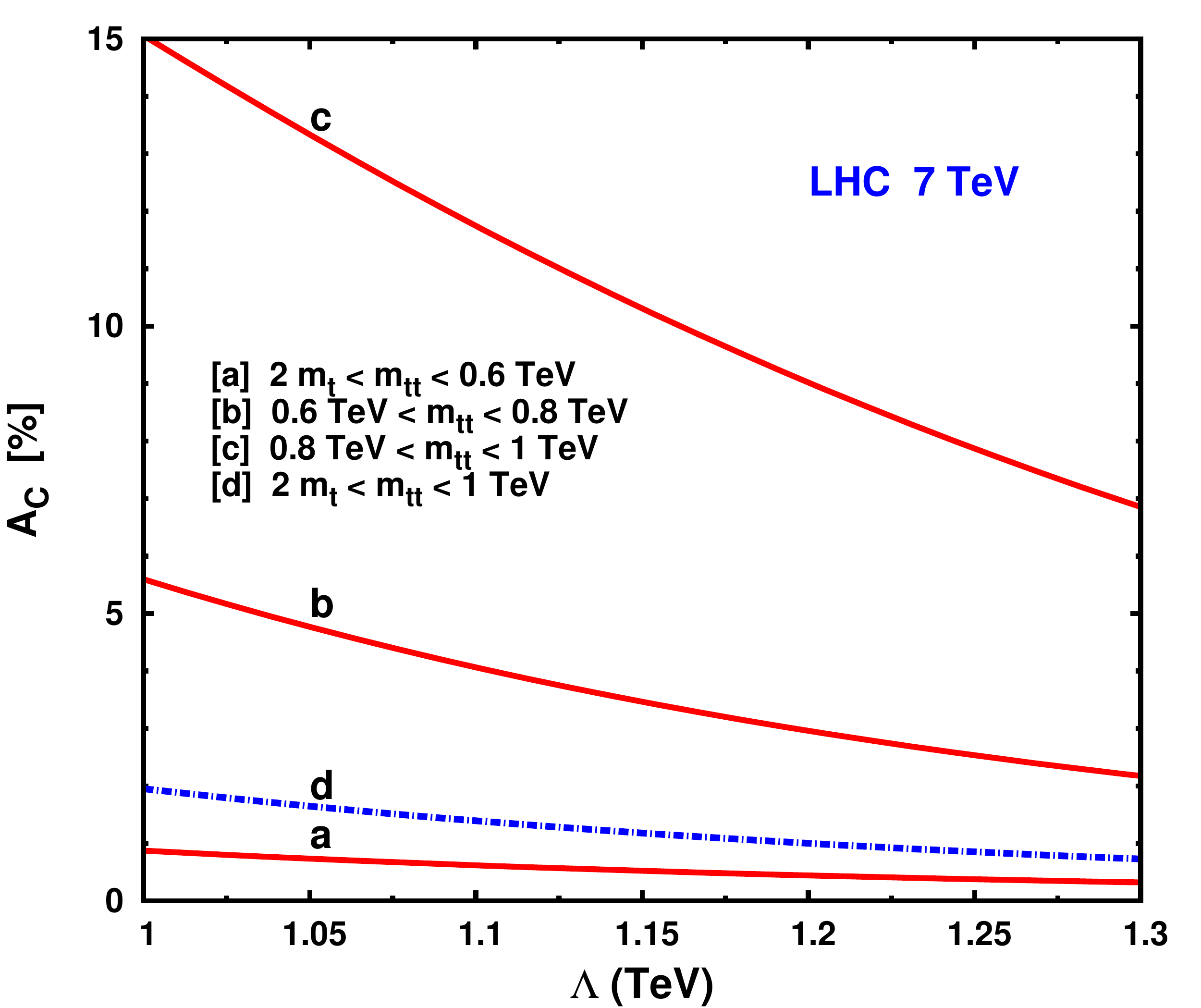}
\includegraphics[width=0.34\textwidth, angle=-90]{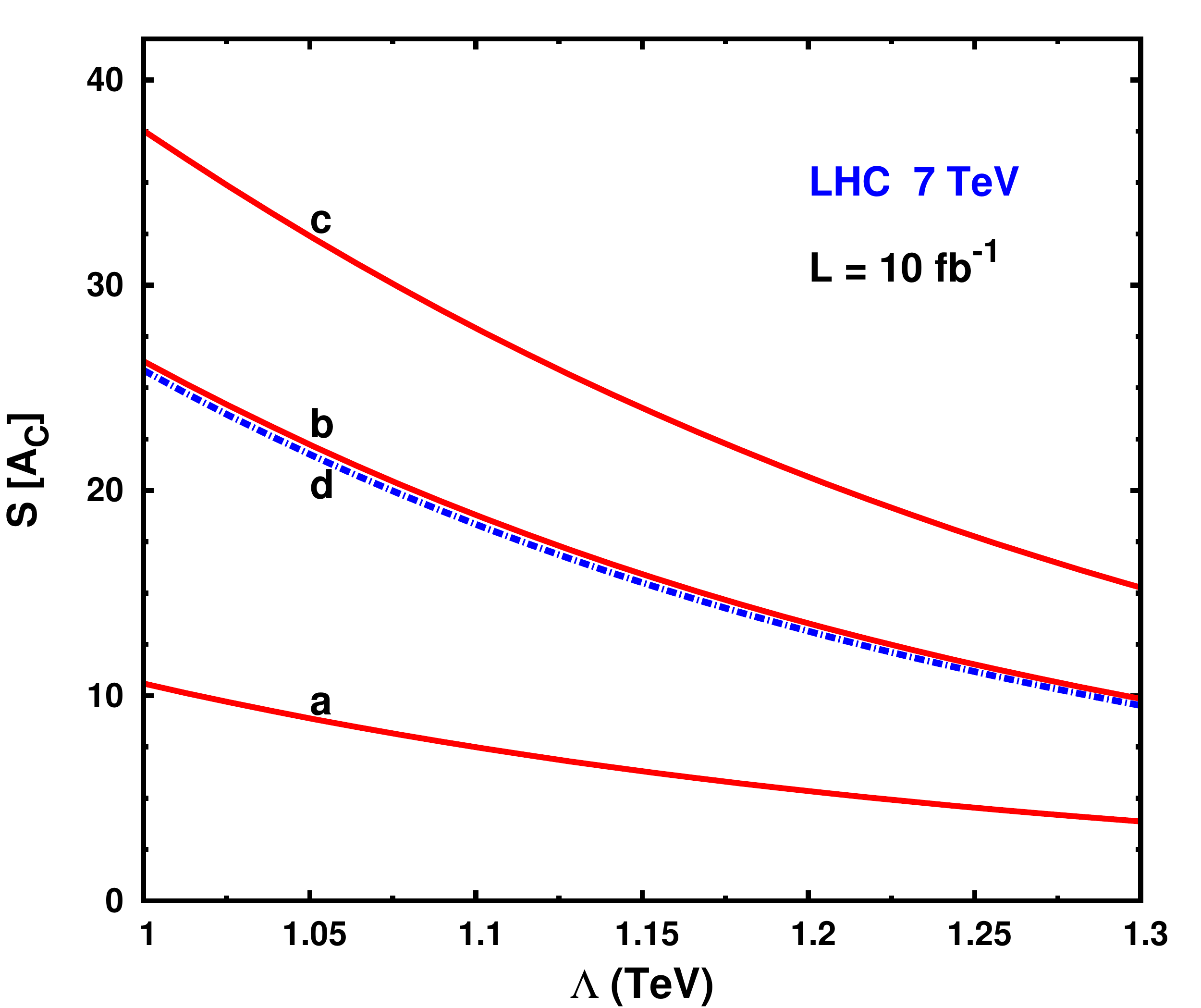}
\caption{The cut-independent  $t\bar t$ charge
asymmetry $A_{C}$ in percentage  (left plots) and corresponding
statistical significance $S[A_C]$ (right plots) at LHC with $pp$
center of mass energy $\sqrt{s}$ = 7 TeV and integrated luminosity
$L=10\ {\rm fb}^{-1}$, with $m_t=172$ GeV, as a function of the
scale $\Lambda$ in TeV, for several regions ([a-d]) of $t\bar t$
invariant mass $m_{t\bar{t}}$.} \label{effective_ttbar_A_fig:AC}
\end{center}
\end{figure*}

The numerical values of $\Delta \sigma$
are plotted in Fig.~\ref{effective_ttbar_A_fig:deltasigma} for the 7~TeV and 14~TeV LHC
as function of the new physics scale $\Lambda$  for several kinematic
regions [a-e] of $m_{t\bar{t}}$ as indicated in the Figure.
Regarding the choice of ranges of $m_{t\bar{t}}$,
we adopted the criterion that the maximum value of $m_{t\bar{t}}$ satisfies the
condition
$m_{t\bar{t}}^{\rm max} \leq \Lambda$ as required by the validity range of
the low energy limit in the $g_A$
form factor.  The cross sections have been evaluated at leading order (LO) in QCD.
The picture that emerges
from these results is clear. If we analyse ranges of
$m_{t\bar{t}}$ below 1 TeV, curve [a],
the expected variation in the total cross section is quite
small, below 5\% for both the 7 TeV and 14 TeV LHC energies.
This is due to the fact that requiring $m_{t\bar{t}}$ to be below 1 TeV,
the $g_A$ contribution at low energy is still largely
screened by the gluon fusion mechanism,
which is the dominant mechanism of top-quark pair production
at the LHC. Clearly, by increasing the $m_{t\bar{t}}$ mass range the NP effect could be
amplified and larger deviations could be observed in the cross sections,
due to the quark-antiquark production mechanism.
These results could also be used to set lower bounds on the scale $\Lambda$
by requiring that no excess in the total cross section is observed with
respect to the SM predictions. For instance, by limiting the deviations
on total cross section below 20 \%, one can see from curve [e] that a
lower bound $\Lambda >4 $ TeV could be obtained at LHC 7 TeV. Clearly,
by increasing the $m_{t\bar{t}}$ range, the statistical error on the cross section
also increases, and a more accurate analysis, which is beyond
the purpose of the present work, would be necessary.

\begin{figure*}[t]
\begin{center}
\includegraphics[width=0.34\textwidth, angle=-90]{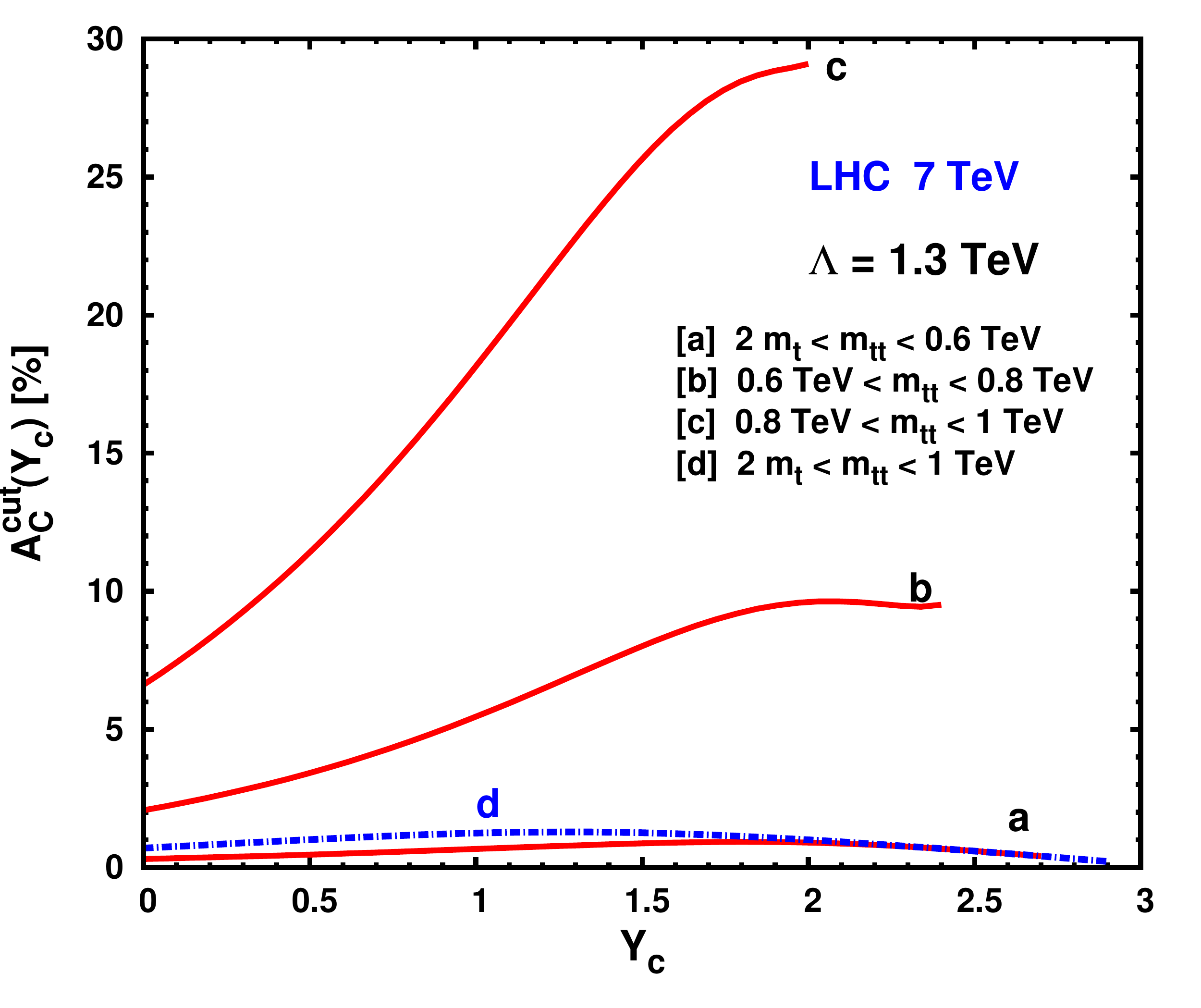}
\includegraphics[width=0.34\textwidth, angle=-90]{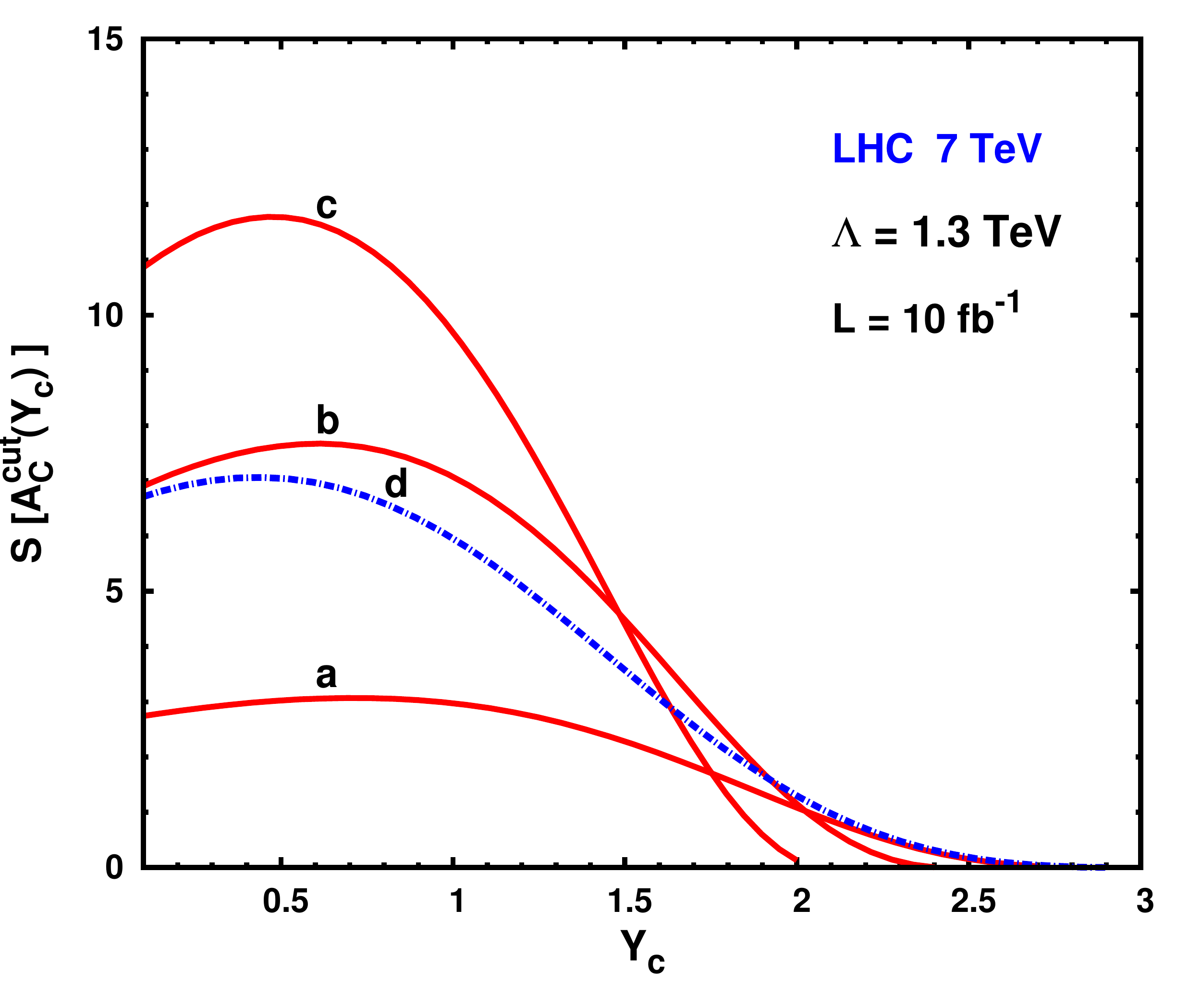}
\vspace{-0.3cm} \caption{Integrated pair charge asymmetry $A^{\rm
cut}_{C}(Y_c)$ in percentage  (left plots) and corresponding
statistical significance $S[A_C^{\rm cut}(Y_c)]$ (right plots) at
LHC with $pp$ center of mass energy $\sqrt{s}$ = 7 TeV and
integrated luminosity $L=10\ {\rm fb}^{-1}$, with $m_t=172$ GeV, as
a function of the cuts on the mean rapidity $|Y|> Y_c$ (in the lab
frame), for several regions ([a-d]) of $t\bar t$ invariant mass
$m_{t\bar{t}}$. Plots correspond to the scale $\Lambda=1.3$ TeV. }
\label{effective_ttbar_A_fig:ACcut}
\end{center}
\end{figure*}

Our numerical results for the $g_A$ contributions to the charge
asymmetries at the 7~TeV LHC defined in Eqs. (\ref{effective_ttbar_A_CA})
and (\ref{effective_ttbar_A_CAcut}) are shown in the left panels
of Fig.~\ref{effective_ttbar_A_fig:AC} and
Fig.~\ref{effective_ttbar_A_fig:ACcut}, respectively. In order to
estimate whether the asymmetries can be measured at the LHC or
not, we
 have used approximate expressions for the statistical significance $S[A_C]$~\cite{Ferrario:2008wm}
\begin{eqnarray}
S[A_C] \simeq  A_C \sqrt{L\, \sigma^{\rm NP+SM}}\, ,
\label{effective_ttbar_A_signf}
\end{eqnarray}
where $L$ stands for the integrated luminosity and $\sigma^{\rm
NP+SM}$ is the total cross section including the SM and NP
 contributions. In $\sigma^{\rm NP+SM}$ we have used the LO cross sections
multiplied by a rescaling factor $K$, which is obtained by simply
rescaling the total cross section evaluated at LO in QCD to its value 
corrected to next-to-next-to-leading order (NNLO) in QCD.
Those results are presented in the right
panels of Fig.~\ref{effective_ttbar_A_fig:AC} and
Fig.~\ref{effective_ttbar_A_fig:ACcut}. 
Notice that the
significance in Eq.(\ref{effective_ttbar_A_signf}) is a simple
theoretical estimation of the true one, since it does not take
into account efficiencies, acceptance, resolution, and
systematics.

In Fig.~\ref{effective_ttbar_A_fig:AC} we plot the cut-independent
$t\bar t$ charge asymmetry $A_{C}$ (left plot) and corresponding
significance (right plot) as a function of the scale $\Lambda$ in
the range [1-1.3] TeV. This is the definition adopted by the ATLAS
and CMS collaborations to measure the top-quark charge asymmetry
at the LHC. The ranges of $m_{t\bar{t}}$ [a-d] appearing in
Figs.~\ref{effective_ttbar_A_fig:AC} and
\ref{effective_ttbar_A_fig:ACcut} are given by
\begin{eqnarray}
[a] &=& 2 m_t < m_{t\bar{t}} < 0.6~  {\rm TeV}, \label{effective_ttbar_A_ranges1} \\
\nonumber
[b] &=& 0.6~{\rm TeV} < m_{t\bar{t}} < 0.8~ {\rm TeV},\\
\nonumber
[c] &=& 0.8~{\rm TeV} < m_{t\bar{t}} < 1~ {\rm TeV},\\
\nonumber
[d] &=& 2 m_t < m_{t\bar{t}} < 1~ {\rm TeV}.
\end{eqnarray}
In the following, we will refer to the $m_{t\bar{t}}$ ranges [a-d] as
the ones defined in Eq.(\ref{effective_ttbar_A_ranges1}).

As seen from these results, the $g_A$ contribution to $A_C$ is quite large
when measured at $m_{t\bar{t}}$ masses close to the value of the
scale $\Lambda$. In particular, for range [c], $A_C$ could be of
order of 15\% or 7\%, for scales
$\Lambda=1$ TeV or $\Lambda=1.3$ TeV respectively, while the
corresponding significance can reach values of 37 and 15 respectively.
On the other hand, when integrated over a large $m_{t\bar{t}}$ range, see for instance
the curve relative to range [b], $A_{C}$ turns out to be quite
small (below 2 \%) and comparable (although a bit larger) to the SM result~\cite{Kuhn:2011ri}.

In Table~\ref{effective_ttbar_A_table} we give some numerical
results for the values of the $t\bar{t}$ {\it cut-independent}
charge asymmetry $A_C$ (in percentage) evaluated at $\Lambda=1$
TeV and at LHC energies $\sqrt{s}=7$ TeV and $\sqrt{s}=8$ TeV,
integrated over the following kinematical ranges for $m_{t\bar{t}}$ and
$Y_{tt}$~\cite{effectivettbarA:cmscomm}:
\begin{eqnarray}
\Delta_1 &=&\{2 m_t < m_{t\bar{t}} < 420~ {\rm GeV} \}\, , \nonumber\\
\Delta_2 &=&\{420~ {\rm GeV} < m_{t\bar{t}} < 512~ {\rm GeV} \}\, , \nonumber\\
\Delta_3 &=&\{512~ {\rm GeV}  < m_{t\bar{t}} < 1~ {\rm  TeV} \}\, , \nonumber\\
\Delta_4 &=&\{0 < |Y_{tt}| < 0.41~ \}\, , \nonumber\\
\Delta_5 &=&\{0.41~  < |Y_{tt}| < 0.9~  \}\, , \nonumber\\
\Delta_6  &=&\{ |Y_{tt}| > 0.9\}\, ,
\label{effective_ttbar_A_ranges2}
\end{eqnarray}
where $Y_{tt}$ is defined as the rapidity of the 4-momentum
$p_{t\bar{t}}\equiv p_t+p_{\bar{t}}$, where $p_t$ and $p_{\bar t}$ are
the top- and antitop-quark 4-momenta in the LHC proton-proton
center of mass system respectively. For each range $\Delta_{1-6}$
in Eq.~(\ref{effective_ttbar_A_ranges2}), the other kinematic
variables are integrated over the whole allowed kinematical range.
In Table~\ref{effective_ttbar_S_table} we give some numerical
results for the values of the significances (see Eq.~(\ref{effective_ttbar_A_signf})) 
of the corresponding {\it
cut-independent} $t\bar{t}$ charge asymmetries $A_C$ computed in
Table \ref{effective_ttbar_A_table}. Results in Table 
\ref{effective_ttbar_S_table} are evaluated according to our definition 
of significance as given in Eq.~(\ref{effective_ttbar_A_signf}), with total
cross sections $\sigma^{\rm SM+NP}$ evaluated at NNLO in QCD, 
and correspond to an integrated 
luminosity  $L=10~{\rm fb}^{-1}$. For the case at $\sqrt{s}=8$ TeV,
we have used the same NNLO QCD rescaling factor $K$ 
as for $\sqrt{s}=7$ TeV.

\begin{table} \begin{center}
\begin{tabular}{|c||c|c|c|c|c|c|}
\hline $\sqrt{S}$
       & $A^{\%}_C(\Delta_1)$
       & $A^{\%}_C(\Delta_2)$
       & $A^{\%}_C(\Delta_3)$
       & $A^{\%}_C(\Delta_4)$
       & $A^{\%}_C(\Delta_5)$
       & $A^{\%}_C(\Delta_6)$
\\ \hline \hline 7 TeV
       &    0.32
       &    0.90
       &    4.9
       &    2.2
       &    1.4
       &    2.1
\\ \hline   8 TeV
       &    0.28
       &    0.78
       &    4.3
       &    1.9
       &    1.3
       &    2.0
\\ \hline \end{tabular}

\caption[]{Numerical values of the {\it cut-independent}
$t\bar{t}$ charge asymmetries $A_C$ in percentage, evaluated at
$\Lambda=1$ TeV and for LHC energies $\sqrt{s}=7$ TeV and
$\sqrt{s}=8$ TeV, integrated in the kinematical ranges
$\Delta_{1-6}$ given in Eq.(\ref{effective_ttbar_A_ranges2}).}
\label{effective_ttbar_A_table}
\end{center}
\end{table}

\begin{table} \begin{center}
\begin{tabular}{|c||c|c|c|c|c|c|}
\hline $\sqrt{S}$
       & $S_{A_C}(\Delta_1)$
       & $S_{A_C}(\Delta_2)$
       & $S_{A_C}(\Delta_3)$
       & $S_{A_C}(\Delta_4)$
       & $S_{A_C}(\Delta_5)$
       & $S_{A_C}(\Delta_6)$
\\ \hline \hline 7 TeV
       &    1.5
       &    3.9
       &    21
       &    8.2
       &    5.4
       &    11
\\ \hline   8 TeV
       &    1.5
       &    4.1
       &    23
       &    8.6
       &    6.0
       &    13
\\ \hline \end{tabular}

\caption[]{Numerical values of the significance of the {\it
cut-independent} $t\bar{t}$ charge asymmetry $A_C$, evaluated at
$\Lambda=1$ TeV and for LHC energies $\sqrt{s}=7$ TeV and
$\sqrt{s}=8$ TeV, corresponding to an integrated luminosity
$L=10~{\rm fb}^{-1}$ and integrated in the kinematical ranges
$\Delta_{1-6}$ given in Eq.(\ref{effective_ttbar_A_ranges2}).}
\label{effective_ttbar_S_table}
\end{center}
\end{table}

In Fig.~\ref{effective_ttbar_A_fig:ACcut} we show the integrated
pair charge asymmetry $A^{\rm cut}_{C}(Y_c)$ as defined in
Eq.(\ref{effective_ttbar_A_CAcut}) and corresponding
significance, as a function of  the cut $Y_c$ on
the mean rapidity $Y$ defined as $Y=(y_t+y_{\bar t})/2$, for the
representative value of $\Lambda=1.3$ TeV. In the case of lower
values of $\Lambda$, the NP effects are just more pronounced and
we will also include the case of $\Lambda = 1$ TeV in our
discussion below. As we can see from these results, $A^{\rm
cut}_{C}(Y_c)$ turns out to the most sensitive probe of our
scenario. In particular, at $\Lambda=1.3$ TeV, for $m_{t\bar{t}}$ in
range [c], the value of $A^{\rm cut}_{C}(Y_c)$ can vary, as a
function of $Y_c$, from 8\%  up to 25\% for $\Lambda=1.3$ TeV and
can increase from 15\% up to 55\% for $\Lambda=1$ TeV. The value
of $A^{\rm cut}_{C}(Y_c)$ in range [c] evaluated for
$Y_c=0.3$, where its significance is maximized, is of order of
10\% for $\Lambda=1.3$ TeV and can reach values of 20\% for
$\Lambda=1$ TeV. However, even at $Y_c=1.5$ the integrated
charge asymmetry is still quite large. As shown in
Fig.~\ref{effective_ttbar_A_fig:ACcut}, for $\Lambda=1.3$ TeV, the
value of $A^{\rm cut}_{C}(1.5)$ is of order  25\% and 8\% for
$m_{t\bar{t}}$ ranges [c] and [b] respectively, with a corresponding
significance of order 5 in both cases. For $\Lambda=1$ TeV, the
value of $A^{\rm cut}_{C}(1.5)$ can increase up to 50\% and 20\%
if integrated in the ranges [c] and [b] respectively, while the
corresponding significances are above 10.

\begin{figure*}[t]
\begin{center}
\includegraphics[width=0.34\textwidth, angle=-90]{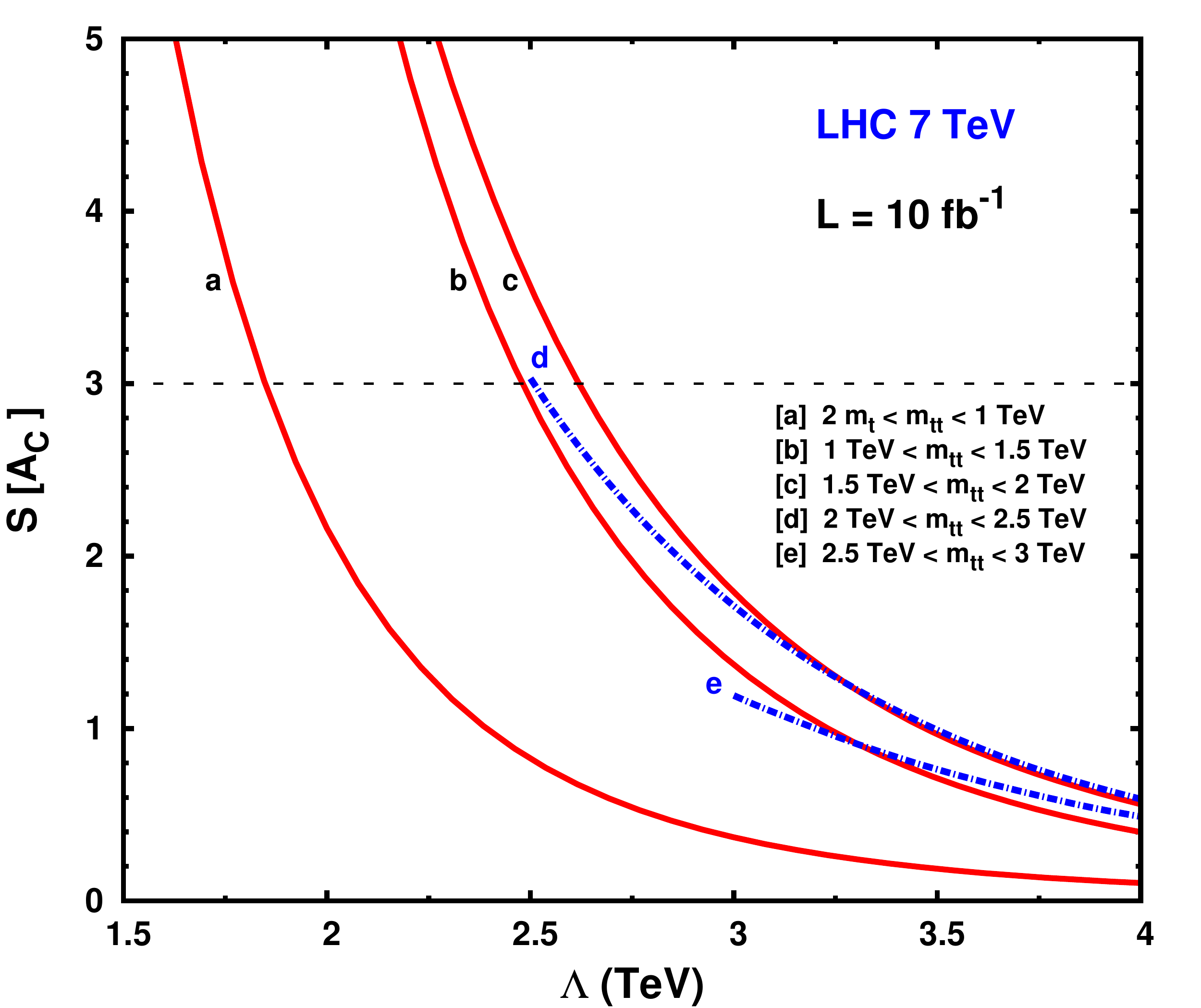}
\includegraphics[width=0.34\textwidth, angle=-90]{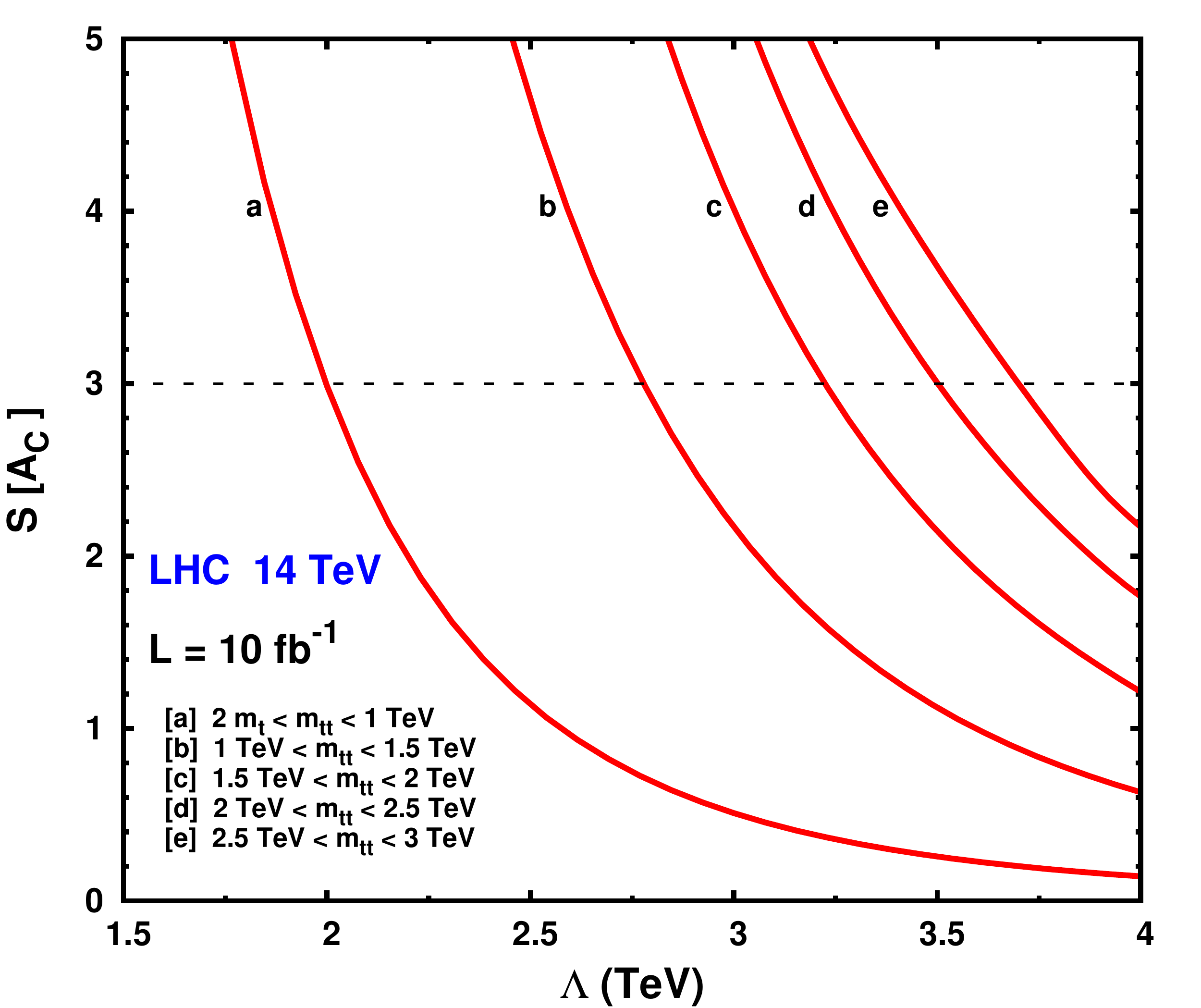}
\vspace{-0.3cm} \caption{Statistical significance $S[A_C]$ of the
cut-independent $t-\bar t$ charge asymmetry $A_C$ at LHC with $pp$
center of mass energies $\sqrt{s}$ = 7 TeV (left plots) and
$\sqrt{s}$ = 14 TeV (right plot) with  integrated luminosity $L=10\
{\rm fb}^{-1}$, with $m_t=172$ GeV, as a function of the scale
$\Lambda$ in TeV, for several regions ([a-e]) of $t\bar t$
invariant mass $m_{t\bar{t}}$.} \label{effective_ttbar_A_fig:S}
\end{center}
\end{figure*}

In Fig.~\ref{effective_ttbar_A_fig:S} we analyze the statistical
significance of the cut-independent charge asymmetry $A_C$ versus
the scale $\Lambda$ in the range of $\Lambda$=[1.5-4] TeV and for
several ranges of $m_{t\bar{t}}$ as indicated in the Figure.  The left
and right plots correspond to LHC center of mass energies of
$\sqrt{s}=$ 7 TeV and $\sqrt{s}=$ 14 TeV respectively. We have
used the same convention for the $\Lambda$ and $m_{t\bar{t}}$
ranges as adopted in Fig.~\ref{effective_ttbar_A_fig:deltasigma}.
Assuming that no significant deviations from SM predictions on the
charge asymmetry $A_C$ will be observed, from results in
Fig.~\ref{effective_ttbar_A_fig:S} one can derive lower bounds on
the scale $\Lambda$. For example, by requiring that $S[A_C] < 3$,
we get for $L=10 {\rm fb}^{-1}$, the following (strongest) lower
bounds:
\begin{itemize}
\item
$\Lambda  > 2.6~ {\rm TeV}$, for
$m_{t\bar{t}}\in$ [1.5-2] TeV at LHC at 7 TeV;
\item
$\Lambda > 3.7$ TeV, for $m_{t\bar{t}}\in$ [2.5-3] TeV at LHC at 14 TeV\,.
\end{itemize}
As can be seen from the dashed (blue) curves in the left plot of
Fig.~\ref{effective_ttbar_A_fig:S}, for LHC at 7 TeV there is no
advantage, concerning the lower bounds of $\Lambda$, in going to
higher bin ranges in $m_{t\bar{t}} > 2$ TeV, due to the loss of
statistics.

\section{CONCLUSIONS}
We have studied the gluon effective axial-vector coupling induced top charge asymmetry at the LHC.
We have compared rapidity cut-dependent and independent asymmetries and shown that the former are
more sensitive to new physics than the latter. We have also studied the asymmetries and variations of
total $t\bar t$ cross sections at different invariant masses of the $t\bar t$ system and shown that
it would be necessary to measure those quantities as functions of $m_{t\bar{t}}$ at the LHC.
A first measurement has been performed by the CMS collaboration~\cite{CMS-PAS-TOP-11-030}, 
finding agreement with both the SM and the considered new physics scenario.
With more data and a better control of the systematic uncertainties the LHC might either 
confirm the Tevatron
top charge asymmetry anomaly or rule it out in the context of the considered new physics
scenario. In the latter case the LHC will be able to put stringent constraint on the new physics
scale Lambda.

\section*{ACKNOWLEDGEMENTS}
We thank T. Chwalek, T. Peiffer,
G. Rodrigo, and J. Wagner-Kuhr for several communications.
This work was supported by the ESF grants  8090,  MTT59, MTT60, JD164,  by the recurrent financing SF0690030s09 project
and by  the European Union through the European Regional Development Fund.

\AddToContent{E.~Gabrielli, A.~Giammanco,  A.~Racioppi, M.~Raidal}
\renewcommand{\thesection}{\arabic{section}}

\chapter{Searching for sgluons in multitop production at the LHC}

{\it Samuel Calvet, Benjamin Fuks, Philippe Gris, Adrien
Renaud, Lo\"ic Val\'ery, Dirk Zerwas}

\begin{abstract}
In this report, we present the results of our activities addressing the
investigation of signatures containing multiple top quarks at the Large Hadron
Collider. We use the framework of a simplified effective theory
containing color-octet scalar particles decaying to top quarks and light jets
and explore final state signatures with either four top quarks or two top quarks
and two light jets. We investigate the sensitivity of both a same-sign
dilepton plus jets
and a single lepton plus jets analysis to the possible presence of new physics
events, and we find that signal cross sections down to a few tens of fb can be
already probed at the Large Hadron Collider, assuming 10 fb$^{-1}$ of collected
data at a center-of-mass energy of 7 TeV.
\end{abstract}

\section{INTRODUCTION}
Among all the alternatives proposed to extend the Standard Model of particle
physics, supersymmetric theories, and more particularly their minimal version, the
so-called Minimal Supersymmetric Standard Model (MSSM) \cite{Nilles:1983ge,
Haber:1984rc}, are theoretically and phenomenologically well-motivated. 
Assuming a symmetry between the fermionic and bosonic degrees of freedom of the
theory, they predict fermionic (bosonic) partners for each
bosonic (fermionic) Standard Model particle. They feature
a solution to the hierarchy problem \cite{Witten:1981nf}, 
gauge coupling unification at high energies \cite{Dimopoulos:1981yj} and
contain, as the lightest supersymmetric particle, a candidate to solve the dark
matter problem in the universe \cite{Ellis:1983ew}. However, supersymmetric
partners of the Standard Model particles have not been observed yet. Therefore,
and in order to remain a viable solution to the hierarchy problem, supersymmetry
has to be softly broken, which shifts the masses of the superpartners around the
TeV scale. 

This explains why searches for the supersymmetric counterparts of the Standard Model
particles are among the main experimental topics at the Large Hadron Collider
(LHC)
experiments at CERN. Direct searches are currently under way and limits on the
supersymmetric masses are set to higher and higher scales by the ATLAS and CMS
experiments \cite{atlas, cms}. However, most of the results are derived 
under the hypothesis of the constrained MSSM,
\textit{i.e.}, a framework in which the ${\cal O}(100)$ new free parameters of the
theory are
reduced to four parameters and one sign. Consequently, there exist alternative
realizations of supersymmetry with properties 
different from those expected in the 
constrained MSSM and thus not covered by the current experimental analyses.
Therefore, their investigation at the LHC deserves dedicated studies in order to
prepare the interpretation of data within these models. 
As an example, we have chosen in this work to focus on $N=1/N=2$ hybrid
supersymmetric theories \cite{Fayet:1975yi, AlvarezGaume:1996mv, Choi:2008pi,
Choi:2008ub, choi:2010gc, Choi:2010an, Schumann:2011ji} and $R$-symmetric
supersymmetric theories \cite{Salam:1974xa, Fayet:1974pd, Kribs:2007ac} which
both predict, among others, the existence of a new color-octet scalar field.
In these models, the $SU(3)_c$ vector supermultiplet of the MSSM is supplemented
by an additional chiral supermultiplet lying in the adjoint representation of
the $SU(3)_c$ gauge group, forming hence a complete gauge supermultiplet of an $N=2$
supersymmetry. The latter contains a vector degree of freedom which corresponds
to the gluon field, two (two-component) fermionic fields mixing to form a Dirac
gluino field, in contrast to the Majorana counterpart included in the MSSM, and
one color-octet scalar particle commonly dubbed sgluon \cite{Plehn:2008ae,
Choi:2009jc}. Let us note that another type of color-octet scalar particles,
called hyperpions, also appear in vectorlike confining theories at the TeV scale
\cite{Kilic:2008pm, Kilic:2008ub, Kilic:2009mi, Kilic:2010et, Dicus:2010bm,
Sayre:2011ed}. While some signatures are similar, the supersymmetric theories
can lead to additional final states. In the following we will therefore only
consider the supersymmetric case.

Being sensitive to the strong interaction, sgluons (strongly) couple to gluons and 
quarks and are therefore expected to be copiously pair-produced at the LHC.
Then, their observation could be possible through their allowed decays
into a pair of light jets, a pair of top quarks, or even an associated pair of a
top quark and a light jet. In this work,
rather than handling the complete model, which also demands a careful 
design of a theoretically motivated and not experimentally excluded 
benchmark scenario, we follow an alternative approach where
we construct a simplified model allowing for sgluon pair hadroproduction and
decay either into two top quarks and two light jets, or into four top
quarks. Assuming a sgluon with a mass lying in the TeV range, together with
sensible couplings to quarks and gluons, we analyze the sensitivity
of the LHC to the possible observation of sgluon particles within multitop events
containing either two top quarks and two additional light jets or four top
quarks. Both the single lepton plus jets and same-sign dilepton plus jets channels are
investigated.

The outline of this contribution is as follows: In Section \ref{sec:theo}, we
describe the effective theory which we have constructed modeling sgluon pair
production and decays at the LHC and define four benchmark scenarios for our
phenomenological studies. In Section \ref{sec:analysis}, we present the details
of our Monte Carlo simulations and the results of our two phenomenological
analyses. Our conclusions are given in Section \ref{sec:conclusions}.

\section{AN EFFECTIVE THEORY FOR SGLUON PRODUCTION AND DECAY AT THE
LHC AND BENCHMARK SCENARIOS}\label{sec:theo}
As stated above, sgluons can be pair-produced at the LHC through their couplings
to gluons and quarks. The same interactions also allow for sgluon
decays, in a flavor conserving fashion or not, to light jets and top quarks.
These properties can be embedded into an effective gauge-invariant  Lagrangian
describing the dynamics of a complex sgluon field $\sigma$,
\begin{equation}\label{eq:lag}\begin{split}
 {\cal L} =&\ {\cal L}_{SM} + \Big(
   \sigma^a \bar d T_a \Big[ a_d P_L + b_d P_R \Big] d + 
   \sigma^a \bar u T_a \Big[ a_u P_L + b_u P_R \Big] u + 
   a_g d_a{}^{bc} \sigma^a g_{\mu\nu b} g^{\mu\nu}{}_c +
       {\rm h.c.} \Big)\\ &\  +   D_\mu \sigma^\dag_a D^\mu \sigma^a - m_\sigma
       \sigma^a \sigma^\dag_a\  .
\end{split}\end{equation}
This effective Lagrangian contains the Standard Model Lagrangian
${\cal L}_{SM}$, the sgluon kinetic and mass ($m_\sigma$) 
terms included in the last two terms of the equation above and the
sgluon 
gauge interactions, which 
are embedded through the $SU(3)_c$ covariant derivative in the adjoint
representation,
\begin{equation}
  D_\mu \sigma^a = \partial_\mu \sigma^a + g_s f^a{}_{bc} g_\mu^b \sigma^c \ ,
\end{equation}
where $g_\mu^b$ denotes the gluon field, $g_s$ the strong coupling constant and
$f^a{}_{bc}$ the structure constants of $SU(3)_c$. In complete hybrid or
$R$-symmetric supersymmetric models, additional interactions of sgluons to quarks
and gluons are generated at the one-loop level. The sgluon can hence also singly
couple to two, three or four gluons or to a pair of quarks. In our effective
theory,  this option is allowed through the Lagrangian terms given in the 
parenthesis of the first line of Eq.\ \eqref{eq:lag}. In the latter, flavor
indices, as well as (anti)fundamental color indices, are understood, and we have
introduced the left-handed and right-handed chirality projectors $P_L$ and
$P_R$. In addition, $g_{\mu\nu a}$ stands for the gluon field strength tensor,
whilst $T_a$ and $d^a{}_{bc}$ are the fundamental representation matrices and
the symmetric structure constants of $SU(3)$, respectively. The relative
coupling strengths are left as free parameters and consist in a set of four
complex $3\times 3$ matrices in flavor space $a_d$, $b_d$, $a_u$,
$b_u$ and a complex number $a_g$.

To investigate the main features of sgluon signatures at the LHC, we consider
two simplified scenarios I and II, each of them split into two categories (1 and
2) which differ only in the mass assumed for the sgluon. For
scenarios I.1 and II.1, the sgluon is assumed to have a mass of 400 GeV,  a rather
collider-friendly value, while for scenarios I.2 and II.2, it is assumed to be a
1000 GeV particle, yielding a phase-space suppressed but still visible sgluon
pair-production cross section at the LHC. The difference between the two
sets of scenarios lies in the allowed decay channels for the
color-octet scalar particle. For scenarios I.1 and I.2, sgluons are only allowed
to decay to a top-antitop pair with a branching ratio equal to
unity. In contrast, for scenarios II.1 and II.2, they can decay, with an equal
rate, to a top or an antitop quark in association with a light jet originated from
an up or charm quark. Let us note that the total branching
ratio of sgluon to this $tj$ channel is again assumed equal to unity.
Moreover, we are looking for moderately low mass objects. Therefore,
kinematically, the parton densities of the proton tell us that gluon fusion
mechanism is dominant over the quark-antiquark scattering processes for these
masses. The production cross section then depends on the mass of the sgluon and
the sgluon-gluon-gluon coupling $a_g$.

The values of the non-zero parameters of the Lagrangian of Eq.\ \eqref{eq:lag}
are given, for our four benchmark scenarios, in Table \ref{tab:params}, whilst
all the Standard Model parameters, with the exception of the top quark mass $m_t$ which is chosen
equal to 172 GeV, follow the conventions of Ref.\ \cite{Christensen:2009jx}.
The numerical values of the non-zero sgluon coupling strengths have been chosen
so that the cross sections related to sgluon pair production and decays are
visible at the LHC.

\begin{table}[!t]
\begin{center}
\begin{tabular}{l | r r r r}
\hline\hline
Parameter &  Scenario I.1 & Scenario I.2 & Scenario II.2 &
  Scenario II.1\\
\hline
$a_g$  & 1. & 1. & 1. & 1.\\
$(a_u)_{tt}$ & 0.085 & 0.085 & 0 & 0\\
$(a_u)_{ct} = (a_u)_{tc}$ & 0 & 0 & 0.04 & 0.04\\
$(a_u)_{ut} = (a_u)_{tu}$ & 0 & 0 & 0.04 & 0.04\\
$m_\sigma$ & 400 GeV & 1000 GeV & 400 GeV & 1000 GeV\\
$m_t$ & 172 GeV & 172 GeV & 172 GeV & 172 GeV\\
 \hline\hline
\end{tabular}
\caption{\label{tab:params} Non-zero input parameters for our four benchmark
scenarios. All the other Standard Model parameters follow the conventions of
Ref.\ \cite{Christensen:2009jx}.}
\end{center}
\end{table}

\section{PHENOMENOLOGICAL INVESTIGATIONS}\label{sec:analysis}
\subsection{Monte Carlo simulations and object definitions} \label{sec:tools}

Event simulation is performed for the LHC at $\sqrt{s} = 7$ TeV and for an
integrated luminosity of 10 fb$^{-1}$. Both for signal and background events,
the hard scattering process has been described with the multipurpose matrix
element generator {\sc MadGraph} 5 \cite{Alwall:2011uj}. Neglecting all quark
masses but the top mass, we employ the leading order set of the CTEQ6 parton
density fit \cite{Pumplin:2002vw} and identify both the renormalization and
factorization scales as the partonic center-of-mass energy. Concerning the
simulation of the sgluon signals, we have implemented the Lagrangian
of Eq.\ \eqref{eq:lag} in the {\sc FeynRules} package
\cite{Christensen:2008py, Duhr:2011se} and used the dedicated UFO-interface to
convert the Feynman rules to the UFO format \cite{Degrande:2011ua}, which has
been adopted as the standard {\sc MadGraph} model format. Using this framework, 
our estimates for the (leading order) cross section for sgluon pair production
(without decays) are 1.86 pb, 0.82 fb, 1.86 pb and 0.82 fb for our four scenarios
I.1, I.2, II.1 and II.2, respectively. The cross sections of the different
generated background processes are presented in Table \ref{tab:xsec}.

\begin{table}
\begin{center}
\begin{tabular}{lr}
\hline
\hline
Sources of background & Cross sections [pb] \\
\hline
$W\to l\nu + \text{jets}$ 			& 14350\\
$\gamma^{*}/Z\to l\bar{l} + \text{jets}$ 	& 1672\\
$t\bar{t} + \text{jets}$ (semileptonic) 	& 17.1 \\
$t\bar{t} + \text{jets}$ (dileptonic) 		& 3.2 \\
single top (leptonic) + \text{jets} [$s$-channel] 	& 0.49\\
single top (leptonic) + \text{jets} [$t$-channel] 	& 10.8\\
single top (semileptonic) + \text{jets} [$tW$-channel] 	& 2.21\\
single top (dileptonic) + \text{jets} [$tW$-channel] 	& 0.41\\
$WW+ \text{jets}\to$ at least one lepton 	& 9 \\
$WZ+ \text{jets}\to$ at least one lepton	& 0.48 \\
$ZZ+ \text{jets}\to$ at least two leptons 	& 2.8 \\
$t\bar{t}W  + \text{jets}$ & 0.03\\
$t\bar{t}Z  + \text{jets}$ & 0.02\\
$t\bar{t}WW + \text{jets}$ & 0.21\\
$t\bar{t}t\bar{t} + \text{jets}$ & 3.7 $10^{-4}$\\
\hline
\hline
\end{tabular}
\caption{Cross sections for the different genererated sources of background.}
\label{tab:xsec}
\end{center}
\end{table}

In order to have a
description of the collisions as accurate as possible, we have matched the {\sc
MadGraph} event samples with parton showering and hadronization as provided by
the {\sc Pythia} program \cite{Sjostrand:2006za, Sjostrand:2007gs}.
We have then performed a fast detector simulation with the program {\sc Delphes}
1.9 \cite{Delphes}, using the publicly available ATLAS detector card.

Jets are reconstructed using the anti-$k_{t}$ algorithm with a radius parameter
set to $R=0.4$, and only those with a transverse momentum $p_T > 10$ GeV are
kept in our analysis. However, jets defined with such a low radius, and
particularly those with a low transverse momentum, may suffer from large bias in
energy measurement due to the possibly significant fraction of the energy
not being reconstructed. We have corrected this effect by an 
\textit{ad-hoc} calibration based on test samples of dijet events issued from
the decay of a $Z$- or a $Z'$-boson, the latter having a mass ranging from 200
GeV to 1 TeV. After this calibration, we only retain, at the analysis level, 
jets with a transverse momentum $p_T> 20$ GeV and a pseudorapidity $|\eta|<2.5$.
In addition, we estimate a $b$-tagging efficiency of about 60\%, together with a
charm and light flavor mistagging rate of 10\% and 1\%, respectively.

Leptons are required to have a transverse momentum $p_T > $ 20 GeV. The
pseudorapidity of electrons must be less than 2.47 and for muons less than 2.5. 
Finally, we require isolation criteria and do not consider in our analysis jets
at a relative distance $\Delta R<0.2$ of an electron and muons at a relative
distance $\Delta R<0.2$ from a jet.

\subsection{Event selection} 

Considering the four tops or two tops plus two light jets signatures described in
Section \ref{sec:theo}, the reconstructed final state can be more or less rich in
leptons and jets after the decays of the top quarks. We design
two different analyses, the first one asking for a same-sign lepton pair with a
moderate number of jets and the second one for exactly one lepton produced in
association with a large number of jets. The first analysis is expected to
be sensitive to our four scenarios, whilst the second one is especially designed
to use the sgluon mass reconstruction to improve the sensitivity 
for scenarios of type II.

In the following, we define the sensitivity of our analyses as in Ref.\
\cite{Schumann:2011ji}, \textit{i.e.}, as the statistical
error on the background predictions (for a luminosity of 10 fb$^{-1}$) divided by
the efficiency and multiplied by 1.64, similarly to a one-sided Gaussian
counting experiment limit. In other words, we estimate the signal cross section 
which can be excluded at the 95\% confidence level in the absence of
a new physics observation.

\subsubsection{Same-sign dilepton plus jets signature} 
In order to probe the same-sign dilepton plus jets signature, our
event selection criteria consist in rejecting events which do not contain
exactly two charged leptons (electrons or muons) with the same electric charge
and a transverse momentum $p_T> 20$ GeV. We require at least four jets, one or more
being tagged as 
$b$-jets, with a reconstructed transverse energy of $E_{T}>25$ GeV. In addition,
we impose that the invariant-mass of the lepton pair is above $20~$GeV to reject
hadronic resonances and we 
demand a missing transverse energy of at least 40 GeV. It is
important to note that requiring same sign leptons allows to reject a
significant fraction of the Standard Model background. We then obtain a
sensitivity of 110 fb, 61 fb, 380 fb and 209 fb for scenarios I.1, I.2, 
II.1 and II.2, respectively. 

In the case of scenarios of type I, the analysis can be further improved by
taking advantage of the larger expected jet multiplicity in the
final state due to the decays of the four top quarks. Requiring 
eight
reconstructed jets with a transverse energy $E_{T}> 25$ GeV, two or more 
of them being $b$-tagged, we
find that the LHC is now sensitive to signal cross sections of 17 fb and 6
fb for the scenarios I.1 and I.2, \textit{i.e.}, for 400 GeV and 1 TeV sgluons,
respectively. 

The important hadronic activity in the final state suggests a last
selection cut on the $H_T$ variable which we define as the scalar sum of the
transverse momentum of all the reconstructed leptons and jets, together with the
missing transverse energy. Our results for the four scenarios, assuming an
optimal cut on the $H_T$ variable, are given in Table
\ref{table_sensitivity_dilepton} together with the value of this cut.

\begin{table}
\begin{center}
\begin{tabular}{c|c c c}
\hline
\hline
  Scenario & Sensitivity before $H_T$ cut [fb]&$H_T$ cut [GeV]& Sensitivity
after $H_T$ cut [fb]  \\
\hline
  I.1    & \phantom{0}17               & \phantom{0}850    & 11.8   \\
  I.2    & \phantom{00}6               & 1750                      & 0.5   \\
  II.1   & 380                                 & \phantom{0}650    & 216  \\
  II.2   & 209                                 & 1550                      & 15   \\ 
\hline
\hline
\end{tabular}
\caption{Sensitivity of our dilepton analysis to a sgluon signal at the LHC
assuming the indicated optimal cuts on the $H_T$ variable.}
\label{table_sensitivity_dilepton}
\end{center}
\end{table}


\subsubsection{Single lepton plus jets signature} 
Turning to our second analysis focusing on signatures containing a single lepton
produced in 
association with a large number of jets, we select events containing at least
one electron or one muon with a transverse momentum $p_T> 25$ GeV and no other
charged lepton with a transverse energy $E_{T}> 20$ GeV. These requirements
ensure the orthogonality with the dilepton analysis above. In addition, we
demand at least six jets with a reconstructed transverse energy $E_{T}> 25$ GeV,
with one or more of them being $b$-tagged. Since the reconstructed lepton comes
from the decay of 
a $W$-boson, together with an unobserved neutrino, we drastically reduce the
multijet background by requiring missing transverse energy $\slashed{E}_T > 40$ GeV
and a moderate $W$-boson transverse mass $M_T^W>25$ GeV, the latter being
defined as
\begin{equation}
  M_T^W = \sqrt{ 2\ p_T^{\rm lepton} \slashed{E}_T 
    \Big[ 1 - \cos\big( \Delta\phi({\rm lepton}, \slashed{E}_T)\big) \Big] } \ .
\end{equation}
One obtains a sensitivity to a sgluon signal of 152 fb, 131 fb, 334 fb
and 228 fb for scenarios I.1, I.2, II.1 and II.2, respectively.

The single lepton analysis allows to reconstruct the sgluon mass
which is used to improve the sensitivity for ditop signal events.
Indeed, combinatorics
render this task extremely difficult for final states containing four top quarks.
Thus, we focus, in the rest of this report, only on the scenarios of
type II which have
been designed for a signature containing two top quarks and two light jets.
We choose to reconstruct the sgluon mass by combining objects so that we
minimize a quantity $\chi^2$ defined as 
\begin{equation}\label{eq:chi2}\begin{split}
  \chi^2  =&\ 
    \left[ \frac{m_{jj}-m^{(r)}_W}{\sigma^{(r)}_W} \right]^{2} 
  + \left[ \frac{\big(m_{jjj}-m_{jj}\big) - m^{(r)}_{tW}}
      {\sigma^{(r)}_{tW}} \right]^{2} 
  + \left[ \frac{m_{l\nu j}-m^{(r)}_{tl}}{\sigma^{(r)}_{tl}} \right]^{2} \\ 
&\ + \left[\frac{(m_{l\nu j,j}-m_{l\nu j})-(m_{jjj,j}-m_{jjj})}
      {\sigma_{\sigma t}^{(r)} \big[ 
       (m_{l\nu j,j}-m_{l\nu j})+(m_{jjj,j}-m_{jjj}) \big] } \right]^{2} \ .
\end{split}\end{equation}
For scenarios of type II, the final state contains one
hadronically
decaying top quark, one leptonically decaying top quark and two additional 
light jets. Hence, two of the reconstructed jets are  
the decay products of a $W$-boson. We estimate the expected mass of the
reconstructed $W$-boson $m^{(r)}_W$ and the associated uncertainty
$\sigma^{(r)}_W$, corresponding to the detector resolution, with the help of the
Monte Carlo truth, fitting the reconstructed $W$-mass with a Gaussian.
We find the values given in Table \ref{tab:chi2}. These Gaussian
parameters are then used in the $\chi^2$ minimization function above, where
among all the possible combinations of jets, we select the one with the
invariant mass $m_{jj}$ closest to $m^{(r)}_W$ as indicated by the first term of
Eq.\ \eqref{eq:chi2}. 

\begin{table}
\begin{center}
\begin{tabular}{c c|c c|c c| c}
\hline
\hline
  $m^{(r)}_W$ & $\sigma^{(r)}_W$ & $m^{(r)}_{tW} $ & 
  $\sigma^{(r)}_{tW}$ & $m^{(r)}_{tl}$ & $\sigma^{(r)}_{tl}$ &
  $\sigma_{\sigma t}^{(r)}$ \\
\hline
  85.5 GeV & 13.2 GeV & 90.5 GeV & 17.3 GeV & 169 GeV & 30.8 GeV & 0.090\\
\hline
\hline
\end{tabular}
\caption{Values of the free parameters entering the $\chi^2$ minimization
function of Eq.\ \eqref{eq:chi2}.}
\label{tab:chi2}\end{center}
\end{table}

Similary, using the Monte Carlo truth, one can extract from the events
the additional $b$-jet
necessary to reconstruct the hadronic top quark.
In order to include this information in the minimization function, one must
however first subtract the $W$-boson reconstructed mass, the latter being
strongly correlated to the reconstructed top mass. The obtained values for the
mass difference $m^{(r)}_{tW}$ and the width $\sigma^{(r)}_{tW}$
are again presented in Table \ref{tab:chi2}.
 These last parameters are used in the second term of
the $\chi^2$ function of Eq.\ \eqref{eq:chi2}, where we select the
right combination of three jets ($b$-tagged or not) of invariant mass $m_{jjj}$
yielding a reconstructed hadronically decaying top quark.

The third term of Eq.\ \eqref{eq:chi2} constrains the reconstruction of the
leptonic top quark. Using the information included in the Monte Carlo truth, one
derives the parameters $m_{tl}^{(r)}$ and $\sigma_{tl}^{(r)}$, the average mass
and the width of the reconstructed (leptonically decaying) top quark. They are
obtained in a similar fashion as above and given in Table \ref{tab:chi2}.
We include them in the third term of the minimization function,
optimizing  
the selection of the jet (whether it is $b$-tagged or not) to be matched with
the final state lepton and missing energy so that a top quark mass is
reconstructed. As above, the difference between the
invariant mass $m_{l\nu j}$ and $m_{tl}^{(r)}$ is minimized. Let us note that
the neutrino four-momentum, necessary for the computation of $m_{l\nu j}$, is
infered from the reconstructed missing transverse energy and the $W$-boson mass
taken equal to 80.4 GeV \cite{Nakamura:2010zzi}\footnote{In this way, a
constraint on the $W$-boson mass is also imposed at the same time.}.

In the last term of Eq.\ \eqref{eq:chi2}, we use the fact that two
sgluons are produced, which allows us to require two similar reconstructed masses
(without assuming anything on the corresponding
numerical value). Combining each of the
reconstructed top quarks with an extra jet, one then requires $m_{l\nu j,j}
\approx m_{jjb,j}$. However, in order to decouple this constraint from those
related to the reconstruction of the top masses, 
we subtract the reconstructed top masses $m_{jjj}$ and
$m_{l\nu j}$ from the two reconstructed sgluon masses, similarly to
what was done to the second term of Eq.\ \eqref{eq:chi2}.
Moreover, in
order to remove a too strong dependence on the sgluon mass, which holds
especially for heavy sgluons, we add an extra factor, at the denominator, 
of $(m_{l\nu j,j}-m_{l\nu j})+(m_{jjj,j}-m_{jjj})$ in the fourth term of
the minimization function. As for
the other terms, the combined width on the sgluon and top mass difference
$\sigma^{(r)}_{\sigma t}$ is extracted from a Gaussian fit based on the Monte
Carlo truth and given in Table \ref{tab:chi2}.

\begin{figure}
\begin{minipage}{0.49\linewidth}
\includegraphics[width=\textwidth]{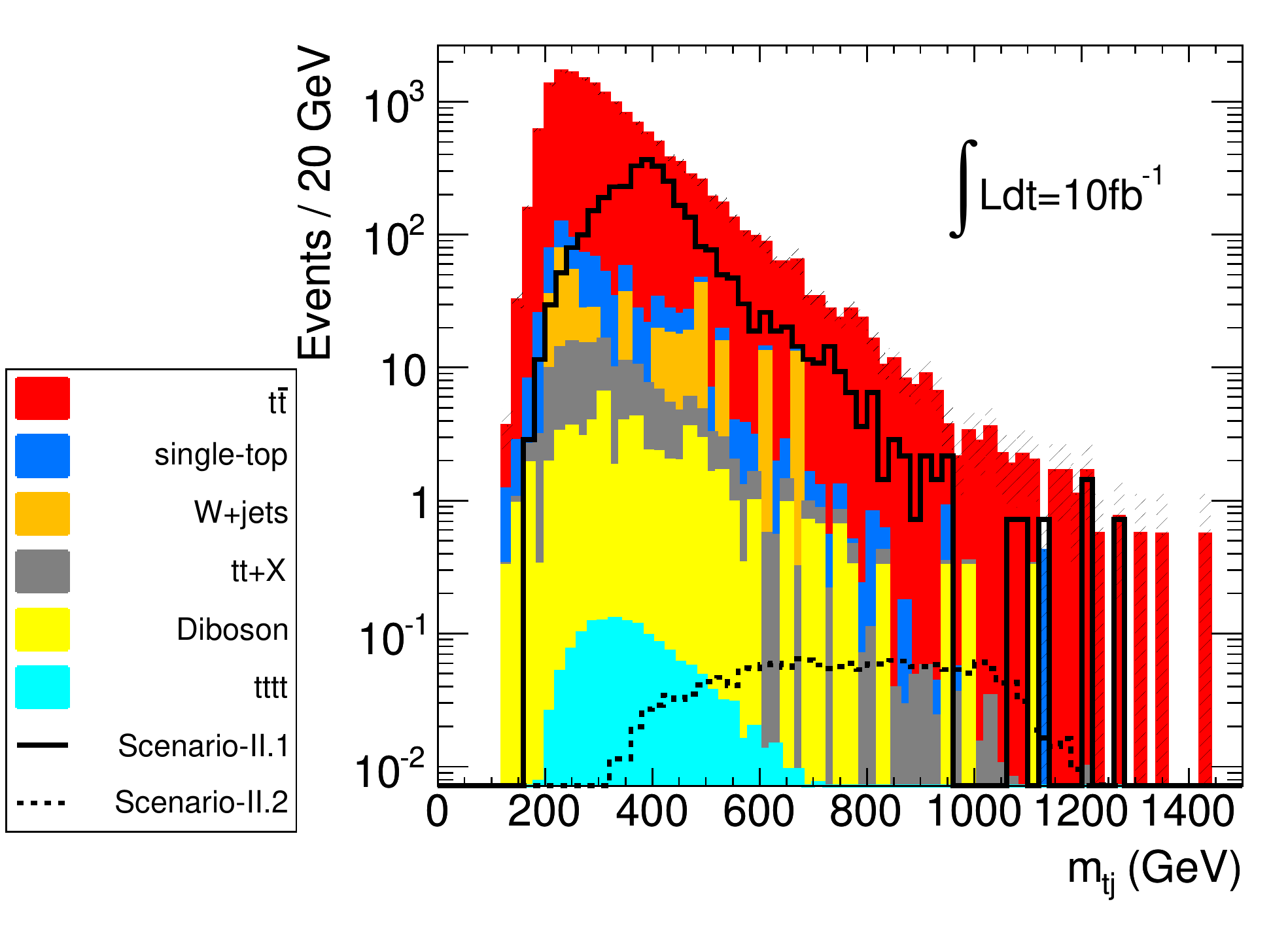}
\end{minipage}
\begin{minipage}{0.49\linewidth}
\includegraphics[width=\textwidth]{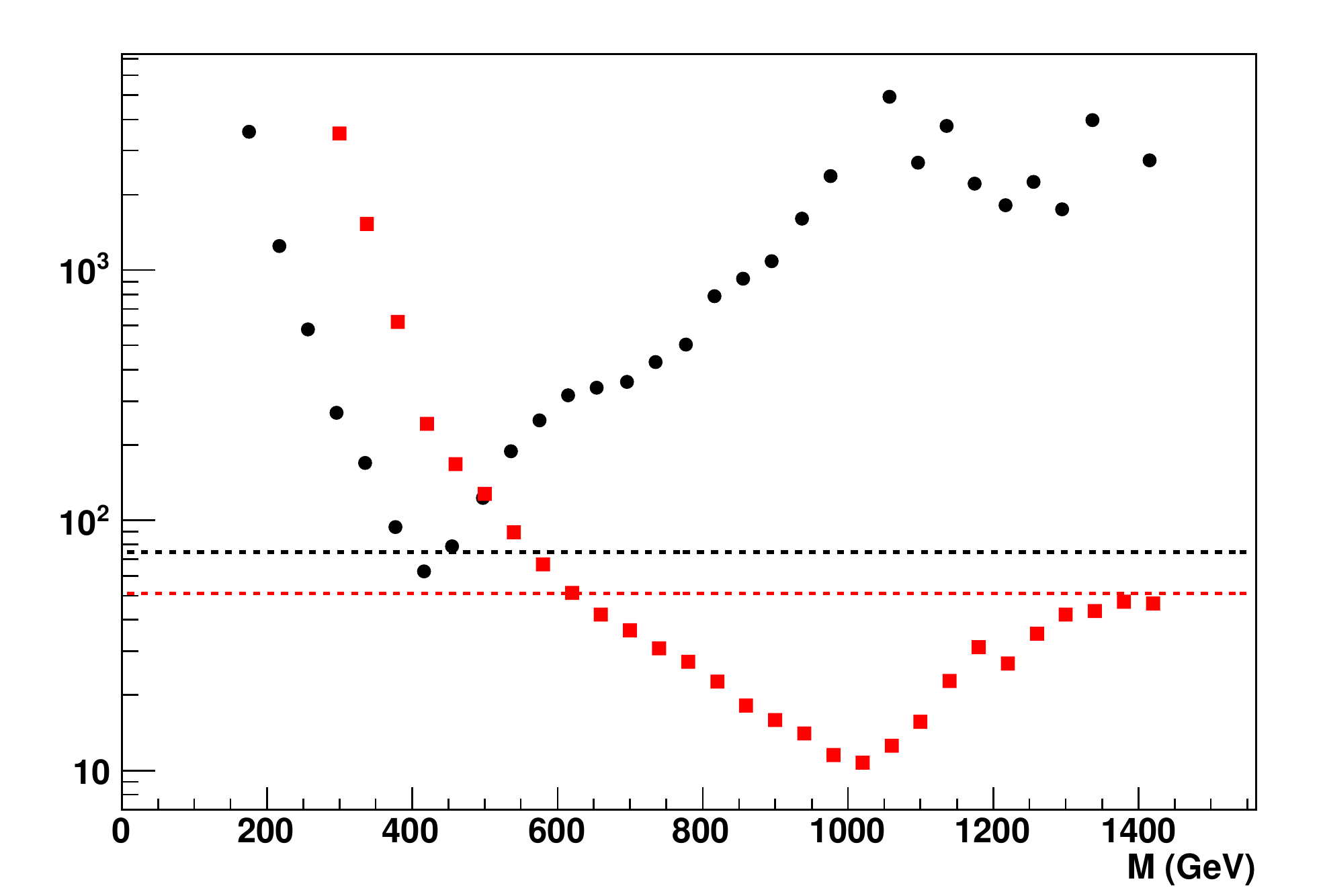}
\end{minipage}  
\caption{Left: reconstructed top-jet invariant mass $m_{tj}$, 
  for both signal events corresponding to our scenarios II.1 (solid) and II.2
  (dashed) and the different sources of backgrounds (see the description in the
  text). Right: assuming a test mass $M$ and retaining events for which $m_{tj}$
  lies in a $[M-10\%, M+10\%]$ window, we present the dependence of the 
  sensitivity of our analysis on $M$ for signal events corresponding to our
  scenario II.1 (a 400 GeV sgluon, in black) and II.2 (a 1 TeV sgluon, in red).
  For comparison purposes, the global sensitivities, without using the
  reconstructed $m_{tj}$ mass, are indicated by dotted lines.}
\label{ReconstructedMass}
\end{figure}

Finally, to reduce combinatorial issues, we demand that the two leading jets
belong to the object combinations reconstructing the mass of the two produced
sgluons. The reconstructed mass $m_{tj}$ distribution is presented in Figure
\ref{ReconstructedMass} (left panel) 
both for the signal and all possible sources of
backgrounds. Signal events for sgluons of 400 GeV and 1 TeV, \textit{i.e.},
scenarios II.1 and II.2, are shown as solid and dashed curves, respectively,
whilst the main
sources of background coming from $t\bar t$ (red), single top (blue) and $W$
plus jets (orange) events are indicated. For the sake of completeness, the other, subleading,
possible sources of background are shown as three groups, the first one,
$t\bar t+X$ (grey) being related to $t\bar t + Z$, $t\bar t + W$ and $t\bar t +
WW$ events, the second one to all diboson events (yellow) and the last one to
the Standard Model production of four top quarks (turquoise). 

Assuming a sgluon
test mass $M$ and retaining events for which the reconstructed invariant mass
$m_{tj}$ lies within a $[M-10\%, M+10\%]$ window around the test mass, the LHC
sensitivity to a sgluon signal, shown in Figure  \ref{ReconstructedMass} (right
panel) is found to improve with respect to the global case discussed above.
For optimal choices of the test mass, the sensitivity reaches even 278 fb and
48 fb for our scenarios II.1 and II.2, respectively, which has to be compared
to the previously found values of 334 fb and 228 fb. Therefore, our analysis is
more sensitive to high sgluon mass, provided enough signal events are
produced.

\section{CONCLUSIONS}\label{sec:conclusions}
In this paper, we have presented a strategy for searches of new color-octet
scalar particles, such as those predicted by hybrid $N=1 / N=2$ or $R$-symmetric
supersymmetric theories. We have used a simplified effective field theory
describing the production and decays of such particles at the LHC and have
investigated two different multitop channels, 
a first one with four tops in the final state
and a second one with two top quarks, possibly carrying the same electric
charge, produced in association with two light jets. 

We have studied the sensitivity of the LHC collider, running at a center-of-mass
energy of 7 TeV to such signals, in two dedicated analyses. In one case, we
consider same-sign dilepton signatures with a moderate number of jets, whislt in the
second case, we only require a single lepton, but ask,  in addition, for a larger
number of jets. We have shown that cross sections as low as a few
tens of fb can be probed, assuming some basic cuts on jet and lepton
multiplicity. Moreover, for ditop analyses, we have presented a way to
reconstruct the top-jet invariant mass, \textit{i.e.}, the sgluon invariant mass
in the case of signal events. We have shown how this observable can be used to
improve the sensitivity to signal events for the corresponding analysis,
especially for heavier sgluon particles. However, one must note that pile-up of
events at the LHC, not considered here, could decrease the sensitivity.

\section*{ACKNOWLEDGEMENTS}
BF, AR and DZ would like to thank the Les Houches school for physics for their
hospitality during the workshop where some of the work contained herein was
performed, as well as the worskhop organizers. BF acknowledges
support from the Theory-LHC France-initiative of the CNRS/IN2P3.



\AddToContent{S.~Calvet, B.~Fuks, P.~Gris, A.~Renaud, L.~Val\'ery, D.~Zerwas}
\renewcommand{\thesection}{\arabic{section}}

\superpart{ Long-lived Particles }



\chapter{Metastable charginos and LHC searches}

{\it G. Polesello, T. G. Rizzo}

\begin{abstract}
In the framework of the Minimal Supersymmetric Standard Model,
the parameter space yielding long lived  charginos is studied.
For the case where the charginos decay outside the detector,
a limit on the production of metastable charginos in the MSSM
is set, on the basis of the results of a recent ATLAS search
for long-lived charged particles.
\end{abstract}

\section{INTRODUCTION}\label{chargino_secintro}

The very successful 2010/2011 data-taking at the LHC has 
allowed the ATLAS and CMS experiments to perform several
searches for supersymmetric particles. 
These searches to date have yielded no discovery, 
and squark/gluino masses in the TeV range have been excluded
for a  very constrained set of models based on the assumptions 
of large mass gaps within sparticles and very simple decay chains.\par
Several detailed studies are ongoing to assess the 
coverage of these analyses in the framework of a generic 
phenomenological Minimal Supersymmetric Model defined
in terms of 19 SUSY-breaking parameters (pMSSM) \cite{Djouadi:2002ze}. 
The aim of these studies is to identify `difficult' models
which may have escaped the first round of LHC analysis.
The `SUSY without prejudice' study \cite{Conley:2011nn,Conley:2010du} 
is based on a very extensive
scan of the 19-parameters space, featuring 
the simulation of $\sim$71k model points, 
and has identified classes of  models which escape the 
available LHC searches based on the requirement of a large
imbalance of transverse momentum ($\etmiss$) in the events. 
A large class of such models includes situations
in which the lightest chargino is approximately degenerate
with the neutralino and is long lived, thus appearing 
as a heavy stable muon-like particle in the detector.\par 
The ATLAS experiment \cite{Aad:2008zzm} has published a search for long lived 
heavy particles (LLP) based on the measurement of the momentum 
and of the velocity $\beta$ of particles reaching the 
muon detector \cite{Aad:2011hz}. The results of the analysis are interpreted 
in terms of long-lived sleptons in the framework GMSB SUSY models,
and in terms of R-hadrons appearing in models with very heavy 
quarks and light gluinos. \par
The purpose of the present study is to reinterpret the results 
of this analysis as limits on the chargino/neutralino sector 
of the pMSSM. 

\section{THE MODEL}\label{chargino_secmodel}
In a generic  MSSM the partners of the gauge and Higgs bosons
of the model, the bino, wino and Higgsinos 
($\tilde{B}, \tilde{W}, \tilde{H_u}, \tilde{H_d}$) mix to yield two 
charginos and four neutralinos.
At tree level the chargino and neutralino mixing matrices 
are determined  by four parameters,  the mass
of the bino $M_1$, the mass of the wino $M_2$, the 
common Higgsino mass $\mu$,
and $\tan\beta$, the ratio of the vacuum expectation values 
of the two Higgs doublets in the model.
The broad features of the chargino/neutralino mass spectra 
are defined by the hierarchy of $M_1$, $M_2$, and $\mu$.
Of special interest for our study is the situation where 
the hierarchy is such that the Next to Lightest
Supersymmetric Particle (NLSP) is the chargino which has a sufficiently 
small mass splitting with the LSP to render it stable on LHC detector
time-scales. 

\begin{figure}[htb]
\begin{center}
\dofigs{0.4\textwidth}{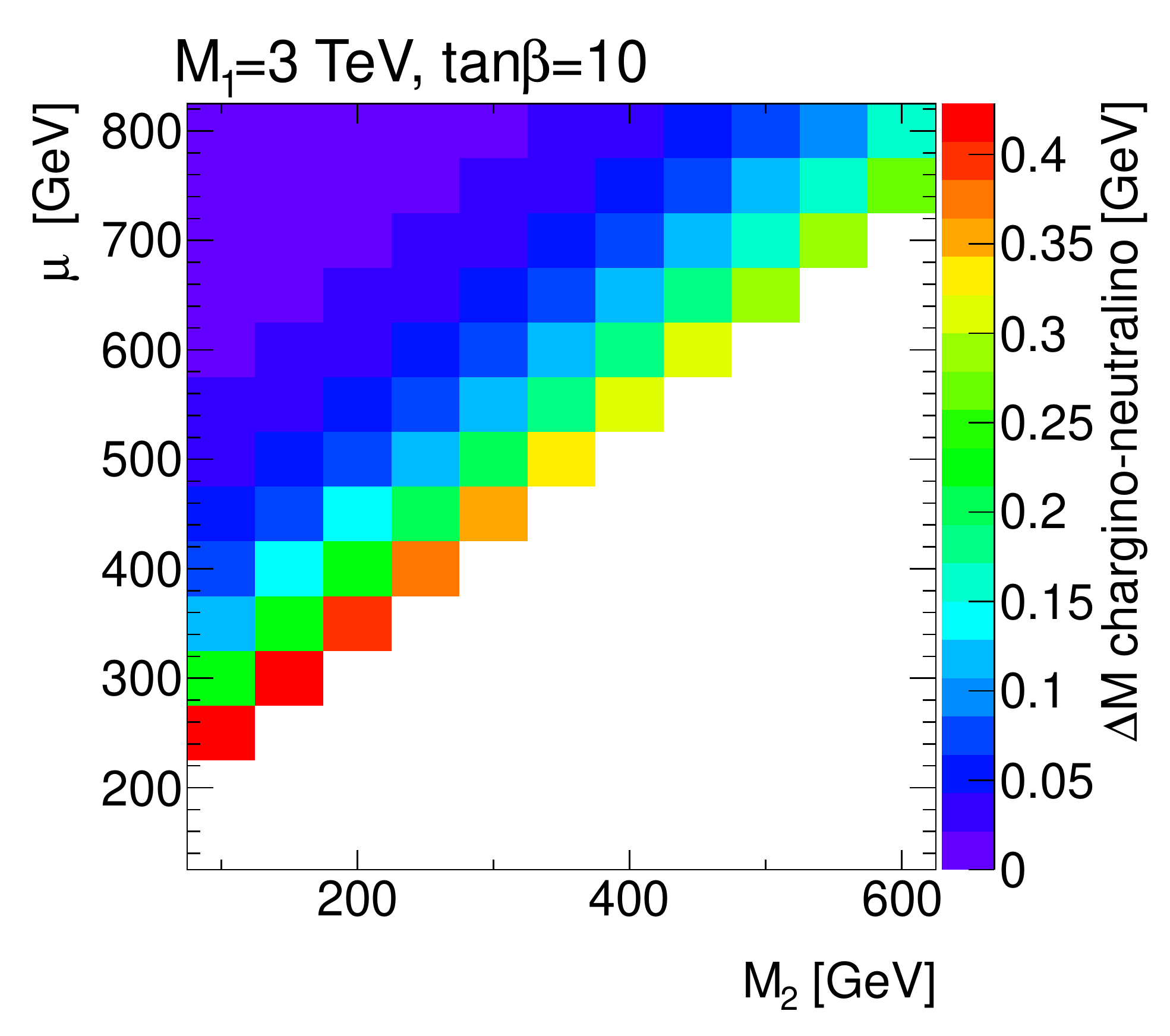}{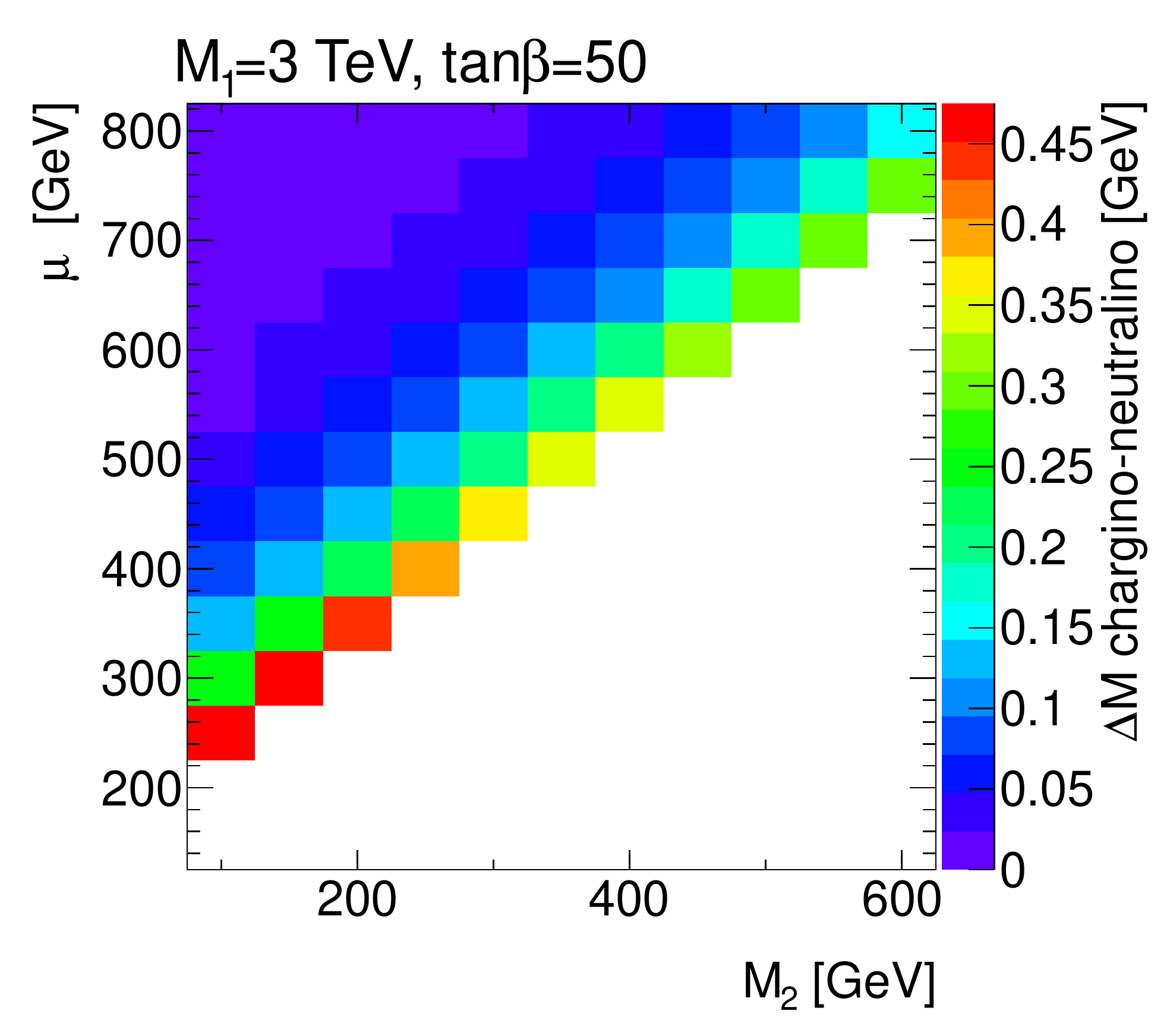}
\caption{Mass difference $\tilde{\chi}^{\pm}_1-\lsp$ 
as a function of $M_2$ and $\mu$ for two values of $\tan\beta$:
10 and 50. $M_1$ is set at 3 TeV.}
\label{chargino_fig1}
\end{center}
\end{figure}

An example of this situation occurs when $M_2<(M_1,\mu)$.  In this case 
both the lightest 
chargino  ($\tilde{\chi}^{\pm}_1$) and the lightest neutralino 
($\lsp$) are predominantly winos,
and both have a mass which is essentially given by $M_2$.
For sufficiently high $\Delta M(M_1,M_2)$ and $\Delta M(\mu,M_2)$  
both $\tilde{\chi}^{\pm}_1$ and $\lsp$ are pure winos and
the mass difference may become smaller than ~100~MeV. A similar situation 
occurs in the opposite limit when $|\mu|<(M_{1,2})$ in which case one 
obtains an Higgsino LSP with a slightly more massive chargino which is also 
Higgsino-like. In such cases, if the $\tilde{\chi}^{\pm}_1$ is NLSP,
 the only available decay chain for 
the $\tilde{\chi}^{\pm}_1$ would be the three-body decay 
$\tilde{\chi}^{\pm}_1\rightarrow e\nu\lsp$
which proceeds through the exchange of a virtual $W$ or selectron, and 
the particle will then appear  in the ATLAS detector as a stable charged 
penetrating particle. 
For the wino case  we studied the mass difference of $\tilde{\chi}^{\pm}_1$  
and $\lsp$ is shown in  Fig.~\ref{chargino_fig1} as a function of 
$M_2$ and $\mu$ with the SUSY-HIT program incorporating the 
loop  corrections to the chargino/neutralino  masses. 
The masses of all of the sfermions and of the gluino where put at 3~TeV,
the mass of the pseudoscalar Higgs to 1 TeV and the trilinear couplings to 
zero.  The required level of mass degeneracy is 
achieved for $\Delta M(\mu,M_2)$ in excess of $\sim250$~GeV. 
The dependence on $M_1$ was also studied, and it was found that 
for $\Delta M(M_1,M_2)$ in excess of $\sim100$~GeV the
mass difference is approximately independent of $M_1$.
Loop corrections are important to determine such a small mass 
splitting, so this result is valid in the limit
of the very high fixed value of the sfermion
masses used in this study.  
However, in an independent study addressing a very broad
range of parameters, where  these corrections are included using, e.g., 
the SUSPECT SUSY spectrum generator \cite{Djouadi:2002ze}, such small mass 
splittings still occur with a 
reasonable frequency \cite{Conley:2011nn,Conley:2010du}

\section{SIGNAL SIMULATION AND ANALYSIS}

In order to explore the wino-like LSP model described in the previous section 
in more detail, a series of model points were defined based on 
SUSY-HIT \cite{Djouadi:2006bz} with the
following parameters: $M_1$=3~TeV, $\mu$=1~TeV, $\tan\beta$=10.
The $M_2$ parameter was varied in steps of 20~GeV between 100 
and 200~GeV, and in steps of 100~GeV between 200 and 500~GeV.
The gluino mass and all the sfermion masses were set to 3~TeV,
and the trilinear couplings to zero.\par
For each point a set of  events were generated 
for the electroweak direct  chargino-chargino production with the 
Herwig++ \cite{Gieseke:2006ga,Gigg:2007cr} shower Monte Carlo generator. 
The production cross-section were normalised to the NLO cross-sections
calculated with the PROSPINO program \cite{Beenakker:1996ed}.
{\footnote {We note that for wino-like charginos, in addition to 
$s$-channel $\gamma,Z$ exchange, a $t$-channel squark exchange graph is also 
present which interferes destructively with those in the $s$-channel. For the 
Higgsino case, such $t$-channel exchanges are suppressed due to small Yukawa 
couplings.}} 
From the particle tree in output of the generator the charginos 
were extracted and passed to a simplified parametrised simulation of
the response of the ATLAS detector.\par
\begin{figure}[htb]
\begin{center}
\dofigs{0.4\textwidth}{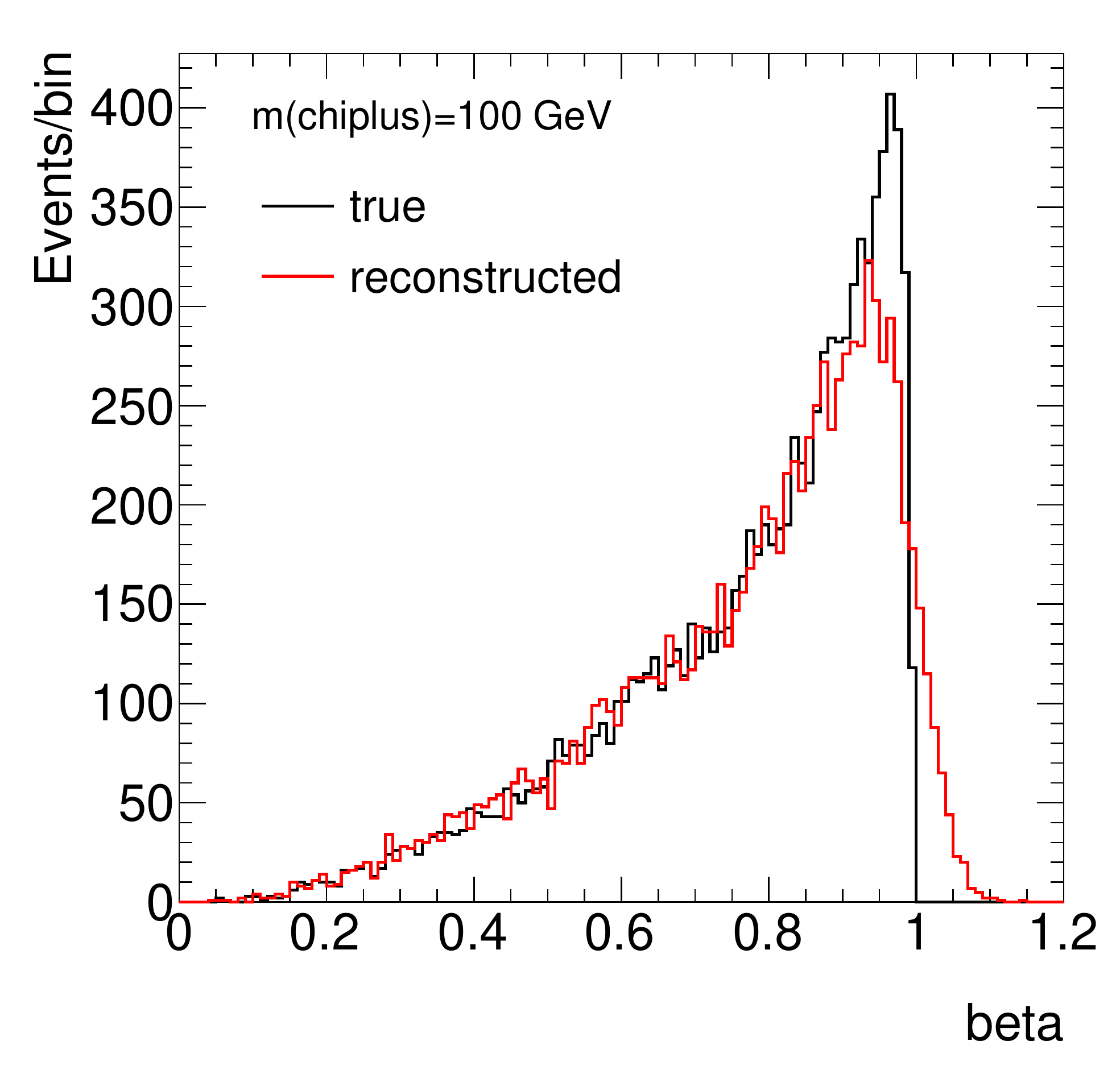}{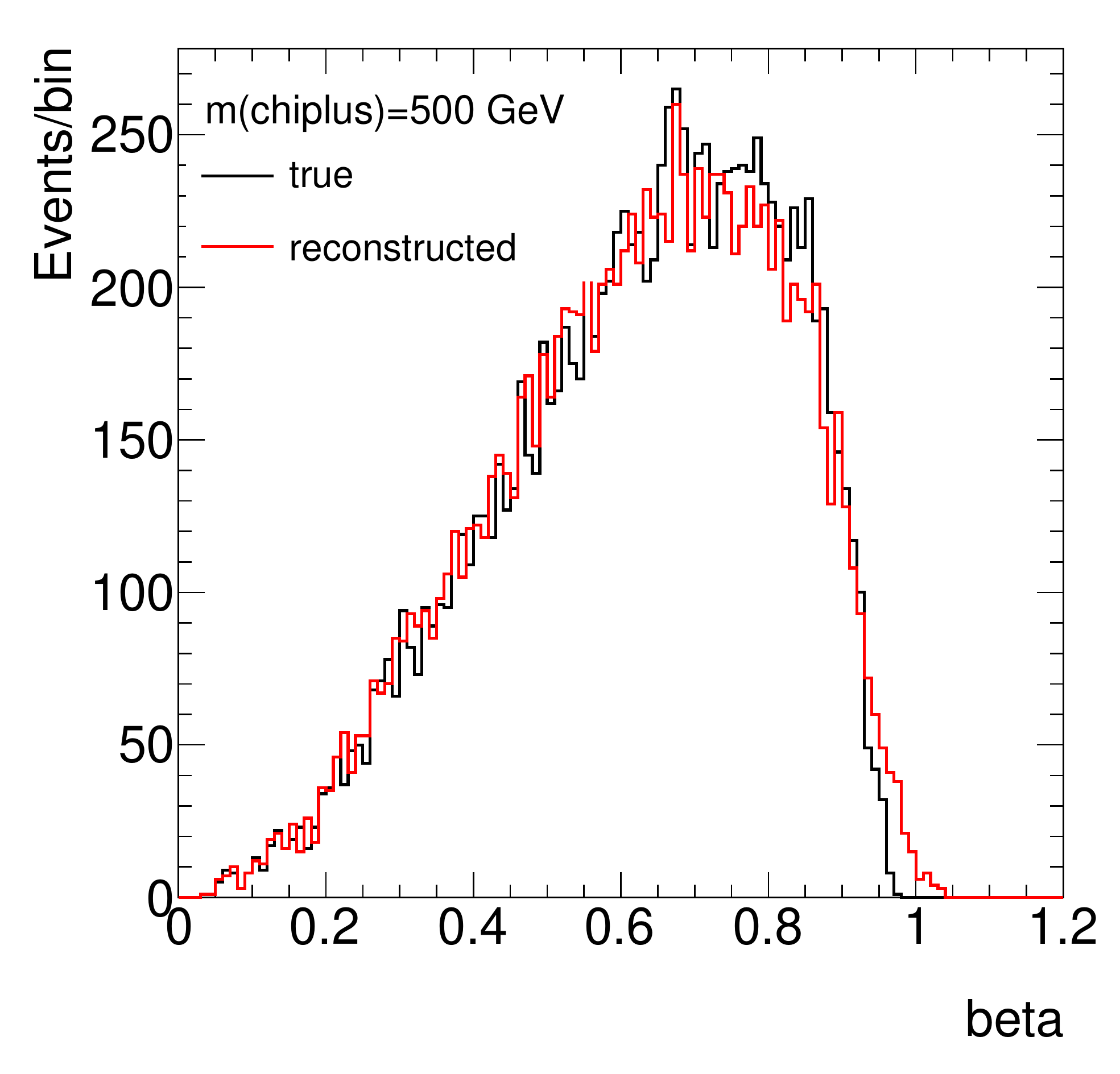}
\caption{Distribution of the $\beta$ variable for charginos of respectively
100 (left) and 500 (right) GeV. The black line shows the 
generated $\beta$, the red line the measured one.}
\label{chargino_fig2}
\end{center}
\end{figure}

\begin{figure}[htb]
\begin{center}
\dofig{0.6\textwidth}{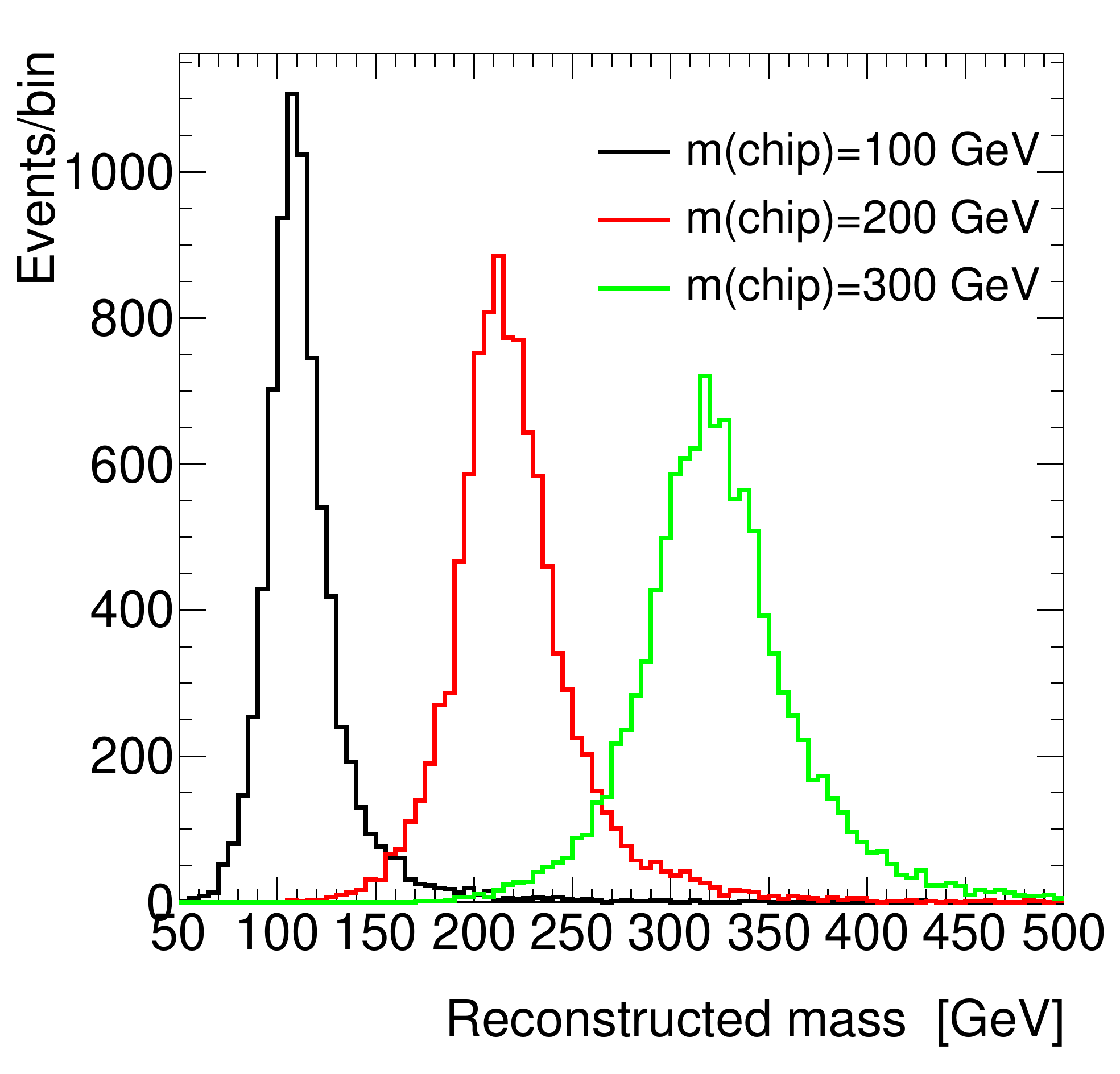}
\caption{Distribution of the reconstructed values for chargino 
masses from $\beta$ and momentum measurements. The distribution 
is shown for charginos of mass 100, 200 and 300 GeV.}
\label{chargino_fig3}
\end{center}
\end{figure}

 
For each chargino the momentum was smeared according to a parametrisation
of the momentum resolution of muons in the ATLAS detector 
as described in \cite{Aad:2008zzm}.
The velocity distribution $\beta$ was smeared 
according to the measured distribution for muons as shown 
in \cite{Aad:2011hz}. A simple gaussian smearing does not completely 
describe the tails of the experimental resolution as shown in 
the auxiliary plots to \cite{Aad:2011hz}. 
The correct simulation of the complete shape of the distribution 
is extremely important for the evaluation of the background 
coming from the tails of the $\beta$ and momentum measurement 
of real muons, and no attempt is done to estimate the
background, which is taken directly from the ATLAS paper. 
For signal, where the real $\beta$ is different from 1, only
the correct evaluation of the distribution width is relevant.
The real and smeared $\beta$ distributions for a ~110 GeV and 
a 300 GeV chargino are shown in Fig.~\ref{chargino_fig2}.
From the smeared beta and momentum the measured sparticle mass
is calculated. The distribution of measured masses for different
values of the chargino mass is shown in Fig.~\ref{chargino_fig3}.\par

The reconstruction efficiency for LLP candidates is a function
of $\beta$, as particles with too low a $\beta$ arrive late 
in the muon detector and cannot be reconstructed. 
No measurement is performed above $|\eta|<2.5$, because of the
coverage of the ATLAS tracking detectors.
For LLP candidates with $p_T>20$~GeV and $|\eta|<2.5$,
the efficiency is zero for $\beta<0.4$, and it reaches a plateau of
0.85 for $\beta\sim 0.65$, determined by the geometrical coverage 
of the muon detectors. These results were obtained in 
\cite{Aad:2011hz,Tarem:2009zz} for GMSB sleptons. 
We use them for charginos although they 
might be somewhat different due to different $p_T$ and $\eta$ 
distribution for our signal.\par
For each candidate in the ATLAS analysis a quality cut
on the $\beta$ measurement is applied. The efficiency of this 
cut can be extracted from Table~3 of \cite{Aad:2011hz} and 
is taken to be 77\% independent of the particle mass and
$\beta$.\par
Following the ATLAS analysis, we require each event to 
contain two candidates with \mbox{$p_T>20$~GeV} within 
\mbox{$|\eta|<2.5$},
reconstructed as a muon. Any candidate which 
combined with a second muon gives a mass
within 10~GeV of the $Z$ mass is rejected.  
We then require that each of the 
events has at least one candidate passing the following 
cuts:
\begin{itemize}
\item
$p_T>40$~GeV and $|\eta|<2.5$;
\item
$\beta$ quality cuts (in our case the efficiency factor defined above);
\item
$\beta>0.95$.
\end{itemize}
The final step is a cut on the mass of the reconstructed 
candidate. We show in Table~\ref{chargino_tab1} for each of the sample 
models considered the cut on reconstructed mass, the number
of expected events after cuts and the efficiency evaluated 
with our approximate simulation.\par
\begin{table}[htb]
\begin{center}
\begin{tabular}{lllllll}
mass (GeV) & $\sigma_{NLO}$(pb) & cut (GeV) &  $\epsilon$ & Expected & 
Limit & Limit \\
&&&& signal & expected & measured \\
\hline
107.4 & 3.29 & 90 & 0.29 & 35.8 & 11.7 & 8.8\\
128.8 & 1.61 & 120 &  0.30 & 17.6 & 7.9 & 5.3\\
150.2 & 0.882 & 130 & 0.35 & 11.3 & 6.2 & 5.0\\
171.4 & 0.518 & 130 &  0.37 & 7.1 & 6.2 & 5.0 \\
192.6 & 0.322 & 130 &  0.38 & 4.5 & 6.2 & 5.0\\
213.7 & 0.209 & 130 &  0.39 & 3.0 & 6.2 & 5.0\\
318.3 & 0.0348 & 130 &  0.41 & 0.52 & 6.2 & 5.0\\
421.8 & 0.0081 & 130 &  0.42 & 0.12 & 6.2 & 5.0\\
524.0 & 0.00226 & 130 &  0.41 & 0.034 & 6.2 & 5.0\\
\hline
\end{tabular}
\caption{As a function of the wino-like chargino mass: NLO cross-section, 
applied cut on the reconstructed candidate mass, analysis efficiency,
number of expected signal events in 37~pb$^{-1}$, 
expected and measured ATLAS 95\% CL upper limit
on the number of signal events.}
\label{chargino_tab1}
\end{center}
\end{table}
The efficiency increases as a function of the
chargino mass, from 30\% for a chargino mass of $\sim$100~GeV
to 40\% for a chargino mass of $\sim$200~GeV. 
These efficiencies in \cite{Aad:2011hz},  vary between 40 and 50\% 
for a comparable range of masses, but for a  model yielding 
different $\eta$ and momentum distribution for the candidate LLPs.\par

\section{Results and outlook}
\begin{figure}[htb]
\begin{center}
\dofig{0.6\textwidth}{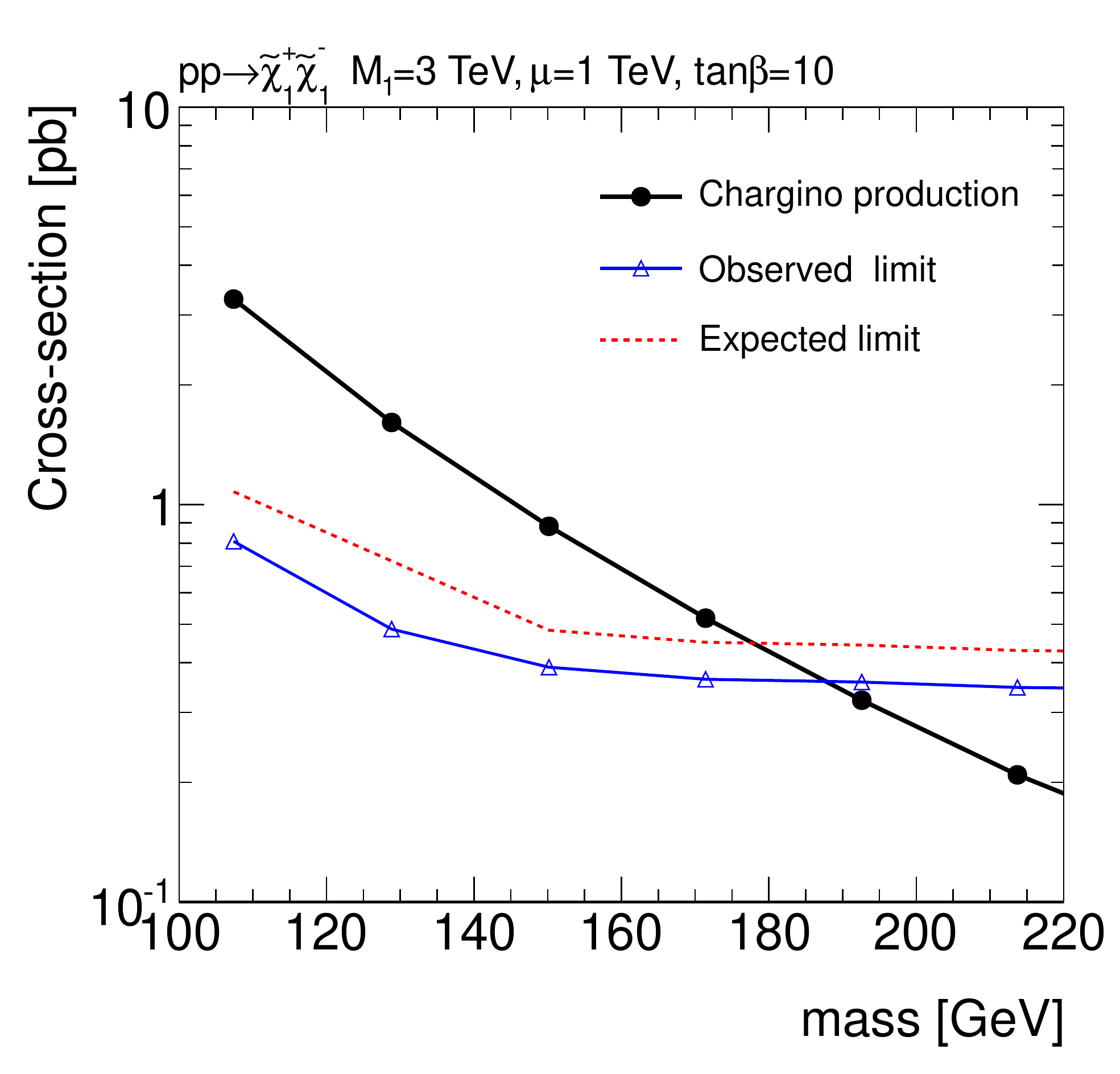}
\caption{The expected production cross-section for wino-like charginos, and the
cross-section upper limit at 95\% CL as a function of the chargino mass.}
\label{chargino_fig4}
\end{center}
\end{figure}

Based on the cross section and efficiency values calculated in the previous
section, and on the number of expected events excluded
by ATLAS at 95\% CL which can be calculated from \cite{Aad:2011hz},
as reported in Table~\ref{chargino_tab1}, an approximate limit 
on the wino-like chargino mass can be evaluated.
The final exclusion plot as a function of the chargino mass 
is given in Fig.~\ref{chargino_fig4}. 
Stable wino-like charginos with masses below approximately 190~GeV
are excluded based on the ATLAS analysis of the 2010 LHC data.  (A  
somewhat weaker result  is expected for Higgsino-like charginos which
have for the same mass a production cross-section which is approximately a 
factor 3.5 lower.) 
From the  consideration of the detailed scan of the parameter space 
discussed in Section~\ref{chargino_secmodel}  this corresponds to excluding
$M_2$ up to $\sim180$~GeV
for  models with the hierarchy $M_2<(M_1,\mu)$ where $M_1$ and $\mu$
are sufficiently heavier than $M_2$.\par
The LHC experiments have collected in 2011 $\sim$5~fb$^{-1}$ of data.
From the numbers of Table~\ref{chargino_tab1} this corresponds to 
$\sim$5 events for a wino-like chargino mass of 520~GeV. Harder cuts 
on $\beta$ and reconstructed mass will be needed for the 2011 analysis
in order to bring down the SM background to few events.
We have checked that for the cuts  $\beta>0.9$ and mass$>400$~GeV,
the efficiency for the chargino is reduced from 0.41 to 0.38.
The 2011 data analysis should therefore be sensitive to 
masses in the proximity of 500(420)~GeV for wino(Higgsino)-like charginos. 
\section{CONCLUSIONS}
The results of an ATLAS search for charged long-lived particles
based on the data collected in 2010 at the LHC
have been interpreted in the framework of a pMSSM model
with degenerate lightest chargino and neutralino.
For such models the ATLAS analysis yields a limit of 
approximately 190 GeV on the wino-like chargino mass.
The more than hundredfold increase in statistics in 2011
should allow ATLAS to probe chargino masses in the 400-500~GeV range.
\section*{ACKNOWLEDGMENTS}
We would like to warmly thank Shikma Bressler who has been very 
helpful in  answering our questions on  the ATLAS analysis.

\AddToContent{G.~Polesello and T.~Rizzo}
\renewcommand{\thesection}{\arabic{section}}


\chapter{Displaced vertex signatures at the LHC from B-L heavy neutrinos and MSSM FIMPs.}

{\it L. Basso, A. Belyaev, E. Dobson, M. C. Thomas, I. Tomalin, S. M. West, A. J. Williams}

\begin{abstract}

In the context of two distinct classes of well-motivated BSM models we
investigate the detection prospects for collider signatures containing
long-lived particles that decay  after travelling up to a meter from
the interaction point of a collision. The first is a $U(1)_{B-L}$
extension of the SM with a dynamical see-saw mechanism, where the RH
neutrino is typically a long-lived particle interacting with the
scalar bosons. 
The second is a version of the MSSM extended by the inclusion of an
extra singlet superfield that is feebly coupled to the MSSM. \\ 
We show that displaced vertices from the models under study provide a
prominent, essentially background-free set of signals, 
which would allow for the identification of these BSM models even if
standard backgrounds, which do not contain displaced vertices, 
are large.
We also comment on the tools used to generate the events including
some useful warnings about the details of handling displaced vertices
in $\tt PYTHIA$.  
\end{abstract}

\section{INTRODUCTION}

Displaced vertices can be common in a number of theories Beyond the
Standard Model (BSM). There are many examples, in the literature, of
models containing long-lived particles that can potentially lead to
displaced vertices at the
LHC~\cite{Langacker:1984dc,Strassler:2006im,Strassler:2006ri,Brooijmans:2010tn,ArkaniHamed:2004fb,Basso:2008iv,Basso:2010yz,Hall:2009bx,Hall:2010jx}.

In this contribution, we study two different classes of models that
contain long-lived states that lead to displaced vertices at the
LHC. We first consider a minimal $U(1)_{B-L}$ extension of the SM at
the TeV scale~\cite{Basso:2008iv}, in which the SM particle content is
extended to include one right-handed (RH) neutrino per generation and
a further singlet scalar, $\chi$, responsible of the spontaneous symmetry
breaking of the extra $U(1)_{B-L}$ group. The interaction between RH
neutrinos and the singlet scalar allows for a dynamical implementation
of the see-saw mechanism as $\chi$ gets a vacuum expectation value
(VEV), allowing at the same time for the Higgs bosons to decay into
pairs of heavy ($\simeq 100$ GeV) Majorana neutrinos (the heavy
partners of the SM-like neutrinos). The mixing of heavy neutrinos with
the SM ones enables the decay of the former into a lepton and a SM
gauge boson. Being suppressed by the see-saw mechanism, the corresponding decay
width is typically very small, thus making the heavy neutrino a
long-lived particle. This will give rise to detectable displaced
vertices which we study in the context of the Higgs boson decays into
multileptons final states  
via these long-lived neutrinos. We study these signatures for two
benchmark scenarios with light Higgs masses relatively small, hence
suitable for the LHC at $\sqrt{s}=7$ TeV, by simulating Monte Carlo
events up to a fast detector simulation employing $\tt
CalcHEP$~\cite{Pukhov:2004ca}, $\tt PYTHIA6$~\cite{Sjostrand:2006za}
and $\tt Delphes$~\cite{Ovyn:2009tx}. 

The second BSM model under study is motivated by dark matter models
that employ the freeze-in mechanism to generate the dark matter relic
abundance \cite{Hall:2009bx, Hall:2010jx}. These models arise by
extending the MSSM with the addition of a singlet superfield, which is
feebly coupled to the MSSM states. This extra superfield is
assumed to contain the Lightest Supersymmetric Particle (LSP), and
consequently the Next to Lightest Supersymmetric Particle (NLSP) will
only be able to decay to the LSP via the feeble coupling leading to
long-lived NLSPs. We investigate one example of this type of model
with a number of benchmark parameter choices. Monte Carlo event
simulations for these benchmarks are generated using a combination of
$\tt MadGraph5$ \cite{Alwall:2011uj}, $\tt BRIDGE$ \cite{Meade:2007js}
and $\tt PYTHIA8$ \cite{Sjostrand:2006za,Sjostrand:2007gs}.

\section{Benchmark models}
\subsection{Long-lived neutrinos in the B-L model}
\label{sec_BLmodel}

The first model under study is the so-called ``pure'' or ``minimal''
$B-L$ model (see ref.~\cite{Basso:2008iv,Basso:2010yz} for conventions
and references) since it has vanishing mixing between the $U(1)_{Y}$
and the new gauged $U(1)_{B-L}$. 
This model extends the SM with three
right-handed neutrinos (needed for the cancellation of the anomalies
related to the new $U(1)$ group) and an additional complex Higgs
singlet, responsible for giving mass to the additional $Z'$ gauge boson
(see \cite{Basso:2008iv,Basso:2010jm} for details). By extending the
fermion and the scalar sectors, this model allows for a
dynamical generation of the neutrino masses. The scalar sector is now
made of two real {\it CP}-even scalars, $h_1$ and $h_2$, with
$m_{h_1}\leq m_{h_2}$, remnants of the Higgs doublet
and singlet fields after electroweak symmetry breaking (EWSB) 
that mix together: 
\begin{equation}\label{displ_scalari_autostati_massa}
\left( \begin{array}{c} h_1\\h_2\end{array}\right) =
  \left( \begin{array}{cc}
    \cos{\alpha}&-\sin{\alpha}\\ \sin{\alpha}&\cos{\alpha} 
	\end{array}\right) \left( \begin{array}{c} h\\h'\end{array}\right) \, .
\end{equation}
The mixing angle 
$-\frac{\pi}{2}\leq \alpha \leq \frac{\pi}{2}$ can be expressed as
$\displaystyle \tan{2\alpha} = \frac{\lambda _3 xv}{\lambda _1 v^2 -
  \lambda _2 x^2} \, ,$ 
where $v$ and $x$ denote the VEV of the doublet and singlet fields,
respectively, and 
$\lambda_{1,2,3}$ are the coefficients of the quartic
terms in the scalar potential. 
Notice that the presence of the new particles (neutrinos and a $Z'$
boson) also alter the properties of the Higgs bosons. Moreover, the
scalar mixing angle $\alpha$ is a free parameter of the model, and the
light (heavy) Higgs boson couples to the new matter content
proportionally to $\sin{\alpha}$ ($\cos{\alpha}$), i.e., with the
complementary angle with respect to the interactions with the SM
content, as in the traditional literature of singlet scalar extended
SM. 

Neutrino mass eigenstates, obtained after applying the see-saw
mechanism, will be called $\nu_l$ and $\nu_h$, where the former are the
SM-like ones. As the Higgs fields develop a VEV, the relevant Yukawa
interactions $\mathcal{L}^\nu_Y= -y^{\nu}_{jk}\overline {l_{jL}} \nu
_{kR}\widetilde H  
	         -y^M_{jk}\overline {(\nu _R)^c_j} \nu _{kR}\chi +  {\rm 
h.c.}$ include
the traditional type-I see-saw mass matrix, that
can be  diagonalised
by a rotation about an angle $\alpha _\nu$, such that $\displaystyle
\tan{2 \alpha_\nu} = -\frac{2m_D}{M}$, where $m_D =
\frac{y^{\nu}}{\sqrt{2}} \, v$ and $M = \sqrt{2} \, y^{M} \, x$. 
We also require the neutrinos to be degenerate in mass.
Thus, $\nu_{L,R}$ can be written as the following linear combination
 of Majorana mass eigenstates $\nu_{l,h}$ :
\begin{equation}\label{nu_mixing} 
\left( \begin{array}{c} \nu_L\\ \nu_R \end{array} \right) = 
\left( \begin{array}{cc} 
\cos{\alpha _\nu} & -\sin{\alpha_\nu} \\ 
\sin{\alpha _\nu} &\cos{\alpha _\nu} 
\end{array} \right) \times \left( \begin{array}{c}
  \nu_l\\ \nu_h \end{array} \right)\, .  
\end{equation} 

 With a reasonable choice of Yukawa couplings, the heavy neutrinos can
 have masses $m_{\nu_h} \sim \mathcal{O}(100)$ GeV $\ll M_{Z'_{B-L}}$.  
In the following, we describe the features relevant to the present
analysis: the pair production of heavy neutrinos via a light Higgs
boson, at the LHC for $\sqrt{s}=7$ TeV.  

\begin{figure}[htb]
\centering
\includegraphics[scale=0.3]{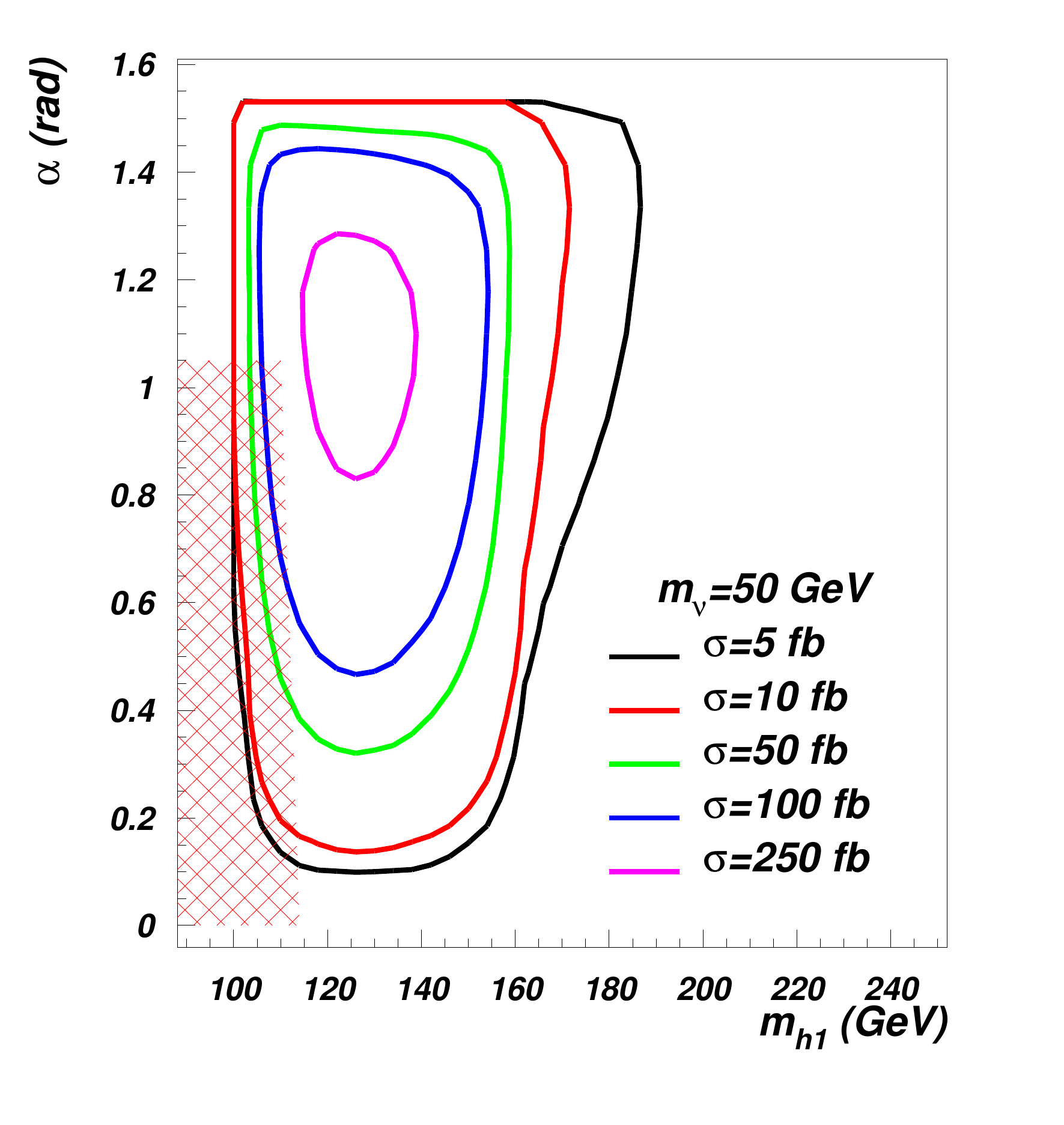}%
\;\;\;
\includegraphics[scale=0.3]{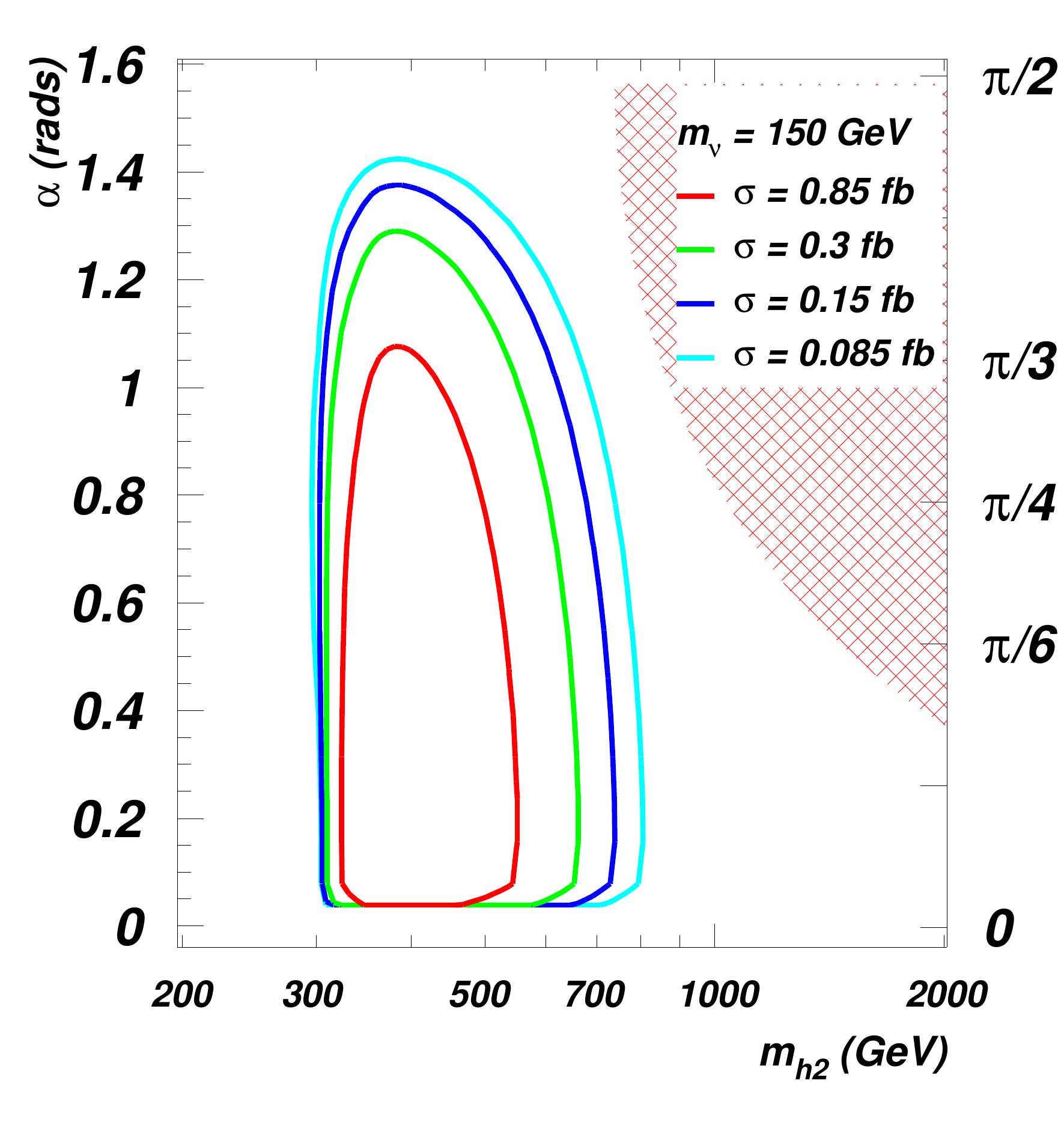}
\caption{Cross sections for (left) $pp\to h_1\to \nu_h \nu_h$ for
  $\sqrt{s}=7$ TeV and for (right) $pp\to h_2\to \nu_h \nu_h$ for
  $\sqrt{s}=14$ TeV, as a function of the intermediate Higgs mass and
  scalar mixing angle~\cite{Basso:2010yz}. (Cross sections summed over
  all the neutrino generations.) The areas with red shade are the regions
  excluded experimentally by LEP (left) and theoretically by Unitarity
  (right)~\cite{Basso:2010jt}.} 
\label{figdispl_BLcross_section}
\end{figure}

Figure~\ref{figdispl_BLcross_section} shows the production cross
sections for the process $pp \to h_{1(2)} \to \nu_h \nu_h$, for the
case of a light and a heavy Higgs boson, $h_1$ and $h_2$, respectively,
as a function of the scalar mixing angle and with suitable
heavy neutrino masses. It is clear that large enough cross sections to
be relevant at
the LHC with $\sqrt{s}=7$ TeV are possible only when the two Higgs
bosons mix strongly, peaking for a value of the mixing angle
around $\pi/3$, and for heavy neutrino masses below $M_W$. The former,
in turn, means that all the SM-like processes (production and decay)
are suppressed by a factor about  $(\cos{\pi/3})^2= 1/4$ 
as compared to the SM Higgs boson. On the other hand, when decaying
into heavy neutrinos, just the production is suppressed (by
$(\cos{\pi/3})= 1/2$), while 
 $BR(h_1\to \nu_h\nu_h)$ can be up to $30\%$, which is
quite a substantial value. 

\begin{figure}[h]
\begin{center}
\includegraphics[scale=0.55]{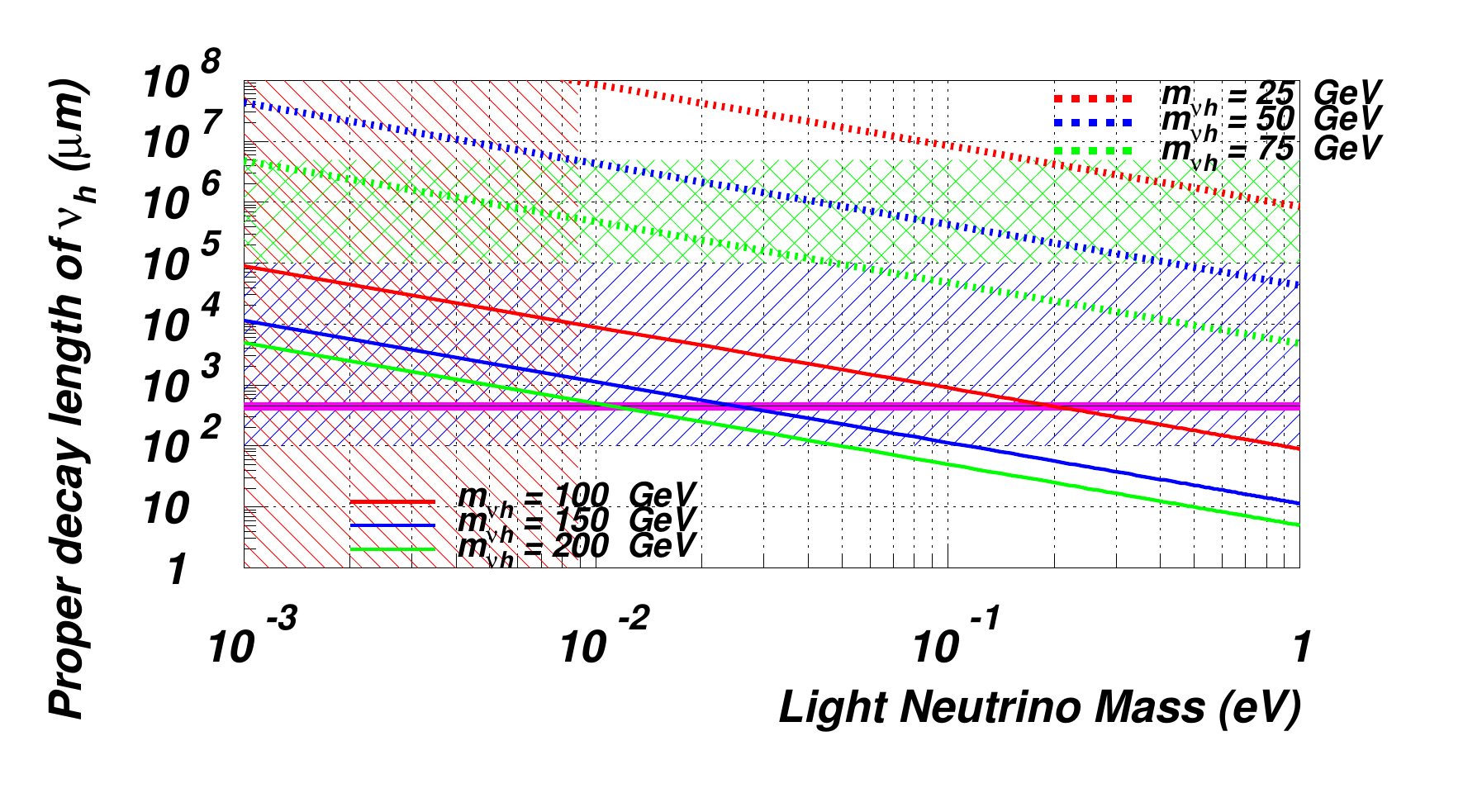}
\caption{Proper decay length of the heavy neutrinos (in their rest
  frame) for a choice of their mass and as a function of the light
  neutrino mass. The purple band presents the proper decay length of
  the $b$-quark. The blue (right) band indicates the range of a
  typical micro-vertex detector, while the green (crossed) band
  indicates the typical detector dimension. The red shaded are on the left
  shows the region excluded by neutrino oscillation direct 
measurements.}
\label{figdispl_BLlifetime}
\end{center}
\end{figure}

The process is interesting because of some peculiar features: first of
all, this decay mode is maximised for a range of masses of the Higgs
boson that has the lowest overall sensitivity at the LHC for SM
processes. It is therefore complementary to stantard light Higgs
searches at the LHC. Second, each heavy
neutrino would decay into a lepton and a SM gauge boson with standard
branching ratios~\cite{Basso:2008iv}: $BR(\nu _h \to \ell ^\pm W^\mp)
\sim \frac{2}{3}$, $BR(\nu _h \to \nu_l Z) \sim \frac{1}{3}$. 
Including the $W$ and $Z$ boson decays, spectacular multi-lepton
decays of the Higgs boson arise. We will focus here on a light Higgs
boson. The relevant cross sections are displayed on the left-hand plot in
figure~\ref{figdispl_BLcross_section}. In this case, the total cross
section is at maximum of $250$~fb  when all neutrino
generations are added up. 
Naively, it will be totally covered by the background. However,
a very efficient way to uncover the signal is to look for displaced
vertices, that are a prominent feature of the heavy
neutrinos. Figure~\ref{figdispl_BLlifetime} shows the lifetime of the
heavy neutrinos: we clearly see that, over a large portion of the
parameter space and depending on the light neutrino mass, 
the heavy neutrinos are long-lived particles.
Hence, the measurement of the
mean lifetime of the heavy Majorana neutrinos and of its mass is of
crucial importance, allowing for a determination of the absolute mass
of the SM-like neutrinos~\cite{Basso:2008iv}. 
We selected $2$ benchmark points to be covered in the present
analysis as summarised in table~\ref{displaced_MC_BL}.  

\begin{table}[h]
\begin{center}
\begin{tabular}{l|l|l|l|l|l|l}\hline
Sample            & $M_{h_1}$ & $M_{\nu_h}$ & $\alpha$ &Final state &
Proper              & $\sigma$ \\ 
                  &           &            &         &            &
decay length        &      \\ \hline 
$B-L$ Benchmark 1 & $120$ GeV  & $50$ GeV & $1.1$ rads & all decay
modes  &  43 cm& $137.9$ fb \\ \hline 
$B-L$ Benchmark 2 & $140$ GeV  & $50$ GeV & $1.1$ rads  & all decay
modes  & 43 cm & $168.5$ fb \\ \hline 
\end{tabular}
\caption{\label{displaced_MC_BL}
Description of Monte Carlo samples used in the analysis. 
The cross sections refer to the sum of the first $2$ generations of
heavy neutrinos only. 
The proper decay length is equal to $ c\cdot \tau$
where is $\tau$ is the heavy neutrino life-time.}
\end{center}
\end{table}

\begin{figure}[h]
\centering
\includegraphics[width=0.85\textwidth]{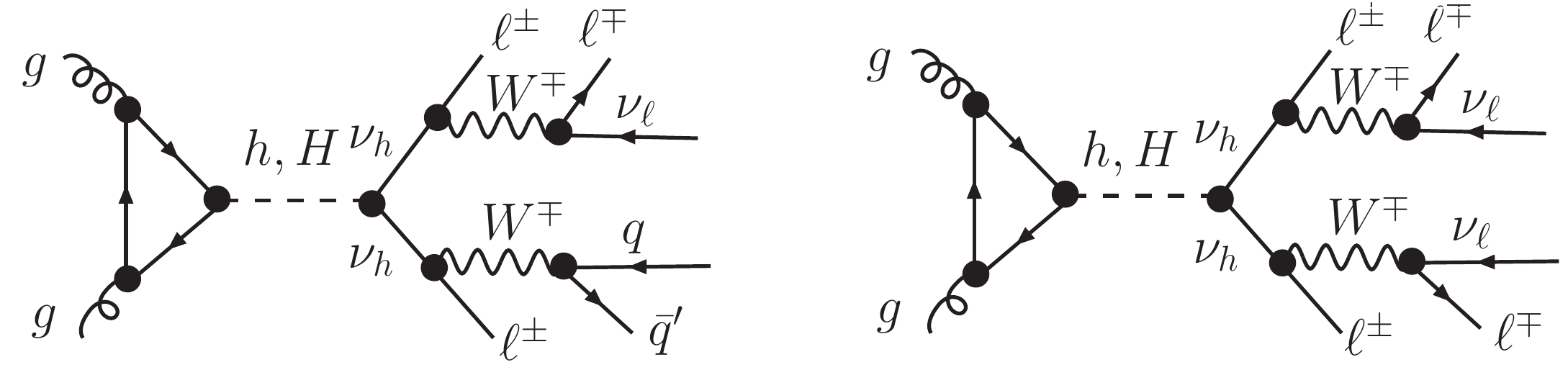}
\caption{Production and cascade decay for (left-hand)
  eq.~(\ref{eq_hnu-semilep}) and for (right-hand)
  eq.~(\ref{eq_hnu-lept}).} 
\label{displaced_fig_BLgraphs}
\end{figure}

Among the possible decay patterns of the heavy neutrinos
(see~\cite{Basso:2011hn}), those that maximise the observation chances
involve leptonic decays of the heavy neutrinos: 
\begin{eqnarray}\label{eq_hnu-semilep}
pp \to h_1 \to \nu_h \nu_h &\to& 3\ell\; 2j\; \mbox{MET}\, ,
\\ \label{eq_hnu-lept} 
pp \to h_1 \to \nu_h \nu_h &\to& 4\ell\; \mbox{MET}\, ,
\end{eqnarray}
the so-called ``tri-lepton'' and ``fully leptonic'' decay modes, whose
pictorial representations are given in
figure~\ref{displaced_fig_BLgraphs}. A channel without missing energy
would also be present (namely, $pp \to h_1 \to \nu_h \nu_h \to 2\ell\,
4j$), even though this is the channel with the largest cross section, 
the reconstruction of $4$ very soft jets dramatically
reduces its efficiency.
In the tri-lepton case one heavy neutrino decays into $2\ell + $ MET
and the second one decays into $\ell + 2j$. The fully leptonic decay
(when both heavy neutrinos decay into $2\ell + $ MET) is preferred for
displaced vertices searches, while the semileptonic decay allows for
higher total cross sections and for an efficient kinematic analysis,
especially when $m_{\nu_h}>M_W$~\cite{Basso:2008iv}. Notice that,
without using displaced vertices techniques, the neutrino mass
reconstruction would be possible only in the tri-lepton case.

It is worth commenting that other scenarios, not included in the present analysis, might be of interest.
First, a very light neutrino, with a mass of around 20~GeV: it would have escaped detection at LEP (because its coupling to the Z boson are heavily see-saw suppressed\footnote{For this mass, $BR(Z\to \nu_l \nu_h)\sim 10^{-11}$ and $BR(Z\to \nu_h \nu_h)\sim 10^{-20}$.}) and it has a very long lifetime, hence resulting at most in a single displaced vertex accompanied by missing momentum from the non-observed heavy neutrino. The efficiency to observe this signal would decrease in inverse proportion to the lifetime.
Second, the decay into pairs of heavy neutrinos could be the only possibility to find the heavy Higgs boson, if weakly mixed with the light partner (therefore feebly coupled also to SM particles, evading the current LHC searches).
It is important to stress that the scalar mixing angle $\alpha$ cannot
vanish, since the most efficient production mechanisms of the heavy
Higgs boson are the typical SM channels at the
LHC~\cite{Basso:2010yz}. Also notice that the couplings to heavy
neutrinos are maximised in the weak mixing limit. 
Given that the cross sections are $\mathcal{O}(1)$ fb, this scenario
can be studied at the LHC with full design luminosity even if a
SM-like light Higgs boson is found, albeit weakly mixed with the heavy partner.

\subsection{Long Lived SUSY states in MSSM models with FIMPs}
\label{section_fimpmodel}

The second class of models we consider is inspired by dark matter
theories that employ the freeze-in mechanism to generate the dark
matter relic abundance \cite{Hall:2009bx, Hall:2010jx}. This mechanism
relies on the dark matter particle never reaching thermal equilibrium
with the visible (e.g the SM or MSSM) sector states.  In order for
this to occur, only very small (we call them feeble) couplings between the hidden sector dark matter states and the visible sector states are allowed. These hidden sector states are referred to as Feebly Interacting Massive Particles (FIMPs).

%

Let us consider a connection between hidden sector FIMP states and the MSSM in the form of a SUSY particle, $Y$, that we will take an MSSM neutralino or chargino, decaying to the hidden sector state, $X$ via $Y \rightarrow X + \ldots$, where the ellipses represent some SM sector states. Here we assume that both $Y$ and $X$ are odd under R-parity. The relic abundance of $X$ states generated by the freeze-in mechanism reads $\Omega h^2 \sim 10^{24}\Gamma_Y m_X/m_Y^2$  \cite{Hall:2009bx}, where $\Gamma_Y$ is the decay rate of $Y$ with $X$ final state. For all particle masses of interest, in order to get $\Omega h^2 \sim 0.1109 \pm 0.0056$~\cite{Komatsu:2010fb}, $\Gamma_Y$ is required to be very small. A consequence of which is that if we produce the particle $Y$ in a detector it will travel a significant distance (greater than a few mm) before decaying. Moreover, the decay width of this state, which could be reconstructed from measurements of a number of such decays, could have a direct link to the relic abundance of dark matter as frozen-in via the decays of $Y$ in the early Universe. Finally, it may also be possible to determine the masses of $Y$ and $X$ using kinematical techniques and therefore test the freeze-in expression for the relic abundance.

Beyond this simple picture, more complicated scenarios are possible and the exact predictions for the decay length and the mass of the $X$ dark matter state can vary~\cite{Cheung:2010gk}. Therefore, 
we consider a range of parameters and masses to remain as general as possible so that this study can be easily applied to a specific picture of a visible and hidden sector feebly coupled with the the dark matter living in the hidden sector. Going beyond the freeze-in mechanism such a scenario can occur in a number of different theories, for example in the context of asymmetric dark matter (see e.g. \cite{Hooper:2004dc, Kaplan:2009ag}). 


Given this motivation, we consider a supersymmetric model with a lightest ordinary supersymmetric particle (LOSP) that has a small decay width to some hidden sector state. This means that in every event in which a pair of supersymmetric particles are produced, there will be two long-lived LOSPs that will decay into hidden sector states. In this analysis, we take a phenomenological approach and optimise the signal by choosing by hand the decay length such that the majority of the decays occur in the tracker thereby allowing for a better reconstruction and a maximisation of the sensitivity of the search. In this way we are looking for the best possible scenario from a detector point of view, which will allow for a better idea of what could be achieved in less ideal scenarios. 

We follow the basic set up outlined in both \cite{Chang:2009sv} and \cite{Cheung:2010gk} where a singlet superfield is added to the MSSM and a survey of the potential operators that can be written down is made. We start these analyses and look in more detail at one of these operators. 
We assume that this extra singlet superfield is even under R-parity and consider only renormalisable operators. Under these assumptions we have chosen to analyse the superpotential Higgs portal operator $\Delta \mathcal{W}= \lambda XH_uH_d$. This choice has been made in order to initiate the study rather than fully cover all possibilities. An immediate extension of this work is to consider all other operators, including the non-renormalisable ones and the case of an R-odd singlet. The consequences of this operator has also been previously studied in \cite{Martin:2000eq}, where an additional singlet superfield was added to the MSSM in order to solve the ``$\mu$-problem". The low energy theory studied in \cite{Martin:2000eq} has many features in common with the set-up we consider here.

In order for this operator to give rise to a long lived LOSP, the coupling $\lambda$ has to be small. 
At order $\lambda$ this term induces the interactions $\widetilde{\chi}^-_i\; X_f \;W^+$ and $\widetilde{\chi}^0_i\; X_f \;Z$ (and the Hermitian conjugates), where $\widetilde{\chi}^-_i$ and $\widetilde{\chi}^0_i$ are the MSSM charginos and neutralinos, respectively, and $X_f$ is the fermionic component of the $X$ superfield.
We look for final-state muons coming from a displaced vertex as this is a particularly powerful way to search for displaced vertices due to low backgrounds. In each event, we require only one cascade, resulting from the production and decay of a SUSY particle, to yield a displaced muon(s). 
\begin{table}[t]
\begin{center}
\begin{tabular}{l|l|l|l|l|l}\hline
Sample & Hard process & Prompt  & Proper decay       & $M_{\mathrm{FIMP}}$ & $\sigma$/pb\\
       &              & objects               &  length /cm &      /GeV      &\\  \hline
FIMP BM1 &  & 2 hard Ws & 25 & 1  & 0.103\\
Neutralino LOSP  & $M_{\tilde{\chi}^+_1}=235$ GeV & & 50 &1 &\\
& $M_{\tilde{\chi}^0_1}=123$ GeV  & & 100 & 1 &\\ 
 &  &  & 50 & 0.1  &  \\ 
 &  & & 50 & 25  &  \\
 \hline
FIMP BM2 & Chargino pair prod. & 2 soft Ws & 25 & 1 & 1.21\\
Neutralino LOSP & $M_{\tilde{\chi}^+_1}=133$ GeV & & 50 &1  &\\
 & $M_{\tilde{\chi}^0_1}=123$ GeV & & 100 & 1 &\\ \hline
FIMP BM3 & Chargino pair prod. & None & 25 & 1 & 0.677\\
Chargino LOSP & $M_{\tilde{\chi}^+_1}=193$ GeV  & & 50 &1 & \\
 &  & & 100 & 1 &\\ \hline
SM sample 1 & $W\mu\nu$ & $W\mu\nu$ & 0 & & 10.46\\ 
SM sample 1 & $Z\mu\mu$ & $Z\mu\mu$ & 0 & & 1.07 \\ 
SM sample 2 & QCD multijet & QCD multijet & 0 & & 1$\times 10^7$\\ \hline
\end{tabular}
\caption{
Description of Monte Carlo samples used in the analysis. The cross sections quoted for the FIMP benchmarks are for the chargino pair production only, see text for more details. \label{displaced_table1_12}}
\end{center}
\vspace{-5mm}
\end{table}
The SUSY production channel considered is through of chargino pairs. In order for this to be the dominant production channel, we consider a region of parameter space with heavy squarks and gluinos. This choice has been made in order to remove the complication of applying existing limits on squarks and gluinos and also to leave the events as clean as possible. 

\begin{figure}[h] 
\begin{center}
\vspace{-5mm}
\includegraphics[scale=0.12]{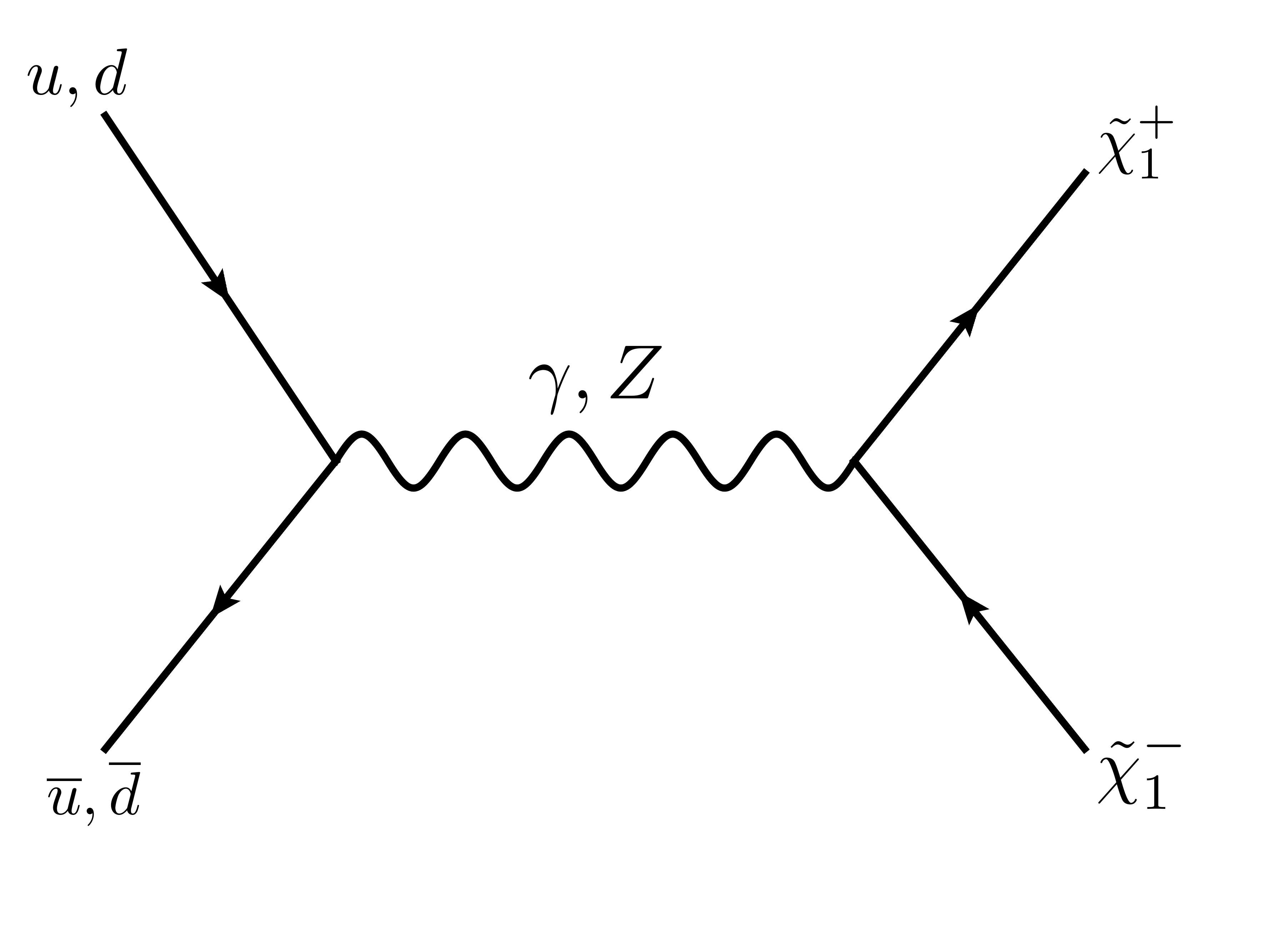} 
\includegraphics[scale=0.12]{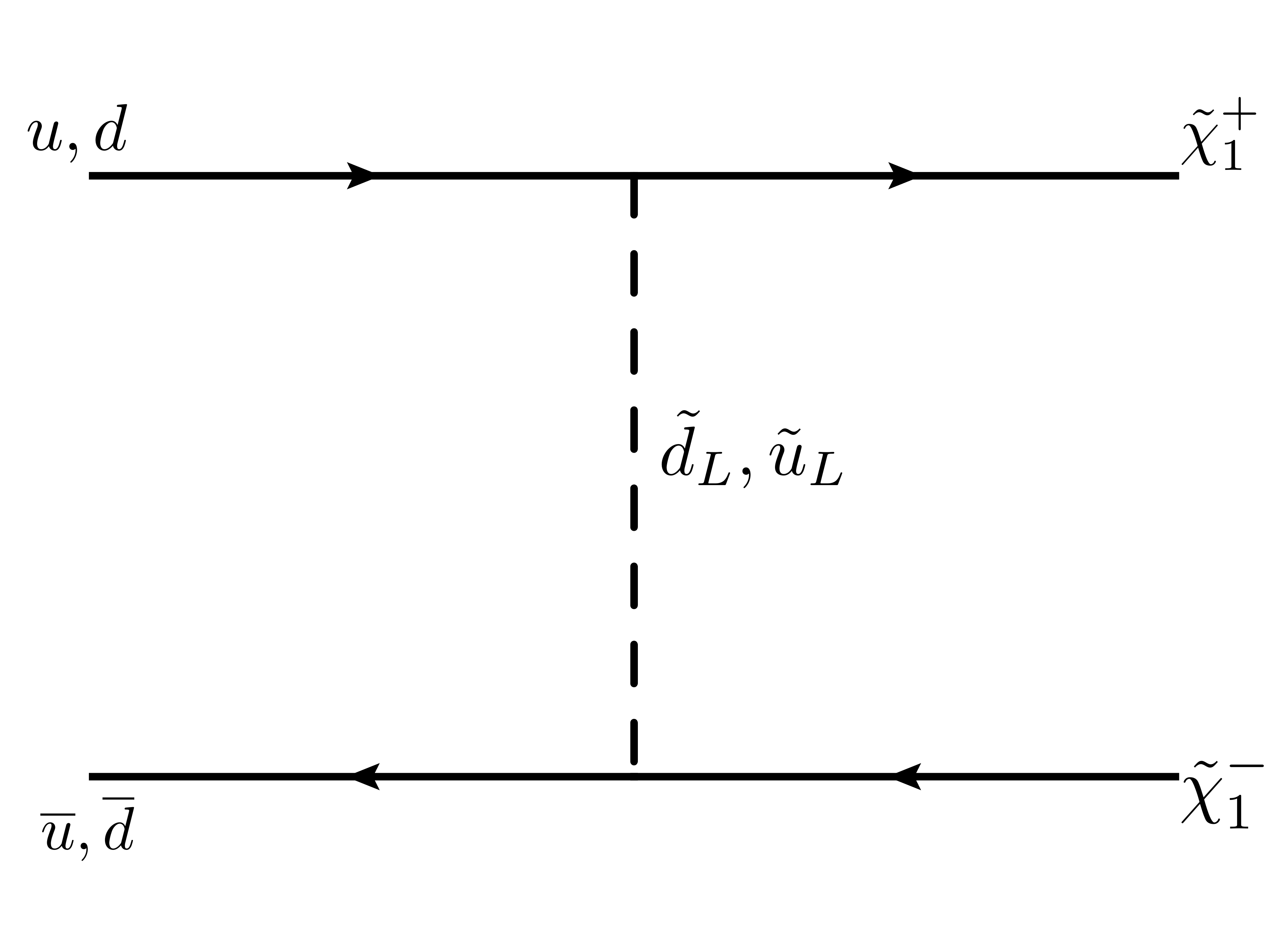} \hspace{3mm}
\includegraphics[scale=0.13]{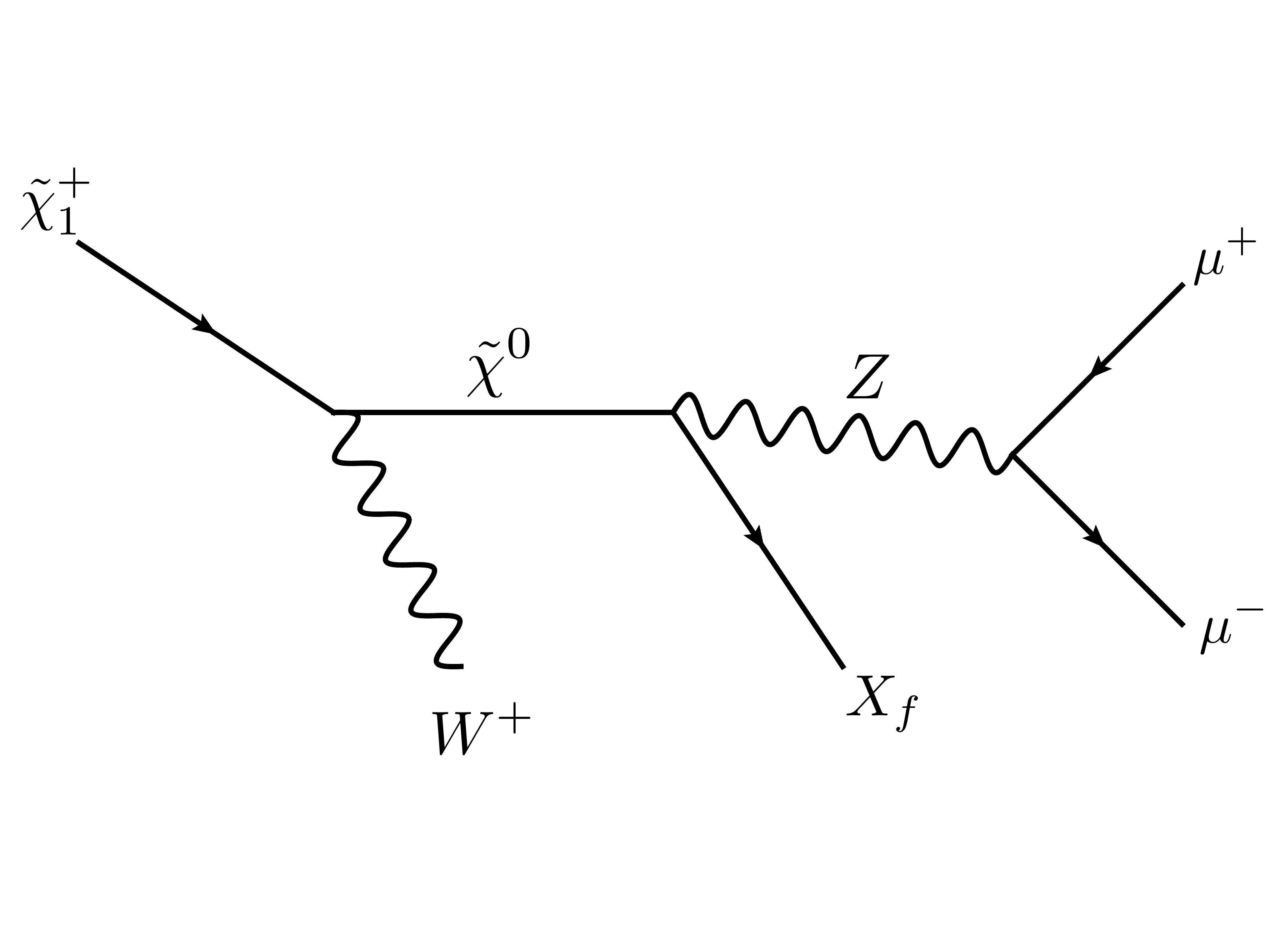} 
\caption{Production for all benchmarks with the cascade decays for BM1 and BM2.}
\label{displaced_fig2}
\end{center}
\vspace{-5mm}
\end{figure}

Some details of the benchmark models we have studied are listed in Table~\ref{displaced_table1_12}. For each benchmark model we generate 100k events using the tools described in section \ref{sec_FIMP_tools}. Benchmark 1 (BM1) looks at a possible point in the MSSM parameter space with a neutralino LOSP. The chargino pair production cross section for this point is calculated using both $\tt MadGraph5$ \cite{Alwall:2011uj} and $\tt CalcHep$ \cite{Pukhov:2004ca} to be $\sim 0.1$ pb.  The diagrams for the production of a pair of charginos are depicted in Fig. \ref{displaced_fig2}. Due to the large masses of the squarks, the t-channel process is sub-dominant compared with the s-channel diagrams. The chargino will decay to a neutralino via a $W$, and with the large mass difference between the chargino and neutralino in BM1, this $W$ can be produced on shell leading to hard jets or a hard single charged lepton. In our sample we allow the $W$ to decay to all possible final states. The neutralino, as the LOSP, has two possible decay modes, both yielding an $X_f$ in the final state. In our sample we only include events in which the neutralino decays to an $X_f$ and an on-shell $Z$, with the $Z$ decaying to muons as shown in Fig. \ref{displaced_fig2}. The other decay mode is to the light Higgs plus $X_f$ and has a small branching ratio for the chosen benchmark points.

The mass for the $X_f$ state has been chosen to be $1$ GeV for the majority of the benchmark scenarios. The reason for this is twofold but ultimately arbitrary. One motivation comes from the fact that we want to produce an on-shell $Z$ in the decay of the neutralino so that we can reconstruct its mass. The other is that in the majority of asymmetric dark matter models the mass needed for the dark matter state is around a 1 GeV. In BM1 we also vary the mass of the $X_f$ state to cover more scenarios and to investigate what effect this has on the kinematics and whether we can use this variation to determine the mass of the hidden sector state and of the LOSP. 

As stated above, we only generate Monte Carlo events in which \textit{both} SUSY cascades end with a pair of oppositely charged muons produced in a displaced vertex. We determine what fraction of these will pass cuts imposed on the displaced vertex (and other aspects of the event) in section \ref{displaced_experimentfimp_43}. Analyses of real data might require only one rather than two displaced pair of oppositely charged muons. A further analysis would be required in that case to accurately estimate the efficiency. In this analysis, we have simply counted the total number of displaced muon pairs that pass the cuts. As there are two of these displaced muon pairs per event there will be a correlation between the displaced muons produced in the same event but as a first pass this estimate gives us a good ball park figure with which to work.  
Given $BR(Z\rightarrow \mu\mu)\sim 0.034$, an estimate, with the chargino pair production cross section of $0.1$ pb, of the number of events with at least one displaced pair of oppositely charged muons is $N \sim 8 \left(\sigma/0.1{\rm pb}\right)\left(L/1 {\rm fb}^{-1}\right)$,
where $L$ is the integrated Luminosity, which we have set at a value of $1 ~{\rm fb}^{-1}$. 

Benchmark 2 (BM2) contains a lighter chargino and the resulting production cross section is larger, $1.21$~pb. Due to the lower chargino mass, the $W$ produced in the decay from the chargino to the neutralino will be soft. Otherwise, BM2 is the same as BM1 and with the increased cross section we expect over 80 events with at least one pair of oppositely charged muons from a displaced vertex. 

The lightest chargino can be lighter than the neutralinos if for low $\tan\beta$; sign$(M_1)\neq$ sign$(M_2) = $sign$(\mu)$, see e.g. \cite{Kribs:2008hq}. In Benchmark 3 (BM3) the chargino still delivers the dominant SUSY production cross section (as our squarks are still heavy), but it is now the LOSP, long-lived and will decay directly to $X_f$ plus a $W$. We then consider the decay of the $W$ into a muon plus muon neutrino thereby producing a displaced muon. The obvious difference in this case is that the LOSP has charged track, which produces a kink when it decays to the muon. The number of events expected in this case is much higher due to the larger branching ratio, $BR(W\rightarrow \mu\nu_{\mu})\sim 0.11$, as compared to the $Z\rightarrow \mu^+\mu^-$ decay.  An estimate for the number of events with at least one displaced muon is close to 150 for $1$ fb$^{-1}$.


\section{Tools}
\label{section_tools}
{\bf B-L model}\label{sec_BM_BLmodel}
Monte Carlo event files have been produced for the present analysis, according to the $2$ benchmark points described in Table~\ref{displaced_MC_BL}. For each benchmark point, $20$k parton level events have been generated in $\tt CalcHEP$~\cite{Pukhov:2004ca}, with full $3$-body decays of the heavy neutrinos, hence properly taking into account $W$ and $Z$ boson off-shellness. Subsequently, the full samples have been hadronized/parton showered with $\tt Pythia$ v.~$6.4.20$ and, finally, fast detector simulation has been added by running $\tt Delphes$ v.~$1.9$ on the samples. Anti-$k_T$  with ``cone size'' $0.5$ has been chosen as jet algorithm to match CMS definitions. We have also modified the standard CMS card to match its isolation criteria.

\label{sec_FIMP_tools}
For the {\bf FIMP model} the Monte Carlo events generated are outlined in Table~\ref{displaced_table1_12}. A $\tt MadGraph5$ \cite{Alwall:2011uj} model file was produced using $\tt FeynRules$ \cite{Christensen:2008py} based on the existing MSSM model file \cite{Duhr:2011se}. The widths and branching ratios of the SUSY particles were calculated using $\tt BRIDGE$ \cite{Meade:2007js} and passed to $\tt Pythia8$ \cite{Sjostrand:2006za,Sjostrand:2007gs} via the SLHA \cite{Skands:2003cj} decay tables. Although $\tt BRIDGE$ is capable of decaying the events in full, it forces all particles in a decay chain to be exactly on-shell (i.e. $\tt BRIDGE$ does not sample widths from the full Breit-Wigner distribution) so we decayed the events using $\tt Pythia8$ instead. For benchmark $3$ the initial scattering is performed in $\tt Pythia8$ using the $\tt MadGraph5$ matrix element as a semi-internal process. This was done to counteract a problem with passing events between $\tt MadGraph5$ and $\tt Pythia8$ where long-lived states are produced in the initial scattering, see section~\ref{sec_pythia_problem} for more details. The subsequent decay chain and showering was performed in $\tt Pythia8$ using the decay tables obtained from $\tt BRIDGE$ for the chargino and neutralino decays. For comparison, Standard Model samples were also produced and these were generated using only $\tt Pythia8$.
\subsection{Displaced vertices in $\tt Pythia$} \label{sec_pythia_problem}
A number of problems had to be overcome to generate displaced vertices correctly using $\tt Pythia$. The calculation of the decay lifetime $\tau$ from the particle width specified in the SLHA header of the event file does not seem to be performed automatically. As a solution, the lifetime of the long-lived particles can be set by hand during initialisation. 

After setting the particle lifetime and decaying the event in $\tt Pythia$ we found that all particles have their production coordinates set at the interaction point. Outputting events in the Les Houches Event (LHE) format~\cite{Alwall:2006yp} is not affected, as only the particles lifetime and momentum is recorded, but outputting events to $\tt HepMC$~\cite{Dobbs:2001ck} format or as root trees will generate events with incorrect production vertices. 
This problem can be overcome by producing the hard process in the LHE format and running the event file through $\tt Pythia$ a second time, since, with this method, the production vertices are calculated correctly from the information in the LHE file. However, producing fully showered events including final state radiation (FSR) seems to be problematic with this procedure as displaced particles in the final state that emit a photon are moved back to the interaction point. A solution is to calculate the production vertices for each displaced particle and set them after the full event has been generated by $\tt Pythia$. This was done recursively for all daughter particles from each displaced vertex before writing the event to disk. 

A further complication arises in handling event files where the long-lived particle is produced in the initial scattering process. For example, in benchmark 3 for the FIMP model, the charginos are long lived and are produced directly via $p p \rightarrow \widetilde{\chi}_1^+ \widetilde{\chi}_1^-$. If this event is generated by $\tt MadGraph5$ and then passed to $\tt Pythia$ to decay, it always decays the charginos at the interaction point even if the lifetime has been correctly set. To avoid this for benchmark 3 the scattering process was calculated entirely in $\tt Pythia$.

Finally, in $\tt Pythia6$, the ``colour string reconnection'' option must be disabled by setting the $\tt Pythia6$ option MSTP$(95)=0$. This is because when deciding which partons to reconnect colour strings to, $\tt Pythia6$ does not explicitly check if the partons are at the same spatial coordinate. With colour reconnection enabled, it therefore sometimes reconnects colour strings between prompt partons and partons from the long-lived particle decay.

\section{Review of existing measurements of displaced vertex signatures}
\label{section_public}
This section presents the status of existing searches for displaced vertex signatures at hadron colliders.

{\bf ATLAS displaced vertex study.}
\label{sec_ATLASpublic}
In Ref.~\cite{Aad:2011zb}, a search in $33$~pb$^{-1}$ of integrated luminosity collected in 2010 with the ATLAS detector for displaced vertices is presented. They study the decay of charged hadrons in association with a high-momentum muon. No signal is observed, and limits are placed on the product of di-squark production cross section and decay-chain branching fraction in a SUGRA scenario in which the lightest neutralino produced in the primary-squark decay undergoes R-parity violating decay into a muon and two quarks. The upper limits at $95\%$~C.L. on the production cross section times branching fraction are at the level 0.5-100~pb for high-mass particles with lifetimes and velocities such that they decay at radial distances between $4$~mm and $180$~mm from the $pp$ interaction point. 

{\bf CMS displaced vertex study.}
\label{sec_CMSpublic}
In Ref.~\cite{CMSNote}, a search in $1.1$~fb$^{-1}$ of integrated luminosity collected in 2011 with the CMS detector for displaced vertices is presented. The search is for Higgs bosons decaying to two long-lived, massive, neutral particles. The analysis searches for these long-lived particles via their decay to dileptons (both electron and muon) within the volume of the CMS tracker. No significant excess was observed above the Standard Model background and upper limits were placed typically in the range $0.003-0.03$ pb, for $X$ bosons whose lifetime is such that their mean transverse decay length is less than $1$~m. 

{\bf CDF displaced vertex study.}
\label{sec_CDFpublic}
In Ref.~\cite{Scott:2004wz}, the results of a search for new particles with long lifetime that decay to a $Z$ boson were presented in $163$~pb$^{-1}$ of data recorded with the CDF detector. Dimuons with invariant mass near the $Z$ peak are vertexed and the decay length distribution studied. No evidence of a long-lived component was found, and cross section limits were presented on a fourth generation quark model. For a long-lived particle of mass 150 GeV, limits were set at $95\%$~C.L. in the range $3-50$ pb for a decay length between 1~mm and 1~m. 

{\bf D0 displaced vertex study.}
\label{sec_D0public}
In Ref.~\cite{Abazov:2006as}, a search for a neutral particle, pair produced in $p\overline{p}$ collisions at a centre-of-mass energy $\sqrt{s}=1.96$~TeV, which
decays into two muons and lives long enough to travel at least $5$~cm before decaying, was presented. The analysis used 380 pb$^{-1}$ of data recorded with the D0 detector. No candidates were observed, and limits were set on the pair-production cross section times branching fraction into dimuons+X for such particles. For a mass of 10 GeV and lifetime of 4$\times 10^{11}$s , values greater than 0.14 pb were excluded at $95\%$~C.L.

\section{Experimental signatures}
This section presents the kinematic and other properties of the two theoretical models presented in 
Sections~\ref{sec_BLmodel} and \ref{section_fimpmodel}. 
It also proposes an experimental analysis strategy, which exploits these properties to  
distinguish the signals from SM backgrounds at the LHC. Furthermore, in the
hypothesis that a significant signal were to be seen, it considers how one might confirm that it is 
indeed due to one of these models, and explores how one might measure the model parameters.

The FIMP model of Section~\ref{section_fimpmodel} predicts long-lived particles decaying to dileptons plus
missing energy. This is also one of the possible decay modes of the long-lived particles of
the $B-L$ model of Sect.~\ref{sec_BLmodel}. The analysis strategies for the two models will thus be 
very similar, with the methods used for the FIMP model tending to be a subset of those used for the
$B-L$ model. In the two sub-sections that follow, results will be presented first for the $B-L$ model,
and then for the FIMP model, emphasizing for the latter just the differences with respect to the former.
\subsection{Phenomenology of Long-Lived Heavy Neutrino Decays in the B-L Model}
\label{sec_nuH_phenom}

This sub-section examines the properties of the two benchmarks described in Section~\ref{sec_BLmodel}, in which the light Higgs boson decays to a pair of long-lived, heavy neutrinos $\nu_h$.
We focus on the two most accessible experimentally $\nu_h$ decay modes, namely
$\nu_h\to \ell^+ \ell'^- \nu$ and $\nu_h\to q\bar q' \ell^\pm$. 
Approximately $17$\% of $\nu_h$ decay to the $\ell^+ \ell'^- \nu$ mode,
via both intermediate channels $\nu_h\to \ell^\pm W^\mp$ and $\nu_h\to Z \nu_l$.
Approximately $45$\% of $\nu_h$ decay to the $q\bar q' \ell^\pm$ mode. (Here $\ell$ or $\ell'$ indicate the sum of electron or muon modes.) The studies were done by
simulating the events with $\tt CalcHep/Pythia$ and simulating the detector response using $\tt Delphes$, configured to model the CMS detector, as described in section~\ref{sec_BM_BLmodel}. $\tt Delphes$ was configured
to assume that electrons, muons and jets can be reconstructed down to $p_T$ of $2$, $4$ and $20$~GeV respectively. No efficiency loss was simulated arising from the triggers,
since it is assumed that new triggers would need to be developed for this channel. It should be noted that $\tt Delphes$ does not record spatial coordinates, and so does not
simulate loss of efficiency caused by the difficulty in reconstructing displaced tracks. All histograms shown in this section are normalised
to one generated signal event. The statistical information shown in the histograms always corresponds to the $140$~GeV Higgs mass.

\begin{figure}[htbp]
\begin{center}
\includegraphics[width=0.3\textwidth]{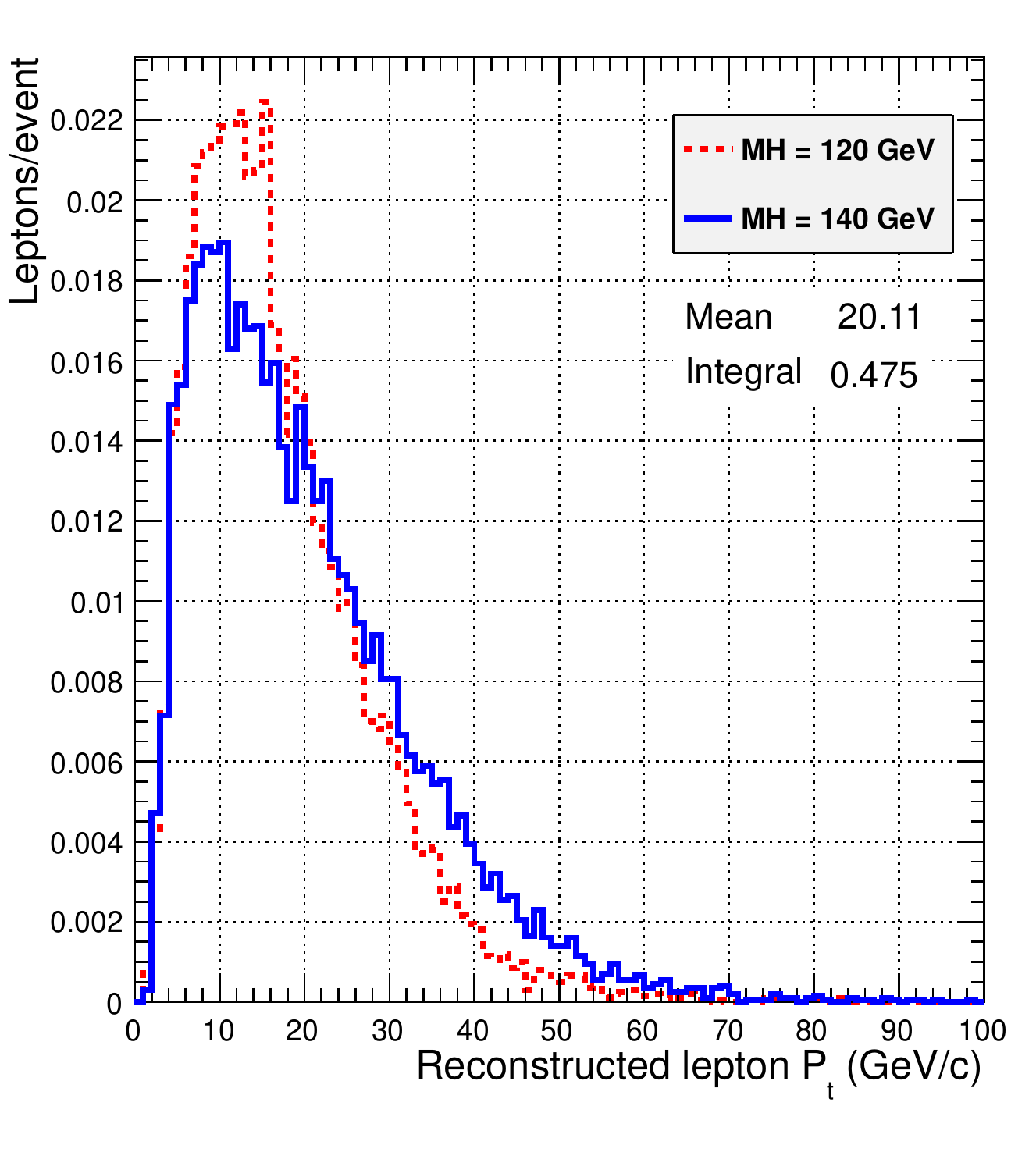}%
\includegraphics[width=0.3\textwidth]{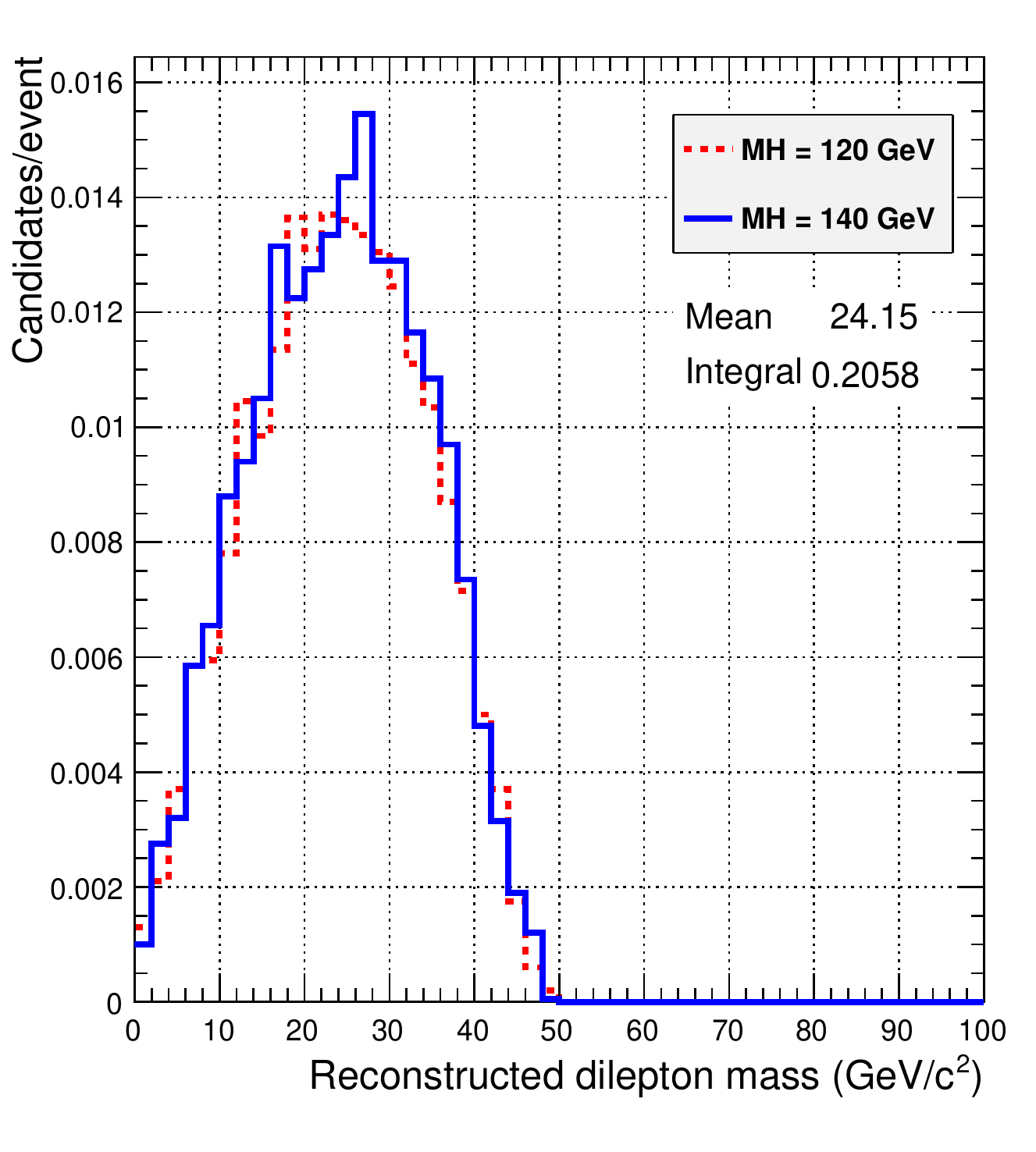}%
\includegraphics[width=0.3\textwidth]{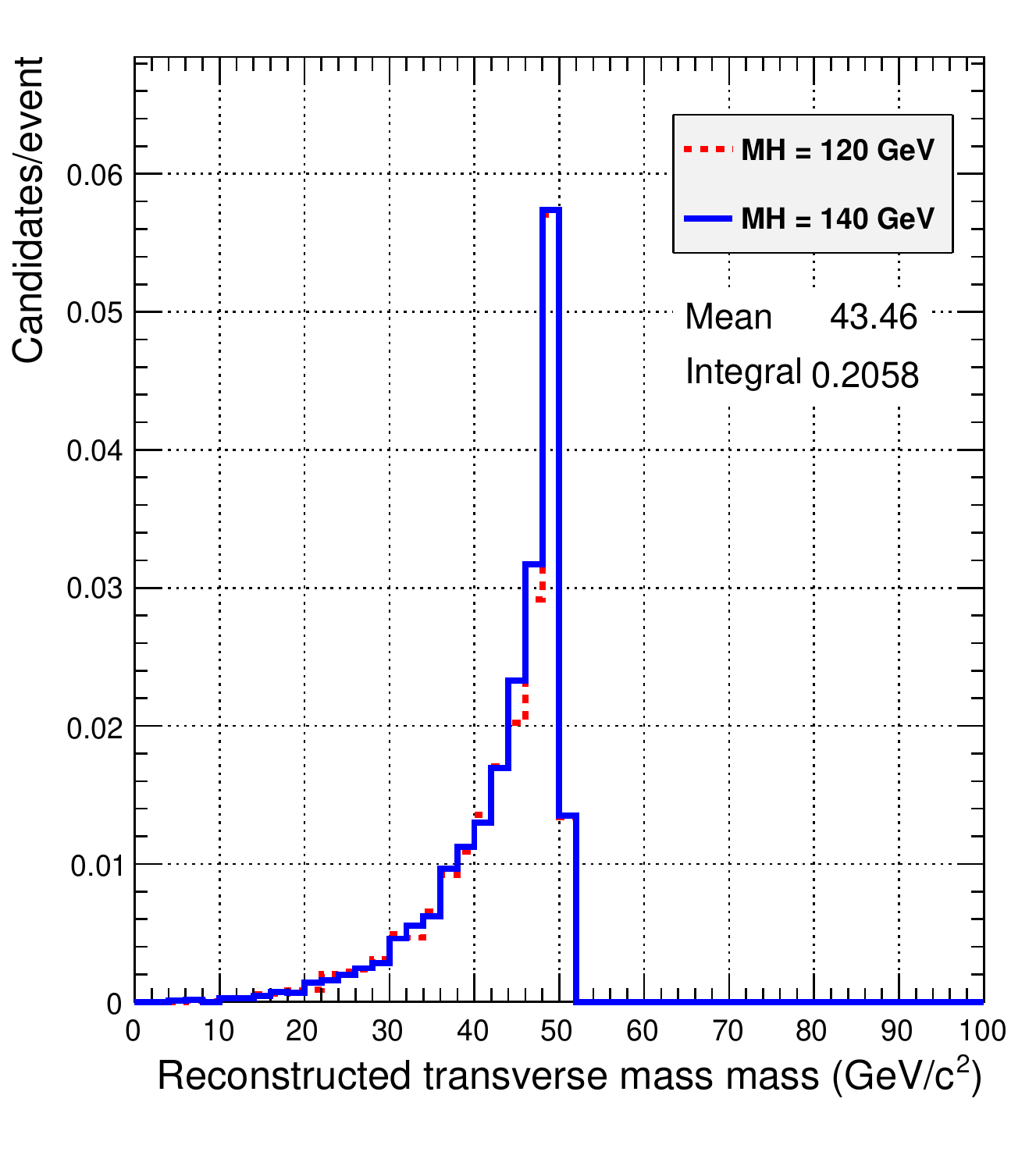} 
\
\caption{For the $\ell^+ \ell'^- \nu$ mode: the left-hand figure shows the $p_T$ of the reconstructed leptons from the heavy neutrino decay; the middle figure shows the reconstructed dilepton mass spectrum; and the right-hand figure the reconstructed 
$\nu_h$ mass, as determined using the more sophisticated method described in the text.}
\label{fig_LLNu}
\end{center}
\end{figure}

The CMS displaced lepton search (Section~\ref{sec_CMSpublic}) indicates that searching for a pair of oppositely charged leptons that are consistent with originating from a displaced 
vertex is a powerful way of searching for this kind of signal, reducing the background to very low levels. The leptons facilitate triggering. Thus the best
way to search for heavy neutrinos should be via the $\ell^+ \ell'^- \nu$ mode. If necessary, the background events could be reduced further by 
searching for the $e^\pm \mu^\mp \nu$ mode or alternatively by requiring the presence of an additional displaced lepton from the second heavy neutrino.
Figure~\ref{fig_LLNu}a shows the $p_T$ of the leptons from the heavy neutrino decay. Because of the low Higgs mass, these are relatively soft. In their publication (Section~\ref{sec_CMSpublic}),
CMS used triggers with lepton $p_T$ thresholds of $23$~GeV ($33$~GeV) for the dimuon (dielectron) channels, respectively. Unless these trigger thresholds can
be reduced, (e.g. by requiring the presence of three leptons at trigger level),  the trigger efficiency for heavy neutrinos would be poor (e.g. for the $120$~GeV
mass Higgs case, only $23$\% ($7$\%) of reconstructed leptons have $p_T > 23$ ($33$)~GeV). The generated transverse decay length distribution of the $\nu_h$ is approximately 
exponential in shape, with a mean of 25 (32)~cm for the 120 (140)~GeV Higgs mass. Since CMS has some efficiency to reconstruct tracks from long-lived exotics with 
transverse decay lengths of up to $50$~cm (see Section~\ref{sec_CMSpublic}), at least some of $\nu_h$ decay products would be found. 


If an excess of displaced dileptons is seen, one would wish to confirm if it were caused by heavy neutrinos and to measure the model parameters. Figure~\ref{fig_LLNu}b
shows the reconstructed invariant mass of the dileptons from the exotic decay. The end-point of this distribution can be used to estimate the heavy neutrino mass.
Figure~\ref{fig_LLNu}c shows a more sophisticated and accurate method to estimate the heavy neutrino mass. In this case, it is assumed that the heavy neutrino
flight direction can be precisely determined from the vector between the reconstructed primary vertex and the displaced vertex. Defining the momentum component of the 
dilepton system perpendicular to the heavy neutrino flight direction as $\vec{p}_T^{\;\mathrm{flight}}(\ell\ell)$, momentum conservation dictates that the corresponding
quantity for the unobserved neutrino must be $\vec{p}_T^{\;\mathrm{flight}}(\nu) = -\vec{p}_T^{\;\mathrm{flight}}(\ell\ell)$. 
The transverse energy of the dilepton system with respect to the flight direction
is then $E_T^{\mathrm{flight}}(\ell\ell) = \sqrt{M(\ell\ell)^2 + p_T^{\mathrm{flight}}(\ell\ell)^2}$ whilst that of the light neutrino (since it is massless with a very good degree of approximation) is
$E_T^{\mathrm{flight}}(\nu) = p_T^{\mathrm{flight}}(\nu)$.
Hence the transverse mass of the heavy neutrino decay products, where `transverse' is defined as meaning perpendicular to the flight direction, is 
$M_T^{\mathrm{flight}} = \sqrt{ [E_T^{\mathrm{flight}}(\ell\ell) + E_T^{\mathrm{flight}}(\nu)]^2 
- [\vec{p}_T^{\;\mathrm{flight}}(\ell\ell) + \vec{p}_T^{\;\mathrm{flight}}(\nu)]^2 }
= E_T^{\mathrm{flight}}(\ell\ell) + E_T^{\mathrm{flight}}(\nu)$. 
This is analogous to the transverse mass
commonly used in hadronic colliders to reconstruct the mass of a resonance that decays to a neutrino plus visible decay products. It differs in that here, it is
the component of the light neutrino momentum parallel to the flight direction that is unknown, whereas conventionally it is the component along the beam-axis that 
is unknown. As shown in Figure~\ref{fig_LLNu}c, the quantity $M_T^{\mathrm{flight}}$ is always less than the heavy neutrino mass (aside from resolution effects), but has far 
more entries near the end-point of the spectrum that the simpler method, thus allowing for a more precise measurement. It ought to be noted that the end-point technique works here only because the heavy neutrino is lighter than the SM gauge bosons, therefore decaying via effective 3-body decays. Instead the more refined transverse mass reconstruction works regardless of the heavy neutrino mass.
 
The displaced lepton vertex position would provide an approximate measurement of the heavy neutrino lifetime, although this would be complicated by the presence
of the light neutrino amongst the heavy neutrino decay products, which would prevent a precise determination of its momentum. From the statistical information
in Figure~\ref{fig_LLNu}c, it can be seen that $0.21$ dilepton candidates/event are reconstructed (for the $140$~GeV Higgs mass), indicating only a small efficiency loss with respect to the $2\times 0.17 = 0.34$ generated.
We emphasize here that the combined measurement of the heavy neutrino lifetime and mass would enable to infer the related light neutrino mass~\cite{Basso:2008iv}.

\begin{figure}[h]
\begin{center}
\includegraphics[width=0.3\textwidth]{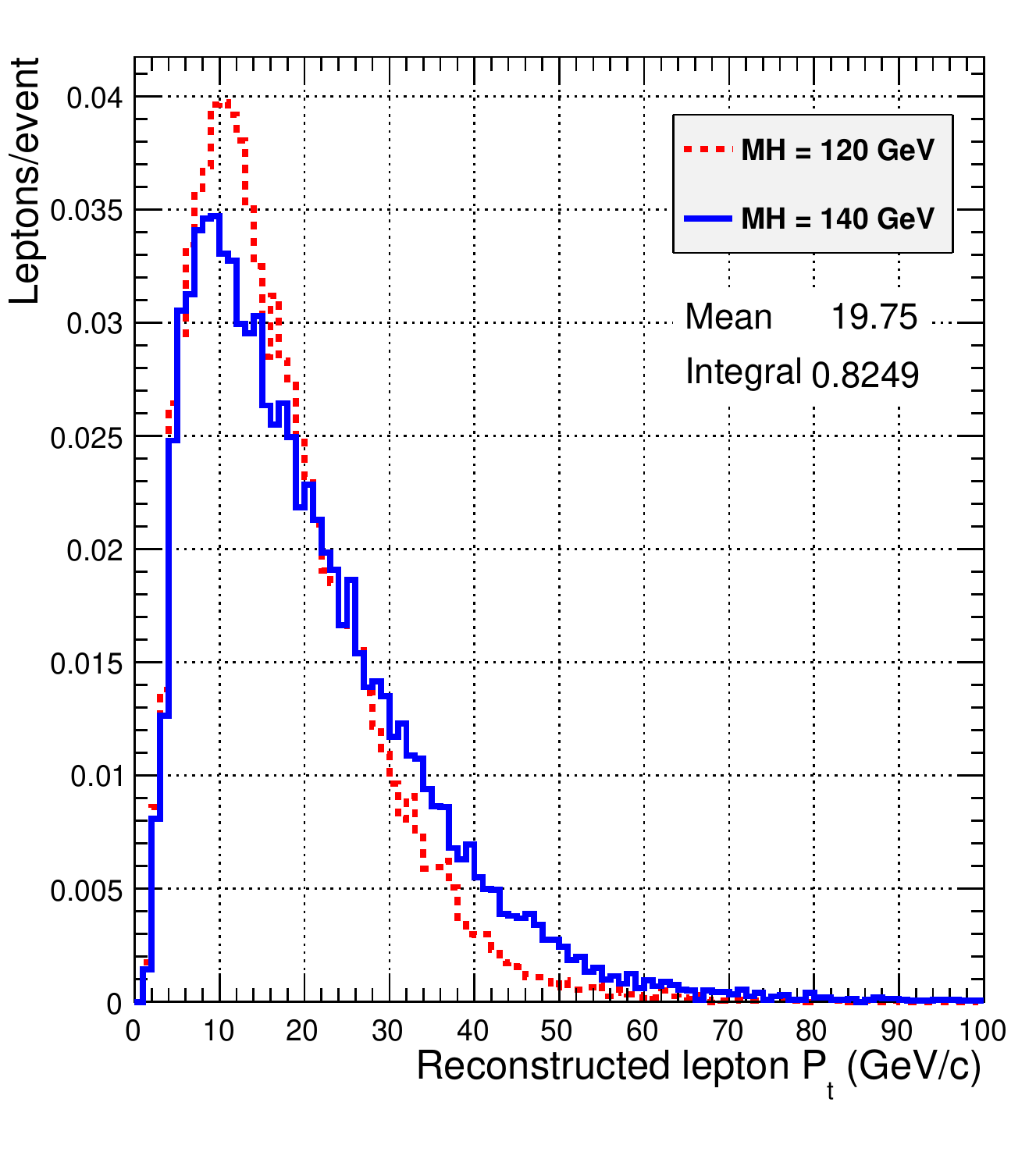} 
\includegraphics[width=0.3\textwidth]{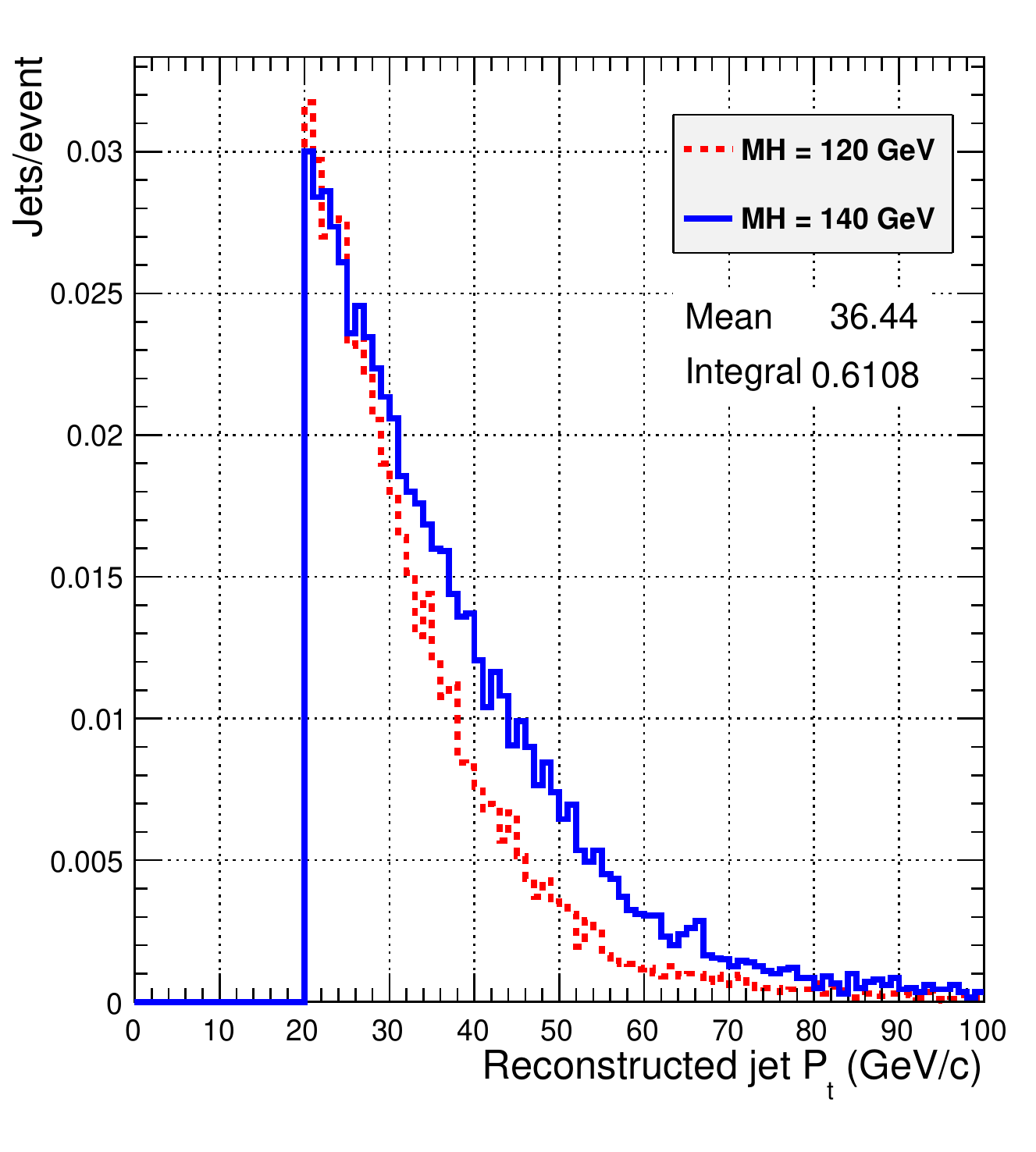}
\includegraphics[width=0.3\textwidth]{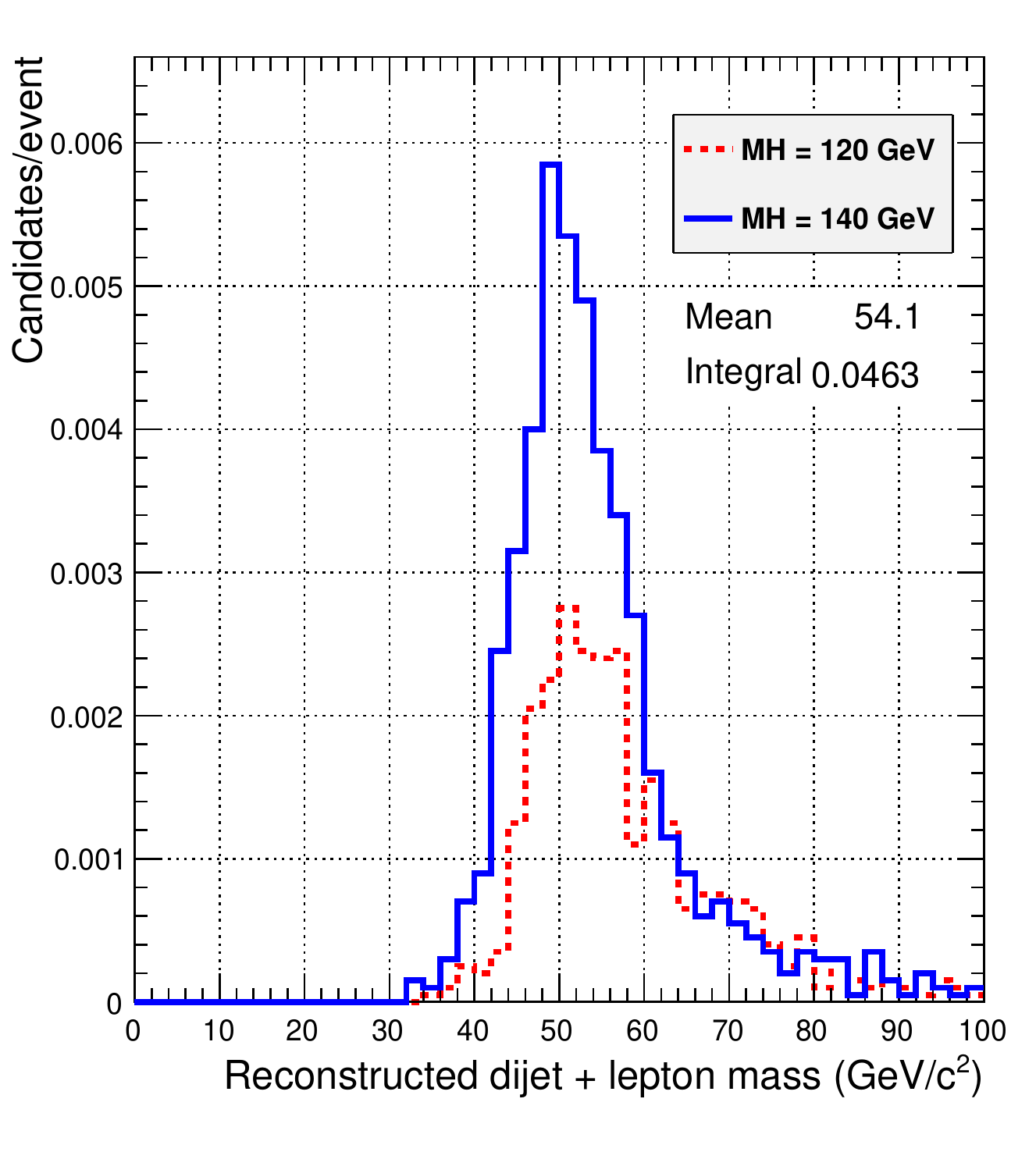}  
\caption{For the $q\bar q' \ell^\pm$ mode: the left-hand figure shows the $p_T$ of the reconstructed lepton and the central figure shows the $p_T$ of the reconstructed jets from the heavy neutrino decay. The right-hand figure shows the reconstructed invariant mass of the lepton plus dijet from the heavy neutrino decay.}
\label{displaced_fig_QQL}
\end{center}
\end{figure}

In events containing a displaced dilepton vertex, one could search for a second heavy neutrino decaying through the $\nu_h\to q\bar q' \ell^-$ channel.  This could be
done by searching for a third displaced lepton and investigating if it formed a displaced vertex with a number of reconstructed charged hadrons. 
(ATLAS successfully found displaced vertices containing a lepton + hadrons in their analysis of Section~\ref{sec_ATLASpublic}.)
Doing so would allow one to measure the relative branching ratio of the two heavy neutrino decay modes. As an additional step, one could then search for a pair
of jets, arising from the $q\bar q'$ pair from the heavy neutrino decay, whose directions would be compatible with these hadrons. Figures~\ref{displaced_fig_QQL}a and \ref{displaced_fig_QQL}b show the 
$p_T$ of the reconstructed lepton and jets from the heavy neutrino in this channel. The soft jet spectrum represents a major challenge. Even assuming
that one can reconstruct jets down to $p_T$ of $20$~GeV, which is optimistic, particularly in the presence of pile-up, the efficiency to reconstruct
these jets is poor. However, if one succeeds, one can then estimate the heavy neutrino mass by plotting the reconstructed lepton + dijet invariant mass,
as shown in Figure~\ref{displaced_fig_QQL}c. (It should also be noted that in the cases in which the $q\bar q'$ pair is reconstructed, simulation indicates
that $\approx 80$\% of the time they are reconstructed as two separate jets, as opposed to a single merged jet. One can thus require the presence of two
reconstructed jets from the displaced vertex. Doing so improves the mass resolution.)
From the statistical information in this histogram, it can be seen that $0.046$ lepton + dijet candidates/event are reconstructed (for the $140$~GeV Higgs mass), 
compared with the $2\times 0.45 = 0.90$ generated $\nu_h\to q\bar q' \ell^\pm$ decays per event. The poor efficiency is due to the difficulty in reconstructing
such soft jets.

\begin{figure}[htbp]
\begin{center}
\includegraphics[width=0.3\textwidth]{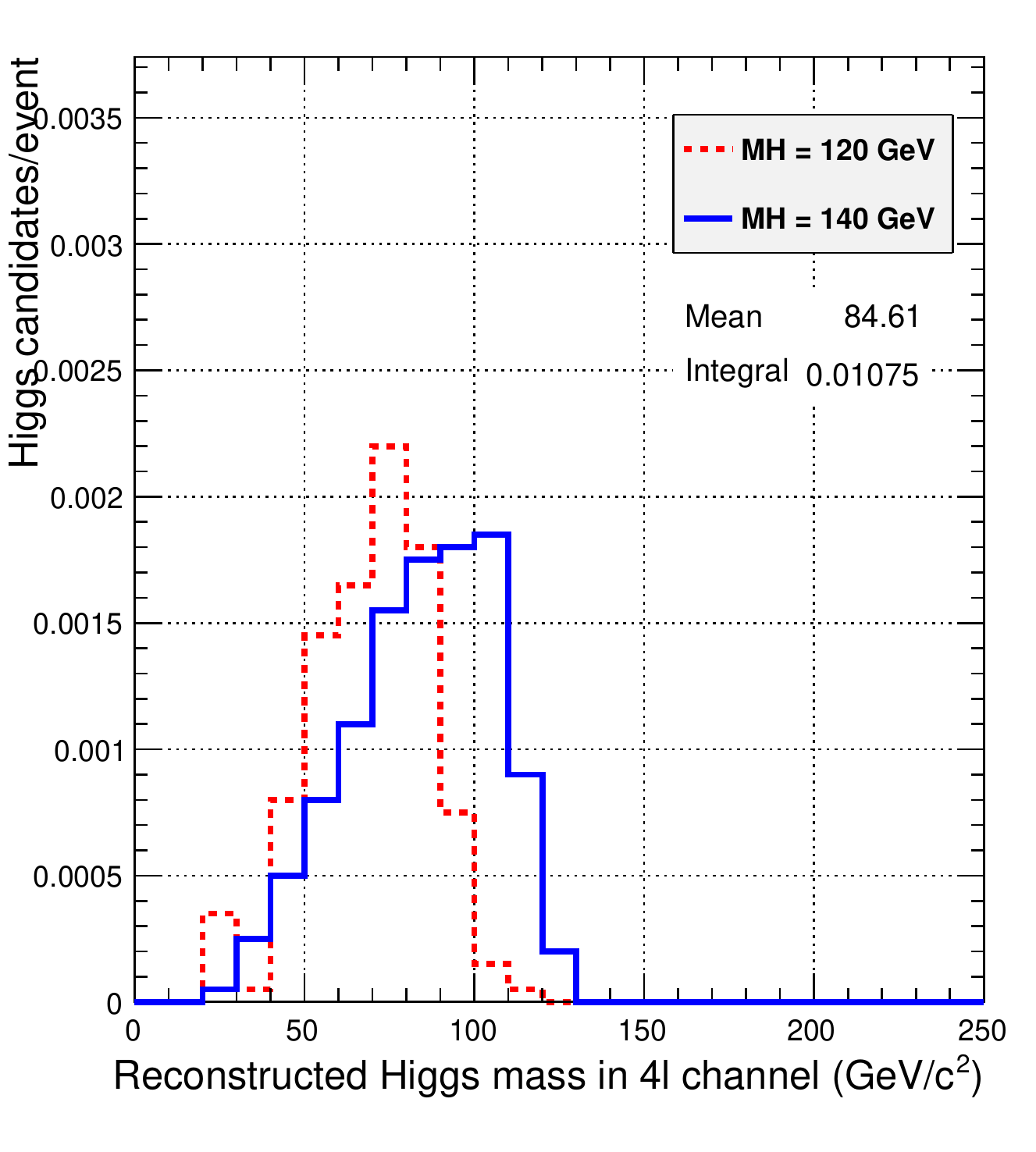} 
\includegraphics[width=0.3\textwidth]{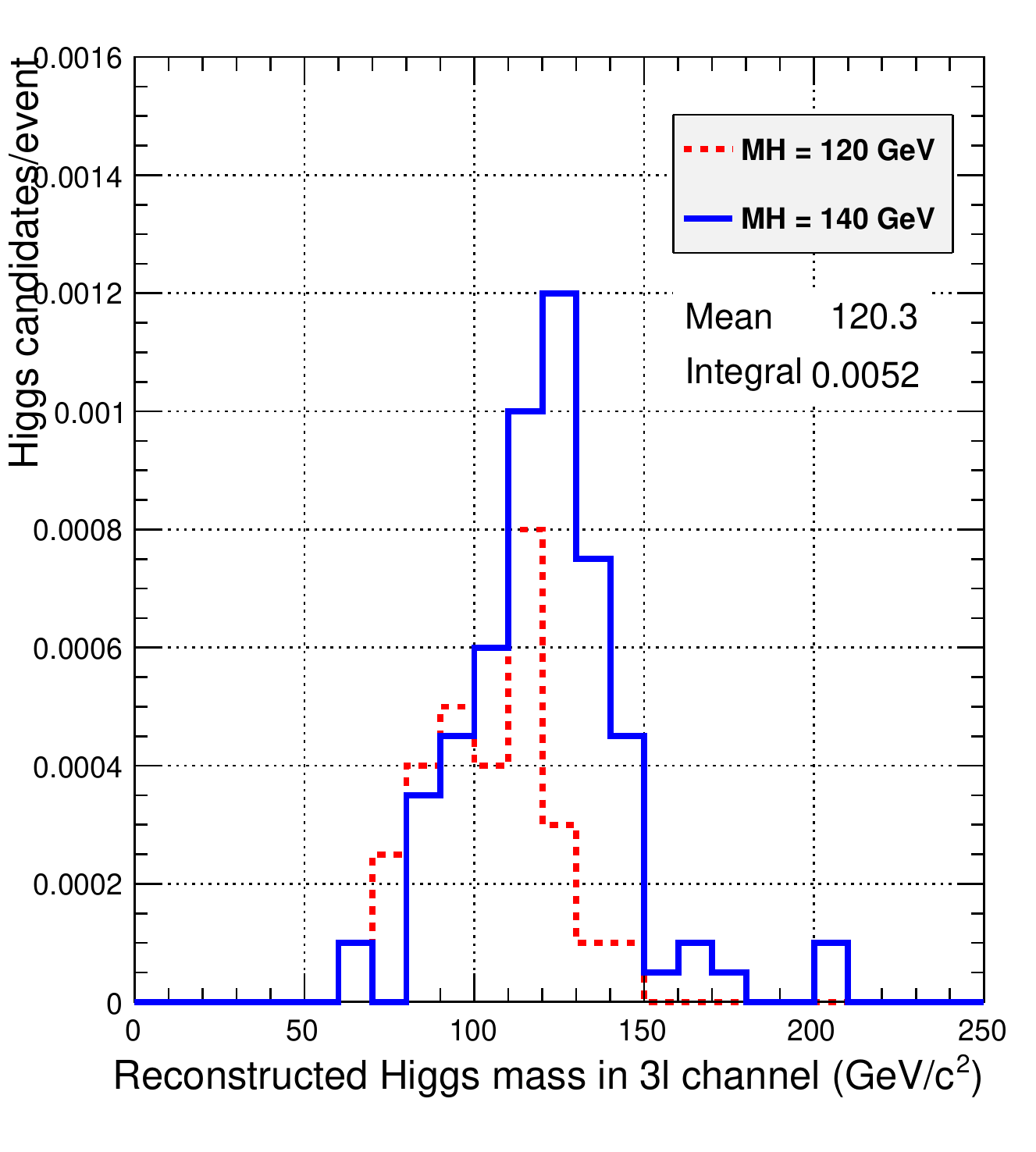} 
\includegraphics[width=0.3\textwidth]{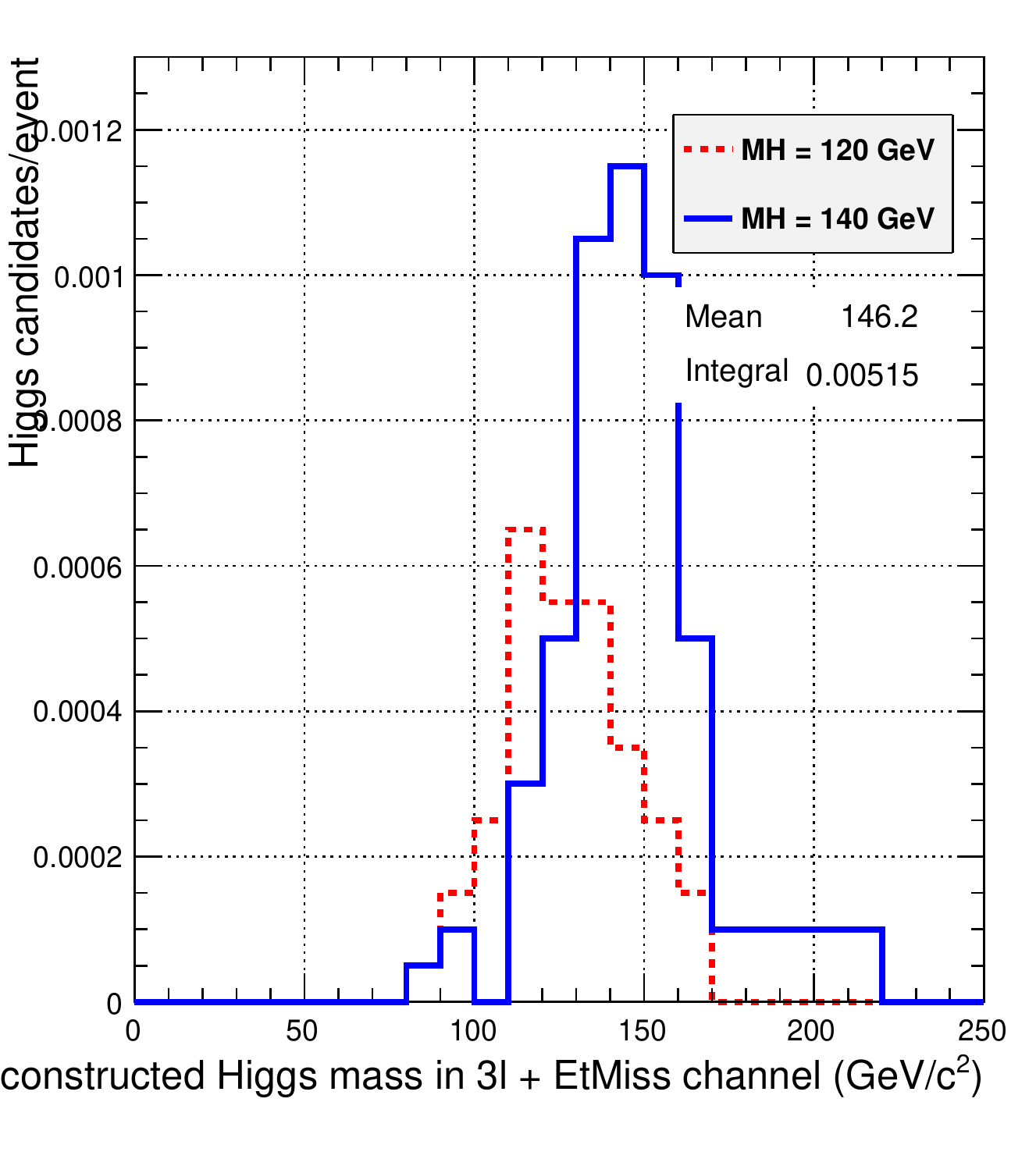} 
\caption{The reconstructed Higgs mass from the visible decay products in the 4-lepton (left) and tri-lepton channels (central), and from the visible decay products and missing transverse momentum in the tri-lepton channel (right).}
\label{displaced_fig_Higgs}
\end{center}
\end{figure}

To measure the Higgs boson mass, one could use either the tri-lepton final state (as in eq.~(\ref{eq_hnu-semilep})) or the four-lepton final state (as in eq.~(\ref{eq_hnu-lept})).
In both cases, the end-point of the spectrum would allow one to infer the Higgs mass.
 Figures~\ref{displaced_fig_Higgs}a and \ref{displaced_fig_Higgs}b
show the Higgs mass reconstructed from its visible decay products in these two channels. Furthermore, as can be seen from the statistical information in the histograms, the efficiency
to reconstruct both heavy neutrinos in an event is at the percent level. As before, more sophisticated techniques can be employed to reconstruct the Higgs mass,
exploiting the reconstructed flight directions of the heavy neutrinos, the known heavy neutrino mass (assumed its mass has already been measured using the
techniques described above) and the measured missing momentum transverse to the beam-axis $p_T(miss)$. In the tri-lepton final state, one
can exploit the missing transverse momentum to construct the
transverse mass of the Higgs decay products (with respect to the beam axis) $M_T = \sqrt{[E_T(vis) + p_T(miss)]^2 - [\vec{p}_T(vis) + \vec{p}_T(miss)]^2}$,
where $E_T(vis) = \sqrt{M(vis)^2 + p_T(vis)^2}$ and $vis$ represents the visible Higgs decay products. This quantity $M_T$ is plotted in Figure~\ref{displaced_fig_Higgs}c.
Were it not for the finite jet and missing energy resolutions, it would always be less than the Higgs mass, and its end-point could be used to estimate the latter.

\subsection{Phenomenology of FIMP model}\label{displaced_experimentfimp_43}
This section describes the properties of the benchmarks described in Section \ref{section_fimpmodel}. The event samples were made according to the procedure discussed in Section \ref{section_tools}. All benchmarks considered predict a long-lived particle decaying into a fermionic dark matter particle in addition to a $W$ or $Z$ boson. In this analysis we consider the subsequent decay of the boson in the muon decay channel. 

\begin{figure}[hbtp]
\begin{center}
\includegraphics[scale=0.35]{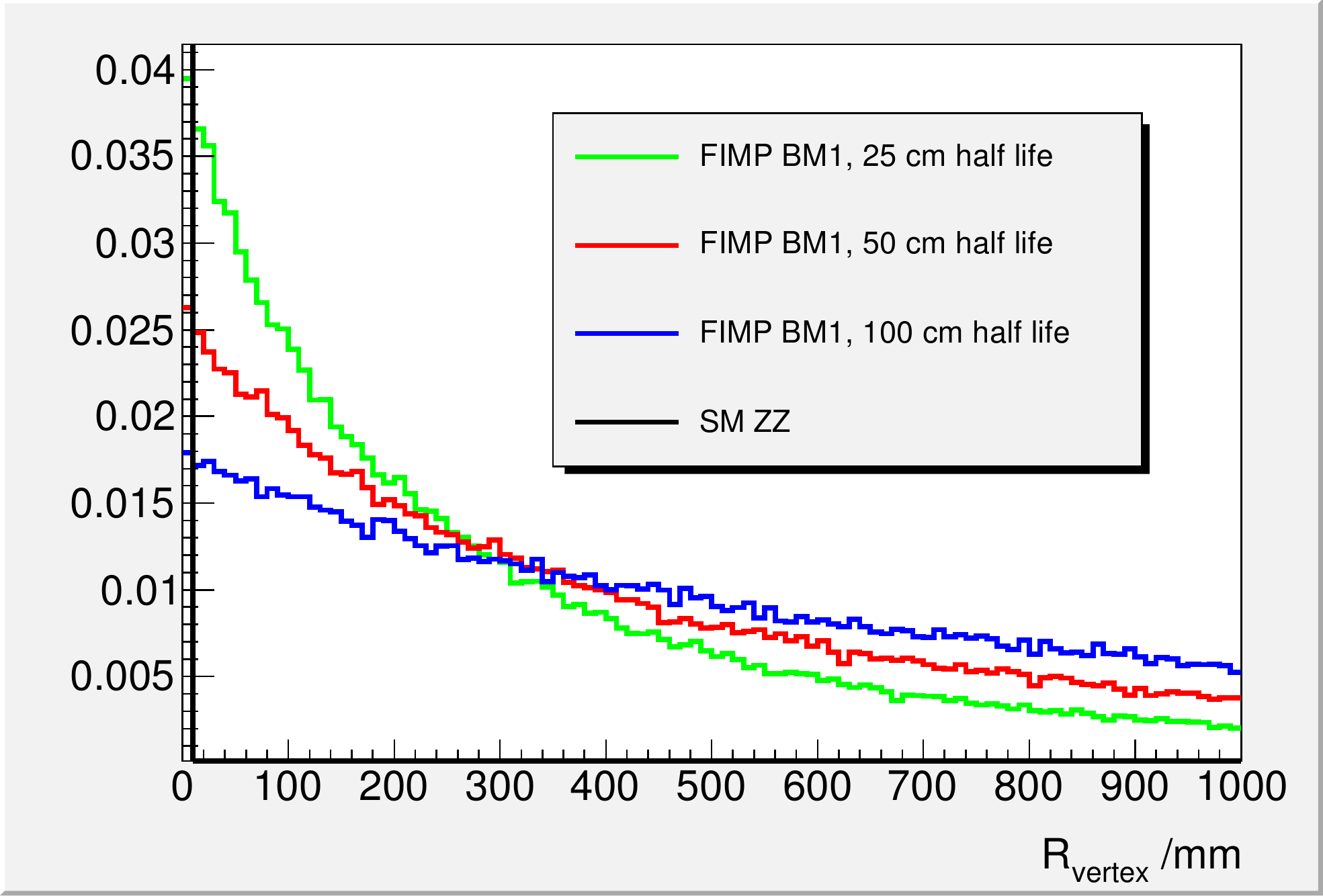}
\includegraphics[scale=0.35]{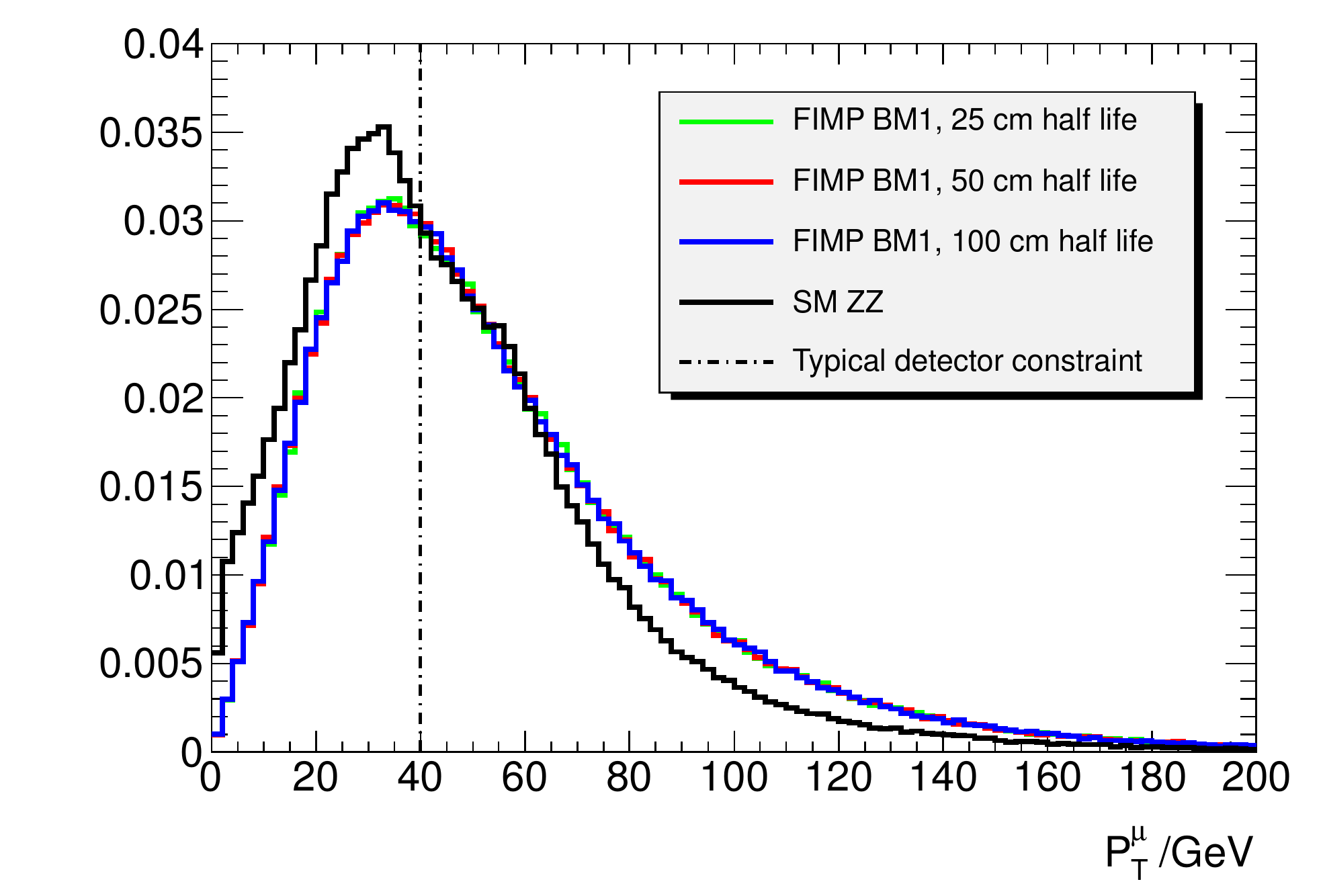} 
\includegraphics[scale=0.35]{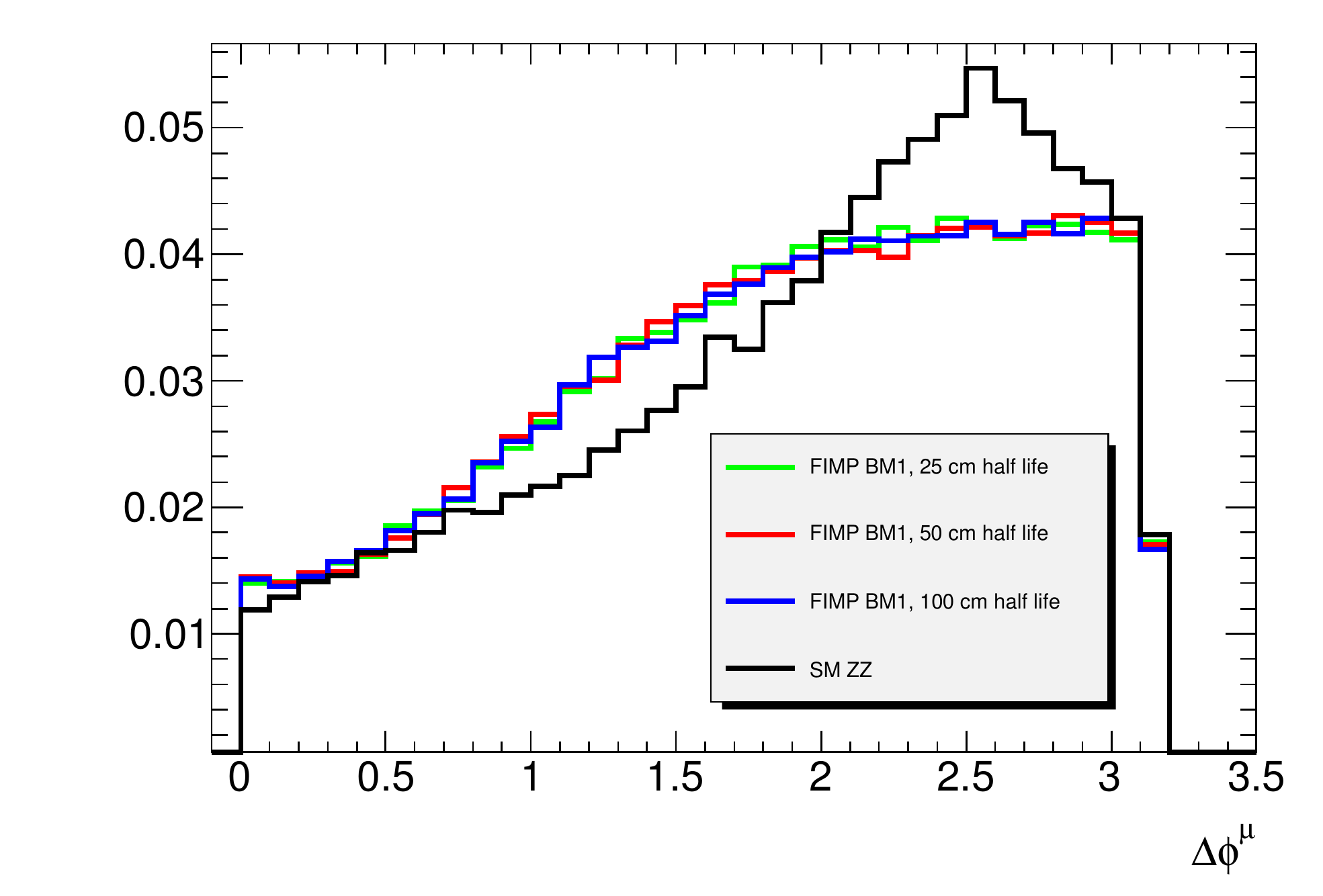} 
\includegraphics[scale=0.35]{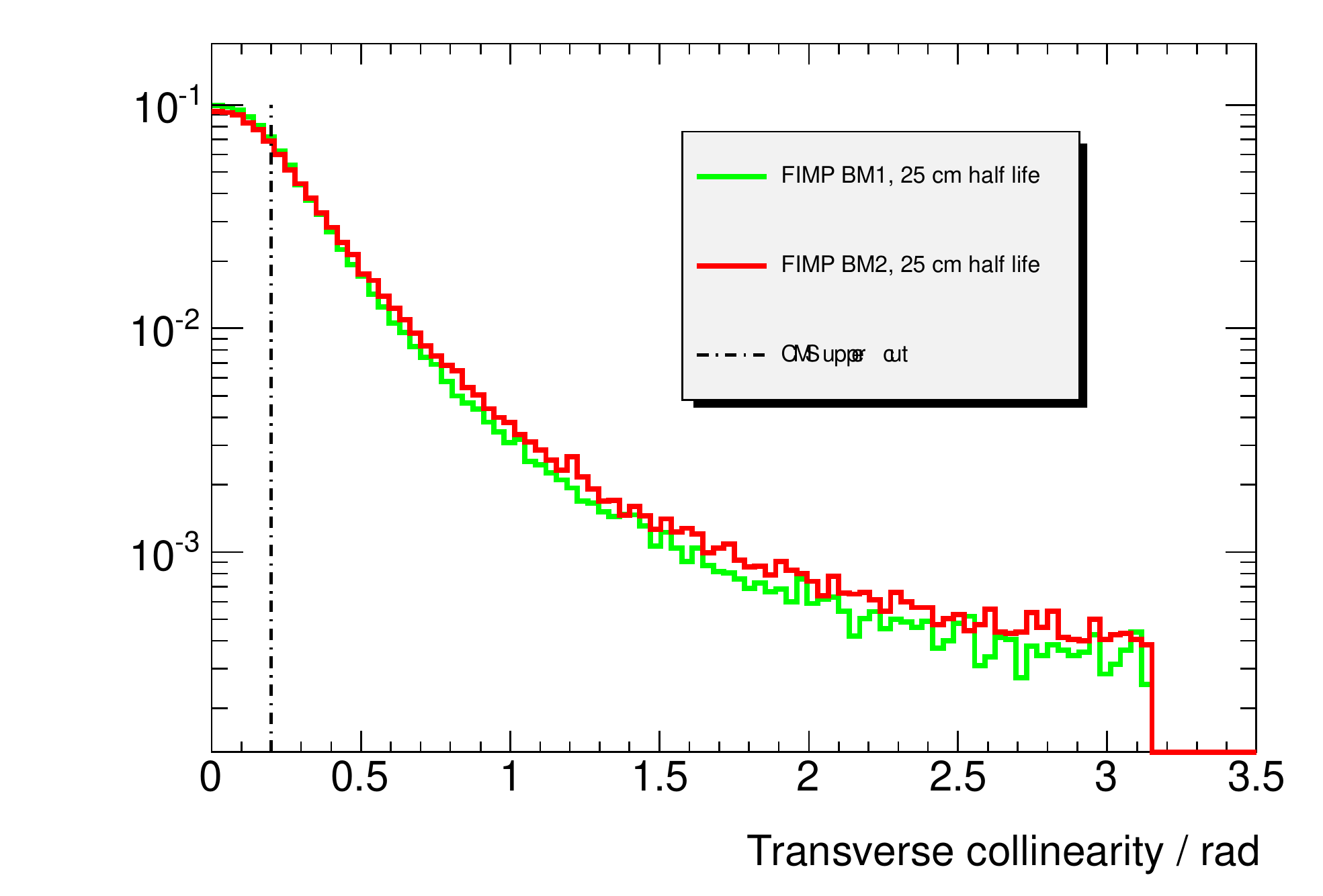}
\caption{Radial displacement (top left), transverse muon momenta (top right), transverse angular muon separation (bottom left) and transverse collinearity (bottom right) of displaced $Z$ decay. Standard Model Drell-Yan kinematics from a prompt $Z$ decay are also shown for comparison. Existing cuts on these variables in displaced vertex searches are shown on the plot.}
\label{fig_displacedMuons}
\end{center}
\end{figure}

Figure \ref{fig_displacedMuons} shows the transverse momentum and angular separation (if applicable) of the muons originating from the heavy boson decay in all three benchmarks. The corresponding distribution for the SM Drell-Yan background is also shown for comparison. The kinematic constraints imposed by existing detector searches are marked on the plot. 
In the interests of maximising acceptance, the optimal lepton $p_T$ requirement should be as low as possible. A single or di-muon trigger is envisaged to record the events studied here and, due to the muon vertex displacement, the trigger is expected not to rely on inner detector tracking information. As already described, trigger prescaling, needed due to the high luminosity of the LHC, means that the trigger threshold is likely to be at the lowest $\simeq$ 20 GeV and may well be higher. However, the muons originating from these models are more energetic than in the $B-L$ model and thus a higher yield of events is envisaged. 
Some cut on the angular separation between the muons -- either transverse or otherwise -- would be necessary to avoid background contamination from cosmic rays. The looser this cut is, the more sensitivity the analysis will have to these models. The D0, and to a lesser extent the CMS, analyses will have reduced sensitivity to the FIMP model as their requirements on angular separation are sufficiently tight that they would remove many candidates. 

Searching for a muon pair displaced from the interaction point is a powerful way of discriminating these models from Standard Model background signatures. Figure \ref{fig_displacedMuons} shows the typical $R$ displacement predicted by the FIMP models when the mean lifetime is set at 25, 50 and 100~cm in BM1. As in the previous model, a measured $R$ displacement represents the lifetime of the chargino lifetime and thus its coupling to the FIMP. For any displaced particle, the most likely point of decay is (0, 0, 0) and then an exponential decay from that point is observed, its mean lifetime governed by the decay length of the particle. It can be seen in the CMS and ATLAS studies that the efficiency for reconstructing these vertices also drops sharply as the displacement is increased. 

The displaced vertex represents a decay into a $Z$ boson and a fermionic dark matter particle which will be observed as missing transverse energy in the detector. Unlike the $B-L$ model, the model considered here does not allow for two leptons of different flavour to originate from the same vertex and thus we only consider decays to two, oppositely charged muons. The invariant mass of muon pairs, observed at a radial displacement $R$, will follow the same distribution as the invariant mass in Standard Model $Z$ decays. A cut around the $Z$ mass peak of the vertex could prove powerful in the proposed analysis, as there are no Standard Model physics backgrounds in which a high mass displaced $\mu \mu$ vertex would be seen in the detector. 

The transverse collinearity angle between the reconstructed momentum vector of the $Z$ boson and the vector from the primary to the secondary vertex -- the neutralino direction of flight -- is also shown in figure \ref{fig_displacedMuons}. It is seen that the most likely direction of the Z is collinear to the neutralino although there is a significant portion of events in which this is not the case. The collinearity cut imposed in the CMS analysis  -- an upper limit of 0.2 radians -- would be likely to remove a significant number of FIMP decays, should they be present, from their selected events. 

\begin{figure}[hbtp]
\begin{center}
\includegraphics[scale=0.35]{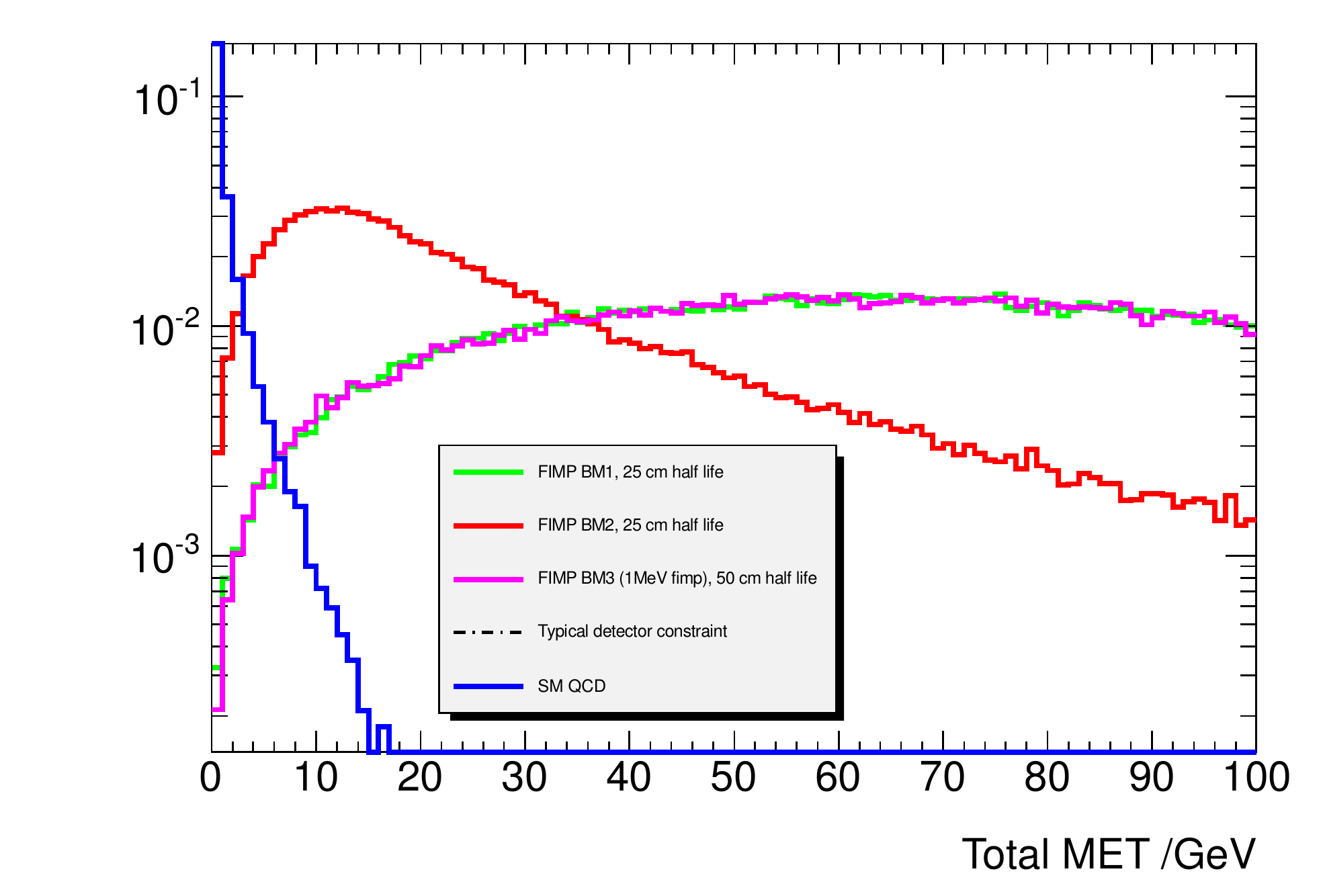} 
\includegraphics[scale=0.35]{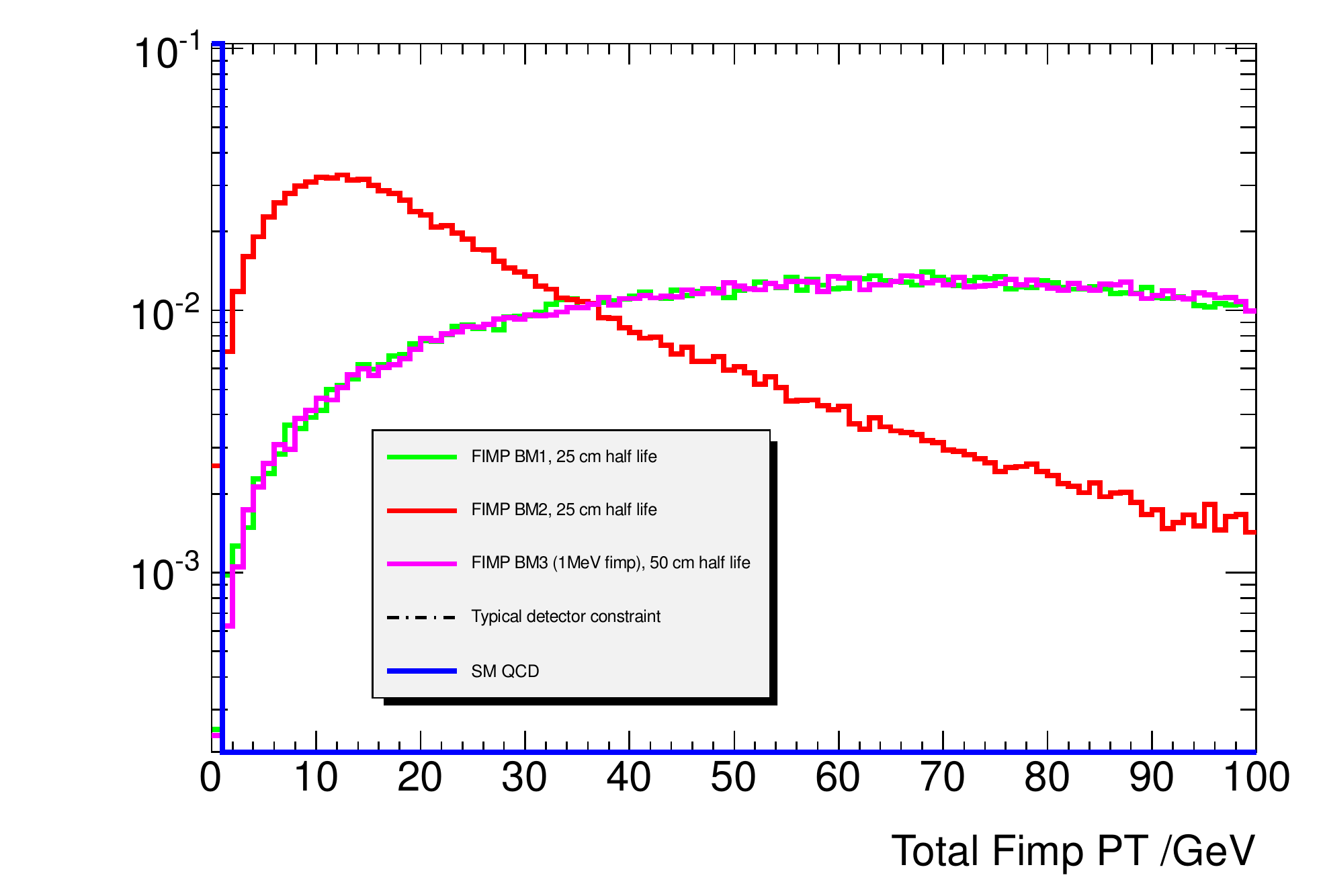} 
\caption{Missing transverse energy from combination of neutrinos and FIMPS (left), and from FIMPS only (right) in all 3 benchmarks. The distribution expected from Standard Model $W\mu\nu$ is shown for comparison.}
\label{fig_missing}
\end{center}
\end{figure}

The presence of missing transverse energy in the event is also considered. Figure \ref{fig_missing} shows the total missing transverse energy from both the neutrinos and the FIMPs and from only the FIMPs in the event, respectively. All benchmarks are shown (with 25 cm mean life) as well as SM $W\rightarrow\mu\nu$ events. Only BM3, due to the additional contribution from the displaced $W$ decay, predicts larger missing transverse energy than the SM $W$ sample. This implies that the transverse missing energy contribution provided by a 1 GeV FIMP is comparable to that from the neutrino in a $W\rightarrow\mu\nu$ decay, and for this reason a cut on transverse missing energy may not be a powerful discriminant for these models.

\subsection{Sensitivity of current displaced vertex searches to models}
Among the current searches for displaced vertex signatures at hadron colliders, the CMS search has the most sensitivity to the model under consideration. The D0 analysis would have limited sensitivity to the model due to the cut on the angular separation between the muons, and the limits set by the CDF analysis (limited by the luminosity available) do not reach the cross sections for the range of FIMP models considered. The ATLAS search is not relevant for the FIMP model as the event selection is on a different final state topology (muon+hadrons). 

To estimate the potential sensitivity of a CMS-like search for FIMP decays, a similar event selection is run -- at truth level -- for the benchmarks. The requirement applied is two energetic (transverse momenta $>$ 25 GeV) and fiducial ($|\eta|<2.4$) muons associated with the same vertex. A veto is applied to muons that are back-to-back (transverse angular separation $<$ 2.8). The transverse collinearity cut employed by the CMS analysis, which would remove many FIMP events, is not applied. The expected yields per fb of data for each benchmark is summarised in table \ref{table_selection}, having taken into account the branching ratio, $\Gamma(\mu)$, for having at least one of the legs decaying muonically. No physics background was seen to pass the proposed event selection, and it is assumed that there is zero machine or detector background given the high mass of the displaced vertex. 

There will be additional losses to the number of events seen to pass the fiducial selection at truth level, due to imperfect detector reconstruction of the event. As has been discussed, the largest losses will be in triggering and reconstruction of the displaced vertex. A functional form of efficiency -- although it must be noted this is very detector dependent -- to reconstruct a muon as a function of radial displacement was assumed from studying the publicly-available CMS efficiency plots. The form was taken as linear fall off from unity at zero displacement to zero at 40 cm displacement. As would be expected, the longer the life-time of the particle is, the more events are lost due to imperfect efficiency modelling in the detector. The results here suggest that certain models (BM2) may very well already be ruled out by existing displaced vertex searches, but others (BM1) may not yet be visible with the current detector limitations and integrated luminosity available.

The $p_T$ cut used on the leptons in the CMS analysis significantly reduces its efficiency in a search for low mass Higgs bosons -- CMS quotes limits only for Higgs masses $>$ 200 GeV. Their publication does not therefore rule out the benchmarks presented in Section~\ref{sec_BLmodel}. One should note that the heavy neutrinos in the $B-L$ model have only a $2.9\%$ branching ratio to $\mu\mu\nu$ (when summed up over the three generations). 
And that the probability of both of these muons being reconstructed with transverse momenta exceeding the cut used by CMS and of the dimuon system being collinear with the exotica flight direction to within 0.2 radians (also required by CMS) is only $2.7\%$. If one attempted to reinterpret the CMS publication in terms of limits on the $B-L$ cross section, the result would be a limit of approximately $0.01 \rm{pb}/(0.029\times 0.027$) = 10 pb. This is well above the theoretically expected cross section.

\begin{table}[h]
\begin{center}
\begin{tabular}{l|l|l|l|l}\hline
Selection & $N_{BM1}\times\Gamma(\mu)$ & $N_{BM2}\times\Gamma(\mu)$ & $N_{BM3}\times\Gamma(\mu)$ & $N_{SM}\times\Gamma(\mu)$ \\ \hline
W & 1 & 0 & 0 & $\simeq{5000000}$ \\
Z & 0.05 & 0 & 0 & $\simeq{100000}$ \\
Displaced Vertex & 4 & 42 & 0.01 & $\simeq{0}$ \\
\hline
\end{tabular}
\caption{\label{table_selection}
Expected yield of FIMP events for 1 fb$^{-1}$ of luminosity passing Standard Model and CMS-style displaced vertex selections at truth level.  }
\end{center}
\end{table}

Due to the cascade decays inherent in the FIMP models, such events can be quite noisy with many hard leptons and jets produced at the interaction point. Standard $W$ and $Z$ event topology selections were applied to the samples. It was found that the expected contribution from the models considered here would be well below the current fractional uncertainty on measured $W$ and $Z$ cross sections\cite{Aad:2011dm}. This suggests that if these events were produced at the LHC, they would not be detected in ATLAS or CMS SM analyses if event yields or basic kinematics alone are considered. 


The strategies developed in Section~\ref{sec_nuH_phenom} for determining the mass of long-lived particles would require some modification of the FIMP analysis. The method illustrated in Figure \ref{fig_LLNu}b of measuring the endpoint of the dilepton mass spectrum would only work if the $Z$ boson is off mass-shell, since otherwise, the dilepton invariant mass would simply be equal to the $Z$ boson mass. Thus this method is only applicable if the mass difference between the gaugino and the FIMP is less than the $Z$ mass. The more sophisticated method illustrated in Figure \ref{fig_LLNu}c, which uses the measured long-lived particle flight direction as a constraint, does not suffer from this limitation. However, it does assume that the invisible particle is massless, so could only be employed for the FIMP analysis if the FIMP has a much lower mass than the long-lived particle.

\section{Conclusions}
We have investigated two classes of models, which are quite different from the theoretical point of view. 
Both models predict long-lived particles -- heavy Majorana neutrinos (from Higgs boson decay) in the case of $B-L$ model 
and charginos/neutralinos in the case of an MSSM model extended by a singlet superfield that is feebly coupled to the standard MSSM states (FIMP model). The decay of these long-lived particles provide spectacular displaced, multi-lepton 
(or lepton + jet) signatures, which are virtually background-free.

For selected benchmark points of the $B-L$ and FIMP models, we performed an analysis showing that experimental 
observation of these displaced
vertices at the LHC should be possible. We also demonstrated how one could reconstruct the masses of the long-lived 
particles, profiting from the reconstructed vertex positions to constrain the long-lived particle flight direction.
Similar analysis techniques could be applied to other BSM models featuring long-lived particles.
In the case of the $B-L$ model, the measurement of the mean lifetime and mass of the Majorana neutrinos 
could be important ingredients, allowing one to deduce the
absolute mass of SM-like neutrinos~\cite{Basso:2008iv}.
We showed that published searches from the Tevatron and the LHC experiments have poor sensitivity to these 
models.  


\section*{ACKNOWLEDGEMENTS}
L.B and A.B.thank the NExT Institute and Royal Society 
for partial financial support. 
L.B. has also been partially supported by the Deutsche Forschungsgemeinschaft through the Research Training Group GRK\,1102 \textit{Physics of Hadron Accelerators}. SMW and AJW thank the Higher Education Funding Council for England and the Science and Technology Facilities Council for financial support under the SEPNet Initiative and the IPPP for financial support. 



\AddToContent{L.~Basso, A.~Belyaev, E.~Dobson, M.~C.~Thomas, I.~Tomalin, S.~M.~West, A.~J.~Williams}
\renewcommand{\thesection}{\arabic{section}}

\superpart{ Dark Matter }

\chapter{Impact of semi-annihilations on dark matter phenomenology -- an example of $Z_N$ symmetric scalar dark matter}

{\it G. B\'elanger, K. Kannike, A. Pukhov, M. Raidal}

\begin{abstract}
We study the impact of semi-annihilations $\chi\chi\leftrightarrow\chi X,$ where $\chi$ is dark matter and $X$ is any standard model particle,  
on dark matter phenomenology. 
We formulate scalar dark matter models with minimal field content that predict non-trivial dark matter phenomenology for different discrete Abelian symmetries  $Z_N,$ $N>2,$ and contain semi-annihilation processes. We implement such an example model in micrOMEGAs and show that
semi-annihilations modify the phenomenology of this type of models.
\end{abstract}

\section{INTRODUCTION}

The origin of dark matter  of the Universe is not known. In popular models with new particles beyond the standard model particle content, such as the minimal supersymmetric standard model, an additional discrete $Z_2$ symmetry is introduced~\cite{Farrar:1978xj}. As a result, the lightest new $Z_2$-odd particle, $\chi,$ is stable and is a good candidate for dark matter. The phenomenology of this type of models is studied extensively.

The discrete symmetry that stabilises dark matter must be the discrete remnant of a broken gauge group \cite{Krauss:1988zc}, because global discrete symmetries are broken by gravity. The most natural way for the discrete symmetry to arise is from breaking of a $U(1)_X$ embedded in a larger gauge group, e.g. $SO(10)$ \cite{Fritzsch:1974nn}. The latter contains gauged $B-L$ as a part of the symmetry, and the existence of dark matter can be related to the neutrino masses, leptogenesis and, in a broader context, to the existence of leptonic and baryonic  matter~\cite{Martin:1992mq,Kadastik:2009dj,Kadastik:2009cu}.  

Obviously, the discrete remnant of $U(1)_X$ need not to be $Z_2$ -- in general it can be any $Z_N$ Abelian symmetry. The possibility that dark matter may exist due to $Z_N,$ $N > 2,$ is a known \cite{Ibanez:1991hv,Agashe:2004ci,Agashe:2004bm,Dreiner:2005rd,Ma:2007gq,Agashe:2010gt,Agashe:2010tu,Batell:2010bp,DEramo:2010ep} but much less studied scenario\footnote{Phenomenology of $Z_3$-symmetric dark matter in supersymmetric models has been studied in Refs.~\cite{Ibanez:1991hv,Dreiner:2005rd} and in extra dimensional models in Refs.~\cite{Agashe:2004ci,Agashe:2004bm}.}.
 Model independently, it has been pointed out in Ref.~\cite{DEramo:2010ep} 
  that  in $Z_N$ models the dark matter annihilation processes
 contain new topologies with different number of dark matter particles in the initial and final states -- called semi-annihilations --, for example $\chi\chi\leftrightarrow\chi X,$ where $X$ can be any standard model particle. It has been argued that those processes 
may significantly change the predictions for generation of dark matter relic abundance  in thermal freeze-out.  
However, no detailed studies have been performed that compare dark matter phenomenology of different $Z_N$ models. 
This is difficult also because presently the publicly available tools for computing dark matter relic abundance do not  include  
the possibility of imposing $Z_N$  discrete symmetry instead of $Z_2$.

The aim of this work is to formulate the minimal scalar dark matter model that predicts different non-trivial scalar potentials for different $Z_N$ symmetries
and to  study their phenomenology. In particular we are interested in quantifying the possible effects of semi-annihilation processes 
$\chi\chi\leftrightarrow\chi X$ on generating the  dark matter relic abundance. In order to perform quantitatively precise analyses we implement 
minimal  $Z_{3}$ symmetric scalar dark matter models that contain one singlet and one extra doublet in micrOMEGAs~\cite{Belanger:2004yn,Belanger:2006is}.
 Using this tool we show that, indeed, the semi-annihilations affect the dark matter phenomenology  and should be taken into account in a quantitatively precise way
 in studies of any particular model.

\section{$Z_{N}$ LAGRANGIANS}

\subsection{$Z_{N}$ symmetry}

Under an Abelian $Z_{N}$ symmetry, where $N$ is a positive integer, addition of charges is modulo $N$. Thus the possible values of $Z_{N}$ charges can be taken to be $0, 1, \ldots, N-1$ without loss of generality. A field $\phi$ with $Z_N$ charge $X$ transforms under a $Z_N$ transformation as $\phi \to \omega^X \phi$, where $\omega^N = 1$, that is $\omega = \exp (i 2 \pi/N)$.

A $Z_{N}$ symmetry can arise as a discrete gauge symmetry from breaking a $U(1)_{X}$ gauge group with a scalar whose $X$-charge is $N$ \cite{Krauss:1988zc,Martin:1992mq}.

For larger values of $N$ the conditions the $Z_{N}$ symmetry imposes on the Lagrangian approximates a $U(1)$ symmetry for two reasons. First, assuming renormalizability, the number of possible Lagrangian terms is limited and will be exhausted for some small finite $N$, though they may come up in different combinations for different values of $N$. Second, if the $Z_{N}$ symmetry arises from some $U(1)_{X}$, the $X$-charges of particles cannot be arbitrarily large, because that would make the model nonperturbative -- if $N$ is larger than the largest charge in the model, the restrictions on the Lagrangian are the same as in the unbroken $U(1)$.

We shall see below that for the large number of possible assignments of $Z_N$ charges to the fields, the number of possible distinct potentials is much smaller.

\subsection{Field content of the minimal model}

In order to study how different discrete $Z_{N}$ symmetries impact dark matter phenomenology the example model must contain more than one type of dark matter candidates.  The minimal dark matter model that possesses such properties  contains, in addition to the standard model fermions and the standard model Higgs boson $H_{1}$, one extra scalar doublet $H_{2}$  and one extra complex scalar singlet $S$~\cite{Kadastik:2009dj}. 
In the case of $Z_2$ symmetry, as proposed in~\cite{Kadastik:2009dj}, those new fields can be identified with the well known
inert doublet $H_{2}$ \cite{Deshpande:1977rw,Ma:2006km,Barbieri:2006dq,LopezHonorez:2006gr} and the complex singlet $S$ \cite{McDonald:1993ex,Barger:2007im,Barger:2008jx,Burgess:2000yq,Gonderinger:2009jp}.
The phenomenology of those models is well studied. However, when they are put together,  qualitatively new features concerning dark matter phenomenology, electroweak symmetry breaking and collider phenomenology occur~\cite{Kadastik:2009cu,Kadastik:2009dj,Kadastik:2009ca,Kadastik:2009gx,Huitu:2010uc}.
The field content of the minimal scalar $Z_{N}$ model is summarized in Table~\ref{tab:fields}.

\begin{table}[htdp]	
\caption{Scalar field content of the low energy theory with the components of the standard model Higgs $H_{1}$ in the Feynman gauge. The value of the Higgs VEV is $v = 246$~GeV. }
\begin{center}
\begin{tabular}{cccccc}
Field & $SU(3)$ & $SU(2)_{L}$ & $T^{3}$ & $Y/2$ & $Q = T^{3} + Y/2$
\\
\hline
$H_{1} = \begin{pmatrix} G^{+} \\ \frac{v+h+iG^{0}}{\sqrt{2}} \end{pmatrix}$ & $\bf 1$ & $\bf 2$ & $\begin{pmatrix} \frac{1}{2} \\ -\frac{1}{2} \end{pmatrix}$ & $\frac{1}{2}$ & $\begin{pmatrix} 1 \\ 0 \end{pmatrix}$ 
\\[3ex]
$H_{2} = \begin{pmatrix}
    H^{+}\\ \frac{H^{0} + i A^{0}}{\sqrt{2}}
    \end{pmatrix}$ & $\bf 1$ & $\bf 2$ & $\begin{pmatrix} \frac{1}{2} \\ -\frac{1}{2} \end{pmatrix}$ & $\frac{1}{2}$ & $\begin{pmatrix} 1 \\ 0 \end{pmatrix}$  
\\[1ex]
$S = \frac{S_{H} + i S_{A}}{\sqrt{2}}$ & $\bf 1$ & $\bf 1$ & $0$ & $0$ & $0$ 
\\[1ex]
\end{tabular}
\end{center}
\label{tab:fields}
\end{table}%

\subsection{Constraints on charge assignments}

The assignments of $Z_{N}$ charges have to satisfy
\begin{equation}
\begin{split}
  X_{S} &> 0, \\
  X_{1} &\neq X_{2}, \\
  -X_{\ell} + X_{1} + X_{e} &= 0 \mod N, \\
  -X_{q} + X_{1} + X_{d} &= 0 \mod N, \\
  -X_{q} - X_{1} + X_{u} &= 0 \mod N.
\end{split}
\label{eq:constraints:L}
\end{equation}
The first and second conditions arise from avoiding the $|H_{1}|^{2} S$ term and from avoiding Yukawa terms for $H_{2}$, respectively, and the rest from requiring Yukawa terms for $H_{1}$.

The choice of $Z_{N}$ charges for standard model fermions, the standard model Higgs $H_{1}$, the inert doublet $H_{2}$ and the complex singlet $S$ must be such that there are no Yukawa terms for $H_{2}$ and no mixing between $H_{1}$ and $H_{2}$: only annihilation and semiannihilation terms for $H_{2}$ and $S$ are allowed.

All possible scalar potentials contain a common piece because the terms where each field is in pair with its Hermitian conjugate are allowed under any $Z_{N}$ and charge assignment. We denote it by $V_{\text{c}}$, where the `c' stands for `common':
\begin{equation}
\begin{split}
V_{\text{c}}&= \mu_{1}^{2} |H_{1}|^{2} + \lambda_{1} |H_{1}|^{4} + \mu_{2}^{2} |H_{2}|^{2} + \lambda_{2} |H_{2}|^{4} + \mu_{S}^{2} |S|^{2} + \lambda_{S} |S|^{4} \\
&+ \lambda_{S1} |S|^{2} |H_{1}|^{2}
+ \lambda_{S2} |S|^{2} |H_{2}|^{2} + \lambda_{3} |H_{1}|^{2} |H_{2}|^{2} + \lambda_{4} (H_{1}^{\dagger} H_{2}) (H_{2}^{\dagger} H_{1}).
\end{split}
\label{eq:V:c}
\end{equation}

\subsection{The $Z_{2}$ scalar potential}

There are 256 ways to assign $0,1$ to the standard model and dark sector fields. Of these, 8 satisfy Eq.~(\ref{eq:constraints:L}); among them, there are 2 different assignments to the dark sector fields, both giving rise to the unique scalar potential
\begin{equation}
\begin{split}
    V &= V_{\rm c} + \frac{\mu_{S}^{\prime 2}}{2} ( S^{2}
    + S^{\dagger 2} ) + \frac{\lambda_{5}}{2} \left[(H_{1}^{\dagger} H_{2})^{2} + (H_{2}^{\dagger} H_{1})^{2} \right]\\
        &+ \frac{\mu_{S H}}{2} (S^{\dagger} H_{1}^{\dagger} H_{2} + S H_{2}^{\dagger} H_{1})
    + \frac{\mu'_{S H}}{2} (S H_{1}^{\dagger} H_{2} + S^{\dagger} H_{2}^{\dagger} H_{1})
    \\
    &+ \frac{ \lambda'_{S}}{2} (S^{4} + S^{\dagger 4})
    + \frac{ \lambda''_{S} }{2} |S|^{2} (S^{2} + S^{\dagger 2})
    \\
    &+ \lambda_{S1} |S|^{2} |H_{1}|^{2}
    + \lambda_{S2} |S|^{2} |H_{2}|^{2} \\
    &+ \frac{ \lambda'_{S1} }{2}
    |H_{1}|^{2} (S^{2} + S^{\dagger 2} )
    + \frac{ \lambda'_{S2} }{2}
    |H_{2}|^{2} (S^{2} + S^{\dagger 2} ).
\end{split}
\label{eq:V:Z2}
\end{equation}

\subsection{$Z_{3}$ scalar potentials}

There are 6561 ways to assign $0,1,2$ to the fields. Of these, 108 satisfy Eq.~(\ref{eq:constraints:L}); among them, there are 12 different 
assignments to the dark sector fields, giving rise to 2 different scalar potentials. The example potential we choose to work with  is
\begin{equation}
\begin{split}
V_{Z_3} &= V_{\text{c}} + \frac{\mu''_{S}}{2} (S^{3} + S^{\dagger 3}) + \frac{\lambda_{S12}}{2} (S^{2} H_{1}^{\dagger} H_{2} + S^{\dagger 2} H_{2}^{\dagger} H_{1}) \\
&+ \frac{\mu_{SH}}{2} (S H_{2}^{\dagger} H_{1} + S^{\dagger} H_{1}^{\dagger} H_{2}),
\end{split}
\label{eq:Lag:Z:3}
\end{equation}
which induces the semi-annihilation processes we are interested in.
The second one is obtained from Eq.~(\ref{eq:Lag:Z:3}) by changing $S \to S^\dagger$ (with $\mu_{SH} \to \mu'_{SH}$ and $\lambda_{S12} \to \lambda_{S21}$).

\subsection{$Z_{4}$ scalar potentials}

There are 65536 ways to assign $0,1,2,3$ to the fields. Of these, 576 satisfy Eq.~\eqref{eq:constraints:L}; among them, there are 36 different assignments to the dark sector fields, giving rise to 5 different scalar potentials.
Among those the only potential that contains semi-annihilation terms is
\begin{equation}
\begin{split}
V_{Z_4}^1 &= V_{\text{c}} + \frac{\lambda'_{S}}{2} (S^{4} + S^{\dagger 4}) + 
\frac{\lambda_{5}}{2} \left[(H_{1}^{\dagger} H_{2})^{2} + (H_{2}^{\dagger} H_{1})^{2} \right] \\
&+ \frac{\lambda_{S12}}{2} (S^{2} H_{1}^{\dagger} H_{2} + S^{\dagger 2} H_{2}^{\dagger} H_{1}) + \frac{\lambda_{S21}}{2} (S^{2} H_{2}^{\dagger} H_{1} + S^{\dagger 2} H_{1}^{\dagger} H_{2}).
\end{split}
\label{eq:Lag:3:Z:4}
\end{equation}
The other four scalar potentials can be formally obtained from the $Z_2$-invariant potential Eq.~(\ref{eq:V:Z2}) by setting all the new terms added to $V_{\rm c}$ to zero, with the exception of the
1) $\lambda'_S$, $\mu_{SH}$,
2) $\lambda'_S$, $\mu'_{SH}$,
3) $\mu'_S$, $\lambda'_S$, $\lambda''_S$, $\lambda'_{S1}$, $\lambda'_{S2}$,
4) $\mu'_S$, $\lambda'_S$, $\lambda''_S$, $\lambda'_{S1}$, $\lambda'_{S2}$, 
  $\mu_{SH}$, $\mu'_{SH}$ terms.

\section{RELIC DENSITY IN CASE OF THE $Z_3$ SYMMETRY}

\subsection{Evolution equations}

Consider the  $Z_3$-symmetric theory.
The imposed $Z_3$ symmetry implies, as usual, just one dark matter candidate.
This is because the $Z_3$ charges $1$ and $-1$ correspond to particle and anti-particle. 
The new feature is that processes of the type $\chi\chi\rightarrow \chi X$, where $X$ is any standard model particle, also contribute to dark matter annihilation. 
The equation for the number density reads
\begin{equation}
\frac{dn}{dt}=-v\sigma_{\chi\chi\rightarrow  XX}  \left(n^2-n_{\rm eq}^2 \right) -\frac{1}{2} v\sigma_{\chi\chi\rightarrow \chi X}  
\left(n^2-n \, n_{\rm eq} \right) -3H n.
\end{equation}
We define 
\begin{equation}
\sigma_v=  v\sigma_{\chi\chi\rightarrow XX}  +\frac{1}{2}v\sigma_{\chi\chi\rightarrow \chi X} v \quad {\rm and} \quad
\alpha=\frac{1/2 \sigma_{\chi\chi\rightarrow \chi X}}{\sigma_v},
\end{equation}
which means that $0\leq \alpha\leq 1$. In terms of the abundance, $Y=n/s,$ where $s$ is the entropy density, we obtain
\begin{equation}
\frac{dY}{dt}=-s \sigma_v \left(Y^2-\alpha Y Y_{\rm eq} -(1-\alpha) Y_{\rm eq}^2 \right) 
\end{equation}
or 
\begin{equation}
\frac{dY}{ds}=\frac{\sigma_v}{H} \left(Y^2-\alpha Y Y_{\rm eq} -(1-\alpha) Y_{\rm eq}^2 \right).
\end{equation}
To solve this equation we follow the usual procedure~\cite{Belanger:2004yn}. Writing $Y=Y_{\rm eq}+\Delta Y$ we find the starting point for the numerical solution of this equation using the Runge-Kutta method using
\begin{equation}
\frac{dY_{\rm eq}}{ds}=\frac{\sigma_v}{H} Y\Delta Y \left( 2-\alpha \right),
\end{equation} 
where $\Delta Y\ll Y$. This
is similar to the standard case except that  $\Delta Y$ increases by a factor $1/(1-\alpha/2)$.
Furthermore, when solving numerically the evolution equation, the decoupling condition
$Y^2\gg Y^2_{\rm eq}$ is modified to 
\begin{equation}
Y^2\gg \alpha Y Y_{\rm eq} + (1-\alpha) Y^2_{\rm eq}.
\end{equation}
This implies that freeze-out starts at an earlier time and lasts until
a later time as compared with the standard case.
This modified evolution equation is implemented in micrOMEGAs~\cite{Belanger:2006is,Belanger:2010gh}.

\subsection{Numerical results with micrOMEGAs}

Using the scalar potential defined in Eq.~(\ref{eq:Lag:Z:3}) we have implemented in micrOMEGAs the scalar model with a $Z_3$ symmetry.  The scalar sector is composed of one additional  complex scalar doublet and one complex singlet. The neutral component of the doublet mixes with the singlet, the lightest component  $\tilde{h}_1$ is therefore the dark matter candidate, while  the heavy component $\tilde{h}_2$ can decay into $\tilde{h}_1h$, where $h$ is the standard model like Higgs boson. 
The $Z_3$ charge of $\tilde{h}_1,\tilde{h}_2,\tilde{h}^+$ is $1$.

We then compute the dark matter relic density as well as  the elastic scattering cross section on nuclei. Here we average over dark matter and anti-dark matter cross section assuming that they have the same density. the main contribution here comes from the $Z$-exchange diagram because there is a $\tilde{h}_1\tilde{h}_1^* Z$ coupling\footnote{In the inert doublet model with a $Z_2$ symmetry~\cite{Deshpande:1977rw,Barbieri:2006dq}, a $\lambda_5$ term splits the complex doublet into a scalar and a pseudoscalar, when the mass splitting is small such coupling leads to inelastic scattering.}.
Furthermore one can easily show that the scattering amplitudes are not the same for protons and neutrons. 
Since the current experimental bounds on $\sigma_{\chi n}^{\rm SI}$ are extracted from experimental results assuming that the couplings to protons ($f_p$) and neutrons ($f_n$) are equal, we define the normalized cross section on a point-like nucleus~\cite{Belanger:2010cd}:
\begin{eqnarray}\label{NormalizedCross}
\sigma_{\tilde{h}_1N}^{\rm SI} = \frac{\mu_{\psi_1}^2}{\pi}\, \frac{[Zf_p + (A-Z)f_n]^2}{A^2}\,.
\end{eqnarray}
This quantity can directly be compared with the limit on $\sigma_{\chi n}^{\rm SI}$.
\begin{figure}
\begin{center}
\includegraphics[width=0.9\textwidth]{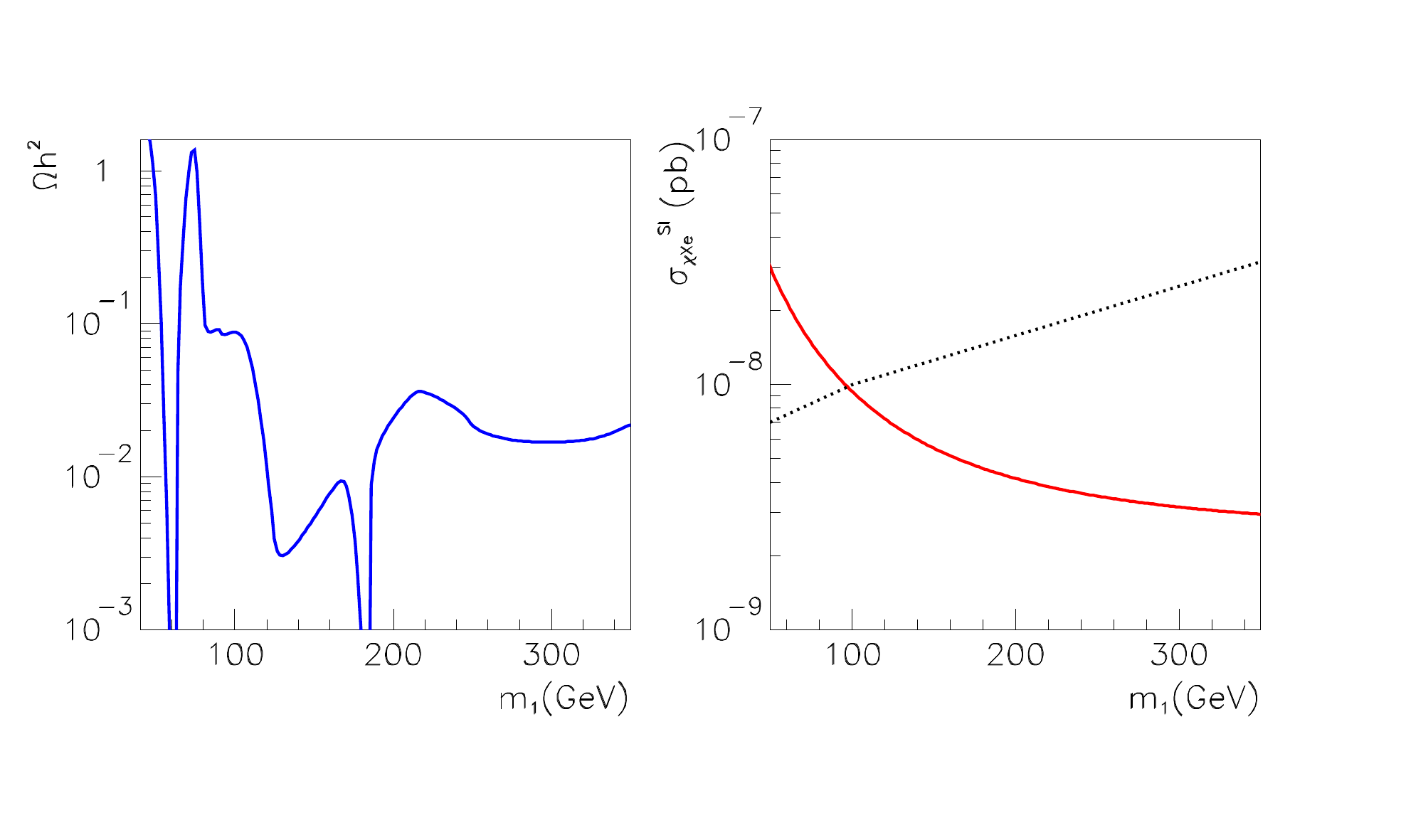}
 \caption{$\Omega h^2$ as a function of the dark matter mass (left panel) and 
$\sigma_{\tilde{h}_1 Xe}^{\rm SI}$. 
The experimental limit from XENON100~\cite{Aprile:2011hi} is also displayed (dashed line.) 
}
\label{fig:omega}
\end{center}
\end{figure}

To illustrate the behavior of the relic density of dark matter, we choose the parameters $\lambda_2=0.73,\lambda_3=0.16, \lambda_4=0.45,\lambda_S=0.93$, $\lambda_{S1}=0.06$, $\lambda_{S2}=0.03,\lambda_{S21}=0.69$, $\mu''_S=427.5$~GeV. Furthermore we fix the mass of the Higgs sector to be $M_h=125$~GeV, 
$M_{h_2}=371.2$~GeV and let $M_{\tilde{h}_1}$ vary. We also assume that $\tilde{h}_1$ is almost singlet taking $\sin\theta_h=0.9995$. 

The variation of $\Omega h^2$ with the dark matter mass is displayed in 
Fig.~\ref{fig:omega}. When the dark matter mass is around $50$~GeV the main annihilation channel is through $Z$-exchange, note that this, however, leads to a very large direct detection rate. 
As $m_1=m_{\tilde{h}_1}$ increases above $m_Z/2$, the relic density increases dramatically. At this point $\alpha$ is very small since the process $\tilde{h}_1 \tilde{h}_1\rightarrow \tilde{h}_1^* Z$ is kinematically forbidden. 
After the threshold for $W$ pair production, the relic density starts to drop, the process $\tilde{h}_1 \tilde{h}_1\rightarrow \tilde{h}_1^* Z$ soon becomes kinematically accessible and $\alpha$ increases rapidly. The contribution of this channel rises rapidly leading to a drop of $\Omega h^2$ to values below the preferred region extracted form WMAP measurements~\cite{Larson:2010gs}. For yet larger dark matter masses the annihilation process $\tilde{h}_1 \tilde{h}_1\rightarrow h_2^*\rightarrow \tilde{h}_1^* h$ occurs near the $h_2$ resonance, this is the second drop in $\Omega h^2$ displayed in Fig.~\ref{fig:omega}. As the dark matter mass increases further, several processes involving one non-standard particle in the final state can take place (notably $W^+H^-,h\tilde{h}_1^*,h\tilde{h}_2^*,Z\tilde{h}_2^*$); such processes completely dominate and $\alpha \approx 1$. 

The value of the SI cross section for the same parameters is displayed in Fig.~\ref{fig:omega} right panel. The region where $\Omega h^2 \approx 0.1$ just satisfies the Xenon100 bound~\cite{Aprile:2011hi}. Note that to achieve that it was necessary 
to have an almost pure singlet dark matter.  Heavier dark matter satisfy easily the direct detection bounds, although their abundance is too low to explain all of the dark matter.

\section{CONCLUSIONS}
We have formulated scalar dark matter models with the minimal particle content in which dark matter stability is due to the discrete $Z_N$ symmetry, $N>2.$
Already the minimal models containing one extra scalar singlet and doublet possess non-trivial dark matter phenomenology.
In particular, the annihilation processes with new topologies like $\chi\chi\rightarrow \chi X$, where $\chi$ is the dark matter and $X$ is any standard model particle,
change the dark matter freeze-out process and must be taken into account when calculating the dark matter relic abundance.
We have performed an example study of  semi-annihilations in the $Z_3$ symmetric scalar dark matter model by implementing the model to micrOMEGAs and 
studying the impact of semi-annihilations to the relic abundance and on the predictions of dark matter direct detection.
We conclude that in this type of models semi-annihilations may significantly affect the phenomenology and must be taken into account 
in numerical analyses quantitatively exactly.

\section*{ACKNOWLEDGEMENTS}
Part of this work was performed in the  Les Houches 2011 summer institute. K.K. and M.R. were supported by the ESF grants 8090, 8499, 8943, MTT8, MTT59, MTT60, MJD140, by the recurrent financing SF0690030s09 project
and by the European Union through the European Regional Development Fund.
A.P. was supported by the Russian foundation for Basic Research, grant RFBR-10-02-01443-a. The work of A.P. and G.B. was supported in part by the GDRI-ACPP of CNRS.

\AddToContent{G.~B\'elanger, K.~Kannike, A.~Pukhov, M.~Raidal}
\renewcommand{\thesection}{\arabic{section}}

\superpart{ From Results to Interpretation }

\def\describeRec{Provide a clear,
explicit description of the analysis in publications.  In particular,
the most crucial information such as basic object definitions and
event selection should be clearly displayed in the publications,
preferably in tabular form, and kinematic variables utilised should be
unambiguously defined.  Further information necessary to reproduce the
analysis should be provided, as soon as it becomes available for
release, on a suitable common platform.}
\def\databaseRec{The community should identify, develop and adopt a 
common platform to store analysis databases, collecting object
definitions, cuts, and all other information, including
well-encapsulated functions, necessary to reproduce or use the
results of the analyses, and as required by other recommendations.}
\def\efficiencyRec{Provide histograms or functional forms of efficiency maps 
wherever possible in the auxiliary information, along with precise definitions 
of the efficiencies, and preferably provide them in standard electronic forms 
that can easily be interfaced with simulation or analysis software.}
\def\simulatorRec{The community should take responsibility for providing, 
validating and maintaing a simplified simulation code for public use, reproducing 
the basic response of the LHC detectors. The validation and tuning of this tool 
should be based on comparisons with actual performance plots, and/or other 
inputs, made available by the experiments along the lines of Recommendation~2a. 
Limits of validity should be investigated and clearly documented.}
\def\analysisNumbersRec{Provide all crucial numbers regarding the results of the analysis, 
preferably in tabulated form in the publication itself. Further relevant information, like fit 
functions or distributions, should be provided as auxiliary material.}
\def\multiBinRec{For multi-bin results, provide an ensemble of sets of the numbers $B$,
$\delta B$, $\cal{L}$, $\delta\cal{L}$, $Q$, $k$, etc in the auxiliary information.
These would be created by sampling from the various experiment-specific systematic effects,
such as the jet energy scale, jet energy resolution, etc. 
Results should be quoted without inclusion of systematic/theoretical uncertainties
external to the experiment.} 
\def\likelihoodRec{When feasible, 
provide a mathematical description of the \underline{final} likelihood 
function in which experimental data and parameters are clearly distinguished, 
either in the publication or the auxiliary information. 
Limits of validity should always be clearly specified.}
\def\roostatsRec{Additionally provide a digitized implementation of the
likelihood that is consistent with the mathematical description.}
\def\interpretRec{In the interpretation of experimental results,  
preferably provide the final likelihood function (following Recommendations~3b/3c). 
When this is not possible or desirable, provide a grid of confidence levels over the parameter space.  
The expected constraints should be given in addition to the observed ones, and 
whatever sensitivity measure is applied must be precisely defined. 
Modeling of the acceptance needs to be precisely described.}
\def\higgsRec{For Higgs searches, provide all relevant information on a channel-by-channel 
basis for both production and decay processes.}
\def\designRec{When relevant, 
design analyses and signal regions that are based on  disjoint sets of events.}

\newcounter{recom}
\setcounter{recom}{0}
\chapter{Searches for New Physics: Les Houches Recommendations 
for the Presentation of LHC Results}

{\it S.~Kraml, 
B.C.~Allanach, 
M.~Mangano, 
H.B.~Prosper, 
S.~Sekmen (editors),\\
C.~Balazs, 
A.~Barr,
P.~Bechtle, 
G.~Belanger, 
A.~Belyaev,
K.~Benslama, 
M.~Campanelli, 
K.~Cranmer, 
A.~De~Roeck, 
M.J.~Dolan, 
J.R.~Ellis, 
M.~Felcini, 
B.~Fuks, 
D.~Guadagnoli,
J.F.~Gunion, 
S.~Heinemeyer, 
J.~Hewett, 
A.~Ismail, 
M.~Kadastik,
M.~Kr\"amer,
J.~Lykken,
F.~Mahmoudi,
S.P.~Martin,
T.~Rizzo, 
T.~Robens, 
M.~Tytgat, 
A.~Weiler}



\begin{abstract}
We present a set of recommendations for the presentation of LHC 
results on searches for new physics, which are aimed at providing a more efficient flow of scientific information between the experimental collaborations and the rest of the high energy physics community, and at facilitating the interpretation of the results in a wide class of models. Implementing these recommendations would aid the full exploitation of the physics potential of the LHC.  
\end{abstract}

\vspace*{1mm}

\section{INTRODUCTION}

The LHC has very successfully begun to explore the TeV energy scale, 
and will be the energy frontier machine for the foreseeable future. Everyone who has had
a hand in bringing this scientific and technological
marvel to fruition deserves considerable credit and our thanks: 
the physicists and engineers who
conceived, designed, and built it; those who operate the machine and its experiments;
those who produce experimental results; those who try to understand them, and the public 
and its representatives whose generous support has enabled all this to happen. 

The LHC was designed as a machine of discovery.
There are high hopes that groundbreaking discoveries will indeed occur and 
shed light on  
electroweak symmetry breaking (be it via the Higgs mechanism or some other new
dynamics) and new physics beyond the Standard Model (SM) of electroweak and strong
interactions.  It is of highest priority to our community to
exploit fully the physics potential of the LHC.
One aspect of this exploitation is the interpretation of LHC results in the contexts 
of different models of new physics. This is crucial if we are to unravel the correct 
new physics model, determine its parameters, and move beyond the SM. 

The ATLAS and CMS collaborations are providing detailed
experimental results~\cite{Atlastwiki,Cmstwiki} of searches in many different channels. 
They are also providing interpretations in terms of popular models, such as the 
CMSSM\footnote{Constrained Minimal Supersymmetric Standard Model, see e.g.~\cite{AbdusSalam:2011fc}.}, 
or in terms of Simplified Models\footnote{Simplified Models are designed as an effective-Lagrangian description 
of a small number of accessible new particles. This approach has a long heritage; for a recent paper
advocating it see e.g.~\cite{Alves:2011wf}.}. 
These results are being used  
to test as large a variety of beyond-the-SM (BSM) scenarios as possible. 
For example, the searches for supersymmetry (SUSY), 
including~\cite{Chatrchyan:2011zy,Aad:2011cwa,Aad:2011ib,ATLAS:2011ad,CMS:SS,CMS:OS},
were interpreted in a number of 
different SUSY-breaking schemes, see e.g.\ \cite{Farina:2011bh,Buchmueller:2011sw,Fowlie:2011mb,Allanach:2011qr,Kats:2011qh,Grellscheid:2011ij}, 
as well as in the weak-scale ``phenomenological'' MSSM~\cite{Sekmen:2011cz,Arbey:2011un}. 
The sensitivity to light stops 
was investigated 
in~\cite{Papucci:2011wy,Bi:2011ha,Desai:2011th}, while implications of
compressed SUSY spectra were analyzed in~\cite{LeCompte:2011fh}. 
Interpretations were also made for non-SUSY models, for instance for the minimal 
universal extra dimension (UED) model in~\cite{Chang:2011aa}.  
Similar non-collaboration efforts to interpret Higgs search results~\cite{ATLAS:2012si,Chatrchyan:2012tx} 
in a large variety of BSM scenarios are also underway.  
These examples illustrate the community's interest in the LHC
experimental results---interest that will surely grow as results become more 
comprehensive and readily available.

A systematic way of presenting LHC results will
also greatly facilitate the comparison and combination of analyses
within and across the LHC collaborations, as well as the assessment of
the physics potential of future facilities. Furthermore, agreement on a set
of recommendations and their implementation would be
a further step towards a more comprehensive approach to the storage,
persistence and future use of LHC results.

In this contribution, we therefore propose 
a set of recommendations for the presentation of LHC results
aimed at maximizing its scientific return.  Many of the
experimental publications already implement several of the basic
recommendations we make. But, as we shall see,  our
recommendations go substantially beyond current practice. Our wish is to work towards an agreement on
a common standard for the presentation of results. The goal is 
to help the community make the most of an extraordinary scientific
opportunity.

\section{EXECUTIVE SUMMARY}

We here summarize our recommendations, which we present in four broad categories:
analysis description, detector modeling, analysis dissemination and analysis design. 
Moreover, we include some recommendations regarding the interpretation of the results. 
Where appropriate, we split our recommendations into options: 
\begin{description}
\item[] {\bf (a)} ``crucial" (mandatory) recommendations, defined as actions that we believe should be undertaken, following 
revision motivated by feedback from the experiments, and 
\item[] {\bf (b), (c)} ``desirable steps'', i.e.\ actions that would help, but whose implementation is recognized as either being controversial, and thus needing more debate, or requiring major efforts and a longer timescale. 
\end{description}
Recommendations without such sub-division are understood as  ``crucial". 
An extensive discussion of these recommendations will be published as an independent document.

\begin{enumerate}
\vskip 1cm
\item \begin{enumerate} 
      \item \describeRec \vskip 0.2cm
      \item \databaseRec 
      \end{enumerate}
\vskip 1cm
\item \begin{enumerate} 
      \item \efficiencyRec \vskip 0.2cm 
      \item \simulatorRec 
      \end{enumerate}
\vskip 1cm
\item \begin{enumerate} 
      \item \analysisNumbersRec \\
      \emph{Addendum}:\\ \multiBinRec \vskip 0.2cm
      \item \likelihoodRec \vskip 0.2cm
      \item \roostatsRec
      \end{enumerate}
\vskip 0.7cm
\item\interpretRec
\vskip 0.4cm
\item \higgsRec
\vskip 0.4cm
\item \designRec
\end{enumerate}


\section{CONCLUSIONS}

This document presents a set of recommendations for the presentation of LHC  
results on searches for new physics, which are aimed at providing a more efficient flow 
of scientific information and at facilitating the interpretation of the results in wide classes of models. 
It originated from discussions at this Les Houches ``Physics at TeV Colliders 2011'' 
workshop and was thoroughly discussed and refined, with valuable input from 
representatives of the ATLAS and CMS collaborations, in a dedicated miniworkshop organized 
by the LHC Physics Centre at CERN~\cite{feb13}. 
The target of
these recommendations are physicists both within and outside the LHC
experiments, interested in the best exploitation of the BSM search analyses.  

The added value for the experiments, and the whole HEP community, in
extending the scope of the information made available about the
experimental results, is a faster and more precise feedback
on the implications of these results for a broad range of theoretical
scenarios. Correlations and consistency checks among the findings of
different experiments, at the LHC and elsewhere, will be facilitated,
and will provide crucial input in the choice of the best research
directions in both the near and far future, at the LHC and
elsewhere. Improving the way the results of the LHC searches are
documented and stored furthermore provides a forum to
explore alternative approaches to long-term data archiving. 

The tools needed to provide extended experimental information will
require some dedicated efforts in terms of resources and manpower, to
be supported by both the experimental and the theory
communities. Practical solutions towards the development of these
tools and the implementation of the proposed recommendations will be
addressed in dedicated Workshops and working groups.

\section*{ACKNOWLEDGEMENTS}
We are grateful to the ATLAS and CMS search group conveners for helpful and supportive advice. 
We are also grateful to the ATLAS and CMS physics coordinators, in particular Richard Hawkings and Greg Landsberg, for very constructive discussions on these recommendations.  
BCA~would like to thank other members of the Cambridge SUSY Working Group for
helpful suggestions. 

This work has been partially supported by IN2P3, 
the Royal Society, STFC, and the ARC Centre of Excellence for Particle Physics at the Terascale.
The work of JE is supported in part by the London Centre for Terauniverse Studies (LCTS), using funding from the European Research Council 
via the Advanced Investigator Grant 267352. 
JFG is supported by US DOE grant DE-FG03-91ER40674.  
HBP is supported in part by US DOE grant DE-FG02-97ER41022..

\AddToContent{S.~Kraml, B.C.~Allanach, M.~Mangano, 
H.B.~Prosper,S.~Sekmen et al.}
\renewcommand{\thesection}{\arabic{section}}




\chapter{High Energy Physics Model Database -- HEPMDB:  
Towards decoding of the underlying theory at the LHC.}

{\it 
Maksym~Bondarenko,
Alexander~Belyaev,
Lorenzo Basso,
Edward Boos,
Vyacheslav Bunichev,
R. Sekhar Chivukula,
Neil D. Christensen,
Simon Cox,
Albert De Roeck,
Stefano Moretti,
Alexander Pukhov,
Sezen Sekmen,
Andrei Semenov,
Elizabeth H. Simmons,
Claire Shepherd-Themistocleous,
Christian Speckner
}



\begin{abstract}
We present here the first stage of development of the High Energy Physics Model Data-Base (HEPMDB) which is a
convenient centralized storage environment for HEP models, and can
accommodate, via web interface to the HPC cluster, the validation of
models, evaluation of LHC predictions and event generation-simulation
chain.
The ultimate goal of HEPMDB is to perform an effective LHC data interpretation
isolating  the most successful theory for explaining
LHC observations.

\end{abstract}

\section{The idea behind HEPMDB}

The year 2010 has marked the start of real data taking at the CERN Large Hadron Collider (LHC). The central task of the LHC is to find  signals from new physics and to reveal  the underlying dynamics of Nature responsible for such signals. This is a highly non-trivial problem since many promising  models could and do  lead to similar signatures. For the next decades research in High Energy Physics (HEP) will be concentrated on 
interpreting the data from the LHC.  
We propose to create a High Energy Physics Model Data-Base (HEPMDB) which will be an  unparallelled tool to isolate  the most successful theory in explaining
LHC observations. This project is aimed at boosting the activity of HEP groups around the world and will play an important role in the LHC data interpretation. 
 
The first phase of the HEPMDB prototype is accessible  at \verb|https://hepmdb.soton.ac.uk|. The  HEPMDB was created as a result of 
 ideas discussed in the context of the ``Dictionary of LHC signatures" activity~\cite{Belyaev:2008pk},
at the FeynRules workshop~\cite{Feynrules2010} and at the ``Mini-Workshop on Dynamical Symmetry Breaking models 
and tools''~\cite{minibsm-at-soton}. Following the ``Dictionary of LHC signatures" activities it was agreed 
that the identification of the underlying 
theory from LHC signals can only be  sensibly realised via a flexible database which collects models and their specific signatures.

\noindent The HEPMDB  is aimed to:
\begin{enumerate}
\item Collect HEP models for various  multipurpose Matrix Element (ME) generators like
{\sf CalcHEP}~\cite{Pukhov:2004ca}, 
{\sf CompHEP}~\cite{Pukhov:1999gg,Boos:2004kh},
{\sf FeynArts}~\cite{Kublbeck:1992mt,Hahn:2000kx},
{\sf MadGraph/MadEvent}~\cite{Maltoni:2002qb,Alwall:2011uj,deAquino:2011ub,Degrande:2011ua}, {\sf AMEGIC ++/COMIX} within {\sf 
SHERPA}~\cite{Gleisberg:2003xi,Gleisberg:2008ta}.
and {\sf WHIZARD}~\cite{Kilian:2007gr}.
Under ``HEP models'' we denote the set of particles, Feynman rules and parameters written 
in the format specific for a given package.
\item Collect model sources which can be used to generate HEP models for various ME generators
using {\sf FeynRules}~\cite{Christensen:2008py} or {\sf LanHEP}~\cite{Semenov:2008jy}, which automate the process of generating Feynman Rules, particle spectra, 
etc.. Under ``model source" we denote the model (lagrangian etc.)
written in the form of input  for  {\sf FeynRules} or  {\sf LanHEP}.
For the moment, {\sf FeynRules} interfaces to {\sf CompHEP, CalcHEP,  FeynArts, GoSam}~\cite{Cullen:2011ac}, 
{\sf MadGraph/MadEvent},  {\sf SHERPA}  and {\sf WHIZARD}~\cite{Christensen:2010wz} are available. {\sf LanHEP}  works with  {\sf CalcHEP, CompHEP, FeynArts} and {\sf GoSam}. Also, the latest {\sf LanHEP v.3.15} has an option (under test) of outputting the model in UFO format~\cite{Degrande:2011ua} which 
provides a way to interface it with  {\sf MadGraph/MadEvent}.
\item Allow users to upload their models onto a server in order to perform
evaluation of HEP processes and event  generation for their own  models using the
full power of the High Performance Computing (HPC) cluster standing behind the
HEPMDB itself. The HPC cluster at Southampton University, IRIDIS3 is the
state-of-the-art  fastest university-owned HPC resource in the UK, with 1008
8-core compute nodes (Intel Nehalem 2.26 GHz), at least 24GB of memory per node, all
connected with fast infiniband network for parallel communication. This is one of
the very powerful features of the HEPMDB: it  provides  a web interface to various
ME generators which can then also be run directly on the HPC cluster. 
This way, users can preform calculations for any model
from the HEPMDB (including their own models which they can upload),
avoiding problems related to installing the actual software, which can
sometimes be quite cumbersome.
\item 
 Cross check and validate  models for different ME generators. We should note
 that similar functionality is also provided by the FeynRules web validation
 framework~\cite{FR-website} which is also presented in these proceedings. However, the
 FeynRules web validation is mainly geared towards comparing FeynRules models and
 can use its knowledge of the model format to provide a highly
 automatized test procedure for those, while the HEPMDB works in a more generic way and will  provide access to more model formats at the price of slightly less automatization.
Also, one should stress that  uploads and evaluations at  the HEPMDB
are available to {\it all users}. This is an important new feature of
the HEPMDB as compared to the FeynRules website.

%

\item Collect predictions and specific features of various models in the
form of a (sub)database of signatures and perform comparisons of various
model predictions with experimental data. There are a lot of different
aspects related to this problem, details of which are outside the scope
of the current short contribution.  We would like to mention though,
that this task includes a comprehensive development of a database of
signatures as well as development of the format of presentation of
these signatures.
This format will be   consistent with the format
which will be used by the experimentalists for the presentation of the LHC data, 
discussed  at the workshop in the context of the ``Les Houches Recommendations for
the Presentation of LHC Results" activity.

\item 
Trace the history of model modifications (in case modifications
take place), and makes available all versions of the models.
Through this application, we stress the importance of reproducibility
of the results coming from the HEPMDB or from a particular model
downloaded from the HEPMDB.

\end{enumerate}
\vskip0.25cm
\noindent

\noindent
We would like to stress that the HEPMDB is an important project
which has no analogy at the moment.
It is different from projects which sound quite similar
but in fact serve a different purpose.
For example, the ``Database of Numerical HEP scattering cross sections"\\
(\verb|http://durpdg.dur.ac.uk/HEPDATA/REAC|) collects various particle scattering 
process cross-sections which are connected to experimental searches of different reactions.
On the other hand the ``Signatures of New Physics at the LHC" web-site\\
(\verb|http://www.lhcnewphysics.org/|) collects various BSM signatures, classifies them and the related papers.
The part of the HEPMDB related to signatures definitely has a lot in common
with ``Signatures of New Physics at the LHC" but the global aim of the HEPMDB is 
generically  different.
The HEPMDB is aimed to provide an important connection between theory (in the form of theoretical models which users can upload), phenomenology (in the form of cross sections and distributions which users can evaluate at the HEPMDB) and experiment (in the form of collections of  model-specific signatures and simulated events which can be used by experimental collaborations).

\section{The current status of the HEPMDB}

\begin{figure}[htb]
\centering
\subfigure[]{
\includegraphics[width=0.50\textwidth]{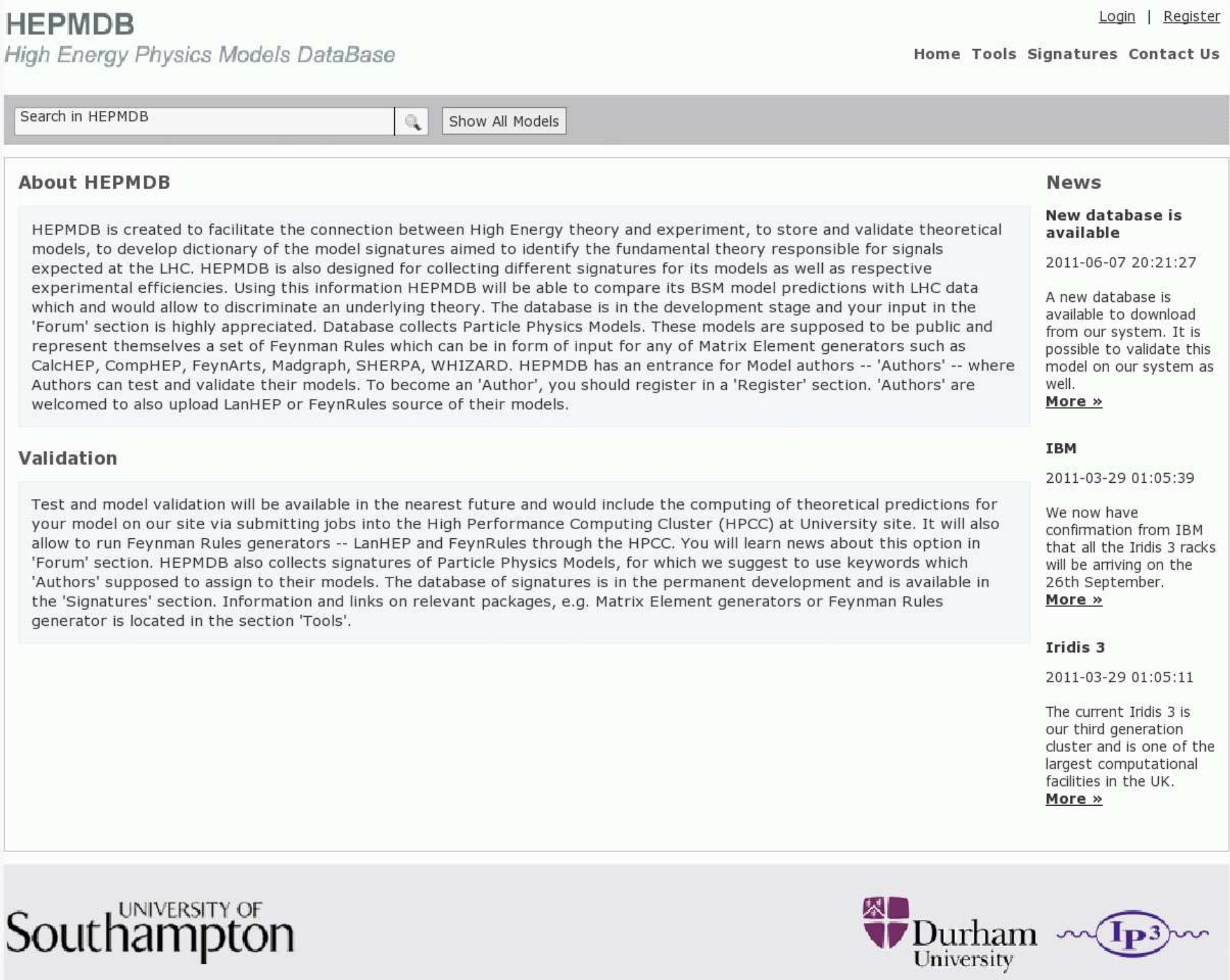}
\label{fig:hepmdb-main}
}%
\subfigure[]{
\includegraphics[width=0.50\textwidth]{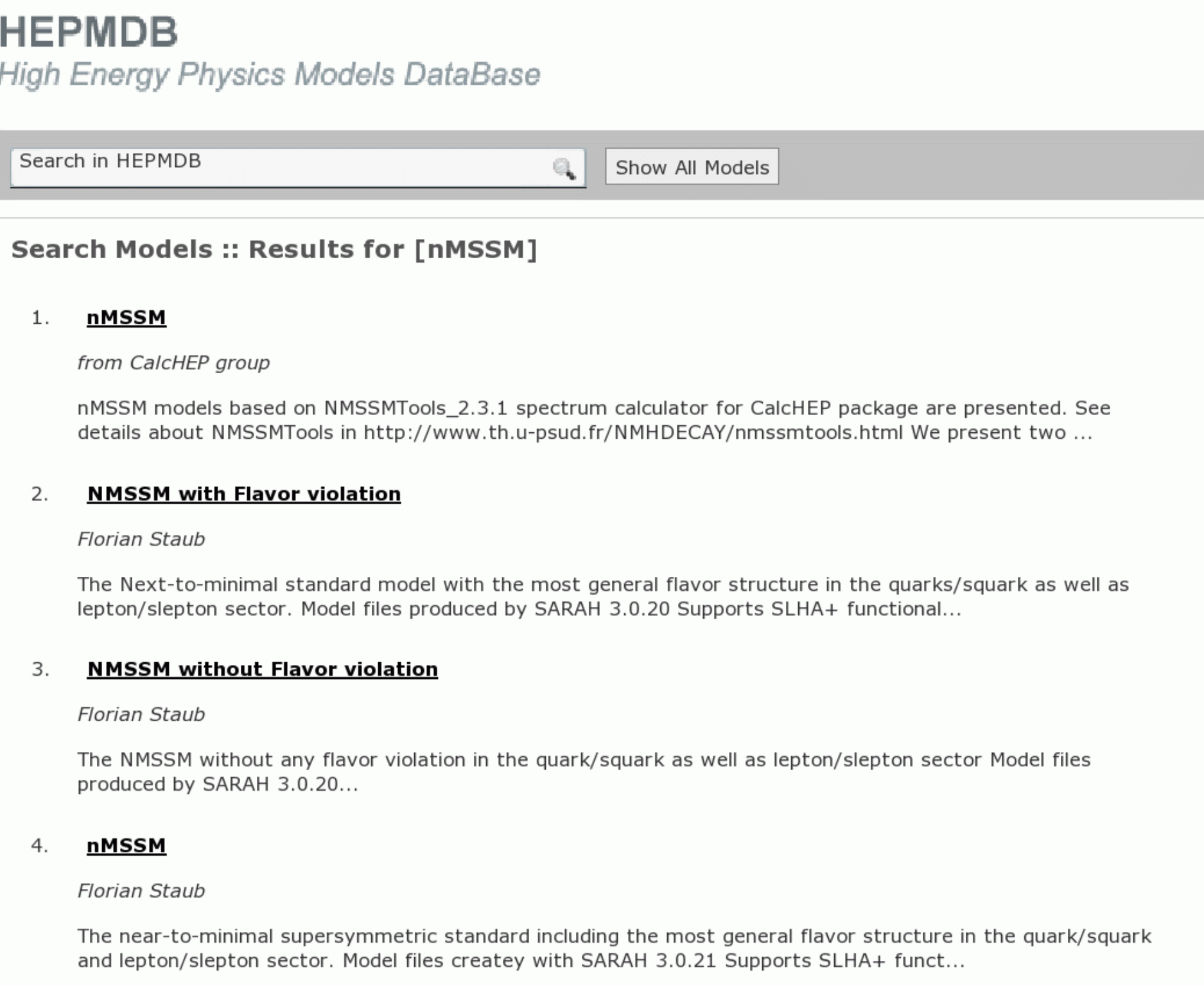}
\label{fig:hepmdb-search}
}
\\
\subfigure[]{
\includegraphics[width=0.50\textwidth]{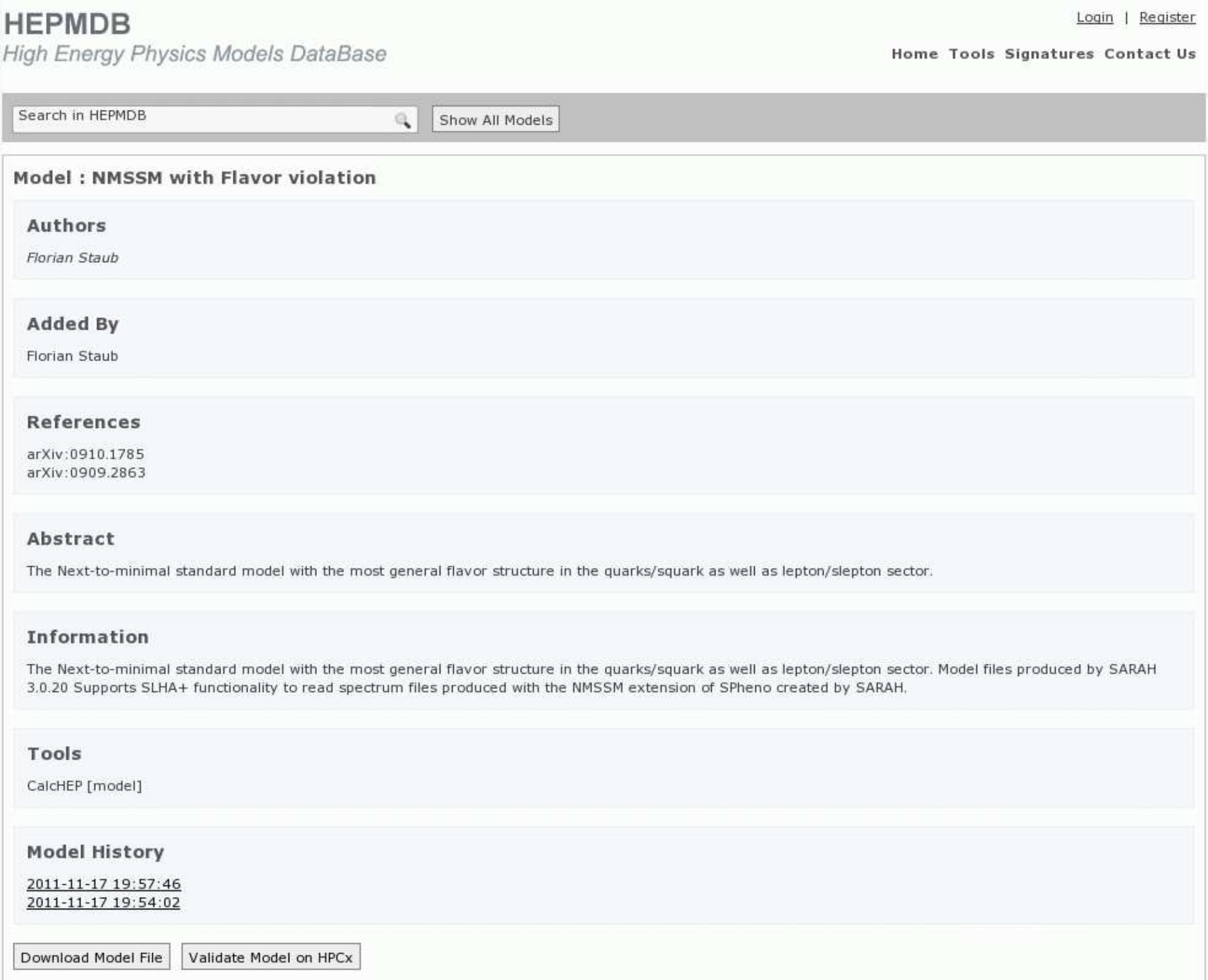}
\label{fig:hepmdb-model}
}%
\subfigure[]{
\includegraphics[width=0.50\textwidth]{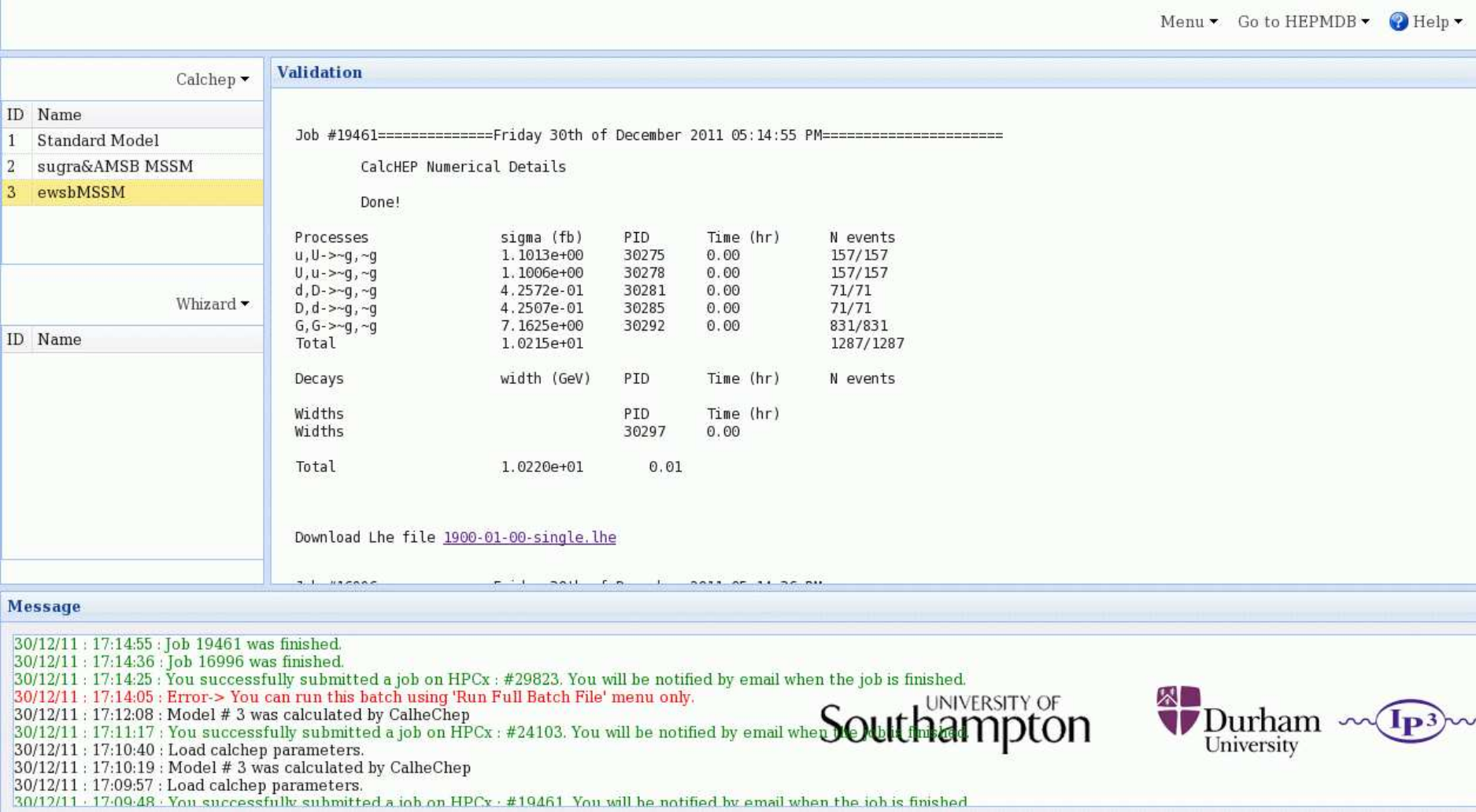}
\label{fig:hepmdb-calcs}
}
\caption{HEPMDB - (a) main page, 
                  (b)  example of the search pattern and results, 
 	            (c)  model details,
	            (d)  ME evaluation.}
\label{fig:hepmdb-figs}
\end{figure}

The HEPMDB project  main page is 
presented 
in Fig.~\ref{fig:hepmdb-main}. At the moment the HEPMDB allows the user to perform the following operations.
\begin{enumerate}
\item Find  and download an existing HEP model from the database.
  The search engine checks patterns in the following fields:
{\sf Model}, {\sf Authors}, {\sf References}, {\sf Abstract}, {\sf Signatures}
and {\sf Information} (see Fig.~\ref{fig:hepmdb-search}).

\item Upload a new model (upon user registration). The model can be uploaded in the format of any ME generator.
Also, a user can upload the  model {\it source} in {\sf FeynRules} or {\sf LanHEP} formats.
An example of the  model entries as well as the model history is given in Fig.~\ref{fig:hepmdb-model}.

\item Perform the evaluation of cross sections for user-defined processes for the  chosen model from the  HEPMDB  
and produce a corresponding  LHE file  with generated parton-level events. 
This file becomes available for download once the process is finished. Currently, the HEPMDB allows the user to perform these calculations (using the aforementioned HPC) for {\sf CalcHEP} and {\sf WHIZARD} models only.
An example of such an evaluation is given in Fig.~\ref{fig:hepmdb-calcs}.

\item Update or add   features and respective signatures 
specific to each model.  These features and signatures
can be  used in the future to distinguish 
the model from others and connect it  to the LHC signatures.

\item Keep track of model changes, providing reproducibility of the results obtained with 
previous versions of the models uploaded to the HEPMDB.
\end{enumerate}

\section{HEPMDB: conclusions and future prospects}

The
present stage of the HEPMDB is a first step towards realizing a systematic
comparison of various model predictions with experimental data and
decrypting of the underlying theory.

The following improvements and additions will take place in the HEPMDB in the coming months.
\begin{enumerate}
\item The {\sf LanHEP} and {\sf FeynRules packages} will be added to provide event generation from model sources.
\item The {\sf MadGraph/MadEvent} and {\sf CompHEP} packages will be added. 
\item A systematic model validation process will be started and the respective pages will be added.  
\item The possibility to study events beyond parton level will be carefully considered, up to detector simulation. One concrete possibility would be the chain \\ \verb|LHE events| $\to$ \verb|HEPMC events| $\to$  \verb|FASTSIM events (ROOT format)|. For the last step, {\sf Delphes}~\cite{Ovyn:2009tx} seems a promising candidate.
\item The structure of the database of signatures will be extended to deal with correlated signatures (i.e.~whereby 
multiple signatures, or lack thereof, must be accounted for simultaneously).
\item The possibility to produce plots for various kinematic distributions will be added.
\end{enumerate}

On the longer time scale (of the order of one year) we plan to install the {\sf MicrOMEGAs} package for evaluation of the dark matter relic density as
well as to provide a possibility for scans of various model parameter
spaces.
During this stage, a development of the format for model predictions consistent with 
the format for presentation of the LHC data by experimentalists is planned.
The question about including automatic tools for NLO evaluations is under discussion
and will be developed further at the later stages of the HEPMDB development.

To conclude, we hope that
starting from the present stage, the HEPMDB development will be boosted
further via involvement of the HEP community. The HEPMDB is already a
convenient centralized storage environment for HEP models, and can
accommodate, via web interface to the HPC cluster, validation of
models, evaluation of LHC predictions and event generation and simulation.
We think that in the near future the HEPMDB will
also become a powerful tool for isolation of  the most  successful  theory in
explaining the LHC data.

\section*{ACKNOWLEDGEMENTS}
AB and MB acknowledge financial support of HEPMDB at its initial stage 
by the University of Southampton and IPPP Institute.
AB, LB and SM thank the NExT Institute and SEPnet
for partial financial support. AB also does so with the Royal Society.
CS and (partially) LB have been supported by the Deutsche Forschungsgemeinschaft
 through the Research Training Group GRK\,1102
\textit{Physics of Hadron Accelerators}.
NC was partially supported by the United States National Science
Foundation under grant NSF-PHY-0705682 and partially by
the PITTsburgh Particle Physics, Astrophysics and Cosmology Center
(PITT PACC).

\AddToContent{M.~Bondarenko et al.}
\renewcommand{\thesection}{\arabic{section}}

\clearpage

\bibliography{LH_NewPhysics_Biblio}

\end{document}